\documentstyle[12pt,amssymb]{article}
\newcommand{\text}[1]{\mbox{\rm #1}}
\newenvironment{arablist}{\begin{list}{(\theenumi)}
{\usecounter{enumi}}}{\end{list}}

\newenvironment{romanlist}{\begin{list}{(\theenumiii)}
{\usecounter{enumiii}}}{\end{list}}

\newenvironment{letterlist}{\begin{list}{(\alph{enumiv})}
{\usecounter{enumiv}}}{\end{list}}
\renewcommand{\thesection}{\arabic{section}}

\renewcommand{\theequation}{\arabic{section}\Alph{subsection}.\arabic{equation}}
\newtheorem{theorem}{THEOREM}
\newtheorem{proposition}[theorem]{PROPOSITION}
\newtheorem{corollary}[theorem]{COROLLARY}
\newtheorem{lemma}[theorem]{LEMMA}
\renewcommand{\thetheorem}
{\arabic{section}\Alph{subsection}.\arabic{theorem}}
\newtheorem{gssm}{GSSM}
\newenvironment{definition}[1]
{\mbox{ } \\ \begin{bf} #1 \end{bf} 
    \begin{list}{}{\setlength{\leftmargin}{\parindent}}\item}
{\end{list}\begin{flushright} End of Dfn.\end{flushright}}
\newenvironment{nosubsec}
{\renewcommand{\theequation}{\arabic{section}.\arabic{equation}}}
{\renewcommand{\theequation}{\arabic{section}\Alph{subsection}.\arabic{equation}}}
\newcounter{subeq}
\newenvironment{abc}
{\setcounter{subeq}{\value{equation}} \setcounter{equation}{0}
    \renewcommand{\theequation}
    {\arabic{section}\Alph{subsection}.\arabic{subeq}\alph{equation}}
    \stepcounter{subeq}}
{\renewcommand{\theequation}{\arabic{section}\Alph{subsection}.\arabic{equation}}
    \setcounter{equation}{\value{subeq}}
    \setcounter{subeq}{0}}
\newcommand{\proof}{\noindent{\em Proof:~}}
\newcommand{\proofs}{\noindent{\em Proofs:~}}
\newcommand{\cheers}{\hfill {\em End~of~proof.}}
\newcommand{\A}{{\mathcal A}}
\newcommand{\B}{{\mathcal B}}
\newcommand{\C}{{\mathcal C}}
\newcommand{\D}{{\mathcal D}}
\newcommand{\E}{{\mathcal E}}
\newcommand{\F}{{\mathcal F}}
\newcommand{\G}{{\mathcal G}}
\newcommand{\I}{{\mathcal I}}
\newcommand{\J}{{\mathcal J}}
\newcommand{\K}{{\mathcal K}}

\newcommand{\M}{{\mathcal M}}
\newcommand{\N}{{\mathcal N}}
\newcommand{\Q}{{\mathcal Q}}

\renewcommand{\S}{{\mathcal S}}

\newcommand{\Y}{{\mathcal Y}}
\newcommand{\Z}{{\mathcal Z}}
\newcommand{\bgreek}[1]{\mbox{\boldmath{$#1$}}}			
\newcommand{\bscript}[1]{\mbox{\boldmath{${\mathcal #1}$}}}	
\newcommand{\x}{{\mathbf x}}
\newcommand{\ba}{{\mathbf a}}
\newcommand{\bb}{{\mathbf b}}

\newcommand{\bu}{{\mathbf u}}
\newcommand{\bv}{{\mathbf v}}
\newcommand{\bw}{{\mathbf w}}
\newcommand{\bA}{{\mathbf A}}
\newcommand{\bC}{{\mathbf C}}

\newcommand{\bH}{{\mathbf H}}

\newcommand{\bK}{{\mathbf K}}

\newcommand{\bP}{{\mathbf P}}

\newcommand{\bS}{{\mathbf S}}
\newcommand{\bU}{{\mathbf U}}
\newcommand{\bV}{{\mathbf V}}
\newcommand{\bsD}{\bscript{D}}
\newcommand{\bsF}{\bscript{F}}
\newcommand{\bsG}{\bscript{G}}
\newcommand{\bsI}{\bscript{I}}

\newcommand{\bsM}{\bscript{M}}
\newcommand{\bsN}{\bscript{N}}
\newcommand{\bsP}{\bscript{P}}
\newcommand{\bsQ}{\bscript{Q}}
\newcommand{\bsR}{\bscript{R}}
\newcommand{\bsS}{\bscript{S}}
\newcommand{\bsT}{\bscript{T}}
\newcommand{\bsU}{\bscript{U}}
\newcommand{\bpartial}{\bgreek{\partial}}
\newcommand{\bchi}{\bgreek{\chi}}
\newcommand{\bdelta}{\bgreek{\delta}}
\newcommand{\bnu}{\bgreek{\nu}}
\newcommand{\bphi}{\bgreek{\phi}}
\newcommand{\bpsi}{\bgreek{\psi}}
\newcommand{\bsigma}{\bgreek{\sigma}}
\newcommand{\btheta}{\bgreek{\theta}}
\newcommand{\bvarphi}{\bgreek{\varphi}}
\newcommand{\bLambda}{\bgreek{\Lambda}}
\newcommand{\bPhi}{\bgreek{\Phi}}
\newcommand{\bPsi}{\bgreek{\Psi}}
\newcommand{\subbsQ}{\mbox{\boldmath ${\scriptstyle \Q}$}}
\newcommand{\subbsF}{\mbox{\boldmath ${\scriptstyle \F}$}}
\newcommand{\subbsI}{\mbox{\boldmath ${\scriptstyle \I}$}}
\newcommand{\bbE}{{\Bbb E}}
\newcommand{\sgn}[1]{{\mathrm sgn}#1}
\newcommand{\nbd}[1]{{\mathrm nbd}#1}
\renewcommand{\Re}[1]{{\mathrm Re}~#1}
\renewcommand{\Im}[1]{{\mathrm Im}~#1}
\newcommand{\dom}[1]{{\mathrm dom}~#1}
\newcommand{\tr}[1]{{\mathrm tr}~#1}
\newcommand{\domE}{D}
\newcommand{\s}{(\x_{0},\I^{(3)},\I^{(4)})}
\newcommand{\triple}{(\x_{0},\x_{1},\x_{2})}
\renewcommand{\u}{\underline}
\newcommand{\zip}[1]{{#1}$_{0}$}
\newcommand{\predot}[1]{{#1}^{\sqcup}}
\newcommand{\postdot}[1]{{#1}^{\sqcap}}

\begin{document}
\title{Group structure of the solution manifold of the hyperbolic
Ernst equation---general study of the subject and detailed elaboration
of mathematical proofs}
\author{I.\ Hauser\thanks{Home address: 4500 19th Street, \#342,
	Boulder, CO 80304} and F.\ J.\ Ernst\thanks{E-mail: gravity@slic.com} \\
	FJE Enterprises, 511 County Route 59, Potsdam, NY 13676, 
	USA\thanks{Homepage URL: http://www.slic.com/gravity}}
\date{September 9, 1998}
\maketitle

\begin{abstract}
Having defined a set $\S_{\E}^{3}$ that includes all $\bC^{4}$ solutions 
$\E$ of the hyperbolic Ernst equation in a certain specified domain, we 
formulate and prove, using a new HHP and equivalent integral equations of
the Alekseev and Fredholm types, a generalized Geroch conjecture concerning
the transformation of one member of $\S_{\E}^{3}$ into another member.
(The original Geroch conjecture, as interpreted by Kinnersley and Chitre,
concerned the transformation of one {\em analytic} solution of the elliptic
Ernst equation into another analytic solution.)  A unique feature of these
notes is that we provide a mathematical proof for every nontrivial statement
that we make.

It should be noted that some of the propositions and theorems that are 
included in these notes were developed before we knew exactly what would be
absolutely necessary in order to accomplish our principal objectives most
expeditiously.
\end{abstract}


\newpage
\part{Basic properties of various potentials; initial value problem}

\section{Introduction}
Some years ago, when we were concerned with axis-accessible stationary 
axisymmetric vacuum fields governed by an elliptic Ernst equation
we published a homogeneous Hilbert problem (HHP) on a closed contour\footnote{
I.~Hauser and F.~J.~Ernst, {\em A homogeneous Hilbert problem for the
Kinnersley--Chitre transformations}, J.\ Math.\ Phys.\ {\bf 21}, 1126-1140
(1980). \label{HHP}} with which we claimed we could effect a Kinnersley--Chitre
(K--C) transformation from one such spacetime to another.  To expedite our 
demonstration, we accepted (as has everyone who has ever devised a formalism 
for treating the same physical problem) certain unproven ``working hypotheses'' 
such as, for example, the existence of a solution of the HHP, and the 
differentiability of the solution of the HHP.  Some people improperly refer to 
such assumptions as ``axioms,'' thereby suggesting that they require no proof.
We, however, proved {\em all\/} our working hypotheses together
with a conjecture of Geroch several years later.\footnote{I.~Hauser and
F.~J.~Ernst, {\em A new proof of an old conjecture}, in Gravitation and
Geometry, Eds.\ Rindler and Trautman, Bibliopolis, Naples (1987).}

\subsection{The set $\S_{\E}\triple$ of $\E$-potentials}

Many methods have been introduced for the purpose of generating solutions of
the vacuum field equations for those spacetimes whose matrix tensors are
locally expressible in the form
\begin{equation}
\begin{array}{rcl}
g(x^{3},x^{4}) & = & dx^{a} \otimes dx^{b} g_{ab}(x^{3},x^{4})
	+ dx^{i} \otimes dx^{j} g_{ij}(x^{3},x^{4}), \\ & &
	\text{ where } a,b \in \{1,2\};i,j \in \{3,4\}, \label{G1.1}
\end{array}
\end{equation}
and which, therefore, have two commuting Killing vectors, $\bpartial_{1}$ 
and $\bpartial_{2}$.  Most of these methods have {\em not\/} been concerned with
the group structure of the solution manifold, but rather with the practical
problem of generating explicit {\em analytic\/} solutions corresponding to
given initial data (in the hyperbolic case) or given axis data (in the
elliptic case).  Therefore, there has been a certain, perhaps excusable, 
tendency to state theorems in a rather casual manner, and to provide only 
what might be called plausibility arguments.  Afterall, once one has an 
analytic solution, it can be checked by verifying directly that the Einstein 
equations are satisfied, and its domain of validity can be extended using 
analytic continuation.  

Such a cavalier attitude toward the statement of theorems and the formulation
of their proofs was not desirable when in 1980 we investigated the group 
structure of the finite transformations, the infinite-dimensional algebra 
of which had been developed in 1977 by Kinnersley and Chitre.  There, in
particular, we identified at the start the domains of the analytic 
$\E$-potentials and other fields that we introduced.  When one leaves the 
realm of analytic functions, as we propose to do in this set of notes, it 
becomes absolutely essential to be precise in stating theorems and
mathematically rigorous in proving them, whether one is attempting to study
the group structure of the solution manifold or simply attempting to 
generate specific solutions of the field equations corresponding to given
initial data.

In fact, we intend to motivate and to introduce a substantial enlargement of 
the family of K--C transformations of those vacuum metrics (\ref{G1.1}) for 
which the Killing vectors $\bpartial_{1}$ and $\bpartial_{2}$ are both spacelike. 
When $\bpartial_{1}$ and $\bpartial_{2}$ are spacelike, 
$g_{11}g_{22}-(g_{12})^{2} > 0$
and there exist coordinates $x^{3}=r$ and $x^{4}=s$ such that
\begin{equation}
g(r,s) = dx^{a} \otimes dx^{b} g_{ab}(r,s)
	+ (dr \otimes ds + ds \otimes dr) g_{34}(r,s).
\label{G1.2}
\end{equation}
The null coordinates $r,s$ are related to Weyl canonical coordinates 
\begin{equation}
\rho := \sqrt{g_{11}g_{22}-(g_{12})^{2}}
\end{equation}
and $z$ by
\begin{equation}
z = \frac{1}{2}(s+r), \quad \rho = \frac{1}{2}(s-r) > 0.
\end{equation}
The problem of solving the vacuum field equations for (\ref{G1.2}) is
reducible to the problem of solving the hyperbolic Ernst equation
\begin{equation}
f d(\rho \star d\E) - \rho d\E \star d\E = 0, \label{G1.5}
\end{equation}
where $\star$ is a two-dimensional duality operator such that
\begin{equation}
\star dr = dr, \quad \star \, ds = - ds, \label{G1.6}
\end{equation}
$d(\rho \star d\E) = d \wedge (\rho \star d\E)$ and
$d\E \star d\E = d\E \wedge (\star d\E)$ in accordance with the convention
of deleting $\wedge$ from exterior derivatives and exterior products of
differential forms, and $\E$ is related to $g_{ab}$ by
\begin{equation}
\E = f+i\chi, \quad f := \Re{\E} = -g_{22},
\end{equation}
and
\begin{equation}
d\omega = \rho f^{-2} \star d\chi, \quad \omega = g_{12}/g_{22}.
\end{equation}

Corresponding to the usual Ernst potential $\E = f + i\chi$, one
can alternatively introduce a $2 \times 2$ matrix potential
\begin{equation}
\bbE := I f + J \chi , \quad 
I := \left( \begin{array}{cc}
1 & 0 \\ 0 & 1
\end{array} \right) , \quad
J := \left( \begin{array}{cc}
0 & 1 \\ -1 & 0
\end{array} \right) .
\end{equation}

\begin{abc}
\begin{theorem}[Ernst equation $\Leftrightarrow d\Delta - \Delta\Delta = 0$]
\label{1.1A}
\mbox{ } \\
The Ernst equation (\ref{G1.5}) is equivalent to the matrix equation
\begin{equation}
\label{1.4a}
d\Delta - \Delta \Delta = 0 ,
\end{equation}
where
\begin{equation}
\Delta := - \left( \frac{\tau-z-\rho\star}{\tau-z+\rho\star}
\right)^{1/2} \frac{d\bbE}{2f} \sigma_{3} + \frac{d\chi}{2f} J 
\label{1.4b}
\end{equation}
depends not only upon $z$ and $\rho$ but also upon a complex spectral
parameter $\tau$.
\end{theorem}
\end{abc}
\begin{abc}
\proof
It will be understood that
\begin{equation}
\left( \frac{\tau-z-\rho\star}{\tau-z+\rho\star} \right)^{1/2} dr
= \left( \frac{\tau-s}{\tau-r} \right)^{1/2} dr , \quad
\left( \frac{\tau-z-\rho\star}{\tau-z+\rho\star} \right)^{1/2} ds
= \left( \frac{\tau-r}{\tau-s} \right)^{1/2} ds ,
\label{1.5a}
\end{equation}
such that the above $1$-forms have the branchcut $[r,s]$ in the
$\tau$-plane and have the values $dr$ and $ds$, respectively, at
$\tau=\infty$.  Moreover,
$\sigma_{3} := \left( \begin{array}{cc}
1 & 0 \\ 0 & -1
\end{array} \right)$ is one of the Pauli matrices.

Employing the easily established property
\begin{equation}
(\star \alpha) \beta = - \alpha (\star \beta) 
\end{equation}
of the two-dimensional duality operator, we obtain
\begin{equation}
\Delta \Delta = -2 \left( \frac{df}{2f} \right) \left( \frac{d\chi}{2f}
\right) J + 2 \left[ \left( \frac{\tau-z-\rho\star}{\mu} \right)
\frac{d\bbE}{2f} \right] \left( \frac{d\chi}{2f} \right) J \sigma_{3},
\end{equation}
using which it is straightforward to prove that Eq.\ (\ref{1.4a}) is
equivalent to the pair of equations
\begin{equation}
\label{1.5d}
d \left( \frac{d\chi}{2f} \right) = - 2 \left( \frac{df}{2f} \right)
\left( \frac{d\chi}{2f} \right) 
\end{equation}
and
\begin{equation}
\label{1.5e}
d \left[ \left( \frac{\tau-z-\rho\star}{\mu} \right) 
\frac{d\bbE}{2f} \right] = 2 \left( \frac{d\chi}{2f} \right)
\left[ \frac{\tau-z-\rho\star}{\mu} \right] \frac{d\bbE}{2f} J .
\end{equation}
Since Eq.\ (\ref{1.5d}) is satisfied identically, Eq.\ (\ref{1.4a})
is equivalent to the single equation (\ref{1.5e}), which, in turn, can
be expressed as 
\begin{equation}
\frac{f}{\mu} d(\rho\star d\bbE) 
- \left[ d\left(\frac{\tau-z}{\mu} \right) 
+ \rho \star d\left(\frac{1}{\mu}\right) \right] f d\bbE
= \frac{\rho}{\mu} d\bbE \star d\bbE,
\end{equation}
where 
\begin{equation}
\mu := \sqrt{(\tau-z)^{2}-\rho^{2}}, \quad
\lim_{\tau\rightarrow\infty} \left(\frac{\mu}{\tau}\right) = 1.
\label{mu}
\end{equation}
Because
$$
d \left( \frac{\tau-z}{\mu} \right) + \rho \star d \left( \frac{1}{\mu}
\right) = 0
$$
identically, we have proved that Eq.\ (\ref{1.4a}) is equivalent to
the Ernst equation (\ref{G1.5}).
\cheers
\end{abc}

We shall focus attention on those solutions of Eq.\ (\ref{G1.5}) that have the
rectangular or truncated rectangular domains (See Fig.~1.)
\begin{equation}
\domE := \{ \x=(r,s):r_{1}<r<r_{2},
s_{2}<s<s_{1} \text{ and } r<s \},
\end{equation}
where $\x_{1}=(r_{1},s_{1})$ is the upper left vertex of the rectangle,
with the understanding that $r_{1}=-\infty$ and/or $s_{1}=+\infty$ are
admissible,
$\x_{2}=(r_{2},s_{2})$ is the lower right vertex, and the lower left
vertex $(r_{1},s_{2})$ and upper right vertex $(r_{2},s_{1})$ have
$\rho \ge 0$, i.e., satisfy
\begin{equation}
r_{1} \le s_{2}, \quad r_{2} \le s_{1}. 
\label{G1.10}
\end{equation}
We shall frequently employ the notations
\begin{equation}
I^{(3)} := \{r: r_{1}<r<r_{2}\}, \quad I^{(4)} := \{s: s_{2}<s<s_{1}\}
\end{equation}
for the open real intervals involved in the definition of the domain $\domE$.

It must be noted that the concepts and results of this set of notes are all
easily extendable, by inspection, to domains that are the unions of $\domE$
with certain sections of its boundary.  There are two important cases.  One
of these is
\begin{abc}
\begin{equation}
\domE \cup \{(r_{1},s):s_{2}<s\le s_{1}\} \cup
\{(r,s_{1}):r_{1}\le r<r_{2}\}. \label{G.12a}
\end{equation}
The other is possible only when $s_{2}<r_{2}$ and it is
\begin{equation}
\domE \cup \{(z,z):s_{2}<z<r_{2}\}, \label{G1.12b}
\end{equation}
where the added boundary section lies on $\rho=0$.  We shall not complicate
our exposition by considering these domains with boundary sections as we
develop our formalism; but it will be a simple exercise for the reader to do 
so.
\end{abc}

We shall also select a {\em reference point\/} $\x_{0}=(r_{0},s_{0})$ within
each domain $\domE$ and, as can always be done, choose the coordinates and the 
arbitrary additive constant in $\chi=\Im{\E}$ so that 
\begin{equation}
\E(\x_{0})=-1.
\end{equation}
The triples $(\x_{0},\x_{1},\x_{2})$ for which it is true that, for every 
$\x \in \domE$, the null line segments $\{(r,s_{0}):r_{1}<r<r_{2}\}$ 
and $\{(r_{0},s):s_{2}<s<s_{1}\}$ both lie in $\domE$, will be called
{\em type\/} A.  This definition is equivalent to the statement that [see 
(\ref{G1.10})]
\begin{equation}
r_{1} < r_{0} \le s_{2} \text{ and } r_{2} \le s_{0} < s_{1} \quad
\text{(type A)} \label{G1.14} 
\end{equation}
For the domains (\ref{G.12a}), we choose $\x_{0} = \x_{1}$ and classify
$(\x_{0},\x_{1},\x_{2})$ as type A.  For the domains (\ref{G1.12b}), it
will be understood that $(\x_{0},\x_{1},\x_{2})$ is type A before one
adds $\{(z,z):s_{2} < z < r_{2}\}$ to the domain.

In this set of notes we shall grant that $\triple$ is type A, for the
concepts and proofs for triples that are not type A are generally much
more cumbersome than the ones for type A.  Accordingly, we shall focus
attention on sets 
\begin{eqnarray}
\S_{\E}\triple & := & \text{ the set of all complex-valued functions $\E$
such that} \nonumber \\ & & 
\text{ $\dom{\E} = \domE$, the derivatives $\E_{r}(\x)$, $\E_{s}(\x)$ and}
\nonumber \\ & &
\text{ $\E_{rs}(\x)$ exist and are continuous at all $\x \in \domE$, }
\\ & &
\text{ $f := \Re{\E} > 0$ and $\E$ satisfies Eq.\ (\ref{G1.5}) thoughout}
\nonumber \\ & &
\text{ $\domE$, the triple $\triple$ is type A, and $\E(\x_{0})=-1$.}
\nonumber
\end{eqnarray}

Our choice of rectangular or truncated rectangular domains came
at the end of a long investigation, which started by considering
all solutions of the hyperbolic Ernst equation with convex open domains
in the half-plane $\{(r,s):r < s\}$, and which accepted the premise that
the existence theorem for solutions of the new HHP is valid.  To accomodate
an arbitrary choice of the reference point $\x_{0}$ in the convex open
domain we formulated the new HHP in a slightly more general way than we 
shall do in the present set of notes.

The result of this investigation was that any solution $\E$
with a convex open domain has an extension $\bar{\E}$
which is also a solution and which has a domain that is of the
rectangular or truncated rectangular form $\domE$.
Moreover, the same HHP which is used to effect the
transformation $\E^{M} \rightarrow \E$ also effects $\E^{M}
\rightarrow \bar{\E}$.  There was clearly no point to
considering convex open domains other than those of the form $\domE$.

With regard to the domains $\domE$ for
which $r_{1}=s_{2}$ and/or $r_{2}=s_{1}$, there is the annoying
fact that no point $\x_{0} \in \domE$
exists for which $(\x_{0},\x_{1},\x_{2})$ is type A.  Nevertheless,
an Ernst potential domain for which $r_{1}=s_{2}$ and/or
$r_{2}=s_{1}$ may still be treated as the limit of a one-parameter
family of Ernst potential domains that do have type A triples.  
For example, suppose $\dom{\E} = \{\x:r_{1}=s_{2} < r < s < r_{2}
= s_{1}\}$.  Let $\epsilon$ denote any positive real number
less than $(r_{2}-s_{2})/2$, let $\E_{\epsilon}$ denote the 
restriction of $\E$ to the domain $\{\x:r_{1}+\epsilon \le r <
s \le s_{1}-\epsilon\}$ and let $\x_{1\epsilon} := (r_{1}+\epsilon,
s_{1}-\epsilon)$.  Then $(\x_{1\epsilon},\x_{1\epsilon},\x_{2})$
is type A; and the limit of $\dom{\E_{\epsilon}}$ as $\epsilon
\rightarrow 0$ is $\dom{\E}$ [in the sense that $\dom{\E_{\epsilon}}
\subset \dom{\E_{\epsilon'}}$ when $\epsilon < \epsilon'$, and
the union over all $\epsilon$ of the sets $\dom{\E_{\epsilon}}$ 
is $\dom{\E}$].

The above discussion shows that no irreparable loss of generality
occurs if one considers only type A triples, since no Ernst potential
domain of the form $\domE$ is thereby
actually excluded from treatment by our formalism.

\setcounter{equation}{0}
\setcounter{theorem}{0}
\subsection{Incompleteness of Kinnersley--Chitre transformations}

That the K--C transformations are incomplete is clear, for
any member of $\S_{\E}\triple$ that is obtained from $\E^{M}$ by a K--C
transformation is analytic, whereas the solutions of the hyperbolic (as
opposed to the elliptic) Ernst equation are generally not analytic even
when they are $\bC^{\infty}$.

Moreover, the general member of the set $\S_{\E}\triple$ is a functional
of {\em four\/} independent real-valued $\bC^{1}$ functions of a single real
variable, and these functions can be chosen arbitrarily.\footnote{This
observation applies also to the elliptic case when the axis is inaccessible,
a point that did not elude G.\ A.\ Alekseev.  See Ref.~\ref{Alekseev}.}  
However, the member of $\S_{\E}\triple$ that is obtained from the general
K--C transformation of its Minkowski space member $\E^{M} \in \S_{\E}\triple$,
where
\begin{equation}
\E^{M}(\x)=-1 \text{ for all } \x \in \domE,
\end{equation}
is a functional of only {\em two\/} independent real-valued {\em analytic\/}
functions of a single real variable.

As we embark upon this study, designed to rectify the limitations of the
K--C group described above, let us warn the reader that our analysis will
involve the use of several HHP's that should not be confused with one another.
One of these HHP's is of earlier vintage,\footnote{I.~Hauser and F.~J.~Ernst, 
{\em Initial value problem for colliding gravitational plane waves-I/II\/}, 
J.\ Math.\ Phys.\ {\bf 30}, 872-887, 2322-2336 (1989). \label{IVP12}} and 
it is used to effect the solution of the initial value problem, in which the
complex Ernst potential is specified upon two null lines.  In this set of
notes this HHP will be used to establish certain relatively elementary
properties of the various potentials with which we shall be concerned, and
to explore the limits of certain of these potentials as one approaches 
certain important cuts in the complex plane.  Using the knowledge gained by 
this investigation, we shall be able to formulate a new HHP on arcs and 
identify a group of mappings of one solution of the hyperbolic Ernst 
equation within a certain set $\S_{\E}^{3}$ (that includes all $\bC^{4}$
solutions) into another such solution, concerning which we shall formulate 
a generalized Geroch conjecture.\footnote{Initially we considered the simpler
case of $\bC^{\infty}$ solutions.}

Logically this work consists of three principal parts.  The first part may
be considered to be prologue, and involves mostly elementary propositions 
concerning various potentials and solving the initial value problem; the 
second part concerns everything up to and including the statement of our new
HHP on arcs and the statement of our generalized Geroch conjecture; the third 
part comprises all the steps that we have to go through, using that new HHP,
in order to prove our generalized Geroch conjecture and certain other claims.

Whenever we are aware of (and have seen the proof of) a standard theorem, 
the exploitation of which will help us achieve our objectives, we shall
provide a reference for the proof of that theorem; whenever we find it
necessary first to generalize a standard theorem, we shall provide a 
reference for the standard theorem and point out why we could not use the
standard theorem itself.\footnote{We would appreciate from readers any
additional references to published proofs of more powerful mathematical
theorems of which we appear to be unaware.}

As a result of this work we have identified a Geroch-like group $\K^{3}$ 
the general element of which is a functional of six arbitrary real-valued
$\bC^{3}$ functions of a real variable.  The group $\K^{3}$ can generate, 
in principle, a set $\S_{\E}^{3}$ of solutions of the Ernst equation such 
that the general member of $\S_{\E}^{3}$ is a functional of {\em four} of
the real-valued functions on which the general member of $\K^{3}$ depends.
This is in accord with the initial value problem, where the initial data 
comprise the complex Ernst potentials $\E^{(3)}$ and $\E^{(4)}$ on two 
null lines.  The family of elements of $\K^{3}$ which generate a given
member of $\S_{\E}^{3}$ can be computed from $\E^{(3)}$ and $\E^{(4)}$,
and vice versa.

It is sufficient to assume that $\E^{(3)}$ and $\E^{(4)}$ are $\bC^{1}$ 
in order to derive from our new HHP an equation analogous to the singular
integral equation that G.\ A.\ Alekseev\footnote{G.~A.\ Alekseev, {\em
The method of the inverse scattering problem and singular integral
equation for interacting massless fields}, Dokl.\ Akad.\ Nauk SSSR
{\bf 283}, 577--582 (1985) [Sov.\ Phys.\ Dokl.\ (USA) {\bf 30}, 565
(1985)], {\em Exact solutions in the general theory of relativity}, 
Trudy Matem.\ Inst.\ Steklova {\bf 176}, 215--262 (1987). \label{Alekseev}}
introduced in 1985 in his treatment of the initial value problem for
the case when the initial data are analytic.  Moreover, it is sufficient
to assume a certain H\"{o}lder condition in addition to $\bC^{2}$ 
differentiability of $\E^{(3)}$ and $\E^{(4)}$ in order to derive
(from our Alekseev-type equation) a Fredholm integral equation of the
second kind that can also be used to handle the initial value problem.
However, our proof that our new HHP, our Alekseev-like integral equation,
and our Fredholm integral equation are equivalent to one another, and
our proof that the solution of each of these equivalent equations 
actually yields a solution of the hyperbolic Ernst equation on the
prescribed domain are only valid if one assumes that $\E^{(3)}$ and
$\E^{(4)}$ are initial data for a member of $\S_{\E}^{3}$.

\begin{figure}[htbp] 
\begin{picture}(300,300)(-60,60)
\put(75,75){\line(1,1){175}}
\put(125,85){the line $r=s$ (or $\rho=0$)}
\put(125,90){\vector(-1,1){15}}
\put(250,125){\vector(0,1){40}}
\put(250,125){\vector(1,0){40}}
\put(248,170){s}
\put(295,122){r}
\put(75,250){\line(1,0){100}}
\put(75,120){\line(1,0){45}}
\put(75,120){\line(0,1){130}}
\put(175,175){\line(0,1){75}}
\put(75,250){\circle*{3}}
\put(75,120){\circle*{3}}
\put(175,250){\circle*{3}}
\put(175,120){\circle*{3}}
\put(65,255){${\mathbf x}_{1}$}
\put(175,110){${\mathbf x}_{2}$}
\put(50,105){$(r_{1},s_{2})$}
\put(180,255){$(r_{2},s_{1})$}
\put(100,220){\circle*{3}}
\put(105,210){${\mathbf x}_{0}$}
\multiput(75,220)(10,0){10}{\line(1,0){5}}
\multiput(100,120)(0,10){13}{\line(0,1){5}}
\end{picture}
\caption{An $\E$-potential domain $\domE$ and choice of $\x_{0}$ for 
which $s_{2} < r_{2}$ and $\triple$ is type~A is illustrated.  The null 
line segments through $\x_{0}$ are represented by the vertical and 
horizontal dashed lines.}
\end{figure}
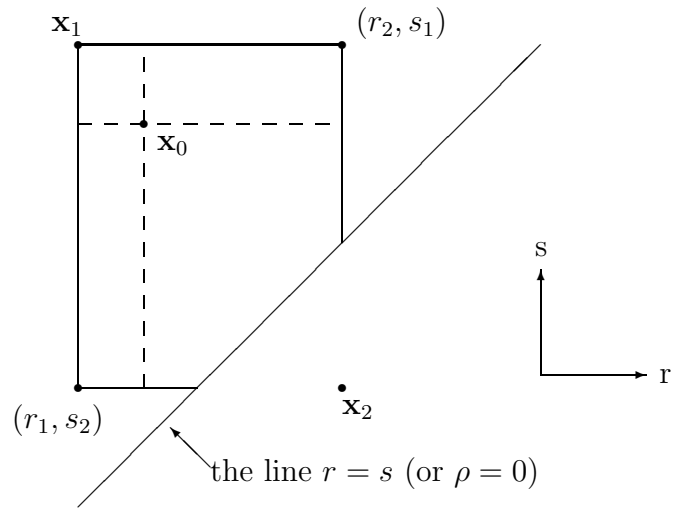

\newpage

\setcounter{equation}{0}
\setcounter{theorem}{0}
\section{Properties of and relations between the solutions $\Q$ and
$\F$ of two associated linear systems $d\Q=\Delta \Q$ and
$d\F=\Gamma \F$ of the hyperbolic Ernst equation \label{Sec_1}}

\begin{nosubsec}

The linear systems $d\Q = \Delta \Q$ and $d\F = \Gamma \F$ are very similar 
to ones introduced many years ago by G.\ Neugebauer\footnote{G.~Neugebauer, 
{\em B\"{a}cklund transformations of axially symmetric stationary 
gravitational fields}, Phys.\ Lett.~A {\bf 12}, L67 (1979). \label{Neu}} 
and by W.\ Kinnersley and D.\ M.\ Chitre,\footnote{W.~Kinnersley and 
D.~M.~Chitre, {\em Symmetries of the stationary Einstein-Maxwell field
equations, III}, J.\ Math.\ Phys.\ {\bf 19}, 1926--1931 (1978).} 
respectively.  Our objective in Sec.~\ref{Sec_1} is to establish (for 
given Ernst potential $\E \in \S_{\E}$), using an approach that we
developed earlier in connection with our treatment of the initial value 
problem\footnote{I.~Hauser and F.~J.~Ernst, {\em Initial value problem
for colliding gravitational plane waves-III/IV\/}, J.~Math.~Phys.\ 
{\bf 31}, 871--881 (1990), {\bf 32}, 198--209 (1991). \label{IVP34}}
for colliding gravitational plane waves, some important properties of the
respective solutions $\Q$ and $\F$ of these two linear systems.

Initially we shall consider fields $\bsQ$ and $\bsF$ that involve two 
points $\x$ and $\x'$, and only later set $\x'$ equal to $\x_{0}$ to
construct the fields $\Q$ and $\F$.\footnote{In this set of notes 
two-point functions will always be denoted by bold type.  On the other
hand, when $\x'$ is replaced by a fixed point $\x_{0}$, the resulting 
function of the remaining variables will be denoted by ordinary type.} 

\end{nosubsec}


\subsection{Notations for frequently encountered sets}

In the hyperbolic case we employ null coordinates $x^{3}$ and
$x^{4}$ such that
\begin{equation}
x^{3} = r := z - \rho, \quad x^{4} = s := z + \rho.
\label{1.2}
\end{equation}
Then 
\begin{equation}
\star dx^{i} = (-1)^{7-i} dx^{i}. \; (i \in \{3,4\})
\label{1.3}
\end{equation}

\begin{abc}
\begin{definition}{Dfns.\ of the subspaces $C^{+}$, $C^{-}$, $\bar{C}^{+}$
and $\bar{C}^{-}$ of $C$\label{def01}} 
Let 
\begin{equation}
C^{\pm} := \{ \tau \in C-\{\infty\}: \pm\Im{\tau}>0 \}
\end{equation}
and
\begin{equation}
\bar{C}^{\pm} := C^{\pm} \cup R^{1} \cup \{\infty\},
\end{equation}
where $R^{1}$ is the real axis.  Arguments of $\tau \in
\bar{C}^{\pm}-\{0,\infty\}$ will always be chosen so that
\begin{equation}
0 \le \pm \arg{\tau} \le \pi
\end{equation}
\end{definition}
\end{abc}

\begin{abc}
\begin{definition}{Dfns.\ of the closed disks $d(\x)$ and $d(\x,\x')$
\label{def02}}
For all $\x, \x' \in D$,
\begin{eqnarray}
d(\x) & := & \text{the closed disk in $C$ that is centered on the origin}
\nonumber \\
& & \text{and that has radius } \sup{\{|x^{3}|,|x^{4}|\}}, 
\label{d1} \\
d(\x,\x') & := & \text{the closed disk in $C$ that is centered on the origin}
\nonumber \\
& & \text{and that has radius } 
\sup{\{|x^{3}|,|x^{4}|,|x^{\prime 3}|,|x^{\prime 4}|\}}. 
\label{d2} 
\end{eqnarray}
\end{definition}
\end{abc}

\begin{definition}{Dfn.\ of the set $|x,x'|$\label{def03}}
If $x$ and $x' \in R^{1}$, then
\begin{equation}
|x,x'| := \{ \lambda x + (1-\lambda) x': 0 \le \lambda \le 1 \}; 
\end{equation}
i.e., $|x,x'|$ is the closed interval with end points $x$ and $x'$.
\end{definition}

\begin{definition}{Dfns.\ of the sets $\bsI^{(i)}(\x,\x')$,
$\bar{\bsI}^{(i)}(\x,\x')$, $\grave{\bsI}^{(i)}(\x,\x')$, $\bsI(\x,\x')$,
$\bar{\bsI}(\x,\x')$, $\grave{\bsI}(\x,\x')$ and $\hat{\bsI}(\x,\x')$
\label{def04}}
Let\footnote{We shall employ bold type for sets that are determined
by specifying {\em two} arbitrary points $\x$ and $\x'$ in 
$\domE := \dom{\E}$.}
\begin{abc}
\begin{eqnarray}
\bsI^{(i)}(\x,\x') & := & \text{the open interval with end points
$x^{i}$ and $x^{\prime i}$}, \label{1.6q} \\
\bar{\bsI}^{(i)}(\x,\x') & := & \text{the closure of $\bsI^{(i)}(\x,\x')$}, \\
\grave{\bsI}^{(i)}(\x,\x') & := & |x^{i},x^{\prime i}|, \label{1.6w} \\[1ex]
\bsI(\x,\x') & := & \bsI^{(3)}(\x,\x') \cup \bsI^{(4)}(\x,\x'), \\
\bar{\bsI}(\x,\x') & := & \bar{\bsI}^{(3)}(\x,\x') \cup \bar{\bsI}^{(4)}(\x,\x'), \\
\grave{\bsI}(\x,\x') & := & \grave{\bsI}^{(3)}(\x,\x') \cup \grave{\bsI}^{(4)}(\x,\x'),
\\
\hat{\bsI}(\x,\x') & := & \text{the minimal closed real axis interval} 
\nonumber \\
& & \text{that contains } \{x^{3},x^{\prime 3},x^{4},x^{\prime 4}\},
\end{eqnarray}
where $\x := (x^{3},x^{4}) = (r,s)$ and $\x' := (x^{\prime 3},x^{\prime 4})
= (r',s')$ are arbitrary points in $\domE := \dom{\E}$.
Finally, we introduce the abbreviations
\end{abc}
\begin{abc}
\begin{eqnarray}
\I^{(i)}(\x) & := & \bsI^{(i)}(\x,\x_{0}), \\
\bar{\I}^{(i)}(\x) & := & \bar{\bsI}^{(i)}(\x,\x_{0}), \\
\grave{\I}^{(i)}(\x) & := & \grave{\bsI}^{(i)}(\x,\x_{0}), \\
\I(\x) & := & \bsI(\x,\x_{0}), \\
\bar{\I}(\x) & := & \bar{\bsI}(\x,\x_{0}), \\
\grave{\I}(\x) & := & \grave{\bsI}(\x,\x_{0}), \\
\hat{\I}(\x) & := & \hat{\bsI}(\x,\x_{0}).
\end{eqnarray}
\end{abc}
\end{definition}
Note that $\grave{\bsI}^{(i)}(\x,\x') = \{x^{i}\}$ whereas 
$\bar{\bsI}^{(i)}(\x,\x')$ is empty when $x^{i}=x^{\prime i}$.
Also, note that 
\begin{equation}
\hat{\bsI}(\x,\x') = \bar{\bsI}(\x,\x') \cup [x^{\prime 3},x^{\prime 4}] = 
\grave{\bsI}(\x,\x') \cup [x^{3},x^{4}]. 
\end{equation}


\setcounter{theorem}{0}
\setcounter{equation}{0}
\subsection{The Neugebauer-like linear system and the sets 
$\S_{\subbsQ^{\pm}}\triple$}  

Equation (\ref{1.4a}) is the integrability condition for the linear system 
\begin{equation}
d\Q(\x,\tau) = \Delta(\x,\tau) \Q(\x,\tau), 
\label{G2.15a}
\end{equation}
where $\Q$ is a $2 \times 2$ matrix function which satisfies the
``initial condition'' $\Q(\x_{0},\tau)=I$.

\begin{abc}
\begin{definition}{Dfns.\ of the functions $\M^{\pm}$ and  $\mu^{\pm}$
\label{def05}}
Let $\M^{\pm}$ denote the function whose domain is $\bar{C}^{\pm}$
and whose values are $\M^{\pm}(0) := 0$, $\M^{\pm}(\infty) = \infty$
and, for each $\tau \in \bar{C}^{\pm}-\{0,\infty\}$,
\begin{equation}
\M^{\pm}(\tau) := \sqrt{|\tau|} \exp \left( \frac{i}{2}\arg{\tau}\right),
\label{1.6d}
\end{equation}
where $\sqrt{|\tau|}$ always denotes the non-negative square root of $\tau$.
Furthermore, $\mu^{\pm}$ will denote the function with the domain
$\{\x:r \le s\} \times \bar{C}^{\pm}$ and the values
\begin{equation}
\mu^{\pm}(\x,\tau) := \M^{\pm}(\tau-r) \M^{\pm}(\tau-s).
\label{1.6e}
\end{equation}
It is to be understood that $\tau^{-1}\mu^{\pm}(\x,\tau)$ has the value
$1$ at $\tau=\infty$.
\end{definition}
Note that $\M^{\pm}$ and $\mu^{\pm}$ are continuous.  The restriction of
$\M^{\pm}$ to $C^{\pm}$ is equal to the restriction of the principal
branch of $\tau^{1/2}$ to $C^{\pm}$.  Under complex conjugation,
\begin{equation}
\M^{-}(\tau) = [\M^{+}(\tau^{*})]^{*}
\label{1.6f}
\end{equation}
and, for $\sigma\in R^{1}$,
\begin{equation}
\M^{\pm}(\sigma) = \left\{ \begin{array}{cc}
\sqrt{\sigma} \text{ if } \sigma \ge 0, \\
\pm i \sqrt{|\sigma|} \text{ if } \sigma < 0.
\end{array} \right.
\end{equation}
\end{abc}

\begin{abc}
\begin{definition}{Dfn.\ of the index $\kappa$\label{def06}}
Let $\kappa$ be an index whose value is determined (not always
uniquely) by $(\x,\sigma)$ as follows: 
\begin{equation}
\kappa = \left\{ \begin{array}{rl}
1 & \text{ if } s \le \sigma, \\
-1 & \text{ if } \sigma \le r, \\
0 & \text{ if } r \le \sigma \le s
\end{array} \right.
\label{1.6h}
\end{equation}
\end{definition}
Then
\begin{equation}
\mu^{\pm}(\x,\sigma) = e^{\pm i \frac{\pi}{2} (1-\kappa)}
\sqrt{|\sigma-r||\sigma-s|}.
\label{1.6i}
\end{equation}
For fixed $\x=(r,s)$, we shall say that $\mu^{\pm}(\x,\tau)$ is
{\em holomorphic on} $\bar{C}^{\pm}-\{r,s,\infty\}$, which means
that the function whose domain is $\bar{C}^{\pm}-\{r,s,\infty\}$
and which equals $\mu^{\pm}(\x,\tau)$ on that domain has a holomorphic
extension to at least one open subset of $C$.  

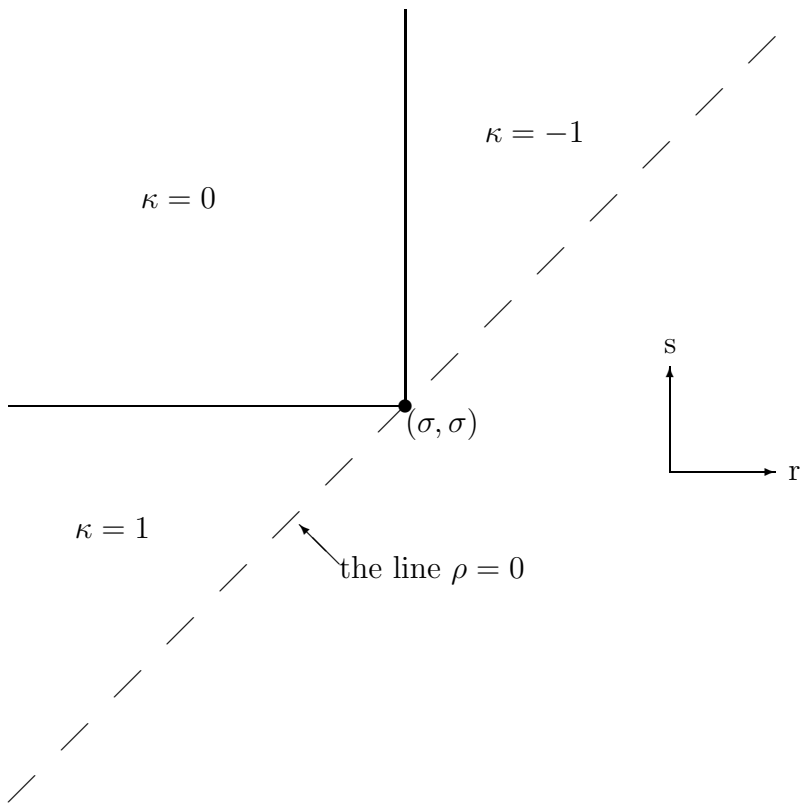
\begin{figure}[htbp] 
\begin{picture}(300,300)(-60,0)
\multiput(0,0)(20,20){15}{\line(1,1){10}}
\put(150,150){\circle*{5}}
\put(150,150){\line(0,1){150}}
\put(0,150){\line(1,0){150}}
\put(250,125){\vector(0,1){40}}
\put(250,125){\vector(1,0){40}}
\put(248,170){s}
\put(295,122){r}
\put(150,140){$(\sigma,\sigma)$}
\put(50,225){$\kappa=0$}
\put(180,250){$\kappa=-1$}
\put(25,100){$\kappa=1$}
\put(125,85){the line $\rho=0$}
\put(125,90){\vector(-1,1){15}}
\end{picture}
\vspace{0.3in}
\caption{The three subspaces of $\{\x:r<s\}$ corresponding to a given
value of $\sigma$ and to $\kappa=1$, $0$ and $-1$.  The point
$(\sigma,\sigma)$ lies on the diagonal $\rho=(s-r)/2=0$.  The horizontal
ray $\{(r,\sigma):r<\sigma\}$ is a subset of both the $\kappa=1$ and the
$\kappa=0$ subspaces.  The vertical ray $\{(\sigma,s):\sigma<s\}$ is
a subset of both the $\kappa=0$ and the $\kappa=-1$ subspaces.}
\end{figure}

The function $\mu$ defined by Eq.\ (\ref{mu}) can be defined equivalently
in terms of $\mu^{\pm}$ as follows:
\begin{equation}
\mu(\x,\tau) := \mu^{\pm}(\x,\tau) \text{ for all } \x=(r,s)
\text{ and } \tau \in \bar{C}^{\pm}-[r,s].
\label{1.6j}
\end{equation}
We recall that $\mu(\x,\tau)$ is a holomorphic function of $\tau$
throughout $C-([r,s]\cup\{\infty\})$ and has a pole of order one at
$\tau=\infty$.
\end{abc}

\begin{abc}
\begin{definition}{Dfns.\ of $\bnu_{i}^{\pm}$\label{def07}} 
Let $\bnu_{i}^{\pm} \; (i \in \{3,4\})$ denote the function
whose domain is
\begin{equation}
\dom{\bnu_{i}^{\pm}} := \{ (x^{i},x^{\prime i},\tau) \in R^{2} \times 
\bar{C}^{\pm}:\tau \ne x^{i} \text{ and } \tau \ne x^{\prime i}\}
\label{1.6l}
\end{equation}
and whose values are
\begin{equation}
\bnu_{i}^{\pm}(x^{i},x^{\prime i},\tau) := \frac{\M^{\pm}(\tau-x^{\prime i})}
{\M^{\pm}(\tau-x^{i})}
\label{1.6m}
\end{equation}
for finite $\tau$, and
\begin{equation}
\bnu_{i}^{\pm}(x^{i},x^{\prime i},\infty) = 1.
\label{1.6n}
\end{equation}
\end{definition}
\end{abc}

Note that $\bnu_{i}^{\pm}$ is finite-valued and continuous throughout its
domain (\ref{1.6l}) and, for fixed $(x^{i},x^{\prime i})$,
$\bnu_{i}^{\pm}(x^{i},x^{\prime i},\tau)$ is a holomorphic function of $\tau$
on $\bar{C}^{\pm}-\{x^{i},x^{\prime i}\}$.  

{}From Eqs.\ (\ref{1.6q}), (\ref{1.6w}) and (\ref{1.6m}), one obtains
\begin{equation}
\bnu_{i}^{\pm}(x^{i},x^{\prime i},\sigma) = \left\{ 
\begin{array}{l}
\sqrt{\frac{\sigma-x^{\prime i}}{\sigma-x^{i}}} \text{ if }
\sigma \in R^{1}-\grave{\bsI}^{(i)}(\x,\x') \\
\pm \sgn{(x^{\prime i}-x^{i})}i \sqrt{\frac{\sigma-x^{\prime i}}{x^{i}-\sigma}}
\text{ if } \sigma \in \bsI^{(i)}(\x,\x').
\end{array} \right.
\label{1.6r}
\end{equation}

\begin{abc}
\begin{definition}{Dfns.\ of $\grave{\bnu}_{i}$ and $\bar{\bnu}_{i}$
\label{def08}}
Equations (\ref{1.6n}) and (\ref{1.6r})
permit us to define functions\footnote{In this set of notes we shall often
employ on functions accents that suggest the $\tau$-plane domains of the
functions.  Typically, one has $\dom{\hat{f}} \subset \dom{\grave{f}}
\subset \dom{\bar{f}}$.}
$\grave{\bnu}_{i}$ and $\bar{\bnu}_{i}$ with
\begin{equation}
\begin{array}{rcl}
\dom{\grave{\bnu}_{i}} & := & \{(x^{i},x^{\prime i},\tau):(x^{i},x^{\prime i})
\in R^{2} \text{ and } \tau \in C - \grave{\bsI}^{(i)}(\x,\x')\}, \\
\dom{\bar{\bnu}_{i}} & := & \{(x^{i},x^{\prime i},\tau):(x^{i},x^{\prime i}) 
\in R^{2} \text{ and } \tau \in C-\bar{\bsI}^{(i)}(\x,\x')\}
\label{1.6t}
\end{array}
\end{equation}
and values
\begin{equation}
\begin{array}{rcl}
\grave{\bnu}_{i}(x^{i},x^{\prime i},\tau) & = & 
\bar{\bnu}_{i}(x^{i},x^{\prime i},\tau) 
\nonumber \\
& := & \bnu_{i}^{\pm}(x^{i},x^{\prime i},\tau) \text{ when } 
\tau \in \bar{C}^{\pm} - \grave{\bsI}^{(i)}(\x,\x'),
\label{1.6u} \\
\bar{\bnu}_{i}(x^{i},x^{i},x^{i}) & := & 1, 
\text{ whereas $\grave{\bnu}_{i}(x^{i},x^{i},x^{i})$ is not defined}.
\nonumber
\end{array}
\end{equation}
\end{definition}
As is well known, $\grave{\bnu}_{i}$ is finite-valued and continuous, while, for
fixed $(x^{i},x^{\prime i})$, $\bar{\bnu}_{i}(x^{i},x^{\prime i},\tau)$ is a
holomorphic function of $\tau$ throughout $C - \bar{\bsI}^{(i)}(\x,\x')$.
\end{abc}

\begin{abc}
\begin{definition}{Dfns.\ of $\bnu^{\pm}$, $\grave{\bnu}$ and $\bar{\bnu}$
\label{def09}}
Let $\bnu^{\pm}$ denote the function with
\begin{equation}
\dom{\bnu^{\pm}} := \{(\x,\x',\tau) \in R^{2} \times R^{2} \times
\bar{C}^{\pm}:r \le s, r' \le s' \text{ and } \tau \notin \{r,r',s,s'\}\}
\label{1.6o}
\end{equation}
and values
\begin{equation}
\bnu^{\pm}(\x,\x',\tau) := \bnu_{3}^{\pm}(r,r',\tau) \bnu_{4}^{\pm}(s,s',\tau).
\label{1.6p}
\end{equation}
Let $\grave{\bnu}$ and $\bar{\bnu}$ denote the functions whose domains are
\begin{equation}
\begin{array}{rcl}
\dom{\grave{\bnu}} & := & \{(\x,\x',\tau) \in R^{2} \times R^{2} \times C: r \le s,
r' \le s', \tau \in C - \grave{\bsI}(\x,\x')\}, \\
\dom{\bar{\bnu}} & := & \{(\x,\x',\tau) \in R^{2} \times R^{2} \times C: r \le s,
r' \le s', \tau \in C - \bar{\bsI}(\x,\x')\}, 
\end{array}
\end{equation}
and whose values are given by
\begin{equation}
\begin{array}{rcl}
\grave{\bnu}(\x,\x',\tau) & := & \grave{\bnu}_{3}(r,r',\tau) 
\grave{\bnu}_{4}(s,s',\tau), \\
\bar{\bnu}(\x,\x',\tau) & := & \bar{\bnu}_{3}(r,r',\tau) 
\bar{\bnu}_{4}(s,s',\tau).
\end{array}
\label{1.6x}
\end{equation}
\end{definition}
\end{abc}

The properties of $\bnu^{\pm}$, $\grave{\bnu}$ and $\bar{\bnu}$ can be
easily obtained from those of $\bnu_{i}^{\pm}$, $\grave{\bnu}_{i}$ and
$\bar{\bnu}_{i}$.  Thus, $\bnu^{\pm}$, $\grave{\bnu}$ and $\bar{\bnu}$
are finite-valued, $\bnu^{\pm}$ and $\grave{\bnu}$ are continuous,
$\bnu^{\pm}(\x,\x',\tau)$ is a holomorphic function of $\tau$ on 
$\bar{C}^{\pm}-\{r,r',s,s'\}$ and $\bar{\bnu}(\x,\x',\tau)$
is a holomorphic function of $\tau$ throughout $C - \bar{\bsI}(\x,\x')$.
{}From Eqs.\ (\ref{1.6l}), (\ref{1.6m}) and (\ref{1.6p}) we obtain
the useful relation
\begin{abc}
\begin{equation}
\bnu^{\pm}(\x,\x',\tau) = \frac{\mu^{\pm}(\x',\tau)}{\mu^{\pm}(\x,\tau)}
\text{ for all } (\x,\x',\tau) \in \dom{\bnu^{\pm}}.
\label{1.6y}
\end{equation}
{}From Eqs.\ (\ref{1.6j}), (\ref{1.6u}) and (\ref{1.6x}) we then obtain
\begin{equation}
\bar{\bnu}(\x,\x',\tau) = \frac{\mu(\x',\tau)}{\mu(\x,\tau)} \text{ for all }
(\x,\x',\tau) \in \dom{\bar{\bnu}}.
\label{1.6z}
\end{equation}
\end{abc}

\begin{abc}
\begin{definition}{Dfn.\ of $\Delta^{\pm}$\label{def10}}
For each $\E \in \S_{\E}\triple$, let $\Delta^{\pm}$ denote the $2 \times 2$
matrix $1$-form such that
\begin{equation}
\dom{\Delta^{\pm}} := \{(\x,\tau):\x\in\domE \text{ and } \tau \in
\bar{C}^{\pm} - \{r,s\}\}
\label{1.7a}
\end{equation}
and $\Delta^{\pm}(\x,\tau)$ is given by Eq.\ (\ref{1.4b}) provided 
we make the substitutions
\begin{equation}
\left( \frac{\tau-z-\rho\star}{\tau-z+\rho\star} \right)^{1/2}
\rightarrow \frac{\M^{\pm}(\tau-z-\rho\star)}{\M^{\pm}(\tau-z+\rho\star)}
\label{1.7b}
\end{equation}
and
\begin{equation}
\left( \frac{\tau-s}{\tau-r} \right)^{1/2} \rightarrow
\frac{\M^{\pm}(\tau-s)}{\M^{\pm}(\tau-r)}, \quad
\left( \frac{\tau-r}{\tau-s} \right)^{1/2} \rightarrow
\frac{\M^{\pm}(\tau-r)}{\M^{\pm}(\tau-s)},
\label{1.7c}
\end{equation}
in Eqs.\ (\ref{1.4b}) and (\ref{1.5a}).  It is to be understood 
that the ratios in Eq.\ (\ref{1.7b}) and (\ref{1.7c}) equal $1$ when 
$\tau=\infty$.
\end{definition}
\end{abc}

\begin{abc}
\begin{proposition}[Properties of $\Delta^{\pm}$]
\label{1.2A}
\mbox{ } \\ \vspace{-3ex}
\begin{romanlist}
\item 
\begin{equation}
\tr{\Delta^{\pm}} = 0 .
\label{1.8a}
\end{equation}
\item 
For all $(\x,\tau) \in \dom{\Delta^{+}}$,
\begin{equation}
[\Delta^{+}(\x,\tau)]^{*} = \Delta^{-}(\x,\tau^{*}) .
\label{1.8b}
\end{equation}
\item 
If $\tau=\sigma$ is real,
\begin{equation}
J^{1-\kappa} \Delta^{-}(\x,\sigma) J^{\kappa-1} = \Delta^{+}(\x,\sigma) .
\label{1.8c}
\end{equation}
\item \mbox{ } 
\begin{equation}
d\Delta^{\pm} - \Delta^{\pm} \Delta^{\pm} = 0 .
\label{1.8d}
\end{equation}
\end{romanlist}
\end{proposition}
\proofs
Equations (\ref{1.8a}) to (\ref{1.8c}) follow by inspection from
Eqs.\ (\ref{1.4b}), (\ref{1.5a}), (\ref{1.6f}), (\ref{1.6i}) and 
(\ref{1.7b}). Equation (\ref{1.8d}) is readily deduced for $\Delta^{+}$
from the Ernst equation (\ref{G1.5}) by using the derivation in the proof of
Thm.\ \ref{1.1A}, with the understanding that 
$(\tau-\alpha)^{1/2} = \M^{+}(\tau-\alpha)$ 
and that $\mu$ is to be replaced by $\mu^{+}$ in the derivation.  The
equation for $\Delta^{-}$ then follows from Eq.\ (\ref{1.8b}).
\cheers
\end{abc}

\begin{abc}
\begin{definition}{Dfn.\ of $\S_{\subbsQ^{\pm}}\triple$\label{def11}}
Let $\S_{\subbsQ^{\pm}}\triple$ denote the set of all $2 \times 2$ matrix
functions $\bsQ^{\pm}$ such that 
\begin{equation}
\dom{\bsQ^{\pm}} := \domE^{2} \times \bar{C}^{\pm}
\label{1.9a}
\end{equation}
and such that the following three conditions hold for all
$(\x',\tau) \in \domE \times \bar{C}^{\pm}$:
\begin{arablist}
\item 
\begin{equation}
\bsQ^{\pm}(\x',\x',\tau)=I.
\label{1.9b}
\end{equation}
\item 
The function of $\x$ given by $\bsQ^{\pm}(\x,\x',\tau)$ is finite-valued
and continuous throughout $\domE$.
\item 
The same function of $\x$ is differentiable throughout
$\{\x\in\domE:(\tau-r)(\tau-s) \ne 0\}$ and there exists
$\E \in \S_{\E}\triple$ such that
\begin{equation}
d\bsQ^{\pm}(\x,\x',\tau) = \Delta^{\pm}(\x,\tau) \bsQ^{\pm}(\x,\x',\tau).
\label{1.9c}
\end{equation}
\end{arablist}
\end{definition}
We shall henceforth suppress the argument `$\triple$' whenever there is no 
likely danger of ambiguity.
\end{abc}

\begin{abc}
\begin{proposition}[Properties of $\bsQ^{\pm} \in \S_{\subbsQ^{\pm}}$]
\label{1.3A}
\mbox{ } \\ \vspace{-3ex}
\begin{romanlist}
\item 
\begin{equation}
\det{\bsQ^{\pm}} = 1.
\label{1.10a}
\end{equation}

\item 
Each $\E \in \S_{\E}$ has no more than one corresponding $\bsQ^{+} \in 
\S_{\subbsQ^{+}}$ and no more than one corresponding $\bsQ^{-} \in 
\S_{\subbsQ^{-}}$.

\item 
Corresponding to each $\E \in \S_{\E}$, $\bsQ^{-}$ exists if and only if $\bsQ^{+}$
exists, and, for each $(\x,\x',\tau) \in \dom{\bsQ^{+}}$,
\begin{equation}
\bsQ^{-}(\x,\x',\tau^{*}) = [\bsQ^{+}(\x,\x',\tau)]^{*}.
\end{equation}

\item 
For all $\x$, $\x''$, $\x' \in \domE$ and $\tau \in \bar{C}^{\pm}$,
\begin{equation}
\bsQ^{\pm}(\x,\x'',\tau)\bsQ^{\pm}(\x'',\x',\tau)=\bsQ^{\pm}(\x,\x',\tau)
\label{1.10c}
\end{equation}
and, as a special case,
\begin{equation}
[\bsQ^{\pm}(\x,\x',\tau)]^{-1} = \bsQ^{\pm}(\x',\x,\tau).
\label{1.10d}
\end{equation}

\item 
For all $\x$, $\x' \in \domE$, 
\begin{equation}
\begin{array}{r}
\bsQ^{+}(\x,\x',\sigma) =
\bsQ^{-}(\x,\x',\sigma) \\
\text{and is real if } \sigma \in R^{1} - \hat{\bsI}(\x,\x'),
\end{array}
\label{1.10f}
\end{equation}
while
\begin{equation}
\begin{array}{c}
-J \bsQ^{+}(\x,\x',\sigma) J = \bsQ^{-}(\x,\x',\sigma) \\
\text{ and is unitary if } 
\bsI^{(3)}(\x,\x') < \sigma < \bsI^{(4)}(\x,\x').
\end{array}
\label{1.10g}
\end{equation}

\item 
For all $\x, \x' \in \domE$,
\begin{equation}
\bsQ^{+}(\x,\x',\infty) = \bsQ^{-}(\x,\x',\infty) = e^{-\sigma_{3}\psi(\x)}
\left( \begin{array}{cc}
1 & -\chi(\x)+\chi(\x') \\ 0 & 1
\end{array} \right)
e^{\sigma_{3}\psi(\x')},
\label{1.10h}
\end{equation}
where $\psi := \frac{1}{2} \ln (-f)$.

\item 
The restriction of $\bsQ^{\pm}$ to $\{(\x,\x',\tau):\x,\x'\in\domE \text{ and }
\tau \in \bar{C}^{\pm} - \hat{\bsI}(\x,\x')\}$ is finite-valued and 
continuous, and, for fixed $(\x,\x')$, the function of $\tau$
given by $\bsQ^{\pm}(\x,\x',\tau)$ is holomorphic on $\bar{C}^{\pm} -
\hat{\bsI}(\x,\x')$ [which means that the function has a holomorphic
extension to at least one open subset of $C$].
\end{romanlist}
\end{proposition}
\end{abc}

\proofs
\begin{romanlist}
\item 
For each $(\x',\tau) \in (\dom{\E} \times \bar{C}^{\pm})$, Eqs.\ (\ref{1.9c}) 
and (\ref{1.8a}) imply 
$$d[\det{\bsQ^{\pm}}(\x,\x',\tau)] = 0$$
at all $\x \in \dom{\E}$ such that $(\tau-r)(\tau-s) \ne 0$.  If $\tau \in
C^{\pm}$ or if $\tau=\infty$, then $\det{\bsQ^{\pm}}(\x,\x',\tau)$
is independent of $\x$ throughout $\dom{\E}$.  So, from Eq.\ (\ref{1.9b}),
we obtain $\det{\bsQ^{\pm}}(\x,\x',\tau) = \det{\Q^{\pm}}(\x',\x',\tau) = 1$ 
at all $\x \in \dom{\E}$.

If $\tau=\sigma \in R^{1}$, the proof is less simple.  The statement
that $d[\det{\bsQ^{\pm}}(\x,\x',\sigma)] = 0$ at all $\x$ such that
$(\sigma-r)(\sigma-s) \ne 0$ implies that
$\det{\bsQ^{\pm}}(\x,\x',\sigma)$ retains the same value as $\x$
moves along any segmentally smooth path that lies in the interior
of any of the three regions in Fig.~2.  The null lines through
$(\sigma,\sigma)$ cannot be crossed by the path.  Therefore [for
fixed $(\x',\sigma)$ and $\pm$], there exist complex numbers
$N^{(\kappa)} \; (\kappa=1,-1,0)$ such that
$\det{\bsQ^{\pm}}(\x,\x',\sigma) = N^{(\kappa)}$ whenever $(\x,\sigma)$
has the index $\kappa$ as defined by Eq.\ (\ref{1.6h}).

However, we now consider the continuity condition (ii) in the definition
of $\bsQ^{\pm}$.  This condition implies that
$\det{\bsQ^{\pm}}(\x,\x',\tau)$ is a continuous function of $\x$
throughout $\dom{\E}$, and this is possible only if $N^{(1)} = 
N^{(-1)} = N^{(0)}$.  The rest of the proof is obvious.
\cheers

\item 
Since $d\bsQ^{\pm} = \Delta^{\pm} \bsQ^{\pm}$,
\begin{abc}
\begin{equation}
d(\bsQ^{\pm})^{-1} = - (\bsQ^{\pm})^{-1} \Delta^{\pm} .
\label{1.10i}
\end{equation}
Therefore, if $\bsQ^{\pm}$ and $\bsQ^{\prime\pm}$ both satisfy all of the
defining conditions of the $\bsQ^{\pm}$-potential for a given Ernst
potential $\E$, then $d\{[\bsQ^{\pm}(\x,\x',\tau)]^{-1}
\bsQ^{\prime\pm}(\x,\x',\tau)\} = 0$ for each $(\x',\tau)$ in 
$\dom{\E} \times \bar{C}^{\pm}$ and for all $\x \in \dom{\E}$ such that
$(\tau-r)(\tau-s) \ne 0$.  The remainder of the proof uses Eq.\ (\ref{1.9b})
and is similar to the proof just given of Prop.~\ref{1.3A}(i).  The conclusion
is that $[\bsQ^{\pm}(\x,\x',\tau)]^{-1} \bsQ^{\prime\pm}(\x,\x',\tau)
= I$ at all $\x \in \dom{\E}$.  So, $\bsQ^{\pm} = \bsQ^{\prime\pm}$.
\cheers

\item 
Apply Eq.\ (\ref{1.8b}) to Eq.\ (\ref{1.9c}).  Then use the uniqueness 
theorem, Prop.~\ref{1.3A}(ii), that we just proved.
\cheers

\item 
By (\ref{1.9c}) and (\ref{1.10i}), the differential of
$[\bsQ^{+}(\x,\x',\tau)]^{-1}$ $\bsQ^{+}(\x,\x'',\tau)$ $\bsQ^{+}(\x'',\x',\tau)$ 
vanishes for each $(\x'',\x',\tau) \in (\dom{\E})^{2} \times \bar{C}^{+}$ 
and at all $\x \in \dom{\E}$ such that $(\tau-r)(\tau-s) \ne 0$.  The rest
of the proof uses Eq.\ (\ref{1.9b}) and parallels the proofs of 
Prop.~\ref{1.3A}(i) and Prop.~\ref{1.3A}(ii).
\cheers

\item 
Note that $\kappa=1$, $-1$ and $0$ when $\bsI^{(4)}(\x,\x') < \sigma$,
$\sigma < \bsI^{(3)}(\x,\x')$ and 
$\bsI^{(3)}(\x,\x') < \sigma < \bsI^{(4)}(\x,\x')$,
respectively, and that the value of $\kappa$ does not change as one moves
along a straight line from $\x'$ to $\x$ in Fig.~2.  Therefore, application
of Eqs.\ (\ref{1.8c}) and (\ref{1.9b}) to Eq.\ (\ref{1.9c}) yields
\begin{equation}
J^{1-\kappa} \bsQ^{-}(\x,\x',\sigma) J^{\kappa-1} = \bsQ^{+}(\x,\x',\sigma).
\label{1.10j}
\end{equation}
The reality of $\bsQ^{\pm}(\x,\x',\sigma)$ when $|\kappa|=1$ and the
unitariness of $\bsQ^{\pm}(\x,\x',\sigma)$ when $\kappa=0$ then follow
from Eq.\ (\ref{1.10j}) and the fact that $M^{-1}=-J M^{T} J$ for any 
$2 \times 2$ matrix $M$ with unit determinant.
\cheers

\item 
Equations (\ref{1.4b}), (\ref{1.7b}) and (\ref{1.7c})
yield
\begin{equation}
\Delta^{\pm}(\x,\infty) = -d\psi \sigma_{3} 
- d\chi \left( \begin{array}{cc}
0 & 1 \\ 0 & 0
\end{array} \right) e^{-2\psi},
\end{equation}
whereupon the solution of Eqs.\ (\ref{1.9b}) and (\ref{1.9c})
at $\tau=\infty$ is Eq.\ (\ref{1.10h}).
\cheers
\end{abc}

\item 
For any given $\x, \x' \in \domE$,
\begin{abc}
\begin{equation}
L(\x,\x') := \{ \alpha \x + (1-\alpha) \x': 0 \le \alpha \le 1\}
\end{equation}
lies entirely in $\domE$ and is the straight line path from $\x'$
to $\x$.  [Here $\alpha \x + (1-\alpha) \x' := (\alpha r + (1-\alpha)r',
\alpha s + (1-\alpha) s')$.]  Letting $\Delta_{3}^{\pm}$ and 
$\Delta_{4}^{\pm}$ be defined by the equation
\begin{equation}
\Delta^{\pm}(\x,\tau) = \sum_{i} dx^{i} \Delta_{i}^{\pm}(\x,\tau)
\end{equation}
and letting\footnote{We shall underline the symbol for an auxiliary 
function that involves a parametrized curve.  The parameter will be
the first argument of the function, and that parameter will be in the
closed interval $[0,1]$.}
\begin{equation}
\u{\bsQ}^{\pm}(\alpha,\x,\x',\tau) := \bsQ^{\pm}(\alpha \x + (1-\alpha) \x',\x',\tau)
\label{1.10n}
\end{equation}
and
\begin{equation}
\u{\Delta}^{\pm}(\alpha,\x,\x',\tau) := \sum_{i}
(x^{i}-x^{\prime i}) \Delta_{i}^{\pm}(\alpha \x + (1-\alpha) \x',\tau),
\end{equation}
one finds from Eqs.\ (\ref{1.9b}) and (\ref{1.9c}) that
\begin{equation}
\u{\bsQ}^{\pm}(0,\x,\x',\tau) = I
\end{equation}
and that, on the interval $0 \le \alpha \le 1$,
\begin{equation}
\frac{\partial \u{\bsQ}^{\pm}(\alpha,\x,\x',\tau)}{\partial \alpha} =
\u{\Delta}^{\pm}(\alpha,\x,\x',\tau) \u{\bsQ}^{\pm}(\alpha,\x,\x',\tau).
\end{equation}
Moreover, from Eq.\ (\ref{1.10n}),
\begin{equation}
\u{\bsQ}^{\pm}(1,\x,\x',\tau) = \bsQ^{\pm}(\x,\x',\tau).
\label{1.10r}
\end{equation}
It is understood, of course, that
\begin{equation}
\dom{\u{\bsQ}^{\pm}} = \dom{\u{\Delta}^{\pm}} := \{(\alpha,\x,\x',\tau):0 \le \alpha \le 1,
(\x,\x') \in \domE^{2} \text{ and } \tau \in \bar{C}^{\pm} 
- \hat{\bsI}(\x,\x')\}.
\label{1.10s}
\end{equation}
Since $\tau \notin \hat{\bsI}(\x,\x')$, it is easy to show no
singularity is encountered by $\u{\Delta}^{\pm}(\alpha,\x,\x',\tau)$ on the
interval $0 \le \alpha \le 1$.  In fact, $\u{\Delta}^{\pm}$ is continuous
throughout its domain (\ref{1.10s}), and, for fixed $(\alpha,\x,\x')$,
the function of $\tau$ that is given by $\u{\Delta}^{\pm}(\alpha,\x,\x',\tau)$
is holomorphic on $\bar{C}^{\pm} - \hat{\bsI}(\x,\x')$.  Therefore, from
two standard theorems\footnote{The standard theorems to which we here
refer are discussed in {\em Differential Equations: Geometric Theory}, by
S.\ Lefschetz, 2nd Ed., (Dover, NY, 1977), Ch.\ II and III. \label{Lefsch}}
on ordinary differential equations, $\u{\bsQ}^{\pm}$ is continuous [as well as 
$\bC^{1}$ in the variable $\alpha$] throughout the domain (\ref{1.10s}), and,
for fixed $(\alpha,\x,\x')$, the function of $\tau$ given by
$\u{\bsQ}^{\pm}(\alpha,\x,\x',\tau)$ is holomorphic on $\bar{C}^{\pm} - 
\hat{\bsI}(\x,\x')$.  The conclusion of the theorem now follows from
Eq.\ (\ref{1.10r}).
\cheers
\end{abc}
\end{romanlist}

We still have to prove that, for every given $\E \in \S_{\E}$, the 
corresponding $\bsQ^{\pm} \in \S_{\subbsQ^{\pm}}$ exists.  This will be done
in Sec.~\ref{Sec_3} of these notes, where we shall also establish that
$\bsQ^{\pm}$ is continuous and has certain differentiability properties.


\setcounter{theorem}{0}
\setcounter{equation}{0}
\subsection{The K--C linear system and the sets $\S_{\subbsF^{\pm}}\triple$} 

In many of our earlier papers we have described the K--C linear system
at great length.  
\begin{abc}
\begin{definition}{Dfns.\ of $\omega$, $A$ and $h$\label{def12}}
Corresponding to each $\E = f+i\chi \in \S_{\E}$, let $\omega$ denote
the real-valued function such that $\dom{\omega} := \domE$, $d\omega$
exists,
\begin{equation}
d\omega := \rho f^{-2} \star d\chi
\label{1.11a}
\end{equation}
and the gauge condition $\omega(\x_{0}) = 0$ holds, and $A$ and $h$
denote the real $2 \times 2$ matrix functions
\begin{equation}
A := \left( \begin{array}{cc}
1 & \omega \\ 0 & 1
\end{array} \right) \left( \begin{array}{cc}
1/\sqrt{-f} & 0 \\ 0 & \sqrt{-f}
\end{array} \right)
\label{1.11b}
\end{equation}
and
\begin{equation}
h := A \left( \begin{array}{cc}
\rho^{2} & 0 \\ 0 & 1
\end{array} \right) A^{T}.
\label{1.11c}
\end{equation}
\end{definition}
\end{abc}

In the spacetime determined by $\E$,
\begin{abc}
\begin{equation}
h = \left( \begin{array}{cc}
g_{11} & g_{12} \\ g_{21} & g_{22}
\end{array} \right),
\label{1.11d}
\end{equation}
where $g_{ab} = \bpartial_{a} \cdot \bpartial_{b}$,
and $\bpartial_{1}$ and $\bpartial_{2}$ are
two commuting spacelike Killing vectors.  In our gauge, note that 
\begin{equation}
A(\x_{0}) = I, \quad h(\x_{0}) = \left( \begin{array}{cc}
\rho_{0}^{2} & 0 \\ 0 & 1
\end{array} \right),
\label{1.11e}
\end{equation}
where $\rho_{0} := \frac{1}{2}(s_{0} - r_{0})$.
\end{abc}

\begin{abc}
\begin{definition}{Dfn.\ of $\S_{H}\triple$\label{def13}}
Let $\S_{H}\triple$ denote the set of all $2 \times 2$ matrix
functions\footnote{W.~Kinnersley, {\em Symmetries of the stationary
Einstein-Maxwell field equations}, J.\ Math.\ Phys.\ {\bf 18}, 1529--1537
(1977).  See also Ref.~\ref{HHP}.} $H$ with $\dom{H} := \domE$ such that 
there exists $\E \in \S_{\E}$ for which
\begin{equation}
\Re{H} = -h,
\label{1.12a}
\end{equation}
$d(\Im{H})$ exists and satisfies
\begin{equation}
\rho d(\Im{H}) = - h J \star dh,
\label{1.12b}
\end{equation}
and the gauge condition
\begin{equation}
H(\x_{0}) = - \left( \begin{array}{cc}
\rho_{0}^{2} & 0 \\ 2iz_{0} & 1
\end{array} \right)
\label{1.12c}
\end{equation}
holds. 
\end{definition}
\end{abc}

\begin{proposition}[Properties of $H \in \S_{H}$ (for given $\E \in \S_{\E}$)]
\label{1.1B}
\mbox{ } \\
The Ernst equation (\ref{G1.5}), as is well known, guarantees that the right sides
of Eqs.\ (\ref{1.11a}) and (\ref{1.11b}) are closed $1$-forms.  Hence, there
exists exactly one $H \in \S_{H}$ corresponding to each $\E \in \S_{\E}$;
and $dH$ and $\partial^{2}H/\partial r \partial s$ exist and are
continuous throughout $\domE$.  Moreover,
\begin{equation}
H_{22} = \E, \quad H-H^{T} = 2z\Omega,
\label{1.12d}
\end{equation}
where $\Omega := iJ$ and the matrices $H_{r}\Omega$ and $H_{s}\Omega$
are idempotent.\footnote{See Ref.~\ref{HHP}.}
\end{proposition}

\begin{definition}{Dfn.\ of $\Gamma$\label{def14}}
Corresponding to each $H \in \S_{H}$, let $\Gamma$ denote the $2 \times 2$
matrix $1$-form whose domain is $\{(\x.\tau):\x \in \domE \text{ and }
\tau \in C - \{r,s\}\}$ and whose values are given by
\begin{equation}
\Gamma(\x,\tau) := \frac{1}{2} (\tau-z+\rho\star)^{-1} dH \Omega.
\label{1.13a}
\end{equation}
\end{definition}
The function $\Gamma$ satisfies\footnote{See Ref.~\ref{HHP}.}
\begin{equation}
d\Gamma - \Gamma \Gamma = 0,
\label{1.13b}
\end{equation}
which is the integrability condition for a linear system that will be
considered in Secs.~\ref{Sec_1}C.

The conventional Minkowski space member of $\S_{\E}$ is denoted by $\E^{M}$
and has the values
\begin{abc}
\begin{equation}
\E^{M}(\x) := -1 \text{ for all } \x \in \domE.
\label{1.14a}
\end{equation}
The corresponding potentials $A$ and $h$ have the values
\begin{equation}
A^{M}(\x) = I, \quad h^{M}(\x) = \left( \begin{array}{cc}
\rho^{2} & 0 \\ 0 & 1
\end{array} \right),
\label{1.14b}
\end{equation}
for all $\x \in \domE$, and the corresponding member of $\S_{H}$ has the values
\begin{equation}
H^{M}(\x) = - \left( \begin{array}{cc}
\rho^{2} & 0 \\ 2iz & 1
\end{array} \right)
\label{1.14c}
\end{equation}
for all $\x \in \domE$.

{}Eqs.\ (\ref{1.14a}) and (\ref{1.4b}), $\Delta^{\pm} = \left( \begin{array}{cc}
0 & 0 \\ 0 & 0
\end{array} \right)$ when $\E=\E^{M}$.  Therefore, from Eqs.\ (\ref{1.9b})
and (\ref{1.9c}), the member of $\S_{\subbsQ^{\pm}}$ that corresponds 
to $\E^{M}$ has the value
\begin{equation}
\bsQ^{M\pm}(\x,\x',\tau) = I
\label{1.15}
\end{equation}
at all $(\x,\x',\tau)$ in $\domE^{2} \times \bar{C}^{\pm}$.

The matrix $1$-form $\Gamma$ for which $H = H^{M}$ will be denoted by
$\Gamma^{M}$.  Useful solutions of the linear equations
\begin{equation}
dP^{M\pm} = \Gamma^{M} P^{M\pm} \text{ and }
dP^{M} = \Gamma^{M} P^{M}
\label{1.16a}
\end{equation}
such that
\begin{equation}
\dom{P^{M\pm}} := \{(\x,\tau) \in \domE \times \bar{C}^{\pm}: \tau \notin
\{r,s\}\}
\label{1.16b}
\end{equation}
and
\begin{equation}
\dom{P^{M}} := \{(\x,\tau) \in \domE \times C: \tau \notin [r,s]\}
\label{1.16c}
\end{equation}
are given by
\begin{equation}
P^{M\pm}(\x,\tau) = \left( \begin{array}{cc}
1 & -i (\tau-z) \mu^{\pm}(\x,\tau)^{-1} \\
0 & \mu^{\pm}(\x,\tau)^{-1}
\end{array} \right) \frac{1}{\sqrt{2}} \left( \begin{array}{cc}
1 & i \\ -i & -1
\end{array} \right)
\label{1.16d}
\end{equation}
and $P^{M}(\x,\tau)$ equals the same expression as above except that
$\mu^{\pm}$ is replaced by $\mu$.  From Eq.\ (\ref{1.6j}),
\begin{equation}
P^{M}(\x,\tau) = P^{M\pm}(\x,\tau) \text{ for all }
\x \in \domE \text{ and } \tau \in \bar{C}^{\pm} - [r,s].
\label{1.16e}
\end{equation}
When $\tau \ne \infty$, Eq.\ (\ref{1.16d}) has the convenient factorized
form
\begin{equation}
P^{M\pm}(\x,\tau) = \left( \begin{array}{cc}
1 & -i(\tau-z) \\ 0 & 1
\end{array} \right) \left( \begin{array}{cc}
1 & 0 \\ 0 & \mu^{\pm}(\x,\tau)^{-1}
\end{array} \right) \frac{1}{\sqrt{2}} (\sigma_{3}+\Omega).
\label{1.16f}
\end{equation}
\end{abc}

\begin{proposition}[Properties of $P^{M\pm}$ and $P^{M}$]
\label{1.1C}
\mbox{ } \\ \vspace{-3ex}
\begin{romanlist}
\item 
The functions $P^{M\pm}$ and $P^{M}$ are finite-valued and continuous
throughout their respective domains.

\item 
For fixed $\x$, $P^{M\pm}(\x,\tau)$ is holomorphic on $\bar{C}^{\pm}
- \{r,s\}$ [which means that the function of $\tau$ whose domain is
$\bar{C}^{\pm} - \{r,s\}$ and whose values on that domain are
$P^{M\pm}(\x,\tau)$ has a holomorphic extension to at least one open
subset of $C$].  For fixed $\x$, $P^{M}(\x,\tau)$ is a holomorphic
function of $\tau$ throughout $C - [r,s]$.

\item 
Recall the definition of the closed disk $d(\x)$, Eq.\ (\ref{d1}).
\begin{abc}
\begin{equation}
F^{M\pm}(\x,\tau) := P^{M\pm}(\x,\tau) \frac{1}{\sqrt{2}}
\left( \begin{array}{cc}
2\tau & 0 \\ 0 & 1
\end{array} \right)
\label{1.17a}
\end{equation}
and its inverse are finite-valued and continuous functions of $(\x,\tau)$
throughout $\{(\x,\tau):\x \in \domE \text{ and } \tau \in \bar{C}^{\pm}
- d(\x)\}$, and, for fixed $\x$, they are holomorphic functions of $\tau$
on $\bar{C}^{\pm} - d(\x)$.  Also,
\begin{equation}
F^{M\pm}(\x,\infty) = \Omega.
\label{1.17b}
\end{equation}

\item 
Recall the definition of the closed disk $d(\x,\x')$, Eq.\ (\ref{d2}).
Let $\bsF^{M\pm}$ denote the function whose domain is
\begin{equation}
\dom{\bsF^{M\pm}} := \{(\x,\x',\tau) \in \domE^{2} \times \bar{C}^{\pm}:
	\tau \notin \{r,r',s,s'\}\}
\label{1.18a}
\end{equation}
and whose values are given by
\begin{equation}
\bsF^{M\pm}(\x,\x',\tau) :=
P^{M\pm}(\x,\tau) [P^{M\pm}(\x',\tau)]^{-1} \text{ when } \tau \ne \infty
\label{1.18b}
\end{equation}
and by
\begin{equation}
\bsF^{M\pm}(\x,\x',\tau) :=
F^{M\pm}(\x,\tau) [F^{M\pm}(\x',\tau)]^{-1} \text{ when } \tau \in
\bar{C}^{\pm} - d(\x,\x').
\label{1.18c}
\end{equation}
Then $\bsF^{M\pm}$ is finite-valued and
continuous; and, for all $(\x,\x',\tau)$ in the above domain (\ref{1.18a}),
\begin{equation}
\bsF^{M\pm}(\x,\x',\infty) = I
\label{1.18d}
\end{equation}
and, if $\tau \ne \infty$,
\begin{equation}
\bsF^{M\pm}(\x,\x',\tau) = \left( \begin{array}{cc}
1 & -i(\tau-z) \\ 0 & 1
\end{array} \right) \left( \begin{array}{cc}
1 & 0 \\ 0 & \bnu^{\pm}(\x,\x',\tau)
\end{array} \right) \left( \begin{array}{cc}
1 & i(\tau-z') \\ 0 & 1
\end{array} \right),
\label{1.18e}
\end{equation}
where $\bnu^{\pm}$ is defined by Eq.\ (\ref{1.6p}).  For each $(\x,\x')
\in \domE^{2}$, $\bsF^{M\pm}(\x,\x',\tau)$ is a holomorphic function of $\tau$
on $\bar{C}^{\pm} - \{r,s,r',s'\}$.

\item 
For each $\x \in \domE$ and $\sigma \in R^{1} - \{r,s\}$,
\begin{equation}
P^{M-}(\x,\sigma) = P^{M+}(\x,\sigma) \Omega^{1-\kappa},
\label{1.19a}
\end{equation}
where $\kappa$ is defined by Eq.\ (\ref{1.6h}).  For each $(\x,\x') \in
\domE^{2}$,
\begin{equation}
\bsF^{M-}(\x,\x',\sigma) = \bsF^{M+}(\x,\x',\sigma) \text{ for all }
\sigma \in R^{1} - \grave{\bsI}(\x,\x'). 
\label{1.19b}
\end{equation}

\item 
Let $\grave{\bsF}^{M}$ denote the function with the domain
\begin{equation}
\dom{\grave{\bsF}^{M}} := \{(\x,\x',\tau):(\x,\x') \in \domE^{2} \text{ and }
\tau \in C - \grave{\bsI}(\x,\x')\}
\label{1.20a}
\end{equation}
and with the values
\begin{equation}
\grave{\bsF}^{M}(\x,\x',\tau) := \bsF^{M\pm}(\x,\x',\tau) \text{ for all }
(\x,\x',\tau) \in \dom{\grave{\bsF}^{M}} \text{ such that } \tau \in \bar{C}^{\pm},
\label{1.20b}
\end{equation}
where we are employing Eq.\ (\ref{1.19b}) when $\tau \in R^{1}$ and
Eq.\ (\ref{1.17b}) when $\tau=\infty$.  Then $\grave{\bsF}^{M}$ is finite-valued 
and continuous, $\grave{\bsF}^{M}(\x,\x',\tau)$ is a holomorphic function of
$\tau$ throughout $C - \bar{\bsI}(\x,\x')$,
\begin{equation}
\grave{\bsF}^{M}(\x,\x',\infty) = I
\label{1.20c}
\end{equation}
and, when $\tau \ne \infty$, $\grave{\bsF}^{M}(\x,\x',\tau)$ is given by the
expression (\ref{1.18e}) after $\bnu^{\pm}(\x,\x',\tau)$ is replaced
by $\grave{\bnu}(\x,\x',\tau)$ as defined by Eq.\ (\ref{1.6p}).  Let $\bar{\bsF}^{M}$
denote the extension of $\grave{\bsF}^{M}$ to
\begin{equation}
\dom{\bar{\bsF}^{M}} := \{(\x,\x',\tau):(\x,\x')\in\domE^{2} \text{ and }
\tau \in C - \bar{\bsI}(\x,\x')\}
\label{1.20d}
\end{equation}
such that, when $\tau \ne \infty$, $\bar{\bsF}^{M}(\x,\x'_{0},\tau)$ is
given by Eq.\ (\ref{1.18e}) with $\bnu^{\pm}(\x,\x',\tau)$
replaced by $\bar{\bnu}(\x,\x',\tau)$ as defined by Eq.\ (\ref{1.6x}).
Then, $\bar{\bsF}^{M}(\x,\x',\tau)$ is a holomorphic function of $\tau$
throughout $C - \bar{\bsI}(\x,\x')$.
\end{abc}
\end{romanlist}
\end{proposition}

\proofs
\begin{romanlist}
\item 
This follows from the facts that $\tau \mu^{\pm}(\x,\tau)^{-1}$ and
$\tau \mu(\x,\tau)^{-1}$ are finite-valued and continuous functions
of $(\x,\tau)$ throughout the domains (\ref{1.16b}) and (\ref{1.16c}),
respectively.
\cheers

\item 
Moreover, for fixed $\x$, $\tau \mu^{\pm}(\x,\tau)^{-1}$ is a
holomorphic function of $\tau$ on $\bar{C}^{\pm} - \{r,s\}$, and
$\tau \mu(\x,\tau)^{-1}$ is a holomorphic function of $\tau$ 
throughout $C - [r,s]$.
\cheers

\item 
After substituting (\ref{1.16d}) into (\ref{1.17a}) and then
multiplying the four factors to form a single $2 \times 2$ matrix,
it can be seen that $F^{M\pm}(\x,\tau)$ and its inverse are both
expressible as series in inverse powers of $2\tau$ such that these
series converge throughout the open neighborhood $\bar{C}^{\pm} - d(\x)$
of $\infty$.  The remainder of the proof is left for the reader.
\cheers

\item 
The proof employs the preceding theorem, Prop.~\ref{1.1C}(iii), and the
properties of $\bnu^{\pm}$ that are detailed in Sec.~\ref{Sec_1}B.  Substitution
from Eq.\ (\ref{1.16d}) into the product (\ref{1.18b}) yields the
product (\ref{1.18e}), which (after multiplying the three factors to
form a single $2 \times 2$ matrix) is easily shown to have a unique
continuous extension to the domain (\ref{1.18a}) such that (\ref{1.18d})
and the conclusion respecting holomorphy hold.
\cheers

\item 
Equation (\ref{1.19a}) follows from Eqs.\ (\ref{1.16d}) and (\ref{1.6i}).
Equation (\ref{1.19b}) then follows from Eqs.\ (\ref{1.18e}), 
(\ref{1.6i}) and (\ref{1.6y}).
\cheers

\item 
{}From Eqs.\ (\ref{1.20b}), (\ref{1.18d}), (\ref{1.18e}), (\ref{1.6u}) and
(\ref{1.6x}), we obtain Eq.\ (\ref{1.20c}) and Eq.\ (\ref{1.18e}) with
$\bnu^{\pm}$ replaced by $\grave{\bnu}$.  The remainder of the proof employs the
properties of $\grave{\bnu}$ and $\bar{\bnu}$ that are detailed in
Sec.~\ref{Sec_1}B.
\cheers
\end{romanlist}

\begin{abc}
\begin{definition}{Dfn. of $\bvarphi^{\pm}:\S_{\subbsQ^{\pm}} \rightarrow 
\S_{\subbsF^{\pm}}$\label{def15}}
Let $\bvarphi^{+}$ and $\bvarphi^{-}$ denote the mappings with the
domains $\S_{\subbsQ^{+}}$ and $\S_{\subbsQ^{-}}$, respectively, where the value
$\bvarphi^{\pm}(\bsQ^{\pm})$ corresponding to $\bsQ^{\pm} \in \S_{\subbsQ^{\pm}}$ 
is the $2 \times 2$ matrix function $\bsF^{\pm}$ with the domain
\begin{equation}
\dom{\bsF^{\pm}} := \{(\x,\x',\tau) \in \domE^{2} \times \bar{C}^{\pm}:
\tau \notin \{r,s,r',s'\}\}
\label{1.21a}
\end{equation}
such that, for each $(\x,\x',\tau) \in \dom{\bsF^{\pm}}$,
\begin{eqnarray}
\bsF^{\pm}(\x,\x',\tau) & := & A(\x) P^{M\pm}(\x,\tau) \bsQ^{\pm}(\x,\x',\tau)
[P^{M\pm}(\x',\tau)]^{-1} [A(\x')]^{-1} \nonumber \\
& & \text{ when } \tau \ne \infty,
\label{1.21b}
\end{eqnarray}
where $A$ is defined by Eq.\ (\ref{1.11b}) and is computed from the member
of $\S_{\E}$ that corresponds to $\bsQ^{\pm}$,
and
\begin{eqnarray}
\bsF^{\pm}(\x,\x',\tau) & := & A(\x) F^{M\pm}(\x,\tau)
\left( \begin{array}{cc}
(2\tau)^{-1} & 0 \\ 0 & 1
\end{array} \right) \bsQ^{\pm}(\x,\x',\tau)
\left( \begin{array}{cc}
2\tau & 0 \\ 0 & 1
\end{array} \right) \nonumber \\ & &
[F^{M\pm}(\x',\tau)]^{-1} [A(\x')]^{-1}
\text{ when } \tau \in \bar{C}^{\pm} - d(\x,\x').
\label{1.21c}
\end{eqnarray}
\end{definition}
\end{abc}

The above equations (\ref{1.21b}) and (\ref{1.21c}) are mutually consistent
as can be seen from Prop.\ \ref{1.1C}(iii) and the following proposition, 
which is evident by inspection of Props.\ \ref{1.3A}(vi) and~(vii):
\begin{equation}
\begin{array}{l}
\left( \begin{array}{cc}
(2\tau)^{-1} & 0 \\ 0 & 1
\end{array} \right) \bsQ^{\pm}(\x,\x',\tau) \left( \begin{array}{cc}
2\tau & 0 \\ 0 & 1 
\end{array} \right) \text{ is a finite-valued and continuous function of } \\
(\x,\x',\tau) \text{ throughout } \{(\x,\x',\tau) \in \domE^{2} \times \bar{C}^{\pm}:
\tau \notin \hat{\bsI}(\x,\x'), \tau \ne 0\}, \text{ and, } \\
\text{ for fixed $(\x,\x')$, is a holomorphic function of $\tau$ on } 
\bar{C}^{\pm} - [\hat{\bsI}(\x,\x') \cup \{0\}].
\end{array}
\label{1.21e}
\end{equation}

\begin{definition}{Dfn.\ of $\S_{\subbsF^{\pm}}\triple$\label{def16}}
\begin{equation}
\S_{\subbsF^{\pm}} := \{ \bvarphi^{\pm}(\bsQ^{\pm}):\bsQ^{\pm}\in\S_{\subbsQ^{\pm}}\}.
\end{equation}
\end{definition}

\begin{theorem}[Properties of $\bsF^{\pm} \in \S_{\subbsF^{\pm}}$]
\label{1.1D}
\mbox{ } \\ \vspace{-3ex}
\begin{romanlist}
\item 
$\bsF^{M\pm} = \bvarphi^{\pm}(\bsQ^{M\pm}) \in \S_{\subbsF^{\pm}}$.

\item 
At each $(\x,\x',\tau) \in \dom{\bsF^{\pm}}$, $d\bsF^{\pm}(\x,\x',\tau)$
exists and
\begin{abc}
\begin{equation}
d\bsF^{\pm}(\x,\x',\tau) = \Gamma(\x,\tau) \bsF^{\pm}(\x,\x',\tau).
\label{1.22}
\end{equation}

\item 
For all $(\x,\x',\tau) \in \dom{\bsF^{\pm}}$,
\begin{equation}
\det{\bsF^{\pm}(\x,\x',\tau)} = \bnu^{\pm}(\x,\x',\tau).
\label{1.23}
\end{equation}

\item 
For all $\x, \x'', \x' \in \domE$ and $\tau \in \bar{C}^{\pm} - 
\{r,s,r'',s'',r',s'\}$,
\begin{equation}
\bsF^{\pm}(\x,\x'',\tau) \bsF^{\pm}(\x'',\x',\tau) = \bsF^{\pm}(\x,\x',\tau). 
\end{equation}
For each $\x \in \domE$ and $\tau \in \bar{C}^{\pm} - \{r,s\}$,
\begin{equation}
\bsF^{\pm}(\x,\x,\tau) = I. 
\label{1.24b} 
\end{equation}
For all $\x,\x'' \in \domE$ and $\tau \in \bar{C}^{\pm} - \{r,s,r'',s''\}$,
\begin{equation}
\left[\bsF^{\pm}(\x,\x'',\tau)\right]^{-1} = \bsF^{\pm}(\x'',\x,\tau).
\end{equation}

\item 
The restriction of $\bsF^{\pm}$ to $\{(\x,\x',\tau) \in \dom{\bsF^{\pm}}:
\tau \notin \grave{\bsI}(\x,\x')\}$ is finite-valued and continuous.  For
fixed $(\x,\x')$, the function of $\tau$ that is given by 
$\bsF^{\pm}(\x,\x',\tau)$ is holomorphic on
$\bar{C}^{\pm} - \grave{\bsI}(\x,\x')$.

\item 
For all $\x,\x' \in \domE$,
\begin{equation}
\bsF^{+}(\x,\x',\infty) = \bsF^{-}(\x,\x',\infty) = I.
\label{1.25}
\end{equation}

\item 
Let
\begin{equation}
\A(\x,\tau) := (\tau-z) \Omega + \Omega h(\x) \Omega.
\label{1.26a}
\end{equation}
Then, for all $(\x,\x',\tau) \in \dom{\bsF^{\pm}}$ such that $\tau \ne \infty$,
\begin{equation}
[\bsF^{\mp}(\x,\x',\tau^{*})]^{\dagger} \A(\x,\tau) \bsF^{\pm}(\x,\x',\tau)
= \A(\x',\tau)
\end{equation}
and, in view of the preceding Thm.~\ref{1.1D}(vi), the above equation 
divided through by $\tau$ holds at $\tau = \infty$.

\item 
For all $(\x,\x') \in \domE^{2}$ and $\sigma \in R^{1}-\grave{\bsI}(\x,\x')$,
\begin{equation}
\bsF^{+}(\x,\x',\sigma) = \bsF^{-}(\x,\x',\sigma).
\label{1.27}
\end{equation}
\end{abc}
\end{romanlist}
\end{theorem}

\proofs
\begin{romanlist}
\item 
This follows from Eq.\ (\ref{1.14b}), Eq.\ (\ref{1.15}) and Prop.~\ref{1.1C}(iv).
\cheers

\item 
For the given member $\bsF^{\pm} = \bvarphi^{\pm}(\bsQ^{\pm})$ of the set
$\S_{\subbsF^{\pm}}$, there exists $\E \in \S_{\E}$ such that (by definition of 
$\S_{\subbsQ^{\pm}}$)
Eq.\ (\ref{1.9c}) holds, and it is a simple exercise to prove that this
member of $\S_{\E}$ is uniquely determined by $\bsQ^{\pm}$.  From $\E$, one can
compute the corresponding $H \in \S_{H}$ by using the definition given in
Sec.~\ref{Sec_1}C.  It is well known\footnote{We shall give our own proof of this
statement in Sec.~\ref{Sec_3}.} that $H$ exists, is unique and (like $\E$) is 
$\bC^{1}$ and that the mixed partial derivative $d\star dH$ exists and
is continuous throughout $\domE$.  From Eqs.\ (\ref{1.12a}) and (\ref{1.12b}),
\begin{abc}
\begin{equation}
dH = - \left( I + \rho^{-1} h \Omega \star \right) dh,
\label{1.28a}
\end{equation}
where recall that $\Omega := iJ$.  From Eqs.\ (\ref{1.11a}) and (\ref{1.11b}),
\begin{equation}
A^{-1} dA = - \left( \begin{array}{cc}
0 & 1 \\ 0 & 0
\end{array} \right)
\frac{\rho \star d\chi}{f} - \sigma_{3} \frac{df}{2f}.
\label{1.28b}
\end{equation}
Substitute the relations (\ref{1.11c}) and (\ref{1.14b}) into Eq.\
(\ref{1.28a}) and then use Eqs.\ (\ref{1.28a}) (for $H^{M}$) and
(\ref{1.28b}) [and the identity $M^{T} \Omega M = \Omega (\det{M})$ for
any $2 \times 2$ matrix $M$] to obtain
\begin{equation}
A^{-1} dH \Omega A = dH^{M} \Omega - \frac{d\E}{f} \left( I \rho \star
+ h^{M} \Omega \right) \sigma_{3}.
\label{1.28c}
\end{equation}
Use Eq.\ (\ref{1.16d}) to compute $P^{M\pm}\sigma_{3}-\sigma_{3}P^{M\pm}$
and $P^{M\pm}J-J P^{M\pm}$.  Then, after regrouping terms, one obtains
from Eqs.\ (\ref{1.4b}), (\ref{1.13a}), (\ref{1.28b}) and (\ref{1.28c}),
\begin{equation}
P^{M\pm}\Delta^{\pm} = \left( A^{-1} \Gamma A - \Gamma^{M} - A^{-1} dA
\right) P^{M\pm}.
\end{equation}
Therefore, from Eqs.\ (\ref{1.9c}) and (\ref{1.16a}),
$$
d(A P^{M\pm} \bsQ^{\pm}) = \Gamma (A P^{M\pm} \bsQ^{\pm})
$$
or, more explicitly,
\begin{equation}
\begin{array}{c}
d[A(\x)P^{M\pm}(\x,\tau) \bsQ^{\pm}(\x,\x',\tau)] = \Gamma(\x,\tau)
\left[ A(\x) P^{M\pm}(\x,\tau) \bsQ^{\pm}(\x,\x',\tau) \right] \\
\text{ for all } (\x,\x') \in \domE^{2} \text{ and }
\tau \in \bar{C}^{\pm} - \{r,s\}.
\end{array}
\label{1.28e}
\end{equation}
Also, from Prop.~\ref{1.1C}(iii),
\begin{equation}
\begin{array}{c}
d\left[A(\x) F^{M\pm}(\x,\tau) \sqrt{2} \left( \begin{array}{cc}
(2\tau)^{-1} & 0 \\ 0 & 1 
\end{array} \right) \bsQ^{\pm}(\x,\x',\tau) \right] \nonumber \\
= \Gamma(\x,\tau) \left[ A(\x) F^{M\pm}(\x,\tau) \sqrt{2}
\left( \begin{array}{cc}
(2\tau)^{-1} & 0 \\ 0 & 1
\end{array} \right) \bsQ^{\pm}(\x,\x',\tau) \right] \\
\text{for all } (\x,\x') \in \domE^{2} \text{ and }
\tau \in \bar{C}^{\pm} - d(\x).
\end{array}
\label{1.28f}
\end{equation}
The conclusion (\ref{1.22}) now follows from Eqs.\ (\ref{1.21b}),
(\ref{1.21c}), (\ref{1.28e}) and (\ref{1.28f}).
\cheers

\item 
{}From Eqs.\ (\ref{1.11b}), (\ref{1.16d}) and (\ref{1.17a}),
\begin{equation}
\det{A(\x)}=1, \quad \det{P^{M\pm}(\x,\tau)}=-[\mu^{\pm}(\x,\tau)]^{-1}
\text{ and } \det{F^{M\pm}(\x,\tau)}=-\tau[\mu^{\pm}(\x,\tau)]^{-1}.
\label{1.29}
\end{equation}
Hence, from Eqs.\ (\ref{1.10a}), (\ref{1.21b}), (\ref{1.21c})
and (\ref{1.6y}), we find $\det{\bsF^{\pm}(\x,\x',\tau)} = 
\bnu^{\pm}(\x,\x',\tau)$ for all $(\x,\x',\tau) \in \dom{\bsF^{\pm}}$.
\cheers

\item 
These follow from Prop.~\ref{1.3A}(iv), Eqs.\ (\ref{1.21b}), (\ref{1.21c})
and (\ref{1.23}) [which implies that $\bsF^{\pm}(\x,\x'',\tau)$ is
invertible].
\cheers

\item 
The proof employs Eqs.\ (\ref{1.22}) and (\ref{1.24b}) as well as 
Thms.~\ref{1.1D}(ii) and (iv).  The proof is an exact parallel
of the proof of Prop.~\ref{1.3A}(vii).  To obtain the proof of the present
theorem, simply make the following replacements in the proof of 
Prop.~\ref{1.3A}(vii):  $\Delta^{\pm} \rightarrow \Gamma$, $\bsQ^{\pm} 
\rightarrow \bsF^{\pm}$ and $\hat{\bsI}(\x,\x') \rightarrow \grave{\bsI}(\x,\x')$.
No other replacements are needed.  [Note that the present theorem holds
even when $\tau$ is real and $\grave{\bsI}^{(3)}(\x,\x') < \tau < 
\grave{\bsI}^{(4)}(\x,\x')$.]
\cheers

\item 
{}From the preceding theorem, Thm.~\ref{1.1D}(v), $\bsF^{\pm}(\x,\x',\infty)$
is finite and is a continuous function of $(\x,\x')$ throughout $\domE$.
{}From Eq.\ (\ref{1.13a}), $\Gamma(\x,\infty) = \left( \begin{array}{cc}
0 & 0 \\ 0 & 0
\end{array} \right)$.  Therefore, from Eqs.\ (\ref{1.22}) and (\ref{1.24b}),
$\bsF^{\pm}(\x,\x',\infty)=I$ throughout $\domE$.
\cheers

\item 
First employ Eqs.\ (\ref{1.16d}), (\ref{1.6e}), (\ref{1.6f}), (\ref{1.11c})
and (\ref{1.14b}), and the fact that $A^{T} \Omega A = I$, to prove that
\begin{equation}
[P^{M\mp}(\x,\tau^{*})]^{\dagger} [A(\x)]^{T} \A(\x,\tau) A(\x)
P^{M\pm}(\x,\tau) = \Omega \text{ if } \tau \ne \infty.
\end{equation}
Therefore, from Props.~\ref{1.3A}(i) and~(iii),
\begin{equation}
[A(\x) P^{M\mp}(\x,\tau^{*}) \bsQ^{\mp}(\x,\x',\tau^{*})]^{\dagger}
\A(\x,\tau) [A(\x) P^{M\pm}(\x,\tau) \bsQ^{\pm}(x,\x',\tau)] = \Omega
\text{ if } \tau \ne \infty.
\end{equation}
Therefore, from Eqs.\ (\ref{1.21b}, (\ref{1.6e}), (\ref{1.6f}) and
(\ref{1.11c}),
\begin{equation}
\begin{array}{r}
[\bsF^{\mp}(\x,\x',\tau^{*}]^{\dagger} \A(\x,\tau) \bsF^{\pm}(\x,\x',\tau)
A(\x')^{T} 
\left( \begin{array}{cc}
1 & 0 \\ -i(\tau-z') & \mu^{\pm}(\x',\tau)
\end{array} \right) \sigma_{3} \\
\left( \begin{array}{cc}
1 & i(\tau-z') \\ 0 & \mu^{\pm}(\x',\tau)
\end{array} \right) A(\x') = \A(\x',\tau)
\end{array}
\end{equation}
for all $(\x,\x',\tau) \in \dom{\bsF^{\pm}}$ such that $\tau \ne \infty$.
\cheers

\item 
Apply Eq.\ (\ref{1.6h}), Prop.~\ref{1.3A}(v) [Eqs.\ (\ref{1.10f}) and
(\ref{1.10g})] and Prop.~\ref{1.1C} [Eqs.\ (\ref{1.19a}) to Eq.\ (\ref{1.21b})].
\cheers
\end{abc}
\end{romanlist}

\begin{proposition}[More properties of $\bsQ^{\pm} \in \S_{\subbsQ^{\pm}}$] 
\label{1.2D}
\mbox{ } \\
The restriction of $\bsQ^{\pm}$ to $\{(\x,\x',\tau):\x,\x'\in\domE \text{ and } 
\tau \in \bar{C}^{\pm} - \grave{\bsI}(\x,\x')\}$ is finite-valued and continuous.
For each $(\x,\x') \in \domE^{2}$, the function of $\tau$ that is given by
$\bsQ^{\pm}(\x,\x',\tau)$ is holomorphic on $\bar{C}^{\pm} - \grave{\bsI}(\x,\x')$;
and, as a consequence, the functions of $\sigma$ [see Prop.~\ref{1.3A}(v)]
that are given by $\bsQ^{+}(\x,\x',\sigma)$ and $\bsQ^{-}(\x,\x',\sigma)$ are
analytic throughout $R^{1} - \grave{\bsI}(\x,\x')$.
\end{proposition}
\proof
This follows by inspection from Eq.\ (\ref{1.21b}), Thm.~\ref{1.1D}(v)
and Props.~\ref{1.1C}(i) and~(ii).

Note:  Henceforth each function that we consider will be understood to
be finite-valued throughout its domain.


\setcounter{theorem}{0}
\setcounter{equation}{0}
\subsection{The sets $\S_{\grave{\subbsF}}\triple$, 
$\S_{\bar{\subbsF}}\triple$ and $\S_{\hat{\subbsQ}}\triple$}

We shall first define $\S_{\grave{\subbsF}}$.  Then, before we define 
$\S_{\bar{\subbsF}}$ and $\S_{\hat{\subbsQ}}$, we shall define the functions 
$\Gamma_{i}$, $\grave{\bsF}^{(i)}$ and $\bar{\bsF}^{(i)}$.

\begin{abc}
\begin{definition}{Dfn.\ of $\S_{\grave{\subbsF}}\triple$\label{def17}}
Let $\S_{\grave{\subbsF}}$ denote the set of all $2 \times 2$ matrix functions
$\grave{\bsF}$ with
\begin{equation}
\dom{\grave{\bsF}} := \{ (\x,\x',\tau):\x,\x' \in \domE \text{ and }
\tau \in C - \grave{\bsI}(\x,\x')\}
\label{1.31a}
\end{equation}
such that, for each $\x \in \domE$ and $\tau \in C - \{r,s\}$, 
\begin{equation}
\grave{\bsF}(\x,\x,\tau) = I
\label{1.31b}
\end{equation}
and there exists $H \in \S_{H}$ such that, for each $(\x,\x',\tau) \in 
\dom{\grave{\bsF}}$, $d\grave{\bsF}(\x,\x',\tau)$ exists and 
\begin{equation}
d\grave{\bsF}(\x,\x',\tau) = \Gamma(\x,\tau) \grave{\bsF}(\x,\x',\tau).
\label{1.31c}
\end{equation}
\end{definition}
\end{abc}

\begin{abc}
\begin{definition}{Dfn.\ of $\Gamma_{i}$ (for a given $H \in \S_{H}$)
\label{def18}}
Let $\Gamma_{3}$ and $\Gamma_{4}$ denote the $2 \times 2$ matrix functions
such that $\dom{\Gamma_{i}} := \{(\x,\tau):\x\in\domE$ and $\tau\in C-\{x^{i}\}$
and
\begin{equation}
\Gamma_{i}(\x,\tau) := \frac{[\partial H(\x)/\partial x^{i}] \Omega}
{2(\tau-x^{i})}
\label{1.31d}
\end{equation}
for all $(\x,\tau) \in \dom{\Gamma_{i}}$.  So, from Eqs.\ (\ref{1.13a})
and (\ref{1.2}),
\begin{equation}
\Gamma(\x,\tau) = \sum_{i} dx^{i} \Gamma_{i}(\x,\tau)
\label{1.31e}
\end{equation}
for all $(\x,\tau) \in \dom{\Gamma}$.
\end{definition}
\end{abc}

As regards the following definition, recall that $I^{(3)} := \{r:r_{1} <
r < r_{2}\}$ and $I^{(4)} := \{s:s_{2} < s < s_{1}\}$.

\begin{abc}
\begin{definition}{Dfn.\ of $\grave{\bsF}^{(i)}$ (for a given $H \in \S_{H}$)
\label{def19}}
Let $\grave{\bsF}^{(3)}$ and $\grave{\bsF}^{(4)}$ denote the $2 \times 2$
matrix functions with
\begin{eqnarray}
\dom{\grave{\bsF}^{(3)}} & := & \{(r,r',s,\tau):r,r'\in \I^{(3)}, s\in \I^{(4)},
\nonumber \\ & & \mbox{ }
r<s, r'<s \text{ and } \tau\in C - \grave{\bsI}^{(3)}(\x,\x')\}, 
\label{1.31f} \\ 
\dom{\grave{\bsF}^{(4)}} & := & \{(s,s',r,\tau):s,s'\in \I^{(4)}, r\in \I^{(3)},
\nonumber \\ & & \mbox{ }
r<s, r<s' \text{ and } \tau\in C - \grave{\bsI}^{(4)}(\x,\x')\},
\label{1.31g}
\end{eqnarray}
respectively, such that
\begin{equation}
\grave{\bsF}^{(i)}(x^{i},x^{i},x^{7-i},\tau) := I \text{ for each } 
x^{i} \in \I^{(i)} \text{ and } \tau\in C - \{x^{i}\}
\label{1.31h}
\end{equation}
and such that, for all $(x^{i},x^{\prime i},x^{7-i},\tau) \in 
\dom{\grave{\bsF}^{(i)}}$, 
$\partial\grave{\bsF}^{(i)}(x^{i},x^{\prime i},x^{7-i},\tau)/\partial x^{i}$
exists and
\begin{equation}
\frac{\partial\grave{\bsF}^{(i)}(x^{i},x^{\prime i},x^{7-i},\tau)}{\partial x^{i}}
= \Gamma_{i}(\x,\tau) \grave{\bsF}^{(i)}(x^{i},x^{\prime i},x^{7-i},\tau),
\label{1.31i}
\end{equation}
where $\x := (r,s) = (x^{3},x^{4})$.
\end{definition}
\end{abc}

\begin{abc}
\begin{proposition}[Properties of $\grave{\bsF}^{(i)} \in \dom{\grave{\bsF}^{(i)}}$]
\label{1.0E} 
\mbox{ } \\ \vspace{-3ex}
\begin{romanlist}
\item 
\mbox{ } \\ \vspace{-3ex}
\begin{equation}
\begin{array}{l}
\grave{\bsF}^{(i)} \text{ exists, is unique and is continuous; and, for fixed $\x$
and $x^{\prime i}$,} \\
\grave{\bsF}^{(i)}(x^{i},x^{\prime i},x^{7-i},\tau) \text { is a holomorphic
function of $\tau$ throughout } C - \grave{\bsI}^{(i)}(\x,\x').
\end{array}
\end{equation}
\item 
Moreover, $\partial\grave{\bsF}^{(i)}(x^{i},x^{\prime i},x^{7-i},\tau)/\partial 
x^{7-i}$ exists and
\begin{eqnarray}
\frac{\partial\grave{\bsF}^{(i)}(x^{i},x^{\prime i},x^{7-i},\tau)}{\partial x^{7-i}}
& = & \Gamma_{7-i}(\x,\tau) \grave{\bsF}^{(i)}(x^{i},x^{\prime i},x^{7-i},\tau) 
\nonumber \\ & & \mbox{ } 
- \grave{\bsF}^{(i)}(x^{i},x^{\prime i},x^{7-i},\tau) \left[ \Gamma_{7-i}(\x,\tau)
\right]_{x^{i}=x^{\prime i}}
\label{1.31j}
\end{eqnarray}
at all $(x^{i},x^{\prime i},x^{7-i},\tau) \in \dom{\grave{\bsF}^{(i)}}$.
\end{romanlist}
\end{proposition}
\end{abc}

\begin{abc}
\proof
\begin{romanlist}
\item 
The function of $(\x,\tau)$ given by $\Gamma_{i}(\x,\tau)$ is clearly
continuous throughout $\dom{\Gamma_{i}}$ and is (for fixed $\x$) a
holomorphic function of $\tau$ throughout $C - \{x^{i}\}$.  Hence,
statement (i) follows from standard theorems\footnote{See Ref.~\ref{Lefsch}.}
on ordinary differential equations.
\item 
We shall supply the proof only from $i=4$, i.e., for the ordinary
linear differential equation
\begin{equation}
\frac{\partial\grave{\bsF}^{(4)}(s,s',r,\tau)}{\partial s} = \Gamma_{4}(\x,\tau)
\grave{\bsF}^{(4)}(s,s',r,\tau).
\label{1.311j}
\end{equation}
subject to the initial condition
\begin{equation}
\grave{\bsF}^{(4)}(s',s',r,\tau) = I.
\label{1.312j}
\end{equation}
The proof for $i=3$ is similar.

{}From Eq.\ (\ref{1.31c}) and Prop.~\ref{1.1B}, $\partial
\Gamma_{4}(\x,\tau)/\partial r$ exists and is a continuous function of
$(\x,\tau)$ throughout $\dom{\Gamma_{4}}$.  So, from a standard 
theorem\footnote{See Sec.~4, Ch.~II, of Ref.~\ref{Lefsch}.} on the
differentiability of solutions with respect to a parameter,
$\partial\grave{\bsF}^{(4)}(s,s',r,\tau)/\partial r$ also exists and is a 
continuous function of $(s,s',r,\tau)$ throughout $\dom{\grave{\bsF}^{(4)}}$.
It then follows from Eq.\ (\ref{1.311j}) that
$$\frac{\partial}{\partial r} \left[ \frac{\partial\grave{\bsF}^{(4)}(s,s',r,\tau)}
{\partial s} \right]$$
exists and is a continuous function of $(s,s',r,\tau)$ throughout
$\dom{\grave{\bsF}^{(4)}}$.  Since the first partial derivatives of
$\grave{\bsF}^{(4)}$ with respect to $r$ and $s$ also exist and are continuous
throughout $\dom{\grave{\bsF}^{(4)}}$, the following partial derivatives both
exist and are equal as indicated.\footnote{See Ref.~\ref{HHP}.}
$$\frac{\partial^{2}\grave{\bsF}^{(4)}(s,s',r,\tau)}{\partial r \partial s}
= \frac{\partial^{2}\grave{\bsF}^{(4)}(s,s',r,\tau)}{\partial s \partial r}.$$
Furthermore, Eq.\ (\ref{1.13b}) is expressible in the form
$$\frac{\partial\Gamma_{4}}{\partial r} -
\frac{\partial\Gamma_{3}}{\partial s} -
\left( \Gamma_{3}\Gamma_{4} - \Gamma_{4}\Gamma_{3} \right) = 0.$$
Therefore, upon differentiating both sides of Eq.\ (\ref{1.311j})
with respect to $r$, one can replace $\partial\Gamma_{4}/\partial r$
by $\left( \partial\Gamma_{3}/\partial s \right) + \Gamma_{3}\Gamma_{4}
- \Gamma_{4}\Gamma_{3}$; and one then obtains, after replacing
$\Gamma_{4}\grave{\bsF}^{(4)}$ by $\partial\grave{\bsF}^{(4)}/\partial s$ and
regrouping terms,
\begin{equation}
\frac{\partial}{\partial s} \left( \frac{\partial\grave{\bsF}^{(4)}}{\partial r}
- \Gamma_{3} \grave{\bsF}^{(4)} \right) = \Gamma_{4} \left(
\frac{\partial\grave{\bsF}^{(4)}}{\partial r} - \Gamma_{3} \grave{\bsF}^{(4)} \right).
\label{1.313j}
\end{equation}
Next, let the $2 \times 2$ matrix function $U$ be defined by the equation
\begin{equation}
\frac{\partial\grave{\bsF}^{(4)}}{\partial r} - \Gamma_{3} \grave{\bsF}^{(4)} =
\grave{\bsF}^{(4)} U.
\label{1.314j}
\end{equation}
Substitution from the above equation into Eq.\ (\ref{1.313j}) then yields,
with the aid of Eq.\ (\ref{1.311j}), $\partial U/\partial s = 0$.  Since
$U$ is independent of $s$, one obtains, after setting $s=s'$ in Eq.\
(\ref{1.314j}) and then employing Eq.\ (\ref{1.312j}),
$$U(r,s',\tau) = - \Gamma_{3}((r,s'),\tau).$$
So, Eq.\ (\ref{1.314j}) becomes Eq.\ (\ref{1.31j}) for the case $i=4$;
viz., 
\begin{equation}
\frac{\partial\grave{\bsF}^{(4)}(s,s',r,\tau)}{\partial r} = \Gamma(\x,\tau)
\grave{\bsF}^{(4)}(s,s',r,\tau) - \grave{\bsF}^{(4)}(s,s',r,\tau)
\Gamma_{3}((r,s'),\tau).
\label{1.315j}
\end{equation}
\end{romanlist}
\cheers
\end{abc}

If $\grave{\bsF} \in \S_{\grave{\subbsF}}$ exists such that Eq.\ (\ref{1.31c})
holds for a given $H \in \S_{H}$, then the reader can easily prove from
Eqs.\ (\ref{1.31a}) to (\ref{1.31i}) that $\grave{\bsF}^{(3)}$ and
$\grave{\bsF}^{(4)}$ are extensions of the functions of $(r,r',s,\tau)$
and $(s,s',r,\tau)$ that are given by $\grave{\bsF}(\x,(r',s),\tau)$ and
$\grave{\bsF}(\x,(r,s'),\tau)$, respectively.  Specifically,
\begin{abc}
\begin{equation}
\grave{\bsF}(\x,(r',s),\tau) = \grave{\bsF}^{(3)}(r,r',s,\tau) \text { for all }
(r,r',s,\tau) \in \dom{\grave{\bsF}^{(3)}} \text{ such that } \tau \ne s
\label{1.31k} 
\end{equation}
and
\begin{equation}
\grave{\bsF}(\x,(r,s'),\tau) = \grave{\bsF}^{(4)}(s,s',r,\tau) \text{ for all }
(s,s',r,\tau) \in \dom{\grave{\bsF}^{(4)}} \text{ such that } \tau \ne r.
\label{1.31m}
\end{equation}
\end{abc}

\begin{abc}
\begin{definition}{Dfn.\ of $\bar{\bsF}^{(i)}$ (for given $H \in \S_{H}$)
\label{def20}}
Let $\bar{\bsF}^{(3)}$ and $\bar{\bsF}^{(4)}$ denote the extensions of 
$\grave{\bsF}^{(3)}$ and $\grave{\bsF}^{(4)}$ to [compare with Eqs.\ 
(\ref{1.31f}) and (\ref{1.31g})]
\begin{eqnarray}
\dom{\bar{\bsF}^{(3)}} & := & \{(r,r',s,\tau):r,r'\in \I^{(3)}, s\in \I^{(4)},
\nonumber \\ & & \mbox{ }
r<s, r'<s \text{ and } \tau\in C - \bar{\bsI}^{(3)}(\x,\x')\}, \\
\dom{\bar{\bsF}^{(4)}} & := & \{(s,s',r,\tau):s,s'\in \I^{(4)}, r\in \I^{(3)},
\nonumber \\ & & \mbox{ }
r<s, r<s' \text{ and } \tau\in C - \bar{\bsI}^{(4)}(\x,\x')\},
\end{eqnarray}
respectively, such that
\begin{equation}
\bar{\bsF}^{(i)}(x^{i},x^{i},x^{7-i},x^{i}) := I.
\label{1.31o}
\end{equation}
\end{definition}
\end{abc}

One can readily see from (\ref{1.31h}) and Prop.~\ref{1.0E}(i) that, 
for fixed $(x^{i},x^{\prime i},x^{7-i})$,
\begin{equation}
\bar{\bsF}^{(i)}(x^{i},x^{\prime i},x^{7-i},\tau) \text{ is a holomorphic
function of $\tau$ throughout } C - \bar{\bsI}^{(i)}(\x,\x').
\label{1.31p}
\end{equation}
[CAUTION:  However, $\bar{\bsF}^{(i)}$ is not generally continuous at 
those points $(x^{i},x^{\prime i},x^{7-i},\tau)$ of its domain
at which $x^{i} = x^{\prime i} = \tau$.]

We shall now take advantage of Eqs.\ (\ref{1.31k}) and (\ref{1.31m}),
together with the above definition of $\bar{\bsF}^{(i)}$ (for given $H \in
\S_{H}$), to help us define an extension of each member of $\S_{\grave{\subbsF}}$.
To grasp the following definition, it is essential to note that, for
all $(\x,\x',\tau)$ such that $x, x' \in \domE$ and $\tau \in C - 
\bar{\bsI}(\x,\x')$, the triple $(x,\x',\tau)$ is a member of $\dom{\grave{\bsF}}$
if and only if it is not true that $s = s' = \tau$ and it is not true
that $r = r' = \tau$.

\begin{abc}
\begin{definition}{Dfn.\ of $\S_{\bar{\subbsF}}\triple$\label{def21}}
Corresponding to each member of $\S_{\grave{\subbsF}}$ that is denoted 
by $\grave{\bsF}$, we shall let $\bar{\bsF}$ denote that extension of
$\grave{\bsF}$ to the domain [compare with Eq.\ (\ref{1.31a})]
\begin{equation}
\dom{\bar{\bsF}} := \{(\x,\x',\tau):\x,\x'\in\domE \text{ and }
\tau \in C - \bar{\bsI}(\x,\x')\}
\label{1.31q}
\end{equation}
such that
\begin{eqnarray}
\bar{\bsF}(\x,(r',s),s) & := & \bar{\bsF}^{(3)}(r,r',s,s), \text{ and }
\label{1.31r} \\
\bar{\bsF}(\x,(r,s'),r) & := & \bar{\bsF}^{(4)}(s,s',r,r).
\label{1.31s}
\end{eqnarray}
Also, let
\begin{equation}
\S_{\bar{\subbsF}} := \{\bar{\bsF}:\grave{\bsF} \in \S_{\grave{\subbsF}}\};
\label{1.31t}
\end{equation}
and, for each member of $\S_{\bar{\subbsF}}$ that is denoted by $\bar{\bsF}$, we
shall let $\grave{\bsF}$ denote the restriction of $\bar{\bsF}$ to the domain
(\ref{1.31a}).
\end{definition}
\end{abc}

{}From Eqs.\ (\ref{1.31b}), (\ref{1.31h}), (\ref{1.31o}), (\ref{1.31r})
and (\ref{1.31s}), note that
\begin{equation}
\bar{\bsF}(\x,\x,\tau) = I \text{ for all } \x \in \domE \text{ and }
\tau \in C.
\end{equation}

\begin{proposition}[Properties of $\bar{\bsF} \in \S_{\bar{\subbsF}}$]
\label{1.1E}
\mbox{ } \\ \vspace{-3ex}
\begin{romanlist}
\item 
$\grave{\bsF}$ is continuous; and, for each $(\x,\x') \in \domE^{2}$,
$\bar{\bsF}(\x,\x',\tau)$ is a holomorphic function of $\tau$ throughout
$C - \bar{\bsI}(\x,\x')$.

\item 
For all $(x,\x',\tau) \in \dom{\bar{\bsF}}$,
\begin{abc}
\begin{equation}
\det{\bar{\bsF}(\x,x',\tau)} = \bar{\bnu}(\x,\x',\tau).
\label{1.32}
\end{equation}

\item 
For all $(x,\x',\tau) \in \dom{\bar{\bsF}}$, $[\bar{\bsF}(\x,\x',\tau)]^{-1}$ exists.
Furthermore, for all $(\x,\x',\tau) \in \dom{\grave{\bsF}}$, 
$d[\grave{\bsF}(\x,\x',\tau)^{-1}]$ exists and
\begin{equation}
d[\grave{\bsF}(\x,\x',\tau)^{-1}] = - \grave{\bsF}(\x,\x',\tau)^{-1} \Gamma(\x,\tau).
\label{1.33}
\end{equation}

\item 
There is no more than one $\bar{\bsF} \in \S_{\bar{\subbsF}}$ corresponding to
each $H \in \S_{H}$ for which Eq.\ (\ref{1.31c}) holds throughout 
$\dom{\grave{\bsF}}$.

\item 
For all $\x,\x',\x'' \in \domE$ and $\tau \in C - [\bar{\bsI}(\x,\x') \cup
\bar{\bsI}(\x',\x'')]$,
\begin{equation}
\bar{\bsF}(\x,\x',\tau) \bar{\bsF}(\x',\x'',\tau) = \bar{\bsF}(\x,\x'',\tau).
\label{1.34a}
\end{equation}
For all $(\x,\x',\tau) \in \dom{\bar{\bsF}}$,
\begin{equation}
[\bar{\bsF}(\x,\x',\tau)]^{-1} = \bar{\bsF}(\x',\x,\tau).
\label{1.34b}
\end{equation}

\item 
For each $(\x,\x') \in \domE^{2}$, the expansion of $\bar{\bsF}(\x,\x',\tau)$ in 
inverse powers of $2\tau$ throughout $C - d(\x,\x')$ is such that
\begin{equation}
\bar{\bsF}(\x,\x',\tau) = I + (2\tau)^{-1} [H(\x)-H(\x')]\Omega
+ O(\tau^{-2}).
\end{equation}

\item 
For all $(\x,\x',\tau) \in \dom{\bar{\bsF}}$ such that $\tau \ne \infty$,
\begin{equation}
\bar{\bsF}^{\dagger}(\x,\x',\tau) \A(\x,\tau) \bar{\bsF}(\x,\x',\tau) = \A(\x',\tau),
\label{1.36}
\end{equation}
where $\bar{\bsF}^{\dagger}(\x,\x',\tau) := [\bar{\bsF}(\x,\x',\tau^{*})]^{\dagger}$.
In view of the preceding Prop.~\ref{1.1E}(vi), Eq.\ (\ref{1.36}) 
with $\A(\x,\tau)$ and $\A(\x',\tau)$ replaced by $\tau^{-1}\A(\x,\tau)$ 
and $\tau^{-1}\A(\x',\tau)$, respectively, holds at $\tau = \infty$.
\end{abc}
\end{romanlist}
\end{proposition}

\proofs
Henceforth, we shall tacitly employ the obvious proposition that the
mapping $\bar{\bsF} \mapsto \grave{\bsF}$ is a bijection of $\S_{\bar{\subbsF}}$ 
onto $\S_{\grave{\subbsF}}$; and, in the proofs of the above Prop.~\ref{1.1E}(ii),
(iv), (v) and (vii), we shall tacitly employ the easily proved proposition 
that, for each $\x' \in \domE$ and $\tau \in C - \{r',s'\}$,
\begin{eqnarray}
S(\x',\tau) & := & \{\x\in\domE:(\x,\x',\tau) \in \dom{\grave{\bsF}}\} \nonumber \\
& = & \{\x\in\domE:\tau \in C - \grave{\bsI}(\x,\x')\}
\label{1.37a}
\end{eqnarray}
is a convex open subset of $R^{2}$ such that $\x' \in S(\x,\tau)$.
Specifically, $S(\x',\tau) = \domE$ if $\tau \notin \, ]r_{1},s_{1}[$; and,
if $\tau=\sigma \in \, ]r_{1},s_{1}[$, then $S(\x',\tau)$ is the intersection
of $\domE$ with one of three regions that are labeled $\kappa=1$, $0$ and $-1$
in Fig.~2 except that none of the points on the horizontal line $s=\sigma$
and on the vertical line $r=\sigma$ are in $S(\x',\tau)$.  In accordance
with Eqs.\ (\ref{1.6h}), $\kappa=1$ if $s'<\sigma$, $\kappa=0$ if
$r'<\sigma<s'$, and $\kappa=-1$ if $\sigma<r'$.

We shall now give the proofs of Prop.~\ref{1.1E}(i) to (vii).
\begin{romanlist}
\begin{abc}
\item 
We first consider the proof of the proposition that $\grave{\bsF}$ is continuous
and that, for each fixed $(\x,\x') \in \domE^{2}$, $\grave{\bsF}(\x,\x',\tau)$ is
a holomorphic function of $\tau$ throughout $C - \grave{\bsI}(\x,\x')$.
The proof employs Eqs.\ (\ref{1.31a}) to (\ref{1.31c}) in the
definition of $\S_{\grave{\subbsF}}$.  The proof is an exact parallel of the proof
of Prop.~\ref{1.3A}(vii).  To obtain the proof of the present proposition,
simply make the following replacements in the proof of Prop.~\ref{1.3A}(vii):
$\bar{C}^{\pm} \rightarrow C$, $\hat{\bsI}(\x,\x') \rightarrow \grave{\bsI}(\x,\x')$, 
$\bsQ^{\pm} \rightarrow \grave{\bsF}$ and $\Delta^{\pm} \rightarrow \Gamma$.  No
other replacements are needed.

There remains only the proof that, for each fixed $(\x,\x') \in \domE^{2}$
such that $s=s'$ or/and $r=r'$, $\bar{\bsF}(\x,\x',\tau)$ is a holomorphic
function of $\tau$ throughout $C - \bar{\bsI}(\x,\x')$.  This proof has
already been given [see Prop.~\ref{1.0E}(i) and Eqs.\ (\ref{1.31k}) to 
(\ref{1.31p})] since we have previously demonstrated that the function
of $\tau$ given by $\grave{\bsF}(\x, (r',s),\tau)$ on the domain 
$C - (\grave{\bsI}^{(3)}(\x,\x') \cup \{s\})$ and by
$\grave{\bsF}(\x,(r,s'),\tau)$ on the domain
$C - (\grave{\bsI}^{(4)}(\x,\x') \cup \{r\})$ have holomorphic extensions
$\bar{\bsF}(\x,(r',s),\tau) := \bar{\bsF}^{(3)}(r,r',s,\tau)$ and 
$\bar{\bsF}(\x,(r,s'),\tau) := \bar{\bsF}^{(4)}(s,s',r,\tau)$ that cover the
isolated points $\tau=s$ and $\tau=r$, respectively.
\cheers

\item 
For any $2 \times 2$ matrix $M$ with complex elements,
\begin{equation}
M^{T} \Omega M = \Omega \det{M}.
\label{1.37b}
\end{equation}
We compute $d(\grave{\bsF}^{T} \Omega \grave{\bsF})$ with the aid of Eq.\ 
(\ref{1.31c}), the definition of $\Gamma$ by Eq.\ (\ref{1.13a}), the second of
Eqs.\ (\ref{1.12d}), Eq.\ (\ref{1.2}) and Eq.\ (\ref{1.3}).  From the definition
of $\grave{\bnu}$ by Eqs.\ (\ref{1.6t}) through (\ref{1.6x}), we obtain
$d[\det{\grave{\bsF}(\x,\x',\tau)}] = d\grave{\bnu}(\x,\x',\tau)$, whereupon 
Eq.\ (\ref{1.32}) follows from Eq.\ (\ref{1.31b}) and the fact that 
$\grave{\bnu}(\x,\x,\tau) = 1$.

Equation (\ref{1.32}) is proved by subjecting both sides of 
$\det{\grave{\bsF}(\x,\x',\tau)} = \grave{\bnu}(\x,\x',\tau)$ to holomorphic
extension in the $\tau$-plane when $s=s'$ and/or $r=r'$.  We leave details 
for the reader.
\cheers

\item 
Since $\bar{\bnu}(\x,\x',\tau) \ne 0$ for all $(\x,\x',\tau) \in \dom{\bar{\bsF}}$,
the preceding Prop.~\ref{1.1E}(ii) implies that $[\bar{\bsF}(\x,\x',\tau)]^{-1}$ 
exists for all $(\x,\x',\tau) \in \dom{\bar{\bsF}}$.  Equation (\ref{1.33}) then 
follows from Eq.\ (\ref{1.31c}) and the fact that
$d\left[\grave{\bsF}(\x,\x',\tau)^{-1} \grave{\bsF}(\x,\x',\tau)\right] = 0$.
\cheers

\item 
If $\bar{\bsF}'$ and $\bar{\bsF}$ are both members of $\S_{\bar{\subbsF}}$
corresponding to the same $H \in \S_{H}$, then Eqs.\ (\ref{1.31c}) and
(\ref{1.33}) imply that $d\left[\grave{\bsF}(\x,\x',\tau)^{-1}
\grave{\bsF}'(\x,\x',\tau)\right] = 0$ at all $(\x,\x',\tau) \in
\dom{\grave{\bsF}}$.  The rest of the proof employs Eq.\ (\ref{1.31b})
and is straightforward.
\cheers

\item 
Note that, for all $\x,\x',\tau \in \domE$,
\begin{equation}
C - (\grave{\bsI}(\x,\x') \cup \grave{\bsI}(\x',\x'')) \subset C 
- \grave{\bsI}(\x,\x'').
\label{1.37c}
\end{equation}
Therefore, from Eq.\ (\ref{1.31a}) and the above Eq.\ (\ref{1.37c}),
the triples $(\x',\x'',\tau)$, $(\x,\x',\tau)$ and $(\x,\x'',\tau)$
are members of $\dom{\grave{\bsF}}$ for all $\tau \in C - (\grave{\bsI}(\x,\x') \cup 
\grave{\bsI}(\x',\x''))$; and, from Eqs. (\ref{1.31c}) and (\ref{1.33}),
\begin{equation}
d\left[\grave{\bsF}(\x,\x'',\tau)^{-1} \grave{\bsF}(x,\x',\tau) 
\grave{\bsF}(x',x'',\tau)\right]=0.
\end{equation}
Hence, for all $\x',\x'' \in \domE$ and $\tau \in C - \grave{\bsI}(\x',\x'')$,
$\grave{\bsF}(\x,\x'',\tau)^{-1} \grave{\bsF}(\x,\x',\tau)
\grave{\bsF}(\x',\x'',\tau)$ has the same value for all $\x \in S(\x',\tau)$,
defined by Eq.\ (\ref{1.37a}); i.e., it has the same value for all $\x \in \domE$
such that $\tau \in (C - \grave{\bsI}(\x',\x'')) - \grave{\bsI}(\x,\x')$.
Therefore, from Eq.\ (\ref{1.31b}),
\begin{equation}
\begin{array}{r}
\grave{\bsF}(\x,\x',\tau) \grave{\bsF}(\x',\x'',\tau) = \grave{\bsF}(\x,\x'',\tau) \\
\text{for all } \x,\x',\x''\in\domE \text{ and} \\
\tau \in C - (\grave{\bsI}(\x,\x') \cup \grave{\bsI}(\x',\x'')).
\end{array}
\label{1.37e}
\end{equation}
Equation (\ref{1.34a}) now follows by holomorphic extension (for fixed
$\x$, $\x'$, $\x''$) of both sides of the above Eq.\ (\ref{1.37e}) to 
the $\tau$-plane domain $C - [\bar{\bsI}^{(3)}(\x,\x') \cup
\bar{\bsI}^{(4)}(\x,\x')]$; and Eq.\ (\ref{1.34b}) follows by setting 
$\x''=\x$ in Eq.\ (\ref{1.34a}).
\cheers

\item 
Noting that $\bar{\bsF}(\x,\x',\tau) = \grave{\bsF}(\x,\x',\tau)$ when $\tau \in 
d(\x,\x')$, apply Eqs.\ (\ref{1.31c}), (\ref{1.13a}) and (\ref{1.31b})
to the expansion of $\bar{\bsF}(\x,\x',\tau)$ in inverse powers of $2\tau$. 
The details are straightforward.
\cheers
\end{abc}

\item 
{}From Eqs.\ (\ref{1.11b}), (\ref{1.11c}) and (\ref{1.37b}), and from the
fact that $\Omega := iJ$,
\begin{abc}
\begin{equation}
h \Omega h = \rho^{2} \Omega.
\label{1.37f}
\end{equation}
{}From Eq.\ (\ref{1.12a}), Eq.\ (\ref{1.12b}) is expressible in the form
\begin{equation}
\rho d(i\Im{H}) = h \Omega \star d(\Re{H}),
\label{1.37g}
\end{equation}
which, with the aid of Eq.\ (\ref{1.37f}) and the fact that $\star\star=1$,
is seen to be equivalent to the equation 
\begin{equation}
\rho d(\Re{H}) = h \Omega \star d(i\Im{H}).
\label{1.37h}
\end{equation}
Upon combining the above Eqs.\ (\ref{1.37g}) and (\ref{1.37h}), we
obtain
\begin{equation}
\rho dH = h \Omega \star dH;
\label{1.37i}
\end{equation}
and, from the definitions of $\Gamma$ and $\A$ by Eqs.\ (\ref{1.13a})
and (\ref{1.26a}), respectively, the above Eq.\ (\ref{1.37i}) is seen
to be equivalent to the equation
\begin{equation}
\frac{1}{2} \Omega dH(\x) \Omega = \A(\x,\tau) \Gamma(\x,\tau).
\label{1.37j}
\end{equation}
Upon multiplying Eq.\ (\ref{1.31c}) through on the left by $\A(\x,\tau)$
and using the above Eq.\ (\ref{1.37j}), we obtain
\begin{equation}
\A(\x,\tau) d\grave{\bsF}(\x,\x',\tau) = \frac{1}{2} \Omega dH(\x) \Omega
\grave{\bsF}(\x,\x',\tau).
\label{1.37k}
\end{equation}
We further note, with the aid of Eqs.\ (\ref{1.12a}) and (\ref{1.12d}),
that (\ref{1.26a}) is expressible in the form
\begin{equation}
\A(\x,\tau) = \tau \Omega - \frac{1}{2} \Omega (H+H^{\dagger}) \Omega.
\label{1.37l}
\end{equation}
It is now simple to prove that the above Eqs.\ (\ref{1.37k}) and 
(\ref{1.37l}) imply
\begin{equation}
d\left[\grave{\bsF}^{\dagger}(\x,\x',\tau) \A(\x,\tau)
\grave{\bsF}(\x,\x',\tau)\right]=0.
\end{equation}
Therefore, from Eq.\ (\ref{1.31b}),
\begin{equation}
\grave{\bsF}^{\dagger}(\x,\x',\tau) \A(\x,\tau) \grave{\bsF}(\x,\x',\tau)
= \A(\x',\tau)
\label{1.37n}
\end{equation}
The remainder of the proof employs holomorphic extension in the
$\tau$-plane and is straightforward.
\cheers
\end{abc}
\end{romanlist}

\begin{proposition}[Relation between $\S_{\subbsF^{\pm}}$ and $\S_{\grave{\subbsF}}$]
\label{1.2E}
\mbox{ } \\ \vspace{-3ex}
\begin{romanlist}
\item 
For each $\bsF^{+} \in \S_{\subbsF^{+}}$ and $\bsF^{-} \in \S_{\subbsF^{-}}$,
the $2 \times 2$
matrix function $\grave{\bsF}$ whose domain is given by Eq.\ (\ref{1.31a}) and whose 
values are 
\begin{equation}
\begin{array}{c}
\grave{\bsF}(\x,\x',\tau) := \bsF^{\pm}(\x,\x',\tau) \text{ for all } \\
\x,\x' \in \domE \text{ and } \tau \in \bar{C}^{\pm} - \grave{\bsI}(\x,\x')
\end{array}
\label{1.38}
\end{equation}
[where we are employing Eq.\ (\ref{1.25}) when $\tau=\infty$ and
Eq.\ (\ref{1.27}) when $\tau \in R^{1} - \grave{\bsI}(\x,\x')$] is the
member of $\S_{\grave{\subbsF}}$ corresponding to the $H \in \S_{H}$ for which
Eq.\ (\ref{1.22}) holds.

\item 
The function $\grave{\bsF}^{M}$ that is defined in Prop.~\ref{1.1C}(vi), Eqs.\
(\ref{1.20a}) and (\ref{1.20b}), is the member of $\S_{\grave{\subbsF}}$ for which 
$H = H^{M}$.  Its extension $\bar{\bsF}^{M}$ to the domain (\ref{1.20d}) is,
of course, the corresponding member of $\S_{\bar{\subbsF}}$.
\end{romanlist}
\end{proposition}

\proofs
\begin{romanlist}
\item 
Use Eq.\ (\ref{1.22}) in Thm.~\ref{1.1D}(ii) and Eq.\ (\ref{1.24b})
in Thm.~\ref{1.1D}(iv) to prove that $\grave{\bsF}$ satisfies Eqs.\ (\ref{1.31c})
and (\ref{1.31b}), respectively.
\cheers

\item 
Use the preceding Prop.~\ref{1.2E}(i) together with Eqs.\ (\ref{1.20a})
and (\ref{1.20b}).
\cheers
\end{romanlist}

\begin{abc}
\begin{definition}{Dfn.\ of the mapping $\bvarphi:\S_{\bar{\subbsF}} 
\rightarrow \S_{\hat{\subbsQ}}$\label{def22}}
As regards the following definition, recall that $\hat{\bsI}(\x,\x')
:= \bar{\bsI}(\x,\x') \cup [r',s'] = \grave{\bsI}^{(3)}(\x,\x') \cup
\grave{\bsI}^{(4)}(\x,\x') \cup [r,s]$; and recall that 
$\bar{\bsF}(\x,\x',\tau) = \grave{\bsF}(\x,\x',\tau)$ when $\tau \in 
C - (\grave{\bsI}^{(3)}(\x,\x') \cup \grave{\bsI}^{(4)}(\x,\x'))$.

Let $\bvarphi$ denote the mapping whose domain is $\S_{\bar{\subbsF}}$ and whose
value $\bvarphi(\bar{\bsF})$ corresponding to each $\bar{\bsF} \in
\S_{\bar{\subbsF}}$ is the $2 \times 2$ matrix function $\hat{\bsQ}$ such that
\begin{equation}
\dom{\hat{\bsQ}} := \{(\x,\x',\tau):\x,\x'\in\domE \text{ and }
\tau \in C - \hat{\bsI}(\x,\x')\}
\label{1.39a}
\end{equation}
and, for all $(\x,\x',\tau) \in \dom{\hat{\bsQ}}$,
\begin{equation}
\hat{\bsQ}(\x,\x',\tau) := [A(\x)P^{M}(\x,\tau)]^{-1} \bar{\bsF}(\x,\x',\tau)
A(\x') P^{M}(\x',\tau).
\label{1.39b}
\end{equation}
\end{definition}
\end{abc}
Note the similarity of the above Eq.\ (\ref{1.39b}) to Eq.\ (\ref{1.21b}).

\begin{definition}{Dfn.\ of $\S_{\hat{\subbsQ}}\triple$\label{def23}}
Let
\begin{equation}
\S_{\hat{\subbsQ}} := \{\bvarphi(\bar{\bsF}):\bar{\bsF} \in \S_{\bar{\subbsF}}\}.
\label{1.40}
\end{equation}
\end{definition}

\begin{proposition}[Properties of $\hat{\bsQ} \in \S_{\hat{\subbsQ}}$]
\label{1.3E}
\mbox{ } \\ \vspace{-3ex}
\begin{romanlist}
\item 
\begin{abc}
\begin{equation}
\det{\hat{\bsQ}} = 1.
\end{equation}

\item 
For each $\x' \in \domE$ and $\tau \in C - [r',s']$,
\begin{equation}
\hat{\bsQ}(\x',\x',\tau) = I.
\end{equation}

\item 
The function $\hat{\bsQ}$ is continuous.  For each $(\x,\x') \in \domE^{2}$,
the function of $\tau$ given by $\hat{\bsQ}(\x,\x',\tau)$ is holomorphic 
throughout $C - \hat{\bsI}(\x,\x')$.

\item 
For each $(\x,\x') \in \domE^{2}$ such that $\grave{\bsI}^{(3)}(\x,\x')$ and
$\grave{\bsI}^{(4)}(\x,\x')$ are disjoint and for each $\sigma \in R^{1}$ such that 
$\grave{\bsI}^{(3)}(\x,\x') < \sigma < \grave{\bsI}^{(4)}(\x,\x')$, the two limits 
($\zeta \in C^{+}$)
\begin{equation}
\hat{\bsQ}(\x,\x',\sigma^{\pm}) := \lim_{\zeta\rightarrow 0}
\hat{\bsQ}(\x,\x',\sigma\pm\zeta)
\label{1.43a}
\end{equation}
exist and are both analytic functions of $\sigma$ and satisfy
\begin{equation}
\hat{\bsQ}(\x,\x',\sigma^{-}) = -J \hat{\bsQ}(\x,\x',\sigma^{+}) J
\label{1.43b}
\end{equation}
throughout the open interval between $\grave{\bsI}^{(3)}(\x,\x')$ and 
$\grave{\bsI}^{(4)}(\x,\x')$.
\end{abc}
\end{romanlist}
\end{proposition}

\proofs
\begin{romanlist}
\item 
This follows from Eqs.\ (\ref{1.39b}), (\ref{1.11b}), (\ref{1.16e}),
(\ref{1.29}), (\ref{1.6j}), (\ref{1.6z}) and (\ref{1.32}).
\cheers

\item 
Use Eqs.\ (\ref{1.11e}), (\ref{1.39b}) and (\ref{1.31b}).
\cheers

\item 
Use Eq.\ (\ref{1.39b}), Eq.\ (\ref{1.11b}), Props.~\ref{1.1C}(i) and~(ii),
and Prop.~\ref{1.1E}(i).
\cheers

\item 
{}From Eqs.\ (\ref{1.16b}), (\ref{1.16c}), (\ref{1.16e}), (\ref{1.31a}),
(\ref{1.39a}) and (\ref{1.39b}) together with Props.\ \ref{1.1C}(i) and~(ii),
and Prop.~\ref{1.1E}(i), the two limits exist, are given by
\begin{equation}
\hat{\bsQ}(\x,\x',\sigma^{\pm}) = [A(\x)P^{M\pm}(\x,\sigma)]^{-1} 
\bar{\bsF}(\x,\x',\sigma) A(\x')P^{M\pm}(\x',\sigma),
\label{1.43c}
\end{equation}
and are analytic functions of $\sigma$ throughout $\grave{\bsI}^{(3)}(\x,\x')
< \sigma < \grave{\bsI}^{(4)}(\x,\x')$.  Equation (\ref{1.43b}) then follows
from Eq.\ (\ref{1.19a}) (for $\kappa=0$) and the above Eq.\ (\ref{1.43c}).
\cheers
\end{romanlist}

\begin{proposition}[Relation between $\S_{\subbsQ^{\pm}}$ and $\S_{\hat{\subbsQ}}$]
\label{1.4E}
\mbox{ } \\ \vspace{-3ex}
\begin{romanlist}
\item 
For each $\bsQ^{+} \in \S_{\subbsQ^{+}}$ and $\bsQ^{-} \in \S_{\subbsQ^{-}}$,
the $2 \times 2$ matrix function $\hat{\bsQ}$ whose domain is given by Eq.\
(\ref{1.39a}) and whose values are
\begin{abc}
\begin{equation}
\hat{\bsQ}(\x,\x',\tau) := \bsQ^{\pm}(\x,\x',\tau) \text{ for all }
\x,\x'\in\domE \text{ and } \tau \in \bar{C}^{\pm} - \hat{\bsI}(\x,\x')
\label{1.44a}
\end{equation}
[where we are employing Eq.\ (\ref{1.10f}) when $\tau \in R^{1} - 
\hat{\bsI}(\x,\x')$ and Eq.\ (\ref{1.10h}) when $\tau=\infty$] is a
member of $\S_{\hat{\subbsQ}}$; and, if $\bsF^{\pm} := \bvarphi^{\pm}(\bsQ^{\pm})$
and $\grave{\bsF}$ is the member of $\S_{\grave{\subbsF}}$ that is defined in terms of
$\bsF^{\pm}$ by Eq.\ (\ref{1.38}), then $\hat{\bsQ} = \bvarphi(\bar{\bsF})$.
Moreover, as regards the limits (\ref{1.43a}) in Prop.~\ref{1.3E}(iv), 
\begin{equation}
\hat{\bsQ}(\x,\x',\sigma^{\pm}) = \bsQ^{\pm}(\x,\x',\sigma) \text{ for all }
\grave{\bsI}^{(3)}(\x,\x') < \sigma < \grave{\bsI}^{(4)}(\x,\x').
\label{1.44b}
\end{equation}

\item 
Recalling Prop.~\ref{1.2E}(ii) and the definition (\ref{1.40}) of
$\S_{\hat{\subbsQ}}$, one sees that
\begin{equation}
\hat{\bsQ}^{M} := \bvarphi(\bar{\bsF}^{M})
\end{equation}
is the member of $\S_{\hat{\subbsQ}}$ corresponding to $H = H_{M}$.  The value of
$\hat{\bsQ}^{M}$ is, for all $(\x,\x',\tau)$ in its domain,
\begin{equation}
\hat{\bsQ}^{M}(\x,\x',\tau) = I.
\label{1.45b}
\end{equation}
\end{abc}
\end{romanlist}
\end{proposition}

\proofs
\begin{romanlist}
\item 
For the given $\bsQ^{\pm} \in \S_{\subbsQ^{\pm}}$, let $\bsF^{\pm} :=
\bvarphi^{\pm}(\bsQ^{\pm})$, whereupon Eqs.\ (\ref{1.21b}) and
(\ref{1.44a}) yield
$$
\hat{\bsQ}(\x,\x',\tau) = [A(\x)P^{M\pm}(\x,\tau)]^{-1} \bsF^{\pm}(\x,\x',\tau)
A(\x') P^{M\pm}(\x',\tau)
$$
for all $\x, \x' \in \domE$ and $\tau \in \bar{C}^{\pm} - \hat{\bsI}(\x,\x')$.
Theorem~\ref{1.2E}(i), Eq.\ (\ref{1.16e}) and the definitions (\ref{1.39a}),
(\ref{1.39b}) and (\ref{1.40}) then supply the conclusion $\hat{\bsQ} = 
\bvarphi(\bar{\bsF}) \in \S_{\hat{\subbsQ}}$, where $\bar{\bsF}$ is defined 
in terms of $\bsF^{\pm}$ by Eq.\ (\ref{1.38}).  Equation (\ref{1.44b})
follows from Eq.\ (\ref{1.44a}) and Prop.~\ref{1.2D}.
\cheers

\item 
{}From the definition (\ref{1.20b}) of $\grave{\bsF}^{M}$, Eq.\ (\ref{1.16e}), 
Thm.~\ref{1.1D}(i), the definition of $\bvarphi^{\pm}$ in Sec.~\ref{Sec_1}C 
[in particular, Eq.\ (\ref{1.21b})] and the definition of $\bvarphi$
[in particular, Eq.\ (\ref{1.39b})], the values of $\hat{\bsQ}^{M} := 
\bvarphi(\bar{\bsF}^{M})$ are given by
\begin{equation}
\hat{\bsQ}^{M}(\x,\x',\tau) = \bsQ^{\pm}(\x,\x',\tau)
\end{equation}
for all $x, x' \in \domE$ and $\tau \in \bar{C}^{\pm} - \hat{\bsI}(\x,\x')$.
Equation (\ref{1.45b}) then follows from Eq.\ (\ref{1.15}).
\cheers
\end{romanlist}


\setcounter{theorem}{0}
\setcounter{equation}{0}
\subsection{The sets $\S_{\Q^{\pm}}\triple$ and $\S_{\F^{\pm}}\triple$} 

\begin{abc}
\begin{definition}{Dfn.\ of $\S_{\Q^{\pm}}\triple$\label{def24}}
Let $\S_{\Q^{\pm}}$ denote the set of all $2 \times 2$ matrix functions
$\Q^{\pm}$ such that
\begin{equation}
\dom{\Q^{\pm}} := \domE \times \bar{C}^{\pm}
\label{1.47a}
\end{equation}
and such that the following three conditions hold for all $\tau \in
\bar{C}^{\pm}$:
\begin{arablist}
\item 
\begin{equation}
\Q^{\pm}(\x_{0},\tau) = I.
\label{1.47b}
\end{equation}

\item 
The function of $\x$ given by $\Q^{\pm}(\x,\tau)$ is continuous throughout
$\domE$.

\item 
The same function of $\x$ is differentiable throughout
$\{\x\in\domE:(\tau-r)(\tau-s)\ne 0\}$ and there exists $\E \in \S_{\E}$
such that
\begin{equation}
d\Q^{\pm}(\x,\tau) = \Delta^{\pm}(\x,\tau) \Q^{\pm}(\x,\tau).
\label{1.47c}
\end{equation}
\end{arablist}
\end{definition}
The reader should compare the above definition to that of $\S_{\subbsQ^{\pm}}$
in Sec.~\ref{Sec_1}B, Eqs.\ (\ref{1.9a}) to (\ref{1.9c}).  In the same way
that we proved $\det{\bsQ^{\pm}} = 1$ [Prop.~\ref{1.3A}(i)], one proves
that $\det{\Q^{\pm}} = 1$.  This fact is used below.
\end{abc}

\begin{proposition}[Relation between $\S_{\Q^{\pm}}$ and $\S_{\subbsQ^{\pm}}$]
\label{1.1F}
\mbox{ } \\ \vspace{-3ex}
\begin{romanlist}
\item 
Corresponding to each $\Q^{\pm} \in \S_{\Q^{\pm}}$, there exists exactly
one $\bsQ^{\pm} \in \S_{\subbsQ^{\pm}}$ such that
\begin{abc}
\begin{equation}
\Q^{\pm}(\x,\tau) = \bsQ^{\pm}(\x,\x_{0},\tau)
\label{1.48}
\end{equation}
for all $(\x,\tau) \in \dom{\Q^{\pm}}$.

\item 
Corresponding to each $\bsQ^{\pm} \in \S_{\subbsQ^{\pm}}$, there exists exactly
one $\Q^{\pm} \in \S_{\Q^{\pm}}$ such that
\begin{equation}
\bsQ^{\pm}(\x,\x',\tau) = \Q^{\pm}(\x,\tau) [\Q^{\pm}(\x',\tau)]^{-1}
\label{1.49}
\end{equation}
for all $(\x,\x',\tau) \in \dom{\bsQ^{\pm}}$.
\end{abc}
\end{romanlist}
\end{proposition}

\proofs
\begin{romanlist}
\item 
Suppose $\Q^{\pm} \in \S_{\Q^{\pm}}$.  Since $\det{\Q^{\pm}} = 1$, 
$(\Q^{\pm})^{-1}$ exists.  Let $\bsQ^{\pm}$ denote the $2 \times 2$
matrix function whose domain is $\domE^{2} \times \bar{C}^{\pm}$ and
whose values are given by
\begin{abc}
\begin{equation}
\bsQ^{\pm}(\x,\x',\tau) := \Q^{\pm}(\x,\tau) [\Q^{\pm}(\x',\tau)]^{-1}.
\label{1.50a}
\end{equation}
Then, from Eqs.\ (\ref{1.47a}) to (\ref{1.47c}) and from Eqs.\ 
(\ref{1.9a}) to (\ref{1.9c}), $\bsQ^{\pm} \in \S_{\subbsQ^{\pm}}$ and
Eq.\ (\ref{1.48}) holds for all $(\x,\tau) \in \dom{\Q^{\pm}}$.

As regards uniqueness, suppose that $\bsQ_{1}^{\pm} \in \S_{\subbsQ^{\pm}}$
such that $\Q^{\pm}(\x,\tau) = \bsQ_{1}^{\pm}(\x,\x_{0},\tau)$ for all
$(\x,\tau) \in \dom{\Q^{\pm}}$.  Then, from Prop.~\ref{1.3A}(iv) and
Eq.\ (\ref{1.50a}),
\begin{equation}
\bsQ_{1}^{\pm}(\x,\x',\tau) = \Q^{\pm}(\x,\tau) [\Q^{\pm}(\x',\tau)]^{-1}
= \bsQ^{\pm}(\x,\x',\tau)
\end{equation}
for all $(\x,\x',\tau) \in \domE^{2} \times \bar{C}^{\pm}$.  So,
$\bsQ_{1}^{\pm} = \bsQ^{\pm}$.
\cheers

\item 
Suppose $\bsQ^{\pm} \in \S_{\subbsQ^{\pm}}$.  Let $\Q^{\pm}$ denote the
$2 \times 2$ matrix function whose domain is $\domE \times \bar{C}^{\pm}$
and whose values are given by
\begin{equation}
\Q^{\pm}(\x,\tau) := \bsQ^{\pm}(\x,\x_{0},\tau).
\label{1.51}
\end{equation}
Then, from Eqs.\ (\ref{1.47a}) to (\ref{1.47c}) and from Eqs.\ (\ref{1.9a})
to (\ref{1.9c}), $\Q^{\pm} \in \S_{\Q^{\pm}}$.  Moreover, from
Prop.~\ref{1.3A}(iv) and Eq.\ (\ref{1.51}), Eq.\ (\ref{1.49}) holds for
all $(\x,\x',\tau) \in \dom{\bsQ^{\pm}}$.  The proof of uniqueness is
left for the reader.
\cheers
\end{abc}
\end{romanlist}

Other properties of members of $\S_{\Q^{\pm}}$ are easily obtained by
inspection from the above theorem and Props.~\ref{1.3A} and~\ref{1.2D}.
We shall tacitly use some of these properties below.

\begin{abc}
\begin{definition}{Dfn.\ of $\S_{\F^{\pm}}\triple$\label{def25}}
Let $\S_{\F^{\pm}}$ denote the set of all $2 \times 2$ matrix functions
$\F^{\pm}$ with
\begin{equation}
\dom{\F^{\pm}} := \{(\x,\tau) \in \domE \times \bar{C}^{\pm}:
\tau \notin \{r,s,r_{0},s_{0}\}\}
\label{1.52a}
\end{equation}
for which $\Q^{\pm} \in \S_{\Q^{\pm}}$ exists such that
\begin{equation}
\begin{array}{c}
\F^{\pm}(\x,\tau) := A(\x) P^{M\pm}(\x,\tau) \Q^{\pm}(\x,\tau)
[P^{M\pm}(\x_{0},\tau)]^{-1} \\
\text{ when } \x \in \domE \text{ and } \tau \in \bar{C}^{\pm} - \{r,s,
r_{0},s_{0},\infty\}
\end{array}
\label{1.52b}
\end{equation}
and
\begin{equation}
\begin{array}{c}
\F^{\pm}(\x,\tau) := A(\x) F^{M\pm}(\x,\tau) \left( \begin{array}{cc}
(2\tau)^{-1} & 0 \\ 0 & 1
\end{array} \right) \Q^{\pm}(\x,\tau) \left( \begin{array}{cc}
2\tau & 0 \\ 0 & 1
\end{array} \right) [F^{M\pm}(\x_{0},\tau)]^{-1} \\
\text{ when } \x \in \domE \text{ and } \tau \in \bar{C}^{\pm} -
d(\x,\x_{0}).
\end{array}
\label{1.52c}
\end{equation}
\end{definition}
The above definition should be compared with the definition of
$\S_{\subbsF^{\pm}}$ in Eqs.\ (\ref{1.21a}) to (\ref{1.21e}).
\end{abc}

\begin{proposition}[Relation between $\S_{\F^{\pm}}$ and $\S_{\subbsF^{\pm}}$]
\label{1.2F}
\mbox{ } \\ \vspace{-3ex}
\begin{romanlist}
\item 
Corresponding to each $\F^{\pm} \in \S_{\F^{\pm}}$, there exists exactly
one $\bsF^{\pm} \in \S_{\subbsF^{\pm}}$ such that
\begin{abc}
\begin{equation}
\F^{\pm}(\x,\tau) = \bsF^{\pm}(\x,\x_{0},\tau)
\label{1.53}
\end{equation}
for all $(\x,\tau) \in \dom{\F^{\pm}}$.

\item 
Corresponding to each $\bsF^{\pm} \in \S_{\subbsF^{\pm}}$, there exists exactly
one $\F^{\pm} \in \S_{\F^{\pm}}$ such that
\begin{equation}
\bsF^{\pm}(\x,\x',\tau) = \F^{\pm}(\x,\tau) [\F^{\pm}(\x',\tau)]^{-1}
\end{equation}
for all $(\x,\x') \in \domE^{1}$ and $\tau \in \bar{C}^{\pm} - \{ r,s,
r',s',r_{0},s_{0}\}$.
\end{abc}
\end{romanlist}
\end{proposition}

\proofs
\begin{romanlist}
\item 
Suppose $\F^{\pm} \in \S_{\F^{\pm}}$.  From the definition of $\S_{\F^{\pm}}$,
there exists $\Q^{\pm} \in \S_{\Q^{\pm}}$ such that (\ref{1.52a}) and
(\ref{1.52b}) hold.  From Prop.~\ref{1.1F}(i), there then exists exactly
one $\bsQ^{\pm} \in \S_{\subbsQ^{\pm}}$ such that
\begin{abc}
\begin{equation}
\begin{array}{c}
\F^{\pm}(\x,\tau) = A(\x) P^{M\pm}(\x,\tau) \bsQ^{\pm}(\x,\x_{0},\tau)
[P^{M\pm}(\x_{0},\tau)]^{-1} \\
\text{ when } \x \in \domE \text{ and }
\tau \in \bar{C}^{\pm} - \{r,s,r_{0},s_{0},\infty\}
\end{array}
\label{1.55a}
\end{equation}
and
\begin{equation}
\begin{array}{c}
\F^{\pm}(\x,\tau) = A(\x) F^{M\pm}(\x,\tau) \left( \begin{array}{cc}
(2\tau)^{-1} & 0 \\ 0 & 1
\end{array} \right) \bsQ^{\pm}(\x,\x_{0},\tau) \left( \begin{array}{cc}
2\tau & 0 \\ 0 & 1 
\end{array} \right) [F^{M\pm}(\x_{0},\tau)]^{-1} \\
\text{ when } \x \in \domE \text{ and } \tau \in \bar{C}^{\pm} - 
d(\x,\x_{0}).
\end{array}
\label{1.55b}
\end{equation}
Comparison of the above Eqs.\ (\ref{1.55a}) and (\ref{1.55b})
with Eqs.\ (\ref{1.21b}) and (\ref{1.21c}), taken together with
the fact that $A(\x_{0}) = I$ [Eq.\ (\ref{1.11e})], now yield
\begin{equation}
\F^{\pm}(\x,\tau) = \bsF^{\pm}(\x,\x_{0},\tau) \text{ for all }
(\x,\tau) \in \dom{\F^{\pm}},
\label{1.55c}
\end{equation}
where
\begin{equation}
\bsF^{\pm} := \bvarphi^{\pm}(\bsQ^{\pm}) \in \S_{\subbsF^{\pm}}
\label{1.55d}
\end{equation}
according to the definition (\ref{1.21e}) of $\S_{\subbsF^{\pm}}$.  That
completes the existence part of the proof.

As regards uniqueness, suppose $\bsF_{1}^{\pm} \in \S_{\subbsF^{\pm}}$
such that
\begin{equation}
\F^{\pm}(\x,\tau) = \bsF_{1}^{\pm}(\x,\x_{0},\tau) \text{ for all }
(\x,\tau) \in \dom{\F^{\pm}}.
\end{equation}
Let $\bsQ_{1}^{\pm}$ be the member of $\S_{\subbsQ^{\pm}}$ for which
\begin{equation}
\bsF_{1}^{\pm} = \bvarphi^{\pm}(\bsQ_{1}^{\pm}).
\label{1.56b}
\end{equation}
Then, from the above Eqs.\ (\ref{1.55c}) to (\ref{1.56b}), and from
the definition of $\bvarphi^{\pm}$ in Sec.~\ref{Sec_1}C,
\begin{equation}
\begin{array}{c}
\bsQ_{1}^{\pm}(\x,\x_{0},\tau) = \bsQ^{\pm}(\x,\x_{0},\tau)
\text{ for all } \\
\x \in \domE \text{ and } \tau \in \bar{C}^{\pm} - \{r,s,r_{0},s_{0}\}.
\end{array}
\label{1.57a}
\end{equation}
Let $\E_{1}$ and $\E$ be the members of $\S_{\E}$ corresponding to
$\bsQ_{1}^{\pm}$ and $\bsQ^{\pm}$, respectively; and let $\Delta_{1}^{\pm}$
and $\Delta^{\pm}$ be the corresponding $2 \times 2$ matrix $1$-forms
as defined by Eqs.\ (\ref{1.7a}) to (\ref{1.7c}).  Then, from
Eqs.\ (\ref{1.9c}) and (\ref{1.57a}),
\begin{equation}
\begin{array}{c}
\Delta_{1}(\x,\tau) = \Delta(\x,\tau) \text{ for all } \\
\x \in \domE \text{ and } \tau \in \bar{C}^{\pm} - \{r,s,r_{0},s_{0}\}.
\end{array}
\end{equation}
Therefore, since $\E_{1}(\x_{0}) = \E(\x_{0}) = -1$, we have $\E_{1} =
\E$, whereupon $\bsQ_{1}^{\pm} = \bsQ^{\pm}$ according to Prop.~\ref{1.3A}(ii);
and $\bsF_{1}^{\pm} = \bsF^{\pm}$ from Eqs.\ (\ref{1.55d}) and (\ref{1.56b}).
\cheers

\item 
Suppose $\bsF^{\pm} \in \S_{\subbsF^{\pm}}$; i.e., there exists $\bsQ^{\pm} 
\in \S_{\subbsQ^{\pm}}$ such that $\bsF^{\pm} = \bvarphi^{\pm}(\bsQ^{\pm})$.
{}From Prop.~\ref{1.1F}(ii), there exists exactly one $\Q^{\pm} \in
\S_{\Q^{\pm}}$ such that Eq.\ (\ref{1.49}) holds for all $(\x,\x',\tau)
\in \dom{\bsQ^{\pm}}$.  It then follows from the definition of
$\bvarphi^{\pm}$ by Eqs.\ (\ref{1.21a}) to (\ref{1.21c}) that
\begin{equation}
\begin{array}{c}
\bsF^{\pm}(\x,\x',\tau) = \F^{\pm}(\x,\tau) [\F^{\pm}(\x',\tau)]^{-1}
\text{ for all } \\
(\x,\x') \in \domE^{2} \text{ and } \tau \in \bar{C}^{\pm} - \{r,s,
r',s',r_{0},s_{0}\},
\end{array}
\end{equation}
where $\F^{\pm}(\x,\tau)$ is defined in terms of $\Q^{\pm}$ by Eqs.\ 
(\ref{1.52a}) to (\ref{1.52c}).  That completes the existence part of
the proof.  The proof of uniqueness is simple and is left for the reader.
\cheers
\end{abc}
\end{romanlist}


\setcounter{theorem}{0}
\setcounter{equation}{0}
\subsection{The sets $\S_{\grave{\F}}\triple$, $\S_{\bar{\F}}\triple$,
$\S_{\hat{\Q}}\triple$}

The following definitions of $\grave{\F}^{(i)}$ and $\bar{\F}^{(i)}$ are tentative
and will be used only in Sec.~\ref{Sec_1}F and Sec.~\ref{Sec_2}A\@, while a 
subtly different employment of the symbols `$\grave{\F}^{(i)}$' and
`$\bar{\F}^{(i)}$' will occur in Sec.~\ref{Sec_2}B.

\begin{abc}
\begin{definition}{Dfns.\ of $\grave{\F}^{(i)}$ and $\bar{\F}^{(i)}$ (for a given
$H \in \S_{H}$)\label{def26}}
Let $\grave{\F}^{(i)}$ denote the function whose domain is
\begin{equation}
\dom{\grave{\F}^{(i)}} := \{(x^{i},\tau):x^{i} \in \I^{(i)} \text{ and }
\tau \in C - \grave{\I}^{(i)}(\x)\}
\label{1.59a}
\end{equation}
and whose values are given by
\begin{equation}
\grave{\F}^{(i)}(x^{i},\tau) := \grave{\bsF}^{(i)}(\x^{i},x_{0}^{i},x_{0}^{7-i},\tau).
\label{1.59b}
\end{equation}
Let $\bar{\F}^{(i)}$ denote the function whose domain is
\begin{equation}
\dom{\bar{\F}^{(i)}} := \{(x^{i},\tau):x^{i} \in \I^{(i)} \text{ and }
\tau \in C - \bar{\I}(\x)\}
\label{1.59c}
\end{equation}
and whose values are given by
\begin{equation}
\bar{\F}^{(i)}(x^{i},\tau) := \bar{\bsF}^{(i)}(x^{i},x_{0}^{i},x_{0}^{7-i},\tau).
\end{equation}
Equivalently, $\bar{\F}^{(i)}$ is the extension of $\grave{\F}^{(i)}$ to the
domain (\ref{1.59c}) such that
\begin{equation}
\bar{\F}^{(i)}(x_{0}^{i},x_{0}^{i}) := I.
\end{equation}
\end{definition}
The existence, uniqueness and some properties of $\grave{\F}^{(i)}$ and of
$\bar{\F}^{(i)}$ can be obtained by inspection of Prop.~\ref{1.0E}(i)
and Eqs.\ (\ref{1.31k}) through (\ref{1.31p}) and will be tacitly 
employed in this section and in Sec.~\ref{Sec_2}A\@.
\end{abc}

\begin{abc}
\begin{definition}{Dfns.\ of $\S_{\grave{\F}}\triple$ and $\S_{\bar{\F}}\triple$
\label{def27}}
Let $\S_{\grave{\F}}$ denote the set of all $2 \times 2$ matrix functions
$\grave{\F}$ with the domain
\begin{equation}
\dom{\grave{\F}} := \{(\x,\tau):\x\in\domE \text{ and } \tau \in
C - \grave{\I}(\x)\}
\label{1.59f}
\end{equation}
such that 
\begin{equation}
\grave{\F}(\x_{0},\tau) := I \text{ for all } \tau \in C - \{r_{0},s_{0}\},
\label{1.59g}
\end{equation}
$d\grave{\F}$ exists throughout $\dom{\grave{\F}}$ and there exists $H \in \S_{H}$
such that
\begin{equation}
d\grave{\F}(\x,\tau) = \Gamma(\x,\tau) \grave{\F}(\x,\tau)
\label{1.59h}
\end{equation}
for all $(\x,\tau) \in \dom{\grave{\F}}$.  Corresponding to each member of
$\S_{\grave{\F}}$ that is denoted by $\grave{\F}$, we shall let $\bar{\F}$ 
denote that extension of $\grave{\F}$ to the domain
\begin{equation}
\dom{\bar{\F}} := \{(\x,\tau):\x\in\domE \text{ and } \tau \in C -
\bar{\I}(\x)\}
\label{1.59i}
\end{equation}
such that 
\begin{eqnarray}
\bar{\F}((r,s_{0}),s_{0}) & := & \bar{\F}^{(3)}(r,s_{0}) \text{ for all } 
r \in \I^{(3)}, \text{ and} 
\label{1.59j} \\
\bar{\F}((r_{0},s),r_{0}) & := & \bar{\F}^{(4)}(s,r_{0}) \text{ for all }
s \in \I^{(4)};
\label{1.59k}
\end{eqnarray}
and let
\begin{equation}
\S_{\bar{\F}} := \{\bar{\F}:\grave{\F} \in \S_{\grave{\F}}\}.
\label{1.59l}
\end{equation}
\end{definition}
Recall from the definition of a type~A triple $\triple$ that
$\grave{\I}^{(3)}(\x) < \grave{\I}^{(4)}(\x)$
for all $\x\in\domE$.
\end{abc}

The set $\S_{\bar{\F}}$ can be equivalently (but less explicitly) defined
as the set of all $2 \times 2$ matrix functions $\bar{\F}$ for which
$$
\dom{\bar{\F}} := \{(\x,\tau):\x \in \domE \text{ and } \tau \in C -
\bar{\I}(\x)\},
$$
and there exists $H \in \S_{H}\triple$ such that, at all $\x\in\domE$ and
$\tau \in [C - \bar{\I}(\x)]-\{r_{0},s_{0}\}$, $d\F(\x,\tau)$ exists and
$$
d\F(\x,\tau) = \Gamma(\x,\tau) \F(\x,\tau)
$$
subject to the initial condition
$$
\F(\x_{0},\tau) = I;
$$
and, for each $r \in I^{(3)}$ and $s \in I^{(4)}$, $\F((r,s_{0}),\tau)$
and $\F((r_{0},s),\tau)$ are continuous functions of $\tau$ at
$\tau = s_{0}$ and at $\tau = r_{0}$, respectively.  

\begin{proposition}[Relation between $\S_{\bar{\F}}$ and $\S_{\bar{\subbsF}}$]
\label{1.1G} \mbox{ } \\
Suppose $\bar{\bsF} \in \S_{\bar{\subbsF}}$.  Then the function $\bar{\F}$ whose
domain is given by Eq.\ (\ref{1.59i}) and whose values are given by
\begin{equation}
\bar{\F}(\x,\tau) := \bar{\bsF}(\x,\x_{0},\tau) \text{ for all } (\x,\tau)
\in \dom{\bar{\F}}
\label{1.59m}
\end{equation}
is the member of $\S_{\bar{\F}}$ that corresponds to the same member of $\S_{H}$
as $\bar{\bsF}$.  [See Prop.~\ref{1.1E}(iii).]
\end{proposition}

\proof This follows by inspection from the definitions of $\S_{\grave{\subbsF}}$
by Eqs.\ (\ref{1.31a}) to (\ref{1.31c}), of $\S_{\grave{\F}}$ by Eqs.\ (\ref{1.59a})
to (\ref{1.59c}), of $\S_{\bar{\subbsF}}$ by Eqs.\ (\ref{1.31q}) to (\ref{1.31t}) and
of $\S_{\bar{\F}}$ by Eqs.\ (\ref{1.59i}) to (\ref{1.59l}).
\cheers

\begin{proposition}[Properties of $\bar{\F} \in \S_{\bar{\F}}$]
\label{1.2G}
\mbox{ } \\ \vspace{-3ex}
\begin{romanlist}
\item 
$\grave{\F}$ is continuous; and, for each $\x \in \domE$, $\bar{\F}(\x,\tau)$ is a
holomorphic function of $\tau$ throughout $C - \bar{\I}(\x)$.

\item 
For all $(\x,\tau) \in \dom{\bar{\F}}$,
\begin{abc}
\begin{equation}
\det{\bar{\F}(\x,\tau)} = \bar{\bnu}(\x,\x_{0},\tau).
\end{equation}

\item 
For all $(\x,\tau) \in \dom{\bar{\F}}$, $[\bar{\F}(\x,\tau)]^{-1}$ exists.
For all $(\x,\tau) \in \dom{\grave{\F}}$, $d[\grave{\F}(\x,\tau)^{-1}]$ exists and
\begin{equation}
d[\grave{\F}(\x,\tau)^{-1}] = - \grave{\F}(\x,\tau)^{-1} \Gamma(\x,\tau).
\end{equation}

\item 
There is no more than one $\bar{\F} \in \S_{\bar{\F}}$ corresponding to each
$H \in \S_{H}$ for which Eq.\ (\ref{1.59h}) holds throughout $\dom{\grave{\F}}$.

\item 
For all $r \in \I^{(3)}$ and $\tau \in C - \bar{\I}^{(3)}(\x)$,
\begin{equation}
\bar{\F}((r,s_{0}),\tau) = \bar{\F}^{(3)}(r,\tau).
\label{1.62a}
\end{equation}
For all $s \in \I^{(4)}$ and $\tau \in C - \bar{\I}^{(4)}(\x)$,
\begin{equation}
\bar{\F}((r_{0},s),\tau) = \bar{\F}^{(4)}(s,\tau).
\label{1.62b}
\end{equation}
For all $(\x,\tau) \in \dom{\bar{\F}}$,
\begin{eqnarray}
\bar{\F}(\x,\tau) [\bar{\F}^{(3)}(r,\tau)]^{-1} = \bar{\bsF}^{(4)}(s,s_{0},r,\tau),
\text{ and } 
\label{1.62c} \\
\bar{\F}(\x,\tau) [\bar{\F}^{(4)}(s,\tau)]^{-1} = \bar{\bsF}^{(3)}(r,r_{0},s,\tau),
\label{1.62d}
\end{eqnarray}

\item 
For each $\x \in \domE$, the expansion of $\bar{\F}(\x,\tau)$ in inverse
powers of $2\tau$ throughout $C-d(\x,\x_{0})$ is such that 
\begin{equation}
\bar{\F}(\x,\tau) = I + (2\tau)^{-1} [H(\x)-H(\x_{0})] \Omega
+ O(\tau^{-2}).
\label{1.63a}
\end{equation}

\item 
For all $(\x,\tau) \in \dom{\bar{\F}}$ such that $\tau \ne \infty$,
\begin{equation}
\bar{\F}^{\dagger}(\x,\tau) \A(\x,\tau) \bar{\F}(\x,\tau) = \A(\x_{0},\tau);
\end{equation}
and, with $\A(\x,\tau)$ and $\A(\x_{0},\tau)$ replaced by 
$\tau^{-1}\A(\x,\tau)$ and $\tau^{-1}\A(\x_{0},\tau)$,
respectively, the above equation holds at $\tau=\infty$.
\end{abc}
\end{romanlist}
\end{proposition}

\proofs
Except for parts (v), the proofs of Prop.~\ref{1.2G} are exactly
like those of Prop.~\ref{1.1E}.  To prove Eqs.\ (\ref{1.62a}) and
(\ref{1.62b}) in Prop.~\ref{1.2G}(v), first use Eqs.\ (\ref{1.31e}) 
and (\ref{1.59h}) to show that
\begin{abc}
\begin{eqnarray}
\frac{\partial\grave{\F}((r,s_{0}),\tau)}{\partial r} & = &
\Gamma_{3}((r,s_{0}),\tau) \grave{\F}((r,s_{0}),\tau), \text{ and}
\label{1.64a} \\
\frac{\partial\grave{\F}((r_{0},s),\tau)}{\partial r} & = &
\Gamma_{4}((r_{0},s),\tau) \grave{\F}((r_{0},s),\tau).
\label{1.64b}
\end{eqnarray}
The above Eqs.\ (\ref{1.64a}) and (\ref{1.64b}), together with
Eqs.\ (\ref{1.59b}), (\ref{1.31i}), (\ref{1.59g}) and (\ref{1.31h}),
yield
\begin{eqnarray}
\grave{\F}((r,s_{0}),\tau) & = & \grave{\F}^{(3)}(r,\tau) \text{ for all }
r \in \I^{(3)} \text{ and } \tau \in C - (\grave{\I}^{(3)}(\x) \cup \{s_{0}\}), 
\label{1.64c} \\
\grave{\F}((r_{0},s),\tau) & = & \grave{\F}^{(4)}(s,\tau) \text{ for all }
s \in \I^{(4)} \text{ and } \tau \in C - (\grave{\I}^{(4)}(\x) \cup \{r_{0}\}).
\label{1.64d}
\end{eqnarray}
Equations (\ref{1.62a}) and (\ref{1.62b}) now follow from Eqs.\ 
(\ref{1.64c}), (\ref{1.64d}), (\ref{1.59j}) and (\ref{1.59k}).

To prove Eqs.\ (\ref{1.62c}) and (\ref{1.62d}), first use Eqs.\ 
(\ref{1.31e}), (\ref{1.59h}), (\ref{1.64c}) and (\ref{1.64d})
to show that, for all $(\x,\tau) \in \dom{\grave{\F}}$,
\begin{equation}
\frac{\partial}{\partial x^{7-i}} \left[
\grave{\F}(\x,\tau) \grave{\F}^{(i)}(x^{i},\tau)^{-1} \right] =
\Gamma_{7-i}(\x,\tau) \left[ \grave{\F}(\x,\tau) \grave{\F}^{(i)}(x^{i},\tau)^{-1}
\right]
\label{1.64e}
\end{equation}
and that, for all $x^{i} \in \I^{(i)}$ and $\tau \in C - (\grave{\I}^{(i)}(\x)
\cup \{x_{0}^{7-i}\})$,
\begin{equation}
\left[ \grave{\F}(\x,\tau) \grave{\F}^{(i)}(x^{i},\tau)^{-1} 
\right]_{x^{7-i}=x_{0}^{7-i}} = I.
\label{1.64f}
\end{equation}
Comparison of the above Eqs.\ (\ref{1.64e}) and (\ref{1.64f}) with
Eqs.\ (\ref{1.31i}) and (\ref{1.31h}) [after interchanging `$i$'
and `$7-i$' in these last two equations] yields
\begin{equation}
\grave{\F}(\x,\tau) \bar{\F}^{(i)}(x^{i},\tau)^{-1} = 
\grave{\bsF}^{(7-i)}(x^{7-i},x_{0}^{7-i},x^{i},\tau) \text{ for all }
(\x,\tau) \in \dom{\grave{\F}}.
\label{1.64g}
\end{equation}
Equation (\ref{1.62c}) and (\ref{1.62d}) now follow after augmenting
the above Eq.\ (\ref{1.64g}) by Eqs.\ (\ref{1.62a}), (\ref{1.62b})
and [with scripts `$i$' and `$7-i$' interchanged] (\ref{1.31h}).
\cheers
\end{abc}

\begin{proposition}[The existence of $\grave{\F} \in \S_{\grave{\F}}$ 
(for a given $H \in \S_{H}$)]
\label{1.21G} \mbox{ } \\
For each given $H \in \S_{H}$, there exists $\grave{\F} \in \S_{\grave{\F}}$
such that $d\grave{\F} = \Gamma \grave{\F}$ with $\Gamma :=
\frac{1}{2}(\tau-z+\rho\star)^{-1} dH \, \Omega$.
\end{proposition}

\begin{abc}
\proof
{}From Prop.~\ref{1.0E}(i) and the definition of $\grave{\F}^{(i)}$ by Eqs.\
(\ref{1.59a}) and (\ref{1.59b}), the functions $\grave{\bsF}^{(i)}$ and
$\grave{\F}^{(i)}$ corresponding to the given $H \in \S_{H}$ exist.  Let 
$\grave{\F}$ denote the function with the domain (\ref{1.59f}) and the values
\begin{equation}
\grave{\F}(\x,\tau) := \grave{\bsF}^{(4)}(s,s_{0},r,\tau) \grave{\F}^{(3)}(r,\tau)
\label{1.641g}
\end{equation}
for all $(\x,\tau) \in \dom{\grave{\F}}$.  From Eqs.\ (\ref{1.31i}) and (\ref{1.59b}),
\begin{equation}
\frac{\partial\grave{\F}^{(3)}(r,\tau)}{\partial r} = \Gamma_{3}((r,s_{0}),\tau)
\grave{\F}^{(3)}(r,\tau).
\label{1.642g}
\end{equation}
Therefore, from Eqs.\ (\ref{1.311j}), (\ref{1.315j}), (\ref{1.641g}) and
(\ref{1.642g}),
\begin{equation}
\frac{\partial\grave{\F}(\x,\tau)}{\partial r} = \Gamma_{3}(\x,\tau)
\grave{\F}(\x,\tau), \quad
\frac{\partial\grave{\F}(\x,\tau)}{\partial s} = \Gamma_{4}(\x,\tau)
\grave{\F}(\x,\tau).
\end{equation}
The right sides of the above two equations are continuous functions of
$(\x,\tau)$ throughout $\dom{\grave{\F}}$.  [See Eq.\ (\ref{1.59b}), Eq.\
(\ref{1.641g}) and Prop.~\ref{1.0E}(i).]  Therefore, $d\grave{\F}$ exists and
equals $\Gamma \grave{\F}$.
\cheers
\end{abc}

\begin{proposition}[Relation between $\S_{\grave{\F}}$ and $\S_{\F^{\pm}}$]
\label{1.3G}
\mbox{ } \\ \vspace{-3ex}
\begin{romanlist}
\item 
Suppose $\F^{+} \in \S_{\F^{+}}$ and $\F^{-} \in \S_{\F^{-}}$;
and $\F^{+}$ and $\F^{-}$ correspond to the same member of $\S_{H}$.
Then the function $\grave{\F}$ whose domain is given by Eq.\ (\ref{1.59f}) and
whose values are given by
\begin{abc}
\begin{equation}
\grave{\F}(\x,\tau) := \F^{\pm}(\x,\tau) \text{ for all } \x \in \domE
\text{ and } \tau \in \bar{C}^{\pm} - \grave{\I}(\x)
\label{1.65}
\end{equation}
[where we are using Prop.~\ref{1.2F}(i), together with Thm.~\ref{1.1D}(vi)
when $\tau=\infty$ and Thm.~\ref{1.1D}(viii) when $\tau \in R^{1} -
\grave{\I}(\x)$] is a member of $\S_{\grave{\F}}$ corresponding to the
same member of $\S_{H}$ as $\F^{\pm}$.

\item 
The member of $\S_{\bar{\F}}$ corresponding to $H = H^{M} \in \S_{H}$ exists
and is the function $\bar{\F}^{M}$ whose domain is given by Eq.\ (\ref{1.59i})
and whose values are
\begin{equation}
\bar{\F}^{M}(\x,\tau) := \bar{\bsF}^{M}(\x,\x_{0},\tau) \text{ for all }
\x \in \domE \text{ and } \tau \in C - \bar{\I}(\x).
\end{equation}
\end{abc}
\end{romanlist}
\end{proposition}

\proofs
See the proofs of Thm.~\ref{1.2E}.
\cheers

\begin{abc}
\begin{definition}{Dfn.\ of the set $\S_{\hat{\Q}}$\label{def28}}
Let $\S_{\hat{\Q}}\triple$ denote the set of all $2 \times 2$ matrix functions
$\hat{\Q}$ with domain
\begin{equation}
\dom{\hat{\Q}} = \left\{ (\x,\tau): \x \in \domE, \tau \in C - \hat{\I}(\x)
\right\}
\label{G2.19}
\end{equation}
and values determined by
\begin{equation}
\bar{\F}(\x,\tau) = A(\x) P^{M}(\x,\tau) \hat{\Q}(\x,\tau) 
[P^{M}(\x_{0},\tau)]^{-1}.
\label{G2.14a}
\end{equation}
\end{definition}
\end{abc}

\begin{proposition}[Properties of $\hat{\Q} \in \S_{\hat{\Q}}$]
\label{1.4G}
\mbox{ } \\ \vspace{-3ex}
\begin{romanlist}
\item 
\begin{abc}
\begin{equation}
\det{\hat{\Q}} = 1.
\end{equation}

\item 
For each $\tau \in C - \{r_{0},s_{0}\}$,
\begin{equation}
\hat{\Q}(\x_{0},\tau) = I.
\end{equation}

\item 
The function $\hat{\Q}$ is continuous.  For each $x \in \domE$, the function of
$\tau$ given by $\hat{\Q}(\x,\tau)$ is holomorphic throughout 
$C - \hat{\I}(\x)$.

\item 
For each $\x \in \domE$ and $\sigma \in R^{1}$ such that
$\bar{\I}^{(3)}(\x) < \sigma < \bar{\I}^{(4)}(\x)$,
the two limits ($\zeta \in C^{+}$)
\begin{equation}
\hat{\Q}(\x,\sigma^{\pm}) := \lim_{\zeta\rightarrow 0} \hat{\Q}(\x,\sigma\pm\zeta)
\end{equation}
exist and are both analytic functions of $\sigma$ and satisfy
\begin{equation}
\hat{\Q}(\x,\sigma^{-}) = - J \hat{\Q}(\x,\sigma^{+}) J
\end{equation}
thoughout the open interval between $\bar{\I}^{(3)}(\x)$ and
$\bar{\I}^{(4)}(\x)$.
\end{abc}
\end{romanlist}
\end{proposition}

\proofs
The proofs are exactly like those of Prop.~\ref{1.3E} with `$\x'$'
replaced by `$\x_{0}$'.
\cheers

Note:  As regards the above Prop.~\ref{1.4G}(iv), recall that the triple
$\triple$ is always chosen so that the intervals $\grave{\I}^{(3)}(\x)$
and $\grave{\I}^{(4)}(\x)$ are disjoint for every $\x \in \domE$.  In 
other words, $\grave{\I}^{(3)}(\x) < \grave{\I}^{(4)}(\x)$ for 
every $\x \in \domE$.  

\begin{proposition}[Relation between $\S_{\hat{\Q}}$ and $\S_{\hat{\subbsQ}}$]
\label{1.5G}
\mbox{ } \\
For each $\hat{\bsQ} \in \S_{\hat{\subbsQ}}$, the function $\hat{\Q}$ with domain
\begin{abc}
\begin{equation}
\dom{\hat{\Q}} := \{(\x,\tau):\x\in\domE \text{ and } \tau \in 
C-\hat{\I}(\x)\}
\label{1.71}
\end{equation}
whose values are given by
\begin{equation}
\hat{\Q}(\x,\tau) := \hat{\bsQ}(\x,\x_{0},\tau)
\label{1.72}
\end{equation}
is a member of $\S_{\hat{\Q}}$ corresponding to the same member of $\S_{H}$
as $\hat{\bsQ}$.
\end{abc}
\end{proposition}

\proof
Suppose $\hat{\bsQ} \in \S_{\hat{\subbsQ}}$.  From the definition of
$\S_{\hat{\subbsQ}}$, there exists $\bar{\bsF} \in \S_{\bar{\subbsF}}$ such that 
Eq.\ (\ref{1.39b}) holds for all $\x,\x'\in\domE$ and 
$\tau \in C-\hat{\bsI}(\x,\x')$.  Upon letting $\x'=\x_{0}$ in Eq.\ 
(\ref{1.39b}), one infers from Eq.\ (\ref{1.11e}) and Prop.~\ref{1.2G} that
there exists $\bar{\F} \in \S_{\bar{\F}}$ corresponding to the same member
of $\S_{H}$ as $\hat{\bsQ}$ such that
\begin{equation}
\hat{\bsQ}(\x,\x_{0},\tau) = [A(\x)P^{M}(\x,\tau)]^{-1} \bar{\F}(\x,\tau)
P^{M}(\x_{0},\tau)
\end{equation}
for all $\x \in \domE$ and $\tau \in C-\hat{\I}(\x)$.  It then follows
from the definition of $\S_{\hat{\Q}}$ that the function $\hat{\Q}$ with
the domain (\ref{1.71}) and the values (\ref{1.72}) is a member of
$\S_{\hat{\Q}}$ corresponding to the same member of $\S_{H}$ as $\hat{\bsQ}$.
\cheers

\begin{proposition}[Relation between $\S_{\hat{\Q}}$ and $\S_{\Q^{\pm}}$]
\label{1.6G}
\mbox{ } \\ \vspace{-3ex}
\begin{romanlist}
\item 
For each $\Q^{+} \in \S_{\Q^{+}}$ and $\Q^{-} \in \S_{\Q^{-}}$, the
function $\hat{\Q}$ whose domain is given by Eq.\ (\ref{1.71}) and whose
values are
\begin{abc}
\begin{equation}
\hat{\Q}(\x,\tau) := \Q^{\pm}(\x,\tau) \text{ for all } \x\in\domE
\text{ and } \tau \in \bar{C}^{\pm} - \hat{\I}(\x)
\end{equation}
[where we are using Prop.~\ref{1.1F}(i), together with Prop.~\ref{1.3A}(v)
when $\tau \in R^{1} - \hat{\I}(\x)$ and Prop.~\ref{1.3A}(vi)
when $\tau=\infty$] is a member of $\S_{\hat{\Q}}$.

\item 
The member of $\S_{\hat{\Q}}$ corresponding to $H = H^{M} \in \S_{H}$ exists
and is given by
\begin{equation}
\hat{\Q}^{M}(\x,\tau) = I
\end{equation}
for all $\x \in \domE$ and $\tau \in C - \hat{\I}(\x)$.
\end{abc}
\end{romanlist}
\end{proposition}

\proofs
See the proofs of Prop.~\ref{1.4E}.
\cheers
\newpage

\setcounter{equation}{0}
\setcounter{theorem}{0}
\section{Initial value problem for $\S_{\E}\triple$, $\S_{H}\triple$ and
$\S_{\bar{\F}}\triple$ \label{Sec_2}}

\begin{nosubsec}

The main theorems in Sec.~\ref{Sec_2} concern the existence of a unique 
$\E \in \S_{\E}$, a unique $H \in \S_{H}$ and a unique $\bar{\F} \in 
\S_{\bar{\F}}$ for any prescribed values of $\E$ on the two null lines
in $\domE$ that pass through the point $\x_{0}$.  Specifically, these 
theorems assert that, for any given complex-valued $\bC^{1}$ functions 
$\E^{(3)}$ and $\E^{(4)}$ which have the domains
\begin{equation}
I^{(3)} := \{r:r_{1} < r < r_{2}\}, \quad
I^{(4)} := \{s:s_{2} < s < s_{1}\},
\label{1.76}
\end{equation}
respectively, and which satisfy
\begin{equation}
\Re{\E^{(i)}(x^{i})} < 0 \text{ for all } x^{i} \in \I^{(i)}
\text{ and } \E^{(3)}(r_{0}) = \E^{(4)}(s_{0}) = -1,
\label{1.77}
\end{equation}
there exist exactly one $\E \in \S_{\E}$, exactly one $H \in \S_{H}$
and exactly one $\bar{\F} \in \S_{\bar{\F}}$ such that $\E = H_{22}$,
\begin{equation}
\E(r,s_{0}) = \E^{(3)}(r) \text{ for all } r \in \I^{(3)}, \quad
\E(r_{0},s) = \E^{(4)}(s) \text{ for all } s \in \I^{(4)};
\label{1.78}
\end{equation}
and
\begin{equation}
d\grave{\F} = \Gamma \grave{\F} = \left[ \frac{1}{2} 
\left( \tau-z+\rho\star \right)^{-1} dH \Omega \right] \grave{\F}.
\label{1.79}
\end{equation}
We shall also present theorems on the differentiability and holomorphy
properties of $H$ and $\bar{\F}$ for any given specification of the 
differentiability classes (or of the analyticities) of the initial 
value functions $\E^{(3)}$ and $\E^{(4)}$.

The proofs of these theorems originate with the authors and employ an 
HHP (homogeneous Hilbert problem) and an equivalent Fredholm integral
equation of the second kind.  Except for some notations and conventions
and except for the domain of the Ernst potential in $\{\x:r<s\}$, the
proofs of most of the theorems in Sec.~\ref{Sec_2} are essentially the same
as those given in two earlier papers by the authors on the IVP (initial 
value problem) for colliding gravitational plane wave pairs.\footnote{
See Ref.~\ref{IVP34}.}  Therefore, instead of repeating those proofs, 
we shall provide the reader with a complete list of notational and
other adjustments which must be made in our earlier papers to bring 
them into accord with our present needs.

\end{nosubsec}


\setcounter{theorem}{0}
\setcounter{equation}{0}
\subsection{Introduction to the HHP adapted to $(\bar{\F}^{(3)},\bar{\F}^{(4)})$}

Our decision as to how to handle the IVP corresponding to any given
$\bC^{1}$ initial value functions $\E^{(3)}$ and $\E^{(4)}$ as
defined by Eqs.\ (\ref{1.76}) and (\ref{1.77}) was motivated by the
following proposition:
\begin{abc}
\begin{proposition}[Selected properties of $\bar{\F} \in \S_{\bar{\F}}$]
\label{2.1A}
\mbox{ } \\ \vspace{-3ex}
\begin{romanlist}
\item 
For each $\x\in \domE$, $\bar{\F}(\x,\tau)$ is a holomorphic function of $\tau$
throughout $C - \bar{\I}(\x)$.
\item 
For each $\x\in \domE$,
\begin{equation}
\bar{\F}(\x,\infty) = I.
\end{equation}
and
\begin{equation}
H(\x) = H(\x_{0}) + \left\{ 2\tau \left[ \bar{\F}(\x,\tau) - I \right] \Omega
\right\}_{\tau=\infty},
\label{2.2}
\end{equation}
where $H(\x_{0})$ is given by Eq.\ (\ref{1.12c}).
\item 
For each $i \in \{3,4\}$ and $\x\in \domE$, the product $\bar{\F}(\x,\tau) 
[\bar{\F}^{(i)}(x^{i},\tau)]^{-1}$ has a holomorphic extension in the 
$\tau$-plane to $C - \bar{\I}^{(7-i)}(\x)$. 
\end{romanlist}
\end{proposition}
\end{abc}

\proof
Statement (i) is simply part of Prop.~\ref{1.2G}(i).  Statement (ii)
is derived from Eq.\ (\ref{1.63a}) in Prop.~\ref{1.2G}(vi).  Statement
(iii) is obtained from (\ref{1.31p}), (\ref{1.62c}) and (\ref{1.62d}).
\cheers

Consideration of the above propositions led us to a four-step procedure
for solving the aforementioned IVP:
\begin{abc}
\begin{arablist}
\item 
The defining equations for $H \in \S_{H}$ in Sec.~\ref{Sec_1}C can be used to
compute 
\begin{eqnarray}
H^{(3)}(r) & := & H(r,s_{0}) \text{ for all } r \in \I^{(3)},
\label{2.3} \\
H^{(4)}(s) & := & H(r_{0},s) \text{ for all } s \in \I^{(4)}
\label{2.4}
\end{eqnarray}
directly from $\E^{(3)}$ and $\E^{(4)}$, respectively.  The expressions
which are obtained by this computation can then be used as new definitions
of $H^{(3)}$ and $H^{(4)}$; and this is exactly what we shall do in 
Sec.~\ref{Sec_2}B.  Unlike the above Eqs.\ (\ref{2.3}) and (\ref{2.4}),
these new definitions do not presuppose the existence of $H$.

\item 
{}From Eqs.\ (\ref{1.31d}), (\ref{1.31h}), (\ref{1.31i}), (\ref{1.59b}),
(\ref{2.3}) and (\ref{2.4}), $\grave{\F}^{(i)}$ is the $2 \times 2$ matrix 
solution on the domain (\ref{1.59a}) of the one-parameter family of 
ordinary differential equations
\begin{equation}
\frac{\partial\grave{\F}^{(i)}(x^{i},\tau)}{\partial x^{i}}
= \Gamma^{(i)}(x^{i},\tau) \grave{\F}^{(i)}(x^{i},\tau)
\label{2.5}
\end{equation}
subject to the condition
\begin{equation}
\grave{\F}^{(i)}(x_{0}^{i},\tau) = I,
\label{2.6}
\end{equation}
where $x_{0}^{3} := r_{0}$, $x_{0}^{4} := s_{0}$ and
\begin{equation}
\Gamma^{(i)}(x^{i},\tau) := \frac{\dot{H}^{(i)}(x^{i}) \Omega}{2(\tau-x^{i})}.
\label{2.7}
\end{equation}
The above pair of Eqs.\ (\ref{2.5}) and (\ref{2.6}) can be used
to define $\grave{\F}^{(i)}$ directly without presupposing [as we did in
Eqs.\ (\ref{2.3}) and (\ref{2.4})] the existence of $H$; and that is exactly
what we shall do in Sec.~\ref{Sec_2}B.  The function $\bar{\F}^{(i)}$
will then be defined as the extension of $\grave{\F}^{(i)}$ to the domain
(\ref{1.59c}) such that $\bar{\F}^{(i)}(x_{0}^{i},x_{0}^{i}) := I$.

\item 
The next step in solving the IVP is to seek a $2 \times 2$ matrix
function $\bar{\F}$ which has the domain (\ref{1.59i}) and which satisfies
the following three conditions for each $\x\in \domE$ and $i\in\{3,4\}$:
\begin{equation}
\bar{\F}(\x,\tau) \text{ is a holomorphic function of } 
\tau \text{ throughout } C - \bar{\I}(\x),
\label{2.8}
\end{equation}
\begin{equation}
\bar{\F}(\x,\infty) = I ,
\end{equation}
\begin{equation}
\bar{\F}(\x,\tau) [\bar{\F}^{(i)}(x^{i},\tau)]^{-1} \text{ has a holomorphic 
extension which covers } \bar{\I}(\x). 
\label{2.10}
\end{equation}
Equation (\ref{1.59i}) together with the above Eqs.\ (\ref{2.8})
to (\ref{2.10}) constitutes what we call ``the HHP adapted to
$(\bar{\F}^{(3)},\bar{\F}^{(4)})$.''  

It is clear that a solution $\bar{\F}$ of this HHP can be sought without
presupposing any prior definition of $\bar{\F}$.  In Sec.~\ref{Sec_2}C we shall 
formally define {\em the } HHP {\em adapted to} $(\bar{\F}^{(i)},\bar{\F}^{(4)})$ 
to be the set of all $2 \times 2$ matrix functions $\bar{\F}$ which have
the domain (\ref{1.59i}) and satisfy the conditions (\ref{2.8}) to
(\ref{2.10}).  Any member of this set will be called a {\em solution}
of the HHP adapted to $(\bar{\F}^{(3)},\bar{\F}^{(4)})$.

\item 
A function $H$ is defined in terms of a given solution $\bar{\F}$ by
Eqs.\ (\ref{2.2}); and one then lets $\E := H_{22}$.  Also, let
$\grave{\F}$ denote the restriction of the given solution to the domain
(\ref{1.59f}).
\end{arablist}
\end{abc}

It will turn out that there exists one and only one solution $\bar{\F}$
of the HHP adapted to $(\bar{\F}^{(3)},\bar{\F}^{(4)})$, that
\begin{abc}
\begin{equation}
\begin{array}{rcl}
\bar{\F}((r,s_{0}),\tau) & = & \bar{\F}^{(3)}(r,\tau) \text{ for all }
(r,\tau) \in \dom{\bar{\F}^{(3)}} \\
\bar{\F}((r_{0},s),\tau) & = & \bar{\F}^{(4)}(s,\tau) \text{ for all } 
(s,\tau) \in \dom{\bar{\F}^{(4)}},
\end{array}
\label{2.11}
\end{equation}
that 
\begin{equation}
H(r,s_{0}) = H^{(3)}(r) \text{ for all } r \in \I^{(3)} \text{ and }
H(r_{0},s) = H^{(4)}(s) \text{ for all } s \in \I^{(4)},
\label{2.12}
\end{equation}
and that 
\begin{equation}
\E(r,s_{0}) = \E^{(3)}(r) \text{ for all } r \in \I^{(3)} \text{ and }
\E(r_{0},s) = \E^{(4)}(s) \text{ for all } s \in \I^{(4)}.
\label{2.13}
\end{equation}
It will also turn out that $\bar{\F} \in \S_{\bar{\F}}$, $H \in \S_{H}$ and 
$\E \in \S_{\E}$; and $H$ is the member of $\S_{H}$ for which 
Eq.\ (\ref{1.79}) holds.  Moreover, it will be evident that $\bar{\F}$ 
is the only member of $\S_{\bar{\F}}$ that satisfies Eqs.\ (\ref{2.11}),
$H$ is the only member of $\S_{H}$ that satisfies Eqs.\ (\ref{2.12})
and $\E$ is the only member of $\S_{\E}$ that satisfies Eqs.\ 
(\ref{2.13}).
\end{abc}


\setcounter{theorem}{0}
\setcounter{equation}{0}
\subsection{The initial value functions $\E^{(i)}$, $H^{(i)}$
and $\bar{\F}^{(i)}$}

\begin{abc}
\begin{definition}{Dfns.\ of $\rho^{(i)}$, $z^{(i)}$, $\rho_{0}$
and $z_{0}$\label{def29}}
Let $\rho^{(i)}$ and $z^{(i)}$ denote the real-valued functions 
such that $\dom{\rho^{(i)}} = \dom{z^{(i)}} = \I^{(i)}$, 
\begin{equation}
\rho^{(i)}(x^{i}) := (-1)^{i-1} \frac{x_{0}^{7-i} - x^{i}}{2}
\label{2.14a}
\end{equation}
and
\begin{equation}
z^{(i)}(x^{i}) := \frac{x_{0}^{7-i}+x^{i}}{2}.
\label{2.14b}
\end{equation}
Observe that Eq.\ (\ref{2.14a}) is equivalent to the pair of equations
\begin{equation}
\rho^{(3)}(r) = \frac{s_{0}-r}{2}, \quad
\rho^{(4)}(s) = \frac{s-r_{0}}{2},
\label{2.14c}
\end{equation}
and Eq.\ (\ref{2.14b}) is equivalent to the pair of equations
\begin{equation}
z^{(3)}(r) = \frac{s_{0}+r}{2}, \quad
z^{(4)}(s) = \frac{s+r_{0}}{2}.
\label{2.14d}
\end{equation}
Comparison of the above Eqs.\ (\ref{2.14c}) and (\ref{2.14d}) with
Eqs.\ (\ref{1.2}) shows that $\rho^{(i)}$ and $z^{(i)}$ are the
restrictions of $\rho$ and $z$, respectively, to the two null lines
in $\domE$ that pass through $\x_{0}$.  Let
\begin{equation}
\rho_{0} := \frac{s_{0}-r_{0}}{2}, \quad
z_{0} := \frac{s_{0}+r_{0}}{2}.
\label{2.14e}
\end{equation}
\end{definition}
\end{abc}

\begin{abc}
\begin{definition}{Dfns.\ of $f^{(i)}$ and $\chi^{(i)}$ (for given $\E^{(i)}$)
\label{def30}}
\begin{eqnarray}
f^{(i)} & := & \Re{\E^{(i)}} < 0 \text{ throughout } \I^{(i)},
\label{2.16b} \\
\chi^{(i)} & := & \Im{\E^{(i)}}.
\label{2.16c}
\end{eqnarray}
\end{definition}
\end{abc}

\begin{abc}
\begin{definition}{Dfn.\ of $\omega^{(i)}$ (for given $\E^{(i)})$
\label{def31}}
Let $\omega^{(i)}$ denote the real-valued function such that
$\dom{\omega^{(i)}} = \I^{(i)}$,
\begin{equation}
\omega^{(i)}(x_{0}^{i}) = 0,
\label{2.17a}
\end{equation}
$d\omega^{(i)}$ exists, and 
\begin{equation}
d\omega^{(i)} = \rho^{(i)} [f^{(i)}]^{-2} \star d\chi^{(i)}.
\label{2.17b}
\end{equation}
\end{definition}
\end{abc}

The above definition should be compared with the definition
of $\omega$ [see Eq.\ (\ref{1.11a})].  The pair of Eqs.\ (\ref{2.17a})
and (\ref{2.17b}) is clearly equivalent to the single equation
\begin{equation}
\omega^{(i)}(x^{i}) = \int_{x_{0}^{i}}^{x^{i}} d\lambda
\left[ \frac{x_{0}^{7-i} - \lambda}{2} \right]
\left[f^{(i)}(\lambda)\right]^{-2} \dot{\chi}^{(i)}(\lambda)
\label{2.17c}
\end{equation}
for all $x^{i} \in \I^{(i)}$, where we have used Eqs.\ (\ref{2.14a})
and (\ref{1.3}).

\begin{abc}
\begin{definition}{Dfns.\ of $A^{(i)}$, $h^{(i)}$ and $g_{ab}^{(i)}$
(for given $\E^{(i)})$\label{def32}}
Let
\begin{equation}
A^{(i)} := \left( \begin{array}{cc}
1 & \omega^{(i)} \\ 0 & 1 
\end{array} \right) \left( \begin{array}{cc}
1/\sqrt{-f^{(i)}} & 0 \\ 0 & \sqrt{-f^{(i)}}
\end{array} \right)
\label{2.18}
\end{equation}
and
\begin{equation}
h^{(i)} := A^{(i)} \left( \begin{array}{cc}
[\rho^{(i)}]^{2} & 0 \\ 0 & 1 
\end{array} \right) [A^{(i)}]^{T}.
\label{2.19}
\end{equation}
Let
\begin{equation}
g_{ab}^{(i)} := \text{ matrix element of } h^{(i)}
\text{ in $a$-th row and $b$-th column.}
\label{2.20}
\end{equation}
\end{definition}
\end{abc}
The above Eqs.\ (\ref{2.18}) to (\ref{2.20}) should be compared with
Eqs.\ (\ref{1.11b}) to (\ref{1.11d}).  Note that
\begin{abc}
\begin{equation}
h^{(i)} \text{ is real, symmetric and positive definite,} 
\end{equation}
\begin{eqnarray}
\det{h^{(i)}} & = & [\rho^{(i)}]^{2}, 
\label{2.21b} \\
g_{12}^{(i)} & = & g_{21}^{(i)} = - f^{(i)} \omega^{(i)},
\label{2.21c} \\
g_{22}^{(i)} & = & - f^{(i)}
\label{2.21d} 
\end{eqnarray}
and, from Eqs.\ (\ref{2.14e}), (\ref{1.77}) and (\ref{2.17a}),
\end{abc}
\begin{equation}
h^{(i)}(x_{0}^{i}) = \left( \begin{array}{cc}
\rho_{0}^{2} & 0 \\ 0 & 1
\end{array} \right)
\label{2.22}
\end{equation}
for both $i=3$ and $i=4$.  From Eq.\ (\ref{2.16b}), Eq.\ (\ref{2.16c})
and the premise that $\E^{(i)}$ is $\bC^{1}$, the real-valued functions
$F^{(i)}$ and $\chi^{(i)}$ are $\bC^{1}$.  Therefore, from Eq.\ (\ref{2.17c}),
$\omega^{(i)}$ is $\bC^{1}$; and, from Eqs.\ (\ref{2.14a}), (\ref{2.18}) and
(\ref{2.19}), $h^{(i)}$ is $\bC^{1}$.  This result is used in the following
definition.

\begin{abc}
\begin{definition}{Dfn.\ of $H^{(i)}$ (for given $\E^{(i)}$)\label{def33}}
Let $H^{(i)}$ denote the $2 \times 2$ matrix function such that
$\dom{H^{(i)}} = \I^{(i)}$,
\begin{equation}
\Re{H^{(i)}} = - h^{(i)},
\label{2.23a}
\end{equation}
$d(\Im{H^{(i)}})$ exists,
\begin{equation}
\rho^{(i)} d(\Im{H^{(i)}}) = - h^{(i)} J \star dh^{(i)}
\end{equation}
and
\begin{equation}
\Im{H^{(i)}(x_{0}^{i})} = \left( \begin{array}{cc}
0 & 0 \\ -2iz_{0} & 0
\end{array} \right).
\label{2.23c}
\end{equation}
\end{definition}
\end{abc}

Equations (\ref{2.23a}) to (\ref{2.23c}) should be compared with
Eqs.\ (\ref{1.12a}) to (\ref{1.12c}).  Recalling that $\Omega=iJ$, one
sees that the above definition of $H^{(i)}$ is equivalent to the single
equation
\begin{equation}
H^{(i)}(x^{i}) = -h^{(i)}(x^{i}) + \left( \begin{array}{cc}
0 & 0 \\ -2iz_{0} & 0 
\end{array} \right) - \int_{x_{0}^{i}}^{x^{i}} d\lambda
\left( \frac{2}{x_{0}^{7-i} - \lambda} \right) h^{(i)}(\lambda)
\Omega \dot{h}^{(i)}(\lambda)
\label{2.24}
\end{equation}
for all $x^{i} \in \I^{(i)}$, where we have used Eqs.\ (\ref{2.14a})
to (\ref{2.14e}) and (\ref{1.3}) to obtain the integrand in the above
integral.

\begin{abc}
\begin{proposition}[Properties of $H^{(i)}$]
\label{2.1B}
\mbox{ } \\ \vspace{-3ex}
\begin{romanlist}
\item 
\begin{equation}
H^{(i)}(x_{0}^{i}) = - \left( \begin{array}{cc}
\rho_{0}^{2} & 0 \\ 2iz_{0} & 1
\end{array} \right).
\label{2.25a}
\end{equation}
\item 
\begin{equation}
H^{(i)} - (H^{(i)})^{T} = 2 z^{(i)} \Omega.
\label{2.25b}
\end{equation}
\item 
\begin{equation}
H_{22}^{(i)} = \E^{(i)}.
\label{2.25c}
\end{equation}
\item 
\begin{equation}
H^{(i)} \text{ is } \bC^{1}.
\end{equation}
\end{romanlist}
\end{proposition}
\end{abc}

\proofs
\begin{romanlist}
\item 
Use Eqs.\ (\ref{2.22}), (\ref{2.23c}) and (\ref{2.24}).
\item 
Use the fact that $\Omega^{T} = - \Omega$ together with Eqs.\ (\ref{2.24}),
(\ref{1.37b}), (\ref{2.21b}), (\ref{2.14a}) and (\ref{2.14b}).
\item 
{}From Eq.\ (\ref{2.17c}) and since $\chi^{(i)}$ and $\omega^{(i)}$ are
$\bC^{1}$,
\begin{equation}
\chi^{(i)}(x^{i}) = \int_{x_{0}^{i}}^{x^{i}} d\lambda
\left( \frac{2}{x_{0}^{7-i} - \lambda} \right) [f^{(i)}(\lambda)]^{2}
\dot{\omega}^{(i)}(\lambda).
\label{2.26}
\end{equation}
Equation (\ref{2.25c}) then follows from the fact that the only non-zero
elements of $\Omega$ are $\Omega_{12}=i$ and $\Omega_{21}=-i$ taken 
together with Eqs.\ (\ref{2.24}), (\ref{2.20}), (\ref{2.21d}), (\ref{2.21c})
and (\ref{2.26}).
\item 
This follows from Eq.\ (\ref{2.24}) and the already noted fact that
$h^{(i)}$ is $\bC^{1}$.
\end{romanlist}
\cheers

\begin{definition}{Dfn.\ of $\Gamma^{(i)}$ (for given $\E^{(i)}$)
\label{def34}}
Let $\Gamma^{(i)}$ denote the $2 \times 2$ matrix function whose domain is
\begin{equation}
\dom{\Gamma^{(i)}} := \left\{ (x^{i},\tau): x^{i} \in \I^{(i)} \text{ and }
\tau \in C - \{x^{i}\}\right\}
\label{2.27}
\end{equation}
and whose values are given by Eq.\ (\ref{2.7}).
\end{definition}

\begin{proposition}[Properties of $\Gamma^{(i)}$]
\label{2.2B}
\mbox{ } \\ \vspace{-3ex}
\begin{romanlist}
\item 
For each $(x^{i},\tau) \in \dom{\Gamma^{(i)}}$,
\begin{equation}
\tr{\Gamma^{(i)}(x^{i},\tau)} = \frac{1}{2(\tau-x^{i})}.
\label{2.28}
\end{equation}
\item 
$\Gamma^{(i)}$ is continuous.  For each $x^{i} \in \I^{(i)}$, $\Gamma^{(i)}$
is a holomorphic function of $\tau$ throughout $C - \{x^{i}\}$ and vanishes
at $\tau=\infty$.
\end{romanlist}
\end{proposition}

\proofs
\begin{romanlist}
\item 
{}From Eqs.\ (\ref{2.14b}) and (\ref{2.25b}), 
$\tr{\left(\dot{H}^{(i)}\Omega\right)} = 1$,
whereupon Eq.\ (\ref{2.28}) follows from Eq.\ (\ref{2.7}).
\item 
This follows from Eq.\ (\ref{2.7}) and Prop.~\ref{2.1B}(iv).
\end{romanlist}
\cheers

\begin{abc}
\begin{definition}{Dfn.\ of $\grave{\F}^{(i)}$ (for a given $\E^{(i)}$)
\label{def35}}
Let $\grave{\F}^{(i)}$ denote any $2 \times 2$ matrix function with
\begin{equation}
\dom{\grave{\F}^{(i)}} := \left\{ (x^{i},\tau): x^{i} \in \I^{(i)} \text{ and }
\tau \in C - \grave{\I}^{(i)}(\x) \right\}
\label{2.29a}
\end{equation}
such that
\begin{equation}
\grave{\F}^{(i)}(x_{0}^{i},\tau) := I \text{ for all } \tau \in C - \{x_{0}^{i}\},
\label{2.29b}
\end{equation}
and $\partial\grave{\F}^{(i)}(x^{i},\tau)/\partial x^{i}$ exists and
\begin{equation}
\frac{\partial\grave{\F}^{(i)}(x^{i},\tau)}{\partial x^{i}} = \Gamma^{(i)}(x^{i},\tau)
\grave{\F}^{(i)}(x^{i},\tau) \text{ for all } (x^{i},\tau) \in \dom{\grave{\F}^{(i)}}.
\label{2.29c}
\end{equation}
\end{definition}
\end{abc}

\begin{abc}
\begin{proposition}[Properties of $\grave{\F}^{(i)}$]
\label{2.3B}
\mbox{ } \\ \vspace{-3ex}
\begin{romanlist}
\item 
For a given $\E^{(i)}$, $\grave{\F}^{(i)}$ exists, is unique and is continuous.
For each $x^{i} \in \I^{(i)}$, the function of $\tau$ given by
$\grave{\F}^{(i)}(x^{i},\tau)$ is holomorphic throughout 
$C - \grave{\I}^{(i)}(\x)$.
\item 
For each $(x^{i},\tau) \in \dom{\grave{\F}^{(i)}}$,
\begin{equation}
\det{\grave{\F}^{(i)}(x^{i},\tau)} = \grave{\bnu}_{i}(x^{i},x_{0}^{i},\tau),
\label{2.30}
\end{equation}
where $\grave{\bnu}_{i}$ is defined by Eqs.\ (\ref{1.6t}) and (\ref{1.6u}).
\item 
For each $(x^{i},\tau) \in \dom{\grave{\F}^{(i)}}$, the inverse of
$\grave{\F}^{(i)}(x^{i},\tau)$ exists,
$\partial\left[\grave{\F}^{(i)}(x^{i},\tau)^{-1}\right]/\partial x^{i}$
exists and
\begin{equation}
\frac{\partial\left[\grave{\F}^{(i)}(x^{i},\tau)^{-1}\right]}{\partial x^{i}}
= - \left[ \grave{\F}^{(i)}(x^{i},\tau)^{-1} \right] \Gamma^{(i)}(x^{i},\tau).
\label{2.31}
\end{equation}
\item 
For each $x^{i} \in \I^{(i)}$, the expansion of $\grave{\F}^{(i)}(x^{i},\tau)$
in inverse powers of $2\tau$ throughout $C$ minus the closed disk 
$d(x^{i},x_{0}^{i})$ with center at the origin and radius
$\sup{\{|x^{i}|,|x_{0}^{i}|\}}$ is such that
\begin{equation}
\grave{\F}^{(i)}(x^{i},\tau) = I + (2\tau)^{-1} \left[ H^{(i)}(x^{i}) -
H^{(i)}(x_{0}^{i}) \right] \Omega + O(\tau^{-2}),
\label{2.32}
\end{equation}
where $H^{(i)}(x_{0}^{i})$ is given by Eq.\ (\ref{2.25a}).
\item 
For all $(x^{i},\tau) \in \dom{\grave{\F}^{(i)}}$ such that $\tau \ne 0$,
\begin{equation}
\grave{\F}^{(i)\dagger}(x^{i},\tau) \A^{(i)}(x^{i},\tau) \grave{\F}^{(i)}(x^{i},\tau)
= \A^{(i)}(x_{0}^{i},\tau) = \left( \begin{array}{cc}
1 & i(\tau-z_{0}) \\ -i(\tau-z_{0}) & \rho_{0}^{2}
\end{array} \right),
\label{2.33a}
\end{equation}
where
\begin{equation}
\A^{(i)}(x^{i},\tau) := (\tau-z^{(i)}) \Omega + \Omega h^{(i)}(x^{i}) \Omega;
\end{equation}
and [in view of Prop.~\ref{2.3B}(iv)] Eq.\ (\ref{2.33a}) with
$\A^{(i)}(x^{i},\tau)$ and $\A^{(i)}(x_{0}^{i},\tau)$ replaced by
$\tau^{-1}\A^{(i)}(x^{i},\tau)$ and $\tau^{-1}\A^{(i)}(x_{0}^{i},\tau)$,
respectively, holds at $\tau = \infty$.
\end{romanlist}
\end{proposition}
\end{abc}

\proofs
In the proofs of (i), (ii) and (iv), we tacitly use the obvious fact that,
for each $\tau \in C -\{x_{0}^{i}\}$,
\begin{abc}
\begin{equation}
S^{(i)}(\tau) := \left\{ x^{i} \in \I^{(i)}: (x^{i},\tau) \in \dom{\grave{\F}^{(i)}}
\right\} = \left\{ x^{i} \in \I^{(i)}: \tau \in C - \grave{\I}^{(i)}(\x) \right\}
\end{equation}
is $I^{(i)}$ or is an open subinterval of $I^{(i)}$ such that
$x_{0}^{i} \in S^{(i)}(\tau)$.
\begin{romanlist}
\item 
This follows from Eq.\ (\ref{2.27}), Prop.~\ref{2.2B}(ii), and standard
theorems\footnote{See Secs.~2, 3 and~5, Ch.~II, of Ref.~\ref{Lefsch}.}
on a family of ordinary differential equations such as the one
given by Eqs.\ (\ref{2.29a}) to (\ref{2.29c}).
\item 
{}From Eq.\ (\ref{2.29c}) and Prop.~\ref{2.2B}(i),
\begin{equation}
\frac{\partial\left[\det{\grave{\F}^{(i)}(x^{i},\tau)}\right]}{\partial x^{i}}
= \frac{1}{2(\tau-x^{i})} \left[ \det{\grave{\F}^{(i)}(x^{i},\tau)} \right];
\label{2.35a}
\end{equation}
and, from Eqs.\ (\ref{1.6t}) and (\ref{1.6u}),
\begin{equation}
\frac{1}{2(\tau-x^{i})} = \frac{1}{\grave{\bnu}_{i}(x^{i},x_{0}^{i},\tau)}
\frac{\partial\grave{\bnu}_{i}(x^{i},x_{0}^{i},\tau)}{\partial x^{i}}
\label{2.35b}
\end{equation}
for all $(x^{i},\tau) \in \dom{\grave{\F}^{(i)}}$.  Equation (\ref{2.30}) now
follows from Eqs.\ (\ref{2.35a}), (\ref{2.35b}) and (\ref{2.29b}).
\item 
{}From Eq.\ (\ref{2.30}) and the fact that $\grave{\bnu}_{i}(x^{i},x_{0}^{i},\tau)
\ne 0$ for all $(x^{i},\tau) \in \dom{\grave{\F}^{(i)}}$, 
$\left[ \grave{\F}^{(i)}(x^{i},\tau) \right]^{-1}$ exists for all 
$(x^{i},\tau) \in \dom{\grave{\F}^{(i)}}$.  The existence of 
$\partial\left[\grave{\F}^{(i)}(x^{i},\tau)^{-1}\right]/\partial x^{i}$ and
Eq.\ (\ref{2.31}) then follows form the existence of
$\partial\grave{\F}^{(i)}(x^{i},\tau)/\partial x^{i}$ and Eq.\ (\ref{2.29c}).
\item 
{}From Props.~\ref{2.3B}(i) and \ref{2.2B}(ii) and Eq.\ (\ref{2.29c}),
$\grave{\F}^{(i)}(x^{i},\tau)$ and $\partial\grave{\F}^{(i)}(x^{i},\tau)/\partial x^{i}$
are both continuous functions of $(x^{i},\tau)$ thoughout $\dom{\grave{\F}^{(i)}}$
and (for fixed $x^{i} \in \I^{(i)}$) are both holomorphic functions of
$\tau$ throughout $C - \grave{\I}^{(i)}(\x)$.  It then follows from well known
theorems on holomorphic functions that there exists an infinite sequence of
$2 \times 2$ matrix functions $\grave{\F}^{(i)}_{n}$ such that $\dom{\grave{\F}^{(i)}_{n}}
= \I^{(i)}$, $\grave{\F}^{(i)}_{n}$ is $\bC^{1}$ and, for all $x^{i} \in \I^{(i)}$
and $\tau \in C - d(x^{i},x_{0}^{i})$,
\begin{equation}
\grave{\F}^{(i)}(x^{i},\tau) = \sum_{n=0}^{\infty} (2\tau)^{-n} \bar{\F}^{(i)}_{n}(x^{i})
\label{2.36a}
\end{equation}
and
\begin{equation}
\frac{\partial\grave{\F}^{(i)}(x^{i},\tau)}{\partial x^{i}} = \sum_{n=0}^{\infty}
(2\tau)^{-n} \dot{\grave{\F}}^{(i)}_{n}(x^{i}).
\label{2.36b}
\end{equation}
Moreover, the above power series are absolutely convergent and are
uniformly convergent on any compact subspace of
$\left\{ (x^{i},\tau): x^{i} \in \I^{(i)} \text{ and } \tau \in
C - d(x^{i},x_{0}^{i}) \right\}$.  Equation (\ref{2.32}) now follows
in a straightforward way after substituting from Eqs.\ (\ref{2.7}),
(\ref{2.36a}) and (\ref{2.36b}) into Eq.\ (\ref{2.29c}) and then using
Eq.\ (\ref{2.29b}).  [Note:  Similar proofs are used for Thm.~\ref{1.1E}(vi)
and Prop.~\ref{1.2G}(vi).]
\item 
The proof is similar to that of Eq.\ (\ref{1.37n}).  Simply replace $\rho$,
$z$, $h$, $H$, $\A(\x,\tau)$ and $\grave{\bsF}(\x,\x',\tau)$ in the proof of 
Thm.~\ref{1.1E}(vii) by $\rho^{(i)}$, $z^{(i)}$, $h^{(i)}$, $H^{(i)}$,
$\A^{(i)}(x^{i},\tau)$ and $\grave{\F}^{(i)}(x^{i},\tau)$, respectively.
\end{romanlist}
\cheers
\end{abc}

\begin{abc}
\begin{definition}{Dfn.\ of $\bar{\F}^{(i)}$ (for a given $\E^{(i)}$)
\label{def36}}
Let $\bar{\F}^{(i)}$ denote the extension of $\grave{\F}^{(i)}$ to the domain
\begin{equation}
\dom{\bar{\F}^{(i)}} := \left\{ (x^{i},\tau): x^{i} \in \I^{(i)} \text{ and }
\tau \in C - \bar{\I}^{(i)}(\x) \right\}
\end{equation}
such that
\begin{equation}
\bar{\F}^{(i)}(x_{0}^{i},x_{0}^{i}) := I.
\label{2.38}
\end{equation}
\end{definition}
\end{abc}

\begin{proposition}[Properties of $\bar{\F}^{(i)}$]
\label{2.4B}
\mbox{ } \\ \vspace{-3ex}
\begin{romanlist}
\item 
For each $x^{i} \in \I^{(i)}$, the function of $\tau$ given by
$\bar{\F}^{(i)}(x^{i},\tau)$ is holomorphic throughout $C -
\bar{\I}^{(i)}(\x)$.
\item 
For each $(x^{i},\tau) \in \dom{\bar{\F}^{(i)}}$,
\begin{equation}
\det{\bar{\F}^{(i)}(x^{i},\tau)} = \bar{\bnu}_{i}(x^{i},x_{0}^{i},\tau)
\end{equation}
\item 
The result of replacing $\grave{\F}^{(i)}$ by $\bar{\F}^{(i)}$ in Eq.\ (\ref{2.33a})
of Prop.~\ref{2.3B}(v) holds for all $(x^{i},\tau) \in \dom{\bar{\F}^{(i)}}$.
\end{romanlist}
\end{proposition}

\proofs
Use Prop.~\ref{2.3B}(i), (ii) and (v) together with Eqs.\ (\ref{1.6u}),
(\ref{2.29b}) and (\ref{2.38}).
\cheers


\setcounter{theorem}{0}
\setcounter{equation}{0}
\subsection{Formulating the HHP adapted to $(\bar{\F}^{(3)},\bar{\F}^{(4)})$
to solve the initial value problem}

\begin{abc}
\begin{definition}{Dfn.\ of the HHP adapted to $(\bar{\F}^{(3)},\bar{\F}^{(4)})$
\label{def37}}
{\em The} HHP {\em adapted to} $(\bar{\F}^{(3)},\bar{\F}^{(4)})$ is the set of
all $2 \times 2$ matrix functions $\bar{\F}$ with
\begin{equation}
\dom{\bar{\F}} = \left\{ (\x,\tau): \x \in \domE \text{ and } \tau \in
C - \bar{\I}(\x) \right\}
\end{equation}
such that, for each $\x \in \domE$ and $i \in \{3,4\}$, the function
of $\tau$ that is given by 
$$\bar{\F}(\x,\tau) \left[ \bar{\F}^{(i)}(x^{i},\tau) \right]^{-1}$$
has a holomorphic extension $\bar{\J}^{(7-i)}(\x,\tau)$ to
the domain $C - \bar{\I}^{(7-i)}(\x)$; and
\begin{equation}
\bar{\F}(\x,\infty) = I.
\label{2.40b}
\end{equation}
Any member $\bar{\F}$ of the HHP adapted to $(\bar{\F}^{(3)},\bar{\F}^{(4)})$ will
be called a {\em solution} of the HHP.  For each solution $\bar{\F}$, we
let $\grave{\F}$ denote the restriction of $\bar{\F}$ to
\begin{equation}
\dom{\grave{\F}} := \left\{ (\x,\tau): \x \in \domE \text{ and } \tau \in
C - \grave{\I}(\x) \right\}
\label{2.41a}
\end{equation}
and $\grave{\J}^{(i)}$ denote the restriction of $\bar{\J}^{(i)}$ to
\begin{equation}
\dom{\grave{\J}^{(i)}} := \left\{ (\x,\tau): \x \in \domE \text{ and }
\tau \in C - \grave{\I}^{(i)}(\x) \right\}.
\label{2.41b}
\end{equation}
\end{definition}
\end{abc}

An immediate consequence of the above definition of the HHP is that,
for each $\x \in \domE$ and $i \in \{3,4\}$,
\begin{abc}
\begin{equation}
\bar{\F}(\x,\tau) = \bar{\J}^{(7-i)}(\x,\tau) \bar{\F}^{(i)}(x^{i},\tau)
\text{ for all } \tau \in C - \bar{\I}(\x)
\end{equation}
and [from Prop.~\ref{2.4B}(i)] the function of $\tau$ given by $\bar{\F}(\x,\tau)$
is holomorphic throughout $C - \bar{\I}(\x)$.  Employing Eq.\
(\ref{2.40b}), one then sees that the function $H$ whose domain is $\domE$
and whose values are defined by
\begin{equation}
H(\x) := H^{M}(\x_{0}) + \left\{ 2\tau
\left[\bar{\F}(\x,\tau)-I\right]\Omega\right\}_{\tau=\infty}
\label{2.42a}
\end{equation}
exists.  We also let
\begin{equation}
\E := H_{22}, \quad h := - \Re{H}.
\label{2.42b}
\end{equation}
It remains, of course, to prove that $\bar{\F}$ exists and is unique.  One
also has to prove that (granting existence) $\bar{\F} \in \S_{\bar{\F}}$, $H \in \S_{H}$
and $\E \in \S_{\E}$.
\end{abc}

\begin{abc}
\begin{theorem}[Properties of a solution of the HHP]
\label{2.1C} \mbox{ } \\
If $\bar{\F}$ is a solution of the HHP adapted to $(\bar{\F}^{(3)},\bar{\F}^{(4)})$,
then the following statements (i) through (vii) hold:
\begin{romanlist}
\item 
For all $r \in \I^{(3)}$ and $\tau \in \bar{\I}^{(3)}(\x)$
\begin{equation}
\bar{\F}((r,s_{0}),\tau) = \bar{\F}^{(3)}(r,\tau);
\end{equation}
and, for all $s \in \I^{(4)}$ and $\tau \in \bar{\I}^{(4)}(\x)$,
\begin{equation}
\bar{\F}((r_{0},s),\tau) = \bar{\F}^{(4)}(s,\tau).
\end{equation}
\item 
For all $\tau \in C$,
\begin{equation}
\bar{\F}(\x_{0},\tau) = I.
\end{equation}
\item 
\begin{equation}
\det{\bar{\F}} = \bar{\nu}.  
\label{2.45}
\end{equation}
Corollary:  $\bar{\F}^{-1}$ exists.
\item 
There is not more than one solution of the HHP adapted to
$(\bar{\F}^{(3)},\bar{\F}^{(4)})$.
\item 
\begin{equation}
\dom{H} = \domE, \quad H(\x_{0}) = - \left( \begin{array}{cc}
\rho_{0}^{2} & 0 \\ 2iz_{0} & 1
\end{array} \right).
\label{2.46a}
\end{equation}
\item 
For all $r \in \I^{(3)}$,
\begin{equation}
H(r,s_{0}) = H^{(3)}(r), \quad \E(r,s_{0}) = \E^{(3)}(r);
\end{equation}
and, for all $s \in \I^{(4)}$,
\begin{equation}
H(r_{0},s) = H^{(4)}(s), \quad \E(r_{0},s) = \E^{(4)}(s).
\end{equation}
\item 
For all $(\x,\tau) \in \dom{\bar{\F}}$ such that $\tau \ne \infty$,
\begin{equation}
\bar{\F}^{\dagger}(\x,\tau) \left\{ \tau\Omega - \frac{1}{2} \Omega
[H(\x)+H^{\dagger}(\x)]\Omega \right\} \bar{\F}(\x,\tau) = 
\tau\Omega + \Omega \left( \begin{array}{cc}
\rho_{0}^{2} & -iz_{0} \\ iz_{0} & 1
\end{array} \right) \Omega;
\label{2.47}
\end{equation}
and the above equation divided through by $\tau$ clearly holds at
$\tau = \infty$.  As a corollary of Eqs.\ (\ref{1.6z}), (\ref{2.45})
and (\ref{2.47}),
\begin{equation}
\det \left\{ \tau\Omega - \frac{1}{2} \Omega [H(\x)+H^{\dagger}(\x)]
\Omega \right\} = - \mu(\x,\tau)^{2}.
\end{equation}

\end{romanlist}
\end{theorem}
\end{abc}

\proof
Parts (i) to (vii) of Thm.~\ref{2.1C} are, except for certain
notations and conventions and except for the domain of the Ernst
potential in $\{\x:r<s\}$, essentially the same as Thms.~1 to~7, 
respectively, in the third of four earlier papers (IVP1, IVP2, IVP3 
and IVP4) by the authors on the initial value problem for colliding
gravitational plane wave pairs.  The papers that are relevant to our 
present subject are IVP3 and IVP4.\footnote{See Ref.~\ref{IVP34}.}
Since the proofs in those papers are quite detailed, the reader should
have no difficulty in adapting them to the current set of notes by 
employing our tables of corresponding notations and conventions\footnote{ 
See Appendix A.}  In particular, see Sec.~VI of IVP3 for the proofs of
Thms.~\ref{2.1C}(i) through (vii).
\cheers

\begin{abc}
\begin{theorem}[More properties of a solution of the HHP]
\label{2.2C} \mbox{ } \\
If $\bar{\F}$ is a solution of the HHP adapted to $(\bar{\F}^{(3)},\bar{\F}^{(4)})$
such that, for both $i=3$ and $i=4$, 
\begin{equation}
\begin{array}{r}
d\grave{\J}^{(i)} \text{ and } \partial^{2}\grave{\J}^{(i)}/\partial r \partial s
\text{ (as well as $\grave{\J}^{(i)}$) exist and are continuous } \\
\text{ functions of $(\x,\tau)$ throughout the domain (\ref{2.41b}),}
\end{array}
\end{equation}
then the following statements (i) to (vi) hold:
\begin{romanlist}
\item 
$\grave{\F}(\x,\tau)$, $d\grave{\F}(\x,\tau)$ and $\partial^{2}\grave{\F}(\x,\tau)/\partial r
\partial s$ exist and are continous functions of $(\x,\tau)$ throughout
the domain (\ref{2.41a}).

Corollary:  $dH$ and $\partial^{2}H/\partial r \partial s$ exist and
are continuous throughout $\domE$.
\item 
Throughout $\dom{\grave{\F}}$,
\begin{equation}
d\grave{\F} = \Gamma \grave{\F},
\end{equation}
where
\begin{equation}
\Gamma(\x,\tau) := \frac{1}{2} (\tau-z+\rho\star)^{-1} dH(\x) \Omega
\end{equation}
and
\begin{equation}
z := \frac{1}{2}(s+r), \quad \rho := \frac{1}{2}(s-r).
\end{equation}
\item 
\begin{equation}
\frac{1}{2}(H+H^{\dagger})\Omega dH = (z-\rho\star) dH.
\end{equation}
\item 
\begin{equation}
\frac{1}{2}(H-H^{T}) = z\Omega.
\end{equation}

Corollary:  $H := - \Re{H}$ is symmetric.
\item 
\begin{equation}
\det{h} = \rho^{2}.
\end{equation}

Corollary:  $h$ is positive definite throughout $\domE$.
\item 
\begin{equation}
h \Omega dH = \rho \star dH.
\label{2.53a}
\end{equation}

Corollary:
\begin{equation}
d(\rho \star dH) = dh \Omega dH.
\label{2.53b}
\end{equation}

Corollary:  $\E$ satisfies the Ernst equation; and
\begin{equation}
\E(r,s_{0}) = \E^{(3)}(r) \text{ for all } r \in \I^{(3)}
\end{equation}
and
\begin{equation}
\E(r_{0},s) = \E^{(4)}(s) \text{ for all } s \in \I^{(4)}.
\end{equation}
\end{romanlist}
\end{theorem}
\end{abc}

\proof
The proofs are essentially the same as the proofs of Thms.~8
through~13, respectively, in Sec.~VI of IVP3.  

The corollary to Thm.~\ref{2.2C}(v) follows easily from the 
corollary to Thm.~\ref{2.2C}(iv), from Thm.~\ref{2.2C}(v) itself, 
the fact that
$h(\x_{0}) = \left( \begin{array}{cc}
\rho_{0}^{2} & 0 \\ 0 & 1
\end{array} \right)$ [see Eqs.\ (\ref{2.42b}) and (\ref{2.46a})]
and the continuity of $h$ [corollary to Thm.~\ref{2.2C}(i)].

As for the second corollary to Thm.~\ref{2.2C}(vi), the fact that 
$\E := H_{22}$ is a solution of the Ernst equation 
$$f d(\rho\star d\E) = \rho d\E \star d\E$$
is derived from the $(2,2)$ matrix elements of Eqs.\ (\ref{2.53a})
and (\ref{2.53b}) by a well known and simple algebraic procedure.
One uses the definition $h := - \Re{H}$ in the derivation.
\cheers


\setcounter{theorem}{0}
\setcounter{equation}{0}
\subsection{An equivalent Fredholm integral equation of the second kind}

Our proofs that the solution of the HHP adapted to
$(\bar{\F}^{(3)},\bar{\F}^{(4)})$
exists and that this solution has certain differentiability-continuity
properties employ a Fredholm integral equation of the second kind that
is equivalent to this HHP.  To help us define this integral equation, let
$\C_{\x}$ denote the set of all
\begin{abc}
\begin{equation}
\Lambda = \Lambda_{3} \cup \Lambda_{4}
\end{equation}
such that $\Lambda_{3}$ and $\Lambda_{4}$ are smooth, simple, positively
oriented and non-intersecting closed contours in the finite complex plane;
and
\begin{equation}
\grave{\I}^{(3)}(\x) \subset \Lambda_{3}^{+}, \quad
\grave{\I}^{(4)}(\x) \subset \Lambda_{4}^{+}, \quad
\grave{\I}^{(3)}(\x) \subset \Lambda_{4}^{-}, \quad
\grave{\I}^{(4)}(\x) \subset \Lambda_{3}^{-},
\end{equation}
where $\Lambda_{i}^{+}$ and $\Lambda_{i}^{-}$ are those open subsets of
$C$ that lie inside and outside $\Lambda_{i}$, respectively, and that
have $\Lambda_{i}$ as their common boundary.  Also, let $\bar{K}_{i} \,
(i=3,4)$ denote the function with the domain
\begin{equation}
\dom{\bar{K}_{i}} := \left\{ (x^{i},\sigma,\tau): x^{i} \in \I^{(i)}
\text{ and } (\sigma,\tau) \in [C - \bar{\I}^{(i)}(\x)]^{2}\right\},
\end{equation}
the values
\begin{equation}
\bar{K}_{i}(x^{i},\sigma,\tau) :=
\frac{\left[\grave{\F}^{(i)}(x^{i},\sigma)\right]^{-1}
\grave{\F}^{(i)}(x^{i},\tau)-I}{\sigma-\tau}
\end{equation}
when $x^{i} \ne x_{0}^{i}$, and the values
\begin{equation}
\bar{K}_{i}(x_{0}^{i},\sigma,\tau) := 0
\end{equation}
for all $(\sigma,\tau) \in C^{2}$.  Let $\grave{K}_{i}$ denote the
restriction of $\bar{K}_{i}$ to the domain
\begin{equation}
\dom{\grave{K}_{i}} := \left\{ (x^{i},\sigma,\tau) \in \dom{\bar{K}_{i}}:
\sigma \ne x_{0}^{i} \text{ and } \tau \ne x_{0}^{i} \right\}.
\end{equation}

The kernel of the integral equation is the function $\bar{K}$ whose domain is
\begin{equation}
\dom{\bar{K}} := \left\{ (\x,\Lambda,\sigma,\tau): \x \in \domE, \Lambda \in \C_{\x},
\sigma \in \Lambda \text{ and } \tau \in C - \bar{\I}(\x) \right\}
\end{equation}
and whose values are
\begin{equation}
\bar{K}(\x,\Lambda,\sigma,\tau) := \bar{K}_{i}(x^{i},\sigma,\tau) \text{ when }
\sigma \in \Lambda_{i}.
\end{equation}
Let $\grave{K}$ denote the restriction of $\bar{K}$ to 
\begin{equation}
\dom{\grave{K}} = \left\{ (\x.\Lambda,\sigma,\tau) \in \dom{\bar{K}}: \tau \ne r_{0}
\text{ and } \tau \ne s_{0} \right\}.
\end{equation}
(When computing $\bar{\F}$ from the integral equation, one uses $\bar{K}$.  When
computing the restriction $\grave{\F}$, one uses $\grave{K}$.)
\end{abc}

\begin{definition}{Dfn.\ of Fredholm equation problem\label{def38}}
The {\em Fredholm equation problem} corresponding to any given
$(\bar{\F}^{(3)},\bar{\F}^{(4)})$, $\x \in \domE$ and $\Lambda \in \C_{\x}$ is the
set of all $2 \times 2$ matrices $\bar{\F}(\x,\tau)$ that are continuous
functions of $\tau$ on $\Lambda$ and satisfy the integral equation
\begin{equation}
\bar{\F}(\x,\tau) - \frac{1}{2\pi i} \int_{\Lambda} d\sigma \, \bar{\F}(\x,\sigma)
\bar{K}(\x,\Lambda,\sigma,\tau) = I
\label{2.57}
\end{equation}
for all $\tau \in \Lambda$.  Any member of this set will be called a
{\em solution} of the Fredholm equation corresponding to $(\bar{\F}^{(3)},\bar{\F}^{(4)})$,
$\x$ and $\Lambda$.
\end{definition}

\begin{theorem}[HHP--Fredholm equation equivalence]
\label{2.1D} 
\mbox{ } \\ \vspace{-3ex}
\begin{romanlist}
\item 
Let $\bar{\F}$ denote a solution of the HHP adapted to $(\bar{\F}^{(3)},\bar{\F}^{(4)})$.
Then, for each choice of $\x \in \domE$ and $\Lambda \in \C_{\x}$,
$\bar{\F}(\x,\tau)$ with $\tau$ restricted to $\Lambda$ is a solution of the
Fredholm equation corresponding to $(\bar{\F}^{(3)},\bar{\F}^{(4)})$, $\x$ and 
$\Lambda$.
\item 
Conversely, suppose that $\bar{\F}(\x,\tau)$ is a solution of the Fredholm
equation corresponding to $(\bar{\F}^{(3)},\bar{\F}^{(4)})$, $\x$ and $\Lambda$.
Then (for fixed $\x$) the holomorphic extension of $\bar{\F}(\x,\tau)$ to
the domain $C - \bar{\I}(\x)$ exists and is obtained simply by
letting $\tau \in C - \bar{\I}(\x)$ on the right hand side of
Eq.\ (\ref{2.57}); and this holomorphic extension is a solution of
the HHP adapted to $(\bar{\F}^{(3)},\bar{\F}^{(4)})$.  Moreover, the solution of
the HHP that is thus obtained is independent of the choice of $\Lambda
\in \C_{\x}$ since there can be no more than one solution of the HHP
according to Thm.~\ref{2.1C}(iv).
\end{romanlist}
\end{theorem}

\proof
The proof of statement (i) is essentially the same as the one given 
in Sec.~VIIA of IVP3, and the proof of statement (ii) is essentially 
the same as the proof of Thm.~14 in Sec.~VIIB of IVP3.
\cheers

\begin{theorem}[Existence and uniqueness, etc.]
\label{2.2D}
\mbox{ } \\ \vspace{-3ex}
\begin{romanlist}
\item 
Let $N_{3}$ and $N_{4}$ denote any positive integers.  Suppose $\E^{(3)}$
is $\bC^{N_{3}}$ and $\E^{(4)}$ is $\bC^{N_{4}}$.  Then the following
statements hold:
\begin{letterlist}
\item 
The functions $H^{(j)}$, $\grave{\F}^{(j)}$ and $\grave{K}_{j}$ are
each $\bC^{N_{j}} \, (j=3,4)$.
\item 
For all $0 \le m \le N_{j}$ and for each given $x^{j} \in \I^{(j)}$,
$\partial^{m}\grave{\F}^{(i)}(x^{j},\tau)/(\partial x^{j})^{m}$ is a holomorphic
function of $\tau$ throughout $C - \grave{\I}^{(j)}(\x)$ and
$\partial^{m}\grave{K}_{j}(x^{j},\sigma,\tau)/(\partial x^{j})^{m}$
is a holomorphic function of $(\sigma,\tau)$ throughout
$[C - \grave{\I}^{(j)}(\x)]^{2}$.
\end{letterlist}
\item 
There exists exactly one solution of the Fredholm integral equation corresponding
to $(\bar{\F}^{(3)},\bar{\F}^{(4)})$, $\x$ and $\Lambda$.  Hence, from Thm.~\ref{2.1D},
there exists exactly one solution of the HHP adapted to $(\bar{\F}^{(3)},\bar{\F}^{(4)})$.
\item 
\mbox{ } \\ \vspace{-3ex}
\begin{letterlist}
\item 
If $\E^{(3)}$ is $\bC^{N_{3}}$ and $\E^{(4)}$ is $\bC^{N_{4}}$, then
for all $0 \le m \le N_{3}$ and $0 \le n \le N_{4}$,
$$\frac{\partial^{m+n}\grave{\F}(\x,\tau)}{(\partial r)^{m} (\partial s)^{n}}$$
exist and are continuous functions of $(\x,\tau)$ throughout $\dom{\grave{\F}}$.
For fixed $\x$, the same partial derivatives are holomorphic functions of
$\tau$ throughout $C - \grave{\I}(\x)$.  Moreover, from Eq.\ (\ref{2.42a})
and the preceding statement concerning the continuity and holomorphy
(for fixed $\x$) of the partial derivatives of $\grave{\F}$, a well known
theorem tells us that the partial derivatives
$$\frac{\partial^{m+n}H(\x)}{(\partial r)^{m} (\partial s)^{n}}
\text{ for } 0 \le m \le N_{3} \text{ and } 0 \le n \le N_{4}$$
also exist and are continuous functions of $\x$ throughout $\domE$.
\item 
If $\E^{(3)}$ and $\E^{(4)}$ are $\bC^{\infty}$, so are $\grave{\F}$ and $H$.
\item 
If $\E^{(3)}$ and $\E^{(4)}$ are holomorphic, so are $\grave{\F}$ and $H$.
\end{letterlist}
\item 
For the same premises and notations as in Thm.~\ref{2.2D}(iii)(a), 
$$\frac{\partial^{m+n} \grave{\J}^{(j)}(\x,\tau)}{(\partial r)^{m} 
(\partial s)^{n}} \text{ for } 0 \le m \le N_{3} \text{ and }
0 \le n \le N_{4}$$
exist and are continuous functions of $(\x,\tau)$ throughout
$\dom{\grave{\J}^{(j)}}$.
\end{romanlist}
\end{theorem}

\proofs
The proofs of (i), (ii), (iii) and (iv) are essentially the same as
the proofs of Thms.~1, 2, 9 and~10 in Secs.~IIE, IIIA, VD and VD, 
respectively, of IVP4.  The reader will find that the proofs of 
Thms.~(ii) and~(iii) are far from trivial.  In particular, the proof
of Thm.~(iii) requires a lengthy and intricate analysis that is 
detailed in Secs.~IV and~V of IVP4.
\cheers

Note:  From the above Thm.~\ref{2.2D}(iv) for the case $N_{3}=N_{4}=1$,
it can be seen that the unique solution $\bar{\F}$ of the HHP adapted to
$(\bar{\F}^{(3)},\bar{\F}^{(4)})$ satisfies the premise of Thm.~\ref{2.2C}.


\setcounter{theorem}{0}
\setcounter{equation}{0}
\subsection{Principal results of Part I} 

\begin{gssm}[Properties of $\S_{\bar{\F}}\triple$]
\label{Thm_3}
\mbox{ } \\ \vspace{-3ex}
\begin{romanlist}
\item 
For each $H \in \S_{H}\triple$, the corresponding $\bar{\F} \in 
\S_{\bar{\F}}\triple$ exists and is unique; and, for each $\x \in \domE$,
$\bar{\F}(\x,\tau)$ is a holomorphic function of $\tau$ throughout 
$C - \bar{\I}(\x)$ and, in at least one neighborhood of $\tau = \infty$,
$$
\bar{\F}(\x,\tau) = I + (2\tau)^{-1} \left[ H(\x) - H(\x_{0}) \right] \Omega
+ O(\tau^{-2}).
$$
\item 
For each $\bar{\F} \in \S_{\bar{\F}}\triple$, there is only one 
$H \in \S_{H}\triple$ for which $d\bar{\F}(\x,\tau) = \Gamma(\x,\tau)
\bar{\F}(\x,\tau)$.
\item 
With the understanding that
$$
\dom{\bar{\nu}} := \{ (\x,\tau): \x \in \domE \text{ and } 
\tau \in C - \bar{\I}(\x)\}
$$
and that $\bar{\nu}(\x,\infty) = 1$, we have
$$
\det{\bar{\F}(\x,\tau)} = \bar{\nu}(\x,\tau) :=
\left( \frac{\tau-r_{0}}{\tau-r} \right)^{1/2}
\left( \frac{\tau-s_{0}}{\tau-s} \right)^{1/2}.
$$
\item 
The member of $\S_{\bar{\F}}\triple$ that corresponds to $\E^{M}$ is given by
\begin{equation}
\bar{\F}^{M}(\x,\tau) = \left( \begin{array}{cc}
1 & -i(\tau-z) \\ 0 & 1
\end{array} \right) \left( \begin{array}{cc}
1 & 0 \\ 0 & \bar{\nu}(\x,\tau)
\end{array} \right) \left( \begin{array}{cc}
1 & i(\tau-z_{0}) \\ 0 & 1
\end{array} \right).
\label{G2.10c}
\end{equation}
\end{romanlist}
\end{gssm}

\proof
The proof of GSSM~\ref{Thm_3}(i) is given by the combined proofs of
Props.~\ref{1.2G}(i), \ref{1.2G}(iv), \ref{1.2G}(vi) and \ref{1.21G}.
GSSM~\ref{Thm_3}(ii) is an obvious consequence of Eq.\ (\ref{1.16a}),
Eq.\ (\ref{1.59h}) and the fact that $\domE$ is connected.  The proof
of GSSM~\ref{Thm_3}(iii) is the proof of Prop.~\ref{1.2G}(ii).  Finally,
GSSM~\ref{Thm_3}(iv) clearly follows from Prop.~\ref{1.1G} and
Prop.~\ref{1.1C}.
\cheers

\begin{gssm}[Properties of $\S_{H}\triple$]
\label{Thm_2}
\mbox{ } \\ \vspace{-3ex}
\begin{romanlist}
\item 
For each $\E \in \S_{\E}\triple$, there is exactly one $H \in \S_{H}\triple$
such that $\E = H_{22}$.
\item 
If, for each $i \in \{3,4\}$, $\E^{(i)}$ is $\bC^{n_{i}}$ ($n_{i} \ge
1$), then, for all $0 \le k < n_{3}$ and $0 \le m \le n_{4}$,
the partial derivatives $\partial^{k+m}H(\x)/\partial r^{k} \partial
s^{m}$ exist and are continuous throughout $\domE$.  If, for each $i \in
\{3,4\}$, $\E^{(i)}$ is analytic, then $H$ is analytic.
\end{romanlist}
\end{gssm}

\proof
The proof of GSSM~\ref{Thm_2}(i) is given by the proof of Prop.~\ref{1.1B}.
The proof of GSSM~\ref{Thm_2}(ii) is contained in the proof of
Thm.~\ref{2.2D}(iii).
\cheers

\begin{gssm}[Property of $\S_{\E}\triple$]
\label{Thm_1}
For each choice of complex valued functions $\E^{(3)}$ and $\E^{(4)}$ 
for which (for $i \in \{3,4\}$) $\dom{\E^{(i)}} = \I^{(i)}$, $\E^{(i)}$ 
is $\bC^{1}$, $f^{(i)} := \Re{\E^{(i)}} < 0$ throughout $\I^{(i)}$, and
$\E^{(3)}(r_{0}) = -1 = \E^{(4)}(s_{0})$, there exists exactly one 
$\E \in \S_{\E}\triple$ such that
$$
\E^{(3)}(r) = \E(r,s_{0}) \text{ and } \E^{(4)}(s) = \E(r_{0},s)
$$
for all $r \in \I^{(3)}$ and $s \in \I^{(4)}$, respectively.
\end{gssm}

\begin{abc}
\proof
{}From Thm.~\ref{2.2D}(i) for the case $N_{3}=N_{4}=1$, $\E := H_{22}$,
$d\E$ and $\partial^{2}\E/\partial r \partial s$ exist and are
continuous throughout $\domE$; and, from the second corollary of 
Thm.~\ref{2.2C}(vi), $\E$ satisfies the Ernst equation and the
initial value conditions
\begin{equation}
\begin{array}{rcl}
\E(r,s_{0}) & = & \E^{(3)}(r) \text{ for all } r \in \I^{(3)}, \\
\E(r_{0},s) & = & \E^{(4)}(s) \text{ for all } s \in \I^{(4)}. 
\end{array}
\label{2.58}
\end{equation}
Since $\E^{(3)}(r_{0}) = \E^{(4)}(s_{0}) = -1$ by definition,
$\E(\x_{0}) = -1$.  Therefore, $\E \in \S_{\E}$.

The only part of GSSM~\ref{Thm_1} that is left to be proved is the assertion
that, for given initial data $(\E^{(3)},\E^{(4)})$, there exists no
more than one $\E \in \S_{\E}$ such that Eqs.\ (\ref{2.58}) hold.
[This is well known, but it is instructive to see how easily the
result is derived with the aid of the HHP adapted to 
$(\bar{\F}^{(3)},\bar{\F}^{(4)})$.] Suppose $\E_{1}, \E_{2} \in \S_{\E}$
such that
\begin{equation}
\begin{array}{rcccl}
\E_{1}(r,s_{0}) & = & \E_{2}(r,s_{0}) & = & \E^{(3)}(r) 
\text{ for all } r \in \I^{(3)}, \\
\E_{1}(r_{0},s) & = & \E_{2}(r_{0},s) & = & \E^{(4)}(s) 
\text{ for all } s \in \I^{(4)}. 
\end{array}
\end{equation}
{}From GSSM~\ref{Thm_2}(i), $\E_{1}, \E_{2}$ uniquely determine $H_{1}, H_{2}
\in \S_{H}$ such that 
\begin{equation}
\E_{1} = (H_{1})_{22}, \quad \E_{2} = (H_{2})_{22};
\label{2.60}
\end{equation}
and, according to GSSM~\ref{Thm_3}(i), $H_{1}, H_{2}$ then uniquely
determine $\bar{\F}_{1}, \bar{\F}_{2} \in \S_{\bar{\F}}$ such that ($b \in \{1,2\}$)
\begin{equation}
\bar{\F}_{b}(\x,\tau) = I + (2\tau)^{-1} \left[H_{b}(\x)-H_{b}(\x_{0})\right]\Omega
+ O(\tau^{-2})
\label{2.61}
\end{equation}
in a neighborhood of $\tau=\infty$.  However, from Prop.~\ref{2.1A},
$\bar{\F}_{1}$ and $\bar{\F}_{2}$ are both solutions of the HHP adapted to
$(\bar{\F}^{(3)},\bar{\F}^{(4)})$.  So, from Thm.~\ref{2.1C}(iv), $\bar{\F}_{1}
= \bar{\F}_{2}$; and Eqs.\ (\ref{2.60}) and (\ref{2.61}) then yield 
$\E_{1} = \E_{2}$.
\cheers
\end{abc}

Throughout the rest of these notes we shall distinguish the theorems that
constitute the principal results of this work by GSSM, which acronym is
based upon the title ``Group Structure of the Solution Manifold of the 
Hyperbolic Ernst Equation.''

\newpage
\part{Quest for a new HHP and a generalized Geroch conjecture}

\setcounter{equation}{0}
\setcounter{theorem}{0}
\section{Existence and continuity of $\Q^{+}$; its employment to
construct other spectral potentials and generalized Abel transforms
$V^{(3)},V^{(4)}$ of initial data $\E^{(3)},\E^{(4)}$
\label{Sec_3}}

\begin{nosubsec}

The proofs of the theorems in Secs.~\ref{Sec_1} and~\ref{Sec_2} were
relatively simple compared to those that we shall now describe.  In the
HHP that we developed\footnote{Ibid.} to effect K--C transformations
among analytic solutions of the Ernst equation, a closed contour 
surrounding the cut $\hat{\I}(\x)$ was involved.  Here, where
we are foregoing the pleasures of analytic functions, we must instead
formulate the HHP on the arcs $\bar{\I}^{(3)}(\x)$ and
$\bar{\I}^{(4)}(\x)$.  However, in order to define the HHP in
a plausible way, we must first figure out what properties should be
demanded of a solution of such an HHP.  For this a comprehensive
study of limits as one approaches $\bar{\I}(\x)$ is necessary,
especially at the end points of the arcs $\bar{\I}^{(3)}(\x)$
and $\bar{\I}^{(4)}(\x)$.  It turns out to be easier to investigate
these limits using the $\Q$-potential than it would be using the 
$\F$-potential, even though the HHP itself is easier to formulate 
in terms of the $\F$-potential.

The chief task of Sec.~\ref{Sec_3} will be to prove that, for each given
$\E \in \S_{\E}$, a unique $\Q^{+} \in \S_{\Q^{+}}$ exists and that
$\Q^{+}$ is continuous.  We shall then be able to construct in terms of
$\Q^{+}$ all the other potentials in which we are interested, and to 
show that those potentials are continuous and unique for any given $\E
\in \S_{\E}$.  Finally, we shall undertake a detailed study of 
$\bsQ^{+}(\x,(\sigma,s),\sigma)$ and $\bsQ^{+}(\x,(r,\sigma),\sigma)$
and introduce the important functions $V^{(3)}$ and $V^{(4)}$.

\end{nosubsec}


\setcounter{theorem}{0}
\setcounter{equation}{0}
\subsection{Introduction to the existence-continuity proofs}

Recall that we have not yet proved the existence and continuity of
the spectral potential $\bsQ^{\pm} \in \S_{\subbsQ^{\pm}}$ corresponding to
any given $\E \in \S_{\E}$.  This potential was defined and discussed
in Sec.~\ref{Sec_1}B, where we proved that [Props.~\ref{1.3A}(iii) and~(iv)]
\begin{abc}
\begin{equation}
\bsQ^{\pm}(\x,\x'',\tau)\bsQ^{\pm}(\x'',\x',\tau) = \bsQ^{\pm}(\x,\x',\tau)
\text{ for all } \x,\x'',\x' \in \domE \text{ and } \tau \in \bar{C}^{\pm}
\label{3.1a}
\end{equation}
and (as a corollary)
\begin{equation}
[\bsQ^{\pm}(\x,\x',\tau)]^{-1} = \bsQ^{\pm}(\x',\x,\tau)
\end{equation}
for each $\bsQ^{\pm} \in \S_{\subbsQ^{\pm}}$.  Furthermore, [Prop.~\ref{1.1F}]
for each given $\E \in \S_{\E}$, the corresponding $\bsQ^{\pm} \in \S_{\subbsQ^{\pm}}$
exists iff the corresponding $\Q^{\pm} \in \S_{\Q^{\pm}}$ exists; and $\Q^{\pm}$
is given in terms of $\bsQ^{\pm}$ by
\begin{equation}
\Q^{\pm}(\x,\tau) = \bsQ^{\pm}(\x,\x_{0},\tau)
\label{3.1c}
\end{equation}
while $\bsQ^{\pm}$ is given in terms of $\Q^{\pm}$ by
\begin{equation}
\bsQ^{\pm}(\x,\x',\tau) = \Q^{\pm}(\x,\tau) [\Q^{\pm}(\x',\tau)]^{-1}.
\label{3.1d}
\end{equation}
Note that Eq.\ (\ref{3.1d}) is implied by Eqs.\ (\ref{3.1a}) to (\ref{3.1c}).
Also, since $\Q^{\pm}(\x_{0},\tau) = I$ by definition (Sec.~\ref{Sec_1}E), Eqs.\
(\ref{3.1a}) to (\ref{3.1c}) are implied by Eq.\ (\ref{3.1d}).  Hence,
\begin{equation}
\text{ Eq.\ (\ref{3.1d}) is equivalent to the 
triad of Eqs.\ (\ref{3.1a}) to (\ref{3.1c}). }
\end{equation}
\end{abc}

We shall now use one of the classical methods of reducing the solving of the
a total differential equation such as $d\Q^{\pm} = \Delta^{\pm} \Q^{\pm}$ to 
the solving of a pair of ordinary linear differential equations along
characteristic lines in $\domE$.  For this purpose let us consider the values
of $\bsQ^{\pm}(\x,\x',\tau)$ along null lines corresponding to fixed $r=r'$ 
or fixed $s=s'$, defining
\begin{abc}
\begin{eqnarray}
\bsQ^{(4)\pm}(s,s',r,\tau) & := & \bsQ^{\pm}(\x,(r,s'),\tau)
\text{ (tentative dfn.) } 
\label{3.2a} \\
\bsQ^{(3)\pm}(r,r',s,\tau) & := & \bsQ^{\pm}(\x,(r',s),\tau)
\text{ (tentative dfn.) }
\label{3.2b}
\end{eqnarray}
for all $(s,s',r,\tau)$ and $(r,r',s,\tau)$, respectively, such that
$\x := (r,s)$, $(r,s')$, $(r',s) \in \domE$ and $\tau \in \bar{C}^{\pm}$.
Then, application of Eqs.\ (\ref{3.1a}) and (\ref{3.1c}) yields
\begin{equation}
\Q^{\pm}(\x,\tau) = \bsQ^{(4)\pm}(s,s_{0},r,\tau) 
\bsQ^{(3)\pm}(r,r_{0},s_{0},\tau)
= \bsQ^{(4)\pm}(s,s_{0},r,\tau) \Q^{(3)\pm}(r,\tau),
\label{3.2c}
\end{equation}
where
\begin{equation}
\Q^{(3)\pm}(r,\tau) := \Q^{\pm}((r,s_{0}),\tau).
\end{equation}
The reason for introducing the above Eq.\ (\ref{3.2c}) is that each of the
two factors on the right side of this equation may be defined as a solution
of an {\em ordinary} homogeneous linear differential equation.  Specifically,
upon letting [see Eqs.\ (\ref{1.7a}) to (\ref{1.7c}) for the definition of
$\Delta^{\pm}$]
\end{abc}
\begin{abc}
\begin{equation}
\Delta^{\pm}(\x,\tau) = dr \Delta_{3}^{\pm}(\x,\tau) +
ds \Delta_{4}^{\pm}(\x,\tau),
\label{3.3a}
\end{equation}
one finds that Eqs.\ (\ref{3.2a}), (\ref{3.2b}), (\ref{1.9b}) and (\ref{1.9c})
yield, for $s \ne \tau$,
\begin{equation}
\frac{\partial\bsQ^{(4)\pm}(s,s',r,\tau)}{\partial s} 
= \Delta_{4}^{\pm}(\x,\tau) \bsQ^{(4)\pm}(s,s',r,\tau)
\label{3.3b}
\end{equation}
and, for $r \ne \tau$,
\begin{equation}
\frac{\partial\bsQ^{(3)\pm}(r,r',s,\tau)}{\partial r} 
= \Delta_{3}^{\pm}(\x,\tau) \bsQ^{(3)\pm}(r,r',s,\tau)
\label{3.3c}
\end{equation}
subject to the initial conditions
\begin{eqnarray}
\bsQ^{(4)\pm}(s',s',r,\tau) & = & I 
\label{3.3d} \\
& \text{and} & \nonumber \\
\bsQ^{(3)\pm}(r',r',s,\tau) & = & I, 
\label{3.3e}
\end{eqnarray}
respectively.  The above Eqs.\ (\ref{3.2c}) to (\ref{3.3e}) suggest the
following plan:
\end{abc}
\begin{enumerate}
\item
Define $\bsQ^{(4)\pm}$ and $\bsQ^{(3)\pm}$ as solutions of the ordinary
differential equations (\ref{3.3b}) and (\ref{3.3c}) subject to the initial
conditions (\ref{3.3d}) and (\ref{3.3e}), respectively, and to certain
simple continuity conditions [instead of defining them in terms of $\bsQ^{\pm}$
by Eqs.\ (\ref{3.2a}) and (\ref{3.2b})].
\item
Then demonstrate the existence, continuity and other interesting properties
of $\bsQ^{(3)\pm}$ and $\bsQ^{(4)\pm}$ by employing the Picard
method of successive approximations and certain well known theorems on
infinite sequences of functions.
\item
Then demonstrate that the right side of Eq.\ (\ref{3.2c}) is indeed
$\Q^{\pm}(\x,\tau)$, whereupon the existence, continuity and other properties
of $\Q^{\pm}$ will automatically follow from the existence, continuity and
other properties of its factors.
\end{enumerate}

The second phase of the above plan is needed, since the usual well known
theorems on the existence, continuity, holomorphy and differentiability
of a solution of an ordinary linear differential equation are not
applicable here.  This is because the functions $\Delta_{i}^{\pm}(\x,\tau)$
that are employed in the differential equations (\ref{3.3b}) and (\ref{3.3c})
do not satisfy all premises of these well known theorems.  Specifically,
$\Delta_{i}^{\pm}(\x,\tau)$ has a non-finite singularity at each point of
the domain of $\bsQ^{(i)\pm}$ at which $x=\tau$; and
$\partial\Delta_{i}^{\pm}(\x,\tau)/\partial y$ has non-finite
singularities at those points of $\dom{\bsQ^{(i)\pm}}$ at which $x=\tau$
and at which $y=\tau$.  These singularities make the proofs of the
existence, continuity and certain other properties of $\bsQ^{(i)\pm}$
somewhat of a challenge.


\setcounter{theorem}{0}
\setcounter{equation}{0}
\subsection{The sets $\domE_{i}$, $\D^{\pm}_{i}$, $\D^{\pm}_{0i}$, 
$\D^{\pm}_{i0}$, $\D^{\pm}_{0i0}$ and ${\mathbf \domE}_{i}$, 
$\bsD^{\pm}_{i}$, $\bsD^{\pm}_{0i}$, $\bsD^{\pm}_{i0}$, $\bsD^{\pm}_{0i0}$}

We shall be employing numerous pairs of related equations such as Eqs.\ 
(\ref{3.2a}) and (\ref{3.2b}), (\ref{3.3b}) and (\ref{3.3c}), and 
(\ref{3.3d}) and (\ref{3.3e}).  The two equations in each of these pairs
can be consolidated into a single equation by employing the free script
$i \in \{3,4\}$ in an appropriate way.  However, to avoid the introduction
of cumbersome explicit superscripts, we shall instead employ (for a fixed 
choice of index $i$) the abbreviations
\begin{equation}
x := x^{i}, \quad x' := x^{\prime i}, \quad x'' := x^{\prime\prime i}, \quad
y := x^{7-i}, \quad y' := x^{\prime 7-i}, \quad y'' := x^{\prime\prime 7-i},
\label{3.9c}
\end{equation}
whenever there is no danger of ambiguity.  We shall also adjust the 
definitions of our functions to the needs of this consolidation program.
In this connection we find it convenient to introduce special notations
for the domains of the functions.

\begin{abc}
\begin{definition}{Dfns.\ of $\domE_{i}$, $\D^{\pm}_{i}$, 
$\D^{\pm}_{0i}$, $\D^{\pm}_{i0}$ and $\D^{\pm}_{0i0}$\label{def39}}

Recalling that $x^{3} := r$ and $x^{4} := s$, we let
\begin{equation}
\domE_{i} := \{(x,y):(r,s) \in \domE\}
\end{equation}
or, equivalently,
\begin{equation}
\begin{array}{rcl}
\domE_{3} & := & \domE, \\
\domE_{4} & := & \{(s,r): (r,s) \in \domE\}.
\end{array}
\end{equation}
We also let\footnote{Observe that the position of the $0$ subscript(s)
is an indication of which value(s) of $\tau$ are to be excluded.}
\begin{eqnarray}
\D^{\pm}_{i} & := & \domE_{i} \times \bar{C}^{\pm}, \\
\D^{\pm}_{0i} & := & \{(x,y,\tau) \in \D^{\pm}_{i}: \tau \ne x\}, \\
\D^{\pm}_{i0} & := & \{(x,y,\tau) \in \D^{\pm}_{i}: \tau \ne y\}, \\
\D^{\pm}_{0i0} & := & \{(x,y,\tau) \in \D^{\pm}_{i}: \tau \notin \{x,y\}.
\end{eqnarray}
\end{definition}
\end{abc}

\begin{abc}
\begin{definition}{Dfns.\ of ${\mathbf \domE}_{i}$, 
$\bsD^{\pm}_{i}$, $\bsD^{\pm}_{0i}$, $\bsD^{\pm}_{i0}$, and $\bsD^{\pm}_{0i0}$
\label{def40}}

Recalling that $x^{\prime 3} := r'$ and $x^{\prime 4} := s'$, we let
\begin{equation}
{\mathbf \domE}_{i} := \{(x,x',y): (x,y) \text{ and }
(x',y') \text{ are both members of } \domE_{i}\}
\end{equation}
or, equivalently,
\begin{equation}
\begin{array}{rcl}
{\mathbf \domE}_{3} & := & \{(r,r',s):(r,s) \in \domE \text{ and } (r',s) \in \domE\}
\\ 
{\mathbf \domE}_{4} & := & \{(s,s',r):(r,s) \in \domE \text{ and } (r,s') \in \domE\}.
\end{array}
\end{equation}
We also let\footnote{We shall employ bold type for the sets that involve
two points $\x$ and $\x'$.}
\begin{eqnarray}
\bsD^{\pm}_{i} & := & {\mathbf \domE}_{i} \times \bar{C}^{\pm}, \\
\bsD^{\pm}_{0i} & := & \{(x,x',y,\tau) \in \bsD^{\pm}_{i}: \tau \ne x\},
\\
\bsD^{\pm}_{i0} & := & \{(x,x',y,\tau) \in \bsD^{\pm}_{i}: \tau \ne y\},
\\
\bsD^{\pm}_{0i0} & := & \{(x,x',y,\tau) \in \bsD^{\pm}_{i}: \tau \notin
\{x,y\} \}.
\end{eqnarray}
\end{definition}
\end{abc}


\setcounter{theorem}{0}
\setcounter{equation}{0}
\subsection{Existence, continuity and differentiability of $\bsQ^{(i)+}$}

\begin{abc}
\begin{definition}{Dfns.\ of $\gamma_{i}^{\pm} \, (i=3,4)$\label{def41}}
Let $\gamma_{i}^{\pm}$ denote the function with domain
\begin{equation}
\dom{\gamma_{i}^{\pm}} := \{(\x,\tau) \in \domE \times \bar{C}^{\pm}:
x \ne \tau \}
\label{3.4a}
\end{equation}
and values
\begin{equation}
\gamma_{i}^{\pm}(\x,\tau) :=
\frac{\M^{\pm}(\tau-y)}{\M^{\pm}(\tau-x)} \text{ when }
\tau \ne \infty
\end{equation}
and
\begin{equation}
\gamma_{i}^{\pm}(\x,\infty) = 1,
\label{3.4c}
\end{equation}
where we recall that $x^{3} := r$, $x^{4} := s$ and $\M^{\pm}$ is
defined by Eq.\ (\ref{1.6d}).
\end{definition}
\end{abc}

\begin{abc}
\begin{proposition}[Properties of $\gamma_{i}^{\pm}$]
\label{3.1B}
\mbox{ } \\ \vspace{-3ex}
\begin{romanlist}
\item 
For all $(\x,\tau) \in \dom{\gamma_{i}^{\pm}}$,
\begin{equation}
\gamma_{i}^{\pm}(\x,\tau) = \frac{\mu^{\pm}(\x,\tau)}{\tau-x},
\label{3.4d}
\end{equation}
$\partial[\mu^{\pm}(\x,\tau)-\tau]/\partial x$ exists, and
\begin{equation}
\frac{\partial[\mu^{\pm}(\x,\tau)-\tau]}{\partial x} =
- \frac{1}{2} \gamma_{i}^{\pm}(\x,\tau),
\label{3.4e}
\end{equation}
where $\mu^{\pm}$ was defined by Eq.\ (\ref{1.6e}).  For each 
$(\x,\tau) \in \dom{\gamma_{i}^{+}}$,
\begin{equation}
\gamma_{i}^{-}(\x,\tau^{*}) = [\gamma_{i}^{+}(\x,\tau)]^{*}.
\end{equation}
\item 
For each $(y,\tau) \in \I^{(7-i)} \times \bar{C}^{\pm}$, the
function of $x$ that is given by $\gamma_{i}^{\pm}(\x,\tau)$
is summable (Lebesgue-integrable) over any bounded subinterval of
$\{x\in \I^{(i)}: r < s\}$.
\item 
$\gamma_{i}^{\pm}$ is continuous.  Note:  All functions that we
have defined and shall define in these notes are, with the sole
exceptions of $\M^{\pm}$, $\mu^{\pm}$, $\mu$ and $[\mu]$, finite-valued
throughout their prescribed domains.
\item 
For each $\tau \in \bar{C}^{\pm}$, the function of $\x$ whose
domain is $\{\x \in \domE: r \ne \tau, s \ne \tau\}$
and whose values are given by $\gamma_{i}^{\pm}(\x,\tau)$ is
analytic and, therefore, also $\bC^{\infty}$.
\item 
For each $\x \in \domE$ and for all non-negative integers $m$ and $n$,
the function of $\tau$ whose domain is $\bar{C}^{\pm}-\{r,s\}$ and
whose values are given by
$$\frac{\partial^{m+n}\gamma_{i}^{\pm}(\x,\tau)}{(\partial r)^{m}
(\partial s)^{n}}$$
is holomorphic [i.e., by definition, this function has an extension
whose domain is an open subset of $C$ and which is holomorphic].
\end{romanlist}
\end{proposition}
\end{abc}

\proof
The above theorem is a part of the theory of elementary functions.
\cheers

\begin{definition}{Dfn.\ of $\Delta_{i}^{\pm}$\label{def42}}
Let $\Delta_{i}^{\pm}$ denote the $2 \times 2$ matrix function
whose domain is the same as that of $\gamma_{i}^{\pm}$ and whose
values are given by
\begin{equation}
\Delta_{i}^{\pm}(\x,\tau) := - \gamma_{i}^{\pm}(\x,\tau)
\frac{\partial \bbE(\x)/\partial x}{2f(\x)} \sigma_{3}
+ \frac{\partial\chi(\x)/\partial x}{2f(\x)} J.
\label{3.5a}
\end{equation}
\end{definition}
Note:  The above definition is consistent with Eq.\ (\ref{3.3a})
and the definition of $\Delta^{\pm}$ by Eqs.\ (\ref{1.7a}) to (\ref{1.7c}).

\begin{abc}
\begin{proposition}[Properties of $\Delta_{i}^{\pm}$]
\label{3.2B}
\mbox{ } \\ \vspace{-3ex}
\begin{romanlist}
\item 
\begin{equation}
\tr{\Delta_{i}^{\pm}} = 0.
\end{equation}
\item 
For each $(\x,\tau) \in \dom{\Delta_{i}^{+}}$,
\begin{equation}
\Delta_{i}^{-}(\x,\tau^{*}) = [\Delta_{i}^{+}(\x,\tau)]^{*}.
\end{equation}
\item 
If $\tau=\sigma$ is real,
\begin{equation}
J^{1-\kappa} \Delta_{i}^{-}(\x,\sigma) J^{\kappa-1} =
\Delta_{i}^{+}(\x,\sigma).
\label{3.5d}
\end{equation}
\item 
At each $(\x,\tau) \in \dom{\Delta_{i}^{\pm}}$ such that $\tau \ne y$,
$\partial\Delta_{i}^{\pm}(\x,\tau)/\partial y$ exists for both
$i=3$ and $i=4$; and
\begin{equation}
\frac{\partial\Delta_{4}^{\pm}(\x,\tau)}{\partial x^{3}} -
\frac{\partial\Delta_{3}^{\pm}(\x,\tau)}{\partial x^{4}} -
\Delta_{3}^{\pm}(\x,\tau) \Delta_{4}^{\pm}(\x,\tau) +
\Delta_{4}^{\pm}(\x,\tau) \Delta_{3}^{\pm}(\x,\tau) = 0.
\end{equation}
\item 
$\Delta_{i}^{\pm}$ is continuous.  Also, $\partial\Delta_{i}^{\pm}(\x,\tau)/
\partial y$ is a continuous function of $(\x,\tau)$ throughout
$\{(\x,\tau) \in \dom{\Delta_{i}^{\pm}}: \tau \ne y\}$.
\item 
For each $\x \in \domE$, the function of $\tau$ whose domain is 
$\bar{C}^{\pm}-\{r,s\}$ and whose values are given by $\Delta_{i}^{\pm}$
is holomorphic.
\end{romanlist}
\end{proposition}
\end{abc}

\proof
For the proofs of the above statements (i), (ii), (iii) and (iv), see
the proofs of the corresponding statements in Prop.~\ref{1.2A}; and, in
the proof of (iv), use the facts that $\E$ (and, therefore, $\bbE$) is
$\bC^{1}$ and that $\partial^{2}\E(\x)/\partial r \partial s$ exists
and is continuous throughout $\domE$.  The proof of statement (v) employs
Props.~\ref{3.1B}(iii) and~(v).  The proof of (vi) employs Prop.~\ref{3.1B}(iv).
\cheers

We now return to the definitions of $\bsQ^{(4)\pm}$ and $\bsQ^{(3)\pm}$
that are to replace the tentative definitions (\ref{3.2a}) and (\ref{3.2b}).

\begin{abc}
\begin{definition}{Dfn.\ of $\check{\I}^{(i)}(y)$\label{def43}}

Corresponding to each $y \in \I^{(7-i)}$, let
\begin{equation}
\check{\I}^{(i)}(y) := \{x \in \I^{(i)}:(x,y) \in \domE_{i}\}
\end{equation}
or, equivalently,
\begin{equation}
\check{\I}^{(3)}(\x) := \{\sigma \in \I^{(3)}:\sigma < s\} \text{ and } 
\check{\I}^{(4)}(\x) := \{\sigma \in \I^{(4)}:r < \sigma\}.
\end{equation}
\end{definition}
\end{abc}

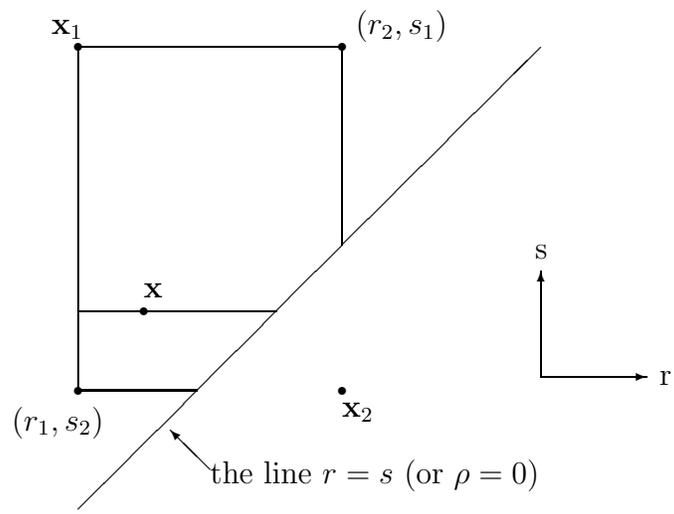
\begin{figure}[htbp] 
\begin{picture}(300,300)(-60,60)
\put(75,75){\line(1,1){175}}
\put(125,85){the line $r=s$ (or $\rho=0$)}
\put(125,90){\vector(-1,1){15}}
\put(250,125){\vector(0,1){40}}
\put(250,125){\vector(1,0){40}}
\put(248,170){s}
\put(295,122){r}
\put(75,250){\line(1,0){100}}
\put(75,120){\line(1,0){45}}
\put(75,120){\line(0,1){130}}
\put(175,175){\line(0,1){75}}
\put(75,250){\circle*{3}}
\put(75,120){\circle*{3}}
\put(175,250){\circle*{3}}
\put(175,120){\circle*{3}}
\put(65,255){${\mathbf x}_{1}$}
\put(175,110){${\mathbf x}_{2}$}
\put(50,105){$(r_{1},s_{2})$}
\put(180,255){$(r_{2},s_{1})$}
\put(100,150){\circle*{3}}
\put(100,155){${\mathbf x}$}
\put(75,150){\line(1,0){75}}
\end{picture}
\caption{The subinterval $\check{\I}^{(3)}(x^{4})$ of the interval $\I^{(3)}$.
In this example, $\check{\I}^{(4)}(x^{3}) = \I^{(4)}$.}
\end{figure}

\pagebreak
\begin{abc}
\begin{definition}{Dfn.\ of $\bsQ^{(i)\pm}$ (for a given $\E \in \S_{\E}$)
\label{def44}}
Let $\bsQ^{(i)+}$ denote any $2 \times 2$ matrix function such that
\begin{equation}
\dom{\bsQ^{(i)+}} := \bsD^{+}_{i}
\label{3.7a}
\end{equation}
and the following conditions are satisfied for each 
$(x',y,\tau) \in \D^{+}_{i}$:
\begin{equation}
\bsQ^{(i)+}(x,x',y,\tau) \text{ is a continuous 
function of $x$ throughout $\check{\I}^{(i)}(y)$,}
\label{3.7b}
\end{equation}
\begin{equation}
\bsQ^{(i)+}(y^{i},x',y,\tau) = I, 
\label{3.7c}
\end{equation}
and, at all $x \in \check{\I}^{(i)}(y)$ at which $x \ne \tau$,
the derivative $\partial \bsQ^{(i)+}(x,x',y,\tau)/\partial
x$ exists and
\begin{equation}
\frac{\partial\bsQ^{(i)+}(x,x',y,\tau)}{\partial x}
= \Delta_{i}^{+}(\x,\tau) \bsQ^{(i)+}(x,x',y,\tau).
\label{3.7d}
\end{equation}
[Note that Eq.\ (\ref{3.7d}) holds for all $(x,x',y,\tau)
\in \bsD^{+}_{0i}$.]  

Corresponding to each $\bsQ^{(i)+}$, we let $\bsQ^{(i)-}$ denote
the $2 \times 2$ matrix function whose domain is
\begin{equation}
\dom{\bsQ^{(i)-}} := \bsD^{-}_{i} 
\end{equation}
and whose values are given by
\begin{equation}
\bsQ^{(i)-}(x,x',y,\tau^{*}) :=
\left[ \bsQ^{(i)+}(x,x',y,\tau) \right]^{*}
\text{ for all } (x,x',y,\tau) \in \bsD^{+}_{i}.
\label{3.7f}
\end{equation}
\end{definition}
\end{abc}

\begin{abc}
\begin{proposition}[Properties of $\bsQ^{(i)+}$ (for a given $\E \in \S_{\E}$)] 
\label{3.3B}
\mbox{ } \\ \vspace{-3ex}
\begin{romanlist}
\item 
If $\bsQ^{(i)+}$ exists,
\begin{equation}
\det{\bsQ^{(i)\pm}} = 1.
\end{equation}
\item 
There is not more than one $\bsQ^{(i)+}$.
\item 
Grant that $\bsQ^{(i)+}$ exists.  Then, for each $(y,\tau)
\in \I^{(7-i)} \times \bar{C}^{+}$ and for all $x$, $x'$
and $x'' \in \check{\I}^{(i)}(y)$,
\begin{equation}
\bsQ^{(i)+}(x,x'',y,\tau)
\bsQ^{(i)+}(x',x',y,\tau)
= \bsQ^{(i)+}(x,x',y,\tau)
\end{equation}
and
\begin{equation}
[\bsQ^{(i)+}(x,x',y,\tau)]^{-1} =
\bsQ^{(i)+}(x',x,y,\tau).
\end{equation}
\item 
Grant that $\bsQ^{(i)+}$ exists.  Then, for each $(x,x',y)
\in {\mathbf \domE}_{i}$,
\begin{eqnarray}
\bsQ^{(i)-}(x,x',y,\sigma) & = &
\bsQ^{(i)+}(x,x',y,\sigma) \text{ [and is, therefore, real] }
\nonumber \\
& & \text{ if } \sigma \in R^{1} - \left(|x,y| \cup
|x',y| \right) \\ 
& & \text{ and if } \sigma=\infty, \nonumber \\
& \text{and} & \nonumber \\
-J \bsQ^{(i)-}(x,x',y,\sigma) J & = &
\bsQ^{(i)+}(x,x',y,\sigma) \text{ and is unitary } \nonumber \\
& & \text{ for all } \sigma \text{ in the real open interval } \\
& & \text{ between } \{y\} \text{ and } |x,x'|. \nonumber
\end{eqnarray}
\end{romanlist}
\end{proposition}
\end{abc}

\proof
The proofs of the above statements (i) to (iv) can be obtained by
formally making the substitutions
$$\bsQ^{\pm}(\x,(r',s),\tau) \rightarrow \bsQ^{(3)\pm}(r,r',s,\tau)
\text{ and }
\bsQ^{\pm}(\x,(r,s'),\tau) \rightarrow \bsQ^{(4)\pm}(s,s',r,\tau)$$
in the proofs of the corresponding statements (i), (ii), (iv) and (v) 
of Prop.~\ref{1.3A}.
\cheers

\begin{abc}
\begin{definition}{Dfn.\ of $\gamma_{0i}^{+}$\label{def45}}
Let $\gamma_{0i}^{+}$ denote the function with domain
\begin{equation}
\dom{\gamma_{0i}^{+}} := \D^{+}_{0i} 
\label{3.9a}
\end{equation}
and values [recalling that $\x := (x^{3},x^{4})$]
\begin{equation}
\gamma_{0i}^{+}(x,y,\tau) := {\gamma}_{i}^{+}(\x,\tau).
\label{3.9b}
\end{equation}
\end{definition}
\end{abc}

Several properties of $\gamma_{0i}^{+}$ can easily be obtained
by inspection from the properties of ${\gamma}_{i}^{+}$ that
are given in Prop.~\ref{3.1B}.  Other properties of interest
are given in the following theorem.

\begin{abc}
\begin{theorem}[Other properties of $\gamma_{0i}^{+}$]
\label{3.1C}
\mbox{ } \\ \vspace{-3ex}
\begin{romanlist}
\item 
For any given closed subintervals
\begin{equation}
[a_{1},a_{2}] \subset \I^{(i)} \text{ and }
[b_{1},b_{2}] \subset \I^{(7-i)}
\label{3.10a}
\end{equation}
such that
\begin{equation}
\bU_{i} := [a_{1},a_{2}]^{2} \times [b_{1},b_{2}] \subset {\mathbf \domE}_{i}
\label{3.10b}
\end{equation}
there exists at least one positive real number $\M(\bU_{i})$ such that
\begin{equation}
\sgn{(x-x')} \int_{x'}^{x} d\lambda |\gamma_{0i}^{+}(\lambda,y,\tau)|
\le \M(\bU_{i}) \text{ for all } (x,x',y,\tau) \in \bU_{i} \times \bar{C}^{+}.
\label{3.10c}
\end{equation}
\item 
Consider any complex-valued function $\bphi^{(i)+}$ whose domain is
\begin{equation}
\dom{\bphi^{(i)+}} := \bsD^{+}_{i} \text{ (resp.\ $\bsD^{+}_{i0}$)}
\label{3.11a}
\end{equation}
such that, for each $(x',x,\tau') \in \D^{+}_{i}$ (resp.\ $\D^{+}_{i0}$), the
function of $x$ that is given by $\bphi^{(i)+}(x,x',y,\tau)$ is continuous
throughout $\check{\I}^{(i)}(y)$.  Then the function $\bpsi^{(i)+}$ whose
domain is the same as that of $\bphi^{(i)+}$ and whose values are
\begin{equation}
\bpsi^{(i)+}(x,x',y,\tau) := \int_{x'}^{x} d\lambda \gamma_{0i}^{+}(\lambda,y,\tau)
\bphi^{(i)+}(\lambda,x',y,\tau) \text{ for all } (x,x',y,\tau) \in \dom{\bphi^{(i)+}}
\label{3.11b}
\end{equation}
exists; and, for each $(x',y,\tau) \in \D^{+}_{i}$ (resp.\ $\D^{+}_{i0}$),
$\bpsi^{(i)+}(x,x',y,\tau)$ is a continuous function of $x$ throughout
$\check{\I}^{(i)}(y)$.  Moreover, $\partial\bpsi^{(i)+}(x,x',y,\tau)/\partial x$
exists and equals
\begin{equation}
\frac{\partial\bpsi^{(i)+}(x,x',y,\tau)}{\partial x} = \gamma_{0i}^{+}(x,y,\tau)
\bphi^{(i)+}(x,x',y,\tau) \text{ for all } (x,x',y,\tau) \in \bsD^{+}_{0i} \text{ 
(resp.\ $\bsD^{+}_{0i0}$)}.
\label{3.11c}
\end{equation}
\item 
Let $\bphi^{(i)+}$ and $\bpsi^{(i)+}$ be defined as in the preceding part (ii) of this
theorem.  If $\bphi^{(i)+}$ is continuous, then $\bpsi^{(i)+}$ is continuous.
\item 
Let $\bphi^{(i)+}$ and $\bpsi^{(i)+}$ be defined as in the preceding part (ii) of this
theorem, and assume that $\bphi^{(i)+}$ is continuous.  If
$\partial\bphi^{(i)+}(x,x',y,\tau)/\partial y$ exists and is a continuous
function of $(x,x',y,\tau)$ thoughout $\bsD^{+}_{i0}$, then
$\partial\bpsi^{(i)+}(x,x',y,\tau)/\partial y$ also exists and is a continuous
function of $(x,x',y,\tau)$ throughout $\bsD^{+}_{i0}$.  Moreover, at all
$(x,x',y,\tau) \in \bsD^{+}_{i0}$,
\begin{eqnarray}
\frac{\partial\bpsi^{(i)+}(x,x',y,\tau)}{\partial y} & = &
\int_{x'}^{x} d\lambda \frac{\partial}{\partial y} \left[
\gamma_{0i}^{+}(\lambda,y,\tau) \bphi^{(i)+}(\lambda,x',y,\tau) \right] 
\label{3.11d} \\
& = & - \frac{\bpsi^{(i)+}(x,x',y,\tau)}{2(\tau-y)} + \int_{x'}^{x} d\lambda
\gamma_{0i}^{+}(\lambda,y,\tau) \left[
\frac{\partial\bphi^{(i)+}(\lambda,x',y,\tau)}{\partial y} \right].
\nonumber
\end{eqnarray}
\end{romanlist}
\end{theorem}
\end{abc}

\proof
\begin{romanlist}
\item 
\begin{abc}
Select a positive real number $R$ which is sufficiently large so that
\begin{equation}
[a_{1},a_{2}] \cup [b_{1},b_{2}] \subset \{\tau:|\tau|<R\}.
\end{equation}
[For example, let $R$ equal twice the supremum of
$\{|\sigma|:\sigma \in [a_{1},a_{2}] \cup [b_{1},b_{2}]\}$.]  Let
\begin{equation}
t:=(2\tau)^{-1}, \; \tau_{1} := \Re{\tau} \text{ when }
|\tau| \le R \text{ and } t_{1} = \Re{t} \text{ when }
|\tau| \ge R.
\end{equation}
{}From Eqs.\ (\ref{3.4a}) to (\ref{3.4c}) and (\ref{3.9a}) to
(\ref{3.9c}), one obtains, for all $(x,x',y) \in \bU_{i}$,
\begin{eqnarray}
\sgn{(x-x')} \int_{x'}^{x} d\lambda |\gamma_{0i}^{+}(\lambda,y,\tau)|
& = & \sgn{(x-x')} \int_{x'}^{x} d\lambda \sqrt{\frac{|\tau-y|}
{|\tau-\lambda|}} \nonumber \\
& \le & \sgn{(x-x')} \sqrt{|\tau-y|} \int_{x'}^{x} d\lambda
\frac{1}{\sqrt{|\tau_{1}-\lambda|}} \nonumber \\
& = & 2 \sgn{(x-x')} \sqrt{|\tau-y|} \nonumber \\
& & \left[ \sgn{(\tau_{1}-x')} \sqrt{|\tau_{1}-x'|} + 
\sgn{(x-\tau_{1})} \sqrt{|\tau_{1}-x|} \right]
\nonumber \\
& & \text{ for all $\tau \in \bar{C}^{+}$ such that $|\tau| \le R$,}
\label{3.12c}
\end{eqnarray}
and
\begin{eqnarray}
\sgn{(x-x')} \int_{x'}^{x} d\lambda |\gamma_{0i}^{+}(\lambda,y,\tau)|
& = & \sgn{(x-x')} \int_{x'}^{x} d\lambda \sqrt{\frac{|1-2ty|}{|1-2t\lambda|}}
\nonumber \\
& \le & \sgn{(x-x')} \sqrt{|1-2ty|} \nonumber \\
& & \mbox{ } \left[ \frac{\sqrt{1-2t_{1}x'}-\sqrt{1-2t_{1}x}}{t_{1}} \right]
\nonumber \\
& & \text{ for all $\tau \in \bar{C}^{+}$ such that $|\tau| \ge R$.}
\label{3.12d}
\end{eqnarray}
Now, the right side of the above Eqs.\ (\ref{3.12c}) and (\ref{3.12d})
are continuous functions of $(x,x',y,\tau)$ and of $(x,x',y,t)$
throughout the compact spaces $\bU_{i} \times \{\tau\in\bar{C}^{+}:|\tau|
\le R\}$ and $\bU_{i} \times \{t: \Im{t} \le 0 \text{ and } |t| \le
(2R)^{-1}\}$, respectively.
Therefore,
\begin{eqnarray}
\M'(\bU_{i}) & := & \text{ the supremum of the set of all right sides of }
\nonumber \\ & & \text{ (\ref{3.12c}) for which } (x,x',y) \in \bU_{i},
\tau \in \bar{C}^{+} 
\\ & & \text{ and } |\tau| \le R, 
\nonumber \\
\M''(\bU_{i}) & := & \text{ the supremum of the set of all right sides of }
\nonumber \\ & & \text{ (\ref{3.12d}) for which } (x,x',y) \in \bU_{i},
\Im{t} \le 0 
\\ & & \text{ and } |t| \le (2R)^{-1} \nonumber
\end{eqnarray}
are both finite (positive real numbers).  Let
\begin{equation}
\M(\bU_{i}) := \sup{\{\M'(\bU_{i}),\M''(:\bU_{i})\}} 
\label{3.12g}
\end{equation}
whereupon (\ref{3.12c}) to (\ref{3.12g}) yield (\ref{3.10c}).
\end{abc}
\item 
For each $(y,\tau) \in \I^{(7-i)} \times \bar{C}^{+}$, the function
of $x$ that is given by $\gamma_{0i}^{+}(x,y,\tau)$ is summable over any
bounded subinterval of $\check{\I}^{(i)}(y)$.  [See Prop.~\ref{3.1B}(ii).]
Therefore, for each $(x',y,\tau) \in \D^{+}_{i}$ (resp.\ $\D^{+}_{i0}$), the
function of $x$ that is given by $\gamma_{0i}^{+}(x,y,\tau) \bphi^{(i)+}(x,x',y,\tau)$
is summable over any bounded subinterval of $\check{\I}^{(i)}(y)$ [since
a summable function times a continuous function is summable].  Therefore,
$\bpsi^{(i)+}$ as defined by Eq.\ (\ref{3.11b}) exists, whereupon from a well-known
theorem, $\bpsi^{(i)+}(x,x',y,\tau)$ is an absolutely continuous function of $x$
on each finite subinterval of $\check{\I}^{(i)}(y)$; and, therefore,
$\bpsi^{(i)+}(x,x',y,\tau)$ is a continuous function of $x$ throughout
$\check{\I}^{(i)}(y)$.

Furthermore, from Prop.~\ref{3.1B}(iii), $\gamma_{0i}^{+}(x,y,\tau)
\bphi^{(i)+}(x,x',y,\tau)$ is a continuous function of $x$ throughout
$\check{\I}^{(i)}(y)-\{\tau\}$.  Therefore, 
$\partial\bpsi^{(i)+}(x,x',y,\tau)/\partial x$ exists at all $x \in \check{\I}^{(i)}(y)$
at which $x \ne \tau$, and Eq.\ (\ref{3.11c}) holds.
\item 
\begin{abc}
Consider any given point $(u,u',v,\xi)$ in $\bsD^{+}_{i}$ (resp.\ $\bsD^{+}_{i0}$). 
Recall that $\xi \ne v$ when $(u,u',v,\xi) \in \bsD^{+}_{i0}$. 

We start by selecting $a_{1}, a_{2} \in \I^{(i)}$ so that
\begin{equation}
a_{1} < u < a_{2}, \quad a_{1} < u' < a_{2}, \quad
(a_{1},v), (a_{2},v) \in \domE_{i},
\end{equation}
whereupon
\begin{equation}
[a_{1},a_{2}] \times \{v\} \subset \domE_{i}.
\end{equation}
We then select a sufficiently small real positive number $\beta$ so that
\begin{equation}
[a_{1},a_{2}] \times [v-\beta,v+\beta] \subset \domE_{i}
\end{equation}
and
\begin{equation}
\xi \notin [v-\beta,v+\beta] \text{ when } \dom{\bphi^{(i)+}} = \bsD^{+}_{i0}.
\end{equation}
With the choices that we have made above,
\begin{equation}
\bU_{i} := [a_{1},a_{2}]^{2} \times [v-\beta,v+\beta] \subset {\mathbf \domE}_{i}
\end{equation}
and
\begin{equation}
\bU_{i} \times \{\xi\} \subset \bsD^{+}_{i} \text{ (resp.\ $\bsD^{+}_{i0}$)}.
\end{equation}

We treat the cases $\xi = \infty$ and $\xi \ne \infty$ in different ways.
When $\xi = \infty$, we select a bounded and closed semi-circular
neighborhood $S_{i}$ of the origin $t=0$ in the space
\begin{equation}
\{ t=(2\tau)^{-1}: \tau \in \bar{C}^{+}\}
\end{equation}
such that the radius $R$ of $S_{i}$ satisfies
\begin{equation}
2R|c| < 1 \text{ for all } c \in [a_{1},a_{2}] \text{ and }
c \in [v-\beta,v+\beta].
\end{equation}
Then, if 
\begin{equation}
A_{i} := \{\tau=(2t)^{-1}:t \in S_{i}\},
\end{equation}
one has
\begin{equation}
\bU_{i} \times A_{i} \subset \bsD^{+}_{i0} \subset \bsD^{+}_{i};
\end{equation}
and
\begin{equation}
\gamma_{0i}^{+}(\lambda,y,\tau) := \frac{\M^{+}(\tau-y)}{\M^{+}(\tau-\lambda)}
= \frac{\M^{+}(1-2ty)}{\M^{+}(1-2t\lambda)}
\end{equation}
is a continuous function of $(\lambda,y,\tau)$ throughout $[a_{1},a_{2}]
\times [v-\beta,v+\beta] \times A_{i}$ [which means that it is a
continuous function of $(\lambda,y,t)$ throughout $[a_{1},a_{2}] \times
[v-\beta,v+\beta] \times S_{i}$].  Therefore, the integrand in Eq.\
(\ref{3.11b}) is a continuous function of $(\lambda,x',y,\tau)$ throughout
$\bU_{i} \times A_{i}$.  Therefore, 
\begin{equation}
\begin{array}{c}
\bpsi^{(i)+} \text{ is continuous at each of the points } \\
(u,u',v,\xi) \in \dom{\bphi^{(i)+}} \text{ for which } \xi = \infty.
\end{array}
\label{3.13a}
\end{equation}
Note:  Recall that we assign the relative topology to each subset of
the extended complex plane.

When $\xi \ne \infty$, we select a bounded and closed rectangular
neighborhood $B_{i}$ of the point $\xi$ in the space $\bar{C}^{+}$
(resp.\ $\bar{C}^{+}-\{v\}$) such that each edge of $B_{i}$ is parallel
to or collinear with the real or the imaginary axis.  Then
$\bU_{i} \times B_{i} \subset \bsD^{+}_{i}$ (resp.\ $\bsD^{+}_{i0}$), and, 
from the multi-variable Weierstrass approximation theorem\footnote{A proof
of the Weierstrass approximation theorem can be found on p.~154 of Ch.~7
of {\em Introduction to Topology and Modern Analysis} by George F.\ Simmons
(McGraw-Hill, New York 1963).  As for the multi-variable version, see
problem 1 on p.~161.},
there exists an infinite sequence of polynomials $\bphi^{(i)+}_{n}$ (with
complex coefficients) each of which has the domain $[C - \{\infty\}]^{5}$
such that, as $n \rightarrow \infty$, $\bphi^{(i)+}_{n}(x,x',y,\Re{\tau},\Im{\tau})
\rightarrow \bphi^{(i)+}(x,x',y,\tau)$ uniformly over the compact space 
$\bU_{i} \times B_{i}$.  In other words,
\begin{equation}
\begin{array}{l}
\text{for each $\epsilon > 0$, there exists a positive integer $N(\epsilon,
\bU_{i} \times B_{i})$ such that,} \\
\text{for all $n > N(\epsilon,\bU_{i} \times B_{i})$ and all
$(x,x',y,\tau) \in \bU_{i} \times B_{i}$}, \\
|\bphi^{(i)+}(x,x',y,\tau) - \bphi^{(i)+}_{n}(x,x',y,\tau_{1},\tau_{2})| < \epsilon, 
\text{ where $\tau_{1} := \Re{\tau}, \; \tau_{2} := \Im{\tau}$}.
\end{array}
\label{3.13b}
\end{equation}
Now let
\begin{equation}
\begin{array}{c}
\bpsi^{(i)+}_{n}(x,x',y,\tau) := \int_{x'}^{x} d\lambda \gamma_{0i}^{+}(\lambda,y,\tau)
\bphi^{(i)+}_{n}(\lambda,x',y,\tau_{1},\tau_{2}) \\
\text{for all $n > 0$ and $(x,x',y,\tau) \in \bU_{i} \times B_{i}$}.
\end{array}
\label{3.13c}
\end{equation}
[The existence of $\bpsi^{(i)+}_{n}$ is demonstrated by the same argument that
was used to establish the existence of $\bpsi^{(i)+}$ in part (ii) of this theorem.]
Then, from the definition of $\bpsi^{(i)+}$ by Eq.\ (\ref{3.11b}) and from
(\ref{3.13b}), for all $n > N(\epsilon,\bU_{i} \times B_{i})$ and
$(x,x',y,\tau) \in \bU_{i} \times B_{i}$, 
\begin{eqnarray*}
|\bpsi^{(i)+}(x,x',\tau,y) - \bpsi^{(i)+}_{n}(x,x',\tau,y)| 
& \le & \sgn(x-x') \int_{x'}^{x} d\lambda |\gamma_{0i}^{+}(\lambda,y,\tau)
|\bphi^{(i)+}(\lambda,x',y,\tau) \\ 
& & \mbox{ } - \bphi^{(i)+}_{n}(\lambda,x',y,\tau_{1},\tau_{2})| \\
& < & \epsilon \, \sgn(x-x') \int_{x'}^{x} d\lambda |\gamma_{0i}^{+}(x,y,\tau)|.
\end{eqnarray*}
Therefore, from part (i) of this theorem [see Eqs.\ (\ref{3.10a}) to
(\ref{3.10c})] there exists a positive real number $\M_{1}(\bU_{i})$ such
that $|\bpsi^{(i)+}(x,x',y,\tau) - \bpsi^{(i)+}_{n}(x,x',y,\tau)| < \epsilon \M_{1}(\bU_{i})$
for all $n > N(\epsilon,\bU_{i} \times B_{i})$ and all $(x,x',y,\tau) \in 
\bU_{i} \times B_{i}$.  So, we have proved that
\begin{equation}
\text{as $n \rightarrow \infty$, $\bpsi^{(i)+}_{n}$ uniformly converges 
to the restriction of } \bpsi^{(i)+} \text{ to } \bU_{i} \times B_{i}.
\label{3.13d}
\end{equation}
Furthermore, from Eqs.\ (\ref{3.13c}), (\ref{3.4d}), (\ref{3.4e}) and
(\ref{3.9b}), 
\begin{eqnarray}
\bpsi^{(i)+}_{n}(x,x',y,\tau) & = & \int_{x'}^{x} d\lambda \gamma_{0i}^{+}(\lambda,y,\tau)
\left[ \bphi^{(i)+}_{n}(\lambda,x',y,\tau_{1},\tau_{2}) -
\bphi^{(i)+}_{n}(\tau,x',y,\tau_{1},\tau_{2}) \right] \nonumber \\
& & - 2 \left[ (\tau-x) \gamma_{0i}^{+}(x,y,\tau) - (\tau-x')
\gamma_{0i}^{+}(x',y,\tau) \right] \bphi^{(i)+}_{n}(\tau,x',y,\tau_{1},\tau_{2})
\nonumber \\
& & \text{ for all } n > 0 \text{ and } (x,x',y,\tau) \in \bU_{i} \times B_{i}.
\label{3.13e}
\end{eqnarray}
Since $(\tau-x) \gamma_{0i}^{+}(x,y,\tau) = \mu^{+}(\x,\tau)$, both terms in
the above Eq.\ (\ref{3.13e}) are continuous functions of $(x,x',y,\tau)$
throughout $\bU_{i} \times B_{i}$.  So we have proved that
\begin{equation}
\bpsi^{(i)+}_{n} \text{ is continuous for all } n > 0.
\label{3.13f}
\end{equation}
{}From the above statements (\ref{3.13d}) and (\ref{3.13f}), the restriction
of $\bpsi^{(i)+}$ to $\bU_{i} \times B_{i}$ is continuous.  Therefore,
\begin{equation}
\bpsi^{(i)+} \text{ is continuous at every point } (u,u',v,\xi) \in \dom{\bphi^{(i)+}}
\text{ for which } \xi \ne \infty.
\label{3.13g}
\end{equation}
Upon combining the statements (\ref{3.13a}) and (\ref{3.13g}), we obtain
the final conclusion; namely, $\bpsi^{(i)+}$ is continuous throughout its domain.
\end{abc}
\item 
\begin{abc}
Let
\begin{eqnarray}
\bsG^{(i)+}(x,x',y,\tau) & := & - \frac{\bphi^{(i)+}(x,x',y,\tau)}{2\M^{+}(\tau-y)}
+ \M^{+}(\tau-y) \left[ \frac{\partial\bphi^{(i)+}(x,x',y,\tau)}{\partial y} \right]
\nonumber \\ & & \text{ for all } (x,x',y,\tau) \in \bsD^{+}_{i0}.
\label{3.14a}
\end{eqnarray}
{}From the given premises,
\begin{equation}
\bsG^{(i)+}(x,x',y,\tau) \text{ exists and is a continuous function of }
(x,x',y,\tau) \text{ throughout } \bsD^{+}_{i0};
\label{3.14b}
\end{equation}
and
\begin{equation}
\frac{\partial}{\partial y} \left[ \gamma_{0i}^{+}(x,y,\tau) \bphi^{(i)+}(x,x',y,\tau)
\right] = \frac{\bsG^{(i)+}(x,x',y,\tau)}{\M^{+}(\tau-x)} \text{ exists at all }
(x,x',y,\tau) \in \bsD^{+}_{0i0}.
\label{3.14c}
\end{equation}
For each given $(x,x',y,\tau) \in \bsD^{+}_{i0}$, one can always select a 
sufficiently small $\epsilon > 0$ so that
\begin{equation}
(x,x',y',\tau) \in \bsD^{+}_{i0} \text{ for all }
y-\epsilon \le y' \le y+\epsilon.
\end{equation}
{}From the statement (\ref{3.14b}), there then exists a positive real
number $\M(x,x',y,\tau,\epsilon)$ such that
\begin{equation}
|\bsG^{(i)+}(\lambda,x',y',\tau)| \le \M(x,x',y,\tau,\epsilon) \text{ for all }
\lambda \in |x,x'| \text{ and } y-\epsilon \le y' \le y+\epsilon.
\label{3.14e}
\end{equation}
Moreover, $[\M^{+}(\tau-\lambda)]^{-1}$ is a summable function of $\lambda$ 
over the interval $|x,x'|$.  Therefore, from Eqs.\ (\ref{3.11b}), (\ref{3.14c})
and (\ref{3.14e}) and a standard theorem\footnote{Sec.~39, Cor.~3.92 in 
{\em Integration}, by Edward J.\ McShane (Princeton University Press, 1944)
p.~217.  In Sec.~39, the author unnecessarily restricts his parameter $t$
to an open interval $(a,b)$.  His proof, however, is valid for all $t \in
[a,b]$. \label{McShane}} on the partial differentiation of a Lebesgue integral
with respect to a parameter, $\partial\bpsi^{(i)+}(x,x',y',\tau)/\partial y'$
exists for all $y-\epsilon < y' < y+\epsilon$ and is given by
\begin{eqnarray*}
\frac{\partial\bpsi^{(i)+}(x,x',y',\tau)}{\partial y'} & = & \int_{x'}^{x}
d\lambda \frac{\partial}{\partial y'} \left[ \gamma_{0i}^{+}(\lambda,y',\tau)
\bphi^{(i)+}(\lambda,x',y',\tau) \right] \\
& = & \int_{x'}^{x} d\lambda
\frac{\bsG^{(i)+}(\lambda,x',y',\tau)}{\M^{+}(\tau-\lambda)}.
\end{eqnarray*}
Since this statement holds for each $(x,x',y,\tau) \in \bsD^{+}_{i0}$,
$\partial\bpsi^{(i)+}(x,x',y,\tau)/\partial y$ exists for
each $(x,x',y,\tau) \in \bsD^{+}_{i0}$ and [from Eq.\ (\ref{3.14a})] is
given by Eq.\ (\ref{3.11d}).  Furthermore, from Eq.\ (\ref{3.11d})
and part (iii) of this theorem, $\partial\bpsi^{(i)+}(x,x',y,\tau)/\partial y$
is a continuous function of $(x,x',y,\tau)$ throughout $\bsD^{+}_{i0}$. 
\end{abc}
\end{romanlist}
\cheers

\begin{corollary}[Mixed partial derivatives of $\bpsi^{(i)}$]
\label{3.2C} \mbox{ } \\
Let $\bphi^{(i)+}$ and $\bpsi^{(i)+}$ be defined as in 
Thm.~\ref{3.1C}(iv), assume that $\bphi^{(i)+}$ is continuous, and that
$\partial\bphi^{(i)+}(x,x',y,\tau)/\partial y$ exists and is a continuous
function of $(x,x',y,\tau)$ throughout $\bsD^{+}_{i0}$.  Then
$\partial^{2}\bpsi^{(i)+}(x,x',y,\tau)/\partial x \partial y$ and 
$\partial^{2}\bpsi^{(i)+}(x,x',y,\tau)/\partial y \partial x$ exist at all
$(x,x',y,\tau) \in \bsD^{+}_{0i0}$, and
\begin{equation}
\frac{\partial^{2}\bpsi^{(i)+}(x,x',y,\tau)}{\partial x \partial y} =
\frac{\partial^{2}\bpsi^{(i)+}(x,x',y,\tau)}{\partial y \partial x} =
\frac{\partial}{\partial y} \left[ \gamma_{0i}^{+}(x,y,\tau)
\bphi^{(i)+}(x,x',y,\tau) \right]
\end{equation}
and is a continuous function of $(x,x',y,\tau)$ throughout $\bsD^{+}_{0i0}$.
\end{corollary}

\proof
Use Thms.~\ref{3.1C}(ii) and (iv).
\cheers

\begin{abc}
\begin{definition}{Dfn.\ of $\Delta_{0i}^{+}$ (for a given $\E \in \S_{\E}$)
\label{def46}}
Let $\Delta_{0i}^{+}$ denote the $2 \times 2$ matrix function with
\begin{equation}
\dom{\Delta_{0i}^{+}} := \D^{+}_{0i} = \dom{\gamma_{0i}^{+}}
\end{equation}
and values
\begin{equation}
\Delta_{0i}^{+}(x,y,\tau) := {\Delta}_{i}^{+}(\x,\tau).
\label{3.16b}
\end{equation}
\end{definition}
\end{abc}

{}From the definition of ${\Delta}_{i}^{+}$ by Eq.\ (\ref{3.5a}),
\begin{abc}
\begin{equation}
\Delta_{0i}^{+}(x,y,\tau) = - \gamma_{0i}^{+}(x,y,\tau)
\frac{\partial \bbE(\x)/\partial x}
{2 f(\x)} \sigma_{3} + \frac{\partial \chi(\x)/\partial x}{2 f(\x)} J
\label{3.16c}
\end{equation}
for all $(x,y,\tau) \in \D^{+}_{0i}$.  Furthermore, Eq.\ (\ref{3.7d}) [in the 
definition of $\bsQ^{(i)+}$ by Eqs.\ (\ref{3.7a}) to (\ref{3.7d})] is 
expressible in the form
\begin{equation}
\frac{\partial \bsQ^{(i)+}(x,x',y,\tau)}{\partial x} = 
\Delta_{0i}^{+}(x,y,\tau) \bsQ^{(i)+}(x,x',y,\tau)
\text{ for all } (x,x',y,\tau) \in \bsD^{+}_{0i}.
\label{3.16d}
\end{equation}
Several properties of $\Delta_{0i}^{+}$ can be seen by inspection of 
Thm.~\ref{3.2B}, Thm.~\ref{3.1C} and Cor.~\ref{3.2C}.  The properties
which are perhaps most needed are given in the following theorem.
\end{abc}

\begin{abc}
\begin{theorem}[Some properties of $\Delta_{0i}^{+}$]
\label{3.3C}
\mbox{ } \\ \vspace{-3ex}
\begin{romanlist}
\item 
$\Delta_{0i}^{+}$ is continuous, and $\partial \Delta_{0i}^{+}(x,y,\tau)/\partial y$
exists, is a continuous function of $(x,y,\tau)$ throughout $\D^{+}_{0i0}$,
and
\begin{eqnarray}
\frac{\partial \Delta_{0[7-i]}^{+}(y,x,\tau)}{\partial x} -
\frac{\partial \Delta_{0i}^{+}(x,y,\tau)}{\partial y} 
+ \Delta_{0[7-i]}^{+}(y,x,\tau) \Delta_{0i}^{+}(x,y,\tau)
\nonumber & & \\
- \Delta_{0i}^{+}(x,y,\tau) \Delta_{0[7-i]}^{+}(y,x,\tau) & = & 0
\label{3.17a}
\end{eqnarray}
at all $(x,y,\tau) \in \D^{+}_{0i0}$.
\item 
Let $\bU_{i}$ be defined by Eqs.\ (\ref{3.10a}) and (\ref{3.10b}) in 
Thm.~\ref{3.1C}.  Then there exists a positive real number $\M_{2}(\bU_{i})$
such that
\begin{equation}
\sgn{(x-x')} \int_{x'}^{x} d\lambda ||\Delta_{0i}^{+}(\lambda,y,\tau)||
\le \M_{2}(\bU_{i}) \text{ for all } (x,x',y,\tau) \in \bU_{i} \times \bar{C}^{+},
\label{3.17b}
\end{equation}
where $||M||$ denotes the norm (by any of the conventional definitions) of
a $2 \times 2$ matrix $M$.  Note:  From Eq.\ (\ref{3.16c}), Eq.\ (\ref{3.9b}),
Prop.~\ref{3.1B}(ii) and the fact that $\E$ (and, therefore, $\bbE$) is
$\bC^{1}$, the Lebesgue integral in (\ref{3.17b}) exists.
\item 
Suppose that $\bPhi^{(i)+}$ is a $2 \times 2$ matrix function whose domain is
\begin{equation}
\dom{\bPhi^{(i)+}} = \bsD^{+}_{i} \text{ (resp.\ $\bsD^{+}_{i0}$) }
\label{3.17c}
\end{equation}
such that, for each $(x',y,\tau) \in \D^{+}_{i}$ (resp.\ $\D^{+}_{i0}$), the
function of $x$ given by $\bPhi^{(i)+}(x,x',y,\tau)$ is continuous throughout
$\check{\I}^{(i)}(y)$.  Then the function $\bPsi^{(i)+}$ whose domain is the same
as that of $\bPhi^{(i)+}$ and whose values are given by
\begin{equation}
\bPsi^{(i)+}(x,x',y,\tau) := \int_{x'}^{x} d\lambda \Delta_{0i}^{+}(\lambda,y,\tau)
\bPhi^{(i)+}(\lambda,x',y,\tau) \text{ for all } (x,x',y,\tau) \in \dom{\bPhi^{(i)+}}
\label{3.17d}
\end{equation}
exists; and, for each $(x',y,\tau) \in \D^{+}_{i}$ (resp.\ $\D^{+}_{i0}$),
$\bPsi^{(i)+}(x,x',y,\tau)$ is a continuous function of $x$ throughout
$\check{\I}^{(i)}(y)$.  Moreover, $\partial \bPsi^{(i)+}(x,x',y,\tau)/\partial x$
exists and equals
\begin{equation}
\frac{\partial \bPsi^{(i)+}(x,x',y,\tau)}{\partial x} = \Delta_{0i}^{+}(x,y,\tau)
\bPhi^{(i)+}(x,x',y,\tau) \text{ for all } (x,x',y,\tau) \in \bsD^{+}_{0i}
\text{ (resp. $\bsD^{+}_{0i0}$)}.
\end{equation}
\item 
Let $\bPhi^{(i)+}$ be any $2 \times 2$ matrix function with the domain 
(\ref{3.17c}).  If $\bPhi^{(i)+}$ is continuous, then the function 
$\bPsi^{(i)+}$ whose domain is the same as that of $\bPhi^{(i)+}$ and
whose values are given by Eq.\ (\ref{3.17d}) is continuous.
\item 
Let $\bPhi^{(i)+}$ and $\bPsi^{(i)+}$ be defined as in the preceding part (ii) 
of this theorem, and assume that $\bPhi^{(i)+}$ is continuous and that $\partial
\bPhi^{(i)+}(x,x',y,\tau)/\partial y$ exists and is a continuous function of
$(x,x',y,\tau)$ throughout $\bsD^{+}_{i0}$.  Then $\partial
\bPsi^{(i)+}(x,x',y,\tau)/\partial y$ also exists and is a continuous function 
of $(x,x',y,\tau)$ throughout $\bsD^{+}_{i0}$; and
\begin{equation}
\frac{\partial \bPsi^{(i)+}(x,x',y,\tau)}{\partial y} = \int_{x`}^{x} d\lambda
\frac{\partial}{\partial y} \left[ \Delta_{0i}^{+}(\lambda,y,\tau) \bPhi^{(i)+}(x,x',y,\tau)
\right] \text{ at all } (x,x',y,\tau) \in \bsD^{+}_{i0}.
\end{equation}
Furthermore, $\partial^{2}\bPsi^{(i)+}(x,x',y,\tau)/\partial x \partial y$ and
$\partial^{2}\bPsi^{(i)+}(x,x',y,\tau)/\partial y \partial x$ exist and are equal
at all $(x,x',y,\tau) \in \bsD^{+}_{0i0}$, and
\begin{equation}
\frac{\partial \bPsi^{(i)+}(x,x',y,\tau)}{\partial x \partial y} = 
\frac{\partial}{\partial y} \left[ \Delta_{0i}^{+}(x,y,\tau)
\bPhi^{(i)+}(x,x',y,\tau) \right].
\end{equation}
\end{romanlist}
\end{theorem}
\end{abc}

\proofs
\begin{romanlist}
\item 
Use Props.~\ref{3.2B}(iv) and (v).
\item 
Use Eq.\ (\ref{3.16c}), the fact that $\E$ is $\bC^{1}$, and
Thm.~\ref{3.1C}(i).
\item 
Use Eq.\ (\ref{3.16c}), the fact that $\E$ is $\bC^{1}$, and
Thm.~\ref{3.1C}(ii).
\item 
Use Eq.\ (\ref{3.16c}), the fact that $\E$ is $\bC^{1}$, and
Thm.~\ref{3.1C}(iii).
\item 
Use Eq.\ (\ref{3.16c}), the fact that $\E$ is $\bC^{1}$, the
fact that $\partial^{2}\E(\x)/\partial r \partial s$ exists
and is continuous throughout $\domE$, Thm.~\ref{3.1C}(iv) and
Cor.~\ref{3.2C}.
\end{romanlist}
\cheers

\begin{theorem}[Picard integral equation for $\bsQ^{(i)+}$]
\label{3.4C} \mbox{ } \\
The definition of $\bsQ^{(i)+}$ in Sec.\ \ref{Sec_3}C is equivalent to the statement
that $\bsQ^{(i)+}$ is a $2 \times 2$ matrix function with the domain
$\bsD^{+}_{i}$ such that the continuity condition (\ref{3.7b}) holds and such
that
\begin{equation}
\bsQ^{(i)+}(x,x',y,\tau) = I + \int_{x'}^{x} d\lambda \Delta_{0i}^{+}(x,y,\tau)
\bsQ^{(i)+}(\lambda,x',y,\tau)
\end{equation}
for all $(x,x',y,\tau) \in \bsD^{+}_{i}$, where the above integral is defined
in the sense of Lebesgue.
\end{theorem}

\proof
Use Thm.~\ref{3.3C}(iii).
\cheers

\begin{theorem}[Existence and continuity of $\bsQ^{(i)}$ (for a
given $\E \in \S_{\E}$)]
\label{3.1D} \mbox{ } \\
For each $\E \in \S_{\E}$, $\bsQ^{(i)+}$ exists and is continuous.
\end{theorem}

\proof
The proof will be given in three phases:
\begin{arablist}
\item 
\begin{abc}
We introduce the following Picard sequence of successive approximations
for each $(x,x',y,\tau)$ in $\bsD^{+}_{i}$:
\begin{equation}
\bPhi_{0}^{(i)+}(x,x',y,\tau) := I
\label{3.19a}
\end{equation}
and, for all $n \ge 0$,
\begin{equation}
\bPhi_{n+1}^{(i)+}(x,x',y,\tau) := I + \int_{x'}^{x} d\lambda
\Delta_{0i}^{+}(\lambda,y,\tau) \bPhi_{n}^{(i)+}(\lambda,x',y,\tau).
\label{3.19b}
\end{equation}
The principle of mathematical induction, the part of Thm.~\ref{3.3C}(iii)
that concerns existence, and Thm.~\ref{3.3C}(iv) yield
\begin{equation}
\bPhi_{n}^{(i)+} \text{ exists and is continuous for all } n \ge 0.
\label{3.19c}
\end{equation}
Equations (\ref{3.19a}) to (\ref{3.19c}) are collectively equivalent,
as one proves by mathematical induction, to the following statement:
\end{abc}
\begin{abc}
\begin{equation}
\bPhi_{n}^{(i)+} = \sum_{k=0}^{n} \bsQ_{k}^{(i)+} \text{ for all }
n \ge 0,
\label{3.19d}
\end{equation}
where, for all $(x,x',y,\tau) \in \bsD^{+}_{i}$ and all $k \ge 0$,
\begin{eqnarray}
\bsQ_{0}^{(i)+}(x,x',y,\tau) & := & I,  
\label{3.19e} \\
\bsQ_{k+1}^{(i)+}(x,x',y,\tau) & := & \int_{x'}^{x} d\lambda \Delta_{0i}^{+}(\lambda,y,\tau)
\bsQ_{k}^{(i)+}(\lambda,x',y,\tau),
\label{3.19f}
\end{eqnarray}
and
\begin{equation}
\bsQ_{k}^{(i)+} \text{ exists and is continuous}.
\end{equation}
{}From Eqs.\ (\ref{3.19e}) and (\ref{3.19f}), one obtains, for all $k \ge 1$,
\end{abc}
\begin{eqnarray*}
||\bsQ_{1}^{(i)+}(x,x',y,\tau)|| & \le & \sgn{(x-x')} \int_{x'}^{x} d\lambda
||\Delta_{0i}^{+}(\lambda,y,\tau)||, \\
||\bsQ_{k+1}^{(i)+}(x,x',y,\tau)|| & \le & \sgn{(x-x')} \int_{x'}^{x} d\lambda
||\Delta_{0i}^{+}(\lambda,y,\tau)|| \, ||\bsQ_{k}^{(i)+}(\lambda,x',y,\tau)||.
\end{eqnarray*}
Therefore, for all $k \ge 1$,
\begin{equation}
||\bsQ_{k}^{(i)+}(x,x',y,\tau)|| = \frac{1}{k!} \left[ \sgn{(x-x')}
\int_{x'}^{x} d\lambda ||\Delta_{0i}^{+}(\lambda,y,\tau)|| \right]^{k}.
\label{3.19h}
\end{equation}
\item 
We next define $\bU_{i}$ as we did in Thm.~\ref{3.1C} by Eqs.\ (\ref{3.10a})
and (\ref{3.10b}), whereupon Eq.\ (\ref{3.19h}) and Thm.~\ref{3.3C}(ii)
imply the existence of a positive real number $\M_{2}(\bU_{i})$ such that
$$
||\bsQ_{k}^{(i)+}(x,x',y,\tau)|| \le \frac{1}{k!} \left[ \M_{2}(\bU_{i}) \right]^{k}
\text{ for all } k \ge 1 \text{ and } (x,x',y,\tau) \in \bsD^{+}_{i}.
$$
Therefore, from Eq.\ (\ref{3.19d}),
\begin{abc}
\begin{equation}
||\bPhi_{n+1}^{(i)+}(x,x',y,\tau)|| \le \sum_{k=0}^{n+1} \frac{1}{k!}
\left[ \M_{2}(\bU_{i}) \right]^{k} \text{ for all } n \ge -1 \text{ and }
(x,x',y,\tau) \in \bsD^{+}_{i}.
\label{3.19i}
\end{equation}
The comparison test of absolute and uniform convergence and the above
Eq.\ (\ref{3.19i}) imply that the restriction of $\bPhi_{n}^{(i)+}$ to
$\bU_{i} \times \bar{C}^{+}$ converges absolutely and uniformly as
$n \rightarrow \infty$.  However, for each point $(u,u',v)$ in the
space ${\mathbf \domE}_{i}$, $\bU_{i}$ can always be selected so that it is a
neighborhood of $(u,u',v)$.  Therefore,
\begin{equation}
\bPhi^{(i)+}(x,x',y,\tau) := \lim_{n \rightarrow \infty}
\bPhi_{n}^{(i)+}(x,x',y,\tau) \text{ exists for all } (x,x',y,\tau)
\in \bsD^{+}_{i}
\end{equation}
and the convergence is absolute throughout $\bsD^{+}_{i}$ and is uniform
on any compact subspace of $\bsD^{+}_{i}$.

Moreover, from (\ref{3.19c}), the restriction of $\bPhi^{(i)+}$ to each
compact subspace of $\bsD^{+}_{i}$ is continuous, because the limit of any
uniformly convergent sequence of continuous functions is also continuous.
Therefore, $\bPhi^{(i)+}$ is continuous.
\end{abc}
\item 
\begin{abc}
It now follows from the existence part of Thm.~\ref{3.3C}(iii) that
$$
\int_{x'}^{x} d\lambda \Delta_{0i}^{+}(\lambda,y,\tau)
\bPhi^{(i)+}(\lambda,x',y,\tau)
$$
exists for all $(x,x',y,\tau) \in \bsD^{+}_{i}$; and, from Eq.\ (\ref{3.19b}),
\begin{eqnarray}
\bPhi_{n+1}^{(i)+}(x,x',y,\tau) & = & I + \int_{x'}^{x} d\lambda
\Delta_{0i}^{+}(\lambda,y,\tau) \bPhi^{(i)+}(\lambda,x',y,\tau) 
\label{3.19l} \\
& & \mbox{ } + \int_{x'}^{x} d\lambda \Delta_{0i}^{+}(\lambda,y,\tau)
\left[ \bPhi_{n}^{(i)+}(\lambda,x',y,\tau)-\bPhi^{(i)+}(\lambda,x',y,\tau)\right]
\nonumber
\end{eqnarray}
for all $n \ge 0$ and $(x,x',y,\tau) \in \bsD^{+}_{i}$.  By employing
Thm.~\ref{3.3C}(ii) and the fact that the restriction of $\bPhi_{n}^{(i)+}$
to $\bU_{i} \times \bar{C}^{+}$ converges uniformly to the restriction of
$\bPhi^{(i)+}$ to $\bU_{i} \times \bar{C}^{+}$, the reader can easily prove
that the second integral on the right side of the above Eq.\ (\ref{3.19l})
converges to zero as $n \rightarrow \infty$.  So, one obtains
\begin{equation}
\bPhi^{(i)+}(x,x',y,\tau) = I + \int_{x'}^{x} d\lambda \Delta_{0i}^{+}(\lambda,x',y,\tau)
\bPhi^{(i)+}(\lambda,x',y,\tau) \text{ for all } (x,x',y,\tau) \in \bsD^{+}_{i}.
\label{3.19m}
\end{equation}
We conclude from Prop.~\ref{3.3B}(ii), Thm.~\ref{3.4C}, the continuity of
$\bPhi^{(i)+}$, and the above Eq.\ (\ref{3.19m}) that a unique
\begin{equation}
\bsQ^{(i)+} := \bPhi^{(i)+} = \sum_{n=0}^{\infty} \bsQ_{n}^{(i)+}
\label{3.19n}
\end{equation}
exists and is continuous.
\end{abc}
\end{arablist}
\cheers

\begin{theorem}[Existence and continuity of partial derivative]
\label{3.1E} \mbox{ } \\
At each $(x,x',y,\tau) \in \bsD^{+}_{i0}$, $\partial \bsQ^{(i)+}(x,x',y,\tau)/\partial
y$ exists, has the value
\begin{equation}
\frac{\partial \bsQ^{(i)+}(x,x',y,\tau)}{\partial y} =
\Delta_{0[7-i]}^{+}(y,x,\tau) \bsQ^{(i)+}(x,x',y,\tau) -
\bsQ^{(i)+}(x,x',y,\tau) \Delta_{0[7-i]}^{+}(y,x',\tau),
\label{3.20}
\end{equation}
and is a continuous function of $(x,x',y,\tau)$ throughout $\bsD^{+}_{i0}$.
\end{theorem}

\proof
The proof will be given in five stages:
\begin{arablist}
\item 
For each $(x,x',y,\tau) \in \bsD^{+}_{i0}$, there clearly exists a positive real
number $b$ such that, if $\bU_{i}$ denotes the rectangle
\begin{abc}
\begin{equation}
\bU_{i} := |x,x'| \times [y-b,y+b],
\end{equation}
then
\begin{equation}
\bU_{i} \subset \domE_{i} \text{ and } \tau \notin [y-b,y+b].
\end{equation}
Equivalently,
\begin{equation}
(\lambda,x',y',\tau) \in \bsD^{+}_{i0} \text{ for every } (\lambda,y') \in \bU_{i}.
\end{equation}
In the stages (2) and (3) of this proof, $(x,x',y,\tau)$ will remain fixed,
and we shall consider the set of all $(\lambda,y') \in \bU_{i}$.
\end{abc}

\item 
{}From Props.~\ref{3.3B}(i) and \ref{3.1D}, $\bsQ^{(i)+}(\lambda,x',y',\tau)^{-1}$
exists and
\begin{abc}
\begin{equation}
\bsQ^{(i)+}(\lambda,x',y',\tau)^{-1} = - J \bsQ^{(i)+}(\lambda,x',y',\tau) J
\text{ for all } (\lambda,y') \in \bU_{i};
\label{3.21d}
\end{equation}
and, from the differential equation (\ref{3.16c}),
\begin{eqnarray}
\frac{\partial \bsQ^{(i)+}(\lambda,x',y',\tau)}{\partial\lambda} & = &
\Delta_{0i}^{+}(\lambda,y',\tau) \bsQ^{(i)+}(\lambda,x',y',\tau), \nonumber \\
\frac{\partial[\bsQ^{(i)+}(\lambda,x',y',\tau)^{-1}]}{\partial\lambda} & = &
- \bsQ^{(i)+}(\lambda,x',y',\tau)^{-1} \Delta_{0i}^{+}(\lambda,y',\tau) \nonumber \\
& & \mbox{ } \text{ for all } (\lambda,y') \in \bU_{i} \text{ at which }
\lambda \ne \tau.
\label{3.21e}
\end{eqnarray}
So, 
\begin{eqnarray}
\lefteqn{\frac{\partial}{\partial\lambda} \left[
\bsQ^{(i)+}(\lambda,x',y,\tau)^{-1} \bsQ^{(i)+}(\lambda,x',y',\tau) \right] = } 
\nonumber \\
& & \bsQ^{(i)+}(\lambda,x',y,\tau)^{-1} \left[
\Delta_{0i}^{+}(\lambda,y',\tau) - \Delta_{0i}^{+}(\lambda,y,\tau) \right]
\bsQ^{(i)+}(\lambda,x',y',\tau) \nonumber \\
& & \mbox{ } \text{ for all } (\lambda,y') \in \bU_{i} \text{ at which }
\lambda \ne \tau.
\label{3.21f}
\end{eqnarray}
Note that, for each $y' \in [y-b,y+b]$, the function of $\lambda$ given by
\begin{equation}
\gamma_{0i}^{+}(\lambda,y',\tau) := \M^{+}(\tau-y')/\M^{+}(\tau-\lambda)
\end{equation}
is continuous on $|x,x'| - \{\tau\}$ and summable on
$|x,x'|$.  As can be seen from the expression (\ref{3.16c}) for
$\Delta_{0i}^{+}$ and the fact that $\E$ is $\bC^{1}$,
\begin{equation}
\begin{array}{l}
\Delta_{0i}^{+}(\lambda,y',\tau) = \gamma_{0i}^{+}(\lambda,y,\tau)
\bLambda_{i}(\lambda,y',\tau,y) \text{ where } \bLambda_{i}(\lambda,y',\tau,y) \\
\text{ is a continuous function of } (\lambda,y') \text{ throughout } \bU_{i}.
\end{array}
\label{3.21h}
\end{equation}
Therefore, since
\begin{equation}
\Delta_{0i}^{+}(\lambda,y',\tau) - \Delta_{0i}^{+}(\lambda,y,\tau) = \gamma_{0i}^{+}(\lambda,y,\tau)
\left[ \bLambda_{i}(\lambda,y',\tau,y) - \bLambda_{i}(\lambda,y,\tau,y) \right],
\label{3.21i}
\end{equation}
and since $\bsQ^{(i)+}$ is continuous, the function of $\lambda$ given (for
fixed $y'$) by the right side of Eq.\ (\ref{3.21f}) is continuous on
$|x,x'| - \{\tau\}$ and summable on $|x,x'|$.  Therefore,
employing the continuity condition (\ref{3.7b}) and the initial condition
(\ref{3.7c}) in the definition of $\bsQ^{(i)+}$, one obtains, after integrating
(\ref{3.21f}) over $\lambda \in |x,x'|$ and then multiplying through
by $\bsQ^{(i)+}(x,x',y,\tau)$,
\begin{eqnarray}
\lefteqn{\bsQ^{(i)+}(x,x',y',\tau) - \bsQ^{(i)+}(x,x',y,\tau) = 
\bsQ^{(i)+}(x,x',y,\tau) } \nonumber \\ & &
\int_{x}^{x'} d\lambda \bsQ^{(i)+}(\lambda,x',y,\tau)^{-1} \left[
\Delta_{0i}^{+}(\lambda,y',\tau) - \Delta_{0i}^{+}(\lambda,y,\tau) \right]
\bsQ^{(i)+}(\lambda,x',y',\tau) \nonumber \\ & & \hspace{10em}
\text{ for all } y' \in [y-b,y+b].
\label{3.21j}
\end{eqnarray}
\end{abc}

\item 
Since the function of $(\lambda,y')$ given by $\bsQ^{(i)+}(\lambda,x',y',\tau)$
is continuous on the compact space $\bU_{i}$, 
\begin{abc}
\begin{eqnarray}
\bsM_{i}(x,x',y,\tau) & := & \text{ the supremum of the set of all }
\nonumber \\ & &
||\bsQ^{(i)+}(\lambda,x',y',\tau)|| \text{ for which } (\lambda,y') \in
\bU_{i} \nonumber \\ & < & \infty. 
\label{3.21k}
\end{eqnarray}
Moreover, since $\partial^{2}\E(\x)/\partial r \partial s$ exists and is a 
continuous function of $\x$ throughout $\domE$, it can be seen from Eqs.\
(\ref{3.16c}) and (\ref{3.21h}) that
\begin{eqnarray*}
& \partial\bLambda_{i}(\lambda,y',\tau,y)/\partial y' 
\text{ exists and is a continuous } & \\
& \text{ function of } (\lambda,y') \text{ throughout } \bU_{i}, \text{ and } &
\end{eqnarray*}
\begin{eqnarray}
\frac{\partial \Delta_{0i}^{+}(\lambda,y',\tau)}{\partial y'} & = & 
\gamma_{0i}^{+}(\lambda,y,\tau) \frac{\partial\bLambda_{i}(\lambda,y',\tau,y)}
{\partial y'} \\ & &
\text{ at all } (\lambda,y') \text{ at which } \lambda \ne \tau. \nonumber
\end{eqnarray}
Hence
\begin{eqnarray}
\bsM_{s}(x,x',y,\tau) & := & \text{ the supremum of the set of all }
\nonumber \\ & &
||\partial\bLambda_{i}(\lambda,y',\tau,y)/\partial y'|| 
\text{ for which } (\lambda,y') \in \bU_{i} \nonumber \\ & < & \infty. 
\end{eqnarray}
and, from Eq.\ (\ref{3.21i}),
\begin{eqnarray}
\left|\left|
\frac{\Delta_{0i}^{+}(\lambda,y',\tau) - \Delta_{0i}^{+}(\lambda,y,\tau)}{y'-y}
\right|\right| & = & \left| \frac{\gamma_{0i}^{+}(\lambda,y,\tau)}{y'-y} \right|
\; \left|\left| \int_{y}^{y'} dy''
\frac{\partial\bLambda_{i}(\lambda,y'',\tau,y)}{\partial y''} \right|\right|
\nonumber \\ & \le & |\gamma_{0i}^{+}(\lambda,y,\tau)| \bsM_{s}(x,x',y,\tau)
\nonumber \\ & & \text{ for all } (\lambda,y') \in \bU_{i}.
\label{3.21n}
\end{eqnarray}
We now see from (\ref{3.21d}), (\ref{3.21k}) and (\ref{3.21n}) that the
integrand in Eq.\ (\ref{3.21j}), divided by $y'-y$, satisfies
\begin{eqnarray}
\lefteqn{
\left|\left|
\bsQ^{(i)+}(\lambda,x',y,\tau)^{-1} \left[
\frac{\Delta_{0i}^{+}(\lambda,y',\tau) - \Delta_{0i}^{+}(\lambda,y,\tau)}{y'-y}
\right] \bsQ^{(i)+}(\lambda,x',y',\tau)
\right|\right|
} \hspace{5em} \nonumber \\
& \le & |\gamma_{0i}^{+}(\lambda,y,\tau)| [\bsM_{i}(x,x',y,\tau)]^{2}
\bsM_{s}(x,x',y,\tau) \nonumber \\
& < & \infty \text{ for all } (\lambda,y') \in \bU_{i}.
\label{3.21o}
\end{eqnarray}
\end{abc}

\item 
So, since $|\gamma_{0i}^{+}(\lambda,y,\tau)|$ is a summable function of $\lambda$
on $|x,x'|$, Lebesgue's theorem on an integral whose integrand
depends on a parameter ($y'$ in our case) and is dominated, as in our 
statement (\ref{3.21o}), asserts that the integral in Eq.\ (\ref{3.21j}),
divided by $y'-y$, converges as $y' \rightarrow y$, and that the limit is
$$
\int_{x'}^{x} d\lambda \bsQ^{(i)+}(\lambda,x',y,\tau)^{-1} \lim_{y' \rightarrow
y} \left\{ \left[ \frac{\Delta_{0i}^{+}(\lambda,y',\tau) 
- \Delta_{0i}^{+}(\lambda,y,\tau)}{y'-y}
\right] \bsQ^{(i)+}(\lambda,x',y',\tau) \right\}.
$$
Therefore, from Eq.\ (\ref{3.21j}),
$$
\partial \bsQ^{(i)+}(x,x',y,\tau)/\partial y \text{ exists and } \hspace{20em}
$$
\begin{eqnarray}
\frac{\partial \bsQ^{(i)+}(x,x',y,\tau)}{\partial y} & = & \bsQ^{(i)+}(x,x',y,\tau)
\label{3.21p} \\ & & \mbox{ }
\int_{x'}^{x} d\lambda \bsQ^{(i)+}(\lambda,x',y,\tau)^{-1}
\frac{\partial \Delta_{0i}^{+}(\lambda,y,\tau)}{\partial y} \bsQ^{(i)+}(\lambda,x',y,\tau).
\nonumber
\end{eqnarray}
\item 
We next use Eq.\ (\ref{3.17a}) to replace $\partial
\Delta_{0i}^{+}(\lambda,y,\tau)/\partial y$ in the integrand of
Eq.\ (\ref{3.21p}), and Eqs.\ (\ref{3.21e}) are then used to obtain
\begin{eqnarray}
\lefteqn{\frac{\partial \bsQ^{(i)+}(x,x',y,\tau)}{\partial y} = 
\bsQ^{(i)+}(x,x',y,\tau)} 
\label{3.21q} \\ & & \mbox{ }
\int_{x'}^{x} d\lambda \frac{\partial}{\partial \lambda} \left[
\bsQ^{(i)+}(\lambda,x',y,\tau)^{-1} \Delta_{0[7-i]}^{+}(y,\lambda,\tau)
\bsQ^{(i)+}(\lambda,x',y,\tau) \right],
\nonumber
\end{eqnarray}
where $\bsQ^{(i)+}(\lambda,x',y,\tau)^{-1} \Delta_{0[7-i]}^{+}(y,\lambda,\tau)
\bsQ^{(i)+}(\lambda,x',y,\tau)$ is [for fixed $(x',y,\tau) \in \D^{+}_{i0}$] a 
continuous function of $\lambda$ throughout $|x,x'|$.  The 
conclusion (\ref{3.20}) then follows from Eqs.\ (\ref{3.7c}) and (\ref{3.21q}).
Finally, (\ref{3.20}) implies that $\partial \bsQ^{(i)+}(x,x',y,\tau)/\partial y$
is a continuous function of $(x,x',y,\tau)$ throughout $\bsD^{+}_{i0}$, because
$\bsQ^{(i)+}$ is continuous throughout $\bsD^{+}_{i}$ and
$\Delta_{0[7-i]}^{+}$ is continuous throughout $\D^{+}_{i0}$.
\end{arablist}
\cheers

\begin{corollary}[Mixed partial derivatives of $\bsQ^{(i)}$]
\label{3.2E} \mbox{ } \\
Both $\partial^{2}\bsQ^{(i)+}(x,x',y,\tau)/\partial x \partial y$ and
$\partial^{2}\bsQ^{(i)+}(x,x',y,\tau)/\partial y \partial x$ exist,
are equal, and are continuous functions of $(x,x',y,\tau)$ throughout
$\bsD^{+}_{0i0}$.
\end{corollary}

\proof
Use Thm.~\ref{3.3C}(i), Eq.\ (\ref{3.16c}) and Eq.\ (\ref{3.20}).
\cheers


\setcounter{theorem}{0}
\setcounter{equation}{0}
\subsection{Schwarz extendability from $\bsD_{i}^{+}$ and
from $\bsD_{0i}^{+}$}

We shall next consider a Neumann series for $\bsQ^{(i)+}(x,x',y,\tau)$
that is different from the one,
$$
\bsQ^{(i)+} = \sum_{n=0}^{\infty} \bsQ_{n}^{(i)+},
$$
that was defined by Eq.\ (\ref{3.19n}) in Thm.~\ref{3.1D}.  The new 
series will provide a better understanding of the dependences of
$\bsQ^{(i)+} (x,x',y,\tau)$ on $y$ and on $\tau$.

\begin{abc}
\begin{definition}{Dfns.\ of ${\theta}_{i}$, $L_{i}$
and $\bsR^{(i)+}$\label{def47}}

Let ${\theta}_{i}$ and $L_{i}$ be the functions with the domains
\begin{equation}
\dom{{\theta}_{i}} = \dom{L_{i}} := \domE_{i}
\label{3.22a}
\end{equation}
and the values
\begin{eqnarray}
{\theta}_{i}(x,y) & := & \int_{x_{0}}^{x} d\lambda \left[
\frac{\partial\chi(\x)/\partial x}{2 f(\x)} \right]_{x=\lambda},
\label{3.22b} \\
L_{i}(x,y) & := & - e^{-2J {\theta}_{i}(x,y)} \left[
\frac{\partial\bbE(\x)/\partial x}{2 f(\x)} \right]
\end{eqnarray}
for each $(x,y) \in \domE_{i}$.  Let $\bsR^{(i)+}$ be the 
function with domain $\bsD^{+}_{i}$ such that 
\begin{equation}
\bsQ^{(i)+}(x,x',y,\tau) = e^{J\theta_{i}(x,y)} 
\bsR^{(i)+}(x,x',y,\tau) e^{-J\theta_{i}(x',y)}.
\label{3.22d}
\end{equation}
\end{definition}
\end{abc}

With the above definitions (\ref{3.22a}) to (\ref{3.22d}), the expression
(\ref{3.16c}) for $\Delta_{0i}^{+}(x,y,\tau)$ becomes
\begin{abc}
\begin{equation}
\Delta_{0i}^{+}(x,y,\tau) = \gamma_{0i}(x,y,\tau) 
e^{2J{\theta}_{i}(x,y)} L_{i}(x,y) \sigma_{3} +
\frac{\partial{\theta}_{i}(x,y)}{\partial x} J;
\end{equation}
and, after a brief calculation, the differential equation (\ref{3.16d}) 
for $\bsQ^{(i)+}$ is seen to be equivalent to the following
differential equation for $\bsR^{(i)+}$:
\begin{equation}
\frac{\partial \bsR^{(i)+}(x,x',y,\tau)}{\partial x} =
\gamma_{0i}(x,y,\tau) L_{i}(x,y) \sigma_{3}
\bsR^{(i)+}(x,x',y,\tau).
\end{equation}
\end{abc}

\begin{abc}
\begin{theorem}[Equivalence theorem]
\label{3.1F} \mbox{ } \\
The definition of $\bsQ^{(i)+}$ in Eqs.\ (\ref{3.7a}) to (\ref{3.7c})
is equivalent to the statement that $\bsR^{(i)+}$ is a $2 \times 2$
matrix function with domain
\begin{equation}
\dom{\bsR^{(i)+}} = \bsD^{+}_{i}
\end{equation}
such that, for each $(x',y,\tau) \in \D^{+}_{i}$, $\bsR^{(i)+}(x,x',y,\tau)$
is a continuous function of $x$ throughout $\check{\I}^{(i)}(y)$; and
\begin{eqnarray}
\bsR^{(i)+}(x,x',y,\tau) & = & I + \int_{x'}^{x} d\lambda
\gamma_{0i}^{+}(\lambda,y,\tau) L_{i}(\lambda,y) \sigma_{3}
\bsR^{(i)+}(\lambda,x',y,\tau) \nonumber \\ & & 
\text{ for all } (x,x',y,\tau) \in \bsD^{+}_{i}.
\label{3.24b}
\end{eqnarray}
\end{theorem}
\end{abc}

\proof
Similar to the proof of Thm.~\ref{3.4C}.  Use Thm.~\ref{3.1C}(ii).
\cheers

Just as we did for $\bsQ^{(i)+}$ in Sec.~\ref{Sec_3}C, we can generate the
solution of Eq.\ (\ref{3.24b}) in the form of a Neumann series,
\begin{abc}
\begin{equation}
\bsR^{(i)+} = \sum_{n=0}^{\infty} \bsR^{(i)+}_{n},
\label{3.25a}
\end{equation}
where
\begin{equation}
\bsR^{(i)+}_{0}(x,x',y,\tau) := I
\label{3.25b}
\end{equation}
and, for all $n \ge 0$,
\begin{equation}
\bsR^{(i)+}_{n+1}(x,x',y,\tau) := \int_{x'}^{x} d\lambda
\gamma_{0i}(\lambda,y,\tau) L_{i}(\lambda,y) \sigma_{3}
\bsR^{(i)+}_{n}(\lambda,x',y,\tau);
\end{equation}
and (just as we did for $\bsQ^{(i)+}$ in Sec.~\ref{Sec_3}C) we can employ
Thm.~\ref{3.1C}(i) and the fact that $L_{i}$ is continuous to
prove that the above series (\ref{3.25a}) is absolutely convergent and is
also uniformly convergent in any compact subspace of $\bsD^{+}_{i}$. 
\end{abc}

Since $L_{i}(\lambda,y)$ is a real linear combination of the
matrices $I$ and $J$, and since $\sigma_{3} J \sigma_{3} = - J = J^{T}$,
the even numbered terms in the Neumann series (\ref{3.25a}) are clearly
given by Eq.\ (\ref{3.25b}) and, for all $n \ge 0$, by 
\begin{abc}
\begin{equation}
\bsR^{(i)+}_{2n+2}(x,x',y,\tau) = (\tau-y) \int_{x'}^{x} d\lambda_{2}
\int_{x'}^{\lambda_{2}} d\lambda_{1} \left\{
\frac{L_{i}(\lambda_{2},y)L_{i}^{T}(\lambda_{1},y)}
{\M^{+}(\tau-\lambda_{2})\M^{+}(\tau-\lambda_{1})} 
\bsR^{(i)+}_{2n}(\lambda_{1},x',y,\tau) \right\};
\label{3.26a}
\end{equation}
and the odd numbered terms are then given, for all $n \ge 0$, by
\begin{equation}
\bsR^{(i)+}_{2n+1}(x,x',y,\tau) = \int_{x'}^{x} d\lambda
\gamma_{0i}^{+}(\lambda,y,\tau) L_{i}(\lambda,y)
\sigma_{3} \bsR^{(i)+}_{2n}(\lambda,x',y,\tau).
\label{3.26b}
\end{equation}
For a further analysis of the above integrals, the following variant of
Thm.~\ref{3.1C} will be useful.
\end{abc}

\begin{abc}
\begin{theorem}[Integral transforms employing the kernel 
$\M^{+}(\tau-\lambda)^{-1}$]
\label{3.2F}
\mbox{ } \\ \vspace{-3ex}
\begin{romanlist}
\item 
Let $\bU_{i} \subset {\mathbf \domE}_{i}$ be defined as in Thm.~\ref{3.1C}(i).  Then
there exists a positive real number $m(\bU_{i})$ such that 
\begin{equation}
\sgn{(x-x')} \int_{x}^{x'} d\lambda |\M^{+}(\tau-\lambda)^{-1}| < m(\bU_{i})
\text{ for all } (x,x',y,\tau) \in \bU_{i} \times \bar{C}^{+}.
\end{equation}
\item 
Consider any complex-valued function $\bphi^{(i)+}$ whose domain is $\bsD^{+}_{i}$
such that, for each $(x',y,\tau) \in \D^{+}_{i}$, the function of $x$ given by
$\bphi^{(i)+}(x,x',y,\tau)$ is continuous throughout $\check{\I}^{(i)}(y)$.
Then the function $\bpsi^{(i)+}$ whose domain is $\bsD^{+}_{i}$ and whose value
for each $(x,x',y,\tau) \in \bsD^{+}_{i}$ is
\begin{equation}
\bpsi^{(i)+}(x,x',y,\tau) := \int_{x'}^{x} d\lambda
\frac{\bphi^{(i)+}(\lambda,x',y,\tau)}{\M^{+}(\tau-\lambda)}
\end{equation}
exists; and, for each $(x',y,\tau)$, $\bpsi^{(i)+}(x,x',y,\tau)$ is a
continuous function of $x$ throughout $\check{\I}^{(i)}(y)$.  Moreover,
$\partial\bpsi^{(i)+}(x,x',y,\tau)/\partial x$ exists and
\begin{equation}
\frac{\partial\bpsi^{(i)+}(x,x',y,\tau)}{\partial x} = 
\frac{\bphi^{(i)+}(x,x',y,\tau)}{\M^{+}(\tau-x)} \text{ for all }
(x,x',y,\tau) \in \bsD^{+}_{0i}.
\end{equation}
\item 
Let $\bphi^{(i)+}$ and $\bpsi^{(i)+}$ be defined as in the preceding
part (ii) of this theorem.  Then, if $\bphi^{(i)+}$ is continuous, so
is $\bpsi^{(i)+}$.
\item 
Let $\bphi^{(i)+}$ and $\bpsi^{(i)+}$ be defined as in the preceding
part (ii), and assume that $\bphi^{(i)+}$ is continuous and that
$\partial\bphi^{(i)+}(x,x',y,\tau)/\partial y$ exists and is a continuous
function of $(x,x',y,\tau)$ throughout $\bsD^{+}_{i}$.  Then
$\partial\bpsi^{(i)+}(x,x',y,\tau)/\partial y$ exists, is a continuous
function of $(x,x',y,\tau)$ throughout $\bsD^{+}_{i}$, and 
\begin{equation}
\frac{\partial\bpsi^{(i)+}(x,x',y,\tau)}{\partial y} = \int_{x'}^{x} d\lambda
\frac{\partial\bphi^{(i)+}(\lambda,x',y,\tau)/\partial
y}{\M^{+}(\tau-\lambda)}.
\end{equation}
Clearly, $\partial^{2}\bpsi^{(i)+}(\x,x',y,\tau)/\partial y\partial x$ and
$\partial^{2}\bpsi^{(i)+}(x,x',y,\tau)/\partial x \partial y$ exist, are equal and
\begin{equation}
\frac{\partial^{2}\bpsi^{(i)+}(x,x',y,\tau)}{\partial x \partial y} =
\frac{\partial\bphi^{(i)+}(x,x',y,\tau)/\partial y}{\M^{+}(\tau-x)}
\text{ at all } (x,x',y,\tau) \in \bsD^{+}_{0i}.
\end{equation}
\end{romanlist}
\end{theorem}
\end{abc}

\proofs
The proofs of the above statements (i), (ii), (iii) and (iv) are obtained
by replacing `$\gamma_{0i}^{+}(\lambda,y,\tau)$' by `$\M^{+}(\tau-\lambda)$'
and `$\M_{1}(\bU_{i})$' by `$m(\bU_{i})$' in the proofs of parts (i), (ii),
(iii) and (iv) of Thm.~\ref{3.1C}, respectively.  [Eqs.\ (\ref{3.12c}) and
(\ref{3.12d}) in those proofs are to be divided through by $\sqrt{|\tau-y|}$.]
\cheers

We shall also be considering integral transforms such as Eq.\ (\ref{3.26b}),
where the kernel is $[\mu^{+}(\lambda_{1},\lambda_{2},\tau)]^{-1} =
[\M^{+}(\tau-\lambda_{1})\M^{+}(\tau-\lambda_{2})]^{-1}$ with 
$\lambda_{1}$, $\lambda_{2} \in |x,x'|$.

\begin{definition}{Dfns.\ of $\mu^{\pm}$, $[\mu]$ and $\mu$\label{def48}}
We let $\mu^{+}$ and $\mu^{-}$ denote those functions whose domains are
$(R^{1})^{2} \times \bar{C}^{+}$ and $(R^{1})^{2} \times \bar{C}^{-}$, 
respectively, such that
\begin{abc}
\begin{equation}
\mu^{\pm}(\alpha,\beta,\tau) := \M^{\pm}(\tau-\alpha)\M^{\pm}(\tau-\beta)
\text{ for all } \alpha, \beta \in R^{1} \text{ and } \tau \in \bar{C}^{\pm}.
\label{3.28a}
\end{equation}
We shall let $[\mu]$ denote that extension of $\mu^{+}$ such that
\begin{equation}
\dom{[\mu]} := (R^{1})^{2} \times C
\end{equation}
and
\begin{equation}
[\mu](\alpha,\beta,\tau) := [\mu^{+}(\alpha,\beta,\tau^{*})]^{*} 
= \mu^{-}(\alpha,\beta,\tau) \text{ for all } \alpha, \beta \in R^{1}
\text{ and } \tau \in \bar{C}^{-}.
\label{3.28c}
\end{equation}
Thus, $[\mu](\alpha,\beta,\tau) = \mu^{\pm}(\alpha,\beta,\tau)$ when
$\tau \in \bar{C}^{\pm}$, and $[\mu]$ is an extension of $\mu^{-}$ as
well as of $\mu^{+}$.  The functions $\mu^{+}$ and $\mu^{-}$ will be
called the {\em components} of the relation $[\mu]$; and we recall that
the restriction of $[\mu]$ to $\{(\alpha,\beta,\tau): \alpha,\beta \in
R^{1}, \tau \in C - |\alpha,\beta|\}$ is denoted by $\mu$.  Note:  {\em
Strictly, $[\mu]$ is not a function since $[\mu](\alpha,\beta,\sigma)$
has the two distinct values}
\begin{equation}
\mu^{\pm}(\alpha,\beta,\sigma) = \pm i \sqrt{(\sigma-\alpha)(\sigma-\beta)}
\text{ when } \alpha \ne \beta \text{ and } \sigma \in |\alpha,\beta|
-\{\alpha,\beta\}.
\label{3.28d}
\end{equation}
\end{abc}
\end{definition}

\begin{abc}
\begin{definition}{Dfn.\ of a function $\bphi^{(i)+}$ that is Schwarz 
extendable from $\bsD^{+}_{i}$ and Dfns.\ of $\bphi^{(i)-}$, $[\bphi^{(i)}]$
and $\bphi^{(i)}$\label{def49}}

Suppose that $\bphi^{(i)+}$ is a complex valued (or matrix valued with
complex elements) function with the domain $\bsD^{+}_{i}$ such that
$\bphi^{(i)+}$ is continuous and, for each $(x,x',y) \in {\mathbf \domE}_{i}$,
$\bphi^{(i)+}(x,x',y,\tau)$ is a holomorphic function of $\tau$
throughout $C^{+}$ and
\begin{equation}
\begin{array}{l}
\bphi^{(i)+}(x,x',y,\sigma) \text{ is real for all }
\sigma \in R^{1} - |x,x'| \text{ (which, because of } \\
\text{the continuity of } \bphi^{(i)+}, \text{ implies that }
\bphi^{(i)+}(x,x',y,\infty) \text{ is real).}
\end{array}
\label{3.28e}
\end{equation}
Then we shall let $\bphi^{(i)-}$ denote the function whose domain is
$\bsD^{-}_{i}$ and whose value at each $(x,x',y,\tau)$ in this
domain is
\begin{equation}
\bphi^{(i)-}(x,x',y,\tau) := [\bphi^{(i)+}(x,x',y,\tau^{*})]^{*};
\end{equation}
and we shall let $[\bphi^{(i)}]$ denote that extension of $\bphi^{(i)+}$
such that
\begin{equation}
\dom{[\bphi^{(i)}]} := {\mathbf \domE}_{i} \times C
\end{equation}
and
\begin{equation}
[\bphi^{(i)}](x,x',y,\tau) := \bphi^{(i)\pm}(x,x',y,\tau)
\text{ when } \tau \in \bar{C}^{\pm}.
\end{equation}
The functions $\bphi^{(i)+}$ and $\bphi^{(i)-}$ will be called the
{\em components} of $[\bphi^{(i)}]$.  Furthermore, $\bphi^{(i)}$
will denote the restriction of $[\bphi^{(i)}]$ to
\begin{eqnarray}
\dom{\bphi^{(i)}} & := & \{(x,x',y,\tau): (x,x',y) \in {\mathbf \domE}_{i}, \tau
\in C - \bar{\bsI}^{(i)}(\x,\x') 
\label{3.28i} \\ & & \text{where $\x=(x,y)$ and $\x'=(x',y)$ if $i=3$}
\nonumber \\ & & \text{~~and $\x=(y,x)$ and $\x'=(y,x')$ if $i=4$} \}. 
\nonumber
\end{eqnarray}
{}From the Schwarz reflection principle,\footnote{H.~A.~Schwarz, J.\ f\"{u}r
Math.\ {\bf 70}, 105-120 (1869).  See, for example, E.\ C.\ Titchmarsh,
{\em Theory of Functions} (1932), p.\ 155.}
$\bphi^{(i)}(x,x',y,\tau)$ is a holomorphic function of $\tau$ throughout 
$C - \bar{\bsI}^{(i)}(\x,\x')$.  A function  $\bphi^{(i)+}$ will henceforth 
be called {\em Schwarz extendable from } $\bsD^{+}_{i}$ iff it satisfies all 
of the conditions given in the sentence which contains (\ref{3.28e}).
\end{definition}
\end{abc}

Note:  The set of all complex (or $2 \times 2$ matrix) valued functions
which are Schwarz extendable from $\bsD^{+}_{i}$ is a commutative ring.

\begin{abc}
\begin{theorem}[Functions which are Schwarz extendable from $\bsD^{+}_{i}$]
\label{3.3F}
\mbox{ } \\ \vspace{-3ex}
\begin{romanlist}
\item 
If $\bphi^{(i)+}$ is Schwarz extendable from $\bsD^{+}_{i}$, then
$\bphi^{(i)-}$ is continuous and, for each $(x,x',y) \in {\mathbf \domE}_{i}$,
$\bphi^{(i)}(x,x',y,\tau)$ is a holomorphic function of $\tau$ throughout
$C^{-}$, and
\begin{equation}
\begin{array}{l}
\bphi^{(i)-}(x,x',y,\sigma) = \bphi^{(i)+}(x,x',y,\sigma)
\text{ and is real for all } \\
\sigma \in R^{1} - |x,x'| \text{ and at } \sigma = \infty,
\sigma = x \text{ and } \sigma = x'.
\end{array}
\label{3.29a}
\end{equation}
Also, if $\zeta$ denotes any point in $C^{+}$,
\begin{equation}
\lim_{\zeta \rightarrow \sigma} \bphi^{(i)}(x,x',y,\sigma \pm \zeta)
= \bphi^{(i)\pm}(x,x',y,\sigma) \text{ at all } 
\sigma \in |x,x'| - \{x,x'\} =: \I^{(i)}(\x,\x').
\end{equation}
\item 
If $\bphi^{(i)+}_{1}$ and $\bphi^{(i)+}_{2}$ are Schwarz extendable
from $\bsD^{+}_{i}$, then the function $\bpsi^{(i)+}$ whose domain is $\bsD^{+}_{i}$
and whose value for each $(x,x',y,\tau) \in \bsD^{+}_{i}$ is
\begin{equation}
\bpsi^{(i)+}(x,x',y,\tau) := \int_{x'}^{x} d\lambda_{2}
\int_{x'}^{\lambda_{2}} d\lambda_{1}
\frac{\bphi^{(i)+}_{2}(\lambda_{2},x',y,\tau)
\bphi^{(i)+}_{1}(\lambda_{1},x',y,\tau)}
{\mu^{+}(\lambda_{1},\lambda_{2},\tau)}
\end{equation}
is also extendable from $\bsD^{+}_{i}$; and
\begin{equation}
\bpsi^{(i)-}(x,x',y,\tau) := \int_{x'}^{x} d\lambda_{2}
\int_{x'}^{\lambda_{2}} d\lambda_{1}
\frac{\bphi^{(i)-}_{2}(\lambda_{2},x',y,\tau)
\bphi^{(i)-}_{1}(\lambda_{1},x',y,\tau)}
{\mu^{-}(\lambda_{1},\lambda_{2},\tau)}.
\end{equation}
Also, $[\bpsi^{(i)}](x,x',y,\tau)$ is obtained when $\bphi^{(i)-}_{1}$,
$\bphi^{(i)-}_{2}$ and $\mu^{-}$ in the above integrand are replaced by
$[\bphi^{(i)}_{1}]$, $[\bphi^{(i)}_{2}]$ and $[\mu]$, respectively; and
$\bpsi^{(i)}(x,x',y,\tau)$ is obtained when $\bphi^{(i)-}_{1}$,
$\bphi^{(i)-}_{2}$ and $\mu^{-}$ are replaced by $\bphi^{(i)}_{1}$,
$\bphi^{(i)}_{2}$ and $\mu$, respectively.
\item 
Suppose that each member $\bphi_{n}^{(i)+}$ of an infinite sequence of
functions is Schwarz extendable from $\bsD^{+}_{i}$ and that, for each choice of
$\bU_{i} \subset {\mathbf \domE}_{i}$ as defined in Thm.~\ref{3.1C}(i), the
restriction of $\bphi^{(i)+}_{n}$ to $\bU_{i} \times \bar{C}^{+}$ uniformly
converges as $n \rightarrow \infty$.  Then there exists exactly one function
$\bphi^{(i)+}$ whose domain is $\bsD^{+}_{i}$ and whose value at each
$(x,x',y,\tau) \in \bsD^{+}_{i}$ is $\bphi^{(i)+}(x,x',y,\tau) := \lim_{n
\rightarrow \infty} \bphi^{(i)+}_{n}(x,x',y,\tau)$; and
$\bphi^{(i)+}$ is Schwarz extendable from $\bsD^{+}_{i}$.
\end{romanlist}
\end{theorem}
\end{abc}

\proofs
\begin{romanlist}
\item 
The proof is elementary.
\item 
Use Thms.~\ref{3.2F}(ii) and~(iii) (twice in succession) to establish the
existence and continuity of $\bpsi^{(i)+}$.  The remainder of the
proof is elementary.
\item 
Use the theorems on any uniformly convergent sequence of continuous functions
and on any uniformly convergent sequence of holomorphic functions.
\end{romanlist}
\cheers

\begin{abc}
\begin{theorem}[The functions $\bsT^{(i)+}$ and $\bsU^{(i)+}$]
\label{3.4F}
\mbox{ } \\ \vspace{-3ex}
\begin{romanlist}
\item 
For each $(x,x',y,\tau) \in \bsD^{+}_{i}$, let
\begin{equation}
\bsS^{(i)+}_{0}(x,x',y,\tau) := I
\label{3.30a}
\end{equation}
and, for each integer $n \ge 0$,
\begin{eqnarray}
\bsS^{(i)+}_{2n+2}(x,x',y,\tau) & := & \int_{x'}^{x} d\lambda_{2}
\int_{x'}^{\lambda_{2}} d\lambda_{1} \frac{L_{i}(\lambda_{2},y)
L_{i}^{T}(\lambda_{1},y)}{\mu^{+}(\lambda_{1},\lambda_{2},\tau)}
\bsS^{(i)+}_{2n}(\lambda_{1},x',y,\tau) \nonumber
\\ & \text{and} & 
\label{3.30b} \\
\bsS^{(i)+}_{2n+1}(x,x',y,\tau) & := & \int_{x'}^{x} d\lambda
\frac{L_{i}^{T}(\lambda,y)}{\M^{+}(\tau-\lambda)}
\bsS^{(i)+}_{2n}(\lambda,x',y,\tau).
\label{3.30c}
\end{eqnarray}
Then, for all $n \ge 0$, the functions $\bsS^{(i)+}_{n}$ exist; and
$\bsS^{(i)+}_{2n}$
is Schwarz extendable from $\bsD^{+}_{i}$.  The function whose domain is $\bsD^{+}_{i}$ and 
whose values are given by $\M^{+}(\tau-x') \bsS^{(i)+}_{2n+1}(x,x',y,\tau)$ is
also Schwarz extendable from $\bsD^{+}_{i}$.

\item 
Moreover, for all $n \ge 0$ and $(x,x',y,\tau) \in \bsD^{+}_{i}$,
$\left[\M^{+}(\tau-x) \frac{\partial}{\partial x}\right]
\bsS^{(i)+}_{n}(x,x',y,\tau)$ and $\partial \bsS^{(i)+}_{n}(x,x',y,\tau)
/\partial y$ exist, are continuous functions of $(x,x',y,\tau)$ throughout 
$\bsD^{+}_{i}$,
and
\begin{eqnarray}
\M^{+}(\tau-x) \frac{\partial \bsS^{(i)+}_{2n+2}(x,x',y,\tau)}{\partial x} & = &
L_{i}(x,y) \bsS^{(i)+}_{2n+1}(x,x',y,\tau) , \nonumber \\
\M^{+}(\tau-x) \frac{\partial \bsS^{(i)+}_{2n+1}(x,x',y,\tau)}{\partial x} & = &
L_{i}(x,y) \bsS^{(i)+}_{2n}(x',x,y,\tau) , \nonumber \\
\bsS^{(i)+}_{n}(x,x,y,\tau) & = & \delta_{n0} I.
\label{3.30d}
\end{eqnarray}

\item 
For all $n \ge 0$ and $(x,x',y,\tau) \in \bsD^{+}_{i}$,
\begin{equation}
\bsR^{(i)+}_{n}(x,x',y,\tau) = [\sigma_{3}\M^{+}(\tau-y)]^{n}
\bsS^{(i)+}_{n}(x,x',y,\tau)
\label{3.30e}
\end{equation}
and is a continuous function of $(x,x',y,\tau)$ throughout $\bsD^{+}_{i}$.

\item 
Let $\bsT^{(i)+}$ and $\bsU^{(i)+}$ denote the functions whose domains are
$\bsD^{+}_{i}$ and whose values for each $(x,x',y,\tau) \in
\bsD^{+}_{i}$ are
\begin{eqnarray}
\bsT^{(i)+}(x,x',y,\tau) & := & 
\sum_{n=0}^{\infty} (\tau-y)^{n} \bsS^{(i)+}_{2n}(x,x',y,\tau) 
\label{3.30f}
\\ & \text{and} & \nonumber \\
\bsU^{(i)+}(x,x',y,\tau) & := & 
\sum_{n=0}^{\infty} (\tau-y)^{n} \bsS^{(i)+}_{2n+1}(x,x',y,\tau).
\label{3.30g}
\end{eqnarray}
Then $\bsT^{(i)+}$ and $\bsU^{(i)+}$ exist and, for each
$(x,x',y,\tau) \in \bsD^{+}_{i}$,
\begin{eqnarray}
\bsT^{(i)+}(x,x',y,\tau) & = & I + \int_{x'}^{x} d\lambda_{2} 
\int_{x'}^{\lambda_{2}} d\lambda_{1}
\label{3.30h} \\ & & \mbox{ }
\frac{(\tau-y)L_{i}(\lambda_{2},y)L_{i}^{T}(\lambda_{1},y)}
{\mu^{+}(\lambda_{1},\lambda_{2},\tau)} \bsT^{(i)+}(\lambda_{1},x',y,\tau),
\nonumber \\
\bsU^{(i)+}(x,x',y,\tau) & = & 
\int_{x'}^{x} d\lambda \frac{L_{i}^{T}(\lambda,y)}
{\M^{+}(\tau-\lambda)} 
\bsT^{(i)+}(\lambda,x',y,\tau)
\label{3.30i}
\\ & \text{and} & \nonumber \\
\bsR^{(i)+}(x,x',y,\tau) & = & \bsT^{(i)+}(x,x',y,\tau)
+ \sigma_{3} \M^{+}(\tau-y) \bsU^{(i)+}(x,x',y,\tau).
\label{3.30j}
\end{eqnarray}
Furthermore, $\bsT^{(i)+}$ is Schwarz extendable from $\bsD^{+}_{i}$, 
$\bsU^{(i)+}$ is continuous, and the function whose domain is $\bsD^{+}_{i}$
and whose value at each $(x,x',y,\tau) \in \bsD^{+}_{i}$ is
$\M^{+}(\tau-x') \bsU^{(i)+}(x,x',y,\tau)$ is Schwarz extendable from
$\bsD^{+}_{i}$.

\item 
For all $(x,x',y) \in {\mathbf \domE}_{i}$,
\begin{equation}
\bsR^{(i)+}(x,x',y,y) = I.
\label{3.30k}
\end{equation}
\end{romanlist}
\end{theorem}
\end{abc}

\proof
\begin{abc}
\begin{romanlist}
\item 
First use the principle of mathematical induction and Thm.~\ref{3.3F}(ii)
to prove that, for all $n \ge 0$, $\bsS_{2n}^{(i)+}$ exists and is Schwarz
extendable from $\bsD^{+}_{i}$; and, also,
\begin{equation}
\begin{array}{l}
\text{the function with domain $\bsD^{+}_{i}$ and values} \\
(\tau-y)^{n} \bS_{2n}^{(i)+}(x,x',y,\tau) \text{ is Schwarz
extendable from $\bsD^{+}_{i}$.}
\label{3.31a}
\end{array}
\end{equation}
The above statement (\ref{3.31a}) will be used later in the proof.

{}From its definition by Eq.\ (\ref{3.30c}) and from Thms.~\ref{3.2F}(ii)
and~(iii), $\bsS_{2n+1}^{(i)+}$ exists and is continuous.  Furthermore,
note that
\begin{eqnarray}
\M^{+}(\tau-x') \bsS_{2n+1}^{(i)+}(x,x',y,\tau) & = &
2 \int_{x'}^{x} d\lambda_{2} \int_{x'}^{\lambda_{2}} d\lambda_{1}
\frac{L_{i}(\lambda_{2},y) \bsS_{2n}^{(i)+}(\lambda_{2},x',y,\tau)}
{\mu^{+}(\lambda_{1},\lambda_{2},\tau)} \nonumber \\ & & \mbox{ }
+ \int_{x'}^{x} d\lambda_{2} L_{i}^{T}(\lambda_{2},y)
\bsS_{2n}^{(i)+}(\lambda_{2},x',y,\tau).
\label{3.31b}
\end{eqnarray}
The second integral on the right side of the above equation clearly defines
a function which is Schwarz extendable from $\bsD^{+}_{i}$.  Therefore, from 
the above Eq.\ (\ref{3.31b}) and from Thm.~\ref{3.3F}(ii), the function
whose domain is $\bsD^{+}_{i}$ and whose values are given by
$\M^{+}(\tau-x') \bsS_{2n+1}^{(i)+}(x,x',y,\tau)$ is Schwarz extendable
from $\bsD^{+}_{i}$.
\cheers

\item 
Apply Thms.~\ref{3.2F}(ii) and~(iv) to Eqs.\ (\ref{3.30a}) to (\ref{3.30c})
to obtain the third of Eqs.\ (\ref{3.30d}).
\cheers

\item 
Apply the principle of mathematical induction to Eqs.\ (\ref{3.25b}),
(\ref{3.26a}), (\ref{3.26b}), and (\ref{3.30a}) to (\ref{3.30c}).
\cheers

\item 
We already know that the infinite series on the right side of Eq.\
(\ref{3.25a}) uniformly converges to $\bsR^{(i)+}(x,x',y,\tau)$ on any
given compact subspace of $\bsD^{+}_{i}$.  Therefore, employing Eq.\
(\ref{3.30e}), one sees that
$$
\sum_{n=0}^{\infty} \bsR_{2n}^{(i)+}(x,x',y,\tau) = \sum_{n=0}^{\infty}
(\tau-y)^{n} \bsS_{2n}^{(i)+}(x,x',y,\tau)
$$
uniformly converges on any given compact subspace of $\bsD^{+}_{i}$.  Therefore,
from the definition of $\bsT^{(i)+}$ by Eq.\ (\ref{3.30f}), the statement
(\ref{3.31a}) and Thm.~\ref{3.3F}(iii), $\bsT^{(i)+}$ exists and is
Schwarz extendable from $\bsD^{+}_{i}$.

Equation (\ref{3.30h}) is obtained by multiplying both sides of Eq.\
(\ref{3.30b}) by $(\tau-y)^{n}$, summing over all $n \ge 0$ and using
Eq.\ (\ref{3.30a}), and then applying the theorem on term by term 
integration of a uniformly convergent series and using the definition
(\ref{3.30f}) of $\bsT^{(i)+}$.

Next, by applying Thms.~\ref{3.2F}(ii) and~(iii) to the following integral,
and then using Eq.\ (\ref{3.30f}) to replace $\bsT^{(i)+}$ by a uniformly
convergent series of continuous terms, and then applying the theorem on
term by term integration of a uniformly convergent series, one obtains
\begin{equation}
\begin{array}{l}
\int_{x'}^{x} d\lambda \frac{L_{i}^{T}(\lambda,y)}{\M^{+}(\tau-\lambda)}
\bsT^{(i)+}(\lambda,x',y,\tau) \text{ exists for all } (x,x',y,\tau) \in
\bsD^{+}_{i}, \\
\text{is a continuous function of $(x,x',y,\tau)$ throughout $\bsD^{+}_{i}$,} \\
\text{and equals }
\sum_{n=0}^{\infty} (\tau-y)^{n} \int_{x'}^{x} d\lambda
\frac{L_{i}^{T}(\lambda,y)}{\M^{+}(\tau-\lambda)}
\bsS_{2n}^{(i)+}(\lambda,x',y,\tau), \\
\text{which (according to the full theorem on term by term integration} \\
\text{of a uniformly convergent series) is also uniformly convergent} \\
\text{on any given compact subspace of $\bsD^{+}_{i}$.}
\end{array}
\label{3.31c}
\end{equation}
So, from the above statement (\ref{3.31c}) and the definition (\ref{3.30c})
of $\bsS_{2n+1}^{(i)+}$, one sees that $\bsU^{(i)}$ as defined by Eq.\
(\ref{3.30g}) exists and is given in terms of $\bsT^{(i)+}$ by Eq.\
(\ref{3.30i}); and, furthermore, the series in Eq.\ (\ref{3.30g}) uniformly
converges to $\bsT^{(i)+}$ on any given compact subspace of $\bsD^{+}_{i}$,
whereupon part (i) of this theorem [the part concerning the Schwarz
extendability of $\M^{+}(\tau-x') \bsT_{2n+1}^{(i)+}(x,x',y,\tau)$] and
Thm.~\ref{3.3F}(iii) imply that the function whose domain is $\bsD^{+}_{i}$ and
whose values are given by $\M^{+}(\tau-x') \bsU^{(i)+}(x,x',y,\tau)$ is
Schwarz extendable from $\bsD^{+}_{i}$.

Since $\bsT^{(i)+}$ is continuous, Eq.\ (\ref{3.30i}) and Thm.~\ref{3.2F}(iii)
imply that $\bsU^{(i)+}$ is continuous.

Finally, Eq.\ (\ref{3.30j}) is obtained from (\ref{3.25a}), (\ref{3.30e}),
(\ref{3.30f}) and (\ref{3.30g}).

\cheers

\item 
Use Eqs.\ (\ref{3.30f}), (\ref{3.30a}) and (\ref{3.30j}).
\cheers
\end{romanlist}
\end{abc}

Note that from the definition of $\bsT^{(i)+}$ and $\bsU^{(i)+}$
by Eqs.\ (\ref{3.30f}) and (\ref{3.30g}) and from Eq.\ (\ref{3.30d}),
\begin{equation}
\bsT^{(i)+}(x,x,y,\tau) = I \text{ and } 
\bsU^{(i)+}(x,x,y,\tau) = 0 \text{ for all }
(x,y,\tau) \in \D^{+}_{i}.
\label{3.31d}
\end{equation}
The above statement can also be deduced from Eqs.\ (\ref{3.7c}), (\ref{3.22b}),
(\ref{3.22d}), (\ref{3.30j}) and the fact that $\bsU^{(i)+}$ is continuous.

\begin{definition}{Dfns.\ of $\bsM^{(i)+}$ and $\bsN^{(i)+}$\label{def50}}
Let $\bsM^{(i)+}$ and $\bsN^{(i)+}$ denote the functions with the domain
$\bsD^{+}_{0i}$ and the values
\begin{eqnarray}
\bsM^{(i)+}(x,x',y,\tau) & := & e^{J\theta_{i}(x,y)} 
\bsT^{(i)+}(x,x',y,\tau) e^{-J\theta_{i}(x',y)} \nonumber \\ & \text{and} & 
\label{3.31e} \\
\bsN^{(i)+}(x,x',y,\tau) & := & e^{-J\theta_{i}(x,y)} 
\bsU^{(i)+}(x,x',y,\tau) e^{-J\theta_{i}(x',y)}. 
\end{eqnarray}
\end{definition}

{}From Eqs.\ (\ref{3.22d}), (\ref{3.30j}) and the above definitions
(\ref{3.31e}),
\begin{abc}
\begin{eqnarray}
\bsQ^{(i)+}(x,x',y,\tau) & = & \bsM^{(i)+}(x,x',y,\tau)
+ \sigma_{3} \M^{+}(\tau-y) \bsN^{(i)+}(x,x',y,\tau) \nonumber \\
& & \text{ for all } (x,x',y,\tau) \in \bsD^{+}_{i}.
\label{3.31f}
\end{eqnarray}
{}From Eqs.\ (\ref{3.31e}) and Thm.~\ref{3.4F}(iv),
\begin{equation}
\begin{array}{l}
\bsN^{(i)+} \text{ is continuous, and } \bsM^{(i)+}
\text{ and the function whose domain is } \bsD^{+}_{i} \\
\text{ and whose values are given by} 
\M^{+}(\tau-x') \bsN^{(i)+}(x,x',\tau,y) \text{ are } \\
\text{ both Schwarz extendable from } \bsD^{+}_{i};
\end{array}
\label{3.31g}
\end{equation}
and, from Eqs.\ (\ref{3.31d}),
\begin{equation}
\bsM^{(i)+}(x,x,y,\tau) = I \text{ and } \bsN^{(i)+}(x,x,y,\tau) = 0
\text{ for all } (x,y,\tau) \in \D^{+}_{i}.
\label{3.31h}
\end{equation}
\end{abc}

\begin{definition}{Dfn.\ of a function which is Schwarz extendable
from $\bsD^{+}_{0i}$\label{def51}}
A complex valued (or matrix valued with complex elements) function
$\bphi^{(i)+}$ will be called {\em Schwarz extendable from} $\bsD^{+}_{0i}$
if its domain is $\bsD^{+}_{0i}$, it is continuous and, for each $(x,x',y)
\in {\mathbf \domE}_{i}$, $\bphi^{(i)+}(x,x',y,\tau)$ is a holomorphic
function of $\tau$ throughout $C^{+}$ and
\begin{equation}
\begin{array}{l}
\bphi^{(i)+}(x,x',y,\sigma) \text{ is real for all } \sigma \in R^{1}
-\grave{\bsI}^{(i)}(\x,\x') \\
\text{[which, because of the continuity of } \bphi^{(i)+}, \text{ implies
that } \\
\bphi^{(i)+}(x,x',y,x') \text{ is real when } x' \ne x \text{ and }
\bphi^{(i)+}(x,x',y,\infty) \text{ is real].}
\end{array}
\label{3.31i}
\end{equation}
$\bphi^{(i)-}$, $[\bphi^{(i)}]$ and $\bphi^{(i)}$ are defined as they
were for a function which is Schwarz extendable from $\bsD^{+}_{i}$; and,
from the Schwarz reflection principle, $\bphi^{(i)+}(x,x',y,\tau)$
is a holomorphic function of $\tau$ throughout $C - \grave{\bsI}^{(i)}(\x,\x')$.
\end{definition}
As the reader can easily prove from Eq.\ (\ref{3.30i}), the fact
that $\bsT^{(i)+}$ is Schwarz extendable from $\bsD^{+}_{i}$, and
Eq.\ (\ref{3.31e}),
\begin{equation}
\begin{array}{l}
\text{the functions with the domains } \bsD^{+}_{0i} \text{ and values } \\
\bsU^{(i)+}(x,x',y,\tau)/\M^{+}(\tau-x) \text{ and }
\bsN^{(i)+}(x,x',y,\tau)/\M^{+}(\tau-x) \\
\text{are Schwarz extendable from } \bsD^{+}_{0i}.
\end{array}
\label{3.31j}
\end{equation}


\setcounter{theorem}{0}
\setcounter{equation}{0}
\subsection{Construction of $\Q^{\pm}$, $\bsQ^{\pm}$, $\bsF^{\pm}$,
$\F^{\pm}$, $\grave{\bsF}$, $\grave{\F}$, $\hat{\bsQ}$ and $\hat{\Q}$
from $\bsQ^{(i)+}$}

\begin{abc}
\begin{theorem}[Existence and continuity of $\Q^{+} \in \S_{\Q^{+}}$
for each $\E \in \S_{\E}$]
\label{3.1G} \mbox{ } \\
Corresponding to each $\E \in \S_{\E}$, the function $\Q^{+}$ whose
domain is $\domE \times \bar{C}^{+}$ and whose value for each
$(\x,\tau)$ in this domain is
\begin{equation}
\Q^{+}(\x,\tau) := \bsQ^{(4)+}(s,s_{0},r,\tau)
\bsQ^{(3)+}(r,r_{0},s_{0},\tau)
\label{3.32a}
\end{equation}
is the member of $\S_{\Q^{+}}$ such that
\begin{equation}
d\Q^{+} = \Delta^{+} \Q^{+},
\end{equation}
where $\Delta^{+}$ is defined in terms of $\E$ by Eqs.\ (\ref{1.7a})
to (\ref{1.7c}).  Moreover, $\Q^{+}$ is continuous.
\end{theorem}
\end{abc}

\begin{abc}
\proof
The sets $\S_{\Q^{\pm}}$ were defined in Sec.~\ref{Sec_3}D.  [See Eqs.\ 
(\ref{1.47a}) to (\ref{1.47c}).]

{}From Eqs.\ (\ref{3.32a}) and (\ref{3.7c}), $\Q^{+}(\x_{0},\tau) = I$ for
all $\tau \in \bar{C}^{+}$.  So, condition (i) in the definition of
$\S_{\Q^{+}}$ is satisfied by $\Q^{+}$.

{}From Thm.~\ref{3.1D}, the left hand factor on the right side of Eq.\ 
(\ref{3.32a}) is a continuous function of $(s,r,\tau)$ throughout
$\domE_{4} \times \bar{C}^{+}$, while the right hand factor is a 
continuous function of $(r,\tau)$ throughout $\I^{(3)} \times \bar{C}^{+}$.
[Note: $\check{\I}^{(i)}(x^{i}_{0}) = \I^{(i)}$, since $r < s_{0}$ for
all $r \in \I^{(3)}$ and $r_{0} < s$ for all $s \in \I^{(4)}$.]  Therefore,
$\A^{+}(\x,\tau)$ is a continuous function of $(\x,\tau)$ throughout
$\dom{\Q^{+}} := \domE \times \bar{C}^{+}$.

It follows from the continuity of $\Q^{+}$ that condition (ii) in
the definition of $\S_{\Q^{+}}$ is satisfied by $\Q^{+}$.

{}From Eq.\ (\ref{3.7d}), Thm.~\ref{3.2B}(v) and Thm.~\ref{3.1D},
$\partial\bsQ^{(4)}(s,s_{0},r,\tau)/\partial s$ exists, and
\begin{equation}
\frac{\partial\bsQ^{(4)}(s,s_{0},r,\tau)}{\partial s} = 
\Delta^{+}_{4}(\x,\tau) \bsQ^{(4)+}(s,s_{0},r,\tau)
\label{3.33a}
\end{equation}
and is a continuous function of $(s,r,\tau)$ throughout
$\{(s,r,\tau) \in \D_{4}^{+}: s \ne \tau\}$; \hfill
\linebreak
and $\partial\bsQ^{(3)+}(r,r_{0},s,\tau)/\partial r$ exists, and
\begin{equation}
\frac{\partial\bsQ^{(3)+}(r,r_{0},s_{0},\tau)}{\partial r} = 
\Delta^{+}_{3}((r,s_{0}),\tau) \bsQ^{(3)+}(r,r_{0},s_{0},\tau)
\end{equation}
and is a continuous function of $(r,\tau)$ throughout
$\{(r,\tau) \in \I^{(3)} \times \bar{C}^{+}: r \ne \tau\}$.  From
Eq.\ (\ref{3.16b}) and Thm.~\ref{3.1E} [see Eq.\ (\ref{3.20})],
$\partial \bsQ^{(4)+}(s,s_{0},r,\tau)/\partial r$ exists, and
\begin{equation}
\frac{\partial\bsQ^{(4)+}(s,s_{0},r,\tau)}{\partial r} =
\Delta^{+}_{3}(\x,\tau) \bsQ^{(4)+}(s,s_{0},r,\tau)
- \bsQ^{(4)+}(s,s_{0},r,\tau) \Delta^{+}_{3}((r,s_{0}),\tau)
\label{3.33c}
\end{equation}
and is a continuous function of $(s,r,\tau)$ throughout 
$\{(s,r,\tau) \in \D_{4}^{+}: r \ne \tau\}$.  It now follows from
the above statements which contain Eqs.\ (\ref{3.33a}) to (\ref{3.33c})
that, at all $(\x,\tau) \in \domE \times \bar{C}^{+}$ at which
$\tau \notin \{r,s\}$, $d\Q^{+}(\x,\tau)$ exists and equals
$\Delta^{+}(\x,\tau) \Q^{+}(\x,\tau)$.  So, all three conditions
in the definition of $\S_{\Q^{+}}$ are satisfied by $\Q^{+}$.
\cheers
\end{abc}

\begin{abc}
\begin{theorem}[Existence and continuity of $\bsQ^{+} \in \S_{\subbsQ^{+}}$
for each $\E \in \S_{\E}$] 
\label{3.2G} \mbox{ } \\
Corresponding to each $\E \in \S_{\E}$, the function $\bsQ^{+}$ whose
domain is $\domE^{2} \times \bar{C}^{+}$ and whose value for each
$(\x,\x',\tau)$ in this domain is
\begin{equation}
\bsQ^{+}(\x,\x',\tau) := \Q^{+}(\x,\tau) [\Q^{+}(\x',\tau)]^{-1}
\label{3.34a}
\end{equation}
is the member of $\S_{\subbsQ^{+}}$ such that
\begin{equation}
d\bsQ^{+} = \Delta^{+} \bsQ^{+},
\end{equation}
where $\Delta^{+}$ is defined in terms of $\E$ by Eqs.\ (\ref{1.7a})
to (\ref{1.7c}).  Moreover, $\bsQ^{+}$ is continuous.  

Note:  From Thm.~\ref{3.3B}(i) and Eq.\ (\ref{3.32a}),
\begin{equation}
\det{\Q^{+}(\x,\tau)} = 1 \text{ for all }
(\x,\tau) \in \dom{\Q^{+}}.
\end{equation}
Therefore, $[\Q^{+}(\x',\tau)]^{-1}$ exists for all $(\x',\tau)
\in \dom{\Q^{+}}$.
\end{theorem}
\end{abc}

\proof
Use the preceding Thm.~\ref{3.1G} to prove that $\bsQ^{+}$ as defined
by Eq.\ (\ref{3.34a}) satisfies all three conditions in the definition
of $\S_{\subbsQ^{+}}$ in Sec.~\ref{Sec_1}B.  [See Eqs.\ (\ref{1.9a})
to (\ref{1.9c}).]  Clearly, $\bsQ^{+}$ is continuous, because $\Q^{+}$
is continuous.
\cheers

\begin{abc}
\begin{corollary}[Existences and continuities of $\bsQ^{-}$ and $\Q^{-}$
for each $\E \in \S_{\E}$]
\label{3.3G} \mbox{ } \\
Corresponding to each $\E \in \S_{\E}$, the functions $\bsQ^{-}$ and
$\Q^{-}$ whose domains are $\domE^{2} \times \bar{C}^{-}$ and
$\domE \times \bar{C}^{-}$, and whose values are given by
\begin{eqnarray}
\bsQ^{-}(\x,\x',\tau) & := & [\bsQ^{+}(\x.\x',\tau^{*})]^{*}
\text{ for all } (\x,\x',\tau) \in \dom{\bsQ^{-}}, 
\label{3.35a} \\
\Q^{-}(\x,\tau) & := & [\Q^{+}(\x,\tau^{*})]^{*} \text{ for all }
(\x,\tau) \in \dom{\Q^{-}}
\label{3.35b}
\end{eqnarray}
are the members of $\S_{\subbsQ^{-}}$ and $\S_{\Q^{-}}$ such that
\begin{equation}
d\bsQ^{-} = \Delta^{-} \bsQ^{-} \text{ and }
d\Q^{-} = \Delta^{-} \Q^{-},
\end{equation}
respectively, where $\Delta^{-}$ is defined in terms of $\E$ by
Eqs.\ (\ref{1.7a}) to (\ref{1.7c}).  Moreover, $\bsQ^{-}$ and
$\Q^{-}$ are continuous.
\end{corollary}
\end{abc}

\proof
Use the preceding Thms.~\ref{3.1G} and~\ref{3.2G} together with
Prop.~\ref{1.2A}.
\cheers

Note that, from Eqs.\ (\ref{3.7f}), (\ref{3.32a}) and (\ref{3.35b}),
\begin{abc}
\begin{equation}
\Q^{\pm}(\x,\tau) = \bsQ^{(4)\pm}(s,s_{0},r,\tau)
\bsQ^{(3)\pm}(r,r_{0},s_{0},\tau) \text{ for all }
(\x,\tau) \in \domE \times \bar{C}^{\pm};
\label{3.36a}
\end{equation}
and, from Eqs.\ (\ref{3.34a}), (\ref{3.35a}) and (\ref{3.35b}),
\begin{equation}
\bsQ^{\pm}(\x,\x',\tau) = \Q^{\pm}(\x,\tau) [\Q^{\pm}(\x',\tau)]^{-1}
\text{ for all } (\x,\x',\tau) \in \dom{\bsQ^{\pm}},
\end{equation}
and, since $Q^{\pm}(\x_{0},\tau) = I$,
\begin{equation}
\Q^{\pm}(\x,\tau) = \bsQ^{\pm}(\x,\x_{0},\tau) \text{ for all }
(\x,\tau) \in \dom{\Q^{\pm}}.
\label{3.36c}
\end{equation}
Several properties of $\bsQ^{\pm}$ are given by Props.~\ref{1.3A}
and~\ref{1.2D}.  We leave it for the reader to prove from the 
definitions of $\S_{\subbsQ^{\pm}}$ and $\bsQ^{(i)\pm}$ in 
Secs.~\ref{Sec_1}B and~\ref{Sec_3}C, respectively, and from the
uniqueness theorems \ref{1.3A}(ii) and \ref{3.3B}(ii) that
\begin{eqnarray}
\bsQ^{(3)\pm}(r,r',s,\tau) & = & \bsQ^{\pm}((r,s),(r',s),\tau)
\text{ for all } (r,r',s,\tau) \in \dom{\bsQ^{(3)\pm}},
\label{3.36d} \\
\bsQ^{(4)\pm}(s,s',r,\tau) & = & \bsQ^{\pm}((r,s),(r,s'),\tau)
\text{ for all } (s,s',r,\tau) \in \dom{\bsQ^{(4)\pm}}.
\label{3.36e}
\end{eqnarray}
Also, from Prop.~\ref{1.3A}(iv) and the above Eqs.\ (\ref{3.36c}
to (\ref{3.36e}),
\begin{equation}
\A^{\pm}(\x,\tau) = \bsQ^{(3)\pm}(r,r_{0},s,\tau)
\bsQ^{(4)\pm}(s,s_{0},r_{0},\tau).
\end{equation}
The reader will note that we could have used the above equation
instead of Eq.\ (\ref{3.32a}) to define $\Q^{+}$.
\end{abc}

The set $\S_{\subbsF^{\pm}}$ was defined by Eqs.\ (\ref{1.21a})
to (\ref{1.21c}), and $\S_{\F^{\pm}}$ was defined by Eqs.\ (\ref{1.52a})
to (\ref{1.52c}).  The reader should review these definitions at this
time.

\begin{theorem}[Existences and continuities of $\bsF^{\pm}$ and
$\F^{\pm}$ for each $\E \in \S_{\E}$]
\label{3.4G} \mbox{ } \\
Corresponding to each $\E \in \S_{\E}$, there exists exactly one
$\bsF^{\pm} \in \S_{\subbsF^{\pm}}$ and exactly one $\F^{\pm} \in
\S_{\F^{\pm}}$, and $\bsF^{\pm}$ and $\F^{\pm}$ are continuous
throughout their respective domains (\ref{1.21a}) and (\ref{1.52a}).
\end{theorem}

\proof
This follows from the definition of $\bsF^{\pm}$ in terms of
$(A,\bsQ^{\pm},P^{M\pm})$, the definition of $\F^{\pm}$ in terms
of $(A,\Q^{\pm},P^{M\pm})$, the existences and continuities of
$\bsQ^{\pm}$ and $\Q^{\pm}$, Thm.~\ref{1.1C}(i) on the continuity
of $P^{M\pm}$, and the continuity of $A$ as is evident from its
definition by Eqs.\ (\ref{1.11a}) and (\ref{1.11b}).
\cheers

The relations between $\bsF^{\pm} \in \S_{\subbsF^{\pm}}$ and
$\F^{\pm} \in \S_{\F^{\pm}}$ corresponding to the same $\E \in \S_{\E}$
were given by Prop.~\ref{1.2F}.  In particular,
\begin{equation}
\F^{\pm}(\x,\tau) = \bsF^{\pm}(\x,\x_{0},\tau) \text{ for all }
(\x,\tau) \in \dom{\F^{\pm}}.
\label{3.37}
\end{equation}
Several properties of $\bsF^{\pm}$ were given by Thm.~\ref{1.1D};
and several properties of $\F^{\pm}$ can be obtained from 
Thms.~\ref{1.1D}(i), (ii), (iii) and (v) by using Eq.\ (\ref{3.37}).
Some pertinent properties of $\F^{\pm}$ are given by the following
theorems.

\begin{abc}
\begin{theorem}[Properties of $\F^{\pm} \in \S_{\F^{\pm}}$]
\label{3.5G}
\mbox{ } \\ \vspace{-3ex}
\begin{romanlist}
\item 
The member of $\S_{\F^{\pm}}$ that corresponds to $\E^{M}$ is given by
\begin{equation}
\F^{M\pm}(\x,\tau) = \left( \begin{array}{cc}
1 & -i(\tau-z) \\ 0 & 1
\end{array} \right) \left( \begin{array}{cc}
1 & 0 \\ 0 & \bnu^{\pm}(\x,\x_{0},\tau)
\end{array} \right) \left( \begin{array}{cc}
1 & i(\tau-z_{0}) \\ 0 & 1 
\end{array} \right),
\end{equation}
where $\bnu^{\pm}$ is defined by Eqs.\ (\ref{1.6l}) to (\ref{1.6n}),
(\ref{1.6o}) and (\ref{1.6p}).
\item 
At each $(\x,\tau) \in \dom{\F^{\pm}}$, $d\F^{\pm}(\x,\tau)$ exists, and
\begin{equation}
d\F^{\pm}(\x,\tau) = \Gamma(\x,\tau) \F^{\pm}(\x,\tau).
\end{equation}
\item 
For all $(\x,\tau) \in \dom{\F^{\pm}}$,
\begin{equation}
\det{\F^{\pm}(\x,\tau)} = \bnu^{\pm}(\x,\x_{0},\tau).
\end{equation}
\item 
For each $\x \in \domE$, $\F^{\pm}(\x,\tau)$ is a holomorphic function
of $\tau$ throughout $C^{\pm}$; and the boundary values of $\F^{\pm}(\x,\tau)$
satisfy
\begin{equation}
\F^{+}(\x,\sigma) = \F^{-}(\x,\sigma) \text{ for all }
\sigma \in R^{1} - \grave{\I}(\x),
\label{3.39a}
\end{equation}
and
\begin{equation}
\F^{+}(\x,\infty) = \F^{-}(\x,\infty) = I.
\label{3.39b}
\end{equation}
\item 
For each $\x \in \domE$, $\alpha \in \{r_{0},s_{0}\}$ and $\beta \in
\{r,s\}$, the following limits exist and are equal as indicated:
\begin{eqnarray}
\lim_{\tau \rightarrow \alpha} \F^{+}(\x,\tau) & = & \lim_{\tau \rightarrow
\alpha} \F^{-}(\x,\tau),
\label{3.40a} \\
\lim_{\tau \rightarrow \beta} [\F^{+}(\x,\tau)]^{-1} & = & 
\lim_{\tau \rightarrow \beta} [\F^{-}(\x,\tau)]^{-1},
\label{3.40b}
\end{eqnarray}
where, of course, $\tau \in \bar{C}^{\pm} - \{r,s,r_{0},s_{0}\}$.
\end{romanlist}
\end{theorem}
\end{abc}

\proofs
\begin{romanlist}
\item 
Use Eqs.\ (\ref{1.18e}) and (\ref{3.37}).
\cheers
\item 
Use Eqs.\ (\ref{1.22}) and (\ref{3.37}).
\cheers
\item 
Use Eqs.\ (\ref{1.23}) and (\ref{3.37}).
\cheers
\item 
Use Thm.~\ref{1.1D}(v) and Eq.\ (\ref{3.37}) to obtain the statement that
$\F^{\pm}(\x,\tau)$ is a holomorphic function of $\tau$ throughout $C^{\pm}$.
Then use Thms.~\ref{1.1D}(viii) and~(vi), together with Eq.\ (\ref{3.37}) to
obtain Eqs.\ (\ref{3.39a}) and (\ref{3.39b}).
\cheers
\item 
This proof will be given in three parts (1), (2) and (3); and the proof will
be given in a form which displays the dependence of $\F^{\pm}$ on the
functions $\bsM^{(i)}$ and $\bsN^{(i)}$.
\begin{arablist}
\begin{abc}
\item 
{}From Eqs.\ (\ref{3.31f}) and (\ref{3.35a}),
\begin{eqnarray}
\bsQ^{(i)\pm}(x,x',y,\tau) & = & \bsM^{(i)\pm}(x,x',y,\tau) + \sigma_{3}
\M^{\pm}(\tau-y) \bsN^{(i)\pm}(x,x',y,\tau) \nonumber \\ & &
\text{for all } (x,x',y) \in {\mathbf \domE}_{i} \text{ and }
\tau \in \bar{C}^{\pm},
\label{3.41a}
\end{eqnarray}
where, for all $(x,x',y) \in {\mathbf \domE}_{i}$ and $\tau \in \bar{C}^{-}$,
\begin{eqnarray}
\bsM^{(i)-}(x,x',y,\tau) & := & [\bsM^{(i)+}(x,x',y,\tau^{*}]^{*},
\label{3.41b} \\
& \text{and} & \nonumber \\
\bsN^{(i)-}(x,x',y,\tau) & := & [\bsN^{(i)+}(x,x',y,\tau^{*}]^{*}.
\label{3.41c}
\end{eqnarray}
It will be convenient to express $\bsM^{(i)\pm}$ and $\bsN^{(i)\pm}$ as
linear combinations of the $2 \times 2$ matrices $I$ and $J$:
\begin{equation}
\begin{array}{rcl}
\bsM^{(i)\pm} & = & I \bsM^{(i)\pm}_{0} + J \bsM^{(i)\pm}_{2}, \\
\bsN^{(i)\pm} & = & I \bsN^{(i)\pm}_{3} + J \bsN^{(i)\pm}_{1},
\end{array}
\label{3.41d}
\end{equation}
where $\bsM^{(i)\pm}_{0}$, $\bsM^{(i)\pm}_{2}$, $\bsN^{(i)\pm}_{3}$ and
$\bsN^{(i)\pm}_{1}$ are complex valued functions, each of which has the
domain $\bsD^{+}_{i}$.  As we know from Thm.~\ref{3.4F}(iv),
\begin{equation}
\bsM^{(i)+}_{0} \text{ and } \bsM^{(i)+}_{2} \text{ are Schwarz
extendable from } \bsD^{+}_{i}, 
\label{3.41e}
\end{equation}
\begin{equation}
\bsN^{(i)\pm}_{3} \text{ and } \bsN^{(i)\pm}_{1} \text{ are continuous},
\label{3.41f}
\end{equation}
and the functions with the domain $\bsD^{+}_{i}$ and the values
\begin{equation}
\begin{array}{l}
\M^{+}(\tau-x') \bsN^{(i)+}_{3}(x,x',y,\tau) \text{ and }
\M^{+}(\tau-x') \bsN^{(i)+}_{1}(x,x',y,\tau) \\
\text{are Schwarz extendable from } \bsD^{+}_{i}.
\end{array}
\end{equation}
Furthermore, from (\ref{3.31j}) and (\ref{3.31h}), the functions with
the domain $\bsD^{+}_{0i}$ and the values
\begin{equation}
\begin{array}{l}
\bsN^{(i)+}_{3}(x,x',y,\tau)/\M^{+}(\tau-x) \text{ and }
\bsN^{(i)+}_{1}(x,x',y,\tau)/\M^{+}(\tau-x) \\
\text{are Schwarz extendable from } \bsD^{+}_{0i};
\end{array}
\label{3.41h}
\end{equation}
and
\begin{equation}
\begin{array}{l}
\bsM^{(i)\pm}_{0}(x',x',y,\tau) = 1 \text{ and } \\
\bsM^{(i)\pm}_{2}(x',x',y,\tau) =
\bsN^{(i)\pm}_{3}(x',x',y,\tau) =
\bsN^{(i)\pm}_{1}(x',x',y,\tau) = 0 \\
\text{for all } (x',y,\tau) \in \D^{+}_{i}.
\end{array}
\end{equation}
\end{abc}

\begin{abc}
\item 
Next, let $\bsF^{(i)\pm}$ denote the function whose domain is 
\begin{equation}
\dom{\bsF^{(i)\pm}} := \{ (x,x',y,\tau): (x,x',y) \in {\mathbf \domE}_{i},
\tau \in \bar{C}^{\pm} - \{x,x'\}\}
\label{3.42a}
\end{equation}
and whose values are
\begin{equation}
\bsF^{(i)\pm}(x,x',y,\tau) := A(\x) P^{M\pm}(\x,\tau)
\bsQ^{(i)\pm}(x,x',y,\tau) [A(\x')P^{M\pm}(\x',\tau)]^{-1}
\label{3.42b}
\end{equation}
where we recall that
\begin{equation}
\x := (r,s) = \left\{ \begin{array}{c}
(x,y) \text{ when } i = 3 \\
(y,x) \text{ when } i = 4
\end{array} \right.
\end{equation}
and it is to be understood that
\begin{equation}
\x' := \left\{ \begin{array}{rcl}
(r',s) & = & (x',y) \text{ when } i = 3 \\
(r,s') & = & (y,x') \text{ when } i = 4.
\end{array} \right.
\end{equation}
{}From the continuity of $A$, the continuity of $P^{M\pm}$ [Thm.~\ref{1.1C}(i)]
and the continuity of $\bsQ^{(i)\pm}$, we obtain
\begin{equation}
\bsF^{(i)\pm} \text{ is continuous.}
\label{3.42e}
\end{equation}
After a straightforward (but not brief) calculation, Eqs.\ (\ref{3.41a}),
(\ref{3.41d}), (\ref{3.42b}), and the expression for $P^{M\pm}(\x,\tau)$
that is given by Eq.\ (\ref{1.16f}), yield
\begin{eqnarray}
& \bsF^{(i)\pm}(x,x',y,\tau) = A(\x) \left( \begin{array}{cc}
1 & -i(\tau-z) \\ 0 & 1 
\end{array} \right) & \nonumber \\ & 
\left( \begin{array}{cc}
\bsM^{(i)\pm}_{0} - i \bsM^{(i)\pm}_{2} &
i(\tau-y) \M^{+}(\tau-x') (\bsN^{(i)\pm}_{3} + i \bsN^{(i)\pm}_{1}) \\
-i(\bsN^{(i)\pm}_{3} - i \bsN^{(i)\pm}_{1})/\M^{+}(\tau-x) &
(\bsM^{(i)\pm}_{0} + i \bsM^{(i)\pm}_{2}) \bnu_{i}^{\pm}(x,x',\tau)
\end{array} \right) & \nonumber \\ &
\left( \begin{array}{cc}
1 & i(\tau-z') \\ 0 & 1
\end{array} \right) A(\x')^{-1}, &
\label{3.42f}
\end{eqnarray}
where we have suppressed the argument symbol `$(x,x',y,\tau)$' that
follows each of the symbols `$\bsM^{(i)\pm}_{0}$', `$\bsM^{(i)\pm}_{2}$',
`$\bsN^{(i)\pm}_{3}$' and `$\bsN^{(i)\pm}_{1}$'.  Above,
\begin{equation}
z = \frac{1}{2}(x+y), \quad z'=\frac{1}{2}(x'+y),
\end{equation}
and we recall that
\begin{equation}
\bnu_{i}^{\pm}(x,x',\tau) := \frac{\M^{\pm}(\tau-x')}{\M^{\pm}(\tau-x)}.
\end{equation}

Now, from the defining equations (\ref{3.42a}) and (\ref{3.42b}) of the 
function $\bsF^{(i)\pm}$ and from Eq.\ (\ref{3.7c}),
\begin{equation}
\bsF^{(i)\pm}(x',x',y,\tau) = I \text{ for all }
(x',y,\tau) \in \D^{+}_{0i}.
\label{3.42i}
\end{equation}
Therefore, with the aid of (\ref{3.29a}), (\ref{3.31i}), (\ref{3.41e})
to (\ref{3.41h}), and the above statement (\ref{3.42i}), one deduces
that the following limits exist for each $(x,x',y) \in {\mathbf \domE}_{i}$
and are equal as indicated:
\begin{equation}
\begin{array}{l}
\lim_{\tau \rightarrow x'} \bsF^{(i)+}(x,x',y,\tau) =
\lim_{\tau \rightarrow x'} \bsF^{(i)-}(x,x',y,\tau), \\
\text{where } \tau \in \bar{C}^{+} - \{x,x'\} 
\text{ in the first of the above limits,} \\
\text{and } \tau \in \bar{C}^{-} - \{x,x'\} 
\text{ in the second limit.}
\end{array}
\label{3.43a}
\end{equation}

Moreover, recalling that the inverse of $\bsQ^{(i)\pm}(x,x',y,\tau)$ is
$\bsQ^{(i)\pm}(x',x,y,\tau)$, one deduces from Eq.\ (\ref{3.42b}) that
\begin{equation}
[\bsF^{(i)\pm}(x,x',y,\tau)]^{-1} = \bsF^{(i)\pm}(x',x,y,\tau)
\text{ for all } (x,x',y,\tau) \in \dom{\bsF^{(i)\pm}}.
\label{3.43b}
\end{equation}
Therefore, from the above statements (\ref{3.43a}) and (\ref{3.43b}),
one deduces that the following limits exist and are equal as indicated:
\begin{equation}
\lim_{\tau \rightarrow x} [\bsF^{(i)+}(x,x',y,\tau)]^{-1} =
\lim_{\tau \rightarrow x} [\bsF^{(i)-}(x,x',y,\tau)]^{-1}.
\label{3.43c}
\end{equation}
Furthermore, since $y \in R^{1} - |x,x'|$, Thm.~\ref{3.3F}(i)
[Eq.\ (\ref{3.29a})] together with (\ref{3.31i}), (\ref{3.41e}),
(\ref{3.41f}), (\ref{3.41h}) and (\ref{3.42e}) enable us to deduce
from Eq.\ (\ref{3.42f}) that the following limits exist and are equal
as indicated:
\begin{equation}
\begin{array}{l}
\bsF^{(i)+}(x,x',y,y) = \lim_{\tau \rightarrow y} \bsF^{(i)+}(x,x',y,\tau) = \\
\lim_{\tau \rightarrow y} \bsF^{(i)-}(x,x',y,\tau) = \bsF^{(i)-}(x,x',y,y);
\end{array}
\label{3.43d}
\end{equation}
and, from Eq.\ (\ref{3.43b}), a like statement holds for
$[\bsF^{(i)\pm}(x,x',y,\tau)]^{-1}$.
\end{abc}

\item 
We next employ Eqs.\ (\ref{3.36a}), (\ref{3.42a}) and (\ref{3.42b}) to
deduce that [recall that $A(\x_{0}) = I$]
\begin{eqnarray}
\F^{\pm}(\x,\tau) & := & A(\x) P^{M\pm}(\x,\tau) \Q^{\pm}(\x,\tau)
[P^{M\pm}(\x_{0},\tau)]^{-1} \nonumber \\
& = & \bsF^{(4)\pm}(s,s_{0},r,\tau) \bsF^{(3)\pm}(r,r_{0},s_{0},\tau)
\label{3.44} \\
& & \text{for all } (\x,\tau) \in \dom{\F^{\pm}}.
\nonumber
\end{eqnarray}
The final conclusions (\ref{3.40a}) and (\ref{3.40b}) readily follow
from the above Eqs.\ (\ref{3.43a}), (\ref{3.43c}), (\ref{3.43d}) and
(\ref{3.44}).
\end{arablist}
\cheers
\end{romanlist}

The set $\S_{\grave{\subbsF}}$ was defined in Sec.~\ref{Sec_1}D [Eqs.\ 
(\ref{1.31a}) to (\ref{1.31c})]; and, for each $\bsF \in \S_{\grave{\subbsF}}$,
an extension $\bar{\bsF}$ of the function $\grave{\bsF}$ was defined by
Eqs.\ (\ref{1.31q}) to (\ref{1.31s}).  The set $\S_{\subbsF}$ of all of
these extensions was then defined by Eq.\ (\ref{1.31t}).  Various properties
of $\grave{\bsF} \in \S_{\grave{\subbsF}}$ and of the corresponding $\bar{\bsF}
\in \S_{\bar{\subbsF}}$ were given by Prop.~\ref{1.1E}; and the relation 
between $\S_{\subbsF^{\pm}}$ and $\S_{\grave{\subbsF}}$ was spelled out by
Prop.~\ref{1.2E}.  Equation (\ref{1.38}) in Prop.~\ref{1.2E} enables us to
construct the member of $\S_{\grave{\subbsF}}$ corresponding to a given
$\E \in \S_{\E}$ from the members of $\S_{\subbsF^{+}}$ and $\S_{\subbsF^{-}}$
that correspond to the same given $\E$.

\begin{theorem}[Uniqueness theorem]
\label{3.6G} \mbox{ } \\
There exists exactly one $\grave{\bsF} \in \S_{\grave{\subbsF}}$ (and,
therefore, exactly one $\bar{\bsF} \in \S_{\bar{\subbsF}}$) corresponding
to each given $\E \in \S_{\E}$.
\end{theorem}

\proof
Use Prop.~\ref{1.2E} and Thm.~\ref{3.4G} together with the uniqueness theorem,
Prop.~\ref{1.1E}(iv).
\cheers

The set $\S_{\grave{\F}}$ and the set of corresponding extensions 
$\S_{\bar{\F}}$ were defined by Eqs.\ (\ref{1.59f}) to (\ref{1.59l}),
the relation between $\S_{\bar{\F}}$ and $\S_{\bar{\subbsF}}$ was
given by Eq.\ (\ref{1.59m}) in Prop.~\ref{1.1G}, various properties
of $\grave{\F} \in \S_{\grave{\F}}$ and of its extension $\bar{\F}$
were given by Prop.~\ref{1.2G}, and the existence of $\grave{\F} \in
\S_{\grave{\F}}$ corresponding to each $H \in \S_{H}$ was established
by Thm.~\ref{1.21G}.  The construction of $\grave{\F} \in \S_{\grave{\F}}$
corresponding to a given $H \in \S_{H}$ in terms of the members
$\F^{\pm} \in \S_{\F^{\pm}}$ corresponding to $\E = H_{22}$ was given
by Eq.\ (\ref{1.65}) in Prop.~\ref{1.3G}.

\begin{abc}
\begin{gssm}[Limits of $\bar{\F}$]
\label{3.7G}
\mbox{ } \\
Let $\zeta \in C^{+}$.  For each $\x \in \domE$ and $i \in \{3,4\}$,
let $\tau \in C - \bar{\I}(\x)$, $\sigma \in \I^{(i)}(\x)$
and $\alpha$ and $\beta$ be points of $\bar{\I}^{(i)}(\x)$
such that $\alpha \in \{r_{0},s_{0}\}$ and $\beta \in \{r,s\}$.  Then
the limits $\lim_{\zeta \rightarrow 0} \bar{\F}(\x,\sigma \pm \zeta)$
exist and
\begin{equation}
\lim_{\zeta \rightarrow 0} \bar{\F}(\x,\sigma \pm \zeta) 
= \F^{\pm}(\x,\sigma).
\label{3.45}
\end{equation}
Also, the following limits all exist and are equal as indicated:
\begin{eqnarray}
\lim_{\sigma \rightarrow \alpha} \F^{\pm}(\x,\sigma) & = &
\lim_{\tau \rightarrow \alpha} \bar{\F}(\x,\tau), 
\label{3.46a} \\
\lim_{\sigma \rightarrow \beta} [\F^{\pm}(\x,\sigma)^{-1}] & = &
\lim_{\tau \rightarrow \beta} [\bar{\F}(\x,\tau)^{-1}].
\label{3.46b}
\end{eqnarray}
\end{gssm}
\end{abc}

\begin{abc}
\proof
Note that $\sigma \in \I^{(i)}(\x)$, $\alpha \in
\bar{\I}^{(i)}(\x)$ and $\beta \in \bar{\I}^{(i)}(\x)$
exist iff $\I^{(i)}(\x)$ is not empty; and, if $\I^{(i)}(\x)$
is not empty, then $\bar{\I}^{(i)}(\x) = \grave{\I}^{(i)}(\x)$
and
\begin{equation}
\begin{array}{r}
\bar{\F}(\x,\tau) = \grave{\F}(\x,\tau) \text{ for all $\tau$ in at least } \\
\text{one open neighborhood of } \bar{\I}^{(i)}(\x).
\end{array}
\label{3.46c}
\end{equation}
{}From Prop.~\ref{1.3G},
\begin{equation}
\grave{\F}(\x,\tau) = \F^{\pm}(\x,\tau) \text{ for all }
\tau \in \bar{C}^{\pm} - \grave{\I}(\x),
\label{3.46d}
\end{equation}
where $\F^{\pm}$ is the member of $\S_{\F^{\pm}}$ that corresponds to
the same $\E \in \S_{\E}$ as does the given $\bar{\F} \in \S_{\bar{\F}}$.
The conclusion (\ref{3.45}) now follows from Eqs.\ (\ref{3.46c}),
(\ref{3.46d}) and the fact that $\F^{+}$ and $\F^{-}$ are continuous
(Thm.~\ref{3.4G}); and the conclusions (\ref{3.46a}) and (\ref{3.46b})
follow from Eq.\ (\ref{3.46c}), Eq.\ (\ref{3.46d}) and Thm.~\ref{3.5G}(v).
\cheers
\end{abc}

The set $\S_{\hat{\subbsQ}}$ was defined by Eqs.\ (\ref{1.39a}), (\ref{1.39b})
and (\ref{1.40}); and several properties of $\hat{\bsQ} \in \S_{\hat{\subbsQ}}$, 
including its continuity, were given by Thm.~\ref{1.3E}.  The existence
and uniqueness of $\hat{\bsQ} \in \S_{\hat{\subbsQ}}$ corresponding to each given
$\E \in \S_{\E}$ is readily deduced from Thm.~\ref{3.6G} and the definition
[Eqs.\ (\ref{1.39a}) and (\ref{1.39b}] of $\hat{\bsQ}$ in terms of $(P^{M},A,
\grave{\bsF})$.  Theorem~\ref{1.4E} provides the method of constructing
$\hat{\bsQ}$ directly from those members of $\S_{\subbsQ^{+}}$ and $\S_{\subbsQ^{-}}$
that correspond to the same member of $\S_{\E}$ as does $\hat{\bsQ}$.  There will
be no need in these notes to derive any properties of $\hat{\bsQ}$ that have not
already been covered in Sec.~\ref{Sec_1}D.

The set $\S_{\hat{\Q}}$ was defined by Eqs.\ (\ref{G2.19}) and (\ref{G2.14a}).

\begin{proposition}[Existence and uniqueness of $\hat{\Q} \in \S_{\hat{\Q}}$]
\label{3.8G} \mbox{ } \\
There exists exactly one $\hat{\Q} \in \S_{\hat{\Q}}$ corresponding to each
given $\E \in \S_{\E}$ [or, equivalently (by GSSM~\ref{Thm_2}), to each
given $H \in \S_{H}$].
\end{proposition}

\proof
Use Thm.~\ref{1.21G} on the existence of $\grave{\F}$ for each given
$H \in \S_{H}$, Thm.~\ref{1.2G}(iv) on the uniqueness of $\grave{\F}$
for each given $H \in \S_{H}$, and the definition of $\hat{\Q}$ in terms
of $(P^{M},A,\grave{\F})$ by Eqs.\ (\ref{G2.19}) and (\ref{G2.14a}).
\cheers

Several properties of $\hat{\Q} \in \S_{\hat{\Q}}$ (including its continuity)
are provided by Prop.~\ref{1.4G}, the construction of $\hat{\Q}$ from the
member of $\S_{\hat{\subbsQ}}$ that corresponds to the same member of
$\S_{\E}$ as does $\hat{\Q}$ is covered by Thm.~\ref{1.5G}, and the 
construction of $\hat{\Q}$ from those members of $\S_{\Q^{+}}$ and 
$\S_{\Q^{-}}$ that correspond to the same member of $\S_{\E}$ as does 
$\hat{\Q}$ is detailed in Prop.~\ref{1.6G}.

\begin{gssm}[Properties of $\S_{\hat{\Q}}$]
\label{3.9G} \mbox{ } \\ \vspace{-3ex}
\begin{romanlist}
\item 
\begin{abc}
The definition of $\S_{\hat{\Q}}$ by Eqs.\ (\ref{G2.19}) and (\ref{G2.14a})
is equivalent to the statement that $\S_{\hat{\Q}}$ is the set of all 
$2 \times 2$ matrix functions $\hat{\Q}$ that have the domain
\begin{equation}
\dom{\hat{\Q}} := \{(\x,\tau):\x \in \domE, \tau \in C -
\hat{\I}(\x)\},
\label{3.47a}
\end{equation}
that satisfy
\begin{equation}
\hat{\Q}(\x_{0},\tau) = I \text{ for all } \tau \in C - [r_{0},s_{0}],
\label{3.47b}
\end{equation}
and that satisfy the conditions that $d\hat{\Q}$ and $\E \in \S_{\E}$ 
exist such that $d\hat{\Q} = \Delta \hat{\Q}$, where $\Delta$ is defined
by Eqs.\ (\ref{1.4b}) and (\ref{1.5a}).
\end{abc}
\item 
\begin{abc}
If $\hat{\Q} \in \S_{\hat{\Q}}$, then
\begin{equation}
\det{\hat{\Q}} = 1
\label{3.48a}
\end{equation}
and
\begin{equation}
\hat{\Q}^{*}(\x,\tau) := [\hat{\Q}(\x,\tau^{*})]^{*} = \hat{\Q}(\x,\tau)
\text{ for all } (\x,\tau) \in \dom{\hat{\Q}}.
\label{3.48b}
\end{equation}
Furthermore, for each $\x \in \domE$, $\hat{\Q}(\x,\tau)$ is a holomorphic
function of $\tau$ throughout $C - \hat{\I}(\x)$, and
\begin{equation}
\hat{\Q}(\x,\infty) = \frac{1}{\sqrt{-f(\x)}} \left( \begin{array}{cc}
1 & - \chi(\x) \\ 0 & - f(\x)
\end{array} \right).
\label{3.48c}
\end{equation}
\end{abc}
\item 
\begin{abc}
Let $\zeta$ denote any point in $C^{+}$.  Then for each $\x \in \domE$ 
and each $\sigma \in R^{1}$, the limits as $\zeta \rightarrow 0$ of
$\hat{\Q}(\x,\sigma+\zeta)$ and $\hat{\Q}(\x,\sigma-\zeta)$ exist, and
are given by
\begin{equation}
\lim_{\zeta \rightarrow 0} \hat{\Q}(\x,\sigma \pm \zeta) = \Q^{\pm}(\x,\sigma).
\label{3.49a}
\end{equation}
Also,
\begin{equation}
\Q^{-}(\x,\sigma) = [\Q^{+}(\x,\sigma)]^{*}
\label{3.49b}
\end{equation}
and $\Q^{+}(\x,\sigma)$ is a continuous function of $\sigma$ throughout
$R^{1}$.
\end{abc}
\item 
Moreover, $\Q^{+}(\x,\sigma) = \Q^{-}(\x,\sigma)$ is a real analytic
function of $\sigma$ at all $\sigma < \bar{\I}^{(3)}(\x)$ and
$\sigma > \bar{\I}^{(4)}(\x)$, while $\Q^{+}(\x,\sigma)$ is an
analytic (but not generally real) function of $\sigma$ in the interval
$\bar{\I}^{(3)}(\x) < \sigma < \bar{\I}^{(4)}(\x)$.
\item 
At all $\sigma \in R^{1} - \{r,s\}$, $d\Q^{+}(\x,\sigma)$ exists and
equals $\Delta^{+}(\x,\sigma) \Q^{+}(\x,\sigma)$, where $\Delta^{+}$
is defined in terms of $\E$ by Eqs.\ (\ref{1.7a}) to (\ref{1.7c}).
\end{romanlist}
\end{gssm}

\proofs
\begin{romanlist}
\item 
\begin{abc}
First suppose that $\hat{\Q} \in \S_{\hat{\Q}}$ is defined as in Eqs.\ 
(\ref{G2.19}) and (\ref{G2.14a}).  Let $\E$ denote the member of $\S_{\E}$
corresponding to the given $\hat{\Q}$.  From Thm.~\ref{3.1G} and
Cor.~\ref{3.3G}, there exist $\Q^{+} \in \S_{\Q^{+}}$ and $\Q^{-} \in
\S_{\Q^{-}}$ such that $d\Q^{\pm}(\x,\tau)$ exists and
\begin{equation}
d\Q^{\pm}(\x,\tau) = \Delta^{\pm}(\x,\tau) \Q^{\pm}(\x,\tau)
\text{ for all } \x \in \domE \text{ and } \tau \in \bar{C}^{\pm}
- \{r,s\},
\label{3.50a}
\end{equation}
where $\Delta^{\pm}$ is defined in terms of $\E$ by Eqs.\ (\ref{1.7a})
to (\ref{1.7c}).  Then, from Thm.~\ref{1.6G}(i) and the uniqueness part
of Prop.~\ref{3.8G},
\begin{equation}
\hat{\Q}(\x,\tau) = \Q^{\pm}(\x,\tau) \text{ for all } \x \in \domE
\text{ and } \tau \in \bar{C}^{\pm} - \hat{\I}(\x).
\label{3.50b}
\end{equation}
The definitions of $\Delta$ by Eqs.\ (\ref{1.4b}) and (\ref{1.5a})
and of $\Delta^{\pm}$ by Eqs.\ (\ref{1.7a}) to (\ref{1.7c}) yield
\begin{equation}
\Delta(\x,\tau) = \Delta^{\pm}(\x,\tau) \text{ for all } \x \in \domE
\text{ and } \tau \in \bar{C}^{\pm} - \hat{\I}(\x).
\label{3.50c}
\end{equation}
Therefore, from Eq.\ (\ref{1.47b}) and the above Eqs.\ (\ref{3.50a})
to (\ref{3.50c}), $\hat{\Q}$ is a $2 \times 2$ matrix function which
has the domain (\ref{3.47a}) which satisfies the condition (\ref{3.47b}),
and which satisfies the conditions that $d\hat{\Q}$ exists and equals
$\Delta\hat{\Q}$.

As regards the converse, suppose that $\hat{\Q}$ is a given $2 \times 2$
matrix function which has the domain (\ref{3.47a}), which satisfies the
condition (\ref{3.47b}), and which satisfies the conditions that $d\hat{\Q}$
exists and that $\E \in \S_{\E}$ exists such that 
\begin{equation}
d\hat{\Q} = \Delta \hat{\Q},
\end{equation}
where $\Delta$ is defined in terms of $\E$ by Eqs.\ (\ref{1.4b}) and
(\ref{1.5a}).  From Thm.~\ref{3.1G} and Cor.~\ref{3.3G}, there exist
$\Q^{\pm} \in \S_{\Q^{\pm}}$ such that
\begin{equation}
d\Q^{\pm}(\x,\tau) = \Delta^{\pm}(\x,\tau) \Q^{\pm}(\x,\tau)
\text{ for all } \x \in \domE \text{ and } \tau \in \bar{C}^{\pm}
- \{r,s\},
\end{equation}
where $\Delta^{\pm}$ is defined in terms of $\E$ by Eqs.\ (\ref{1.7a})
to (\ref{1.7c}); and, from Thm.~\ref{1.6G}(i), the function $\hat{\Q}'$
whose domain is the same as that of $\hat{\Q}$ and whose values are
given by 
\begin{equation}
\hat{\Q}'(\x,\tau) := \Q^{\pm}(\x,\tau) \text{ for all } \x \in \domE
\text{ and } \tau \in \bar{C}^{\pm} - \hat{\I}(\x)
\label{3.50f}
\end{equation}
is a member of $\S_{\hat{\Q}}$.  Recall that $\det{\hat{\Q}'} = 1$
and that, therefore, $\hat{\Q}'(\x,\tau)^{-1}$ exists for all
$(\x,\tau)$ in its domain.  It follows from Eqs.\ (\ref{3.50c}) to
(\ref{3.50f}) that
$$
d[\hat{\Q}'(\x,\tau)^{-1}\Q(\x,\tau)] = 0 \text{ for all }
(\x,\tau) \in \dom{\hat{\Q}}.
$$
However, for each $\tau \in C - [r_{0},s_{0}]$, the space 
$\{ \x \in \domE:\tau \notin \hat{\I}(\x) \}$ is connected.
Hence, upon using Eqs.\ (\ref{3.47b}), (\ref{3.50f}) and (\ref{1.47b}),
we obtain
$$
\hat{\Q}'(\x,\tau)^{-1}\hat{\Q}(\x,\tau) = \hat{\Q}'(\x_{0},\tau)^{-1}
\hat{\Q}(\x_{0},\tau) = I \text{ for all } (\x,\tau) \in \dom{\hat{\Q}}.
$$
So, $\hat{\Q} = \hat{\Q}' \in \S_{\hat{\Q}}$.
\end{abc}
\cheers
\item 
\begin{abc}
Equation (\ref{3.48a}) is the same as Prop.~\ref{1.4G}(i).  Equation
(\ref{3.48b}) follows from Eqs.\ (\ref{3.35b}) and (\ref{3.50b}).
The statement that $\hat{\Q}(\x,\tau)$ is a holomorphic function of
$\tau$ throughout $C - \hat{\I}(\x)$ is contained in
Prop.~\ref{1.4G}(iii).  Equation (\ref{3.48c}) follows from Eq.\ 
(\ref{3.50b}), (\ref{3.36c}) and Prop.~\ref{1.3A}(vi).
\cheers

We shall now give an instructive alternative proof that $\hat{\Q}(\x,\tau)$
is a holomorphic function of $\tau$ throughout $C - \hat{\I}(\x)$.
{}From Eqs.\ (\ref{3.36a}), (\ref{3.35b}), (\ref{3.31f}), (\ref{3.41b}),
(\ref{3.41c}) and (\ref{3.28a}),
\begin{eqnarray}
\Q^{\pm}(\x,\tau) & = & 
\left[ \bsM^{(4)\pm}(s,s_{0},r,\tau) + \sigma_{3}
\mu^{\pm}(r,s,\tau) \bsP^{(4)\pm}(s,s_{0},r,\tau) \right] 
\nonumber \\ & & \mbox{ }
\left[ \bsM^{(3)\pm}(r,r_{0},s_{0},\tau) + \sigma_{3}
\mu^{\pm}(r,s_{0},\tau) \bsP^{(3)\pm}(r,r_{0},s_{0},\tau) \right]
\label{3.51a}
\end{eqnarray}
for all $\x \in \domE$ and $\tau \in \bar{C}^{\pm} - \{r,s\}$, where
\begin{equation}
\bsP^{(i)\pm}(x,x',y,\tau) := 
\frac{\bsN^{(i)\pm}(x,x',y,\tau)}{\M^{\pm}(\tau-x)}.
\end{equation}
Recall that $\bsM^{(i)+}$ is Schwarz extendable from $\bsD_{i}$ [Eq.\ 
(\ref{3.31g})] and $\bsP^{(i)+}$ is Schwarz extendable from $\bsD_{0i}$
[Eq.\ (\ref{3.31j})].  From the definition of $\bsM^{(i)}$ by Eqs.\ 
(\ref{3.28e}) to (\ref{3.28i}), the like definition of $\bsP^{(i)}$ by
Eq.\ (\ref{3.31i}), the definition of $\mu$ by Eqs.\ (\ref{3.28a}) to 
(\ref{3.28c}), and Eqs.\ (\ref{3.50b}) and (\ref{3.51a}),
\begin{eqnarray}
\hat{\Q}(\x,\tau) & = &
\left[ \bsM^{(4)}(s,s_{0},r,\tau) + \sigma_{3}
\mu(r,s,\tau) \bsP^{(4)}(s,s_{0},r,\tau) \right] 
\nonumber \\ & & \mbox{ }
\left[ \bsM^{(3)}(r,r_{0},s_{0},\tau) + \sigma_{3}
\mu(r,s_{0},\tau) \bsP^{(3)}(r,r_{0},s_{0},\tau) \right] 
\label{3.51c}
\end{eqnarray}
for all $\x \in \domE$ and 
\begin{equation}
\tau \in C - \left( [s,s_{0}] \cup [r,s] \cup [r,r_{0}] \cup [r,s_{0}] \right)
= C - \hat{\I}(\x).
\end{equation}
Recall that (for fixed $\x$) $\mu(r,s,\tau)$ is a holomorphic function of
$\tau$ throughout $C - [r,s]$, $\bsM^{(4)}(s,s_{0},r,\tau)$ and 
$\bsP^{(4)}(s,s_{0},r,\tau)$ are holomorphic functions of $\tau$ throughout
$C - \grave{\I}^{(4)}(\x)$, $\mu(r,s_{0},\tau)$ is a holomorphic
function of $\tau$ throughout $C - [r,s_{0}]$, and
$\bsM^{(3)}(r,r_{0},s_{0},\tau)$ and $\bsP^{(3)}(r,r_{0},s_{0},\tau)$ are
holomorphic functions of $\tau$ throughout $C - \grave{\I}^{(3)}(\x)$.
Therefore, $\hat{\Q}(\x,\tau)$ is a holomorphic function of $\tau$ throughout
$C - \hat{\I}(\x)$.
\cheers
\end{abc}
\item 
The existences of the limits of $\hat{\Q}(\x,\sigma \pm \zeta)$ as $\zeta
\rightarrow 0$ and the values (\ref{3.49a}) of these limits are direct
consequences of Eq.\ (\ref{3.50b}) and the fact that $\Q^{+}$ and $\Q^{-}$
are continuous [Thm.~\ref{3.1G} and Cor.~\ref{3.3G}].

Equation (\ref{3.49b}) follows from Eq.\ (\ref{3.35b}); and, since $\Q^{+}$
is continuous, $\Q^{+}(\x,\sigma)$ is a continuous function of $\sigma$
throughout $R^{1}$.
\cheers
\item 
\begin{abc}
{}From Eqs.\ (\ref{3.49b}) and (\ref{3.50b}),
\begin{equation}
\Q^{+}(\x,\sigma) = \Q^{-}(\x,\sigma) = \hat{\Q}(\x,\sigma)
\text{ and is real for all } \sigma \in R^{1} - \hat{\I}(\x);
\end{equation}
and, since $\hat{\Q}(\x,\tau)$ is a holomorphic function of $\tau$
throughout $C - \hat{\I}(\x)$, $\hat{\Q}(\x,\sigma)$ is a
real analytic function of $\sigma$ throughout $R^{1} - \hat{\I}(\x)$.

{}From Eq.\ (\ref{3.51a}), the definitions of Schwarz extendability from
$\bsD_{i}$ and $\bsD_{0i}$, the Schwarz reflection principle and Eq.\
(\ref{3.28d}),
\begin{eqnarray}
\Q^{\pm}(\x,\sigma) & = & 
\left[\bsM^{(4)}(s,s_{0},r,\sigma) \pm i \sigma_{3}
\sqrt{(\sigma-r)(s-\sigma)} \bsP^{(4)}(s,s_{0},r,\sigma)\right] 
\nonumber \\ & & \mbox{ }
\left[\bsM^{(3)}(r,r_{0},s,\sigma) \pm i \sigma_{3}
\sqrt{(\sigma-r)(s_{0}-\sigma)} \bsP^{(3)}(r,r_{0},s_{0},\sigma)\right]
\nonumber \\ & & \mbox{ }
\text{ for all } \grave{\I}^{(3)}(\x) < \sigma <
\grave{\I}^{(4)}(\x);
\label{3.51f}
\end{eqnarray}
and $\Q^{+}(\x,\sigma)$ and $\Q^{-}(\x,\sigma)$ are analytic functions of
$\sigma$ throughout the open interval $\grave{\I}^{(3)}(\x) < 
\sigma < \grave{\I}^{(4)}(\x)$.  (Note:  Of course, more can be
said in the exceptional case when $r = r_{0}$ and $s = s_{0}$.)  The above
Eq.\ (\ref{3.51f}) should be compared with Eq.\ (\ref{3.51c}).
\end{abc}
\cheers
\item 
Obvious.
\end{romanlist}


\setcounter{theorem}{0}
\setcounter{equation}{0}
\subsection{Generalized Abel transforms $V^{(3)}$ and $V^{(4)}$ 
of initial data functions $\E^{(3)}$ and $\E^{(4)}$}

Based upon prior knowledge of the Weyl case\footnote{Our notes on the Weyl
case were archived (\#9303021) at the Los Alamos preprint library in 1993.}
we expected the values of $\bsQ^{(i)+}(x,\sigma,y,\sigma)$ for $\sigma
\in \check{\I}^{(i)}(y)$ to be particularly significant for our
formalism.  From the definition (\ref{3.22d}) of $\bsR^{(i)+}$,
\begin{eqnarray}
\bsQ^{(i)+}(x,\sigma,y,\sigma) & = & 
e^{J\theta_{i}(x,y)} \bsR^{(i)+}(x,\sigma,y,\sigma) 
e^{-J\theta_{i}(\sigma,y)} \text{ for all } (x,y) \in \domE_{i}
\nonumber \\ & & \text{ and } \sigma \in \check{\I}^{(i)}(y) 
\text{ [i.e., for all $(x,\sigma,y) \in {\mathbf \domE}_{i}$].}
\end{eqnarray}
We, therefore, start with an analysis of $\bsR^{(i)+}(x,\sigma,y,\sigma)$
for all $(x,y) \in \domE_{i}$ and $\sigma \in \check{\I}^{(i)}(y)$.
According to Eq.\ (\ref{3.30j}), when $\sigma \ne x$,
\begin{equation}
\bsR^{(i)+}(x,\sigma,y,\sigma) = \bsT^{(i)+}(x,\sigma,y,\sigma)
+ \mu^{+}(x,y,\sigma) \frac{\bsU^{(i)+}(x,\sigma,y,\sigma)}{\M^{+}(\sigma-x)};
\label{3.53a}
\end{equation}
and [Thm.~\ref{3.4F}(iv) and Eq.\ (\ref{3.31j})] since the functions
given by $\bsT^{(i)+}(x,x',y,\tau)$ and
$\bsU^{(i)+}(x,x',y,\tau)/\M^{+}(\tau-x)$ are Schwarz extendable from
$\bsD_{i}$ and $\bsD_{0i}$, respectively,
\begin{equation}
\bsT^{(i)+}(x,\sigma,y,\sigma) \text{ and }
\bsU^{(i)+}(x,\sigma,y,\sigma)/\M^{+}(\sigma-x)
\text{ are real linear combinations of $I$ and $J$.}
\end{equation}

A point which may not be immediately evident is that, although the
denominator in $\bsU^{(i)+}(x,\sigma,y,\sigma)/\M^{+}(\sigma-x)$
vanishes at $\sigma = x$, this ratio is extendable to a continuous
function of $(x,\sigma,y)$ over the entire domain ${\mathbf \domE}_{i}$.
This will soon become obvious.

\begin{abc}
\begin{definition}{Dfns.\ of $\u{\bsT}^{(i)+}$ and $\u{\bsU}^{(i)+}$
\label{def52}}
Let $\u{\bsT}^{(i)+}$ and $\u{\bsU}^{(i)+}$ denote the functions whose
domains are
\begin{equation}
\dom{\u{\bsT}^{(i)+}} = \dom{\u{\bsU}^{(i)+}} := [0,1] \times 
{\mathbf \domE}_{i}
\end{equation}
and whose values are
\begin{eqnarray}
\u{\bsT}^{(i)+}(\xi,x,\sigma,y) & := & 
\bsT^{(i)+}(\sigma-\xi^{2}(\sigma-x),\sigma,y,\sigma), 
\label{3.54b} \\
\u{\bsU}^{(i)+}(\xi,x,\sigma,y) & := & 
-2 \int_{0}^{\xi} d\eta L_{i}^{T}(\sigma-\eta^{2}(\sigma-x),y)
\u{\bsT}^{(i)+}(\eta,x,\sigma,y).
\label{3.54c}
\end{eqnarray}
\end{definition}
\end{abc}

\begin{abc}
\begin{proposition}[Properties of $\u{\bsT}^{(i)+}$ and $\u{\bsU}^{(i)+}$]
\label{3.1H}
\mbox{ } \\ \vspace{-3ex}
\begin{romanlist}
\item 
$\u{\bsT}^{(i)+}$ and $\u{\bsU}^{(i)+}$ are continuous.
\item 
For all $(x,\sigma,y) \in {\mathbf \domE}_{i}$,
\begin{eqnarray}
\u{\bsT}^{(i)+}(0,x,\sigma,y) & = & I, 
\label{3.55a} \\
\u{\bsT}^{(i)+}(1,x,\sigma,y) & = & \bsT^{(i)+}(x,\sigma,y,\sigma), 
\\
\u{\bsU}^{(i)+}(0,x,\sigma,y) & = & 0.
\label{3.55c}
\end{eqnarray}
\item 
For all $(x,\sigma,y) \in {\mathbf \domE}_{i}$ such that $\sigma \ne x$,
\begin{equation}
\u{\bsU}^{(i)+}(1,x,\sigma,y) = \frac{\bsU^{(i)+}(x,\sigma,y,\sigma)}
{\M^{+}(\sigma-x)}.
\label{3.55d}
\end{equation}
\item 
For all $(x,\sigma,y) \in {\mathbf \domE}_{i}$,
\begin{equation}
\bsR^{(i)+}(x,\sigma,y,\sigma) = \u{\bsT}^{(i)+}(1,x,\sigma,y)
+ \sigma_{3} \mu^{+}(x,y,\sigma) \u{\bsU}^{(i)+}(1,x,\sigma,y).
\label{3.56}
\end{equation}
\end{romanlist}
\end{proposition}
\end{abc}

\begin{abc}
\proofs
\begin{romanlist}
\item 
This follows from Eq.\ (\ref{3.54b}), Eq.\ (\ref{3.54c}) and the facts
that $\bsT^{(i)+}$ and $L_{i}$ are continuous.  (The statement that
$\bsT^{(i)+}$ is continuous is part of our definition of Schwarz 
extendability from $\bsD_{i}$.)
\cheers
\item 
These follow from Eq.\ (\ref{3.54b}), Eq.\ (\ref{3.54c}) and the fact
that $\bsT^{(i)+}(x,x,y,\tau) = I$ [which, in turn, derives from Eqs.\ 
(\ref{3.30d}) and (\ref{3.30f})].
\cheers
\item 
In the expression for $\u{\bsU}^{(i)+}(1,x,\sigma,y)$ which is defined
by Eq.\ (\ref{3.54c}), introduce the new integration variable
\begin{equation}
\lambda = \sigma - \eta^{2}(\sigma-x), \quad (\sigma \ne x).
\end{equation}
Then ($\lambda \in |\sigma,x|$)
\begin{equation}
d\lambda = -2(\sigma-x)\eta d\eta = -2(\sigma-x) 
\sqrt{\frac{\sigma-\lambda}{\sigma-x}} d\eta,
\end{equation}
and, since
\begin{equation}
(\sigma-x) \sqrt{\frac{\sigma-\lambda}{\sigma-x}} = \mu^{+}(x,\lambda,\sigma)
\text{ for both } \sigma > x \text{ and } \sigma < x,
\label{3.57c}
\end{equation}
one obtains, with the aid of Eq.\ (\ref{3.54b}),
\begin{equation}
\u{\bsU}^{(i)+}(1,x,\sigma,y) = \int_{\sigma}^{x} d\lambda
\frac{L_{i}^{T}(\lambda,y)}{\mu^{+}(x,\lambda,\sigma)}
\bsT^{(i)+}(\lambda,\sigma,y,\sigma).
\end{equation}
Comparison of the above result with Eq.\ (\ref{3.30i}) nets Eq.\
(\ref{3.55d}).
\cheers
\item 
Employing Eqs.\ (\ref{3.53a}) and (\ref{3.55d}), one sees that Eq.\
(\ref{3.56}) holds for all $(x,\sigma,y) \in {\mathbf \domE}_{i}$
for which $\sigma \ne x$.  However, $\bsR^{(i)+}$ and $\u{\bsU}^{(i)+}$
are both continuous.  Therefore, Eq.\ (\ref{3.56}) holds at $\sigma = x$
as well.
\cheers
\end{romanlist}
\end{abc}

\begin{abc}
\begin{proposition}[Equivalent differential equations]
\label{3.2H}
\mbox{ } \\
For all $(\xi,x,\sigma,y) \in \dom{\u{\bsT}^{(i)+}}$, 
\begin{eqnarray}
\u{\bsT}^{(i)+}(\xi,x,\sigma,y) & = & I + 4(\sigma-x)(\sigma-y) 
\int_{0}^{\xi} d\eta_{2} \int_{0}^{\eta_{2}} d\eta_{1} \,
\label{3.58}
\\ & & \mbox{ } 
L_{i}(\sigma-\eta_{2}^{2}(\sigma-x),y)
L_{i}^{T}(\sigma-\eta_{1}^{2}(\sigma-x),y)
\u{\bsT}^{(i)+}(\eta_{1},x,\sigma,y);
\nonumber
\end{eqnarray}
and the pair of differential equations that are equivalent to the pair
of integral equations (\ref{3.54c}) and (\ref{3.58}) are as follows:
\begin{eqnarray}
\frac{\partial\u{\bsT}^{(i)+}(\xi,x,\sigma,y)}{\partial\xi} & = & 
-2 (\sigma-x)(\sigma-y) L_{i}(\sigma-\xi^{2}(\sigma-x),y)
\u{\bsU}^{(i)+}(\xi,x,\sigma,y), \nonumber \\
& \text{and} & 
\label{3.59a} \\
\frac{\partial\u{\bsU}^{(i)+}(\xi,x,\sigma,y)}{\partial\xi} & = & 
-2 L_{i}^{T}(\sigma-\xi^{2}(\sigma-x),y) \u{\bsT}^{(i)+}(\xi,x,\sigma,y)
\label{3.59b}
\end{eqnarray}
subject to the initial conditions (\ref{3.55a}) and (\ref{3.55c}).
\end{proposition}
\end{abc}

\proof
To derive Eq.\ (\ref{3.58}), introduce the new integration variables
$$
\eta_{1} = \sqrt{\frac{\sigma-\lambda_{1}}{\sigma-x}}
\text{ and }
\eta_{2} = \sqrt{\frac{\sigma-\lambda_{2}}{\sigma-x}}
$$
into Eq.\ (\ref{3.30h}), and use Eqs.\ (\ref{3.54b}) and (\ref{3.57c}).
[Recall that we employ the Lebesgue definition of integrals.]  The proof
of the rest of the proposition is left for the reader.
\cheers

\begin{theorem}[Continuity]
\label{3.3H}
\mbox{ } \\ \vspace{-3ex}
\begin{romanlist}
\item 
Suppose $\E^{(i)}$ is $\bC^{n_{i}} \, (n_{i} \ge 1)$.  Then, for each
$y \in \I^{(7-i)}$, $\u{\bsT}^{(i)+}(\xi,x,\sigma,y)$ and
$\u{\bsU}^{(i)+}(\xi,x,\sigma,y)$ are $\bC^{n_{i}-1}$ functions of
$(\xi,x,\sigma)$ throughout $[0,1] \times [\check{\I}^{(i)}(y)]^{2}$;
and, for each $(x,\sigma,y) \in {\mathbf \domE}_{i}$, these are
$\bC^{n_{i}}$ functions of $\xi$.
\item 
Suppose $\E^{(i)}$ is analytic.  Then, for each $(\xi,y) \in [0,1] \times
I^{(7-i)}$, the functions of $(x,\sigma)$ that are given by 
$\u{\bsT}^{(i)+}(\xi,x,\sigma,y)$ and $\u{\bsU}^{(i)+}(\xi,x,\sigma,y)$
are analytic throughout $[\check{\I}^{(i)}(y)]^{2}$.
\end{romanlist}
\end{theorem}

\begin{abc}
\proofs
\begin{romanlist}
\item 
If $\E^{(i)}$ is $\bC^{n_{i}}$, then it follows from Thm.~\ref{2.2D}(iii)
that $L_{i}(x,y)$ is a $\bC^{n_{i}-1}$ function of $x$ throughout
$\check{\I}^{(i)}(y)$.  The conclusions of this part of the theorem then
follow from the differential equations (\ref{3.59a}) and (\ref{3.59b}),
the initial conditions (\ref{3.55a}) and (\ref{3.55c}), and a standard
theorem on ordinary differential equations.\footnote{See Ref.~\ref{Lefsch},
Ch.~II, Sec.~4.}
\cheers
\item 
If $\E^{(i)}$ is analytic, then it has a holomorphic extension to an open
neighborhood of the real axis interval $\I^{(i)}$ in the space $C$.\footnote{
{\em Several Complex Variables} by S.~Bochner and W.~T.~Martin (Princeton
University Press, Princeton 1948), Ch.~II, Sec.~2, Thm.~6. \label{Bochner}}
Then, from an obvious variant of part (c) of Thm.~\ref{2.2D}(iii), the function
of $x$ given by $L_{i}(x,y)$ has a holomorphic extension to an open neighborhood
$J^{(i)}(y)$ of the real axis interval $\check{\I}^{(i)}(y)$ in the space $C$.

We may then consider the integral equation (\ref{3.58}) for real $\xi \in
[0,1]$, real $y \in \I^{(7-i)}$, and complex $(x,\sigma) \in [J^{(i)}(y)]^{2}$.
In the usual way, one proves that the Neumann series
\begin{equation}
\sum_{n=0}^{N} \u{\bsT}^{(i)+}_{2n}(\xi,x,\sigma,y),
\end{equation}
where
\begin{equation}
\u{\bsT}^{(i)+}_{0}(\xi,x,\sigma,y) := I
\end{equation}
and, for all $n \ge 0$, 
\begin{eqnarray}
\u{\bsT}^{(i)+}_{2n+2}(\xi,x,\sigma,y) & := & 4(\sigma-x)(\sigma-y) 
\int_{0}^{\xi} d\eta_{2} \int_{0}^{\eta_{2}} d\eta_{1} 
\\ & & \mbox{ }
L_{i}(\sigma-\eta_{2}^{2}(\sigma-x),y)
L_{i}^{T}(\sigma-\eta_{1}^{2}(\sigma-x),y)
\u{\bsT}^{(i)+}_{2n}(\eta_{1},x,\sigma,y),
\nonumber
\end{eqnarray}
uniformly converges as $N \rightarrow \infty$ to
\begin{eqnarray}
\u{\bsT}^{\prime(i)+}(\xi,x,\sigma,y) & := & \sum_{n=0}^{\infty}
\u{\bsT}^{(i)+}_{2n}(\xi,x,\sigma,y) \text{ for all } \\
& & \text{for all } \xi \in [0,1], y \in \I^{(7-i)} \text{ and }
(x,\sigma) \in [J^{(i)}(y)]^{2}.
\nonumber
\end{eqnarray}
Moreover, since the solution of the integral equation (\ref{3.58}) is
unique, 
\begin{equation}
\u{\bsT}^{\prime(i)+}(\xi,x,\sigma,y) = \u{\bsT}^{(i)+}(\xi,x,\sigma,y) 
\text{ for all } (\xi,x,\sigma,y) \in [0,1] \times {\mathbf \domE}_{i}.
\label{3.60e}
\end{equation}
Now, it is clear that each term $\u{\bsT}^{(i)+}_{2n}(\xi,x,\sigma,y)$
of the Neumann series is a holomorphic function of $(x,\sigma)$ 
throughout $[J^{(i)}(y)]^{2}$.  Therefore, from the well known theorem
on a uniformly convergent series of holomorphic functions,
$\u{\bsT}^{\prime(i)+}(\xi,x,\sigma,y)$ is a holomorphic function of
$(x,\sigma)$ throughout $[J^{(i)}(y)]^{2}$.  It follows from Eq.\
(\ref{3.60e}) that $\u{\bsT}^{(i)+}(\xi,x,\sigma,y)$ is a real analytic
function of $(x,\sigma)$ throughout $[\check{\I}^{(i)}(y)]^{2}$; and
it then follows from Eq.\ (\ref{3.54c}) that $\u{\bsU}^{(i)+}(\xi,x,\sigma,y)$
is a real analytic function of $(x,\sigma)$ throughout
$[\check{\I}^{(i)}(y)]^{2}$.
\cheers
\end{romanlist}
\end{abc}

Note:  We could have given a considerably abbreviated proof of part (ii)
of the above theorem by using Thm.~(10.3) in Ch.~II, Sec.~5, of the
aforementioned reference by S.~Lefschetz.  However, we felt that a proof
which takes advantage of the fact that the integral equation that we are
considering is linear would be better understood by most readers.  The
Neumann series for the solution of the linear integral equation can also
be used to prove part (i) of the above theorem; but the proof would then
be considerably longer.

We shall now focus attention on the dependences of 
$\u{\bsT}^{(i)}(1,x,\sigma,y)$ and $\u{\bsU}^{(i)}(1,x,\sigma,y)$ on
$\sigma$ (for a fixed $\x$).  Here we shall begin using a (potentially
confusing) notation in which the fixed parameter $\x$ will appear as a
{\em label\/} within parentheses as an integral part of the name of the
function, as in the symbols $C^{(i)}(\x)$ and $D^{(i)}(\x)$ below.  The
value of such a function, say $f(\x)$, at a particular point 
$\sigma \in \dom{f(\x)}$ will be denoted either by $f(\x)(\sigma)$ or
by $f(\x,\sigma)$.  Lest you are not yet sufficiently confused, we shall
often suppress the label $\x$, denoting the value of the function $f(\x)$
at $\sigma \in \dom{f(\x)}$ by $f(\sigma)$.  These same conventions will
be employed frequently in what follows, so be forewarned!

\begin{abc}
\begin{definition}{Dfns.\ of $C^{(i)}(\x)$, $D^{(i)}(\x)$, $A^{(i)}$
$B^{(i)}$ and $V^{(i)}$\label{def53}}

Recall that $\x := (r,s) = (x,y)$ when $i=3$, and $\x = (y,x)$ when
$i=4$.  For each $\x \in \domE$, let $C^{(i)}(\x)$ and $D^{(i)}(\x)$
denote the functions whose domains are
\begin{equation}
\dom{C^{(i)}(\x)} = \dom{D^{(i)}(\x)} := \check{\I}^{(i)}(y)
\label{3.61a}
\end{equation}
and whose values are, for all $\sigma \in \check{\I}^{(i)}(y)$,
\begin{eqnarray}
C^{(i)}(\x,\sigma) & := & C^{(i)}(\x)(\sigma) \nonumber \\ 
& := & e^{J\theta_{i}(x,y)} \u{\bsT}^{(i)}(1,x,\sigma,y)
e^{-J\theta_{i}(\sigma,y)} e^{J\theta_{7-i}(y,\sigma)},
\label{3.61b} \\
D^{(i)}(\x,\sigma) & := & D^{(i)}(\x)(\sigma) \nonumber \\ 
& := & e^{-J\theta_{i}(x,y)} \u{\bsU}^{(i)}(1,x,\sigma,y)
e^{-J\theta_{i}(\sigma,y)} e^{J\theta_{7-i}(y,\sigma)}.
\label{3.61c}
\end{eqnarray}
Let 
\begin{equation}
A^{(i)} := C^{(i)}(\x_{0}), \quad B^{(i)} := D^{(i)}(\x_{0}).
\label{3.62a}
\end{equation}
Note that, since $\check{\I}^{(i)}(y_{0}) = \I^{(i)}$,
\begin{equation}
\dom{A^{(i)}} = \dom{B^{(i)}} = \I^{(i)}.
\end{equation}
Let $V^{(i)}$ denote the function whose domain is
\begin{equation}
\dom{V^{(i)}} := \I^{(i)}
\end{equation}
and whose values are, for all $\sigma \in \I^{(i)}$,
\begin{equation}
V^{(i)}(\sigma) := \bsQ^{(i)+}(x_{0},\sigma,y_{0},\sigma)
\label{3.63b}
\end{equation}
[where recall that $(x_{0},y_{0}) = (r_{0},s_{0})$ if $i=3$, and
$(x_{0},y_{0}) = (s_{0},r_{0})$ if $i=4$].  Equivalently, as one
sees from Eq.\ (\ref{1.10d}) and Eqs.\ (\ref{3.36c}) to (\ref{3.36e}),
\begin{eqnarray}
V^{(3)}(\sigma) & = & \bsQ^{+}(\x_{0},(\sigma,s_{0}),\sigma) \nonumber \\
& = & [\Q^{+}((\sigma,s_{0}),\sigma)]^{-1}, 
\label{G2.21g1} \\
V^{(4)}(\sigma) & = & \bsQ^{+}(\x_{0},(r_{0},\sigma),\sigma) \nonumber \\
& = & [\Q^{+}((r_{0},\sigma),\sigma)]^{-1},
\label{G2.21g2}
\end{eqnarray}
for all $\sigma \in \I^{(3)}$ and $\sigma \in \I^{(4)}$, 
respectively.\footnote{See Appendix~B for the values of $V^{(3)}$ and
$V^{(4)}$ for several familiar metrics and for an explanation of why
we have referred to these functions as generalized Abel transforms of
the initial data.}
\end{definition}
\end{abc}

\begin{abc}
\begin{theorem}[Properties of $C^{(i)}(\x)$ and $D^{(i)}(\x)$]
\label{3.4H} \mbox{ } \\
For each $\x \in \domE$, $i \in \{3,4\}$ and integer $n_{i} \ge 1$:
\begin{romanlist}
\item 
$C^{(i)}(\x)$, $D^{(i)}(\x)$, $A^{(i)}$ and $B^{(i)}$ are real-valued
linear combinations of the $2 \times 2$ matrices $I$ and $J$.
\item 
If $\E^{(i)}$ is $\bC^{n_{i}}$ (resp.\ analytic), then $C^{(i)}(\x)$,
$D^{(i)}(\x)$, $A^{(i)}$ and $B^{(i)}$ are $\bC^{n_{i}-1}$ (resp.\ 
analytic).
\item 
For each $\sigma \in \I^{(i)}$,
\begin{equation}
V^{(i)}(\sigma) = A^{(i)}(\sigma) + \mu^{+}(\x_{0},\sigma) \sigma_{3}
B^{(i)}(\sigma).
\label{3.64}
\end{equation}
\item 
For each $\sigma \in \check{\I}^{(i)}(y)$,
\begin{equation}
\Q^{+}(\x,\sigma) V^{(i)}(\sigma) = C^{(i)}(\x,\sigma) + \mu^{+}(\x,\sigma)
\sigma_{3} D^{(i)}(\x,\sigma).
\label{3.65}
\end{equation}
\item 
If $\E^{(i)}$ is $\bC^{n_{i}}$ (resp.\ analytic), then
${\F}^{\pm}(\x,\sigma)$ is a $\bC^{n_{i}-1}$ (resp.\ analytic)
function of $\sigma$ throughout $\I^{(i)}(\x)$.
\end{romanlist}
\end{theorem}
\end{abc}

\begin{abc}
\proofs
\begin{romanlist}
\item 
Obvious.
\item 
{}From the definition of $\theta_{i}$ by Eqs.\ (\ref{3.22a}) and (\ref{3.22b}),
and from an obvious corollary of Thm.~\ref{2.2D}(iii),
\begin{eqnarray}
\theta_{i}(\sigma,y) \text{ and } \theta_{7-i}(y,\sigma) \text{ are }
\bC^{n_{i}} \text{ (resp.\ analytic) } \nonumber \\
\text{functions of } \sigma \text{ throughout } \check{\I}^{(i)}(y).
\label{3.66a}
\end{eqnarray}
So, from the definitions (\ref{3.61a}) to (\ref{3.62a}), and from
Thm.~\ref{3.3H}, $C^{(i)}(\x)$, $D^{(i)}(\x)$, $A^{(i)}$ and $B^{(i)}$
are $\bC^{n_{i}-1}$ (resp.\ analytic).
\cheers
\item 
{}From the definition of $\theta_{i}$ by Eqs.\ (\ref{3.22a}) and (\ref{3.22b}),
\begin{equation}
\theta_{i}(x_{0},y) = 0 \text{ and } \theta_{7-i}(y_{0},x) = 0.
\label{3.66b}
\end{equation}
So, from Eqs.\ (\ref{3.63b}), (\ref{3.22d}) and (\ref{3.56}),
\begin{eqnarray}
V^{(i)}(\sigma) & = & \bsR^{(i)+}(x_{0},\sigma,y_{0},\sigma)
e^{-J\theta_{i}(\sigma,y_{0})} 
\label{3.67} \\
& = & \left[ \u{\bsT}^{(i)+}(1,x_{0},\sigma,y_{0}) + \mu^{+}(\x_{0},\sigma)
\sigma_{3} \u{\bsU}^{(i)+}(1,x_{0},\sigma,y_{0}) \right]
e^{-J\theta(\sigma,y_{0})}.
\nonumber
\end{eqnarray}
Equation (\ref{3.64}) now follows from Eqs.\ (\ref{3.61b}), (\ref{3.61c}),
(\ref{3.62a}), (\ref{3.66b}) and (\ref{3.67}).
\cheers
\item 
{}From Eqs.\ (\ref{3.36c}), (\ref{3.36d}), (\ref{3.36e}) and (\ref{3.63b}),
and from Prop.~\ref{1.3A}(iv) [Eq.\ (\ref{1.10c})],
\begin{eqnarray*}
\Q^{+}(\x,\sigma) V^{(3)}(\sigma) & = & \bsQ^{+}(\x,\x_{0},\sigma)
\bsQ^{+}(\x_{0},(\sigma,s_{0}),\sigma) \\
& = & \bsQ^{+}(\x,(\sigma,s_{0}),\sigma) \\
& = & \bsQ^{+}((r,s),(\sigma,s),\sigma)
\bsQ^{+}((\sigma,s),(\sigma,s_{0}),\sigma) \\
& = & \bsQ^{(3)+}(r,\sigma,s,\sigma) \bsQ^{(4)+}(s,s_{0},\sigma,\sigma)
\end{eqnarray*}
for all $\sigma \in \check{\I}^{(i)}(s)$.  A like derivation for the
product $\Q^{+}(\x,\sigma) V^{(4)}(\sigma)$ nets for both $i=3$ and
$i=4$
\begin{equation}
\Q^{+}(\x,\sigma) V^{(i)}(\sigma) = \bsQ^{(i)+}(x,\sigma,y,\sigma)
\bsQ^{(7-i)+}(y,y_{0},\sigma,\sigma) \text{ for all }
\sigma \in \check{\I}^{(i)}(y).
\label{3.68a}
\end{equation}
{}From Thm.~\ref{3.4F}(v) [Eq.\ (\ref{3.30k})], Eq.\ (\ref{3.22d}) and
Eq.\ (\ref{3.66b}),
\begin{equation}
\bsQ^{(7-i)+}(y,y_{0},\sigma,\sigma) = e^{J\theta_{i}(y,\sigma)}.
\label{3.68b}
\end{equation}
Also, from Eqs.\ (\ref{3.22d}) and (\ref{3.56}),
\begin{eqnarray}
\lefteqn{\bsQ^{(i)+}(x,\sigma,y,\sigma) = } 
\label{3.68c} \\
& & e^{J\theta_{i}(x,y)} \left[ \u{\bsT}^{(i)+}(1,x,\sigma,y) +
\mu^{+}(\x,\sigma) \sigma_{3} \u{\bsU}^{(i)+}(1,x,\sigma,y) \right]
e^{-J\theta_{i}(\sigma,y)}. \nonumber
\end{eqnarray}
Equation (\ref{3.65}) now follows from Eqs.\ (\ref{3.68a}), (\ref{3.68b}),
(\ref{3.68c}), (\ref{3.61b}) and (\ref{3.61c}).
\cheers
\item 
{}From Eqs.\ (\ref{3.64}) and (\ref{3.65}) and from part (i) of this theorem,
\begin{eqnarray}
\Q^{\pm}(\x,\sigma) & = & 
\left[ C^{(i)}(\x,\sigma) + \mu^{\pm}(\x,\sigma)
\sigma_{3} D^{(i)}(\x,\sigma) \right]
\left[ A^{(i)}(\sigma) + \mu^{\pm}(\x_{0},\sigma)
\sigma_{3} B^{(i)}(\sigma) \right]^{-1} \nonumber \\ & &
\text{ for all } \sigma \in \check{\I}^{(i)}(y),
\label{3.69}
\end{eqnarray}
where we have used Eq.\ (\ref{3.35b}) and (\ref{1.6i}), and we recall
that $M^{-1} = - J M^{T} J$ for any $2 \times 2$ matrix $M$ for which
$\det{M} = 1$.  From Eq.\ (\ref{1.16f}),
\begin{eqnarray}
\F^{M\pm}(\x,\tau) & = & P^{M\pm}(\x,\tau) [P^{M\pm}(\x_{0},\tau)]^{-1}
\nonumber \\
& = & \left( \begin{array}{cc}
1 & -i(\tau-z) \\ 0 & 1
\end{array} \right) \left( \begin{array}{cc}
1 & 0 \\ 0 & \bnu^{\pm}(\x,\x_{0},\tau)
\end{array} \right) \left( \begin{array}{cc}
1 & i(\tau-z_{0}) \\ 0 & 1
\end{array} \right) \nonumber \\
& & \text{ for all } \x \in \domE \text{ and }
\tau \in \bar{C}^{\pm} - \{r,s,r_{0},s_{0}\},
\label{3.70a}
\end{eqnarray}
\begin{eqnarray}
P^{M\pm}(\x,\tau) J [P^{M\pm}(\x,\tau)]^{-1} & = & 
\left( \begin{array}{cc}
-i & 2(\tau-z) \\ 0 & i
\end{array} \right) \nonumber \\
& & \text{ for all } \x \in \domE \text{ and }
\tau \in \bar{C}^{\pm} - \{r,s\},
\label{3.70b}
\end{eqnarray}
and
\begin{eqnarray}
P^{M\pm}(\x,\tau) \mu^{\pm}(\x,\tau) \sigma_{3} [P^{M\pm}(\x,\tau)]^{-1}
& = & \left( \begin{array}{cc}
-\tau+z & -i\rho^{2} \\ -i & \tau-z
\end{array} \right)
\label{3.70c} \\
& & \text{ for all } \x \in \domE \text{ and }
\tau \in \bar{C}^{\pm} - \{r,s\},
\nonumber
\end{eqnarray}
Also, note that $\I^{(i)}(\x) \subset \check{\I}^{(i)}(y) \subset
I^{(i)}(y_{0}) = \I^{(i)}$; and $|x,x_{0}|$ and $|y,y_{0}|$ are disjoint.
Therefore, from the defining equation (\ref{1.52b}) for $\F^{\pm}$
and from the above Eqs.\ (\ref{3.69}) to (\ref{3.70c}),
\begin{eqnarray}
\F^{\pm}(\x,\sigma) & = & 
\left[ \tilde{C}^{(i)}(\x,\sigma) + \left( \begin{array}{cc}
-\sigma+z & -i\rho^{2} \\ -i & \sigma-z
\end{array} \right) \tilde{D}^{(i)}(\x,\sigma) \right] \nonumber \\
& & \F^{M\pm}(\x,\sigma) \left[ \tilde{A}^{(i)}(\sigma) 
+ \left( \begin{array}{cc}
-\sigma+z_{0} & -i\rho_{0}^{2} \\ -i & \sigma-z_{0}
\end{array} \right) \tilde{B}^{(i)}(\sigma) \right]^{-1} \nonumber \\
& & \text{ for all } \x \in \domE, i \in \{3,4\} \text{ and }
\sigma \in \I^{(i)}(\x),
\label{3.71a}
\end{eqnarray}
where
\begin{equation}
\begin{array}{rcl}
\tilde{C}^{(i)}(\x,\sigma) & := & P^{M\pm}(\x,\sigma) C^{(i)}(\x,\sigma)
[P^{M\pm}(\x,\sigma)]^{-1}, \\
\tilde{D}^{(i)}(\x,\sigma) & := & P^{M\pm}(\x,\sigma) D^{(i)}(\x,\sigma)
[P^{M\pm}(\x,\sigma)]^{-1}, \\
\tilde{A}^{(i)}(\sigma) & := & P^{M\pm}(\x_{0},\sigma) A^{(i)}(\sigma)
[P^{M\pm}(\x_{0},\sigma)]^{-1}, \\
\tilde{B}^{(i)}(\sigma) & := & P^{M\pm}(\x_{0},\sigma) B^{(i)}(\sigma)
[P^{M\pm}(\x_{0},\sigma)]^{-1}, 
\end{array}
\label{3.71b}
\end{equation}
and $\F^{M\pm}(\x,\sigma)$ is given by Eq.\ (\ref{3.70a}) with
\begin{eqnarray}
\nu^{\pm}(\x,\sigma) & = & \pm i \sqrt{\frac{\sigma-x_{0}}{x-\sigma}}
\sqrt{\frac{\sigma-y_{0}}{\sigma-y}} \sgn{(\sigma-x)} \nonumber \\
& & \text{ for all } \x \in \domE \text{ and }
\sigma \in \I^{(i)}(\x). 
\label{3.71c}
\end{eqnarray}
The conclusion of this part of the theorem now follows from parts (i)
and (ii) of this theorem taken together with the above Eqs.\ (\ref{3.70a}),
(\ref{3.70b}), (\ref{3.71a}), (\ref{3.71b}) and (\ref{3.71c}).
\cheers
\end{romanlist}
\end{abc}

\begin{definition}{Dfn.\ of $H(\gamma)$\label{def54}}
Suppose $f$ is a real- or complex-valued function the domain of which 
is a union of disjoint intervals of $R^{1}$, and $[a,b]$ is a given
closed subinterval of $\dom{f}$.  Then $f$ is said {\em to obey a
H\"{o}lder condition of index $0 < \gamma \le 1$ on $[a,b]$; i.e.,
to be $H(\gamma)$ on $[a,b]$}, if there exists $M(a,b,\gamma) > 0$
such that $|f(x')-f(x)| \le M(a,b,\gamma) |x'-x|^{\gamma}$ for all
$x, x' \in [a,b]$.\footnote{N.\ I.\ Muskhelishvili, {\em Singular 
Integral Equations}, 2nd Edition, Dover (1992), Ch.~1, Sec.~3 and 
Sec.~6.  \label{Musk}} 
\end{definition}

\begin{theorem}[H\"{o}lder condition for $\bsQ^{(i)+}$]
\label{3.5H} \mbox{ } \\
For each $\x \in \domE$, the function of $\sigma$ that is given by
$\bsQ^{(i)+}(x,\sigma,y,\sigma)$ is $H(1/2)$ on each closed subinterval
of $\check{\I}^{(i)}(y)$.
\end{theorem}

\begin{abc}
\proof
The proof will be given in three parts.
\begin{arablist}
\item 
Each closed subinterval of $\check{\I}^{(i)}(y)$ is contained in at 
least one closed interval $[a,b] \subset \check{\I}^{(i)}(y)$ such tbat
$x \in [a,b]$.  Therefore, to prove this theorem, it is sufficient to
prove that $\bsQ^{(i)+}(x,\sigma,y,\sigma)$ (for fixed $\x$) is $H(1/2)$
on an arbitrary $[a,b] \in \check{\I}^{(i)}(y)$ for which $x \in [a,b]$.
Henceforth in this proof, we shall grant that $x \in [a,b]$; and we
shall let, for each $\x \in \domE$ and each choice of $[a,b]$,
\begin{eqnarray}
M^{(i)}_{1}(\x,a,b) & := & \sup{\{ ||\bsQ^{(i)+}(x,\sigma,y,\sigma)||:
\sigma \in [a,b] \}},
\label{3.72a} \\
M^{(3)}_{2}(\x,a,b) & := & \sup{ \left\{ 
\left| \frac{\partial \E((\lambda,s))/\partial\lambda}{2f((\lambda,s))}
\right|: \lambda \in [a,b] \right\} },
\label{3.72b} \\
M^{(4)}_{2}(\x,a,b) & := & \sup{ \left\{
\left| \frac{\partial \E((r,\lambda))/\partial\lambda}{2f((r,\lambda))}
\right|: \lambda \in [a,b] \right\} },
\label{3.72c} \\
M^{(i)}_{3}(\x,a,b) & := & \sqrt{b-a} + \sup{ \left\{ \sqrt{|a-y|},
\sqrt{|b-y|} \right\} }.
\label{3.72d}
\end{eqnarray}
Since $\bsQ^{(i)+}$ is continuous, $M^{(i)}_{1}(\x,a,b)$ is finite.
Likewise, $M^{(i)}_{2}(\x,a,b)$ and $M^{(i)}_{3}(\x,a,b)$ are finite.

\item 
Since $\bsQ^{(i)+}(x,\sigma,y,\sigma)$ is a continuous function of $\sigma$
throughout $\check{\I}^{(i)}(y)$, the statement that it is $H(1/2)$ on
$[a,b]$ is equivalent to the statement that it is $H(1/2)$ on both $[a,x]$
and $[x,b]$.  

{}From Eqs.\ (\ref{3.7c}) and (\ref{3.16d}), one obtains, for all $\sigma,
\sigma' \in \check{\I}^{(i)}(y)$,
\begin{eqnarray}
\lefteqn{\bsQ^{(i)+}(x,\sigma,y,\sigma) - \bsQ^{(i)+}(x,\sigma',y,\sigma') = }
\nonumber \\
& & \bsQ^{(i)+}(x,\sigma',y,\sigma') \left\{ \int_{\sigma}^{\sigma'}
d\lambda \Delta_{0i}^{+}(\lambda,y,\sigma) \bsQ^{(i)+}(\lambda,\sigma,
y,\sigma) + \right. \\ & & \left.
\int_{\sigma'}^{x} d\lambda \left[ \bsQ^{(i)+}(\lambda,\sigma',y,\sigma')^{-1}
\left( \Delta_{0i}^{+}(\lambda,y,\sigma) - \Delta_{0i}^{+}(\lambda,y,\sigma')
\right) \bsQ^{(i)+}(\lambda,\sigma,y,\sigma) \right] \right\}.
\nonumber
\end{eqnarray}
Therefore, from Eqs.\ (\ref{3.16c}) and (\ref{3.72a}) to (\ref{3.72c}),
\begin{eqnarray}
\lefteqn{||\bsQ^{(i)+}(x,\sigma',y,\sigma') - \bsQ^{(i)+}(x,\sigma,y,\sigma)||
\le } \nonumber \\ & &
M^{(i)}_{1}(\x,a,b) \left\{ M^{(i)}_{1}(\x,a,b)M^{(i)}_{2}(\x,a,b)
\int_{\sigma}^{\sigma'} d\lambda \left( \left| \frac{\M^{+}(\sigma-y)}
{\M^{+}(\sigma-\lambda)} \right| + 1 \right) + \right. \nonumber \\ & &
\left.  M^{(i)}_{1}(\x,a,b)^{2}M^{(i)}_{2}(\x,a,b) \int_{x}^{\sigma'} d\lambda
\left| \frac{\M^{+}(\sigma-y)}{\M^{+}(\sigma-\lambda)} - \frac{\M^{+}(\sigma'-y)}
{\M^{+}(\sigma'-\lambda)} \right| \right\} \nonumber \\ & &
\text{ for all } \sigma, \sigma' \in [x,b] \text{ such that } \sigma' > \sigma.
\label{3.73b}
\end{eqnarray}
One readily shows that, for both $i=3$ and $i=4$,
\begin{eqnarray}
\lefteqn{\int_{\sigma}^{\sigma'} d\lambda \left( \left| \frac{\M^{+}(\sigma-y)}
{\M^{+}(\sigma-\lambda)} \right| + 1 \right) = } \nonumber \\
& & \sqrt{|\sigma-y|} \sqrt{\sigma'-\sigma} 
+ (\sigma'-\sigma) \le M^{(i)}_{3}(\x,a,b) \sqrt{\sigma'-\sigma}
\end{eqnarray}
where $M^{(i)}_{3}(\x,a,b)$ is defined by Eq.\ (\ref{3.72d}).  Therefore,
the inequality (\ref{3.73b}) becomes
\begin{eqnarray}
\lefteqn{||\bsQ^{(i)+}(x,\sigma',y,\sigma') - \bsQ^{(i)+}(x,\sigma,y,\sigma)||
\le } \nonumber \\
& & M^{(i)}_{4}(\x,a,b) \sqrt{\sigma'-\sigma} + M^{(i)}_{5}(\x,a,b)
\Lambda^{(i)}(\x,\sigma,\sigma') \nonumber \\ & &
\text{ for all } \sigma, \sigma' \in [a,b] \text{ such that }
\sigma' > \sigma,
\label{3.74a}
\end{eqnarray}
where
\begin{eqnarray}
M^{(i)}_{4}(\x,a,b) & := & M^{(i)}_{1}(\x,a,b)^{2}M^{(i)}_{2}(\x,a,b)
M^{(i)}_{3}(\x,a,b), \\
M^{(i)}_{5}(\x,a,b) & := & M^{(i)}_{1}(\x,a,b)^{3}M^{(i)}_{2}(\x,a,b),
\end{eqnarray}
and
\begin{eqnarray}
\Lambda^{(i)}(\x,\sigma,\sigma') & := & \int_{x}^{\sigma'} d\lambda
\left| \frac{\M^{+}(\sigma-y)}{\M^{+}(\sigma-\lambda)} -
\frac{\M^{+}(\sigma'-y)}{\M^{+}(\sigma'-\lambda)} \right|
\nonumber \\ & &
\text{ for all } \sigma, \sigma' \in [x,b] \text{ such that }
\sigma' > \sigma.
\end{eqnarray}

\item 
Since, when $i=3$, $x=x^{3}=r$ and $y=x^{4}=s$, we have
\begin{eqnarray}
\Lambda^{(3)}(\x,\sigma,\sigma') & = & \int_{r}^{\sigma} d\lambda 
\left| i \sqrt{\frac{s-\sigma}{\sigma-\lambda}} - 
i \sqrt{\frac{s-\sigma'}{\sigma'-\lambda}} \right| \nonumber \\
& & \mbox{ } + \int_{\sigma}^{\sigma'} d\lambda \left| 
\frac{\M^{+}(\sigma-s)}{\M^{+}(\sigma-\lambda)} +
\frac{\M^{+}(\sigma'-s)}{\M^{+}(\sigma'-\lambda)} \right| \nonumber \\
& \le & \int_{r}^{\sigma} d\lambda \left(
\sqrt{\frac{s-\sigma}{\sigma-\lambda}} -
\sqrt{\frac{s-\sigma'}{\sigma'-\lambda}} \right) \nonumber \\
& & \mbox{ } + \int_{\sigma}^{\sigma'} d\lambda \left(
\left| \frac{M^{+}(\sigma-s)}{M^{+}(\sigma-\lambda)} \right| +
\left| \frac{M^{+}(\sigma'-s)}{M^{+}(\sigma'-\lambda)} \right| \right)
\nonumber \\
& = & 2 \sqrt{s-\sigma}\sqrt{\sigma-r} - 2 \sqrt{s-\sigma'}\sqrt{\sigma'-r}
+ 2 \sqrt{s-\sigma'}\sqrt{\sigma'-\sigma} \nonumber \\
& & \mbox{ } + 2 \sqrt{s-\sigma}\sqrt{\sigma'-\sigma} +
2 \sqrt{s-\sigma'}\sqrt{\sigma'-\sigma}.
\label{3.75a}
\end{eqnarray}
\end{arablist}
The functions of $\sigma$ given by $\sqrt{s-\sigma}$ and $\sqrt{\sigma-r}$
are both $H(1/2)$ on $[x,b]$.  Therefore, the function of $\sigma$ given
by $\sqrt{s-\sigma}\sqrt{\sigma-r}$ is $H(1/2)$ on $[x,b]$.  Therefore,
from (\ref{3.75a}), there exists a finite $M^{(3)}_{6}(\x,a,b) > 0$ such
that
\begin{equation}
\Lambda^{(3)}(\x,\sigma,\sigma') \le M^{(3)}_{6}(\x,a,b) \sqrt{\sigma'-\sigma}
\text{ for all } \sigma, \sigma' \in [x,b] \text{ for which } \sigma' >
\sigma.
\label{3.75b}
\end{equation}
The reader can likewise show that a finite $M^{(4)}_{6}(\x,a,b) > 0$ exists
such that
\begin{equation}
\Lambda^{(4)}(\x,\sigma,\sigma') \le M^{(4)}_{6}(\x,a,b) \sqrt{\sigma'-\sigma}
\text{ for all } \sigma, \sigma' \in [x,b] \text{ for which } \sigma' >
\sigma.
\label{3.75c}
\end{equation}
So, one obtains from (\ref{3.74a}), (\ref{3.75b}) and (\ref{3.75c}),
\begin{eqnarray}
\lefteqn{||\bsQ^{(i)+}(x,\sigma',y,\sigma') - \bsQ^{(i)+}(x,\sigma,y,\sigma)||
\le } \nonumber \\
& & M^{(i)}(\x,a,b) \sqrt{\sigma'-\sigma} \text{ for all } \sigma', \sigma
\in [x,b] \text{ for which } \sigma' > \sigma,
\label{3.76}
\end{eqnarray}
where
\begin{equation}
M^{(i)}(\x,a,b) := M^{(i)}_{4}(\x,a,b) + M^{(i)}_{5}(\x,a,b)
M^{(i)}_{6}(\x,a,b).
\end{equation}
Hence, we have shown that $\bsQ^{(i)+}(x,\sigma,y,\sigma)$ (for fixed $\x$) 
is $H(1/2)$ on $[x,b]$; and we shall leave the similar proof that it is
$H(1/2)$ on $[a,x]$ as an exercise for the reader.
\cheers
\end{abc}

\begin{theorem}[H\"{o}lder condition for $V^{(i)}$]
\mbox{ } \\ \vspace{-3ex}
\label{3.6H}
\begin{romanlist}
\item 
For each $\x \in \domE$, the function of $\sigma$ given by $\Q^{+}(\x,\sigma)
V^{(i)}(\sigma)$ is $H(1/2)$ on each closed subinterval of
$\check{\I}^{(i)}(y)$.
\item 
$V^{(i)}$ is $H(1/2)$ on each closed subinterval of $\I^{(i)}$.
\end{romanlist}
\end{theorem}

\proofs
\begin{romanlist}
\item 
{}From Eqs.\ (\ref{3.68a}) and (\ref{3.68b}),
\begin{equation}
\Q^{+}(\x,\sigma) V^{(i)}(\sigma) = \bsQ^{(i)+}(x,\sigma,y,\sigma)
e^{J\theta_{7-i}(y,\sigma)}.
\label{3.78}
\end{equation}
{}From the preceding Thm.~\ref{3.5H}, the function of $\sigma$ given by
$\bsQ^{(i_+}(x,\sigma,y,\sigma)$ is $H(1/2)$ on each closed subinterval
of $\check{\I}^{(i)}(y)$; and, from (\ref{3.66a}), $\theta_{7-i}(y,\sigma)$
is a continuously differentiable function of $\sigma$ throughout 
$\check{\I}^{(i)}(y)$.  Therefore, the product (\ref{3.78}) is $H(1/2)$
on each closed subinterval of $\check{\I}^{(i)}(y)$.
\cheers
\item 
Use the fact that $\Q(\x_{0},\sigma) V^{(i)}(\sigma) = V^{(i)}(\sigma)$.
\cheers
\end{romanlist}

\newpage

\section{HHP's corresponding to $(\bv,\bar{\F}_{0},\x)$ and to
$(\bv,\bar{\F}_{0})$; generalized Geroch conjecture \label{Sec_4}}


\setcounter{equation}{0}
\setcounter{theorem}{0}
\subsection{Important sets and groups}

Readers who are familiar with our earlier treatment of Kinnersley-Chitre
transformations will recognize that we have not yet revealed what is to
be the analog of those spacetime-independent $SL(2,R)$ matrices $v(\tau)$
(or $u(t)$) that played such an important role in formulating the
Kinnersley--Chitre realization of the Geroch group.  

Our quest for a suitable formulation of the new HHP will begin with the
set
\begin{abc}
\begin{eqnarray}
\S_{\bV}\triple & := & \text{ the set of all ordered pairs } \bV = (V^{(3)},V^{(4)}), 
\nonumber \\
& & \text{ where $V^{(i)}$ is a $2 \times 2$ matrix function with the} 
\nonumber \\
& & \text{ domain $\I^{(i)}$ and there exists } \hat{\Q} \in \S_{\hat{\Q}}\triple
\label{G3.1} \\
& & \text{ such that Eqs.\ (\ref{G2.21g1}) and (\ref{G2.21g2}) hold,} \nonumber
\end{eqnarray}
and the $H(1/2)$ version of the $B$-group of Kinnersley and Chitre; namely,
\begin{eqnarray}
B(\I^{(i)}) & := & \text{ the multiplicative group of all }
\nonumber \\ & &
\exp \left( J \bvarphi^{(i)} \right) = \left( \begin{array}{cc}
	\cos \bvarphi^{(i)} & \sin \bvarphi^{(i)} \\
	-\sin \bvarphi^{(i)} & \cos \bvarphi^{(i)}
	\end{array} \right) \label{G3.2} \\ & &
\text{ such that $\bvarphi^{(i)}$ is any real-valued function} 
\nonumber \\ & &
\text{ that has the domain $\I^{(i)}$ and is $H(1/2)$} 
\nonumber \\ & &
\text{ on every closed subinterval of $\I^{(i)}$.} 
\nonumber
\end{eqnarray}
Specifically, we shall be concerned with the set 
\begin{eqnarray}
k\s & := & \text{ the set of all ordered pairs $(V^{(3)}w^{(3)},
V^{(4)}w^{(4)})$} \nonumber \\ & &
\text{ such that $(V^{(3)},V^{(4)}) \in \S_{\bV}\triple$ and} 
\label{G3.3} \\ & &
\text{ $(w^{(3)},w^{(4)}) \in B(\I^{(3)},\I^{(4)}) :=
B(\I^{(3)}) \times B(\I^{(4)})$.} \nonumber
\end{eqnarray}
For any members $\bv = (v^{(3)},v^{(4)})$ and $\bv' = (v^{(3)\prime},
v^{(4)\prime})$ of $k\s$, we let $\bv\bv' := (v^{(3)}v^{(3)\prime},
v^{(4)}v^{(4)\prime})$.  It is not known whether $k\s$ is a group
with respect to this multiplicative operation; but it is obvious
that $k\s \subset K\s$, where $K\s$ is a group that we shall now
identify.

{}From Eqs.\ (\ref{3.64}), (\ref{G3.2}) and (\ref{G3.3}), one sees that, if 
$\bv = (v^{(3)},v^{(4)}) \in k\s$, then $v^{(i)}$ is $H(1/2)$ on every
closed subinterval of $\I^{(i)}$ and is expressible in the form
\begin{equation}
v^{(i)}(\sigma) = \alpha^{(i)}(\sigma) + \mu^{+}(\x_{0},\sigma)
\sigma_{3} \beta^{(i)}(\sigma),
\end{equation}
where $\alpha^{(i)}(\sigma)$ and $\beta^{(i)}(\sigma)$ are real
linear combinations of the matrices $I$ and $J$ and are continuous
functions of $\sigma$ throughout $\I^{(i)}$.  This suggests our
next step.
\end{abc}

\begin{abc}
\begin{definition}{Dfn.\ of the group $K\s$ and its subgroups\label{def55}}
We let $K(\x_{0},\I^{(i)})$ denote the multiplicative group of
all $2 \times 2$ matrix functions $v^{(i)}$ such that
\begin{equation}
\dom{v^{(i)}} = \I^{(i)}, \quad \det{v^{(i)}} = 1,
\end{equation}
$v^{(i)}$ is $H(1/2)$ on every closed subinterval of $\I^{(i)}$,
and there exist $2 \times 2$ matrix functions $\alpha^{(i)}$ and 
$\beta^{(i)}$ which are each real linear combinations of the matrices 
$I$ and $J$, have $\I^{(i)}$ as their common domain, are $\bC^{0}$, and 
satisfy
\begin{equation}
v^{(i)}(\sigma) = \alpha^{(i)}(\sigma) + \mu^{+}(\x_{0},\sigma)
\sigma_{3} \beta^{(i)}(\sigma) \label{G3.6}
\end{equation}
for all $\sigma \in \I^{(i)}$.  

We shall say that $f$ is $\bC^{n+}$ if its $n$th derivative $D^{n}f$
exists throughout $\dom{f}$ and $D^{n}f$ obeys a H\"{o}lder condition
of arbitrary index on each closed subinterval of $\dom{f}$.  (The index
may be different for different closed subintervals of $\dom{f}$.)  The 
same terminology is used if $f(x)$ is a matrix with real or complex
elements, and $|f(x)|$ is its norm.

We shall be employing certain proper subgroups of $K(\x_{0},\I^{(i)})$.
For each integer $n \ge 1$, $K^{n}(\x_{0},\I^{(i)})$ will denote the
subgroup of all $v^{(i)} \in K(\x_{0},\I^{(i)})$ such that $\alpha^{(i)}$
and $\beta^{(i)}$ are $\bC^{n}$.  $K^{n+}(\x_{0},\I^{(i)})$ will denote
the subgroup of all $v^{(i)} \in K^{n}(\x_{0},\I^{(i)})$ such that
$\alpha^{(i)}$ and $\beta^{(i)}$ are $\bC^{n+}$.  Likewise,
$K^{\infty}(\x_{0},\I^{(i)})$ is the subgroup of all $v^{(i)} 
\in K^{n+}(\x_{0},\I^{(i)})$ for which $\alpha^{(i)}$ and
$\beta^{(i)}$ are $\bC^{\infty}$; and $K^{an}(\x_{0},\I^{(i)})$
will denote the subgroup of all $v^{(i)} \in K^{\infty}(\x_{0},\I^{(i)})$
for which $\alpha^{(i)}$ and $\beta^{(i)}$ are analytic.

Also,
\begin{equation}
K\s := K(\x_{0},\I^{(3)}) \times K(\x_{0},\I^{(4)}),
\label{G3.7}
\end{equation}
and $K^{n}\s$, $K^{n+}\s$, $K^{\infty}\s$ and $K^{an}\s$
will denote the direct products $K^{n}(\x_{0},\I^{(3)}) \times
K^{n}(\x_{0},\I^{(4)})$, etc.  Henceforth, we shall abbreviate our
equations by deleting the arguments `$\triple$' and `$\s$' when there is
little danger of ambiguity; and we shall employ $\Box$ as a generic 
superscript that stands for $n$, $n+$, $\infty$ or `an' (analytic).

Recalling that $k \subset K$, we let
\begin{equation}
k^{\Box} := \{ \bv \in k: \bv \in K^{\Box} \}
\end{equation}
and
\begin{equation}
\S_{\bV}^{\Box} := \{ \bV \in \S_{\bV}: \text{ there exists $\bw \in 
B(\I^{(3)},\I^{(4)})$ for which $\bV\bw \in k^{\Box}$} \}.  
\label{G3.8b}
\end{equation}
The peculiar structure of the above definition (\ref{G3.8b}) is due to
the possibility that $\bV$ and $\bw$ may exist for which $\bV \notin K^{n}$,
$\bw \notin K^{n}$, but $\bV\bw \in K^{n}$.  Though we have never come
across an example of this kind of $(\bV,\bw)$, we must allow for its
existence until a proof to the contrary is devised.
\end{definition}
\end{abc}

It will be found expedient to let $\S_{\E}^{\Box}$ and $\S_{\F}^{\Box}$ denote 
the sets of all $\E \in \S_{\E}$ and $\F \in \S_{\F}$, respectively, for which
the corresponding $\bV$ is a member of $\S_{\bV}^{\Box}$.  Recalling Theorem
\ref{3.9G}, one sees that $\{ \E \in \S_{\E}: \E^{(3)}$ and $\E^{(4)}$ are 
$\bC^{n+1}\} \subset \S_{\E}^{n}$.  In fact, further analysis (on the special
case of the real members of $\S_{\E}$) shows that $\{ \E \in \S_{\E}: \E^{(3)}$
and $\E^{(4)}$ are $\bC^{n+1}\}$ is a proper subset of $\S_{\E}^{n}$ [since
there are cases where $\bV \in \S_{\bV}^{n}$ and the corresponding $\E^{(3)}$
and $\E^{(4)}$ are $\bC^{n}$ or $\bC^{n-1}$, but are not $\bC^{n+1}$ or
$\bC^{n}$, respectively].  In contrast, $\{ \E \in \S_{\E}: \E^{(3)}$ and
$\E^{(4)}$ are $\bC^{\infty}\} = \S_{\E}^{\infty}$, and $\{ \E \in \S_{\E}:
\E^{(3)}$ and $\E^{(4)}$ are analytic $\} = \S_{\E}^{an}$.

By considering real $\E \in \S_{\E}$, we have proved that $k$ is a proper subset 
of $K$.  It is as yet unknown whether $k^{n}=K^{n}$ or $k^{n+} = K^{n+}$ when 
$n=1$ or $n=2$.  Later in these notes, however, we shall deduce the far from
obvious fact that $k^{3}\s = K^{3}\s$.

\begin{abc}
A multiplicative group $\tilde{K}\s$ that is isomorphic to $K\s$ and that
provides an alternative way to identify our extension of the K--C group is
defined as follows.  For each $\bv \in K\s$, we let 
\begin{equation}
v^{(i)+} := v^{(i)}, \quad v^{(i)-} := [v^{(i)}]^{*},
\label{G3.9}
\end{equation}
whereupon Eqs.\ (\ref{1.6i}) and (\ref{G3.6}) yield
\begin{equation}
v^{(i)\pm}(\sigma) = \alpha^{(i)}(\sigma) + \mu^{\pm}(\x_{0},\sigma)
\sigma_{3} \beta^{(i)}(\sigma).  
\label{G3.10}
\end{equation}

Equation (\ref{1.16d}) supplies us with the relations
\begin{eqnarray}
P^{M}(\x,\tau) J [P^{M}(\x,\tau)]^{-1} & = & \left( \begin{array}{cc}
-i & 2(\tau-z) \\ 0 & i
\end{array} \right), 
\label{G3.11a} \\
P^{M}(\x,\tau) \mu(\x,\tau) \sigma_{3} [P^{M}(\x,\tau)]^{-1}
& = & \left( \begin{array}{cc}
-\tau+z & -i\rho^{2} \\ -i & \tau-z
\end{array} \right). 
\label{G3.11b}
\end{eqnarray}
The reader may now use Eqs.\ (\ref{G3.10}), (\ref{G3.11a}) and (\ref{G3.11b}) 
to prove that
\begin{eqnarray}
\tilde{v}^{(i)}(\sigma) & := & 
P^{M+}(\x_{0},\sigma) v^{(i)+}(\sigma)
[P^{M+}(\x_{0},\sigma)]^{-1} \nonumber \\
& = & P^{M-}(\x_{0},\sigma) v^{(i)-}(\sigma)
[P^{M-}(\x_{0},\sigma)]^{-1} \text{ and} 
\label{G3.12} \\
& & \text{ is a $\bC^{\Box}$ function of $\sigma$ throughout $\I^{(i)}$
if $v^{(i)} \in K^{\Box}(\x_{0},\I^{(i)})$,}
\nonumber
\end{eqnarray}
where, of course, $P^{M+}(\x,\sigma)$ and $P^{M-}(\x,\sigma)$ are the
limits of $P^{M}(\x,\tau)$ as $\tau \rightarrow \sigma$ from above
and from below the real axis, respectively.  Even though 
$P^{M\pm}(\x_{0},\sigma)$ does not exist when $\sigma \in 
\{r_{0},s_{0}\}$, the product $P^{M\pm}(\x_{0},\sigma)$ 
$v^{(i)\pm}(\sigma)$ $[P^{M\pm}(\x_{0},\sigma)]^{-1}$ clearly has a
continuous extension to the entire interval $\I^{(i)}$; and this
extension is $\bC^{\Box}$ if $\alpha^{(i)}$ and $\beta^{(i)}$ are 
$\bC^{\Box}$.
\end{abc}

\begin{abc}
\begin{definition}{Dfn.\ of the group $\tilde{K}\s$ and its subgroups
\label{def56}}
For any given $\bv = (v^{(3)},v^{(4)})$ in the group $K\s$, we shall
let $\tilde{v}^{(i)}$ denote the function whose domain is $\I^{(i)}$
and whose values are defined by Eq.\ (\ref{G3.12}); and we shall let
\begin{equation}
\tilde{K}\s := \{\tilde{\bv}: \bv \in K\s\},
\label{G3.13a}
\end{equation}
where
\begin{equation}
\tilde{\bv} := (\tilde{v}^{(3)},\tilde{v}^{(4)}),
\end{equation}
and $\tilde{K}^{\Box}$ will denote the subgroup of all $\tilde{\bv} \in 
\tilde{K}$ such that $\bv \in K^{\Box}$.
\end{definition}
\end{abc}

Since
\begin{abc}
\begin{equation}
\S_{\bV}\triple \subset k\s \subset K\s
\label{G3.14}
\end{equation}
and
\begin{equation}
B(\I^{(3)},\I^{(4)}) \subset k\s,
\end{equation}
our definitions of $\tilde{v}^{(i)}$ and $\tilde{\bv}$ are also applicable
to members of $\S_{\bV}$, $B$ and $k$.
\end{abc}

\begin{abc}
\begin{gssm}[The group $\tilde{K}^{\Box}$]
\label{4.1A}
\mbox{ } \\
The group $\tilde{K}^{\Box}$ defined above is identical to the 
multiplicative group of all ordered pairs 
$\tilde{\bv} = (\tilde{v}^{(3)},\tilde{v}^{(4)})$
of $2 \times 2$ matrix functions such that, for both $i=3$ and $i=4$,
\begin{equation}
\dom{\tilde{v}^{(i)}} = \I^{(i)}, \quad \det{\tilde{v}^{(i)}} = 1, \quad
\tilde{v}^{(i)} \text{ is } \bC^{\Box}
\label{4.1}
\end{equation}
and
\begin{equation}
\tilde{v}^{(i)}(\sigma)^{\dagger} \A^{M}(\x_{0},\sigma) \tilde{v}^{(i)}(\sigma)
= \A^{M}(\x_{0},\sigma) \text{ for all } \sigma \in \I^{(i)},
\label{4.2}
\end{equation}
where $\A(\x,\tau)$ is defined by Eq.\ (\ref{1.26a}), and $h^{M}$ is given
by Eq.\ (\ref{1.14b}).
\end{gssm}
\end{abc}

\proof
The proof will be given in three parts:
\begin{arablist}
\item 
In this first part we assume that $\tilde{\bv} =
(\tilde{v}^{(3)},\tilde{v}^{(4)}) \in \tilde{K}^{\Box}$.  From the definition
of $\tilde{K}^{\Box}$ given above, the three conditions (\ref{4.1}) in the
theorem are all satisfied.  It remains to prove that the condition (\ref{4.2})
is also satisfied.

Equations (\ref{G3.12}) and (\ref{G3.9}) yield, for all $\sigma \in \I^{(i)} -
\{r_{0},s_{0}\}$,
\begin{abc}
\begin{eqnarray}
\tilde{v}^{(i)}(\sigma)^{\dagger} \A^{M}(\x_{0},\sigma) \tilde{v}^{(i)}(\sigma)
& = & \left[ P^{M-}(\x_{0},\sigma) v^{(i)-}(\sigma) P^{M-}(\x_{0},\sigma)^{-1}
\right]^{\dagger} \A^{M}(\x_{0},\sigma)
\nonumber \\ & & \hspace{5em}
\left[ P^{M+}(\x_{0},\sigma) v^{(i)+}(\sigma) P^{M+}(\x_{0},\sigma)^{-1}
\right] \nonumber \\
& = & \left[ P^{M-}(\x_{0},\sigma)^{\dagger}\right]^{-1} v^{(i)}(\sigma)^{T}
\nonumber \\ & & \left[ P^{M-}(\x_{0},\sigma)^{\dagger} \A^{M}(\x_{0},\sigma)
P^{M+}(\x_{0},\sigma) \right] 
\nonumber \\ & & \hspace{5em}
v^{(i)}(\sigma) P^{M+}(\x_{0},\sigma)^{-1}.
\label{4.3a}
\end{eqnarray}
However, from Eqs.\ (\ref{1.26a}), (\ref{1.16b}), (\ref{1.6e}), (\ref{1.6f})
and (\ref{1.14b},
\begin{equation}
P^{M\mp}(\x,\tau^{*})^{\dagger} \A^{M}(\x,\tau) P^{M\pm}(\x,\tau) = \Omega
\text{ for all } \x \in \domE \text{ and } \tau \in \bar{C}^{\pm} - \{r,s\}.
\label{4.3b}
\end{equation}
Also, since $\det{v^{(i)}} = 1$,
\begin{equation}
v^{(i)}(\sigma)^{T} \Omega v^{(i)}(\sigma) = \Omega \text{ for all }
\sigma \in \I^{(i)}.
\label{4.3c}
\end{equation}
Therefore, upon employing (\ref{4.3b}), and then employing (\ref{4.3c}),
and then employing (\ref{4.3b}) once again, Eq.\ (\ref{4.3a}) becomes
$$
\tilde{v}^{(i)}(\sigma)^{\dagger} \A^{M}(\x_{0},\sigma) v^{(i)}(\sigma)
= \A^{M}(\x_{0},\sigma).
$$
Now, it is ture that the above result has so far been deduced only for
$\sigma \in \I^{(i)} - \{r_{0},s_{0}\}$.  However, both sides of this result
are continuous functions of $\sigma$ throughout $\I^{(i)}$.  Therefore, the
condition (\ref{4.2}) is satisfied.
\end{abc}

\item 
The rest of the proof will concern the converse of what we proved in part
(1).  Suppose that $\tilde{\bv} = (\tilde{v}^{(3)},\tilde{v}^{(4)})$ is
a given ordered pair of $2 \times 2$ matrix functions which satisfy the
conditions (\ref{4.1}) and (\ref{4.2}).  For both $i=3$ and $i=4$, let
\begin{abc}
\begin{equation}
v^{(i)}(\sigma) := P^{M+}(\x_{0},\sigma)^{-1} \tilde{v}^{(i)}(\sigma)
P^{M+}(\x_{0},\sigma)
\label{4.4}
\end{equation}
for all $\sigma \in \I^{(i)} - \{r_{0},s_{0}\}$.  From the condition 
(\ref{4.2}) and the definition (\ref{4.4}),
\begin{eqnarray*}
\left[ P^{M+}(\x_{0},\sigma)^{\dagger} \right]^{-1}
v^{(i)}(\sigma)^{\dagger} P^{M+}(\x_{0},\sigma)^{\dagger}
\A^{M}(\x_{0},\sigma) & & \\
P^{M+}(\x_{0},\sigma) v^{(i)}(\sigma) P^{M+}(\x_{0},\sigma)^{-1} & = &
\A^{M}(\x_{0},\sigma)
\end{eqnarray*}
for all $\sigma \in \I^{(i)} - \{r_{0},s_{0}\}$.  Upon applying Eq.\
(\ref{1.19a}) to the above equation, one gets
\begin{eqnarray*}
& \left[ \Omega^{1-\kappa_{0}} v^{(i)}(\sigma)^{\dagger} \Omega^{1-\kappa_{0}}
\right] \left[ P^{M-}(\x_{0},\sigma)^{\dagger} \A^{M}(\x_{0},\sigma) 
P^{M+}(\x_{0},\sigma) \right] v^{(i)}(\sigma) = & \\ &
P^{M-}(\x_{0},\sigma)^{\dagger} \A^{M}(\x_{0},\sigma) P^{M+}(\x_{0},\sigma); &
\end{eqnarray*}
and, therefore, from Eq.\ (\ref{4.3b}),
\begin{eqnarray}
& \left[ \Omega^{1-\kappa_{0}} v^{(i)}(\sigma)^{\dagger} \Omega^{1-\kappa_{0}}
\right] \Omega v^{(i)}(\sigma) = \Omega, & \\ &
\text{ where } \kappa_{0} = \left\{ \begin{array}{l}
1 \text{ when } s_{0} < \sigma, \\
-1 \text { when } \sigma < r_{0}, \\
0 \text{ when } r_{0} < \sigma < s_{0}.
\end{array} \right. & \nonumber
\label{4.5a}
\end{eqnarray}
However, from conditions (\ref{4.1}) and the definition (\ref{4.4}),
\begin{equation}
\det{v^{(i)}(\sigma)} = 1
\label{4.5b}
\end{equation}
for all $\sigma \in \I^{(i)} - \{r_{0},s_{0}\}$; or, equivalently,
\begin{equation}
v^{(i)}(\sigma)^{T} \Omega v^{(i)}(\sigma) = \Omega
\label{4.5c}
\end{equation}
for all $\sigma \in \I^{(i)} - \{r_{0},s_{0}\}$.  Comparison of the above
Eqs.\ (\ref{4.5a}) and (\ref{4.5c}) nets
\begin{equation}
v^{(i)}(\sigma)^{*} = \Omega^{1-\kappa_{0}} v^{(i)}(\sigma) 
\Omega^{1-\kappa_{0}}
\label{4.6}
\end{equation}
for all $\sigma \in \I^{(i)} - \{r_{0},s_{0}\}$.

For each $\sigma \in \I^{(i)} - \{r_{0},s_{0}\}$, let
$\alpha^{(i)}_{1}(\sigma)$, $\alpha^{(i)}_{2}(\sigma)$, 
$\beta^{(i)}_{1}(\sigma)$ and $\beta^{(i)}_{2}(\sigma)$ denote those
complex numbers for which
\begin{equation}
v^{(i)}(\sigma) = \alpha^{(i)}_{1}(\sigma) I + \alpha^{(i)}_{2}(\sigma) J
+ \sigma_{3} \left[ \beta^{(i)}_{1}(\sigma) I + \beta^{(i)}_{2}(\sigma) J
\right].
\label{4.7a}
\end{equation}
It is then clear that the condition (\ref{4.6}) is equivalent to the 
quadruple of conditions
\begin{eqnarray}
& & \alpha_{1}(\sigma)^{*} = \alpha_{1}(\sigma), \quad
\alpha_{2}(\sigma)^{*} = \alpha_{2}(\sigma), \nonumber \\
& & \beta_{1}(\sigma)^{*} = (-1)^{1-\kappa_{0}} \beta_{1}(\sigma), \quad
\beta_{2}(\sigma)^{*} = (-1)^{1-\kappa_{0}} \beta_{2}(\sigma), \nonumber \\
& & \text{ for all } \sigma \in \I^{(i)} - \{r_{0},s_{0}\}. 
\label{4.7b}
\end{eqnarray}
\end{abc}

\item 
We next employ Eq.\ (\ref{4.7a}), the definition (\ref{4.4}), and Eqs.\ 
(\ref{3.70b}) and (\ref{3.70c}) to compute $\tilde{v}^{(i)}(\sigma)$
in terms of $\alpha^{(i)}_{1}(\sigma)$, $\alpha^{(i)}_{2}(\sigma)$,
$\beta^{(i)}_{1}(\sigma)$ and $\beta^{(i)}_{2}(\sigma)$.  The result
of this straightforward calculation is 
\begin{abc}
\begin{eqnarray}
\tilde{v}^{(i)}(\sigma) & = & P^{M+}(\x_{0},\sigma) v^{(i)}(\sigma)
P^{M+}(\x_{0},\sigma)^{-1} \nonumber \\
& = & \alpha^{(i)}_{1}(\sigma) I + \alpha^{(i)}_{2}(\sigma)
\left( \begin{array}{cc}
-i & 2(\sigma-z_{0}) \\ 0 & i
\end{array} \right) \nonumber \\ & &
\mbox{ } + \frac{\beta^{(i)}_{1}(\sigma)}{\mu^{+}(\x_{0},\sigma)} \sigma_{3}
\left( \begin{array}{cc}
- (\sigma-z_{0}) & -i \rho_{0}^{2} \\ i & -(\sigma-z_{0})
\end{array} \right) \nonumber \\ & &
\mbox{ } + \frac{\beta^{(i)}_{2}(\sigma)}{\mu^{+}(\x_{0},\sigma)} \sigma_{3}
\left( \begin{array}{cc}
i (\sigma-z_{0}) & - 2 [\mu^{+}(\x_{0},\sigma)]^{2} - \rho_{0}^{2} \\
1 & i (\sigma-z_{0})
\end{array} \right) \nonumber \\
& & \text{ for all } \sigma \in \I^{(i)} - \{r_{0},s_{0}\}.
\label{4.8}
\end{eqnarray}
Equation (\ref{4.8}) can obviously be expressed in the form
\begin{eqnarray*}
\tilde{v}^{(i)}(\sigma) & = & 
A^{(i)}_{1}(\sigma) I + A^{(i)}_{2}(\sigma)
\left( \begin{array}{cc}
-i & 2(\sigma-z_{0}) \\ 0 & i
\end{array} \right) \\ & &
\mbox{ } + B^{(i)}_{1}(\sigma) \sigma_{3} \left( \begin{array}{cc}
- (\sigma-z_{0}) & -i \rho_{0}^{2} \\ i & -(\sigma-z_{0})
\end{array} \right) \\ & &
\mbox{ } + B^{(i)}_{2}(\sigma) \sigma_{3} \left( \begin{array}{cc}
i (\sigma-z_{0}) & - 2 [\mu^{+}(\x_{0},\sigma)]^{2} - \rho_{0}^{2} \\
1 & i (\sigma-z_{0})
\end{array} \right), 
\end{eqnarray*}
where, for all $\sigma \in \I^{(i)} - \{r_{0},s_{0}\}$,
\begin{equation}
\begin{array}{rcl}
\alpha^{(i)}_{1}(\sigma) & = & A^{(i)}_{1}(\sigma), \\
\alpha^{(i)}_{2}(\sigma) & = & A^{(i)}_{2}(\sigma), \\
\beta^{(i)}_{1}(\sigma) & = & \mu^{+}(\x_{0},\sigma) B^{(i)}_{1}(\sigma), \\
\beta^{(i)}_{2}(\sigma) & = & \mu^{+}(\x_{0},\sigma) B^{(i)}_{2}(\sigma).
\end{array}
\label{4.12a}
\end{equation}
{}From the conditions (\ref{4.1}),
\begin{equation}
\dom{A^{(i)}_{1}} = \dom{A^{(i)}_{2}} = \dom{B^{(i)}_{1}} 
= \dom{B^{(i)}_{2}} = \I^{(i)};
\label{4.10a}
\end{equation}
and 
\begin{equation}
A^{(i)}_{1}, A^{(i)}_{2}, B^{(i)}_{1} \text{ and } B^{(i)}_{2}
\text{ are } \bC^{\Box}.
\label{4.10b}
\end{equation}
Noting that
\begin{equation}
\mu^{+}(\x_{0},\sigma)^{*} = (-1)^{1-\kappa_{0}} \mu^{+}(\x_{0},\sigma),
\label{4.11a}
\end{equation}
we see that 
\begin{equation}
A^{(i)}_{1}(\sigma), A^{(i)}_{2}(\sigma), B^{(i)}_{1}(\sigma) \text{ and }
B^{(i)}_{2}(\sigma) \text{ are real.}
\label{4.12b}
\end{equation}

Now, let
\begin{equation}
\begin{array}{rcl}
A^{(i)} & := & A^{(i)}_{1} + A^{(i)}_{2} J, \\
B^{(i)} & := & B^{(i)}_{1} + B^{(i)}_{2} J,
\end{array}
\end{equation}
whereupon the statements (\ref{4.10a}) and (\ref{4.10b}) are equivalent
to the statements
\begin{equation}
\dom{A^{(i)}} = \dom{B^{(i)}} = \I^{(i)}
\end{equation}
and
\begin{equation}
A^{(i)} \text{ and } B^{(i)} \text{ are } \bC^{\Box}.
\end{equation}
For all $\sigma \in \I^{(i)} - \{r_{0},s_{0}\}$, Eqs.\ (\ref{4.7a})
and (\ref{4.12a}) give us
\begin{equation}
v^{(i)}(\sigma) = A^{(i)}(\sigma) + \sigma_{3} \mu^{+}(\x_{0},\sigma)
B^{(i)}(\sigma)
\label{4.13a}
\end{equation}
and, from (\ref{4.12b}),
\begin{equation}
A^{(i)}(\sigma) \text{ and } B^{(i)}(\sigma) \text{ are real.}
\label{4.13b}
\end{equation}

So far we have proved Eqs.\ (\ref{4.5b}) [$\det{v^{(i)}(\sigma)}=1$],
(\ref{4.13a}) and (\ref{4.13b}) only for all $\sigma \in \I^{(i)}$
such that $\sigma \notin \{r_{0},s_{0}\}$.  The basic reason for the
restriction $\sigma \notin \{r_{0},s_{0}\}$ goes back to the definition
of $v^{(i)}(\sigma)$ by Eq.\ (\ref{4.4}).  The fact is that the definition
(\ref{4.4}) is good only for all $\sigma \in \I^{(i)} - \{r_{0},s_{0}\}$.

However, $A^{(i)}(\sigma)$ and $B^{(i)}(\sigma)$ are defined in terms of
matrix elements of $\tilde{v}^{(i)}(\sigma)$ for all $\sigma \in \I^{(i)}$.
So, we may use the right side of Eq.\ (\ref{4.13a}) to define $v^{(i)}(r_{0})$
and $v^{(i)}(s_{0})$ as 
\begin{equation}
v^{(i)}(r_{0}) := A^{(i)}(r_{0}), \quad v^{(i)}(s_{0}) := B^{(i)}(s_{0}).
\end{equation}
Then $v^{(i)}(\sigma)$ is given by Eq.\ (\ref{4.13a}) for all $\sigma \in
I^{(i)}$; and, since $A^{(i)}$ and $B^{(i)}$ are continuous, the newly
extended $v^{(i)}$ is continuous throughout $\I^{(i)}$, $\det{v^{(i)}(\sigma)}
= 1$ for all $\sigma \in \I^{(i)}$, and $A^{(i)}(\sigma)$ and $B^{(i)}(\sigma)$
are real for all $\sigma \in \I^{(i)}$.  Moreover, since $A^{(i)}$ and $B^{(i)}$
are $\bC^{\Box}$ and since the function of $\sigma$ given by
$\mu^{+}(\x_{0},\sigma)$ is $H(1/2)$ on each closed subinterval of $\I^{(i)}$,
$v^{(i)}(\sigma)$ is also $H(1/2)$ on each closed subinterval of $\I^{(i)}$.

In short, we have proved that $\bv = (v^{(3)},v^{(4)}) \in K^{\Box}$.
\end{abc}
\end{arablist}
\cheers

\begin{corollary}[The group $\tilde{K}$]
\label{4.2A} \mbox{ } \\
The group $\tilde{K}$ as defined above is identical to the multiplicative
group of all ordered pairs $\tilde{\bv} = (\tilde{v}^{(3)},\tilde{v}^{(4)})$
of $2 \times 2$ matrix functions such that, for both $i=3$ and $i=4$,
\begin{equation}
\dom{\tilde{v}^{(i)}} = \I^{(i)}, \quad \det{\tilde{v}^{(i)}} = 1, \quad
\tilde{v}^{(i)} \text{ is } H(1/2) \text{ on each closed subinterval of }
\I^{(i)}
\end{equation}
and the condition (\ref{4.2}) holds.
\end{corollary}

\proof
The proof is the same as that of the preceding theorem except that the 
condition of being $\bC^{\Box}$ is replaced by the more general condition
of being $H(1/2)$ on each closed subinterval of $\I^{(i)}$.
\cheers


\setcounter{equation}{0}
\setcounter{theorem}{0}
\subsection{A theorem that suggested the form of our new HHP}

\begin{abc}
\begin{gssm}[Motivation]
\label{4.1C}
\mbox{ } \\
For all $\bv \in K$ and for all $\S_{\bar{\F}}$ members $\bar{\F}$ and 
$\bar{\F}_{0}$ whose corresponding $\S_{\bV}$ members are $\bV$ and
$\bV_{0}$, respectively, the following statements (i) and (ii) are
equivalent to one another:
\begin{romanlist}
\item 
There exists $\bw \in B(\I^{(3)},\I^{(4)})$ such that 
\begin{equation}
\bv = \bV \bw \bV_{0}^{-1}.
\end{equation}

\item 
For each $\x \in \domE$, $i \in \{3,4\}$ and $\sigma \in \I^{(i)}(\x)$,
\begin{equation}
\F^{+}(\x,\sigma) \tilde{v}^{(i)}(\sigma) [\F_{0}^{+}(\x,\sigma)]^{-1}
= 
\F^{-}(\x,\sigma) \tilde{v}^{(i)}(\sigma) [\F_{0}^{-}(\x,\sigma)]^{-1}.
\label{4.16b}
\end{equation}
Moreover, if $\E^{(i)}$ and $\E_{0}^{(i)}$ are $\bC^{n_{i}}$ (resp.\ 
analytic) and $w^{(i)}$ is $\bC^{n_{i}-1}$ (resp.\ analytic), then the
function of $\sigma$ that equals each side of Eq.\ (\ref{4.16b}) has
a $\bC^{n_{i}-1}$ (resp.\ analytic) extension $\tilde{Y}^{(i)}(\x)$
to the interval
\begin{equation}
\dom{\tilde{Y}^{(i)}(\x)} = \check{\I}^{(i)}(x^{7-i})
\end{equation}
and, if $\bv \in K^{\Box}$ and $\bar{\F}_{0} \in \S_{\bar{\F}}^{\Box}$,
then $\bV \in \S_{\bV}^{\Box}$ and $\bar{\F} \in \S_{\bar{\F}}^{\Box}$.
\end{romanlist}
\end{gssm}
\end{abc}

\proof
The proof will be given in three parts:
\begin{arablist}
\item 
{}From Thms.~\ref{3.4H}(i) and~(iv),
\begin{abc}
\begin{eqnarray}
\Pi^{(i)\pm}(\x,\sigma) & := & \Q^{\pm}(\x,\sigma) V^{(i)\pm}(\sigma)
\nonumber \\ & = & C^{(i)}(\x,\sigma) + \sigma_{3} \mu^{\pm}(\x,\sigma)
D^{(i)}(\x,\sigma) \nonumber \\ & & \text{ for all } \x \in \domE
\text{ and } \sigma \in \check{\I}^{(i)}(x^{7-i}),
\end{eqnarray}
where
\begin{equation}
V^{(i)+} := V^{(i)}, \quad V^{(i)-} := (V^{(i)})^{*}.
\label{4.17b}
\end{equation}
Therefore, from Eqs.\ (\ref{3.70b}), (\ref{3.70c}) and (\ref{3.71b}),
\begin{eqnarray}
\tilde{\Pi}^{(i)}(\x,\sigma) & := & 
P^{M+}(\x,\sigma) \Pi^{(i)+}(\x,\sigma) [P^{M+}(\x,\sigma)]^{-1} \nonumber \\
& = & 
P^{M-}(\x,\sigma) \Pi^{(i)-}(\x,\sigma) [P^{M-}(\x,\sigma)]^{-1} \nonumber \\
& = & \tilde{C}^{(i)}(\x,\sigma) + \left( \begin{array}{cc}
-(\sigma-z) & -i\rho^{2} \\ -i & \sigma+z
\end{array} \right) \tilde{D}^{(i)}(\x,\sigma) \nonumber \\
& & \text{ for all } \x \in \domE \text{ and } \sigma \in
\check{\I}^{(i)}(x^{7-i}),
\label{4.17c}
\end{eqnarray}
where $\tilde{C}^{(i)}$ and $\tilde{D}^{(i)}$ are defined by Eqs.\
(\ref{3.71b}).  As regards the above definition of $\tilde{\Pi}^{(i)}$,
it is true that $P^{M\pm}(\x,\sigma)$ does not exist at $\sigma \in \{r,s\}$;
but the product $P^{M\pm}(\x,\sigma) \Pi^{(i)\pm}(\x,\sigma)
[P^{M\pm}(\x,\sigma)]^{-1}$ clearly has a continuous extension to
$\check{\I}^{(i)}(x^{7-i})$.  In fact, from Eq.\ (\ref{4.17c}), Eqs.\
(\ref{3.71b}), (\ref{3.70c}) and Thm.~\ref{3.4H}(ii), one obtains
\begin{eqnarray}
\text{If $\E^{(i)}$ is $\bC^{n_{i}}$ ($n_{i} \ge 1$) (resp.\ analytic), 
then for each $\x \in \domE$, the functions } \nonumber \\ 
\text{ of $\sigma$ given by $\tilde{\Pi}^{(i)}(\x,\sigma)$ is $\bC^{n_{i}-1}$
(resp.\ analytic) throughout $\check{\I}^{(i)}(x^{7-i})$.}
\label{4.17d}
\end{eqnarray}
\end{abc}

\item 
We next use Eq.\ (\ref{1.52b}) together with Eq.\ (\ref{G3.12}) to obtain
\begin{abc}
\begin{eqnarray*}
\lefteqn{\F^{\pm}(\x,\sigma) \tilde{v}^{(i)}(\sigma)
[\F_{0}^{\pm}(\x,\sigma)]^{-1} = } \\
& & A(\x) P^{M\pm}(\x,\sigma) \Q^{\pm}(\x,\sigma) v^{(i)\pm}(\sigma)
[\Q_{0}^{\pm}(\x,\sigma)]^{-1} [P^{M\pm}(\x,\sigma)]^{-1} A_{0}(\x)^{-1},
\end{eqnarray*}
whereupon Eqs.\ (\ref{4.17b}) and (\ref{4.17c}) give us
\begin{eqnarray}
\lefteqn{\F^{\pm}(\x,\sigma) \tilde{v}^{(i)}(\sigma)
[\F_{0}^{\pm}(\x,\sigma)]^{-1} = } \nonumber \\
& & A(\x) \tilde{\Pi}^{(i)}(\x,\sigma) P^{M\pm}(\x,\sigma) w^{(i)\pm}(\sigma)
[P^{M\pm}(\x,\sigma)]^{-1} [\tilde{\Pi}^{(i)}(\x,\sigma)]^{-1}
A_{0}(\x)^{-1}
\nonumber \\
& & \text{ for each } \x \in \domE, i \in \{3,4\} \text{ and } 
\sigma \in \I^{(i)}(\x),
\label{4.18a}
\end{eqnarray}
where
\begin{equation}
w^{(i)\pm} := [V^{(i)\pm}]^{-1} v^{(i)\pm} V_{0}^{(i)\pm}.
\end{equation}
Recall that $v^{(i)+} := v^{(i)}$ and $v^{(i)-} := v^{(i)*}$ for 
each $\bv \in K$.  Let $\bw := (w^{(3)},w^{(4)})$ be the member of
$K$ given by 
\begin{equation}
w^{(i)} := [V^{(i)}]^{-1} v^{(i)} V_{0}^{(i)}.
\end{equation}
Then
\begin{equation}
w^{(i)+} = w^{(i)} \text{ and } w^{(i)-} = w^{(i)*}.
\end{equation}
With the aid of Eqs.\ (\ref{1.19a}) one now deduces the following statement
from Eq.\ (\ref{4.18a}):
\begin{eqnarray}
& \text{Statement (ii) of the theorem is true iff } & \nonumber \\
& w^{(i)*}(\sigma) = \Omega^{1-\kappa} w^{(i)}(\sigma) \Omega^{1-\kappa} &
\nonumber \\
& \text{for all } \sigma \in \I^{(i)}(\x), \text{ where }
\kappa = \left\{ \begin{array}{l}
1 \text{ if } s < \sigma, \\
-1 \text{ if } \sigma < r, \\
0 \text{ if } s < \sigma < r.
\end{array} \right. &
\label{4.18e}
\end{eqnarray}
However, since $\bw \in K$, $w^{(i)}(\sigma)$ has the form (by definition
of $K$)
\begin{eqnarray}
w^{(i)}(\sigma) & = & a^{(i)}(\sigma) + \mu^{+}(\x_{0},\sigma) \sigma_{3}
b^{(i)}(\sigma) \text{ for all } \sigma \in \I^{(i)} \nonumber \\
& & \text{ where } a^{(i)}(\sigma) \text{ and } b^{(i)}(\sigma)
\text{ are real linear } \nonumber \\
& & \text{ combinations of $I$ and $J$;}
\label{4.19a}
\end{eqnarray}
and, therefore,
\begin{eqnarray}
w^{(i)}(\sigma)^{*} & = & a^{(i)}(\sigma) + (-1)^{1-\kappa_{0}} 
\mu^{+}(\x_{0},\sigma) \sigma_{3} b^{(i)}(\sigma) 
\text{ for all } \sigma \in \I^{(i)} \nonumber \\
& & \text{ where } \kappa_{0} = \left\{ \begin{array}{l}
1 \text{ if } s_{0} \le \sigma, \\
-1 \text{ if } \sigma \le r_{0}, \\
0 \text{ if } s_{0} \le \sigma \le r_{0}. 
\end{array} \right.
\label{4.19b}
\end{eqnarray}
With the aid of the above relations (\ref{4.19a}) and (\ref{4.19b}), one
infers from (\ref{4.18e}) that
\begin{eqnarray}
& \text{Statement (ii) of the theorem is true iff} & \nonumber \\ &
(-1)^{1-\kappa} b^{(i)}(\sigma) = (-1)^{1-\kappa_{0}} b^{(i)}(\sigma)
& \nonumber \\ &
\text{for each choice of } \x \in \domE, i \in \{3,4\} \text{ and }
\sigma \in \I^{(i)}(\x).
\label{4.20a}
\end{eqnarray}
However, from the meanings of the indices $\kappa$ and $\kappa_{0}$ in
Eq.\ (\ref{4.18e}) and (\ref{4.19b}) and from the fact that $|r,r_{0}|
< |s,s_{0}|$,
\begin{equation}
\kappa-\kappa_{0} = 1 \text{ or } -1 \text{ for all } \sigma \in 
\I^{(i)}(\x)
\end{equation}
(as the reader can quickly verify by considering all possible cases).
Therefore, the proposition (\ref{4.20a}) becomes
\begin{eqnarray*}
& & \text{Statement (ii) of the theorem is true iff } b^{(i)}(\sigma) = 0
\\ & & \text{ for each choice of } \x \in \domE, i \in \{3,4\} 
\text{ and } \sigma \in \I^{(i)}(\x).
\end{eqnarray*}
Since one can choose any $\x$ in $\domE$, however, the above proposition
implies
\begin{eqnarray*}
& & \text{Statement (ii) of the theorem is true iff } b^{(i)}(\sigma) = 0
\\ & & \text{ for each } i \in \{3,4\} \text{ and } \sigma \in \I^{(i)} -
\{r_{0},s_{0}\};
\end{eqnarray*}
and, since $b^{(i)}$ is continuous, we finally obtain
\begin{eqnarray}
& & \text{Statement (ii) of the theorem is true iff } b^{(i)}(\sigma) = 0
\text{ for all }
\nonumber \\ & & i \in \{3,4\} \text{ and } \sigma \in \I^{(i)}.
\text{ Equivalently, statement (ii) of the } \nonumber \\
& & \text{ theorem is true iff } \bw \in B(\I^{(3)},\I^{(4)})
\text{ and } \bw = \bV^{-1} \bv \bV_{0}.
\end{eqnarray}
In other words, we have proved that statement (ii) of the theorem is
true iff statement (i) of the theorem is true.
\end{abc}

\item 
Granting that statements (i) and (ii) are true, we now return to Eq.\ 
(\ref{4.18a}), which becomes
\begin{abc}
\begin{eqnarray}
\lefteqn{\F^{\pm}(\x,\sigma) \tilde{v}^{(i)}(\sigma)
[\F_{0}^{\pm}(\x,\sigma)]^{-1} = A(\x) \tilde{\Pi}^{(i)}(\x,\sigma) }
\nonumber \\ & & \left[ I \cos{\phi^{(i)}(\sigma)} + \left( \begin{array}{cc}
-i & 2(\sigma-z) \\ 0 & i 
\end{array} \right) \sin{\phi^{(i)}(\sigma)} \right] \nonumber \\ & & 
[\tilde{\Pi}^{(i)}(\x,\sigma)]^{-1} A_{0}(\x)^{-1},
\label{4.22a}
\end{eqnarray}
where $w^{(i)}$ has been expressed in its exponential form
\begin{equation}
w^{(i)} = \exp{[J\phi^{(i)}]}
\end{equation}
and where we have used Eq.\ (\ref{3.70b}).  From the above Eq.\ (\ref{4.22a})
and from the proposition (\ref{4.17d}), one immediately sees that, if
$\E^{(i)}$ and $\E_{0}^{(i)}$ are $\bC^{n_{i}}$ (resp.\ analytic), and $w^{(i)}$
is $\bC^{n_{i}-1}$ (resp.\ analytic), then the function of $\sigma$ that
equals (for each $\x \in \domE$)
$$
\F^{\pm}(\x,\sigma) \tilde{v}^{(i)}(\sigma) [\F_{0}^{\pm}(\x,\sigma)]^{-1}
$$
has the $\bC^{n_{i}-1}$ (resp.\ analytic) extension
\begin{equation}
\tilde{Y}^{(i)}(\x) = A(\x) \tilde{\Pi}^{(i)}(\x) e^{J\phi^{(i)}}
[\tilde{\Pi}_{0}^{(i)}(\x)]^{-1} A_{0}(\x)^{-1}
\end{equation}
to the domain $\check{\I}^{(i)}(x^{7-i})$.  [It is understood that 
$\tilde{\Pi}^{(i)}(\x)$ is the function with the domain $\check{\I}(x^{7-i})$
such that $\tilde{\Pi}^{(i)}(\x)(\sigma) := \tilde{\Pi}^{(i)}(\x,\sigma)$.]

Finally, suppose $\bv \in K^{\Box}$ and $\bar{\F}_{0} \in
\S_{\bar{\F}}^{\Box}$.  Then, by definition of $\S_{\bar{\F}}^{\Box}$, 
$\bV_{0} \in \S_{\bV}^{\Box}$, i.e., there exists $\bw' \in B(\I^{(3)},\I^{(4)})$
such that $\bV_{0} \bw' \in K^{\Box}$.  Therefore, since
$$
\bV \bw \bw' = \bv \bV_{0} \bw' \in K^{\Box},
$$
it follows that $\bV \in \S_{\bV}^{\Box}$ and (by definition of 
$\S_{\bar{\F}}^{\Box}$) $\bar{\F} \in \S_{\bar{\F}}^{\Box}$.
\end{abc}
\end{arablist}
\cheers


\setcounter{equation}{0}
\setcounter{theorem}{0}
\subsection{The HHP's corresponding to $(\bv,\bar{\F}_{0},\x)$ and to
$(\bv,\bar{\F}_{0})$}

In the following definitions recall our convention that $\bar{\F}(\x)$,
$\F^{\pm}(\x)$, $\bar{\nu}(\x)$, etc., are {\em names} of functions in 
which the $\x$ plays the role of a parameter that distinguishes one
function from another.

\begin{definition}{Dfn.\ of the HHP corresponding to $(\bv,\bar{\F}_{0},\x)$
\label{def57}}
For each $\bv \in K\s$, $\bar{\F}_{0} \in \S_{\bar{\F}}\triple$ and $\x \in
\domE$, {\em the\/} HHP {\em corresponding to} $(\bv,\bar{\F}_{0},\x)$ will
mean the set of all $2 \times 2$ matrix functions $\bar{\F}(\x)$ such that
\begin{arablist}
\item 
$\bar{\F}(\x)$ is holomorphic throughout $\dom{\bar{\F}(\x)} := 
C - \bar{\I}(\x)$, 
\item 
$\bar{\F}(\x,\infty) = I$, 
\item 
$\F^{\pm}(\x)$ exist, and 
\begin{eqnarray}
Y^{(i)}(\x,\sigma) & := & \F^{+}(\x,\sigma) \tilde{v}^{(i)}(\sigma)
[\F_{0}^{+}(\x,\sigma)]^{-1} \nonumber \\
& = & \F^{-}(\x,\sigma) \tilde{v}^{(i)}(\sigma)
[\F_{0}^{-}(\x,\sigma)]^{-1} 
\label{G3.17} \\
& & \text{ for each $i \in \{3,4\}$ and $\sigma \in \I^{(i)}(\x)$,}
\nonumber
\end{eqnarray}
\item 
$\bar{\F}(\x)$ is bounded at $\x_{0}$ and $\bar{\nu}(\x)^{-1} \bar{\F}(\x)$
is bounded at $\x$, and the function $Y(\x)$ whose domain is $\I(\x)$ 
and whose values are given by $Y(\x,\sigma) := Y^{(i)}(\x,\sigma)$ for each
$\sigma \in \I^{(i)}(\x)$ is bounded at $\x_{0}$ and at $\x$. 
\end{arablist}
The members of the HHP corresponding to $(\bv,\bar{\F}_{0},\x)$ will be
called its {\em solutions}.
\end{definition}

Note:  $\G^{+}(\x)$ and $\G^{-}(\x)$ denote the functions that have 
the common domain $\I(\x)$ and the values ($\Im{\zeta} > 0$)
\begin{equation}
\G^{\pm}(\x,\sigma) := \lim_{\zeta \rightarrow 0} \bar{\G}(\x,\sigma\pm\zeta).
\label{G3.19}
\end{equation}
It is understood that $\G^{+}(\x)$ and $\G^{-}(\x)$ exist if and only if
the above limits exist for every $\sigma \in \I(\x)$.

In item (4), $\bar{\nu}(\x)$ denotes the function whose domain is 
$C - \bar{\I}(\x)$ and whose values $\bar{\nu}(\x,\tau)$ are defined
in GSSM~\ref{Thm_3}.  

In the following definition, it is understood that
$\dom{\bar{\G}(\x)} = C - \bar{\I}(\x)$ and $\dom{Z(\x)} = \I(\x)$.

\begin{abc}
\begin{definition}{Dfn.\ of boundedness at $\x_{0}$ and at $\x$\label{def58}}
It is to be understood that $\bar{\G}(\x)$ is {\em bounded at $\x_{0}$}
if there exists a neighborhood $\nbd(\x_{0})$ of the set $\{r_{0},s_{0}\}$
in the space $C$ such that
\begin{equation}
\{ \bar{\G}(\x,\tau):\tau \in \nbd(\x_{0}) \}
\end{equation}
is bounded.  Likewise, $\bar{\G}(\x)$ is said to be {\em bounded at $\x$}
if there exists a neighborhood $\nbd(\x)$ of the set $\{r,s\}$ in the space
$C$ such that
\begin{equation}
\{ \bar{\G}(\x,\tau):\tau \in \nbd(\x) \}
\end{equation}
is bounded.

We say that $\Z(\x)$ is {\em bounded at $\x_{0}$} if there exists a 
neighborhood $\nbd(\x_{0})$ of the set $\{r_{0},s_{0}\}$ in the space 
$R^{1}$ such that
\begin{equation}
\{ \Z(\x,\sigma):\sigma \in \nbd(\x_{0}) \}
\end{equation}
is bounded.  Likewise, $\Z(\x)$ is {\em bounded at $\x$} if there
exists a neighborhood $\nbd(\x)$ of the set $\{r,s\}$ in the space
$R^{1}$ such that
\begin{equation}
\{ \Z(\x,\sigma):\sigma \in \nbd(\x) \}
\end{equation}
is bounded.
\end{definition}
\end{abc}

\begin{definition}{Dfn.\ of the HHP corresponding to $(\bv,\bar{\F}_{0})$
\label{def59}}
{\em The\/} HHP {\em corresponding to} $(\bv,\bar{\F}_{0})$ will mean 
the set of all functions $\bar{\F}$ [which are not presumed to be members 
of $\S_{\bar{\F}}\triple$] such that $\dom{\bar{\F}} = \{(\x,\tau):\x \in
\domE, \tau \in C - \bar{\I}(\x)\}$ and such that, for each $\x \in \domE$, 
the functions $\bar{\F}(\x)$ whose domains are $C - \bar{\I}(\x)$ and whose 
values are $\bar{\F}(\x,\tau)$ is a solution of the HHP corresponding to
$(\bv,\bar{\F}_{0},\x)$.  The members of the HHP corresponding to
$(\bv,\bar{\F}_{0})$ will be called its {\em solutions}.
\end{definition}

\begin{proposition}[Relation of $\bar{\F}_{0}$ and $\bV_{0}$]
\label{4.1D} \mbox{ } \\
For each $\bar{\F}_{0} \in \S_{\bar{\F}}$ whose corresponding member of
$\S_{\bV}$ is $\bV_{0}$, and for each $\bw \in B(\I^{(3)},\I^{(4)})$,
$\bar{\F}_{0}$ is a solution of the HHP corresponding to $(\bV_{0}\bw,
\bar{\F}^{M})$.
\end{proposition}

\begin{abc}
\proof
For each $\x \in \domE$ and $i \in \{3,4\}$,
\begin{arablist}
\item 
Prop.~\ref{2.1A}(i) states that $\bar{\F}_{0}(\x)$ is holomorphic
throughout its domain $C - \bar{\I}(\x)$,

\item 
GSSM~\ref{3.7G} states that $\F^{\pm}(\x)$ exist and, from Thm.~\ref{4.1C}
and the fact that $\bV^{M} = (I,I)$,
\begin{eqnarray}
Y_{0}^{(i)}(\x,\sigma) & := & \F_{0}^{+}(\x,\sigma) \tilde{V}_{0}^{(i)}(\sigma)
\tilde{w}^{(i)}(\sigma) [\F^{M+}(\x,\sigma)]^{-1} \nonumber \\
& = & \F_{0}^{-}(\x,\sigma) \tilde{V}_{0}^{(i)}(\sigma)
\tilde{w}^{(i)}(\sigma) [\F^{M-}(\x,\sigma)]^{-1} \nonumber \\
& & \text{ for all } \sigma \in \I^{(i)}(\x);
\label{4.24}
\end{eqnarray}
and GSSM~\ref{3.7G} and Prop.~\ref{1.2G}(ii) imply that $\bar{\F}_{0}(\x)$
is bounded at $\x_{0}$ and $\bar{\nu}(\x)^{-1} \bar{\F}_{0}(\x)$ is bounded
at $\x$, while the function $Y_{0}(\x)$ whose domain is $\I(\x)$ and whose 
values are given by $Y_{0}^{(i)}(\x,\sigma)$ at each $\sigma \in \I^{(i)}(\x)$
satisfies the condition
\begin{equation}
Y_{0}(\x) \text{ is bounded at $\x$ and at $\x_{0}$.}
\label{4.25}
\end{equation}
Thus, $\bar{\F}_{0}$ is a solution of the HHP corresponding to
$(\bV\bw,\bar{\F}^{M})$.
\end{arablist}
\cheers
\end{abc}

\begin{gssm}[Reduction theorem]
\label{4.2D}
\mbox{ } \\
For each $\x \in \domE$ and $2 \times 2$ matrix function $\bar{\F}(\x)$
with the domain $C - \bar{\I}(\x)$, for each $\bv \in K$ and $\bar{\F}_{0}
\in \S_{\bar{\F}}$ whose corresponding member of $\S_{\bV}$ is $\bV_{0}$,
and for each $\bw \in B(\I^{(3)},\I^{(4)})$, the following two statements are
equivalent to one another:
\begin{arablist}
\item 
The function $\bar{\F}(\x)$ is a solution of the HHP corresponding to
$(\bv,\bar{\F}_{0},\x)$.
\item
The function $\bar{\F}(\x)$ is a solution of the HHP corresponding to
$(\bv\bV_{0}\bw,\bar{\F}^{M},\x)$.
\end{arablist}
\end{gssm}

\begin{abc}
\proof
Suppose that statement (i) is true.  Then $\bar{\F}(\x)$ satisfies all
four conditions (1) through (4) in the definition of the HHP corresponding
to $(\bv,\bar{\F}_{0},\x)$.  In particular, from conditions (3) and (4),
\begin{eqnarray}
Y^{(i)}(\x,\sigma) & := & \F^{+}(\x,\sigma) \tilde{v}^{(i)}(\sigma)
[\F^{+}_{0}(\x,\sigma)]^{-1} \nonumber \\
& = & \F^{-}(\x,\sigma) \tilde{v}^{(i)}(\sigma)
[\F^{-}_{0}(\x,\sigma)]^{-1} \nonumber \\
& & \text{ for all } i \in \{3,4\} \text{ and } \sigma \in \I^{(i)}(\x);
\label{4.26}
\end{eqnarray}
and
\begin{eqnarray}
Y(\x) \text{ is bounded at $\x$ and at $\x_{0}$.}
\label{4.27}
\end{eqnarray}
So, from the preceding Prop.~\ref{4.1D} and Eqs.\ (\ref{4.24}) and
(\ref{4.26}),
\begin{eqnarray}
X^{(i)}(\x,\sigma) & := & \F^{+}(\x,\sigma) \tilde{u}^{(i)}(\sigma)
[\F^{M+}(\x,\sigma)]^{-1} \nonumber \\
& = & \F_{0}^{-}(\x,\sigma) \tilde{u}^{(i)}(\sigma)
[\F^{M-}(\x,\sigma)]^{-1} \nonumber \\
& & \text{ for all } i \in \{3,4\} \text{ and } \sigma \in \I^{(i)}(\x),
\label{4.28a}
\end{eqnarray}
where 
\begin{equation}
\bu := \bv(\bV_{0}\bw)
\label{4.28b}
\end{equation}
(which is a member of $K$, since $\S_{\bV} \subset K$ and $B \subset K$);
and, furthermore,
\begin{equation}
X(\x) = Y(\x) Y_{0}(\x)
\label{4.28c}
\end{equation}
and, from (\ref{4.25}), (\ref{4.27}) and (\ref{4.28c}),
\begin{equation}
X(\x) \text{ is bounded at $\x$ and $\x_{0}$.}
\label{4.28d}
\end{equation}
Therefore, we have proved that statement (ii) is true if statement (i)
is true.

Next, suppose statement (ii) is true.  Then $\bar{\F}(\x)$ satisfies
all four conditions in the definition of the HHP corresponding to
$(\bu,\bar{\F}^{M},\x)$, where $\bu$ is defined by Eq.\ (\ref{4.28b}).
In particular, from conditions (3) and (4), Eq.\ (\ref{4.28a}) and
the statement (\ref{4.28d}) hold.  Since $\det{V_{0}^{(i)}} =
\det{w^{(i)}} = 1$ and since $\det{\bar{\F}_{0}(\x)} = \det{\bar{\F}^{M}(\x)}
= \bar{\nu}(\x)$ [Prop.~\ref{1.2G}(ii)], Eq.\ (\ref{4.24}) yields
$\det{Y^{(i)}(\x)} = 1$.  Therefore, both sides of Eq.\ (\ref{4.24})
are invertible, and
\begin{eqnarray}
[Y_{0}^{(i)}(\x,\sigma)]^{-1} & = & \F^{M+}(\x,\sigma)
[\tilde{V}_{0}^{(i)}(\sigma) \tilde{w}^{(i)}(\sigma)]^{-1}
[\F_{0}^{+}(\x,\sigma)]^{-1} \nonumber \\
& = & \F^{M-}(\x,\sigma)
[\tilde{V}_{0}^{(i)}(\sigma) \tilde{w}^{(i)}(\sigma)]^{-1}
[\F_{0}^{-}(\x,\sigma)]^{-1} \nonumber \\
& & \text{ for all } i \in \{3,4\} \text{ and } \sigma \in \I^{(i)}(\x);
\label{4.29a}
\end{eqnarray}
and, from (\ref{4.25}),
\begin{equation}
Y_{0}(\x)^{-1} \text{ is bounded at $\x$ and at $\x_{0}$.}
\label{4.29b}
\end{equation}
So, by multiplying both sides of Eq.\ (\ref{4.28a}) by the corresponding
sides of Eq.\ (\ref{4.29a}), and then using (\ref{4.28b}), (\ref{4.28d})
and (\ref{4.29b}), we establish that $\bar{\F}$ is a solution of the
HHP corresponding to $(\bv,\bar{\F}_{0},\x)$.
\cheers
\end{abc}

This tells us that to answer the questions of existence and of membership 
in $\S_{\bar{\F}}\triple$ of the solution of the HHP corresponding to 
$(\bv,\bar{\F}_{0})$ for arbitrary $\bv \in K\s$ and $\bar{\F}_{0} \in 
\S_{\bar{\F}}\triple$, one need only answer the same questions regarding
the HHP corresponding to $(\bv,\bar{\F}^{M})$ for arbitrary $\bv \in K\s$.

\begin{gssm}[Properties of HHP solution]
\label{4.3D}
\mbox{ } \\
Suppose that $\bv \in K$, $\bar{\F}_{0} \in \S_{\bar{\F}}$ and 
$\x \in \domE$ exist such that a solution $\bar{\F}(\x)$ of the HHP 
corresponding to $(\bv,\bar{\F}_{0},\x)$ exists.  Then
\begin{romanlist}
\item
${\F}^{+}(\x)$, ${\F}^{-}(\x)$ and $Y(\x)$ are continuous throughout 
$\I(\x)$,
\item
${\F}^{\pm}(\x)$ are bounded at $\x_{0}$, and 
$[\nu^{\pm}(\x)]^{-1} {\F}^{\pm}(\x)$ are bounded at $\x$, 
\item
$\det{\bar{\F}(\x)} = \bar{\nu}(\x), \quad \det{Y(\x)} = 1$, 
\item
the solution $\bar{\F}(\x)$ is unique, and
\item
the solution of the HHP corresponding to $(\bv,\bar{\F}_{0},\x_{0})$
is given by
\begin{equation}
\bar{\F}(\x_{0},\tau) = I := \left( \begin{array}{cc}
1 & 0 \\ 0 & 1
\end{array} \right) 
\label{4.30}
\end{equation}
for all $\tau \in C$.
\end{romanlist}
\end{gssm}

\proofs
\begin{romanlist}
\item 
The statement that $\F^{+}(\x)$ and $\F^{-}(\x)$ are continuous is a
direct consequence of a theorem by P.~Painlev\'{e} which is stated and
proved by N.\ I.\ Muskhelishvili.\footnote{See Ref.~\ref{Musk}, Ch.\ 2,
Sec.~14, pp.\ 33-34.}  The continuity of $Y(\x)$ then follows from its
definition by Eq.\ (\ref{G3.17}), the fact that $\tilde{v}^{(i)}$ is
continuous [Eq.\ (\ref{G3.12})] and the fact that $\F_{0}^{+}$ and
$\F_{0}^{-}$ are continuous [Thm.~\ref{3.4G}].
\cheers

\item 
{}From Eq.\ (\ref{G3.17}),
\begin{abc}
\begin{equation}
\F^{\pm}(\x) = Y^{(i)}(\x) \F_{0}^{\pm}(\x) [\tilde{v}^{(i)}]^{-1}
\label{4.31}
\end{equation}
for each $i \in \{3,4\}$.  The function $Y(\x)$ is bounded at $\x$ and
at $\x_{0}$ according to condition (4) in the definition of the HHP;
and $\tilde{v}^{(i)}$ and its inverse are continuous throughout $\I^{(i)}$.
As regards $\F_{0}^{\pm}(\x)$, Eqs.\ (\ref{1.23}) and (\ref{1.53}) yield
\begin{eqnarray}
\det{\F_{0}^{\pm}(\x,\tau)} & = & \bnu^{\pm}(\x,\x_{0},\tau) \nonumber \\
& & \text{ for all } \tau \in \bar{C}^{\pm} - \{r,s,r_{0},s_{0}\},
\label{4.32a}
\end{eqnarray}
whereupon
\begin{eqnarray}
\F_{0}^{\pm}(\x,\tau)^{-1} & = & - [\bnu^{\pm}(\x,\x_{0},\tau)]^{-1}
J [\F_{0}^{\pm}(\x,\tau)]^{T} J \nonumber \\
& & \text{ for all } \tau \in \bar{C}^{\pm} - \{r,s,r_{0},s_{0}\}.
\end{eqnarray}
Therefore, from Eqs.\ (\ref{3.46a}) and (\ref{3.46b}) in GSSM~\ref{3.7G},
$\F_{0}^{\pm}(\x)$ is bounded at $\x_{0}$, and $[\nu^{\pm}(\x)]^{-1}
\F_{0}^{\pm}(\x)$ is bounded at $\x$.

So, from Eq.\ (\ref{4.31}), $\F^{\pm}(\x)$ is bounded at $\x_{0}$, and
$[\bar{\nu}(\x)]^{-1} \F^{\pm}(\x)$ is bounded at $\x$.
\cheers
\end{abc}

\item 
Let
\begin{abc}
\begin{equation}
Z_{1}(\x) := \det{\bar{\F}(\x)}/\bar{\nu}(\x),
\label{4.33a}
\end{equation}
whereupon Eq.\ (\ref{4.32a}) and conditions (1), (2), (3) and (4) of the
definition of the HHP imply that 
\begin{equation}
Z_{1}(\x) \text{ is holomorphic throughout } C - \bar{\I}(\x),
\label{4.33b}
\end{equation}
\begin{equation}
Z_{1}(\x,\infty) = 1,
\label{4.33c}
\end{equation}
the limits $Z_{1}^{\pm}(\x)$ exist and
\begin{equation}
\det{Y(\x,\sigma)} = Z_{1}^{+}(\x,\sigma) = Z_{1}^{-}(\x,\sigma)
\text{ for all } \sigma \in \I(\x),
\label{4.33d}
\end{equation}
\begin{equation}
\begin{array}{l}
\bar{\nu}(\x) Z_{1}(\x) \text{ is bounded at } \x_{0} \text{ and } \\
\bar{\nu}(\x)^{-1} Z_{1}(\x) \text{ is bounded at } \x,
\end{array}
\label{4.33e}
\end{equation}
and
\begin{equation}
\det{Y(\x)} = Z_{1}^{\pm}(\x) \text{ is bounded at $x$ and at $\x_{0}$.}
\label{4.33f}
\end{equation}

At this point and at several other points in this set of notes, we shall employ
a trilogy of elementary theorems; namely, a theorem of Riemann on analytic
continuation across an arc, a theorem of Riemann on isolated singularities
of holomorphic functions, and a (generalized) theorem of Liouville\footnote{
For the theorem of Riemann on analytic continuation across an arc, see Sec.~24,
Ch.~1, of {\em A Course of Higher Mathematics}, Vol.~III, Part Two, by V.\ I.\
Smirnov (Addison-Wesley, 1964).  For the theorem of Riemann on isolated 
singularities and the generalized Liouville theorem, see Secs.~133 and
167-168, respectively, of {\em Theory of Functions of a Complex Variable},
Vol.~1, by C.\ Caratheodory, 2nd English edition (Chelsea Publishing Company,
1983).\label{trilogy}} on entire functions that do not have an essential
singularity at $\tau=\infty$.

{}From the above statements (\ref{4.33b}) and (\ref{4.33d}) together with
the theorem on analytic continuation across an arc, $Z_{1}(\x)$ has a 
holomorphic extension to the domain $C - \{r,s,r_{0},s_{0}\}$; and, from 
the statements (\ref{4.33e}) and (\ref{4.33f}), together with the theorem 
on isolated singularities of holomorphic functions, $Z_{1}(\x)$ has a 
further holomorphic extension $Z_{1}^{ex}(\x)$ to $C$.  Finally, the 
theorem of Liouville, together with Eq.\ (\ref{4.33c}), then yields 
\begin{equation}
Z_{1}^{ex}(\x,\tau) = 1 \text{ for all } C.
\end{equation}
Thus, we have shown that $\det{\bar{\F}(\x)} = \bar{\nu}(\x)$, whereupon
Eq.\ (\ref{4.33d}) yields $\det{Y(\x)} = 1$.
\cheers
\end{abc}

\item 
Suppose that $\bar{\F}'(\x)$ is also a solution of the HHP corresponding
to $(\bv,\bar{\F}_{0},\x)$.  Since $\det{\bar{\F}(\x)} = \bar{\nu}(\x)$,
$\bar{\F}(\x)$ is invertible.  Let
\begin{abc}
\begin{equation}
Z_{2}(\x) := \bar{\F}'(\x) \bar{\F}(\x)^{-1},
\end{equation}
whereupon conditions (1), (2), (3) and (4) in the definition of the HHP
imply that
\begin{equation}
Z_{2}(\x) \text{ is holomorphic throughout } C - \bar{\I}(\x),
\end{equation}
\begin{equation}
Z_{2}(\x,\infty) = I,
\end{equation}
the limits $Z_{2}^{\pm}(\x)$ exist and 
\begin{eqnarray}
Y'(\x) Y(\x)^{-1} & = & Z_{2}^{+}(\x) \nonumber \\
& = & Z_{2}^{-}(\x) \text{ throughout } \I(\x),
\end{eqnarray}
\begin{equation}
Z_{2}(\x) \text{ is bounded at $\x$ and at $\x_{0}$}
\end{equation}
and
\begin{equation}
Y'(\x) Y(\x)^{-1} = Z_{2}^{\pm}(\x) 
\text{ is bounded at $\x$ and at $\x_{0}$.}
\end{equation}
The same kind of reasoning that was used in the proof of part (iii) of
the theorem nets $Z(\x) = I$.  So $\bar{F}'(\x) = \bar{\F}(\x)$.
\cheers
\end{abc}

\item 
When $x = \x_{0}$, $\I(x)$ and its closure $\bar{\I}(\x)$ are empty.
So, condition (1) of the HHP definition implies that $\bar{\F}(\x_{0})$
is holomorphic throughout $C$, whereupon condition (2) tells us that
$\bar{\F}(\x_{0})$ has the value $I$ throughout $C$.  [$\F^{\pm}(\x)$
are empty sets when $\x = \x_{0}$; and conditions (3) and (4) hold
trivially when $\x = \x_{0}$.]
\cheers
\end{romanlist}


\setcounter{equation}{0}
\setcounter{theorem}{0}
\subsection{Our generalized Geroch conjecture}
We shall conclude Part II of these notes with a statement of our generalized
Geroch conjecture.  First, we conjecture that the set $k^{3}\s$ and the 
group $K^{3}\s$ are, in fact, identical.  [See Eqs.\ (\ref{G3.3}) and
(\ref{G3.7}).]  We shall, accordingly, regard the Kinnersley--Chitre 
realization of the Geroch group and our generalization of that group to be 
groups of permutations of $\S_{\bar{\F}}^{3}$, i.e., one-to-one mappings of
that set onto itself.\footnote{Precisely how the K--C realization of the
Geroch group fits within our larger group is described in Appendix~C.}

Next, we conjecture that there exists, for each $\bv \in K^{3}$, exactly
one solution $\bar{\F}$ of the HHP corresponding to $(\bv,\bar{\F}_{0})$.
This will allow us to define a mapping 
\begin{abc}
\begin{equation}
[\bv]:\S_{\bar{\F}}^{3} \rightarrow \S_{\bar{\F}}^{3}
\end{equation}
such that, for each $\bar{\F}_{0} \in \S_{\bar{\F}}^{3}$,
\begin{equation}
[\bv](\bar{\F}_{0}) = \bar{\F}
\end{equation}
is that unique solution of the HHP corresponding to $(\bv,\bar{\F}_{0})$.
Then we shall be able to define
\begin{equation}
\K^{3}\s := \{[\bv]:\bv\in K^{3}\s\}.
\end{equation}
\end{abc}

Now, the center of $K(\x_{0},\I^{(i)})$ is, as the reader can verify,
\begin{abc}
\begin{equation}
Z^{(i)} := \{ \delta^{(i)},-\delta^{(i)} \},
\end{equation}
where
\begin{equation}
\delta^{(i)}(\sigma) = I \text{ for all $\sigma \in \I^{(i)}$.}
\end{equation}
The center of $K\s$ is thus $Z^{(3)} \times Z^{(4)}$. 
\end{abc}

Finally, our principal goal in Part III of these notes will be to prove
the following conjecture.\footnote{A remarkable feature of our generalized 
Geroch group is described in Appendix~E.}

\noindent
{\bf GENERALIZED GEROCH CONJECTURE}
\mbox{ } \\ \vspace{-3ex}
\begin{romanlist}
\item 
{\em The mapping $[\bv]$ is the identity map on $\S_{\bar{\F}}^{3}$ iff
$\bv \in Z^{(3)} \times Z^{(4)}$.}
\item 
{\em The set $\K^{3}$ is a group of permutations of $\S_{\bar{\F}}^{3}$
such that the mapping $\bv \rightarrow [\bv]$ is a homomorphism of
$K^{3}$ onto $\K^{3}$; and the mapping $\{\bv\bw: \bw \in
Z^{(3)} \times Z^{(4)}\} \rightarrow [\bv]$ is an isomorphism [viz,
the isomorphism of $K^{3}/(Z^{(3)} \times Z^{(4)})$ onto $\K^{3}$].}
\item 
{\em The group $\K^{3}$ is transitive [i.e., for each $\bar{\F}_{0},
\bar{\F} \in \S_{\bar{\F}}^{3}$ there exists at least one element of
$\K^{3}$ that transforms $\bar{\F}_{0}$ into $\bar{\F}$].} 
\end{romanlist}

\newpage
\part{Proof of our generalized Geroch conjecture}

\section{An Alekseev-type singular integral equation that is equivalent
to the HHP when $\bv \in K^{1+}$\label{Sec_5}}

Using an ingenious argument G.\ A.\ Alekseev\footnote{See Ref.~\ref{Alekseev}.}
derived a singular integral equation, supposing that $\bar{\F}(\tau)$ was
analytic in a neighborhood of $\{r,s\}$ except for branch points of index
$1/2$ at $\tau=r$ and $\tau=s$.  We shall now show that the same type integral
equation arises in connection with solutions of our new HHP that need not be 
analytic.

\setcounter{equation}{0}
\setcounter{theorem}{0}
\subsection{Preliminary propositions}

In the following proposition and in later propositions and theorems, it will
be understood that for each $(\x,\x') \in \domE^{2}$ such that
$\bsI^{(3)}(\x,\x') < \bsI^{(4)}(\x,\x')$ and for each complex-valued 
function $F$ whose domain contains $\bsI(\x,\x')$ and which is summable 
on the intervals $\bsI^{(3)}(\x,\x')$ and $\bsI^{(4)}(\x,\x')$,
\begin{abc}
\begin{equation}
\int_{\bar{\subbsI}(\x,\x')} d\sigma F(\sigma) := 
\int_{a^{3}}^{b^{3}} d\sigma F(\sigma) +
\int_{a^{4}}^{b^{4}} d\sigma F(\sigma),
\end{equation}
where
\begin{equation}
a^{3} := \inf{\{r,r'\}}, \quad a^{4} := \inf{\{s,s'\}}, \quad
b^{3} := \sup{\{r,r'\}}, \quad b^{4} := \sup{\{s,s'\}}.
\label{5.2}
\end{equation}
In other words, $\bar{\bsI}^{(i)}(\x,\x') = \bar{\bsI}^{(i)}(\x',\x)$
will henceforth be regarded as an arc whose orientation is in the direction
of increasing $\sigma \in \bar{\bsI}^{(i)}(\x,\x')$.
\end{abc}

\begin{abc}
\begin{definition}{Dfn.\ of an extension of $\F^{\pm}(\x)$ for each solution
$\bar{\F}(\x)$ of the HHP corresponding to $(\bv,\bar{\F}_{0},\x)$\label{def60}}
For each $\bv \in K$, $\bar{\F}_{0} \in \S_{\bar{\F}}$, $\x \in \domE$ and
solution $\bar{\F}(\x)$ of the HHP corresponding to $(\bv,\bar{\F}_{0},\x)$, 
we shall henceforth let $\F^{\pm}(\x)$ denote the function whose domain is
\begin{equation}
\dom{\F^{\pm}(\x)} := \bar{C}^{\pm} - \{r,s,r_{0},s_{0}\}
\end{equation}
and whose values are given by
\begin{equation}
\F^{\pm}(\x,\tau) := \F^{\pm}(\x)(\tau) := \bar{\F}(\x,\tau)
\text{ when } \tau \in \bar{C}^{\pm} - \grave{\I}(\x)
\label{5.3b}
\end{equation}
and by
\begin{equation}
\F^{\pm}(\x,\sigma) := \F^{\pm}(\x)(\sigma) := \lim_{\zeta \rightarrow 0}
\bar{\F}(\x,\sigma \pm \zeta) \text{ (where $\Im{\zeta} > 0$) when }
\sigma \in \I(\x).
\end{equation}
This newly defined $\F^{\pm}(\x)$ is thus an extension of the function
$\F^{\pm}(\x)$ that was defined by Eq.\ (\ref{G3.19}).  Likewise,
$\nu^{\pm}(\x)$ will denote the function whose domain is $\bar{C}^{\pm}
- \{r,s,r_{0},s_{0}\}$ and whose values are given by
\begin{equation}
\nu^{\pm}(\x)(\tau) := \bnu^{\pm}(\x,\x_{0},\tau) \text{ for all }
\tau \in \bar{C}^{\pm} - \{r,s,r_{0},s_{0}\},
\end{equation}
where $\bnu^{\pm}$ is defined by Eqs.\ (\ref{1.6l}) to (\ref{1.6p}).
\end{definition}
\end{abc}

\begin{proposition}[Properties of $\F^{\pm}(\x)$]
\label{5.1A} \mbox{ } \\
For each $\bv \in K$, $\bar{F}_{0} \in \S_{\bar{\F}}$, $\x \in \domE$
and solution $\bar{\F}(\x)$ of the HHP corresponding to
$(\bv,\bar{\F}_{0},\x)$, the following statements hold:
\begin{romanlist}
\item 
\begin{equation}
\F^{+}(\x,\sigma) = \F^{-}(\x,\sigma) = \bar{\F}(\x,\sigma)
\text{ for all } \sigma \in R^{1} - \grave{\I}(\x).
\end{equation}
\item 
$\F^{\pm}(\x)$ is continuous, and the restriction of $\F^{\pm}(\x)$ to
$C^{\pm}$ is holomorphic.
\item 
$\F^{\pm}(\x)$ is bounded at $\x_{0}$ and $[\nu^{\pm}(\x)]^{-1}
\F^{\pm}(\x)$ is bounded at $\x$, which means that there exist neighborhoods
$\nbd(\x_{0})$ and $\nbd(\x)$ of the sets $\{r_{0},s_{0}\}$ and $\{r,s\}$,
respectively, in the space $\bar{C}^{\pm}$ such that 
$$
\{\F^{\pm}(\x,\tau):\tau \in \nbd(\x_{0}) - \{r,s,r_{0},s_{0}\}\}
$$
and
$$
\{\bnu^{\pm}(\x,\x_{0},\tau)^{-1}\F^{\pm}(\x,\tau):\tau \in \nbd(\x)
- \{r,s,r_{0},s_{0}\}\}
$$
are bounded.
\item 
The functions of $\sigma$ whose domains are $R^{1} - \{r,s,r_{0},s_{0}\}$
and whose values are given by $\F^{\pm}(\x,\sigma)$ and by
$\bnu^{\pm}(\x,\x_{0},\sigma)^{-1}\F^{\pm}(\x,\sigma)$ are summable on
each closed subinterval of $R^{1}$.
\end{romanlist}
\end{proposition}

\begin{abc}
\proofs
\begin{romanlist}
\item 
This follows directly from Eq.\ (\ref{5.3b}) in the definition of $\F^{\pm}$.
\cheers
\item 
This follows from GSSM~\ref{4.3D}(i) and the fact that, by definition,
$\bar{\F}(\x,\tau)$ is a holomorphic (and, therefore, also continuous)
function of $\tau$ throughout $C - \bar{\I}(\x)$.
\cheers
\item 
This follows from GSSM~\ref{4.3D}(ii) and the fact that, by definition of
the HHP, $\bar{\F}(\x)$ is bounded at $\x_{0}$, and 
$\bar{\nu}(\x)^{-1} \bar{\F}(\x)$ is bounded at $\x$.
\cheers
\item 
With $\x' = \x_{0}$, let $a^{i}$ and $b^{i} (i=3,4)$ be defined by
Eqs.\ (\ref{5.2}); and let
\begin{eqnarray*}
\epsilon^{3} & := & \inf{\left\{\frac{1}{2}|r-r_{0}|,
\frac{1}{2}|a^{4}-b^{3}|\right\}}, \\
\epsilon^{4} & := & \inf{\left\{\frac{1}{2}|s-s_{0}|,
\frac{1}{2}|a^{4}-b^{3}|\right\}}.
\end{eqnarray*}
Since $\F^{\pm}(\x,\sigma)$ and $\bnu^{\pm}(\x,\x_{0},\sigma)^{-1}
\F^{\pm}(\x,\sigma)$ are defined and are continuous functions of $\sigma$
throughout $R^{1} - \{r,s,r_{0},s_{0}\}$, it is sufficient to prove that
these functions are summable on the closed intervals
\begin{equation}
[r_{0}-\epsilon^{3},r_{0}+\epsilon^{3}], \quad
[r-\epsilon^{3},r+\epsilon^{3}], \quad
[s_{0}-\epsilon^{4},s_{0}+\epsilon^{4}], \quad
[s-\epsilon^{4},s+\epsilon^{4}]
\label{5.5}
\end{equation}
in order to establish part (iv) of this theorem.

{}From parts (ii) and (iii) of this theorem,
\begin{equation}
\begin{array}{l}
\text{$\F^{\pm}(\x,\sigma)$ is a bounded and continuous function of
$\sigma$ on} \\[0pt]
[r_{0}-\epsilon^{3},r_{0}+\epsilon^{3}] - \{r_{0}\} \text{ and on }
[s_{0}-\epsilon^{4},s_{0}+\epsilon^{4}] - \{s_{0}\};
\end{array}
\label{5.6a}
\end{equation}
and
\begin{equation}
\begin{array}{l}
\text{$\bnu^{\pm}(\x,\x_{0},\sigma)^{-1}\F^{\pm}(\x,\sigma)$ is a bounded
and continuous function} \\
\text{of $\sigma$ on $[r-\epsilon^{3},r+\epsilon^{3}] - \{r\}$ and on
$[s-\epsilon^{4},s+\epsilon^{4}] - \{s\}$.}
\end{array}
\label{5.6b}
\end{equation}
However, a well known theorem\footnote{This is a special case of Corollary
22.4s in Ref.~\ref{McShane}.  Also, see Thm.~21.4s in Ref.~\ref{McShane}.}
asserts that the product of any complex-valued function which is summable
on $[a,b] \subset R^{1}$ by a function which is bounded and continuous on
$[a,b] - \text{ (any given finite set)}$ is also summable on $[a,b]$; and
recall that $\bnu^{\pm}(\x,\x_{0},\sigma)$ and
$\bnu^{\pm}(\x,\x_{0},\sigma)^{-1}$ are summable on any closed subinterval
of $R^{1}$.  Therefore, one infers from (\ref{5.6a}) and (\ref{5.6b}) that
$\F^{\pm}(\x,\sigma)$ and $\bnu^{\pm}(\x,\x_{0},\sigma)^{-1}
\F^{\pm}(\x,\sigma)$ are both summable functions of $\sigma$ on all of the
intervals (\ref{5.5}).
\cheers
\end{romanlist}
\end{abc}

Note: We claim no originality for the next theorem or for the two lemmas
associated with that theorem.  We include proofs, however, since we know of 
no references which are easily accessible and which provide proofs with 
exactly the premises that we employ.  For example, unlike some current 
references on singular integral equations, we are not presently assuming
that $\F^{\pm}(\x,\sigma)$ (for fixed $\x$) obeys a H\"{o}lder condition 
on each closed subinterval of $\I(\x)$.

\begin{abc}
\begin{definition}{Dfns.\ of $R(\x,\tau,\sigma)$ and $\C^{\pm}(\sigma,\alpha)$
\label{def61}}
For each $\x \in \domE$, $\tau \in C - \{r_{0},s_{0},r,s\}$ and
$\sigma \in \grave{\I}(\x)$, let
\begin{equation}
R(\x,\tau,\sigma) := \inf{\left\{\frac{1}{2}|b-\sigma|: b \in 
\{r_{0},s_{0},r,s,\tau\} \text{ and } b \ne \sigma \right\}};
\label{5.7a}
\end{equation}
and, for each real number $\alpha$ such that
\begin{equation}
0 < \alpha \le R(\x,\tau,\sigma),
\end{equation}
let
\begin{equation}
\C^{\pm}(\sigma,\alpha) := \{\tau'\in\bar{C}^{\pm}:|\tau'-\sigma|=\alpha\}
\label{5.7c}
\end{equation}
with an assigned counterclockwise orientation.
\end{definition}
\end{abc}

We shall now present two lemmas, the first of which concerns the end points
$\{r_{0},s_{0},r,s\}$ of the closed interval $\bar{\I}(\x)$, and the second
of which concerns the interior points, i.e., $\I(\x)$.

\begin{abc}
\begin{lemma}[End point theorem]
\label{5.2A} \mbox{ } \\
For each $\bv \in K$, $\bar{\F}_{0} \in \S_{\bar{\F}}$, $\x \in \domE$,
solution $\bar{\F}(x)$ of the HHP corresponding to $(\bv,\bar{\F}_{0},\x)$
and 
\begin{equation}
\tau \in C - \{r_{0},s_{0},r,s\},
\end{equation}
the ratios
\begin{eqnarray}
\frac{\F^{\pm}(\x,\tau')}{\tau'-\tau} \text{ and }
\frac{\bnu^{\pm}(\x_{0},\x,\tau') \F^{\pm}(\x,\tau')}{\tau'-\tau} \nonumber \\
\text{are both continuous functions of $\tau'$} \nonumber \\ 
\text{throughout their common domain } \nonumber \\
\bar{C}^{\pm} - \{r_{0},s_{0},r,s\}.
\label{5.8b}
\end{eqnarray}
Furthermore, for each
\begin{equation}
c \in \{r_{0},s_{0},r,s\},
\end{equation}
the punctured closed semi-disk
\begin{equation}
\begin{array}{r}
\{\tau' \in \bar{C}^{\pm}: 0 < |\tau'-c| \le R(\x,\tau,c)\} 
\text{ is a subset } \\
\text{ of the domain } \bar{C}^{\pm} - \{r_{0},s_{0},r,s\};
\end{array}
\label{5.8d}
\end{equation}
and
\begin{eqnarray}
\frac{1}{2\pi i} \int_{\C^{\pm}(c,\alpha)} d\tau'
\frac{\F^{\pm}(\x,\tau')}{\tau'-\tau} & \rightarrow & 0 \text{ as }
\alpha \rightarrow 0,
\label{5.8e} \\
\frac{1}{2\pi i} \int_{\C^{\pm}(c,\alpha)} d\tau'
\frac{\bnu^{\pm}(\x_{0},\x,\tau')\F^{\pm}(\x,\tau')}{\tau'-\tau} 
& \rightarrow & 0 \text{ as } \alpha \rightarrow 0.
\label{5.8f}
\end{eqnarray}
\end{lemma}
As regards the above relations involving $\bnu^{\pm}$, recall that
\begin{equation}
\bnu^{\pm}(\x_{0},\x,\tau') = \bnu^{\pm}(\x,\x_{0},\tau')^{-1}.
\label{5.8g}
\end{equation}
\end{abc}

\begin{abc}
\proof
The proof will be given in four parts.
\begin{arablist}
\item 
Statement (\ref{5.8b}) follows from Prop.~\ref{5.1A}(ii) and the fact that
$\bnu^{\pm}(\x_{0},\x,\tau')$ and $(\tau'-\tau)^{-1}$ are both continuous
functions of $\tau'$ throughout $\bar{C}^{\pm} - \{r_{0},s_{0},r,s\}$.
\item 
{}From Eq.\ (\ref{5.7a}), if $\tau' \in \bar{C}^{\pm}$ such that
\begin{equation}
0 < \alpha := |\tau'-c| \le R(\x,\tau,c),
\label{5.9a}
\end{equation}
we infer that, for each $b \in \{r_{0},s_{0},r,s,\tau\}$ such that $b \ne c$,
\begin{equation}
|\tau'-b| \ge |b-c| - |\tau'-c| \ge 2R(\x,\tau,c)-\alpha
\ge R(\x,\tau,c) > 0.
\label{5.9b}
\end{equation}
Therefore, if $0 < |\tau'-c| \le R(\x,\tau,c)$, then $|\tau'-b| > 0$
for all $b \in \{r_{0},s_{0},r,s,\tau\}$, whereupon statement (\ref{5.8d})
follows.
\item 
{}From statement (\ref{5.8d}) and from Props.~\ref{5.1A}(ii) and~(iii),
we obtain the following two statements:
\begin{equation}
\begin{array}{c}
\text{If $c \in \{r_{0},s_{0}\}$, there exists a positive real number} \\
M(\x,\tau,c) \text{ such that } ||\F^{\pm}(\x,\tau')|| \le M(\x,\tau,c) 
\text{ for} \\ \text{all }
\tau' \in \bar{C}^{\pm} \text{ for which } 0 < |\tau'-c| \le R(\x,\tau,c);
\end{array}
\label{5.10a}
\end{equation}
\begin{equation}
\begin{array}{c}
\text{If $c \in \{r,s\}$, there exists a positive real number } M(\x,\tau,c) \\
\text{ such that } ||\bnu^{\pm}(\x_{0},\x,\tau)\F^{\pm}(\x,\tau)|| \le 
M(\x,\tau,c) \text{ for} \\ \text{all }
\tau' \in \bar{C}^{\pm} \text{ for which } 0 < |\tau'-c| \le R(\x,\tau,c).
\end{array}
\label{5.10b}
\end{equation}
Now, if (\ref{5.9a}) holds and $b \in \{r_{0},s_{0},r,s\}$, then
\begin{equation}
|\tau'-b| \le |b-c| + |\tau'-c| = |b-c| + \alpha.
\label{5.10c}
\end{equation}
Therefore, upon applying the above inequality (\ref{5.10c}) to the
numerators and the inequality (\ref{5.9b}) to the denominators of the
following ratios, and upon using the relations $s-r=2\rho$ and 
$s_{0}-r_{0}=2\rho_{0}$, one obtains the relations
\begin{eqnarray}
\left| \frac{\bnu^{\pm}(\x,\x_{0},\tau')}{\tau'-\tau} \right| & \le &
\frac{2\rho_{0}+\alpha}{\sqrt{\alpha} R(\x,\tau,c)^{3/2}} 
\text{ when $c \in \{r,s\}$} \nonumber \\
& & \text{and } 0 < |\tau'-c| \le R(\x,\tau,c)
\label{5.10d}
\end{eqnarray}
and
\begin{eqnarray}
\left| \frac{\bnu^{\pm}(\x_{0},\x,\tau')}{\tau'-\tau} \right| & \le &
\frac{2\rho+\alpha}{\sqrt{\alpha} R(\x,\tau,c)^{3/2}} 
\text{ when $c \in \{r_{0},s_{0}\}$} \nonumber \\ 
& & \text{and } 0 < |\tau'-c| \le R(\x,\tau,c).
\end{eqnarray}
It is now a simple matter of using (\ref{5.10a}), (\ref{5.10b}), (\ref{5.10d})
and the identity (\ref{5.8g}) to prove that
\begin{eqnarray}
\left|\left| \frac{\F^{\pm}(\x,\tau')}{\tau'-\tau} \right|\right| & \le & 
\left\{ \begin{array}{l}
\frac{M(\x,\tau,c)}{R(\x,\tau,c)} \text{ if } c \in \{r_{0},s_{0}\}, \\
\frac{(2\rho_{0}+\alpha)M(\x,\tau,c)}{\sqrt{\alpha} R(\x,\tau,c)^{3/2}}
\text{ if } c \in \{r,s\}
\end{array} \right. 
\label{5.10f}
\end{eqnarray}
and
\begin{eqnarray}
\left|\left| \frac{\bnu^{\pm}(\x_{0},\x,\tau')\F^{\pm}(\x,\tau')}{\tau'-\tau}
\right|\right| & \le & \left\{ \begin{array}{l}
\frac{M(\x,\tau,c)}{R(\x,\tau,c)} \text{ if } c \in \{r,s\}, \\
\frac{(2\rho+\alpha)M(\x,\tau,c)}{\sqrt{\alpha} R(\x,\tau,c)^{3/2}}
\text{ if } c \in \{r_{0},s_{0}\}
\end{array} \right.
\label{5.10g}
\end{eqnarray}
for all $0 < |\tau'-c| \le R(\x,\tau,c)$.
\item 
The conclusion (\ref{5.8e}) now follows from (\ref{5.10f}) and the familiar
inequality 
$$
\left|\left| \frac{1}{2\pi i} \int_{\C^{\pm}(c,\alpha)} d\tau'
\frac{\F^{\pm}(\x,\tau')}{\tau'-\tau} \right|\right| \le 
\frac{1}{2} \alpha \sup{\left\{ 
\left|\left| \frac{\F^{\pm}(\x,\tau')}{\tau'-\tau} \right|\right|:
\tau' \in \C^{\pm}(c,\alpha) \right\}}. 
$$
The conclusion (\ref{5.8f}) similarly follows from (\ref{5.10g}).
\end{arablist}
\cheers
\end{abc}

\begin{abc}
\begin{lemma}[Interior point theorem]
\label{5.3A}
For each $\bv \in K$, $\bar{\F}_{0} \in \S_{\bar{\F}}$, $\x \in \domE$,
solution $\bar{\F}(\x)$ of the HHP corresponding to $(\bv,\bar{\F}_{0},\x)$
and
\begin{equation}
\sigma \in \I(\x),
\end{equation}
let
\begin{equation}
R(\x,\sigma) := \inf{\left\{\frac{1}{2}|b-\sigma|: b \in \{r_{0},s_{0},r,s\}
\right\}}
\label{5.11b}
\end{equation}
[which is identical to $R(\x,\sigma,\sigma)$ as defined by Eq.\ (\ref{5.7a})].
Then the closed semi-disk
\begin{equation}
\begin{array}{r}
\{\tau' \in \bar{C}^{\pm}: |\tau'-\sigma| \le R(\x,\sigma)\}
\text{ is a subset of} \\
\text{the domain } \bar{C}^{\pm} - \{r_{0},s_{0},r,s\} \text{ of }
\F^{\pm}(\x);
\end{array}
\label{5.11c}
\end{equation}
and, for $0 < \alpha \le R(\x,\sigma)$,
\begin{eqnarray}
\frac{1}{2\pi i} \int_{\C^{\pm}(\sigma,\alpha)} d\tau'
\frac{\F^{\pm}(\x,\tau')}{\tau'-\sigma} & \rightarrow & 
\frac{1}{2} \F^{\pm}(\x,\sigma) \text{ as } \alpha \rightarrow 0,
\label{5.11d} \\
\frac{1}{2\pi i} \int_{\C^{\pm}(\sigma,\alpha)} d\tau'
\frac{\bnu^{\pm}(\x_{0},\x,\tau')\F^{\pm}(\x,\tau')}{\tau'-\sigma}
& \rightarrow & \frac{1}{2} \bnu^{\pm}(\x_{0},\x,\sigma)\F^{\pm}(\x,\sigma)
\text{ as } \alpha \rightarrow 0.
\label{5.11e}
\end{eqnarray}
\end{lemma}
Note: The above integrals exist by virtue of statement (\ref{5.8b}) in
the preceding lemma.
\end{abc}

\begin{abc}
\proof
Statement (\ref{5.11c}) is an obvious consequence of the definition
(\ref{5.11b}).  

We introduce a new integration variable $\varphi$ such that
\begin{equation}
0 \le \varphi \le \pi \text{ and } \tau' = \sigma \pm \alpha e^{i\varphi}
\text{ when } \tau' \in \C^{\pm}(\sigma,\alpha),
\end{equation}
whereupon
\begin{equation}
\frac{1}{2\pi i} \int_{\C^{\pm}(\sigma,\alpha)} d\tau'
\frac{\F^{\pm}(\x,\tau')}{\tau'-\sigma} =
\frac{1}{2\pi} \int_{0}^{\pi} d\varphi 
\F^{\pm}(\x,\sigma\pm\alpha e^{i\varphi}).
\label{5.12b}
\end{equation}
{}From Prop.~\ref{5.1A}(ii) and (\ref{5.11c}), $\F^{\pm}(\x,\tau')$ is a
uniformly continuous function of $\tau'$ throughout the closed semi-disk
$\{\tau' \in \bar{C}^{\pm}: |\tau'-\sigma| \le R(\x,\sigma)\}$.  Therefore,
the limit of the integral (\ref{5.12b}) as $\alpha \rightarrow 0$ exists
and is given by (\ref{5.11d}).  The statement (\ref{5.11e}) is proved in
the same way.
\cheers
\end{abc}

Henceforth, whenever there is no danger of ambiguity, the arguments `$\x$'
and `$\x_{0}$' will be suppressed.  For example, `$\bar{\F}(\tau)$'
and `$\F^{\pm}(\sigma)$' will generally be used as abbreviations for
`$\F(\x,\tau)$' and `$\F^{\pm}(\x,\sigma)$', respectively; and 
`$\bar{\nu}(\tau)$', `$\nu^{\pm}(\sigma)$' and `$\bar{\I}$' will 
generally stand for `$\bar{\bnu}(\x,\x_{0},\tau)$',
`$\bnu^{\pm}(\x,\x_{0},\sigma)$' and `$\bar{\I}(\x)$', respectively.

\begin{abc}
\begin{gssm}[Alekseev preliminaries]
\label{5.4A} \mbox{ } \\ \vspace{-3ex}
\begin{romanlist}
\item
Suppose that the solution $\bar{\F}(\x)$ of the HHP corresponding to
$(\bv,\bar{\F}_{0},\x)$ exists.  Then, for each $\tau \in C - \bar{\I}(\x)$,
\begin{eqnarray}
\frac{{\F}^{+}(\sigma') - {\F}^{-}(\sigma')}{\sigma' - \tau}
\text{ and }
[\nu^{+}(\sigma')]^{-1} \frac{{\F}^{+}(\sigma') + {\F}^{-}(\sigma')}
{\sigma' - \tau} \nonumber \\
\text{are summable over $\sigma' \in \bar{\I}(\x)$, with assigned} 
\nonumber \\
\text{orientation in the direction of increasing $\sigma'$},
\label{5.13a}
\end{eqnarray}
and
\begin{equation}
\bar{\F}(\tau) = I + \frac{1}{2\pi i} \int_{\bar{\I}} d\sigma'
\frac{{\F}^{+}(\sigma') - {\F}^{-}(\sigma')}{\sigma'-\tau},
\label{5.13b}
\end{equation}
while
\begin{equation}
\left[\bar{\nu}(\tau)\right]^{-1} \bar{\F}(\tau) = 
I + \frac{1}{2\pi i} \int_{\bar{\I}} d\sigma'
[\nu^{+}(\sigma')]^{-1} \frac{{\F}^{+}(\sigma') + {\F}^{-}(\sigma')}
{\sigma'-\tau}. 
\label{5.13c}
\end{equation}
\item
Moreover, for each $\sigma \in \I(\x)$, 
\begin{eqnarray}
\frac{{\F}^{+}(\sigma') - {\F}^{-}(\sigma')}{\sigma' - \sigma}
\text{ and }
[\nu^{+}(\sigma')]^{-1} \frac{{\F}^{+}(\sigma') + {\F}^{-}(\sigma')}
{\sigma' - \sigma} \nonumber \\
\text{are summable over $\sigma' \in \bar{\I}(\x)$ in the} \nonumber \\
\text{principal value (PV) sense,}
\label{5.14a}
\end{eqnarray}
and
\begin{equation}
\frac{1}{2} \left\{ {\F}^{+}(\sigma) + {\F}^{-}(\sigma) \right\} = 
I + \frac{1}{2\pi i} \int_{\bar{\I}} d\sigma' 
\frac{{\F}^{+}(\sigma') - {\F}^{-}(\sigma')}{\sigma' - \sigma} , 
\label{5.14b}
\end{equation}
while
\begin{equation}
\frac{1}{2} [\nu^{+}(\sigma)]^{-1} \left\{ {\F}^{+}(\sigma) - {\F}^{-}(\sigma)
\right\} = I + \frac{1}{2\pi i} \int_{\bar{\I}} d\sigma' 
[\nu^{+}(\sigma')]^{-1} \frac{{\F}^{+}(\sigma') + {\F}^{-}(\sigma')}
{\sigma' - \sigma} . 
\label{5.14c}
\end{equation}
\end{romanlist}
\end{gssm}
\end{abc}

\proofs
\begin{romanlist}
\begin{abc}
\item 
Statement (\ref{5.13a}) follows from Prop.~\ref{5.1A}(iv) and the fact
that $(\sigma'-\tau)^{-1}$ is a continuous function of $\sigma'$
throughout $\bar{\I}(\x)$.

Next, select any piecewise smooth, simple positively oriented contours
$\Lambda^{(3)}$ and $\Lambda^{(4)}$ such that (using the plus sign to
designate a union of arcs)
\begin{eqnarray}
\Lambda & := & \Lambda^{(3)} + \Lambda^{(4)} \subset C - \{\infty\},
\label{5.15a} \\
\bar{\I}^{(i)}(\x) & \subset & \Lambda^{(i)}_{int} := \text{that
bounded open subset} \nonumber \\
& & \text{of $C$ whose boundary is $\Lambda^{(i)}$,}
\label{5.15b} \\
\Lambda^{(3)} \cap \Lambda^{(4)} & = & \emptyset, \text{ and }
\label{5.15c} \\
\tau & \in & \Lambda_{ext} := \text{that open subset of $C$ whose}
\nonumber \\
& & \text{boundary is $\Lambda$ and which has $\infty$ as a member.}
\label{5.15d}
\end{eqnarray}
{}From Cauchy's integral formula and the HHP condition $\bar{\F}(\infty)
= I$,
\begin{equation}
\bar{\F}(\tau) = I - \frac{1}{2\pi i} \int_{\Lambda} d\tau'
\frac{\bar{\F}(\tau')}{\tau'-\tau}.
\label{5.15e}
\end{equation}
For the purpose of this proof, it is convenient to let $\Lambda^{(i)}$
be any member of a specific family of rectangular contours.  Recall the
definitions of $a^{i}$ and $b^{i}$ by Eqs.\ (\ref{5.2}) (with $\x'$
set equal to $\x_{0}$).  Since $\bar{\I}(\x)$ is a bounded closed
subset of $C$ and since $\tau \notin \bar{\I}(\x)$,
\begin{equation}
l(\x,\tau) := \inf{\left[
\left\{\frac{1}{4}|\tau-\sigma|: \sigma \in \bar{\I}(\x)\right\} \cup
\left\{\frac{1}{4}|b-c|: (b,c) \in \{r_{0},s_{0},r,s\}^{2} \text{ and }
b \ne c\right\}
\right]}
\end{equation}
is finite and positive.  In the rest of the proof, we let $\Lambda^{(3)}$
and $\Lambda^{(4)}$ be positively oriented contours which are rectangular
and have the following vertices for each $i \in \{3,4\}$:
\begin{equation}
a^{i}-\alpha\pm i\sqrt{2}\alpha, \quad
b^{i}+\alpha\pm i\sqrt{2}\alpha,
\end{equation}
where $\alpha$ is any real number in the range
\begin{equation}
0 < \alpha \le l(\x,\sigma).
\label{5.16c}
\end{equation}
It is easy to show that this choice of $\Lambda^{(3)}$ and $\Lambda^{(4)}$
satisfies the requisite conditions (\ref{5.15a}) to (\ref{5.15d}). 
Consider the subarcs 
\begin{equation}
\Lambda^{(i)\pm} := \{\tau' \in \Lambda^{(i)}: \tau' \in \bar{C}^{\pm}\}
\label{5.16d}
\end{equation}
of $\Lambda^{(i)}$, and let
\begin{equation}
\Lambda^{\pm} := \Lambda^{(3)\pm} + \Lambda^{(4)\pm}.
\label{5.16e}
\end{equation}
Then Eq.\ (\ref{5.15e}) is expressible in the alternative form
\begin{equation}
\bar{\F}(\tau) = I - \frac{1}{2\pi i} \int_{\Lambda^{+}} d\tau'
\frac{\F^{+}(\tau')}{\tau'-\tau} - \frac{1}{2\pi i} \int_{\Lambda^{-}}
d\tau' \frac{\F^{-}(\tau')}{\tau'-\tau}.
\label{5.16f}
\end{equation}
We now recall that well known generalization\footnote{See Remark 2 in Sec.~2,
Ch.~II, of {\em Analytic Functions} by M.\ A.\ Evgrafov (Dover Publications,
1978).} 
of Cauchy's integral theorem which asserts that the integral of a function 
about a simple piecewise smooth contour $\K$ is zero if the given function
is holomorphic throughout $\K_{int}$ and is continuous throughout 
$\K \cup \K_{int}$.  Upon applying this generalization of Cauchy's 
integral theorem to each of the simple positively oriented contours
$$
\Lambda^{(i)+} - \C^{+}(a^{i},\alpha) + [a^{i}+\alpha,b^{i}-\alpha]
- \C^{+}(b^{i},\alpha)
$$
and
$$
\Lambda^{(i)-} - \C^{-}(a^{i},\alpha) + [a^{i}+\alpha,b^{i}-\alpha]
- \C^{-}(a^{i},\alpha),
$$
where $\C^{\pm}(\sigma,\alpha)$ is defined by Eq.\ (\ref{5.7c}), and,
upon using Prop.~\ref{5.1A}(ii) and the statement (\ref{5.8b}) in
Lem.~\ref{5.2A}, one obtains
\begin{eqnarray}
\int_{\Lambda^{(i)\pm}} d\tau' \frac{\F^{\pm}(\tau')}{\tau'-\tau} - 
\int_{\C^{\pm}(a^{i},\alpha)} d\tau' \frac{\F^{\pm}(\tau')}{\tau'-\tau} 
& & \nonumber \\
\pm \int_{a^{i}+\alpha}^{b^{i}-\alpha} d\sigma'
\frac{\F^{\pm}(\sigma')}{\sigma'-\tau} -
\int_{\C^{\pm}(b^{i},\alpha)} d\tau' \frac{\F^{\pm}(\tau')}{\tau'-\tau} 
& = & 0.
\label{5.17a}
\end{eqnarray}
Substitution from the above Eq.\ (\ref{5.17a}) into Eq.\ (\ref{5.16f})
then yields, after commuting and collecting terms,
\begin{eqnarray}
\bar{\F}(\tau) & = & I + \frac{1}{2\pi i} \int_{a^{3}+\alpha}^{b^{3}-\alpha}
d\sigma' \frac{\F^{+}(\sigma') - \F^{-}(\sigma')}{\sigma'-\tau} \nonumber \\
& & \mbox{ } + \frac{1}{2\pi i} \int_{a^{4}+\alpha}^{b^{4}-\alpha}
d\sigma' \frac{\F^{+}(\sigma') - \F^{-}(\sigma')}{\sigma'-\tau} \nonumber \\
& & \mbox{ } - \frac{1}{2\pi i} \int_{\C^{+}(\ba,\bb,\alpha)}
\frac{\F^{+}(\tau')}{\tau'-\tau}
- \frac{1}{2\pi i} \int_{\C^{-}(\ba,\bb,\alpha)}
\frac{\F^{-}(\tau')}{\tau'-\tau},
\label{5.17b}
\end{eqnarray}
where
\begin{equation}
\C^{\pm}(\ba,\bb,\alpha) := \C^{\pm}(a^{3},\alpha) + \C^{\pm}(b^{3},\alpha)
+ \C^{\pm}(a^{4},\alpha) + \C^{\pm}(b^{4},\alpha).
\label{5.17c}
\end{equation}
The above Eqs.\ (\ref{5.17b}) and (\ref{5.17c}) hold for all real $\alpha$
in the interval (\ref{5.16c}).  From a well known theorem\footnote{See
Cor.~27.7 in Ref.~\ref{McShane}.} on Lebesgue integrals,
\begin{equation}
\int_{a^{i}+\alpha}^{b^{i}-\alpha} d\sigma'
\frac{\F^{\pm}(\sigma')}{\sigma'-\tau} \rightarrow
\int_{a^{i}}^{b^{i}} d\sigma'
\frac{\F^{\pm}(\sigma')}{\sigma'-\tau} \rightarrow
\text{ as } \alpha \rightarrow 0.
\label{5.17d}
\end{equation}
Upon applying statement (\ref{5.8e}) in Lem.~\ref{5.2A} and the above
statement (\ref{5.17d}) to Eq.\ (\ref{5.17b}), one obtains the conclusion
(\ref{5.13b}).

The conclusion (\ref{5.13c}) is proved in the same way with the aid of
statement (\ref{5.8f}).  All that one has to do is to replace
`$\bar{\F}(\tau)$', `$\F^{\pm}(\tau')$' and `$\F^{\pm}(\sigma')$'
in the above proof of Eq.\ (\ref{5.13b}) by `$\bar{\nu}(\tau)^{-1}
\bar{\F}(\tau)$', `$[\nu^{\pm}(\tau')]^{-1} \F^{\pm}(\tau')$' and
`$[\nu^{\pm}(\sigma')]^{-1} \F^{\pm}(\sigma')$', respectively, and
then use the fact that $\nu^{-}(\sigma') = - \nu^{+}(\sigma')$
for all $\sigma' \in \I(\x)$.
\cheers
\end{abc}

\begin{abc}
\item 
We select $\Lambda^{(3)}$ and $\Lambda^{(4)}$ as before so that they
satisfy the conditions (\ref{5.15a}) to (\ref{5.15c}); but, this time,
in place of the condition (\ref{5.15d}), 
\begin{equation}
\sigma \in \Lambda_{int} := \Lambda^{(3)}_{int} + \Lambda^{(4)}_{int}.
\end{equation}
So, in place of Eq.\ (\ref{5.15e}), we have
\begin{equation}
0 = I - \frac{1}{2\pi i} \int_{\Lambda} d\tau'
\frac{\bar{\F}(\tau')}{\tau'-\sigma}.
\label{5.18b}
\end{equation}
In the rest of this proof, we let $\Lambda^{(3)}$ and $\Lambda^{(4)}$
be positively oriented contours which are rectangular and which have
the following vertices for each $i \in \{3,4\}$:
\begin{equation}
a^{i}-\alpha\pm i\sqrt{2}\beta, \quad b^{i}+\alpha\pm i\sqrt{2}\beta,
\end{equation}
where $\alpha$ and $\beta$ are any real numbers such that
\begin{equation}
0 < \alpha < \beta \le m(\x,\sigma)
\label{5.18d}
\end{equation}
and
\begin{equation}
m(\x,\sigma) := \inf{\left\{
\frac{1}{4}|b-c|: (b,c) \in \{r_{0},s_{0},r,s,\sigma\}^{2}
\text{ and } b \ne c \right\}}.
\end{equation}
The subarcs $\Lambda^{(i)\pm}$ and $\Lambda^{\pm}$ are defined as
before by Eqs.\ (\ref{5.16d}) and (\ref{5.16e}), whereupon Eq.\ (\ref{5.18b})
is expressible in the form 
\begin{equation}
0 = I - \frac{1}{2\pi i} \int_{\Lambda^{+}} d\tau'
\frac{\F^{+}(\tau')}{\tau'-\sigma}
- \frac{1}{2\pi i} \int_{\Lambda^{-}} d\tau'
\frac{\F^{-}(\tau')}{\tau'-\sigma}.
\label{5.18f}
\end{equation}

In the remainder of this proof, we shall assume that $\sigma \in 
\I^{(3)}(\x)$.  Upon applying the generalized Cauchy integral theorem
to each of the four simple positively oriented contours
\begin{equation}
\Lambda^{(3)\pm} - \C^{\pm}(a^{3},\alpha) \pm [a^{3}+\alpha,\sigma-\beta]
- \C^{\pm}(\sigma,\beta) \pm [\sigma+\beta,b^{3}-\alpha] -
\C^{\pm}(b^{3},\alpha)
\end{equation}
and
\begin{equation}
\Lambda^{(4)\pm} - \C^{\pm}(a^{4},\alpha) \pm [a^{4}+\alpha,b^{4}-\alpha]
- \C^{\pm}(b^{4},\alpha),
\end{equation}
one obtains, with the aid of Prop.~\ref{5.1A}(ii) and statement (\ref{5.8b}),
\begin{eqnarray}
\int_{\Lambda^{(3)\pm}} d\tau' \frac{\F^{\pm}(\tau')}{\tau'-\sigma} -
\int_{\C^{\pm}(a^{3},\alpha)} d\tau' \frac{\F^{\pm}(\tau')}{\tau'-\sigma} 
& & \nonumber \\ 
\pm \int_{a^{3}+\alpha}^{\sigma-\beta} d\sigma'
\frac{\F^{\pm}(\sigma')}{\sigma'-\sigma} -
\int_{\C^{\pm}(\sigma,\beta)} d\tau'
\frac{\F^{\pm}(\tau')}{\tau'-\sigma}
& & \nonumber \\ 
\pm \int_{\sigma+\beta}^{b^{3}-\alpha} d\sigma'
\frac{\F^{\pm}(\sigma')}{\sigma'-\sigma} -
\int_{\C^{\pm}(b^{3},\alpha)} d\tau'
\frac{\F^{\pm}(\tau')}{\tau'-\sigma} & = & 0 
\label{5.19a}
\end{eqnarray}
and
\begin{eqnarray}
\int_{\Lambda^{(4)\pm}} d\tau' \frac{\F^{\pm}(\tau')}{\tau'-\sigma} -
\int_{\C^{\pm}(a^{4},\alpha)} d\tau' \frac{\F^{\pm}(\tau')}{\tau'-\sigma} 
& & \nonumber \\
\pm \int_{a^{4}+\alpha}^{b^{4}-\alpha} d\sigma'
\frac{\F^{\pm}(\sigma')}{\sigma'-\sigma} -
\int_{\C^{\pm}(b^{4},\alpha)} d\tau' 
\frac{\F^{\pm}(\tau')}{\tau'-\sigma} & = & 0. 
\label{5.19b}
\end{eqnarray}
Substitution from the above Eqs.\ (\ref{5.19a}) and (\ref{5.19b}) into
Eq.\ (\ref{5.18f}) yields, after collecting terms and after transforming
the integrals over $\C^{\pm}(\sigma,\beta)$ to the left side of the
equation,
\begin{eqnarray}
\lefteqn{\frac{1}{2\pi i} \int_{\C^{+}(\sigma,\beta)} d\tau'
\frac{\F^{+}(\tau')}{\tau'-\sigma} +
\frac{1}{2\pi i} \int_{\C^{-}(\sigma,\beta)} d\tau'
\frac{\F^{-}(\tau')}{\tau'-\sigma}} \nonumber \\
& = & I + \frac{1}{2\pi i} \int_{a^{3}+\alpha}^{\sigma-\beta} d\sigma'
\frac{\F^{+}(\sigma') - \F^{-}(\sigma')}{\sigma'-\sigma} \nonumber \\
& & \mbox{ } + \frac{1}{2\pi i} \int_{\sigma+\beta}^{b^{3}-\alpha} d\sigma'
\frac{\F^{+}(\sigma') - \F^{-}(\sigma')}{\sigma'-\sigma} \nonumber \\
& & \mbox{ } + \frac{1}{2\pi i} \int_{a^{4}+\alpha}^{b^{4}-\alpha} d\sigma'
\frac{\F^{+}(\sigma') - \F^{-}(\sigma')}{\sigma'-\sigma} \nonumber \\
& & \mbox{ } - \frac{1}{2\pi i} \int_{\C^{+}(\ba,\bb,\alpha)} d\tau'
\frac{\F^{+}(\tau')}{\tau'-\sigma} - 
\frac{1}{2\pi i} \int_{\C^{-}(\ba,\bb,\alpha)} d\tau'
\frac{\F^{-}(\tau')}{\tau'-\sigma},
\label{5.19c}
\end{eqnarray}
where $\C^{\pm}(\ba,\bb,\alpha)$ is defined by Eq.\ (\ref{5.17c}).
The above Eq.\ (\ref{5.19c}) holds for all $\alpha$ and $\beta$ such
that $0 < \alpha < \beta \le m(\x,\sigma)$ [Eq.\ (\ref{5.18d})].
Keeping $\beta$ fixed, we let $\alpha \rightarrow 0$ in the above
Eq.\ (\ref{5.19c}), whereupon statement (\ref{5.8e}) in Lem.~\ref{5.2A}
yields
\begin{eqnarray}
\lefteqn{\frac{1}{2\pi i} \int_{C^{+}(\sigma,\beta)} d\tau'
\frac{\F^{+}(\tau')}{\tau'-\sigma} +
\frac{1}{2\pi i} \int_{C^{-}(\sigma,\beta)} d\tau'
\frac{\F^{-}(\tau')}{\tau'-\sigma}} \nonumber \\
& = & I + \frac{1}{2\pi i} \int_{a^{3}}^{\sigma-\beta} d\sigma'
\frac{\F^{+}(\sigma') - \F^{-}(\sigma')}{\sigma'-\sigma} 
\nonumber \\
& & \mbox{ } + \frac{1}{2\pi i} \int_{\sigma+\beta}^{b^{3}} d\sigma'
\frac{\F^{+}(\sigma') - \F^{-}(\sigma')}{\sigma'-\sigma}
\nonumber \\
& & \mbox{ } + \frac{1}{2\pi i} \int_{a^{4}}^{b^{4}} d\sigma'
\frac{\F^{+}(\sigma') - \F^{-}(\sigma')}{\sigma'-\sigma}.
\label{5.19d}
\end{eqnarray}
[To obtain the above result, one uses the same theorem on Lebesgue
integrals that provided us with Eq.\ (\ref{5.17d}).]

The next step in the proof is to employ statement (\ref{5.11d}) of
Lem.~\ref{5.3A}.  This statement tells us that the left side of
Eq.\ (\ref{5.19d}) converges as $\beta \rightarrow 0$, whereupon
Eq.\ (\ref{5.19d}) tells us that
\begin{eqnarray}
\lefteqn{\frac{1}{2\pi i} \int_{a^{3}}^{b^{3}} d\sigma'
\frac{\F^{+}(\sigma') - \F^{-}(\sigma')}{\sigma'-\sigma}} \\
& := & \lim_{\beta \rightarrow 0} \left[
\frac{1}{2\pi i} \int_{a^{3}}^{\sigma-\beta} d\sigma'
\frac{\F^{+}(\sigma') - \F^{-}(\sigma')}{\sigma'-\sigma} +
\frac{1}{2\pi i} \int_{\sigma+\beta}^{b^{3}} d\sigma'
\frac{\F^{+}(\sigma') - \F^{-}(\sigma')}{\sigma'-\sigma} \right]
\nonumber
\end{eqnarray}
exists; and statement (\ref{5.11d}) and Eq.\ (\ref{5.19d}) then yield
the final conclusion that, when $\sigma \in \I^{(3)}(\x)$,
$[\F^{+}(\sigma') - \F^{-}(\sigma)]/(\sigma'-\sigma)$
is summable over $\sigma' \in \bar{\I}(\x)$ in the PV sense; and
Eq.\ (\ref{5.14b}) holds.  The proof for the case $\sigma \in \I^{(4)}(\x)$
is similar; and we have already provided the substitutions that are to
be made in the wording of the above proof in order to obtain the proof
of Eq.\ (\ref{5.14c}).
\cheers
\end{abc}
\end{romanlist}

\setcounter{equation}{0}
\setcounter{theorem}{0}
\subsection{Derivation of two Alekseev-type singular integral equations}

Proceeding from either equations (\ref{5.13b}) and (\ref{5.14b}) or from
equations (\ref{5.13c}) and (\ref{5.14c}), one can construct a singular
integral equation of the Alekseev type and, if $\bv \in K^{1+}$, a Fredholm
equation of the second kind. 

We begin by observing that Eq.\ (\ref{G2.10c}) implies that, for each
$\sigma \in \I(\x) \cup \{r_{0},s_{0}\}$, 
\begin{abc}
\begin{eqnarray}
\frac{1}{2} \left\{ {\F}^{M+}(\sigma) + {\F}^{M-}(\sigma) \right\} & = &
\left( \begin{array}{cc}
1 & -i(\sigma-z) \\ 0 & 1
\end{array} \right) \left( \begin{array}{cc}
1 & 0 \\ 0 & 0
\end{array} \right) \left( \begin{array}{cc}
1 & i(\sigma-z_{0}) \\ 0 & 1
\end{array} \right),
\end{eqnarray}
and
\begin{eqnarray}
\frac{1}{2} [\nu^{+}(\sigma)]^{-1}
\lefteqn{
\left\{ {\F}^{M+}(\sigma) - {\F}^{M-}(\sigma) \right\} = } 
\nonumber \\ & &
\left( \begin{array}{cc}
1 & -i(\sigma-z) \\ 0 & 1
\end{array} \right) \left( \begin{array}{cc}
0 & 0 \\ 0 & 1
\end{array} \right) \left( \begin{array}{cc}
1 & i(\sigma-z_{0}) \\ 0 & 1
\end{array} \right).
\end{eqnarray}
If $\bar{\F}$ is a solution of the HHP corresponding to $(\bv,\bar{\F}^{M})$,
Eq.\ (\ref{G3.17}) tells us that, for any $\sigma \in \I(\x)$,
\end{abc}
\begin{equation}
{\F}^{\pm}(\sigma) \tilde{v}^{(i)}(\sigma) = Y^{(i)}(\sigma) {\F}^{M\pm}(\sigma),
\end{equation}
and, therefore,
\begin{abc}
\begin{eqnarray}
\lefteqn{\frac{1}{2} \left\{ {\F}^{+}(\sigma) 
+ {\F}^{-}(\sigma) \right\} \tilde{v}^{(i)}(\sigma) = } \nonumber \\
& & Y^{(i)}(\sigma) \left( \begin{array}{cc}
1 & -i(\sigma-z) \\ 0 & 1
\end{array} \right) \left( \begin{array}{cc}
1 & 0 \\ 0 & 0
\end{array} \right) \left( \begin{array}{cc}
1 & i(\sigma-z_{0}) \\ 0 & 1
\end{array} \right), 
\end{eqnarray}
and
\begin{eqnarray}
\lefteqn{
\frac{1}{2} [\nu^{+}(\sigma)]^{-1}
\left\{ {\F}^{+}(\sigma) - {\F}^{-}(\sigma) \right\} \tilde{v}^{(i)}(\sigma) = }
\nonumber \\ & &
Y^{(i)}(\sigma) \left( \begin{array}{cc}
1 & -i(\sigma-z) \\ 0 & 1
\end{array} \right) \left( \begin{array}{cc}
0 & 0 \\ 0 & 1
\end{array} \right) \left( \begin{array}{cc}
1 & i(\sigma-z_{0}) \\ 0 & 1
\end{array} \right).
\end{eqnarray}
This motivates the introduction two new $2 \times 2$ matrices.
\end{abc}

\begin{abc}
\begin{definition}{Dfn.\ of functions $W^{(i)}(\x)$ and $\Y^{(i)}(\x)$
\label{def62}}
For each $\bv \in K$, we let $W^{(i)}(\x)$ denote the function whose domain
is $\I^{(i)}$ and whose value for each $\sigma \in \I^{(i)}$ is 
\begin{eqnarray}
W^{(i)}(\x,\sigma) & := & W^{(i)}(\x)(\sigma) := 
\tilde{v}^{(i)}(\sigma) \left( \begin{array}{cc}
1 & -i(\sigma-z_{0}) \\ 0 & 1
\end{array} \right),
\label{5.22a}
\end{eqnarray}
and, for each solution $\bar{\F}(\x)$ of the HHP corresponding to
$(\bv,\bar{\F}^{M},\x)$, we let $\Y^{(i)}(\x)$ denote the function whose
domain is $\I^{(i)}(\x)$ and whose value for each $\sigma \in \I^{(i)}(\x)$
is
\begin{eqnarray}
\Y^{(i)}(\x,\sigma) & := & \Y^{(i)}(\x)(\sigma) := Y^{(i)}(\x,\sigma) 
\left( \begin{array}{cc}
1 & -i(\sigma-z) \\ 0 & 1
\end{array} \right). 
\label{5.22b}
\end{eqnarray}
\end{definition}
\end{abc}

In terms of these matrices we may write [suppressing `$\x$']
\begin{equation}
{\F}^{\pm}(\sigma) W^{(i)}(\sigma) = \Y^{(i)}(\sigma) \left( \begin{array}{cc}
1 & 0 \\ 0 & \nu^{\pm}(\sigma)
\end{array} \right) 
\label{G4.8}
\end{equation}
as well as
\begin{abc}
\begin{eqnarray}
\frac{1}{2} \left\{ {\F}^{+}(\sigma) + {\F}^{-}(\sigma) \right\} W^{(i)}(\sigma) 
& = & \Y^{(i)}(\sigma) \left( \begin{array}{cc}
1 & 0 \\ 0 & 0 
\end{array} \right), 
\label{5.25a}
\end{eqnarray}
and
\begin{eqnarray}
\frac{1}{2} [\nu^{+}(\sigma)]^{-1}
\left\{ {\F}^{+}(\sigma) - {\F}^{-}(\sigma) \right\} W^{(i)}(\sigma) & = &
\Y^{(i)}(\sigma) \left( \begin{array}{cc}
0 & 0 \\ 0 & 1
\end{array} \right). 
\label{5.25b}
\end{eqnarray}
\end{abc}

We shall here consider the HHP corresponding to $(\bv,\bar{\F}^{M},\x)$,
where $\bv \in K$ and $\x \in \domE$.  Let us recall Eq.\ (\ref{G3.17}),
which is the third condition in the definition of the HHP.  This equation
is as follows when $\bar{\F}_{0} = \bar{\F}^{M}$:
\begin{abc}
\begin{eqnarray}
Y^{(i)}(\x,\sigma) & := & \F^{+}(\x,\sigma) \tilde{v}^{(i)}(\sigma)
[\F^{M+}(\x,\sigma)]^{-1} \nonumber \\
& = & \F^{-}(\x,\sigma) \tilde{v}^{(i)}(\sigma)
[\F^{M-}(\x,\sigma)]^{-1} \nonumber \\
& & \text{for each $i \in \{3,4\}$ and } \sigma \in \I^{(i)}(\x).
\end{eqnarray}
Condition (4) in the definition of the HHP stipulated that the function
$Y(\x)$ whose domain is $\I(\x)$ and whose values are given by
\begin{equation}
Y(\x,\sigma) := Y(\x)(\sigma) := Y^{(i)}(\x,\sigma)
\end{equation}
satisfies
\begin{equation}
Y(\x) \text{ is bounded at } \x_{0} \text{ and at } \x.
\end{equation}
Moreover, from GSSM~\ref{4.3D}(i) and~(iii), 
\begin{equation}
Y(\x) \text{ is continuous throughout } \I(\x)
\end{equation}
and
\begin{equation}
\det{Y(\x)} = 1.
\end{equation}
\end{abc}

\begin{abc}
\begin{definition}{Dfns.\ of $W(\x)$, $\Y(\x)$, $W_{a}(\x)$ and $\Y_{a}(\x)$
\label{def63}}
Let $W(\x)$ and $\Y(\x)$ denote the functions\footnote{We shall frequently
suppress $\x$.} with domain $\I(\x)$ and values
\begin{equation}
\begin{array}{l}
W(\x,\sigma) := W(\x)(\sigma) := W^{(i)}(\x,\sigma) \text{ and } \\
\Y(\x,\sigma) := \Y(\x)(\sigma) := \Y^{(i)}(\x,\sigma) \\
\text{for each $i \in \{3,4\}$ and $\sigma \in \I^{(i)}(\x)$.}
\end{array}
\label{5.22c}
\end{equation}
Moreover, let
\begin{equation}
\begin{array}{l}
W_{a}(\x,\sigma) := a^{th} \text{ column of } W(\x,\sigma) \text{ and } \\ 
\Y_{a}(\x,\sigma) := a^{th} \text{ column of } \Y(\x,\sigma), \text{ where }
a \in \{1,2\}.
\end{array}
\end{equation}
\end{definition}
\end{abc}

\begin{theorem}[Alekseev-type equation]
\label{5.1B} \mbox{ } \\
For each $\bv \in K$, $\x \in \domE$, solution $\bar{\F}(\x)$ of the HHP
corresponding to $(\bv,\bar{\F}^{M},\x)$, $\tau \in C - \bar{\I}(\x)$
and $\sigma \in \I(\x)$, both the following statements (i) and (ii) hold:
\begin{romanlist}
\begin{abc}
\item 
\begin{equation}
\nu^{+}(\sigma') \Y_{2}(\sigma') W_{1}^{T}(\sigma') (\sigma'-\tau)^{-1} 
\text{ is summable over } \sigma' \in \bar{\I}(\x),
\label{5.23a}
\end{equation}
\begin{equation}
\bar{\F}(\tau) = I + \frac{1}{\pi i} \int_{\bar{\I}} d\sigma'
\nu^{+}(\sigma') \Y_{2}(\sigma') 
\frac{W_{1}^{T}(\sigma') J}{\sigma'-\tau},
\label{5.23b}
\end{equation}
\begin{equation}
\begin{array}{r}
\nu^{+}(\sigma') \Y_{2}(\sigma') W_{1}^{T}(\sigma') (\sigma'-\sigma)^{-1}
\text{ is summable over } \sigma' \in \bar{\I}(\x) \\
\text{in the PV sense,}
\end{array}
\label{5.23c}
\end{equation}
\begin{eqnarray}
\Y_{1}(\sigma) & = & W_{1}(\sigma) + \frac{1}{\pi i} \int_{\bar{\I}}
d\sigma' \nu^{+}(\sigma') Y_{2}(\sigma') 
\frac{W_{1}^{T}(\sigma') J W_{1}(\sigma)}{\sigma'-\sigma}, 
\label{5.23d} \\
\text{and} & & \nonumber \\
0 & = & W_{2}(\sigma) + \frac{1}{\pi i} \int_{\bar{\I}} d\sigma' 
\nu^{+}(\sigma') \Y_{2}(\sigma') 
\frac{W_{1}^{T}(\sigma') J W_{2}(\sigma)}{\sigma'-\sigma}.
\label{5.23e}
\end{eqnarray}
\end{abc}
\begin{abc}
\item 
\begin{equation}
[\nu^{+}(\sigma')]^{-1} \Y_{1}(\sigma') W_{2}^{T}(\sigma') (\sigma'-\tau)^{-1} 
\text{ is summable over } \sigma' \in \bar{\I}(\x),
\label{5.24a}
\end{equation}
\begin{equation}
\bar{\nu}(\tau)^{-1}\bar{\F}(\tau) = I + \frac{1}{\pi i} \int_{\bar{\I}} d\sigma'
[\nu^{+}(\sigma')]^{-1} \Y_{1}(\sigma') 
\frac{W_{2}^{T}(\sigma') J}{\sigma'-\tau},
\label{5.24b}
\end{equation}
\begin{equation}
\begin{array}{r}
[\nu^{+}(\sigma')]^{-1} \Y_{1}(\sigma') W_{2}^{T}(\sigma') (\sigma'-\sigma)^{-1}
\text{ is summable over } \sigma' \in \bar{\I}(\x) \\
\text{in the PV sense,}
\end{array}
\label{5.24c}
\end{equation}
\begin{eqnarray}
\Y_{2}(\sigma) & = & W_{2}(\sigma) - \frac{1}{\pi i} \int_{\bar{\I}}
d\sigma' [\nu^{+}(\sigma')]^{-1} Y_{1}(\sigma') 
\frac{W_{2}^{T}(\sigma') J W_{2}(\sigma)}{\sigma'-\sigma},
\label{5.24d} \\
\text{and} & & \nonumber \\
0 & = & W_{1}(\sigma) + \frac{1}{\pi i} \int_{\bar{\I}} d\sigma' 
[\nu^{+}(\sigma')]^{-1} \Y_{1}(\sigma') 
\frac{W_{2}^{T}(\sigma') J W_{1}(\sigma)}{\sigma'-\sigma}.
\label{5.24e}
\end{eqnarray}
\end{abc}
\end{romanlist}
\end{theorem}

\proof
The statements (\ref{5.23a}) to (\ref{5.23c}) and (\ref{5.24a}) to
(\ref{5.24c}) are obtained by using Eqs.\ (\ref{5.25a}) and (\ref{5.25b})
together with the relation
$$
W(\sigma)^{-1} = - J W^{T}(\sigma) J
$$
to replace $\F^{+}(\sigma') - \F^{-}(\sigma')$ and $[\nu^{+}(\sigma')]^{-1}
[\F^{+}(\sigma') + \F^{-}(\sigma)]$ in statements (\ref{5.13a}) to
(\ref{5.14a}) of Thm.~\ref{5.4A}.  The same replacements are to be made in
the integrands on the right sides of Eqs.\ (\ref{5.14b}) and (\ref{5.14c})
in Thm.~\ref{5.4A}.  Equations (\ref{5.23d}) is then obtained by multiplying
both sides of Eq.\ (\ref{5.14b}) by $W_{1}(\sigma)$ and replacing the
product on the left side of the equation with the first column of
(\ref{5.25a}).  Equation (\ref{5.23e}) is obtained by multiplying both
sides of Eq.\ (\ref{5.15b}) by $W_{2}(\sigma)$ and replacing the product
on the left side of the equation with the second column of (\ref{5.25a}).
Equation (\ref{5.24d}) is obtained by multiplying both sides of Eq.\
(\ref{5.14c}) by $W_{2}(\sigma)$ and replacing the product on the left
side with the second column of (\ref{5.25b}) multiplied by
$[\nu^{+}(\sigma)]^{-1}$.  Equation (\ref{5.24e}) is obtained by
multiplying both sides of Eq.\ (\ref{5.14c}) by $W_{1}(\sigma)$ and
replacing the product on the left side with the first column of (\ref{5.25b}).
\cheers

Equation (\ref{5.24e}) has the form of the singular integral equation
which Alekseev obtained in the analytic case; and Eq.\ (\ref{5.23e}) is
an alternative singular integral equation of the same type.

\setcounter{equation}{0}
\setcounter{theorem}{0}
\subsection{Extension of the function $\Y(\x)$ from $\I(\x)$ to $\bar{\I}(\x)$}

Since $C^{T} J C = 0$ (the zero matrix) for any $2 \times 2$ matrix $C$,
Eqs.\ (\ref{5.23d}) and (\ref{5.24d}) are expressible in the following 
forms for each $i \in \{3,4\}$:
\begin{abc}
\begin{eqnarray}
\lefteqn{\Y_{1}^{(i)}(\sigma) = W_{1}^{(i)}(\sigma)} \nonumber \\
& & \mbox{ } + \frac{1}{\pi i} \int_{a^{i}}^{b^{i}} d\sigma'
\nu^{+}(\sigma') \Y_{2}^{(i)}(\sigma') W_{1}^{(i)}(\sigma')^{T} J
\left[ \frac{W_{1}^{(i)}(\sigma) - W_{1}^{(i)}(\sigma')}{\sigma'-\sigma}
\right] \nonumber \\
& & \mbox{ } + \frac{1}{\pi i} \int_{a^{7-i}}^{b^{7-i}} d\sigma'
\nu^{+}(\sigma') \Y_{2}^{(7-i)}(\sigma') W_{1}^{(7-i)}(\sigma')^{T} J
\left[ \frac{W_{1}^{(i)}(\sigma)}{\sigma'-\sigma} \right],
\label{5.30a} \\
\lefteqn{\Y_{2}^{(i)}(\sigma) = W_{2}^{(i)}(\sigma)} \nonumber \\
& & \mbox{ } - \frac{1}{\pi i} \int_{a^{i}}^{b^{i}} d\sigma'
[\nu^{+}(\sigma')]^{-1} \Y_{1}^{(i)}(\sigma') W_{2}^{(i)}(\sigma')^{T} J
\left[ \frac{W_{2}^{(i)}(\sigma) - W_{2}^{(i)}(\sigma')}{\sigma'-\sigma}
\right] \nonumber \\
& & \mbox{ } - \frac{1}{\pi i} \int_{a^{7-i}}^{b^{7-i}} d\sigma'
[\nu^{+}(\sigma')]^{-1} \Y_{1}^{(7-i)}(\sigma') W_{2}^{(7-i)}(\sigma')^{T} J
\left[ \frac{W_{2}^{(i)}(\sigma)}{\sigma'-\sigma} \right],
\label{5.30b}
\end{eqnarray}
for all $\sigma \in \I^{(i)}(\x)$, where recall that $a^{i} :=
\inf{\{x^{i},x^{i}_{0}\}}$ and $b^{i} := \sup{\{x^{i},x^{i}_{0}\}}$.

Now, from Prop.~\ref{5.1A}(iv), Eq.\ (\ref{5.25a}) and Eq.\ (\ref{5.25b}),
\begin{equation}
\begin{array}{l}
\nu^{+}(\sigma') \Y_{2}(\sigma') W_{1}(\sigma')^{T} J \\
\text{and } [\nu^{+}(\sigma')]^{-1} \Y_{1}(\sigma') W_{2}(\sigma')^{T} J \\
\text{are summable over } \sigma' \in \bar{\I}(\x).
\end{array}
\label{5.30c}
\end{equation}
{}From GSSM~\ref{4.1A} and the definition of $W^{(i)}$ by Eq.\ (\ref{5.22a}),
the following statement holds for each $\x \in \domE$ and $i \in \{3,4\}$:
\begin{equation}
\begin{array}{l}
\text{If $\bv \in K^{1}$, then $W^{(i)}$ is $\bC^{1}$ throughout $\I^{(i)}$,}
\\[0pt]
[W^{(i)}(\sigma') - W^{(i)}(\sigma)](\sigma'-\sigma)^{-1}
\text{ is a continuous } \\
\text{function of } (\sigma',\sigma) \text{ throughout } \I^{(i)} \times \I^{(i)},
\text{ and} \\
W^{(i)}(\sigma)(\sigma'-\sigma)^{-1} \text{ is a $\bC^{1}$ function of }
(\sigma',\sigma) \\
\text{throughout } \bar{\I}^{(7-i)}(\x) \times \check{\I}^{(i)}(x^{7-i}).
\end{array}
\label{5.30d}
\end{equation}
\end{abc}
{}From the above statements (\ref{5.30c}) and (\ref{5.30d}), and from the 
theorem that asserts the summability over a finite interval of the product
of a summable function by a continuous function, the extension of
$\Y^{(i)}(\x)$ that is defined below exists.

\begin{abc}
\begin{definition}{Dfn.\ of $\check{\Y}^{(i)}(\x)$ when $\bv \in K^{1}$
\label{def64}}
For each $\bv \in K^{1}$, $\x \in \domE$, solution $\bar{\F}(\x)$ of the
HHP corresponding to $(\bv,\bar{\F}^{M},\x)$ and $i \in \{3,4\}$, let
$\check{\Y}^{(i)}(\x)$ denote the function whose domain is
$\check{\I}(x^{7-i})$ and whose value for each $\sigma \in
\check{\I}^{(i)}(\x^{7-i})$ is given by [suppressing `$\x$']
\begin{eqnarray}
\check{\Y}_{1}^{(i)}(\sigma) & := & \text{ right side of Eq.\ (\ref{5.30a}),}
\label{5.31a} \\
\check{\Y}_{2}^{(i)}(\sigma) & := & \text{ right side of Eq.\ (\ref{5.30b}).}
\label{5.31b}
\end{eqnarray}
\end{definition}
\end{abc}

Such an extension can also be defined when $\bv \in K^{\Box}$, where
$\Box$ is $n \ge 1$, $n+$ (with $n \ge 1$), $\infty$ or `an'.  The next
theorem and three lemmas associated with that theorem pertain to these
cases.

\begin{lemma}[Continuity and differentiability of $W^{(i)}$]
\label{5.1C}
\mbox{ } \\ 
\begin{romanlist}
\item 
If $\bv \in K^{\Box}$, then $W^{(i)}$ is $\bC^{\Box}$ throughout its domain
$\I^{(i)}$, and the function whose domain is $\bar{\I}^{(7-i)}(\x) \times
\check{\I}^{(i)}(x^{7-i})$ and whose values for each $(\sigma',\sigma)$ in
this domain is $W^{(i)}(\sigma)(\sigma'-\sigma)^{-1}$ is also $\bC^{\Box}$.
\item 
If $\bv \in K^{\Box}$, then the function of $(\sigma',\sigma)$ whose domain
is $\I^{(i)} \times \I^{(i)}$ and whose value for each $(\sigma',\sigma)$ in
this domain is $[W^{(i)}(\sigma)-W^{(i)}(\sigma')]/(\sigma'-\sigma)$ is
$\bC^{n-1}$ if $\Box$ is $n \ge 1$, is $\bC^{(n-1)+}$ if $\Box$ is 
$n^{+}$ ($n \ge 1$), is $\bC^{\infty}$ if $\Box$ is $\infty$, and is
$\bC^{an}$ if $\Box$ is `an'.
\end{romanlist}
\end{lemma}

\begin{abc}
\proofs
\begin{romanlist}
\item 
Use GSSM~\ref{4.1A} and the definition of $W^{(i)}$ by Eq.\ (\ref{5.22a}).
\cheers
\item 
The conclusions when $\Box$ is $n$, $\infty$ or `an' are well known.
As regards the case when $\Box$ is $n^{+}$ ($n \ge 1$), we shall use
the relation
\begin{equation}
\frac{W^{(i)}(\sigma) - W^{(i)}(\sigma')}{\sigma-\sigma'} =
\int_{0}^{1} dt (DW^{(i)})(t\sigma+(1-t)\sigma'),
\label{5.32a}
\end{equation}
where $D^{p}W^{(i)}$ ($1 \le p \le n$) denotes the function whose domain
is $\I^{(i)}$ and whose value for each $\sigma \in \I^{(i)}$ is
\begin{equation}
(D^{p}W^{(i)})(\sigma) :=
\frac{\partial^{p}W^{(i)}(\sigma)}{\partial\sigma^{p}};
\end{equation}
and $DW^{(i)} := D^{1}W^{(i)}$.  From Eq.\ (\ref{5.32a}), one obtains,
for all $0 \le m \le n-1$,
\begin{eqnarray}
U_{m}(\sigma',\sigma) & := & \frac{\partial^{n-1}}{\partial\sigma^{m}
(\partial\sigma')^{n-1-m}} \left[
\frac{W^{(i)}(\sigma) - W^{(i)}(\sigma')}{\sigma'-\sigma}
\right] \nonumber \\
& = & \int_{0}^{1} dt \, t^{m} (1-t)^{n-1-m} (D^{n}W^{(i)})
(t\sigma+(1-t)\sigma') \nonumber \\
& & \text{ for all } (\sigma',\sigma) \in \I^{(i)} \times \I^{(i)}.
\label{5.32c}
\end{eqnarray}
By definition of the class $\bC^{n+}$ of functions of a single real
variable, each $[c,d] \subset \I^{(i)}$ has at least one corresponding
real number $0 < \gamma(c,d) \le 1$ and at least one positive real
number $M(c,d)$ such that
\begin{eqnarray}
\left|\left|(D^{n}W^{(i)})(\sigma_{2})-(D^{n}W^{(i)})(\sigma_{1})\right|\right|
& \le & M(c,d) |\sigma_{2}-\sigma_{1}|^{\gamma(c,d)} \nonumber \\
& & \text{for all } (\sigma_{1},\sigma_{2}) \in [c,d]^{2}.
\label{5.32d}
\end{eqnarray}
For each $[c,d] \subset \I^{(i)}$ and $[c',d'] \subset \I^{(i)}$, let
\begin{equation}
\predot{c} := \inf{\{c,c'\}} \text{ and } \postdot{d} := \sup{\{d,d'\}},
\end{equation}
whereupon Eqs.\ (\ref{5.32c}) and (\ref{5.32d}) yield
\begin{eqnarray}
\lefteqn{\left|\left| U_{m}(\sigma',\sigma_{2}) - U_{m}(\sigma',\sigma_{1})
\right|\right|} \nonumber \\
& \le & \int_{0}^{1} dt \, t^{m+1}(1-t)^{n-m-1} M(\predot{c},\postdot{d})
|\sigma_{2}-\sigma_{1}|^{\gamma(\predot{c},\postdot{d})} \nonumber \\
& \le & M(\predot{c},\postdot{d}) 
|\sigma_{2}-\sigma_{1}|^{\gamma(\predot{c},\postdot{d})}
\text{ for all } 0 \le m \le n-1, \nonumber \\
& & (\sigma_{2},\sigma_{1}) \in [c,d]^{2} \text{ and }
\sigma' \in [c',d'];
\end{eqnarray}
and, likewise,
\begin{eqnarray}
\lefteqn{\left|\left| U_{m}(\sigma'_{2},\sigma) - U_{m}(\sigma'_{1},\sigma)
\right|\right|} \nonumber \\
& \le & M(\predot{c},\postdot{d}) 
|\sigma'_{2}-\sigma'_{1}|^{\gamma(\predot{c},\postdot{d})}
\text{ for all } 0 \le m \le n-1, \nonumber \\
& & (\sigma'_{1},\sigma'_{2}) \in [c',d']^{2} \text{ and }
\sigma \in [c,d].
\end{eqnarray}
Therefore,
\begin{eqnarray}
\lefteqn{\left|\left| U_{m}(\sigma'_{2},\sigma_{2}) - 
U_{m}(\sigma'_{1},\sigma_{1}) \right|\right|} \nonumber \\
& \le & M(\predot{c},\postdot{d}) 
\left\{ |\sigma_{2}-\sigma_{1}|^{\gamma(\predot{c},\postdot{d})} +
|\sigma'_{2}-\sigma'_{1}|^{\gamma(\predot{c},\postdot{d})} \right\}
\nonumber \\
& & \text{ for all } 0 \le m \le n-1, \nonumber \\
& & (\sigma_{1},\sigma_{2}) \in [c,d]^{2} \text{ and }
(\sigma'_{1},\sigma'_{2}) \in [c',d']^{2}.
\end{eqnarray}
We have thus shown that, in addition to the fact that
$[W^{(i)}(\sigma') - W^{(i)}(\sigma)](\sigma'-\sigma)^{-1}$ is
(given the premises of our theorem) $\bC^{n-1}$, each of its
partial derivatives of order $n=1$ satisfies a H\"{o}lder
condition\footnote{For the concept of H\"{o}lder conditions for 
functions of several real variables, see Ch.~I, Sec.~4, p.\ 12, 
of Ref.~\ref{Musk}.} on each closed subinterval of the space
$\I^{(i)} \times \I^{(i)}$.
\cheers
\end{romanlist}
\end{abc}

\begin{abc}
\begin{lemma}[Integral of product]
\label{5.2C} \mbox{ } \\
Suppose $[a,b] \subset R^{1}$, $S$ is a connected open subset of
$R^{m}$ ($m \ge 1$), $f$ is a real-valued function defined almost
everywhere on and summable over $[a,b]$, and $g$ is a real-valued
function whose domain is $[a,b] \times S$ and which is continuous.
Let $\sigma := (\sigma^{1},\ldots,\sigma^{m})$ denote any member
of $S$, and let $F$ denote the function whose domain is $S$ and
whose value at each $\sigma \in S$ is
\begin{equation}
F(\sigma) := \int_{a}^{b} d\sigma' f(\sigma') g(\sigma',\sigma).
\label{5.33a}
\end{equation}
Then the following statements hold:
\begin{romanlist}
\item 
$F$ is continuous.
\item 
If $\partial g(\sigma',\sigma)/\partial\sigma^{k}$ exists for each
$k \in \{1,\ldots,m\}$ and is a continuous function of $(\sigma',\sigma)$
throughout $[a,b] \times S$, then $F$ is $\bC^{1}$ and
\begin{equation}
\frac{\partial F(\sigma)}{\partial\sigma^{k}} = \int_{a}^{b} d\sigma'
f(\sigma') \frac{\partial g(\sigma',\sigma)}{\partial\sigma^{k}}
\label{5.33b}
\end{equation}
for all $k \in \{1,\ldots,m\}$ and $\sigma \in S$.  In particular,
$F$ is $\bC^{1}$ if $g$ is $\bC^{1}$.
\item 
Assume (for simplicity) that $m=1$.  
If there exists a positive integer $p$ such that 
$\partial^{p}g(\sigma',\sigma)/(\partial\sigma)^{p}$ exists and is a
continuous function of $(\sigma',\sigma)$ throughout $[a,b] \times S$,
then $F$ is $bC^{p}$ and
\begin{equation}
\frac{\partial^{p}F(\sigma)}{\partial\sigma^{p}} = \int_{a}^{b} d\sigma'
f(\sigma') \frac{\partial^{p}g(\sigma',\sigma)}{\partial\sigma^{p}}
\end{equation}
for all $\sigma \in S$.  In particular, $F$ is $\bC^{p}$ if $g$ is
$\bC^{p}$.
\item 
Assume (for simplicity) that $m=1$.  If $[c,d] \subset S$ and $g$ obeys
a H\"{o}lder condition on $[a,b] \times [c,d]$, then $F$ obeys a H\"{o}lder
condition on $[c,d]$.
\item 
Assume (for simplicity) that $m=1$.  If $g$ is analytic [i.e., if $g$ has
an analytic extension to an open subset $[a-\epsilon,b+\epsilon] \times S$
of $R^{2}$, then $F$ is analytic.
\end{romanlist}
\end{lemma}
\end{abc}

\proofs
\begin{romanlist}
\item 
For each $\sigma \in S$, let
\begin{abc}
\begin{equation}
|\sigma| := \sup{\{|\sigma^{1}|,\ldots,|\sigma^{m}|\}}.  
\end{equation}
For each 
\begin{equation}
[c,d] := [c^{1},d^{1}] \times \cdots \times [c^{m},d^{m}] \subset S,
\end{equation}
$g(\sigma',\sigma)$ is a uniformly continuous function of $(\sigma',\sigma)$
throughout $[a,b] \times [c,d]$.  Therefore, for each $\epsilon > 0$,
there exists $\delta(\epsilon,c,d) > 0$ such that
\begin{equation}
\begin{array}{r}
|g(\sigma',\sigma) - g(\sigma',\xi)| < \epsilon \text{ whenever }
\sigma' \in [a,b], \\
\sigma \in S, \xi \in S \text{ and } |\sigma-\xi| < \delta(\epsilon,c,d).
\end{array}
\end{equation}
Therefore, from Eq.\ (\ref{5.33a}),
\begin{eqnarray}
|F(\sigma) - F(\xi)| < \epsilon \int_{a}^{b} d\sigma' |f(\sigma')|
\text{ whenever} \nonumber \\
\sigma \in S, \xi \in S \text{ and } |\sigma-\xi| < \delta(\epsilon,c,d).
\end{eqnarray}
Therefore, for each choice of $[c,d] \subset S$, $F(\sigma)$ is a
(uniformly) continuous function of $\sigma$ throughout $[c,d]$.  Therefore,
$F$ is continuous.
\cheers
\end{abc}

\item 
For each $[c,d] \subset S$, there exists a positive real number $M(c,d)$
such that
\begin{eqnarray*}
\left| \frac{\partial g(\sigma',\sigma)}{\partial\sigma^{k}} \right|
& < & M(c,d) \text{ for all } k \in \{1,\ldots,m\} \\
& & \text{ and } (\sigma',\sigma) \in [a,b] \times [c,d].
\end{eqnarray*}
Therefore,
\begin{eqnarray*}
\left| f(\sigma') \frac{\partial g(\sigma',\sigma)}{\partial\sigma^{k}}
\right| & < & |f(\sigma')| M(c,d) \text{ for all } k \in \{1,\ldots,m\} \\
& & \text{ and } (\sigma',\sigma) \in [a,b] \times [c,d].
\end{eqnarray*}
However, (since the norm of a summable function is also summable) the
right side of the above inequality is summable over $[a,b]$.  Therefore,
from a well known theorem,\footnote{See footnote \ref{McShane}.}
$\partial F(\sigma)/\partial\sigma^{k}$ exists and is given by Eq.\
(\ref{5.33b}) for all $k \in \{1,\ldots,m\}$ and $\sigma \in [c,d]$;
and, since this is true for every choice of $[c,d] \subset S$, 
$\partial F(\sigma)/\partial\sigma^{k}$ exists and is given by 
Eq.\ (\ref{5.33b}) for all $k \in \{1,\ldots,m\}$ and $\sigma \in S$.
Moreover, from part (i) of our theorem, these partial derivative are
continuous functions of $\sigma$ throughout $S$.
\cheers

\item 
This follows by induction from the preceding part (ii) of our theorem
(specialized to $m=1$).
\cheers

\item 
By definition of the H\"{o}lder condition for functions of two real variables,
there exist positive real numbers $M(c,d)$, $M'(c,d)$, $\gamma(c,d)$ and
$\gamma'(c,d)$ [where we are suppressing the dependence on $(a,b)$] such
that
\begin{abc}
\begin{equation}
0 < \gamma(c,d) \le 1, \quad 0 < \gamma'(c,d) \le 1,
\end{equation}
and
\begin{eqnarray}
|g(\sigma',\sigma) - g(\sigma',\xi)| & \le & M'(c,d) 
|\sigma'-\xi'|^{\gamma'(c,d)} + M(c,d) |\sigma-\xi|^{\gamma(c,d)} \nonumber \\
& & \text{ for all } (\sigma',\xi') \in [a,b]^{2} \text{ and }
(\sigma,\xi) \in [c,d]^{2}.
\end{eqnarray}
Therefore, from Eq.\ (\ref{5.33a}),
\begin{eqnarray}
|F(\sigma) - F(\xi)| & \le & M(c,d) \int_{a}^{b} d\sigma' |f(\sigma')|
\; |\sigma-\xi|^{\gamma(c,d)} \nonumber \\
& & \text{ for all } (\sigma,\xi) \in [c,d]^{2};
\end{eqnarray}
i.e., $F$ obeys a H\"{o}lder condition on $[c,d]$.
\cheers
\end{abc}

\item 
Since $g$ is analytic, it has a holomorphic extension\footnote{See 
Ref.~\ref{Bochner}.} $[g]$ to a connected open neighborhood of $[a,b] 
\times S$ in the space $C^{2}$.  For each point $(\sigma',\x) \in [a,b]
\times S$, there then exists at least one $\delta(\sigma',\xi) > 0$ such 
that
$$
d(\sigma',\delta(\sigma',\xi)) \times d(\xi,\delta(\sigma',\xi))
\subset \dom{[g]},
$$
where $d(\sigma',\delta(\sigma',\xi))$ and $d(\xi,\delta(\sigma',\xi))$
are those open disks in $C$ whose centers are $\sigma'$ and $\xi$,
respectively, and which both have the radius $\delta(\sigma',\xi)$.
Hence
\begin{abc}
\begin{equation}
I(\sigma',\delta(\sigma',\xi)) \times d(\xi,\delta(\sigma',\xi))
\subset \dom{[g]},
\label{5.36a}
\end{equation}
where $I(\sigma',\delta(\sigma',\xi))$ is that open interval in $R^{1}$
whose midpoint is $\sigma'$ and whose length is $2\delta(\sigma',\xi)$.
For fixed $\xi$, the set of all $I(\sigma',\delta(\sigma',\xi))$ such
that $\sigma' \in [a,b]$ covers $[a,b]$.  So, by the Heine-Borel theorem,
there exists a finite set of points $\sigma'_{k} \in [a,b]$ ($k=1,\ldots,N$) 
such that
\begin{equation}
[a,b] \subset \cup_{k=1}^{N} I(\sigma'_{k},\delta(\sigma'_{k},\xi)).
\label{5.36b}
\end{equation}
Now, let $\delta(\xi) := \inf{\{\delta(\sigma'_{k},\xi):1 \le k \le N\}}$.
Then, from (\ref{5.36a}),
\begin{equation}
I(\sigma'_{k},\delta(\sigma'_{k},\xi)) \times d(\xi,\delta(\xi))
\subset \dom{[g]};
\label{5.36c}
\end{equation}
and, from (\ref{5.36b}) and (\ref{5.36c}),
\begin{equation}
[a,b] \times d(\xi,\delta(\xi)) \subset \dom{[g]}.
\label{5.36d}
\end{equation}
So, we have proved that, for each $\xi \in S$, there exists $\delta(\x)
> 0$ such that the above relation (\ref{5.36d}) holds.

As the next step in the proof, for each $\xi \in S$, let
\begin{equation}
G(\xi) := \text{ the restriction of } [g] \text{ to }
[a,b] \times d(\xi,\delta(\xi)),
\end{equation}
and, for each $(\sigma',\tau) \in [a,b] \times d(\xi,\delta(\xi))$,
\begin{eqnarray*}
G(\xi,\sigma',\tau) & := & G(\xi)(\sigma',\tau), \\
G_{1}(\xi,\sigma',\tau) & := & \Re{G(\xi,\sigma',\tau)}, \\
G_{2}(\xi,\sigma',\tau) & := & \Im{G(\xi,\sigma',\tau)},
\end{eqnarray*}
and
\begin{equation}
\sigma := \sigma^{1} := \Re{\tau}, \quad \sigma^{2} := \Im{\tau},
\end{equation}
whereupon, from the definition of $[g]$ as a holomorphic extension of
$g$ from $\sigma := \sigma^{1}$,
\begin{equation}
\begin{array}{l}
G_{1}(\xi,\sigma',\sigma) = g(\sigma',\sigma) \text{ and }
G_{2}(\xi,\sigma',\sigma) = 0 \\
\text{for all } \sigma' \in [a,b] \text{ and }
\xi-\delta(\xi) < \sigma < \xi+\delta(\xi),
\end{array}
\label{5.36f}
\end{equation}
\begin{equation}
\begin{array}{l}
G_{k}(\xi,\sigma',\tau) \text{ (for each $k \in \{1,2\}$) is a continuous } \\
\text{function of } (\sigma',\sigma^{1},\sigma^{2}) \text{ throughout }
[a,b] \times d(\xi,\delta(\xi)),
\end{array}
\label{5.36g}
\end{equation}
\begin{equation}
\begin{array}{l}
\partial G_{k}(\xi,\sigma',\tau)/\partial\sigma^{l} \text{ (for each
$k \in \{1,2\}$ and $l \in \{1,2\}$) exists and is } \\
\text{a continuous function of } (\sigma',\sigma^{1},\sigma^{2})
\text{ throughout } [a,b] \times d(\xi,\delta(\xi))
\end{array}
\label{5.36h}
\end{equation}
and
\begin{equation}
\frac{\partial G_{1}(\xi,\sigma',\tau)}{\partial\sigma^{1}} = 
\frac{\partial G_{2}(\xi,\sigma',\tau)}{\partial\sigma^{2}}, \quad
\frac{\partial G_{1}(\xi,\sigma',\tau)}{\partial\sigma^{2}} = 
-\frac{\partial G_{2}(\xi,\sigma',\tau)}{\partial\sigma^{1}}
\label{5.36i}
\end{equation}
at all $(\sigma',\tau) \in [a,b] \times d(\xi,\delta(\xi))$.  So, upon
letting
\begin{equation}
F_{k}(\xi,\tau) := \int_{a}^{b} d\sigma' f(\sigma') G_{k}(\xi,\sigma',\tau)
\text { for each } k \in \{1,2\} \text{ and } \tau \in d(\xi,\delta(\xi)),
\end{equation}
[which exist because of (\ref{5.36g}) and the summability of $f$] the
statement (\ref{5.36f}) and Eqs.\ (\ref{5.33a}) (for $m=1$) yield
\begin{equation}
F_{1}(\xi,\sigma) = F(\sigma) \text{ and } F_{2}(\xi,\sigma) = 0
\text{ for all } \xi-\delta(\xi) < \sigma < \xi+\delta(\xi),
\label{5.37a}
\end{equation}
the statement (\ref{5.36g}) and part (i) (for $m=2$) of our theorem
yield, for each $k \in \{1,2\}$,
\begin{equation}
F_{k}(\xi,\tau) \text{ is a continuous function of } (\sigma^{1},\sigma^{2})
\text{ throughout } d(\xi,\delta(\xi)),
\end{equation}
the statement (\ref{5.36h}) and part (ii) (for $m=2$) of our theorem 
yield, for each $(k,l) \in \{1,2\}^{2}$,
\begin{equation}
\begin{array}{l}
\partial F_{k}(\xi,\tau)/\partial\sigma^{l} \text{ exist and are
continuous} \\ \text{functions of } (\sigma^{1},\sigma^{2}) 
\text{ throughout } d(\xi,\delta(\xi))
\end{array}
\label{5.37c}
\end{equation}
and Eqs.\ (\ref{5.36i}) and (\ref{5.33b}) then yield the Cauchy-Riemann
equations
\begin{eqnarray}
\frac{\partial F_{1}(\xi,\tau)}{\partial\sigma^{1}} & = & 
\frac{\partial F_{2}(\xi,\tau)}{\partial\sigma^{2}} \text{ and } 
\nonumber \\
\frac{\partial F_{1}(\xi,\tau)}{\partial\sigma^{2}} & = & 
-\frac{\partial F_{2}(\xi,\tau)}{\partial\sigma^{1}} 
\nonumber \\
& & \text{for all } \tau \in d(\xi,\delta(\xi)).
\label{5.37d}
\end{eqnarray}

Finally, (\ref{5.37c}) and (\ref{5.37d}) imply that
$$
F_{1}(\xi,\tau) + i F_{2}(\xi,\tau) \text{ is a holomorphic function of }
\tau \text{ throughout } d(\xi,\delta(\xi)),
$$
whereupon (\ref{5.37a}) enables us to infer that
$$
F(\sigma) \text{ is a real analytic function of } \sigma \text{ for all }
\xi-\delta(\xi) < \sigma < \xi+\delta(\xi);
$$
and, since the above statement holds for all $\xi \in S$, $F$ is analytic.
\cheers
\end{abc}
\end{romanlist}

\begin{lemma}[Generalization of Lem.~\ref{5.2C}]
\label{5.3C} \mbox{ } \\
All of the conclusions of the preceding lemma remain valid when the only
alteration in the premises is to replace the statement that $f$ and $g$
are real valued by the statement that they are complex valued or are
finite matrices (such that the product $fg$ exists) with complex-valued
elements.
\end{lemma}

\proof
Use the definition
$$
\int_{a}^{b} d\sigma' h(\sigma') := \int_{a}^{b} d\sigma' \Re{h(\sigma')}
+ i \int_{a}^{b} d\sigma' \Im{h(\sigma')}
$$
for any complex-valued function $h$ whose real and imaginary parts are
summable over $[a,b]$.  The rest is obvious.
\cheers

\begin{theorem}[Continuity of $\check{\Y}^{(i)}(\x)$]
\label{5.4C} \mbox{ } \\
For each $\bv \in K^{\Box}$, $\x \in \domE$, solution $\bar{\F}(\x)$ of
the HHP corresponding to $(\bv,\bar{\F}^{M},\x)$ and $i \in \{3,4\}$,
$\check{\Y}^{(i)}(\x)$ [see Eqs.\ (\ref{5.31a}) and (\ref{5.31b})] is
$\bC^{n-1}$ if $\Box$ is $n$, is $\bC^{(n-1)+}$ if $\Box$ is $n+$,
is $\bC^{\infty}$ if $\Box$ is $\infty$ and is $\bC^{an}$ if $\Box$
is `an'.
\end{theorem}

\proof
Apply Lemmas~\ref{5.1C}, \ref{5.2C} and~\ref{5.3C} to the definitions
(\ref{5.31a}) and (\ref{5.31b}) of $\check{\Y}_{1}^{(i)}(\x)$ and
$\check{\Y}_{2}^{(i)}(\x)$.  It is then easily shown that the second
term on the right side of each of the Eqs.\ (\ref{5.30a}) and (\ref{5.30b})
[with $\sigma \in \check{\I}^{(i)}(x^{7-i})$] is $\bC^{n-1}$ if $\Box$
is $n$, is $\bC^{(n-1)+}$ if $\Box$ is $n+$, is $\bC^{\infty}$ if
$\Box$ is $\infty$ and is $\bC^{an}$ if $\Box$ is `an'.  The first and
third terms on the right sides of each of the Eqs.\ (\ref{5.30a}) and
(\ref{5.30b}) are, on the other hand, both $\bC^{\Box}$ even when $\Box$
is $n$ or is $n+$.  However, a $\bC^{n}$ function is also a $\bC^{n-1}$
function; and a $\bC^{n+}$ function is also a $\bC^{(n-1)+}$ function.
\cheers

\begin{corollary}[On $\Y(\x)$ when $\bv \in K^{\Box}$]
\label{5.5C} \mbox \\
\begin{romanlist}
\item 
Suppose $\bv \in K^{1}$, $\x \in \domE$ and the solution $\bar{\F}(\x)$
of the HHP corresponding to $(\bv,\bar{\F}^{M},\x)$ exists.  Then $\Y(\x)$
has a unique continuous extension to $\bar{\I}(\x)$.  Henceforth, $\Y(\x)$
will denote this extension.
\item 
If $\bv \in K^{\Box}$, then (the extension) $\Y(\x)$ is $\bC^{n-1}$ if
$\Box$ is $n$, is $\bC^{(n-1)+}$ if $\Box$ is $n+$, is $\bC^{\infty}$
if $\Box$ is $\infty$ and is $\bC^{an}$ if $\Box$ is `an'.
\end{romanlist}
\end{corollary}

\begin{abc}
\proof
An extension is defined by [suppressing $\x$]
\begin{eqnarray}
\Y(\sigma) & := & \check{\Y}^{(3)}(\sigma) \text{ when }
\sigma \in \bar{\I}^{(3)}(\x) \text{ and } \\
\Y(\sigma) & := & \check{\Y}^{(4)}(\sigma) \text{ when }
\sigma \in \bar{\I}^{(4)}(\x),
\end{eqnarray}
whereupon statement (ii) of this corollary follows from Thm.~\ref{5.4C}.
The uniqueness follows, of course, from the fact that a function defined
and continuous on an open subset of $R^{1}$ has no more than one 
continuous extension to the closure of that subset.
\cheers
\end{abc}

\setcounter{equation}{0}
\setcounter{theorem}{0}
\subsection{Equivalence of the HHP to an Alekseev-type equation when 
$\bv \in K^{1+}$}

We are free to use either one of the Alekseev-type singular integral
equations (\ref{5.23e}) or (\ref{5.24e}).  We choose to employ (\ref{5.24e})
in the following theorem and its corollary.
\begin{gssm}[HHP-Alekseev equivalence theorem]
\label{5.1D} \mbox{ } \\
Suppose $\bv \in K^{1+}$ and $\x \in \domE$, and suppose that $\bar{\F}(\x)$
and $\Y_{1}(\x)$ are $2 \times 2$ matrix functions, respectively, such that
\begin{equation}
\dom{\bar{\F}(\x)} = C - \bar{\I}(\x), \quad \dom{\Y_{1}(\x)} = \bar{\I}(\x)
\text{ and } \Y_{1}(\x) \text{ is } \bC^{0+}.
\label{5.39}
\end{equation}
Then the following two statements are equivalent to one another:
\begin{romanlist}
\item 
The function $\bar{\F}(\x)$ is a solution of the HHP corresponding to
$(\bv,\bar{\F}^{M},\x)$, and $\Y_{1}(\x)$ is the function whose 
restriction to $\I(\x)$ is defined in terms of $\F^{+}(\x) + \F^{-}(\x)$
by Eq.\ (\ref{5.25a}) [where $\x$ is suppressed] and whose existence and
uniqueness [for the given $\bar{\F}(\x)$] is asserted by Cor.~\ref{5.5C}
when $\Box$ is $1+$.
\item 
The restriction of $\Y_{1}(\x)$ to $\I(\x)$ is a solution of the singular
integral equation (\ref{5.24e}), and $\bar{\F}(\x)$ is defined in terms
of $\Y_{1}(\x)$ by Eq.\ (\ref{5.24b}) [where $\x$ is suppressed].
\end{romanlist}
\end{gssm}

\proof
That (i) implies (ii) has already been proved. [See Thm.~\ref{5.1B} and
Cor.~\ref{5.5C}.]  The proof that (ii) implies (i) will be given in four
parts:
\begin{arablist}
\item 
Assume that statement (ii) is true.  From the definition of $\bar{\F}(\x)$
by Eq.\ (\ref{5.24b}),
\begin{abc}
\begin{equation}
\bar{\F}(\x) \text{ is holomorphic;} 
\label{5.40a}
\end{equation}
and, from two theorems of Plemelj\footnote{See Secs.~16 and~17 of Ch.~II
of Ref.~\ref{Musk} (pp.\ 37-43).} 
\begin{equation}
\F^{+}(\x) \text{ and } \F^{-}(\x) \text{ exist }
\label{5.40b}
\end{equation}
and, since $\nu^{-}(\sigma) = -\nu^{+}(\sigma)$ for all $\sigma \in \I(\x)$,
\begin{equation}
\frac{1}{2}\left[\F^{+}(\sigma)+\F^{-}(\sigma)\right] =
- \Y_{1}(\sigma) W_{2}^{T}(\sigma) J
\label{5.41a}
\end{equation}
and
\begin{equation}
\frac{1}{2}[\nu^{+}(\sigma)]^{-1}\left[\F^{+}(\sigma)-\F^{-}(\sigma)\right] =
I - \frac{1}{\pi i} \int_{\bar{\I}} d\sigma' [\nu^{+}(\sigma')]^{-1}
\Y_{1}(\sigma') \frac{W_{2}^{T}(\sigma') J}{\sigma'-\sigma}
\label{5.41b}
\end{equation}
for all $\sigma \in \I(\x)$.  Upon multiplying Eqs.\ (\ref{5.41a}) and
(\ref{5.41b}) through by $W_{2}(\sigma)$ on the right, one obtains,
for all $\sigma \in \I(\x)$,
\begin{eqnarray}
\frac{1}{2}\left[\F^{+}(\sigma)+\F^{-}(\sigma)\right] W_{2}(\sigma) & = & \left(
\begin{array}{c}
0 \\ 0
\end{array} \right),
\label{5.42a} \\
\text{and} & & \nonumber \\
\frac{1}{2}[\nu^{+}(\sigma)]^{-1} \left[\F^{+}(\sigma)-\F^{-}(\sigma)\right]
W_{2}(\sigma) & = & \Y_{2}(\sigma),
\label{5.42b}
\end{eqnarray}
where $\Y_{2}(\x)$ has the domain $\bar{\I}(\x)$ and the values
\begin{eqnarray}
\lefteqn{\Y_{2}(\sigma) := W_{2}(\sigma) } \nonumber \\
& & \mbox{ }-\frac{1}{\pi I} \int_{\bar{\I}} d\sigma' [\nu^{+}(\sigma')]^{-1}
\Y_{1}(\sigma') \frac{W_{2}^{T}(\sigma') J [W_{2}(\sigma)-W_{2}(\sigma')]}
{\sigma'-\sigma}
\nonumber \\
& & \text{for all } \sigma \in \bar{\I}(\x).
\end{eqnarray}
{}From Lemmas~\ref{5.1C}(ii), \ref{5.2C}(iv) and \ref{5.3C},
\begin{equation}
\Y_{2}(\x) \text{ is } \bC^{0+}.
\label{5.43}
\end{equation}
Upon multiplying Eqs.\ (\ref{5.41a}) and (\ref{5.41b}) through by
$W_{1}(\sigma)$ on the right, upon using the fact that $\det{W(\sigma)}=1$ 
is equivalent to the equation
\begin{equation}
W_{2}^{T}(\sigma) J W_{1}(\sigma) = - (1), 
\end{equation}
and, upon using Eq.\ (\ref{5.24e}), one obtains, for all $\sigma \in \I(\x)$,
\begin{eqnarray}
\frac{1}{2}\left[\F^{+}(\sigma)+\F^{-}(\sigma)\right] W_{1}(\sigma) & = &
\Y_{1}(\sigma) 
\label{5.45a} \\
\text{and} & & \nonumber \\
\frac{1}{2}\left[\F^{+}(\sigma)-\F^{-}(\sigma)\right] W_{1}(\sigma) & = &
\left( \begin{array}{c}
0 \\ 0
\end{array} \right).
\label{5.45b}
\end{eqnarray}
\end{abc}

\item 
We next note that the four equations (\ref{5.42a}), (\ref{5.42b}),
(\ref{5.45a}) and (\ref{5.45b}) are collectively equivalent to the
single equation
\begin{abc}
\begin{equation}
\F^{\pm}(\sigma) W(\sigma) = \Y(\sigma) \left( \begin{array}{cc}
1 & 0 \\ 0 & \nu^{\pm}(\sigma)
\end{array} \right) \text{ for all } \sigma \in \I(\x),
\label{5.46}
\end{equation}
where $\Y(\sigma)$ is defined to be the $2 \times 2$ matrix whose first
and second columns are $\Y_{1}(\sigma)$ and $\Y_{2}(\sigma)$, respectively.
{}From the definition of $W(\sigma)$ by Eqs.\ (\ref{5.22a}) and (\ref{5.22c}),
and from the expression for $\bar{\F}^{M}(\tau)$ that is given by 
Eq.\ (\ref{G2.10c}), Eq.\ (\ref{5.46}) is equivalent to the statement
\begin{eqnarray}
\F^{+}(\sigma) \tilde{v}^{(i)}(\sigma) [\F^{M+}(\sigma)]^{-1} & = &
\F^{-}(\sigma) \tilde{v}^{(i)}(\sigma) [\F^{M-}(\sigma)]^{-1} 
\nonumber \\
& = & Y(\sigma) \text{ for all } \sigma \in \I^{(i)}(\x),
\label{5.47}
\end{eqnarray}
where
\begin{equation}
Y(\sigma) := \Y(\sigma) \left( \begin{array}{cc}
1 & i(\sigma-z) \\ 0 & 1
\end{array} \right) \text{ for all } \sigma \in \bar{\I}(\x).
\label{5.48}
\end{equation}
{}From the above Eq.\ (\ref{5.48}) and from statements (\ref{5.39}) and
(\ref{5.43}),
\begin{equation}
\begin{array}{l}
\text{the function } Y(\x) \text{ whose domain is } \bar{\I}(\x)
\text{ and whose value for each } 
\\ \sigma \in \bar{\I}(\x) \text{ is }
Y(\x)(\sigma) := Y(\x,\sigma) \text{ is } \bC^{0+}
\text{ and is, therefore, continuous.}
\end{array}
\end{equation}
So, 
\begin{equation}
Y(\x) \text{ is bounded at $\x$ and at $\x_{0}$.}
\label{5.49b}
\end{equation}
\end{abc}

\item 
We now return to the definition of $\bar{\F}(\x)$ in terms of $\Y_{1}(\x)$
by Eq.\ (\ref{5.24b}).  From Lemma~\ref{5.1C}(i) when $\Box$ is $1$, and
from statement (\ref{5.39}) concerning $\Y_{1}(\x)$ being $\bC^{0+}$ on
its domain $\bar{\I}(\x)$, note that the factors in the numerator of the
integrand in Eq.\ (\ref{5.24b}) have the following properties:
\begin{abc}
\begin{equation}
\begin{array}{l}
\Y_{1}(\sigma') [W_{2}(\sigma')]^{T} J \text{ is defined for all }
\sigma' \in \bar{\I}(\x) \\
\text{and obeys a H\"{o}lder condition on } \bar{\I}(\x);
\end{array}
\label{5.50a}
\end{equation}
and
\begin{equation}
\begin{array}{l}
[\nu^{\pm}(\sigma')]^{-1} \text{ is } H(1/2) \text{ on each closed subinterval
of } \I(\x) \\
\text{and converges to zero as } \sigma' \rightarrow r \text{ and as }
\sigma' \rightarrow s.
\end{array}
\label{5.50b}
\end{equation}
Also, recall that
\begin{equation}
\begin{array}{l}
\bar{\nu}(\tau) \text{ is that branch of } (\tau-r_{0})^{1/2}
(\tau-s_{0})^{1/2} (\tau-r)^{-1/2} (\tau-s)^{-1/2} \\
\text{which has the cut } \bar{\I}(\x) \text{ and the value $1$ at
$\tau=\infty$.}
\end{array}
\label{5.50c}
\end{equation}
Several theorems on Cauchy intgrals near the end points of the lines of
inegration are given in Ref.~\ref{Musk}, Sec.~29, Ch.~4.  In particular,
by applying Muskelishvili's Eq.\ (29.4) to our Eq.\ (\ref{5.24b}), one
obtains the following conclusion from the above statements (\ref{5.50a})
and (\ref{5.50b}):
\begin{equation}
\bar{\nu}(\tau)^{-1} \bar{\F}(\tau) \text{ converges as } \tau \rightarrow
r \text{ and as } \tau \rightarrow s.
\label{5.51a}
\end{equation}
Moreover, by applying Muskelishvili's Eqs.\ (29.5) and (29.6) to our
Eq.\ (\ref{5.24b}), one obtains the following conclusion from the above
statements (\ref{5.50a}) to (\ref{5.50c}):
\begin{equation}
\bar{\F}(\tau) \text{ converges as } \tau \rightarrow r_{0} \text{ and as }
\tau \rightarrow s_{0}.
\label{5.51b}
\end{equation}
\end{abc}

\item 
{}From the above statements (\ref{5.40a}), (\ref{5.40b}), (\ref{5.47}),
(\ref{5.49b}), (\ref{5.51a}) and (\ref{5.51b}), all of the
defining conditions for a solution of the HHP corresponding to 
$(\bv,\bar{\F}^{M},\x)$ are satisfied by $\bar{\F}(\x)$ as defined in terms
of $\Y_{1}(\x)$ by Eq.\ (\ref{5.24b}).
\end{arablist}
\cheers

We already know from GSSM~\ref{4.3D}(iv) that there is not more than one
solution of the HHP corresponding to $(\bv,\bar{\F}^{M},\x)$.

\begin{abc}
\begin{corollary}[Uniqueness of $\Y_{1}(\x)$]
\label{5.2D} \mbox{ } \\
For each $\bv \in K^{1+}$ and $\x \in \domE$, there is not more than one
$2 \times 1$ matrix function $\Y_{1}(\x)$ such that
\begin{eqnarray}
\dom{\Y_{1}(\x)} & = & \I(\x), 
\label{5.52a} \\
\Y_{1}(\x) & \text{is} & \bC^{0+}
\label{5.52b}
\end{eqnarray}
and $\Y_{1}(\x,\sigma) := \Y_{1}(\x)(\sigma)$ satisfies the singular
integral equation (\ref{5.24e}) for all $\sigma \in \I(\x)$.
\end{corollary}
\end{abc}

\begin{abc}
\proof
Suppose that $\Y_{1}(\x)$ and $\Y'_{1}(\x)$ are $2 \times 1$ matrix
functions, both of which have domain $\bar{\I}(\x)$, are $\bC^{0+}$
and satisfy Eq.\ (\ref{5.24e}) for all $\sigma \in \I(\x)$ and for
the same given $\bv \in K^{1+}$.  Let $\bar{\F}(\x)$ and $\bar{\F}'(\x)$
be the $2 \times 2$ matrix functions with domain $C - \bar{\I}(\x)$
that are defined in terms of $\Y_{1}(\x)$ and $\Y'_{1}(\x)$, respectively,
by Eq.\ (\ref{5.24b}).  Then, from the preceding GSSM~\ref{5.1D},
$\bar{\F}(\x)$ and $\bar{\F}'(\x)$ are both solutions of the HHP
corresponding to $(\bv,\bar{\F}^{M},\x)$; and, therefore, from
GSSM~\ref{4.3D}(iv),
\begin{equation}
\bar{\F}(\x) = \bar{\F}'(\x);
\end{equation}
and, from Eq.\ (\ref{5.45a}) in the proof of GSSM~\ref{5.1D} and from
statements (\ref{5.52a}) and (\ref{5.52b}),
\begin{equation}
\Y_{1}(\x) = \Y'_{1}(\x).
\end{equation}
\cheers
\end{abc}
\newpage

\setcounter{equation}{0}
\setcounter{theorem}{0}
\section{A Fredholm integral equation of the second kind that is
equivalent to the Alekseev-type singular integral equation when 
$\bv \in K^{2+}$\label{Sec_6}}

If $\bv \in K^{1+}$ and the particular solution $\Y_{1}(\x)$ of Eq.\ 
(\ref{5.24e}) that has a $\bC^{0+}$ extension to $\bar{\I}(\x)$ exists, 
then it can be shown that $\Y_{1}(\x)$ is also a solution of a Fredholm
integral equation of the second kind.


\subsection{Derivation of Fredholm equation from Alekseev-type equation}

\begin{abc}
\begin{definition}{Dfns.\ of $d_{21}$, $L_{1}$, $k_{21}$, $U_{1}(\x)$
and $K_{21}(\x)$\label{def65}}
For each $\bv \in K^{\Box}$ for which $\Box$ is $1+$, $n$ or $n+$ with
$n \ge 2$, $\infty$ or `an', and for each $\x \in \domE$, let $d_{21}$,
$L_{1}$, $k_{21}$, $U_{1}(\x)$ and $K_{21}(\x)$ denote the functions
such that
\begin{eqnarray}
\dom{d_{21}} & = & \dom{L_{1}} = \dom{k_{21}} := \bar{\I}(\x)^{2},
\label{6.1a} \\
\dom{U_{1}(\x)} & := & \bar{\I}(\x),
\label{6.1b} \\
\dom{K_{21}(\x)} & := & 
\bar{\I}(\x) \times [\bar{\I}(\x)-\{r,s\}]
\text{ if } \bv \in K^{1+} \text{ but } \bv \notin K^{2},
\label{6.1c} \\
\dom{K_{21}(\x)} & := & \bar{\I}(\x)^{2} \text{ if } \bv \in K^{2};
\label{6.1d}
\end{eqnarray}
and, for each $(\sigma',\sigma) \in \bar{\I}(\x)^{2}$,
\begin{eqnarray}
d_{21}(\sigma',\sigma) & := & W_{22}(\sigma') W_{11}(\sigma)
- W_{12}(\sigma') W_{21}(\sigma),
\label{6.2a} \\
L_{1}(\sigma',\sigma) & := & \frac{W_{1}(\sigma')-W_{1}(\sigma)}
{\sigma'-\sigma},
\label{6.2b} \\
k_{21}(\sigma',\sigma) & := & \frac{d_{21}(\sigma',\sigma)-1}{\sigma'-\sigma},
\label{6.2c}
\end{eqnarray}
while, for all $\sigma \in \bar{\I}(\x)$, [suppressing $\x$,]
\begin{equation}
U_{1}(\sigma) := U_{1}(\x)(\sigma) := W_{1}(\sigma) - \frac{1}{\pi i}
\int_{\bar{\I}} d\sigma' \nu^{+}(\sigma') L_{1}(\sigma',\sigma).
\label{6.2d}
\end{equation}
Furthermore, for all $(\sigma',\sigma) \in \dom{K_{21}(\x)}$, [suppressing $\x$]
\begin{eqnarray}
\lefteqn{K_{21}(\sigma',\sigma) := K_{21}(\x)(\sigma',\sigma) } \nonumber \\
& := & k_{21}(\sigma',\sigma) - \frac{1}{\pi i} \int_{\bar{\I}} d\sigma''
\nu^{+}(\sigma'') \left[ \frac{k_{21}(\sigma',\sigma'')-k_{21}(\sigma',\sigma)}
{\sigma''-\sigma} \right].
\label{6.2e}
\end{eqnarray}
\end{definition}
\end{abc}

Note:  One should keep in mind that $\bar{\I}(\x)^{2} := \bar{\I}(\x)
\times \bar{\I}(\x)$ is the union of four disjoint Cartesian products
$\bar{\I}^{(i)}(\x) \times \bar{\I}^{(j)}(\x)$ and that to say that a
function $\phi$ is $\bC^{0+}$ on $\bar{\I}(\x)^{2}$ means that it is
$\bC^{0+}$ on each of these four components; and this, in turn, is 
equivalent to the statement\footnote{See Sec.\ 4 of Ref.~\ref{Musk}.}
that there exist positive real numbers $M_{i}(\x,\x_{0})$ and
$\gamma_{i}(\x,\x_{0}) \le 1$ such that, for all
$\sigma'_{1}$, $\sigma'_{2} \in \bar{\I}^{(i)}(\x)$ and all $\sigma_{1}$,
$\sigma_{2} \in \bar{\I}^{(j)}(\x)$, [suppressing $\x$, $\x_{0}$,]
\begin{eqnarray*}
\left| \phi(\sigma'_{1},\sigma)-\phi(\sigma'_{2},\sigma) \right|
& \le & M_{i} |\sigma'_{1}-\sigma'_{2}|^{\gamma_{i}}
\text{ for all } \sigma \in \bar{\I}^{(j)} \\
\text{and} & & \\
\left| \phi(\sigma',\sigma_{1})-\phi(\sigma',\sigma_{2}) \right|
& \le & M_{j} |\sigma_{1}-\sigma_{2}|^{\gamma_{j}}
\text{ for all } \sigma' \in \bar{\I}^{(i)}.
\end{eqnarray*}
Likewise, to say that a function $\phi$ is $\bC^{0+}$ on $\bar{\I}(\x)
\times [\bar{\I}(\x)-\{r,s\}]$ means that, for all $i$ and $j$ and for
each choice of $c^{j} \in \I^{(j)}(\x)$, $\phi$ is $\bC^{0+}$ on
$\bar{\I}^{(i)}(\x) \times |c^{j},x_{0}^{j}|$; and this, in turn,
is equivalent to the statement that there exist positive real numbers
$M_{i}(\x,\x_{0},c^{j})$ and $\gamma_{i}(\x,\x_{0},c^{j}) \le 1$ such
that, for all $\sigma'_{1}$, $\sigma'_{2} \in \bar{\I}^{(i)}(\x)$ and
all $\sigma_{1}$, $\sigma_{2} \in |c^{j},x_{0}^{j}|$,
\begin{eqnarray*}
\left| \phi(\sigma'_{1},\sigma)-\phi(\sigma'_{2},\sigma) \right|
& \le & M_{i} |\sigma'_{1}-\sigma'_{2}|^{\gamma_{i}}
\text{ for all } \sigma \in |c^{j},x_{0}^{j}| \\
\text{and} & & \\
\left| \phi(\sigma',\sigma_{1})-\phi(\sigma',\sigma_{2}) \right|
& \le & M_{j} |\sigma_{1}-\sigma_{2}|^{\gamma_{j}}
\text{ for all } \sigma' \in \bar{\I}^{(i)}(\x).
\end{eqnarray*}
The reader may find it interesting to examine the above concepts in
the cases $r=r_{0}$ or $s=s_{0}$.

\begin{abc}
\begin{proposition}[$U_{1}$ and $K_{21}$]
\label{6.1A} \mbox{ } \\
For each $\sigma \in \bar{\I}(\x)-\{r,s\}$,
\begin{equation}
U_{1}(\sigma) = - \frac{1}{\pi i} \int_{\bar{\I}} d\sigma'
\nu^{+}(\sigma') \frac{W_{1}(\sigma')}{\sigma'-\sigma};
\label{6.3}
\end{equation}
and, for each $(\sigma',\sigma) \in \bar{\I}(\x) \times
[\bar{\I}(\x)-\{r,s\}]$,
\begin{equation}
K_{21}(\sigma',\sigma) = - \frac{1}{\pi i} \int_{\bar{\I}} d\sigma''
\nu^{+}(\sigma'') \frac{d_{21}(\sigma',\sigma'')}{(\sigma''-\sigma)
(\sigma'-\sigma'')}.
\label{6.4}
\end{equation}
\end{proposition}
\end{abc}

\begin{abc}
\proof
Substituting from Eq.\ (\ref{6.2b}) into Eq.\ (\ref{6.2d}), one obtains
\begin{eqnarray*}
\lefteqn{U_{1}(\sigma) = W_{1}(\sigma)} \nonumber \\
& & \mbox{ } + W_{1}(\sigma) \frac{1}{\pi i} \int_{\bar{\I}} d\sigma'
\nu^{+}(\sigma') (\sigma'-\sigma)^{-1} 
- \frac{1}{\pi i} \int_{\bar{\I}} d\sigma' \nu^{+}(\sigma')
\frac{W_{1}(\sigma')}{\sigma'-\sigma},
\end{eqnarray*}
whereupon Eq.\ (\ref{6.3}) follows because of the well known equation
\begin{equation}
\frac{1}{\pi i} \int_{\bar{\I}} d\sigma' \nu^{+}(\sigma')
(\sigma'-\sigma)^{-1} = -1 \text{ for all } 
\sigma \in \bar{\I}(\x)-\{r,s\},
\label{6.5a}
\end{equation}
which one can derive by the usual contour integration technique with
the aid of the facts $\nu^{-}(\sigma)=-\nu^{+}(\sigma)$ and
$\bar{\nu}(\sigma)=1$.  From Eq.\ (\ref{6.5a}), as well as the
identity
\begin{equation}
(\sigma''-\sigma)^{-1}(\sigma'-\sigma'')^{-1} = (\sigma'-\sigma)^{-1}
\left[(\sigma''-\sigma)^{-1}+(\sigma'-\sigma'')^{-1}\right],
\end{equation}
one obtains
\begin{equation}
\frac{1}{\pi i} \int_{\bar{\I}} d\sigma'' \nu^{+}(\sigma'')
(\sigma''-\sigma)^{-1}(\sigma'-\sigma'')^{-1} = 0 \text{ for all }
\sigma \in \bar{\I}(\x)-\{r,s\}.
\label{6.5c}
\end{equation}
Upon substituting the expression for $k_{21}(\sigma',\sigma'')$
that is given by its definition (\ref{6.2c}) into the definition
(\ref{6.2e}), and upon using Eqs.\ (\ref{6.5a}) and (\ref{6.5c}),
one obtains Eq.\ (\ref{6.4}).
\cheers
\end{abc}

\begin{proposition}[Properties of $L_{1}$, $k_{21}$, $U_{1}(\x)$
and $K_{21}(\x)$]
\label{6.2A} \mbox{ } \\
For each $\x \in \domE$ and $\bv \in K^{\Box}$, $L_{i}$, $k_{21}$
and $U_{1}(\x)$ are $\bC^{n-1}$ if $\Box$ is $n$ and $n \ge 2$, are
$\bC^{(n-1)+}$ if $\Box$ is $n+$ and $n \ge 2$, are $\bC^{\infty}$ if
$\Box$ is $\infty$ and are $\bC^{an}$ if $\Box$ is `an'.

If $\bv \in K^{1+}$, then $L_{1}$, $k_{21}$ and $U_{1}(\x)$ are
$\bC^{0+}$; and $K_{21}(\x)$ is also $\bC^{0+}$ \{but, as we recall,
its domain is only $\bar{\I}(\x) \times [\bar{\I}(\x)-\{r,s\}]$\}.
\end{proposition}

\begin{abc}
\proof
Except for the special case of $K_{21}(\x)$ when $\bv \in K^{1+}$ but
$\bv \notin K^{2}$, the proof is effected by straightforward applications
of Lemmas \ref{5.1C}, \ref{5.2C} and \ref{5.3C} to the definitions
(\ref{6.2b}) to (\ref{6.2e}).  One should keep in mind the domains 
defined by Eqs.\ (\ref{6.1a}), (\ref{6.1b}) and (\ref{6.1d}). 

As regards the special case of $K_{21}(\x)$ when $\bv \in K^{1+}$ but
$\bv \notin K^{2}$, note that $k_{21}$ is $\bC^{0+}$ but is not $\bC^{1}$.
Therefore,
$$
\frac{k_{21}(\sigma',\sigma'')-k_{21}(\sigma',\sigma)}{\sigma''-\sigma}
$$
is not a continuous function of $(\sigma'',\sigma',\sigma)$ on the
diagonal square $\sigma''=\sigma$ of the cube $\bar{\I}(\x)^{3}$;
and the Lemma \ref{5.2C}(i) for the case $m=2$ is not, therefore,
applicable.  There are, however, other theorems that we can apply.
{}From Eqs.\ (\ref{6.5a}) and (\ref{6.2e}),
\begin{eqnarray}
K_{21}(\sigma',\sigma) & = & - \frac{1}{\pi i} \int_{\bar{\I}}
d\sigma'' \nu^{+}(\sigma'') \frac{k_{21}(\sigma',\sigma'')}{\sigma''-\sigma}
\nonumber \\
& & \text{for all } (\sigma',\sigma) \in \bar{\I}(\x) \times
[\bar{\I}(\x)-\{r,s\}].
\label{6.6a}
\end{eqnarray}
Note that
\begin{equation}
\begin{array}{l}
\nu^{+}(\sigma'') k_{21}(\sigma',\sigma'') \text{ is a } \bC^{0+}
\text{ function of } (\sigma',\sigma'') \text { on } \nonumber \\
\bar{\I}(\x) \times [\bar{\I}(\x)-\{r,s\}] \text{ and vanishes at }
\sigma'' \in \{r,s\}.
\end{array}
\end{equation}
Therefore, from Sec.~12 in Ref.~\ref{Musk}, the PV integral (\ref{6.6a})
exists for all $(\sigma',\sigma)$ in $\bar{\I}(\x) \times [\bar{\I}(\x)
-\{r,s\}]$; and the theorem in Sec.~20 of that reference enables us to
conclude that $K_{21}(\x)$ is $\bC^{0+}$ on the domain (\ref{6.1c}).
\cheers
\end{abc}

\begin{theorem}[Fredholm equation]
\label{6.3A} \mbox{ } \\
Suppose that, for a given $\bv \in K^{1+}$ and $\x \in \domE$, a solution
$\Y_{1}(\x)$ of the Alekseev-type equation (\ref{5.24e}) exists and is
$\bC^{0+}$ on $\bar{\I}(\x)$.  Then
\begin{eqnarray}
\lefteqn{\Y_{1}(\sigma) - \frac{1}{\pi i} \int_{\bar{\I}} d\sigma'
\nu^{+}(\sigma')^{-1} \Y_{1}(\sigma') K_{21}(\sigma',\sigma) } \nonumber \\
& = & U_{1}(\sigma) \text{ for all } \sigma \in \bar{\I}(\x)
\text{ if } \bv \in K^{2} \nonumber \\
& & \text{and for all } \sigma \in \bar{\I}(\x)-\{r,s\}
\text{ if } \bv \notin K^{2}.
\label{6.7}
\end{eqnarray}
\end{theorem}

\begin{abc}
\proof
We shall be employing a variant of the Poincar\'{e}-Bertrand commutator
theorem in this proof.  Suppose that $L$ is a smooth oriented line or
contour in $C-\{\infty\}$ and $\phi$ is a complex-valued function whose
domain is $L \times L$ and which obeys a H\"{o}lder condition on $L \times L$.
Then the conventional Poincar\'{e}-Bertrand theorem asserts
\begin{eqnarray}
\left[ \frac{1}{\pi i} \int_{L} d\tau'', \frac{1}{\pi i} \int_{L} d\tau'
\right] \frac{\phi(\tau',\tau'')}{(\tau''-\tau)(\tau'-\tau'')}
& = & \phi(\tau,\tau) \text{ for all } \tau \in L \nonumber \\
& & \text{minus its end points,}
\label{6.8a}
\end{eqnarray}
where the above bracketed expression is the commutator of the path integral
operators.  We are, of course, concerned here only with the case
$L=\bar{\I}(\x)$; and our variant asserts that, for any function $\phi$
which is $\bC^{0+}$ on $\bar{\I}(\x)^{2}$,
\begin{eqnarray}
\left[ \frac{1}{\pi i} \int_{\bar{\I}} d\sigma'' \nu^{+}(\sigma''),
\frac{1}{\pi i} \int_{\bar{\I}} d\sigma' \nu^{+}(\sigma')^{-1} \right]
\frac{\phi(\sigma',\sigma'')}{(\sigma''-\sigma)(\sigma'-\sigma'')}
& = & 0 \text{ for all } \nonumber \\
& & \sigma \in \I(\x).
\label{6.8b}
\end{eqnarray}
An elegant and thorough proof of the Poincar\'{e}-Bertrand theorem (\ref{6.8a})
is given by Sec.~23 of Muskhelishvili's treatise, and what we have done is
to construct a proof of (\ref{6.8b}) that parallels his proof step by step.
Our proof of (\ref{6.8b}) will be given in an appendix to these notes and
will include a demonstration that the interated PV integrals in Eq.\
(\ref{6.8b}) exist.

We shall now apply Eq.\ (\ref{6.8b}) to the Alekseev-type equation
(\ref{5.24e}), which we express in the form
\begin{equation}
\frac{1}{\pi i} \int_{\bar{\I}} d\sigma' \nu^{+}(\sigma')^{-1}
\frac{\Y_{1}(\sigma') d_{21}(\sigma',\sigma'')}{\sigma'-\sigma''}
= -W_{1}(\sigma'') \text{ for all } \sigma'' \in \I(\x).
\label{6.9a}
\end{equation}
Since the postulated solution $\Y_{1}(\x)$ is $\bC^{0+}$ on $\bar{\I}(\x)$,
the product $\Y_{1}(\x)d_{21}$ is $\bC^{0+}$ on $\bar{\I}(\x)^{2}$.  Also,
$\det{W(\sigma)} = d_{21}(\sigma,\sigma) = 1$.  Therefore, upon multiplying
both sides of Eq.\ (\ref{6.9a}) by $(\sigma''-\sigma)^{-1}$ and then
applying the PV integral operator
$$
\frac{1}{\pi i} \int_{\bar{\I}} d\sigma'' \nu^{+}(\sigma''),
$$
Eq.\ (\ref{6.8b}) gives us
\begin{eqnarray}
\lefteqn{\Y_{1}(\sigma) + \frac{1}{\pi i} \int_{\bar{\I}} d\sigma''
\nu^{+}(\sigma'') \frac{\Y_{1}(\sigma')d_{21}(\sigma',\sigma'')}
{(\sigma''-\sigma)(\sigma'-\sigma'')} } \nonumber \\
& = & -\frac{1}{\pi i} \int_{\bar{\I}} d\sigma'' \nu^{+}(\sigma'')
\frac{W_{1}(\sigma'')}{\sigma''-\sigma} \text{ for all } \sigma \in \I(\x).
\end{eqnarray}
So, from Prop.~\ref{6.1A} [Eqs.\ (\ref{6.3}) and (\ref{6.4})], Eq.\ (\ref{6.7}) 
holds for all $\sigma \in \I(\x)$.  It remains to prove that Eq.\ (\ref{6.7})
holds for all $\sigma \in \bar{\I}(\x)$ when $\bv \in K^{2}$ and for all
$\sigma \in \bar{\I}(\x)-\{r,s\}$ when $\bv \notin K^{2}$.

In both cases,
\begin{equation}
\Y_{1}(\x) \text{ is continuous on } \bar{\I}(\x);
\label{6.9c}
\end{equation}
and, from Prop.~\ref{6.2A},
\begin{equation}
U_{1}(\x) \text{ is continuous on } \bar{\I}(\x)
\label{6.9d}
\end{equation}
and
\begin{equation}
K_{21}(\x) \text{ is continuous on } \dom{K_{21}(\x)}.
\label{6.9e}
\end{equation}
A brief review of the proof of Lem.~\ref{5.2C} shows that this lemma
remains valid if $S$ is a closed or a semi-closed subinterval of $R^{1}$.
Therefore, from (\ref{6.9e}) and Lem.~\ref{5.2C}(i) [with a closed or a
semi-closed $S \subset R^{1}$], the integral in Eq.\ (\ref{6.7}) is a
continuous function of $\sigma$ throughout $\bar{\I}(\x)$ if $\bv \in K^{2}$,
and throughout $\bar{\I}(\x)-\{r,s\}$ if $\bv \notin K^{2}$; and it then
follows from (\ref{6.9c}) and (\ref{6.9d}) that Eq.\ (\ref{6.7}) holds for
all $\sigma \in \bar{\I}(\x)$ if $\bv \in K^{2}$, and for all $\sigma
\in \bar{\I}(\x)-\{r,s\}$ if $\bv \notin K^{2}$.
\cheers
\end{abc}

\setcounter{equation}{0}
\setcounter{theorem}{0}
\subsection{Equivalence of Alekseev-type equation and Fredholm equation
when $\bv \in K^{2+}$}

The Fredholm equation (\ref{6.7}) generally has a singular kernel and is 
generally {\em not\/} equivalent to the Alekseev-type equation (\ref{5.24e}).
In this section we shall restrict our attention to the case $\bv \in K^{2+}$.

In the proof of the next theorem, we shall employ the following corollaries
of the Poincar\'{e}-Bertrand variant:
\begin{abc}
\begin{eqnarray}
\lefteqn{\left[ \frac{1}{\pi i} \int_{\bar{\I}} d\sigma'' \nu^{+}(\sigma''),
\frac{1}{\pi i} \int_{\bar{\I}} d\sigma' \nu^{+}(\sigma')^{-1} \right]
\frac{\phi(\sigma',\sigma'')}{\sigma''-\sigma} } \nonumber \\
& = & 0 \text{ for all functions } \phi \text{ that are } \bC^{0+}
\nonumber \\
& & \text{on } \bar{\I}(\x)^{2} \text{ and for all } \sigma \in \I(\x);
\label{6.10a}
\end{eqnarray}
and
\begin{eqnarray}
\lefteqn{\frac{1}{\pi i} \int_{\bar{\I}} d\sigma'' \nu^{+}(\sigma'')^{-1}
\frac{1}{\pi i} \int_{\bar{\I}} d\sigma' \nu^{+}(\sigma')
\frac{\psi(\sigma')}{(\sigma''-\sigma)(\sigma'-\sigma'')} } \nonumber \\
& = & \psi(\sigma) \text{ for all } \psi \text{ that are } \bC^{0+}
\nonumber \\
& & \text{on } \bar{\I}(\x) \text{ and for all } \sigma \in \I(\x).
\label{6.10b}
\end{eqnarray}

To prove statement (\ref{6.10a}), set
$$
\phi(\sigma',\sigma'') = \frac{\phi(\sigma',\sigma'')(\sigma'-\sigma'')}
{\sigma'-\sigma''}
$$
in the operand, and then apply (\ref{6.8b}).  To prove (\ref{6.10b}),
first consider that the linear space of all complex-valued operands of
the PV integral operators
$$
\frac{1}{\pi i} \int_{\bar{\I}} d\sigma'' \nu^{+}(\sigma'')
\text{ and }
\frac{1}{\pi i} \int_{\bar{\I}} d\sigma' \nu^{+}(\sigma')^{-1}
$$
[i.e., the linear space of all $\phi(\sigma',\sigma'')(\sigma''-\sigma)^{-1}
(\sigma'-\sigma'')^{-1}$ such that $\phi$ is $\bC^{0+}$ on $\bar{\I}(\x)^{2}$]
is one-to-one mapped onto itself by the transposition of $\x$ and $\x_{0}$,
that this transposition induces $\nu^{+} \rightarrow (\nu^{+})^{-1}$, and
that one consequence is the equivalence of (\ref{6.8b}) with the following
statement:
\begin{eqnarray}
\lefteqn{\left[ \frac{1}{\pi i} \int_{\bar{\I}} d\sigma''
\nu^{+}(\sigma'')^{-1}, \frac{1}{\pi i} \int_{\bar{\I}} d\sigma'
\nu^{+}(\sigma') \right]
\frac{\phi(\sigma',\sigma'')}{(\sigma''-\sigma)(\sigma'-\sigma'')} }
\nonumber \\
& = & \phi(\sigma,\sigma) \text{ for all complex-valued $\phi$ such that }
\nonumber \\
& & \phi \text{ is } \bC^{0+} \text{ on } \bar{\I}(\x)^{2}, \text{ and for
all } \sigma \in \I(\x).
\label{6.10c}
\end{eqnarray}
The statement (\ref{6.10b}) is now easily obtained from the above statement
(\ref{6.10c}) and the following statement [which is derived from Eq.\
(\ref{6.5c}) by transposing $\x$ and $\x_{0}$]:
\begin{equation}
\frac{1}{\pi i} \int_{\bar{\I}} d\sigma'' \nu^{+}(\sigma'')^{-1}
(\sigma''-\sigma)^{-1}(\sigma'-\sigma'')^{-1} = 0 \text{ for all }
\sigma \in \bar{\I}(\x)-\{r_{0},s_{0}\}.
\label{6.10d}
\end{equation}
\end{abc}

\begin{gssm}[Alekseev-Fredholm equivalence theorem]
\label{6.1B} \mbox{ } \\ 
Suppose $\bv \in K^{2+}$, $\x \in \domE$ and $\Y_{1}(\x)$ is a $2 \times 1$
column matrix function with domain $\bar{\I}(\x)$.  Then $U_{1}(\x)$ is
$\bC^{1+}$ and $K_{21}(\x)$ is $\bC^{0+}$.  Also, the following two
statements are equivalent to one another:
\begin{romanlist}
\item 
$\Y_{1}(\x)$ is $\bC^{0+}$ and is the solution of Eq.\ (\ref{5.24e}) for
all $\sigma \in \I(\x)$.
\item 
$\Y_{1}(\x)$ is summable over $\bar{\I}(\x)$ and is a solution of Eq.\
(\ref{6.7}) for all $\sigma \in \bar{\I}(\x)$.
\end{romanlist}
\end{gssm}

\begin{abc}
\proof
{}From Prop.~\ref{6.2A}, $U_{1}(\x)$ is $\bC^{1+}$ and $K_{21}(\x)$ is
$\bC^{0+}$; and Thm.~\ref{6.3A} already asserts that statement (i)
implies statement (ii).  It remains only to prove that statement (ii)
implies statement (i).

Grant statement (ii).  Since $U_{1}(\x)$ is $\bC^{1+}$ and $K_{21}(\x)$
is $\bC^{0+}$ on $\bar{\I}(\x)$ and since $\Y_{1}(\x)$ is summable over
$\bar{\I}(\x)$, Eq.\ (\ref{6.7}) and Lem.~\ref{5.2C}(iv) yield
\begin{equation}
\Y_{1}(\x) \text{ is } \bC^{0+} \text{ on } \bar{\I}(\x).
\label{6.11a}
\end{equation}

Next, use Eqs.\ (\ref{6.3}) and (\ref{6.6a}) to replace $U_{1}(\sigma)$
and $K_{21}(\sigma',\sigma)$ in Eq.\ (\ref{6.7}).  Since $k_{21}$ is
$\bC^{1+}$ by Prop.~\ref{6.2A}, we can use statement (\ref{6.10a}) to
obtain the following equivalent of the Fredholm equation (\ref{6.7}):
\begin{equation}
\Y_{1}(\sigma) + \frac{1}{\pi i} \int_{\bar{\I}} d\sigma' \nu^{+}(\sigma')
\frac{\psi(\sigma')+W_{1}(\sigma')}{\sigma'-\sigma} = 0,
\label{6.11b}
\end{equation}
where
\begin{equation}
\psi(\sigma) := \frac{1}{\pi i} \int d\sigma' \nu^{+}(\sigma')^{-1}
\Y_{1}(\sigma') k_{21}(\sigma',\sigma).
\label{6.11c}
\end{equation}
{}From Lem.~\ref{5.2C}(iv) and (\ref{6.11a}),
\begin{equation}
\psi \text{ is } \bC^{0+} \text{ on } \bar{\I}(\x).
\end{equation}

Next, after replacing `$\sigma$' by `$\sigma''$' in Eq.\ (\ref{6.11b})
and then applying the operator
$$
\frac{1}{\pi i} \int_{\bar{\I}} d\sigma'' \nu^{+}(\sigma'')^{-1}
\frac{1}{\sigma''-\sigma},
$$
one finds from Eq.\ (\ref{6.10b}) that
\begin{equation}
\frac{1}{\pi i} \int_{\bar{\I}} d\sigma'' \nu^{+}(\sigma'')^{-1}
\frac{\Y_{1}(\sigma'')}{\sigma''-\sigma} + \psi(\sigma) + W_{1}(\sigma)
= 0.
\label{6.11e}
\end{equation}
Substitute from Eq.\ (\ref{6.11c}) into the above Eq.\ (\ref{6.11e}),
and then use Eq.\ (\ref{6.2c}) to replace $k_{21}(\sigma',\sigma)$ in
the integrand, whereupon two terms involving $\Y_{1}$ cancel one another;
and the result is the Alekseev-type singular integral equation in the form
\begin{equation}
\frac{1}{\pi i} \int_{\bar{\I}} d\sigma' \nu^{+}(\sigma')^{-1}
\Y_{1}(\sigma') \frac{d_{21}(\sigma',\sigma)}{\sigma'-\sigma}
+ W_{1}(\sigma) = 0 \text{ for all } \sigma \in \I(\x),
\label{6.12}
\end{equation}
where $\Y_{1}(\x)$ is $\bC^{0+}$ according to Eq.\ (\ref{6.11a}).
\cheers
\end{abc}

Let us summarize the results given by GSSM~\ref{5.1D} and GSSM~\ref{6.1B}
when $\bv \in K^{2+}$.

\begin{theorem}[Summary]
\label{6.2B} \mbox{ } \\
Suppose $\bv \in K^{2+}$, $\x \in \domE$, and $\bar{\F}(\x)$ and
$\Y_{1}(\x)$ are $2 \times 2$ and $2 \times 1$ matrix functions,
respectively, such that
\begin{equation}
\dom{\bar{\F}(\x)} = C - \bar{\I}(\x) \text{ and }
\dom{\Y_{1}(\x)} = \bar{\I}(\x).
\end{equation}
Then the following three statements are equivalent to one another:
\begin{romanlist}
\item 
The function $\bar{\F}(\x)$ is the solution of the HHP corresponding to
$(\bv,\bar{\F}^{M},\x)$, and $\Y_{1}(\x)$ is the function whose restriction
to $\I(\x)$ is defined by Eq.\ (\ref{5.25a}) and whose extension to 
$\bar{\I}(\x)$ is then defined by Eqs.\ (\ref{5.23d}) and (\ref{5.24d}).
[The existence and uniqueness of this extension is asserted by
Cor.~\ref{5.5C}.]
\item 
The function $\Y_{1}(\x)$ is $\bC^{0+}$ and its restriction to $\I(\x)$
is a solution of the Alekseev-type equation (\ref{5.24e}) [or (\ref{6.12})];
and $\bar{\F}(\x)$ is defined in terms of $\Y_{1}(\x)$ by Eq.\ (\ref{5.24b}).
\item 
The function $\Y_{1}(\x)$ is summable over $\bar{\I}(\x)$ and is a solution
of the Fredholm equation (\ref{6.7}) for all $\sigma \in \bar{\I}(\x)$.
\end{romanlist}
\end{theorem}

\proof
Directly from GSSM~\ref{5.1D} and GSSM~\ref{6.1B}.
\cheers

\begin{corollary}[Uniqueness of solutions]
\label{6.3B} \mbox{ } \\
When $\bv \in K^{2+}$, each of the solutions defined in (i), (ii) and~(iii)
of the preceding theorem is unique if it exists.
\end{corollary}

\proof
This follows from the preceding theorem and the uniqueness theorem
[GSSM~\ref{4.3D}(iv)] for the HHP.
\cheers

\setcounter{equation}{0}
\setcounter{theorem}{0}
\subsection{Homogenized equations, propositions, theorems, etc.}

By considering a homogeneous version of the Fredholm equation (\ref{6.7}),
we found it possible to employ the Fredholm alternative theorem to establish
the existence of the solution of the HHP corresponding to $(\bv,\bar{\F}^{M})$
when $\bv \in K^{2+}$.  In order to avoid having to restate many of our 
equations, propositions, theorems, corollaries and lemmas for the homogeneous 
case, we shall instead define various conventions.

\begin{definition}{Dfn.\ of \zip{HHP}\label{def66}}
The HHP that is defined as in Sec.~\ref{Sec_4}C except that the condition
(2) is replaced by the condition
\begin{equation}
\bar{\F}(\x,\infty) = 0 \text{ (\zip{HHP} condition)}
\label{6.14}
\end{equation}
will be called the \zip{HHP} {\em corresponding to} $(\bv,\bar{\F}_{0},\x)$.
\end{definition}

Clearly, the $2 \times 2$ matrix function $\bar{\F}(\x)$ with the domain
$C - \bar{\I}(\x)$ and the value $\bar{\F}(\x,\tau)=0$ for all $\tau$ in
this domain is a solution of the \zip{HHP} corresponding to 
$(\bv,\bar{\F}_{0},\bv)$.  It will be called the {\em zero solution}.

\begin{definition}{Dfn.\ of equation number with attached subscript `$0$'
\label{def67}}
Each equation which occurs in these notes from GSSM~\ref{5.4A} to
Thm.~\ref{6.2B}, inclusive, and which has a term that is an integral whose
integrand involves `$\bar{\F}$', `$\F^{\pm}$', `$\Y$' or `$\Y^{(i)}$'
(or one of their columns) will be called an {\em integral equation}.
Each of these integral equations is expressible in the form
$$
F(\bar{\F},\F^{+},\F^{-},\Y,\Y^{(i)}) = J,
$$
where neither side of the above equation is identically zero, $J$ is given
and is independent of the choice of $(\bar{\F},\F^{+},\F^{-},\Y,\Y^{(i)})$
and $F(\bar{\F},\F^{+},\F^{-},\Y,\Y^{(i)})$ is a homogeneous functional
of the first degree; i.e., 
$$
F(\alpha\bar{\F},\alpha\F^{+},\alpha\F^{-},\alpha\Y,\alpha\Y^{(i)}) =
\alpha F(\bar{\F},\F^{+},\F^{-},\Y,\Y^{(i)}) \text{ for all finite
complex numbers } \alpha.
$$
We shall let the corresponding homogeneous integral equation
$$
F(\bar{\F},\F^{+},\F^{-},\Y,\Y^{(i)}) = 0
$$
be designated by the symbol that results when the subscript `$0$' is
attached to the equation number for the inhomogeneous integral equation.
For example, Eqs.\ \zip{(\ref{5.13b})}, \zip{(\ref{5.13c})}, 
\zip{(\ref{5.14b})}, \zip{(\ref{5.14c})}, \zip{(\ref{5.23b})} and 
\zip{(\ref{5.24b})} will designate the results of deleting the
unit matrix terms from Eqs.\ (\ref{5.13b}), (\ref{5.13c}), (\ref{5.14b}),
(\ref{5.14c}), (\ref{5.23b}) and (\ref{5.24b}), respectively; and
\zip{(\ref{6.11b})} will denote the result of deleting `$W_{1}(\sigma')$'
from the integrand in Eq.\ (\ref{6.11b}).
\end{definition}

The first theorems on the HHP corresponding to $(\bv,\bar{\F}_{0},\x)$
are given in Sec.~\ref{Sec_4}C.  

\begin{definition}{Dfn.\ of theorem label (etc.) with attached subscript
`$0$'\label{def68}}
When a {\em new} valid assertion results from subjecting a labelled
assertion to the following substitutions, that new valid assertion will
bear the same label with an attached subscript `$0$'.
\begin{arablist}
\item 
`HHP' $\rightarrow$ `\zip{HHP}'
\item 
$\bar{\F}(\x,\infty)=I$' $\rightarrow$ `$\bar{\F}(\x,\infty)=0$' 
in condition (2) of the HHP
\item 
each integral equation $\rightarrow$ the corresponding homogeneous
integral equation
\item 
each equation number for an integral equation $\rightarrow$ the same
equation number with an attached subscript `$0$'.
\end{arablist}
\end{definition}

With this convention, the complete list of {\em new} valid propositions,
theorems, corollaries and lemmas is as follows:
\zip{\ref{4.2D}}, \zip{\ref{4.3D}}(i)~and~(ii), 
\zip{\ref{5.1A}}, \zip{\ref{5.2A}}, \zip{\ref{5.3A}}, \zip{\ref{5.4A}},
\zip{\ref{5.4C}}, \zip{\ref{5.5C}},
\zip{\ref{6.3A}}, \zip{\ref{6.1B}}, \zip{\ref{6.2B}}. 

\setcounter{equation}{0}
\setcounter{theorem}{0}
\subsection{Existence and uniqueness of HHP solution}

For our immediate purpose, we shall need the following explicit version of
Thm.~\ref{6.2B}:

\begin{theorem}[Theorem \zip{\ref{6.2B}}]
\label{6.1C} \mbox{ } \\
Suppose $\bv \in K^{2+}$, $\x \in \domE$, and $\bar{\F}(\x)$ and $\Y_{1}(\x)$
are $2 \times 2$ and $2 \times 1$ matrix functions, respectively, such that
\begin{equation}
\dom{\bar{\F}(\x)} = C - \bar{\I}(\x) \text{ and }
\dom{\Y_{1}(\x)} = \bar{\I}(\x).
\end{equation}
Then the following three statements are equivalent to one another:
\begin{romanlist}
\item 
The function $\bar{\F}(\x)$ is a solution of the \zip{HHP} corresponding
to $(\bv,\bar{\F}^{M},\x)$; and $\Y_{1}(\x)$ is the continuous function
whose restriction to $\I(\x)$ is defined in terms of $\F^{\pm}(\x)$ by
Eq.\ (\ref{5.25a}), and whose existence and uniqueness are asserted by
Cor.~\ref{5.5C}.
\item 
The function $\Y_{1}(\x)$ is $\bC^{0+}$ and its restriction to $\I(\x)$
is a solution of Eq.\ \zip{(\ref{5.24e})}; and $\bar{\F}(\x)$ is defined
in terms of $\Y_{1}(\x)$ by Eq.\ \zip{(\ref{5.24b})}.
\item 
The function $\Y_{1}(\x)$ is summable over $\bar{\I}(\x)$ and is a 
solution of the homogeneous Fredholm integral equation \zip{(\ref{6.7})}
for all $\sigma \in \bar{\I}(\x)$.
\end{romanlist}
\end{theorem}

\proof
This theorem summarizes Thms.~\zip{\ref{5.1D}} and \zip{\ref{6.1B}} for
the case $\bv \in K^{2+}$.
\cheers

The following theorem is clearly a key one.

\begin{theorem}[Only a zero solution of \zip{HHP}]
\label{6.2C} \mbox{ } \\
For each $\bv \in K$, $\bar{\F}_{0} \in \S_{\bar{\F}}$ and $\x \in \domE$,
the only solution of the \zip{HHP} corresponding to $(\bv,\bar{\F}_{0},\x)$
is its zero solution.
\end{theorem}

\proof
The proof will be given in four parts:
\begin{arablist}
\item 
{}From Thm.~\ref{1.1D}(viii) and Prop.~\ref{1.2F}(i), and from the hypothesis
$\bar{\F}_{0} \in \S_{\bar{\F}}$,
\begin{abc}
\begin{equation}
\left[ \F_{0}^{\mp}(\x,\tau^{*})\right]^{\dagger} \A_{0}(\x,\tau)
\F_{0}^{\pm}(\x,\tau) = \A_{0}(\x_{0},\tau) \text{ for all }
\tau \in \bar{C}^{\pm}-\{r,s,r_{0},s_{0},\infty\},
\label{6.15a}
\end{equation}
where
\begin{equation}
\A_{0}(\x,\tau) := (\tau-z)\Omega + \Omega h_{0}(\x) \Omega
\label{6.15b}
\end{equation}
and $h_{0}(\x)$ is defined in terms of $\E_{0} \in \S_{\E}$ by Eqs.\
(\ref{1.11a}) to (\ref{1.11c}).  Since
\begin{equation}
h_{0}(\x_{0}) := h^{M}(\x_{0}) = \left( \begin{array}{cc}
\rho_{0}^{2} & 0 \\ 0 & 1
\end{array} \right)
\label{6.15c}
\end{equation}
in our gauge,
\begin{equation}
\A_{0}(\x_{0},\tau) = \A^{M}(\x_{0},\tau).
\label{6.15d}
\end{equation}
Equation (\ref{6.15a}) is clearly expressible in the alternative form
\begin{eqnarray}
\lefteqn{\F_{0}^{\pm}(\x,\tau) \left[\A^{M}(\x_{0},\tau)\right]^{-1} 
\left[\F_{0}^{\mp}(\x,\tau^{*})\right]^{\dagger} = 
\left[\A_{0}(\x,\tau)\right]^{-1} } \nonumber \\
& & \text{for all } \tau \in \bar{C}^{\pm}-\{r,s,r_{0},s_{0},\infty\},
\label{6.15e}
\end{eqnarray}
since $[\F_{0}^{\pm}(\x,\tau)]^{-1}$ exists for all $\tau \in \bar{C}^{\pm}
-\{r,s,r_{0},s_{0}\}$, and
\begin{equation}
\left[ \A_{0}(\x,\tau) \right]^{-1} =
\frac{\B_{0}(\x,\tau)}{\rho^{2}-(\tau-z)^{2}},
\label{6.15f}
\end{equation}
where
\begin{equation}
\B_{0}(\x,\tau) := h_{0}(\x) - (\tau-z) \Omega
\label{6.15g}
\end{equation}
exists for all $\tau \in C - \{r,s\}$.
\end{abc}

\item 
Next, condition (3) in the definition of the HHP (and the \zip{HHP}) that
is given in Sec.~\ref{Sec_4}C asserts that $\F^{\pm}(\x)$ exist, and
Eq.\ (\ref{G3.17}) is expressible in the form
\begin{abc}
\begin{equation}
\F^{\pm}(\x,\sigma) = Y^{(i)}(\sigma) \F_{0}^{\pm}(\x,\sigma)
[\tilde{v}^{(i)}(\sigma)]^{-1} \text{ for each } i \in \{3,4\}
\text{ and } \sigma \in \I(\x).
\label{6.16a}
\end{equation}
{}From Cor.~\ref{4.2A},
\begin{equation}
[\tilde{v}^{(i)}(\sigma)]^{-1} [\A^{M}(\x_{0},\sigma)]^{-1}
[\tilde{v}^{(i)}(\sigma)^{\dagger}]^{-1} = \A^{M}(\x_{0},\sigma)^{-1}
\text{ for all } \sigma \in \I^{(i)} - \{r,s\}.
\label{6.16b}
\end{equation}
Therefore, from Eqs.\ (\ref{6.16a}), (\ref{6.16b}) and (\ref{6.15e}),
\begin{eqnarray}
\F^{\pm}(\x,\sigma) [\A^{M}(\x_{0},\sigma)]^{-1}
[\F^{\mp}(\x,\sigma)]^{\dagger} & = & Y(\x,\sigma) [\A_{0}(\x,\sigma)]^{-1}
Y(\x,\sigma)^{\dagger} \nonumber \\
& & \text{for all } \sigma \in \I(\x);
\label{6.16c}
\end{eqnarray}
or, equivalently, with the aid of Eqs.\ (\ref{6.15f}), (\ref{6.15g})
and (\ref{6.15c}),
\begin{eqnarray}
\lefteqn{\left[ \frac{\rho^{2}-(\sigma-z)^{2}}
{\rho_{0}^{2}-(\sigma-z_{0})^{2}} \right]
\F^{\pm}(\x,\sigma) \B^{M}(\sigma) [\F^{\mp}(\x,\sigma)]^{\dagger} = } 
\nonumber \\ 
& & Y(\x,\sigma) \B_{0}(\x,\sigma) Y(\x,\sigma)^{\dagger}
\text{ for all } \sigma \in \I(\x),
\label{6.16d}
\end{eqnarray}
where
\begin{equation}
\B^{M}(\tau) := \left( \begin{array}{cc}
\rho_{0}^{2} & -i(\tau-z_{0}) \\ i(\tau-z_{0}) & 1
\end{array} \right).
\label{6.16e}
\end{equation}
\end{abc}

\item 
Next, let $Z(\x)$ denote the function with the (tentative) domain
$C - \bar{\I}(\x)$ and the values
\begin{abc}
\begin{eqnarray}
Z(\x,\tau) & := & Z(\x)(\tau) \nonumber \\
& := & \bar{\bnu}(\x,\x_{0},\tau)^{-1}
\bar{\F}(\x,\tau) \B^{M}(\tau) 
[\bar{\bnu}(\x,\x_{0},\tau^{*})^{-1}\bar{\F}(\x,\tau^{*})]^{\dagger} 
\nonumber \\
& & \text{for all } \tau \in C - \bar{\I}(\x),
\label{6.17a}
\end{eqnarray}
where note that
\begin{equation}
\bar{\bnu}(\x,\x_{0},\tau)^{-2} =
\frac{(\tau-r)(\tau-s)}{(\tau-r_{0})(\tau-s_{0})} =
\frac{(\tau-z)^{2}-\rho^{2}}{\tau-z_{0})^{2}-\rho_{0}^{2}}.
\label{6.17b}
\end{equation}

We again appeal to the trilogy of elementary theorems due to Riemann and
Liouville.\footnote{See Ref.~\ref{trilogy}.}  Using these, we shall define 
an extension of $Z(\x)$, and we shall let $Z(\x)$ denote this extension as 
well.

{}From condition (1) in the definition of the HHP (and the \zip{HHP}),
and from Eqs.\ (\ref{6.17a}), (\ref{6.16e}) and (\ref{6.14}),
\begin{equation}
Z(\x,\tau) \text{ is a holomorphic function of } \tau 
\text{ throughout } C - \bar{\I}(\x),
\label{6.17c}
\end{equation}
and
\begin{equation}
Z(\x,\infty) = 0.
\label{6.17d}
\end{equation}
Let ($\Im{\zeta}>0$)
\begin{equation}
Z^{\pm}(\x,\sigma) := \lim_{\zeta \rightarrow 0} Z(\x,\sigma \pm \zeta)
\text{ for all } \sigma \in \I(\x),
\label{6.17e}
\end{equation}
which exist according to condition (3) in the definition of the HHP
(and the \zip{HHP}).  Then, from Eqs.\ (\ref{6.17a}), (\ref{6.17b})
and (\ref{6.16d}),
\begin{equation}
Z^{+}(\x,\sigma) = Z^{-}(\x,\sigma) 
= Y(\x,\sigma) \B_{0}(\x,\sigma) Y(\x,\sigma)^{\dagger}
\text{ for all } \sigma \in \I(\x).
\label{6.17f}
\end{equation}
The above equation permits us to define a single valued extension of
$Z(\x)$ to the domain $C - \{r,s,r_{0},s_{0}\}$ by letting
\begin{equation}
Z(\x,\sigma) := Z^{\pm}(\x,\sigma)
= Y(\x,\sigma) \B_{0}(\x,\sigma) Y(\x,\sigma)^{\dagger}
\text{ for all } \sigma \in \I(\x),
\label{6.17g}
\end{equation}
whereupon, from (\ref{6.17c}), (\ref{6.17g}) and the theorem on analytic
continuation across an arc,
\begin{equation}
Z(\x,\tau) \text{ is a holomorphic function of } \tau 
\text{ throughout } C -\{r,s,r_{0},s_{0}\}.
\label{6.17h}
\end{equation}

We next apply condition (4) in the definition of the HHP (and \zip{HHP}).
Since, according to condition (4), $\bar{\nu}(\x)^{-1} \bar{\F}(\x)$
and $Y(\x)$ are both bounded at $\x$, Eqs.\ (\ref{6.17a}) and (\ref{6.17g})
yield
\begin{equation}
\begin{array}{l}
\text{There exists a positive real number } M_{1}(\x) \text{ such that } \\
||Z(\x,\tau)|| < M_{1}(\x) \text{ as } \tau \rightarrow r \text{ and as } 
\tau \rightarrow s \\
\text{through any sequence of points in } C - \{r,s,r_{0},s_{0}\}.
\end{array}
\label{6.17i}
\end{equation}
Since $\bar{\F}(\x)$ and $Y(\x)$ are both bounded at $\x_{0}$, Eqs.\
(\ref{6.17a}), (\ref{6.17b}) and (\ref{6.17g}) yield
\begin{equation}
\begin{array}{l}
\text{There exists a positive real number } M_{2}(\x) \text{ such that } \\
||(\tau-r_{0})(\tau-s_{0})Z(\x,\tau)|| < M_{2}(\x) 
\text{as } \tau \rightarrow r_{0} \text{ and as } \tau \rightarrow s_{0} \\
\text{through any sequence of points in } C - \{r,s,r_{0},s_{0}\}.
\end{array}
\label{6.17j}
\end{equation}
However, since $Y(\x)$ is bounded at $\x_{0}$, Eq.\ (\ref{6.17g}) yields

\begin{equation}
\begin{array}{l}
\text{There exists a positive real number } M_{3}(\x) \text{ such that } \\
||Z(\x,\sigma)|| < M_{3}(\x) \text{ as } \sigma \rightarrow r_{0}
\text{ and as } \sigma \rightarrow s_{0} \\ 
\text{through any sequence of points in } \I(\x).
\end{array}
\label{6.17k}
\end{equation}
The theorem on isolated singularities, together with statements (\ref{6.17h})
to (\ref{6.17k}), now informs us that
\begin{equation}
Z(\x) \text{ has a holomorphic extension [which we also denote by $Z(\x)$]
to C,}
\label{6.17l}
\end{equation}
whereupon Eq.\ (\ref{6.17d}) and the (generalized) theorem of Liouville yield
\begin{equation}
Z(\x,\tau) = 0 \text{ for all } \tau \in C.
\label{6.17m}
\end{equation}
\end{abc}

\item 
Putting (\ref{6.17a}) and (\ref{6.17m}) together, one obtains
\begin{abc}
\begin{equation}
\bar{\F}(\x,\sigma) \B^{M}(\sigma) \bar{\F}(\x,\sigma)^{\dagger} = 0
\text{ for all } \sigma \in C - \bar{\I}(\x).
\label{6.18a}
\end{equation}
Note from Eq.\ (\ref{6.16e}), $\B^{M}(\sigma)$ is hermitian,
\begin{equation}
\begin{array}{rcl}
\tr{\B^{M}(\sigma)} & = & 1 + \rho_{0}^{2} \; \text{ and } \\
\det{\B^{M}(\sigma)} & = & (s_{0}-\sigma)(\sigma-r_{0}).
\end{array}
\end{equation}
Recall that $|r,r_{0}| < |s,s_{0}|$ for any type~A triple $\triple$;
and it is clear that
\begin{equation}
\B^{M}(\sigma) \text{ is hermitian and positive definite for all }
|r,r_{0}| < \sigma < |s,s_{0}|.
\end{equation}
Therefore, Eq.\ (\ref{6.18a}) implies
\begin{equation}
\bar{\F}(\x,\sigma) = 0 \text{ for all } \sigma \text{ such that }
|r,r_{0}| < \sigma < |s,s_{0}|.
\end{equation}
However, $\bar{\F}(\x,\tau)$ is a holomorphic function of $\tau$
throughout $C - \bar{\I}(\x)$, and this domain contains the open
interval between $|r,r_{0}|$ and $|s,s_{0}|$.  So,
\begin{equation}
\bar{\F}(\x,\tau) = 0 \text{ for all } \tau \in C - \bar{\I}(\x).
\end{equation}
\end{abc}
\end{arablist}
\cheers

\begin{theorem}[Only a zero solution of \zip{(\ref{6.7})}]
\label{6.3C} \mbox{ } \\
The only solution of Eq.\ \zip{(\ref{6.7})} [a homogeneous Fredholm
integral equation of the second kind] is its zero solution.
\end{theorem}

\begin{abc}
\proof
Let $\Y_{1}(\x)$, with domain $\bar{\I}(\x)$, denote a solution of
Eq.\ \zip{(\ref{6.7})}; and let $\bar{\F}(\x)$, with domain $C -
\bar{\I}(\x)$, be defined in terms of $\Y_{1}(\x)$ by Eq.\ \zip{(\ref{5.24b})}.
Using Thm.~\ref{6.1C}, one obtains
\begin{equation}
\bar{\F}(\x) \text{ is a solution of the \zip{HHP} corresponding to }
(\bv,\bar{\F}^{M},\x),
\end{equation}
whereupon Thm.~\ref{6.2C} delivers
\begin{equation}
\bar{\F}(\x,\tau) = 0 \text{ for all } \tau \in C - \bar{\I}(\x).
\end{equation}
It follows that
\begin{equation}
\F^{\pm}(\x,\sigma) = 0 \text{ for all } \sigma \in \I(\x),
\end{equation}
whereupon, from Thm.~\ref{6.2C}(i), Eq.\ (\ref{5.25a}) and the continuity
of $\Y_{1}(\x)$,
\begin{equation}
\Y_{1}(\x,\sigma) = 0 \text{ for all } \sigma \in \bar{\I}(\x).
\end{equation}
\cheers
\end{abc}

At this point, we note that Eq.\ (\ref{6.7}) is a regular Fredholm equation
in disguise when $\bv \in K^{2+}$.  The transformation\footnote{See 
Appendix~D.} that is detailed by Eqs.\ (\ref{5.26a}) to (\ref{5.29b}) [see, 
in particular, Thm.~\ref{5.2B}] yields the following equation, which is 
equivalent to Eq.\ (\ref{6.7}) and has a $\bC^{0+}$ kernel and a $\bC^{1+}$
inhomogeneous term:
\begin{abc}
\begin{eqnarray}
\lefteqn{y_{1}(\x,\theta) - \frac{2}{\pi} \int_{\Theta} d\theta'
y_{1}(\x,\theta') \kappa_{21}(\x,\theta',\theta) } \nonumber \\
& = & u_{1}(\x,\theta) \text{ for all } \theta \in \Theta :=
[0,\pi/2] \cup [\pi,3\pi/2],
\label{6.20a}
\end{eqnarray}
where
\begin{eqnarray}
y_{1}(\x,\theta) & := & \Y_{1}(\x,\sigma(\x,\theta)),
\label{6.20b} \\
u_{1}(\x,\theta) & := & U_{1}(\x,\sigma(\x,\theta)),
\label{6.20c} \\
\kappa_{21}(\x,\theta',\theta) & := & q(\x,\theta')
K_{21}(\x,\sigma(\x,\theta'),\sigma(\x,\theta))
\end{eqnarray}
and
\begin{equation}
q(\x,\theta) := \left\{ \begin{array}{l}
(r_{0}-r)\cos^{2}\theta
\sqrt{\frac{s-\sigma(\x,\theta)}{s_{0}-\sigma(\x,\theta)}}
\text{ when } \theta \in [0,\pi/2], \\
(s_{0}-s)\cos^{2}\theta
\sqrt{\frac{\sigma(\x,\theta)-r}{\sigma(\x,\theta)-r_{0}}}
\text{ when } \theta \in [\pi,3\pi/2].
\end{array} \right.
\label{6.20e}
\end{equation}

Equations (\ref{6.2d}) and (\ref{6.2e}) are expressible in the following
forms, in which $\x$ and $x_{0}$ are no longer suppressed:
\begin{eqnarray}
U_{1}(\x,\sigma) & = & W_{1}(\sigma) - \frac{2}{\pi} \int_{\Theta}
d\theta' p(\x,\theta') L_{1}(\sigma(\x,\theta'),\sigma), 
\label{G5.3e} \\
K_{21}(\x,\sigma',\sigma) & = & k_{21}(\sigma',\sigma) - \frac{2}{\pi}
\int_{\Theta} d\theta'' p(\x,\theta'')
\lambda(\sigma',\sigma(\x,\theta''),\sigma),
\label{G5.3f}
\end{eqnarray}
where
\begin{equation}
p(\x,\theta) := \left\{ \begin{array}{l}
(r_{0}-r)\sin^{2}\theta
\sqrt{\frac{s_{0}-\sigma(\x,\theta)}{s-\sigma(\x,\theta)}}
\text{ when } \theta \in [0,\pi/2], \\
(s_{0}-s)\sin^{2}\theta
\sqrt{\frac{\sigma(\x,\theta)-r_{0}}{\sigma(\x,\theta)-r}}
\text{ when } \theta \in [\pi,3\pi/2].
\end{array} \right.
\label{G5.3g}
\end{equation}
\end{abc}

\begin{theorem}[Fredholm determinant not zero]
\label{6.4C} \mbox{ } \\
The Fredholm determinant corresponding to the kernel $\kappa_{21}(\x)$
is not zero.  Therefore, there exists exactly one solution of Eq.\
(\ref{6.20a}) for each given $\bv \in K^{2+}$ and $\x \in \domE$; or,
equivalently, there exists exactly one solution of Eq.\ (\ref{6.7})
for each given $\bv \in K^{2+}$ and $\x \in \domE$.
\end{theorem}

\proof
This follows from Thm.~\ref{6.3C} and the Fredholm alternative.
\cheers

Thus, in summation, we have the following theorem:

\begin{gssm}[Existence and uniqueness of HHP solution]
\label{Thm_13} \mbox{ } \\ \vspace{-3ex}
\begin{romanlist}
\item
If $\bv \in K^{2+}$, then the \zip{HHP} corresponding to 
$(\bv,\bar{\F}^{M},\x)$ is equivalent to the homogeneous Fredholm
equation of the second kind that is obtained from Eq.\ (\ref{6.7}) by
deleting the term $U_{1}(\sigma)$, provided that the term $W_{1}(\sigma)$
is also deleted from the expression (\ref{5.24d}) for $\Y_{2}(\sigma)$. 
\item
For any given $\x \in \domE$, and $\bv \in K$, the \zip{HHP} corresponding
to $(v,\bar{\F}^{M},\x)$ has the unique solution $\bar{\F}(\x,\tau)=0$ for
all $\tau \in C - \I(\x)$.
\item
Therefore, if $\bv \in K^{2+}$, the only solution of the homogeneous
Fredholm equation is the zero solution.  Hence from the Fredholm alternative
theorem, the inhomogeneous Fredholm equation (\ref{6.7}) has exactly one
solution.  We conclude that there exists one and only one solution
of the HHP corresponding to $(\bv,\bar{\F}^{M})$ when $\bv \in K^{2+}$.
\end{romanlist}
\end{gssm}

\proof
Directly from Thms.~\ref{6.1C}, \ref{6.2C}, \ref{6.3C} and \ref{6.4C}.
\cheers

\newpage

\setcounter{equation}{0}
\setcounter{theorem}{0}
\section{Properties of the HHP solution $\bar{\F}(\x,\tau)$\label{Sec_7}}

\subsection{The $2 \times 2$ matrix $H(\x)$ associated with each solution
of the HHP corresponding to $(\bv,\bar{\F}_{0},\x)$ when $\bv \in K$}

\begin{abc}
\begin{proposition}[Properties of $H(\x)$ and $h(\x)$]
\label{7.1A} \mbox{ } \\
For each $\bv \in K$, $\bar{\F}_{0} \in \S_{\bar{\F}}$, $\x \in \domE$
and solution $\bar{\F}(\x)$ of the HHP corresponding to
$(\bv,\bar{\F}_{0},\x)$, there exists exactly one $2 \times 2$ matrix
$H(\x)$ such that
\begin{eqnarray}
\bar{\F}(\x,\tau) & = & I + (2\tau)^{-1}\left[H(\x)-H^{M}(\x)\right]\Omega
+ O(\tau^{-2}) \nonumber \\
& & \text{in at least one neighborhood of } \tau = \infty.
\label{7.1a}
\end{eqnarray}
Moreover,
\begin{eqnarray}
H(\x_{0}) & = & H^{M}(\x_{0}),
\label{7.1b} \\
H(\x)-H(\x)^{T} & = & 2z \Omega,
\label{7.1c} \\
h(\x) & := & -\Re{H(\x)} \text{ is symmetric,}
\label{7.1d}
\end{eqnarray}
and
\begin{equation}
h(\x_{0}) = \left( \begin{array}{cc}
\rho_{0}^{2} & 0 \\ 0 & 1
\end{array} \right).
\label{7.1e}
\end{equation}
\end{proposition}
\end{abc}

\begin{abc}
\proof
{}From conditions (1) and (2) in the definition of the HHP, there exists
exactly one $2 \times 2$ matrix $B(\x)$ such that
\begin{eqnarray*}
\bar{\F}(\x,\tau) & = & I + (2\tau)^{-1} B(\x) + O(\tau^{-2}) \\
& & \text{in at least one neighborhood of } \tau = \infty.
\end{eqnarray*}
Let
$$
H(\x) := H^{M}(\x_{0}) + B(\x) \Omega,
$$
whereupon statement (\ref{7.1a}) follows.  From GSSM~\ref{4.3D}(v)
[Eq.\ (\ref{4.30})], $B(\x_{0}) = 0$, whereupon Eq.\ (\ref{7.1b}) follows.

Next, from GSSM~\ref{4.3D}(iii),
\begin{eqnarray}
\det{\bar{\F}(\x,\tau)} & = & \bar{\bnu}(\x,\x_{0},\tau) \nonumber \\
& = & 1 + (2\tau)^{-1} (r+s-r_{0}-s_{0}) + O(\tau^{-2}) \nonumber \\
& = & 1 + \tau^{-1} (z-z_{0}) + O(\tau^{-2}) \nonumber \\
& & \text{in at least one neighborhood of } \tau = \infty.
\label{7.2a}
\end{eqnarray}
Moreover, from Eq.\ (\ref{1.14c}),
\begin{equation}
H^{M}(\x_{0}) - [H^{M}(\x_{0})]^{T} = 2 z_{0} \Omega.
\label{7.2b}
\end{equation}
For any $2 \times 2$ matrix $M$, $M \Omega M^{T} = \Omega \det{M}$.  In
particular,
\begin{equation}
\bar{\F}(\x,\tau) \Omega \bar{\F}(\x,\tau)^{T} = \Omega
\bar{\bnu}(\x,\x_{0},\tau).
\label{7.2c}
\end{equation}
The next step is to consider Eq.\ (\ref{7.2c}) in at least one neighborhood
of $\tau = \infty$ for which the expansions given by Eqs.\ (\ref{7.1a}) and
(\ref{7.2a}) hold.  The reader can then easily deduce Eq.\ (\ref{7.1c}) by
using Eq.\ (\ref{7.2b}) and the relations $\Omega^{T} = - \Omega$ and
$\Omega^{2} = I$.

The statement (\ref{7.1d}) follows from Eq.\ (\ref{7.1c}) and the relation
$\Omega^{*} = - \Omega$.  Equation (\ref{7.1e}) is derived from Eqs.\
(\ref{7.1b}) and (\ref{1.14c}).
\cheers
\end{abc}

\begin{abc}
\begin{theorem}[Quadratic relation]
\label{7.2A} \mbox{ } \\
For each $\bv \in K$, $\bar{\F}_{0} \in \S_{\bar{\F}}$, $\x \in \domE$
and solution $\bar{\F}(\x)$ of the HHP corresponding to
$(\bv,\bar{\F}_{0},\x)$, let $h(\x)$ be defined as in the preceding
proposition, and let
\begin{equation}
\A(\x_{0},\tau) = (\tau-z) \Omega + \Omega h(\x) \Omega.
\label{7.3a}
\end{equation}
Then
\begin{equation}
\bar{\F}^{\dagger}(\x,\tau) \A(\x,\tau) \bar{\F}(\x,\tau) = \A(\x_{0},\tau)
\text{ for all } \tau \in [C - \bar{\I}(\x)] - \{\infty\},
\label{7.3b}
\end{equation}
where
\begin{equation}
\bar{\F}^{\dagger}(\x,\tau) := [\bar{\F}(\x,\tau^{*})]^{\dagger}
\text{ for all } \tau \in C - \bar{\I}(\x).
\end{equation}
\end{theorem}
\end{abc}

\proof
Note that parts (1) and (2) in the proof of Thm.~\ref{6.2C} remain valid
here.  For the sake of convenience, we repeat below Eq.\ (\ref{6.16d})
from part (2) of that proof.
\begin{abc}
\begin{eqnarray}
[\bar{\bnu}(\x,\x_{0},\sigma)]^{-2} \F^{\pm}(\x,\sigma)
\B^{M}(\sigma) [\F^{\mp}(\x,\sigma)]^{\dagger} & = & Y(\x,\sigma)
\B_{0}(\x,\sigma) Y(\x,\sigma)^{\dagger} \nonumber \\
& & \text{for all } \sigma \in \I(\x),
\label{7.4a}
\end{eqnarray}
where
\begin{eqnarray}
\B^{M}(\tau) & := & \left( \begin{array}{cc}
\rho_{0}^{2} & -i(\tau-z_{0}) \\ i(\tau-z_{0}) & 1
\end{array} \right),
\label{7.4b} \\
\B_{0}(\x,\tau) & := & h_{0}(\x) - (\tau-z) \Omega \nonumber \\
& = & [\rho^{2}-(\tau-z)^{2}] \A_{0}(\x,\tau)^{-1}, \\
\bar{\bnu}(\x,\x_{0},\tau)^{-2} & = &
\frac{(\tau-r)(\tau-s)}{(\tau-r_{0})(\tau-s_{0})} \nonumber \\
& = & \frac{(\tau-z)^{2}-\rho^{2}}{(\tau-z_{0})^{2}-\rho_{0}^{2}}.
\label{7.4d}
\end{eqnarray}
\end{abc}

Next, let $Z(\x)$ denote the function with the (tentative) domain
$[C - \bar{\I}(\x)] - \{\infty\}$ and the values
\begin{abc}
\begin{equation}
Z(\x,\tau) := \bar{\bnu}(\x,\x_{0},\tau)^{-1} \bar{\F}(\x,\tau)
\B^{M}(\tau) \left[ \bar{\bnu}(\x,\x_{0},\tau^{*})^{-1} \bar{\F}(\x,\tau^{*})
\right]^{\dagger}.
\label{7.5a}
\end{equation}
{}From conditions (1) and (2) in the definition of the HHP, and from Eqs.\
(\ref{7.4b}) and (\ref{7.5a}),
\begin{equation}
\begin{array}{l}
Z(\x,\tau) \text{ is a holomorphic function of } \tau \\
\text{throughout } [C - \bar{\I}(\x)] - \{\infty\} \\
\text{and has a simple pole at } \tau = \infty.
\end{array}
\label{7.5b}
\end{equation}
Note that Eq.\ (\ref{7.1e}) enables us to express (\ref{7.4b}) in the form
\begin{equation}
\B^{M}(\tau) = h(\x_{0}) - (\tau-z_{0}) \Omega.
\label{7.5c}
\end{equation}
Also, note that Eqs.\ (\ref{7.1c}) and (\ref{7.1d}) imply that
\begin{equation}
H(\x) + H(\x)^{\dagger} = - 2 h(\x) + 2 z \Omega
\label{7.5d}
\end{equation}
and that Eq.\ (\ref{7.4d}) yields
\begin{eqnarray}
\bar{\bnu}(\x,\x_{0},\tau)^{-2} & = & 1 + 2 \tau^{-1} (z_{0}-z) + 
O(\tau^{-2}) \nonumber \\
& & \text{in at least one neighborhood of } \tau = \infty.
\label{7.5e}
\end{eqnarray}
Upon using the relation $\bar{\bnu}(\x,\x_{0},\tau^{*})^{*} = 
\bar{\bnu}(\x,\x_{0},\tau)$ and upon inserting (\ref{7.1a}), (\ref{7.5c})
and (\ref{7.5e}) into the right side of Eq.\ (\ref{7.5a}), one obtains
the following result with the aid of Eqs.\ (\ref{7.1b}) and (\ref{7.4d}):
\begin{eqnarray}
Z(\x,\tau) & = & - (\tau-z) \Omega + h(\x) + O(\tau^{-1}) \nonumber \\
& & \text{in at least one neighborhood of } \tau = \infty.
\label{7.5f}
\end{eqnarray}
\end{abc}

We again appeal to the trilogy of elementary theorems due to Riemann and
Liouville.\footnote{See Ref.~\ref{trilogy}.}  We let $Z^{\pm}(\x,\sigma)$ 
be defined for all $\sigma \in \I(\x)$ by Eq.\ (\ref{6.17e}), whereupon 
Eqs.\ (\ref{7.4a}) and (\ref{7.5a}) yield
\begin{abc}
\begin{equation}
Z^{+}(\x,\sigma) = Z^{-}(\x,\sigma) = Y(\x,\sigma) \B_{0}(\x,\sigma)
Y(\x,\sigma)^{\dagger} \text{ for all } \sigma \in \I(\x).
\end{equation}
The above equation permits us to define a single valued extension of $Z(\x)$
to the domain $C - \{r,s,r_{0},s_{0},\infty\}$ by letting
\begin{equation}
Z(\x,\sigma) := Z^{\pm}(\x,\sigma) = Y(\x,\sigma) \B_{0}(\x,\sigma)
Y(\x,\sigma)^{\dagger} \text{ for all } \sigma \in \I(\x),
\label{7.6b}
\end{equation}
whereupon (\ref{7.5b}), (\ref{7.6b}) and the theorem on analytic continuation
across an arc tell us that
\begin{equation}
\begin{array}{l}
Z(\x,\tau) \text{ is a holomorphic function of } \tau \text{throughout } \\
C - \{r,s,r_{0},s_{0},\infty\} \text{ and has a simple pole at } \tau = \infty.
\end{array}
\label{7.6c}
\end{equation}
We next use condition (4) in the definition of the HHP, and we obtain the
statements (\ref{6.17i}), (\ref{6.17j}) and (\ref{6.17k}) exactly as we did
in the proof of Thm.~\ref{6.2C}.  The theorem on isolated singularities, 
together with the statements (\ref{7.6c}), (\ref{6.17i}), (\ref{6.17j}) and
(\ref{6.17k}) now inform us that
\begin{equation}
\begin{array}{l}
Z(\x) \text{ has a holomorphic extension [which we also denote } \\
\text{by $Z(\x)$] to } C - \{\infty\} \text{ and has a simple pole at }
\tau = \infty,
\end{array}
\end{equation}
whereupon Eq.\ (\ref{7.5f}) and the theorem on entire functions that do not
have an essential singularity at $\tau = \infty$ yield
\begin{equation}
Z(\x,\tau) = - (\tau-z) \Omega + h(\x) \text{ for all } \tau \in 
C - \{\infty\}.
\label{7.6e}
\end{equation}
\end{abc}

We are now close to completing our proof.  From GSSM~\ref{4.3D}(iii), Eqs.\ 
(\ref{7.5a}), (\ref{7.4b}) and (\ref{7.4d}),
\begin{abc}
\begin{equation}
\det{Z(\x,\tau)} = \rho^{2} - (\tau-z)^{2}.
\label{7.7a}
\end{equation}
Therefore, from Eqs.\ (\ref{7.3a}) and (\ref{7.6e}), the matrix
$-(\tau-z)\Omega+h(\x)$ is invertible when $\tau \notin \{r,s,\infty\}$,
and
\begin{equation}
\left[-(\tau-z)\Omega+h(\x)\right]^{-1} = 
\frac{\A(\x,\tau)}{\rho^{2}-(\tau-z)^{2}}.
\label{7.7b}
\end{equation}
[Above, we have used the fact that $M^{-1} = \Omega M^{T} \Omega/\det{M}$
for any invertible $2 \times 2$ matrix $M$.]

One then obtains from Eqs.\ (\ref{7.5a}), (\ref{7.5c}), (\ref{7.6e}) and
(\ref{7.7b}),
\begin{eqnarray*}
\lefteqn{\bar{\F}(\x,\tau) [\A(\x_{0},\tau)]^{-1} \bar{\F}^{\dagger}(\x,\tau)
= \A(\x,\tau)^{-1} } \nonumber \\
& & \text{for all } \tau \in [C - \bar{\I}(\x)] - \{\infty\},
\end{eqnarray*}
whereupon the conclusion (\ref{7.3b}) follows.
\cheers
\end{abc}

\begin{theorem}[More properties of $h(\x)$]
\label{7.3A} \mbox{ } \\
Grant the same premises as in the preceding two theorems, and let $h(\x)$
be defined as before.  Then
\begin{equation}
\det{h(\x)} = \rho^{2}
\end{equation}
and
\begin{equation}
h(\x) \text{ is positive definite }
\end{equation}
as well as real and symetric.
\end{theorem}

\begin{abc}
\proof
Since $h(\x)$ is symmetric
$$
\det{[h(\x)-(\tau-z)\Omega]} = \det{h(\x)} - (\tau-z)^{2}.
$$
Therefore, Eq.\ (\ref{7.6e}) and (\ref{7.7a}) imply that $\det{h(\x)} =
\rho^{2}$.

{}From Eqs.\ (\ref{7.5a}), (\ref{7.6e}) and (\ref{7.4d}),
\begin{eqnarray}
Z(\x,\sigma) & = & \frac{(\sigma-r)(s-\sigma)}{(\sigma-r_{0})(s_{0}-\sigma)}
\bar{\F}(\x,\sigma) \B^{M}(\sigma) \bar{\F}(\x,\sigma)^{\dagger} \nonumber \\
& = & - (\sigma-z) \Omega + h(\x) \text{ for all } |r,r_{0}| < \sigma <
|s,s_{0}|.
\label{7.9a}
\end{eqnarray}
Equation (\ref{7.4b}) provides us with
\begin{equation}
\det{\B^{M}(\sigma)} = (s_{0}-\sigma)(\sigma-r_{0}) \text{ and }
\tr{\B^{M}(\sigma)} = 1 + \rho_{0}^{2}.
\end{equation}
Therefore,
$$
\frac{(\sigma-r)(s-\sigma)}{(\sigma-r_{0})(s_{0}-\sigma)} \B^{M}(\sigma)
$$
is a positive definite hermitian matrix when $|r,r_{0}| < \sigma < |s,s_{0}|$.
Therefore, the left side of Eq.\ (\ref{7.9a}) is a positive definite hermitian
matrix when $|r,r_{0}| < \sigma < |s,s_{0}|$ and must, therefore, have a real
positive trace when $|r,r_{0}| < \sigma < |s,s_{0}|$.  So,
\begin{equation}
\tr{[-(\sigma-z)\Omega+h(\x)]} = \tr{h(\x)} > 0;
\end{equation}
and, since the determinant of $h(\x)$ is also positive, $h(\x)$ is positive
definite.
\cheers
\end{abc}

We caution the reader that the function $H$ whose domain is $\domE$ and whose
value at each $\x \in \domE$ is given by Eq.\ (\ref{7.1a}), where $\bar{\F}$
is the solution of the HHP corresponding to $(\bv,\bar{\F}_{0})$, is not
necessarily a member of $\S_{H}$.  It is one of the tasks of this section to
prove that $H \in \S_{H}$ when $\bv \in K^{3}$.

\setcounter{equation}{0}
\setcounter{theorem}{0}
\subsection{Fredholm equation solution $\Y_{1}$ corresponding to $\bv \in K^{3}$}

When $\bv \in K^{2+}$, the solution $\bar{\F}$ of the HHP may not be a
member of $\S_{\bar{\F}}$.  However, when $\bv \in K^{3}$, the kernel 
$K_{21}(\x,\sigma',\sigma)$ and the inhomogeneous term $U_{1}(\x,\sigma)$
in the Fredholm equation (\ref{6.7}) are $\bC^{1}$ and $\bC^{2}$ functions
of $(\x,\sigma',\sigma)$ and $(\x,\sigma)$, respectively.  This makes it 
possible to employ techniques introduced earlier by the present authors to
establish that the HHP solution $\bar{\F}$ is indeed a member of
$\S_{\bar{\F}}^{3}$.

We again refer the reader to the mappings $\btheta(\x):\bar{\I}(\x)
\rightarrow \Theta$ and $\bsigma(\x):\Theta \rightarrow \bar{\I}(\x)$
that are defined in Appendix~D, for we shall here be discussing the
solution $y_{1}$ of the Fredholm equation (\ref{6.20a}) with kernel
$\kappa_{21}$ and inhomogeneous term $u_{1}$ rather than the solution
$\Y_{1}$ of the Fredholm equation (\ref{6.7}) with kernel $K_{21}$ and
inhomogeneous term $U_{1}$.

\begin{abc}
\begin{definition}{Dfns.\ of $y_{1}$, $u_{1}$ and $\kappa_{21}$ when
$\bv \in K^{2+}$\label{def69}}
Let $y_{1}$, $u_{1}$ and $\kappa_{21}$ denote the functions such that
\begin{eqnarray}
\dom{y_{1}} = \dom{u_{1}} & := & \domE \times \Theta, \\
\dom{\kappa_{21}} & := & \domE \times \Theta^{2}
\end{eqnarray}
and the values $y_{1}(\x,\theta)$, $u_{1}(\x,\theta)$ and
$\kappa_{21}(\x,\theta',\theta)$ for each $(\x,\theta) \in \domE \times \Theta$
and each $(\x,\theta',\theta) \in \domE \times \Theta^{2}$ are given by Eqs.\
(\ref{6.20b}) to (\ref{6.20e}).  Note:  Thus, for each $\x \in \domE$, 
$y_{1}(\x,\theta)$ is the solution of the Fredholm equation of the second
kind (\ref{6.20a}); and, as we already know, this solution exists and is
unique when $\bv \in K^{2+}$.
\end{definition}
\end{abc}

\begin{abc}
\begin{definition}{Dfn.\ of $\lambda_{21}$ when $\bv \in K^{2+}$\label{def70}}
For each $\bv \in K^{2+}$ and $\x \in \domE$, let $\lambda_{21}$ denote
the function such that
\begin{equation}
\dom{\lambda_{21}} := \bar{\I}(\x)^{3}
\end{equation}
and
\begin{equation}
\lambda_{21}(\sigma',\sigma'',\sigma) := 
\frac{k_{21}(\sigma',\sigma'') - k_{21}(\sigma',\sigma)}{\sigma''-\sigma}
\label{7.11b}
\end{equation}
for each $(\sigma',\sigma'',\sigma) \in \bar{\I}(\x)^{3}$.
\end{definition}
\end{abc}

The following lemma is required for the proof of Thm.~\ref{7.3B}.

\begin{lemma}[Differentiability properties of $u_{1}$ and $\kappa_{21}$
when $\bv \in K^{3}$]
\label{7.1B} \mbox{ } \\
When $\bv \in K^{3}$, $u_{1}$ is $\bC^{2}$ and $\kappa_{21}$ is $\bC^{1}$.
Moreover, $\partial^{2}\kappa_{21}(\x,\theta',\theta)/\partial r \partial s$
exists and is a continuous function of $(\x,\theta',\theta)$ throughout
$\domE \times \Theta^{2}$ [whereupon, from a theorem of the calculus,
$\partial^{2}\kappa_{21}/\partial s \partial r$ also exists and is equal to
$\partial^{2}\kappa_{21}/\partial r \partial s$].
\end{lemma}

\proof
The proof will be given in three parts:
\begin{arablist}
\item 
{}From Eqs.\ (\ref{5.26h}) and (\ref{5.26i}), $\sigma(\x,\theta)$ is a real
analytic function of $\theta$ throughout $\domE \times \Theta$,
\begin{abc}
\begin{eqnarray}
\sigma(\x,\theta) & \in & |r,r_{0}| \text{ when } \theta \in [0,\pi/2], 
\label{7.12a} \\
\sigma(\x,\theta) & \in & |s,s_{0}| \text{ when } \theta \in [\pi,3\pi/2],
\label{7.12b} \\
\frac{\partial\sigma(\x,\theta)}{\partial s} & = & 0 \text{ when }
\theta \in [0,\pi/2], 
\label{7.12c} \\
\text{and} & & \nonumber \\
\frac{\partial\sigma(\x,\theta)}{\partial r} & = & 0 \text{ when }
\theta \in [\pi,3\pi/2].
\label{7.12d}
\end{eqnarray}
Therefore,
\begin{equation}
W(\sigma(\x,\theta)) \text{ is a } \bC^{3} \text{ function of }
(\x,\theta) \text{ throughout } \domE \times \Theta,
\label{7.12e}
\end{equation}
\begin{eqnarray}
\frac{\partial W(\sigma(\x,\theta))}{\partial s} & = & 0 \text{ when }
\theta \in [0,\pi/2], \\
\text{and} & & \nonumber \\
\frac{\partial W(\sigma(\x,\theta))}{\partial r} & = & 0 \text{ when }
\theta \in [\pi,3\pi/2];
\end{eqnarray}
and, from Eqs.\ (\ref{6.2a}) to (\ref{6.2c}),
\begin{equation}
d_{21}(\sigma(\x,\theta'),\sigma(\x,\theta)) \text{ is a } \bC^{3}
\text{ function of } (\x,\theta',\theta) \text{ throughout } \domE \times
\Theta^{2}, 
\label{7.12h}
\end{equation}
and
\begin{equation}
\begin{array}{r}
L(\sigma(\x,\theta'),\sigma(\x,\theta)) \text{ and }
k_{21}(\sigma(\x,\theta'),\sigma(\x,\theta)) \text{ are } \\
\bC^{2} \text{ functions of } (\x,\theta',\theta) \text{ throughout }
\domE \times \Theta^{2},
\end{array}
\label{7.12i}
\end{equation}
where, of course,
\begin{equation}
L(\sigma',\sigma) := \frac{W(\sigma')-W(\sigma)}{\sigma'-\sigma}
\text{ for all } (\sigma',\sigma) \in \bar{\I}(\x)^{2}.
\end{equation}
Furthermore, from Eq.\ (\ref{7.11b}) and statement (\ref{7.12i}),
\begin{equation}
\begin{array}{r}
\lambda_{21}(\sigma(\x,\theta'),\sigma(\x,\theta''),\sigma(x,\theta))
\text{ is a } \bC^{1} \text{ function of } \\
(\x,\theta',\theta'',\theta) \text{ throughout } \domE \times \Theta^{3}.
\end{array}
\label{7.12k}
\end{equation}
\end{abc}

\item 
The next step is to prove that
\begin{eqnarray}
& & \frac{\partial\lambda_{21}\left(\sigma(\x,\theta'),\sigma(\x,\theta''),
\sigma(\x,\theta)\right)}{\partial r \partial s} \nonumber \\
& & \text{exists and is a continuous function of } \nonumber \\
& & (\x,\theta',\theta'',\theta) \text{ throughout } \domE \times \Theta^{3}.
\label{7.13}
\end{eqnarray}
To prove the above statement, we consider three distinct cases, (a), (b)
and (c):
\begin{letterlist}
\item 
Consider the case
\begin{abc}
\begin{equation}
(\theta'',\theta) \in [0,\pi/2] \times [\pi,3\pi/2] \text{ or }
(\theta'',\theta) \in [\pi,3\pi/2] \times [0,\pi/2].
\end{equation}
{}From (\ref{7.12a}) and (\ref{7.12b}), and from the relation $|r,r_{0}| <
|s,s_{0}|$,
\begin{equation}
\begin{array}{r}
\sigma(\x,\theta'') - \sigma(\x,\theta) \ne 0 \text{ for all }
(\x,\theta'',\theta) \in \domE \times [0,\pi/2] \times [\pi,3\pi/2] \\
\text{and for all } (\x,\theta'',\theta) \in \domE \times [\pi,3\pi/2]
\times [0,\pi/2].
\end{array}
\end{equation}
Therefore, from Eq.\ (\ref{7.11b}) and statement (\ref{7.12i}),
\begin{equation}
\begin{array}{l}
\lambda_{21}\left(\sigma(\x,\theta'),\sigma(\x,\theta''),\sigma(\x,\theta)\right)
\text{ is a } \bC^{2} \\
\text{function of } (\x,\theta',\theta'',\theta) \text{ throughout the } \\
\text{union of }
\domE \times \Theta \times [0,\pi/2] \times [\pi,3\pi/2] \\
\text{and } \domE \times \Theta \times [\pi,3\pi/2] \times [0,\pi/2];
\end{array}
\end{equation}
and, in particular,
\begin{eqnarray}
& & \frac{\partial^{2}}{\partial r \partial s} \lambda_{21} 
\left(\sigma(\x,\theta'),\sigma(\x,\theta''),\sigma(\x,\theta)\right) 
\nonumber \\
& & \text{exists and is a continuous function } \nonumber \\
& & \text{of } (\x,\theta',\theta'',\theta)
\text{ throughout the union } \nonumber \\
& & \text{of }
\domE \times \Theta \times [0,\pi/2] \times [\pi,3\pi/2] \nonumber \\
& & \text{and } \domE \times \Theta \times [\pi,3\pi/2] \times [0,\pi/2].
\label{7.14d}
\end{eqnarray}
\end{abc}

\item 
Consider the case
\begin{abc}
\begin{equation}
\begin{array}{l}
(\theta'',\theta) \in [0,\pi/2]^{2} \text{ and }
\theta' \in [\pi,3\pi/2], \text{ or } \\
(\theta'',\theta) \in [\pi,3\pi/2]^{2} \text{ and }
\theta' \in [0,\pi/2].
\end{array}
\label{7.14e}
\end{equation}
{}From (\ref{7.12a}) and (\ref{7.12b}), and from the relation $|r,r_{0}| < 
|s,s_{0}|$,
\begin{equation}
\begin{array}{l}
\sigma(\x,\theta') - \sigma(\x,\theta'') \ne 0 \text{ and }
\sigma(\x,\theta') - \sigma(\x,\theta) \ne 0 \\ 
\text{for all } \x \in \domE \text{ and for all } (\theta',\theta'',\theta)
\text{ which satisfy (\ref{7.14e}). }
\end{array}
\end{equation}
Therefore, from Eq.\ (\ref{6.2c}) and statement (\ref{7.12h}),
\begin{eqnarray}
& &
k_{21}\left(\sigma(\x,\theta'),\sigma(\x,\theta'')\right) \text{ and }
k_{21}\left(\sigma(\x,\theta'),\sigma(\x,\theta)\right) \text{ are } 
\nonumber \\
& & \bC^{3} \text{ functions of } (\x,\theta',\theta'',\theta) 
\text{ throughout the union of } \nonumber \\
& & \domE \times [\pi,3\pi/2] \times [0,\pi/2]^{2}
\text{and } \domE \times [0,\pi/2] \times [\pi,3\pi/2]^{2}.
\label{7.14g}
\end{eqnarray}
Therefore, from Eq.\ (\ref{7.11b}),
\begin{eqnarray}
& & \lambda_{21}\left(\sigma(\x,\theta'),\sigma(\x,\theta''),\sigma(\x,\theta)\right)
\text{ is a } \bC^{2} \text{ function of } \nonumber \\
& & (\x,\theta',\theta'',\theta) \text{ throughout the same union of sets } 
\nonumber \\
& & \text{mentioned above in (\ref{7.14g});}
\end{eqnarray}
and, in particular,
\begin{eqnarray}
& & \frac{\partial^{2}}{\partial r \partial s}
\lambda_{21}\left(\sigma(\x,\theta'),\sigma(\x,\theta''),\sigma(\x,\theta)\right)
\nonumber \\
& & \text{exists and is a continuous function of } \nonumber \\
& & (\x,\theta',\theta'',\theta) \text{ throughoutout the union of}
\nonumber \\
& & \domE \times [\pi,3\pi/2] \times [0,\pi/2]^{2} \text{ and }
\nonumber \\
& & \domE \times [0,\pi/2] \times [\pi,3\pi/2]^{2}.
\label{7.14i}
\end{eqnarray}
\end{abc}

\item 
Consider the case
\begin{abc}
\begin{equation}
(\theta',\theta'',\theta) \in [0,\pi/2]^{3} \text{ or }
(\theta',\theta'',\theta) \in [\pi,3\pi/2]^{3}.
\end{equation}
It is easy to see from Eqs.\ (\ref{7.12c}) and (\ref{7.12d}) that
\begin{eqnarray}
& & \frac{\partial^{2}}{\partial r \partial s}
\lambda_{21}\left(\sigma(\x,\theta'),\sigma(\x,\theta''),\sigma(\x,\theta)\right)
\text{ exists} \nonumber \\
& & \text{and equals zero throughout the union of } \nonumber \\
& & \domE \times [0,\pi/2]^{3} \text{ and } 
\domE \times [\pi,3\pi/2]^{3}.
\label{7.14k}
\end{eqnarray}
That completes our three cases.  The statement (\ref{7.13}) now follows from
(\ref{7.14d}), (\ref{7.14i}) and (\ref{7.14k}).
\end{abc}
\end{letterlist}

\item 
{}From Eqs.\ (\ref{6.20c}) to (\ref{G5.3f}), 
\begin{eqnarray}
u_{1}(\x,\theta) & = & W_{1}(\sigma(\x,\theta)) 
- \frac{2}{\pi} \int_{\Theta} d\theta' p(\x,\theta')
L_{1}\left(\sigma(\x,\theta'),\sigma(\x,\theta)\right) \nonumber \\
& & \text{for all } (\x,\theta) \in \domE \times \Theta
\label{7.15a} \\
\text{and} & & \nonumber \\
\kappa_{21}(\x,\theta',\theta) & = &
q(\x,\theta') \left[ k_{21}\left(\sigma(\x,\theta'),\sigma(\x,\theta)\right)
- \frac{2}{\pi} \int_{\Theta} d\theta'' p(\x,\theta'') 
\lambda_{21}\left(\sigma(\x,\theta'),\sigma(\x,\theta)\right) 
\right] \nonumber \\
& & \text{ for all } (\x,\theta',\theta) \in \domE \times \Theta^{2},
\label{7.15b}
\end{eqnarray}
where $p(\x,\theta)$ is defined by Eq.\ (\ref{G5.3g}), and $q(\x,\theta)$
is defined by Eq.\ (\ref{6.20e}).  From statements (\ref{5.29a}) and
(\ref{5.29b}),
\begin{equation}
\begin{array}{l}
p(\x,\theta) \text{ and } q(\x,\theta) \text{ are real analytic } \nonumber \\
\text{functions of } (\x,\theta) \text{ throughout } \domE \times \Theta.
\end{array}
\label{7.15c}
\end{equation}
{}From statements (\ref{7.12e}), (\ref{7.12i}), (\ref{7.12k}), (\ref{7.13})
and (\ref{7.15c}), it is clear that the functions $u_{1}$ and $\kappa_{21}$
whose values are given by Eqs.\ (\ref{7.15a}) and (\ref{7.15b}), respectively,
satisfy the conclusions of our lemma.
\end{arablist}
\cheers

\begin{definition}{Dfn.\ of a function that is $\bC^{N_{1},\ldots,N_{L}}$
on $X \subset R^{L}$\label{def71}}
Suppose that $X$ is an open subset of $R^{L}$ or a closed or semi-closed
subinterval of $R^{L}$, $x=(x^{1},\ldots,x^{L})$ denotes any point in $X$,
$T$ is a topological space, $t$ denotes any point in $T$, and $N_{1},\ldots,
N_{L}$ are $L$ non-negative integers.  Suppose, furthermore, that
$F:(X \times T) \rightarrow C$ and that, for each $L$-tuple of integers
$(n_{1},\ldots,n_{L})$ such that $0 \le n_{k} \le N_{k}$ for all
$k=1,\ldots,L$,
\begin{equation}
\partial_{1 \cdots L}^{n_{1} \cdots n_{L}} F(x,t)
:= \left( \frac{\partial}{\partial x^{1}} \right)^{n_{1}} \cdots
\left( \frac{\partial}{\partial x^{k}} \right)^{n_{k}} \cdots
\left( \frac{\partial}{\partial x^{L}} \right)^{n_{L}} F(x,t)
\end{equation}
exists and is a continuous function of $(x,t)$ throughout $X \times T$.
[It is understood that $(\partial/\partial x^{k})^{0} = 1$.]  Then, we
shall say that {\em $F$ is $\bC^{N_{1},\ldots,N_{L}}$ on $X$}.  

Also, if $F:X \rightarrow C$ and $\partial_{1 \cdots L}^{n_{1} \cdots n_{L}}
F(x)$ exists and is a continuous function of $x$ throughout $X$ for each
choice of $(n_{1},\ldots,n_{L})$ that satisfies $0 \le n_{k} \le N_{k}$
for all $1 \le k \le L$, then we shall say that {\em $F$ is
$\bC^{N_{1},\ldots,N_{L}}$ on $X$}.
\end{definition}

Note:  If $F$ is $\bC^{N_{1},\ldots,N_{L}}$ on $X$, then a theorem of the
calculus tells us that, for each $(n_{1},\ldots,n_{L})$ satisfying
$0 \le n_{k} \le N_{k}$ for all $1 \le k \le L$, the existence and value
of $\partial_{1 \cdots L}^{n_{1} \cdots n_{L}} F$ are unchanged when the
operator factors $\partial/\partial x^{k}$ are subject to any permutation.

The following lemma is applicable to a broad class of Fredholm integral
equations and is clearly capable of further generalization in several
directions.  A $2 \times 2$ matrix version of the lemma for the case
$L=2$ was covered in a paper by the authors on the initial value problem
for colliding gravitational plane wave pairs.\footnote{I.~Hauser and
F.~J.~Ernst, J.\ Math.\ Phys.\ {\bf 32}, 198 (1991), Sec.~V.}  As regards
the current notes, the lemma will play a key role in the proof of
Thm.~\ref{7.3B}.

\begin{lemma}[Fredholm minor $M$ and determinant $\Delta$]
\label{7.2B} \mbox{ } \\
Let $X$, $x$ and $N_{k} \, (k=1,\ldots,L)$ be assigned the same meanings
as in the preceding definition; and let $Y$ denote a compact, oriented,
$m$-dimensional differentiable manifold, $y$ denote any point in $Y$, and
$dy$ denote a volume element at point $y$ (the value of a distinguished
non-zero $m$-form at $y$).  Suppose that $K:X \times (Y \times Y) \rightarrow
C$ and $K$ is $\bC^{N_{1},\ldots,N_{L}}$ on $X$.  Let us regard $K$ as an
$L$-parameter family of Fredholm kernels that is employed in Fredholm
integral equations of the form
\begin{equation}
f(x,y) - \int_{Y} dy' f(x,y') K(x,y',y) = g(x,y) \text{ for all }
(x,y) \in X \times Y,
\end{equation}
where $X$ is the parameter space.  Then, the corresponding Fredholm minor
$M$ and Fredholm determinant $\Delta$ are $\bC^{N_{1},\ldots,N_{L}}$
on $X$.
\end{lemma}

\proof
The Fredholm construction of $M$ and $\Delta$ are given by
\begin{abc}
\begin{eqnarray}
M(x,y',y) & = & \sum_{n=0}^{\infty} \frac{(-1)^{n}}{n!} M^{(n)}(x,y',y), \\
\Delta(x) & = & \sum_{n=0}^{\infty} \frac{(-1)^{n}}{n!} \Delta^{(n)}(x), 
\end{eqnarray}
where
\begin{eqnarray}
M^{(0)}(x,y',y) & := & K(x,y',y),
\label{7.18c} \\
M^{(n)}(x,y',y) & := & \int_{Y} dy_{1} \cdots \int_{Y} dy_{n}
D^{(n+1)}\left(x \left| \begin{array}{cc}
y & y_{1} \cdots y_{n} \\ y' & y_{1} \cdots y_{n} 
\end{array} \right. \right) \nonumber \\
& & \text{ for all } n > 0,
\label{7.18d} \\
\Delta^{(0)}(x) & := & 1, \\
\Delta^{(n+1)}(x) & := & \int_{Y} dy M^{(n)}(x,y,y) \text{ for all } n \ge 0,
\end{eqnarray}
and
\begin{eqnarray}
D^{(n)}\left(x \left| \begin{array}{c}
y_{1} \cdots y_{n} \\ y'_{1} \cdots y'_{n} 
\end{array} \right. \right) & := & \text{the determinant of that } n \times n
\nonumber \\
& & \text{matrix whose element in the $k$th } \nonumber \\
& & \text{row and $l$th column is } K(x,y'_{k},y_{l}).
\end{eqnarray}
In particular,
\begin{equation}
D^{(0)}\left( x \left| \begin{array}{c}
y \\ y' 
\end{array} \right. \right) := K(x,y',y).
\end{equation}
\end{abc}

For each bounded and closed subspace $U$ of $X$, let
\begin{abc}
\begin{eqnarray}
||K_{u}|| & := & \sup\left\{\left| \partial_{1 \cdots L}^{n_{1} \cdots n_{L}}
K(x,y',y) \right|:(x,y',y) \in U \times Y^{2}, \right. \nonumber \\
& & \left. \text{and } 0 \le n_{k} \le N_{k} \text{ for all }
k=1,\ldots,L\right\}.
\end{eqnarray}
Also let
\begin{equation}
V := \int_{Y} dy.
\end{equation}
Then, from Eqs.\ (\ref{7.18c}) and (\ref{7.18d}), and from a generalization
of Hadamard's inequality that was formulated and proved by the authors in the
aforementioned paper on the initial value problem for colliding gravitational
plane wave pairs [see Thm.~7 in that paper],
\begin{eqnarray}
\left| \partial_{1 \cdots L}^{n_{1} \cdots n_{L}} M^{n}(x,y',y) \right|
& \le & V^{n} ||K_{U}||^{n+1} (n+1)^{N_{1}+\ldots+N_{L}+(n+1)/2}
\nonumber \\
& & \text{for all } (x,y',y) \in U \times Y^{2} \text{ and all } \nonumber \\
& & (n_{1},\ldots,n_{L}) \text{ such that } 0 \le n_{k} \le N_{k} \nonumber \\
& & \text{for each } k=1,\ldots,L.
\end{eqnarray}
It follows that, for each positive integer $N$,
\begin{eqnarray}
\sum_{n=0}^{N} \frac{1}{n!} \left| \partial_{1 \cdots L}^{n_{1} \cdots n_{L}}
M^{(n)}(x,y',y) \right| & \le & \sum_{n=0}^{N} \frac{V^{n}||K_{U}||^{n+1}}
{n!} (n+1)^{N_{1}+\ldots+N_{L}+(n+1)/2} \nonumber \\
& & \text{for all } (x,y',y) \in U \times Y^{2} \text{ and all } \nonumber \\
& & \text{choices (the usual) of } (n_{1},\ldots,n_{l}).
\label{7.19d}
\end{eqnarray}
The application of the ratio test to the series on the right side of the
above inequality (\ref{7.19d}) is straightforward and deomonstrates that
this series converges as $N \rightarrow \infty$.  Hence, from the comparison
test, the series on the left side of (\ref{7.19d}) converges for all
$(x,y',y) \in U \times Y^{2}$ and all choices of $(n_{1},\ldots,n_{L})$.
The theorems\footnote{See Sec.~2, Ch.~IV, of {\em Differential and Integral
Calculus} by R.~Courant (Interscience Publishers, Inc., 1936).\label{Courant}}
of the calculus on the continuity and term-by-term differentiability of a
uniformly convergent infinite series of functions then supply us with the 
following conclusions:
\begin{equation}
\begin{array}{l}
\text{For all choices of } (n_{1},\ldots,n_{L}) \text{ for which }
0 \le n_{k} \le N_{k} (1 \le k \le L), \nonumber \\
\partial_{1 \cdots L}^{n_{1} \cdots n_{L}} M(x,y',y) \text{ exists and is 
a continuous function of } (x,y',y)
\nonumber \\
\text{throughout } X \times Y^{2};
\end{array}
\end{equation}
and
\begin{eqnarray}
\partial_{1 \cdots L}^{n_{1} \cdots n_{L}} M(x,y',y) & = & 
\sum_{n=0}^{\infty} \frac{(-1)^{n}}{n!} 
\left[ \partial_{1 \cdots L}^{n_{1} \cdots n_{L}} M^{(n)}(x,y',y) \right],
\text{ and } \nonumber \\
& & \text{the infinite series converges absolutely } \nonumber \\
& & \text{and converges uniformly on each } \nonumber \\
& & \text{compact subspace of } X \times Y^{2}.
\end{eqnarray}

Hence, $M$ is $\bC^{N_{1},\ldots,N_{L}}$ on $X$.  The proof that $\Delta$
is also $bC^{N_{1},\ldots,N_{L}}$ on $X$ is left for the reader.
\end{abc}
\cheers

The following theorem concerns the solution $y_{1}(\x,\theta)$ of the
Fredholm equation (\ref{6.20a}) for all $(\x,\theta) \in \domE \times \Theta$.

\begin{theorem}[Differentiability properties of $y_{1}$ when $\bv \in K^{3}$]
\label{7.3B} \mbox{ } \\
If $\bv \in K^{3}$, then $y_{1}$ is $\bC^{1,1}$ on $\domE$; i.e., 
$\partial y_{1}(\x,\theta)/\partial r$, $\partial y_{1}(\x,\theta)/\partial s$
and $\partial^{2}y_{1}(\x,\theta)/\partial r \partial s$ exist and are
continuous functions of $(\x,\theta)$ throughout $\domE \times \Theta$.
\end{theorem}

\begin{abc}
\proof
Consider the inhomogenous Fredholm equation of the second kind (\ref{6.20a}).
According to Thm.~\ref{6.4C}, the Fredholm determinant for Eq.\ (\ref{6.20a})
is not zero for all choices of $\x \in \domE$.  Therefore, a unique solution
of the Fredholm equation exists and is given by
\begin{equation}
y_{1}(\x,\theta) = u_{1}(\x,\theta) + \frac{2}{\pi} \int_{\Theta} d\theta'
u_{1}(\x,\theta') R(\x,\theta',\theta)
\label{7.20a}
\end{equation}
for all $(\x,\theta) \in \domE \times \Theta$, where the resolvent kernel
$R(\x,\theta'\theta)$ is the following ratio of the Fredholm minor and
determinant:
\begin{equation}
R(\x,\theta',\theta) = \frac{M(\x,\theta',\theta)}{\Delta(\x)}.
\end{equation}
{}From Lem.~\ref{7.1B}, $\kappa_{21}$ is $\bC^{1}$.  Moreover, $\partial^{2}
\kappa_{21}(\x,\theta',\theta)/\partial r \partial s$ exists and is a
continuous function of $(\x,\theta',\theta)$ throughout $\domE \times
\Theta^{2}$.  Therefore,
\begin{equation}
\kappa_{21} \text{ is } \bC^{1,1} \text{ on } \domE.
\end{equation}
The preceding Lem.~\ref{7.2B} is now applied to the present case, for which
\begin{equation}
X=\domE, \quad Y=\Theta, \quad L=2, \quad m=1, \quad dy=2d\theta/\pi.
\end{equation}
Thereupon, one obtains
\begin{equation}
R \text{ is } \bC^{1,1} \text{ on } \domE.
\label{7.20e}
\end{equation}
Lemma~\ref{7.1B} also tells us that (amongst other things)
\begin{equation}
u_{1} \text{ is } \bC^{1,1} \text{ on } \domE.
\label{7.20f}
\end{equation}
Therefore, from Eq.\ (\ref{7.20a}), statements (\ref{7.20e}) and (\ref{7.20f}),
and the theorems\footnote{See Ref.~\ref{Courant}.} of the calculus on the
continuity and differentiability of an integral with respect to parameters,
\begin{equation}
y_{1} \text{ is } \bC^{1,1} \text{ on } \domE.
\end{equation}
\cheers
\end{abc}

Note that, in terms of standard notation and terminology, $\lambda=1$ for
our particular Fredholm equation; and the statement that $\Delta(\x) \ne 0$
is equivalent to the statement that $1$ is not a characteristic value
(eigenvalue) of our kernel.

\setcounter{equation}{0}
\setcounter{theorem}{0}
\subsection{Concerning the partial derivatives of $\Y$, $\bar{\F}$, $H$ and
$\F^{\pm}$ when $\bv \in K^{3}$}

\begin{definition}{Dfns.\ of $\Y^{(i)}$, $\Y$ and the partial derivatives
of $\Y$\label{def72}}
Henceforth, $\Y^{(i)} \, (i \in \{3,4\})$ will denote the function whose
domain is
\begin{equation}
\dom{\Y} := \{(\x,\sigma):\x \in \domE, \sigma \in \check{\I}(x^{7-i})\}
\end{equation}
and whose values are given by
\begin{equation}
\Y^{(i)}(\x,\sigma) := \check{\Y}(\x)(\sigma),
\end{equation}
where $\check{\Y}(\x)$ is the extension of $\Y^{(i)}(\x)$ that is defined
by Eqs.\ (\ref{5.31a}) and (\ref{5.31b}).  We shall let $\Y$ denote the
function whose domain is
$$
\dom{\Y} := \{(\x,\sigma):\x \in \domE, \sigma \in \bar{\I}(\x)\}
$$
and whose values are given by
$$
\Y(\x,\sigma) := \Y^{(i)}(\x,\sigma) \text{ whenever } \sigma \in
\bar{\I}^{(i)}(\x).
$$
[Thus, $\Y(\x,\sigma) = \Y(\x)(\sigma)$.]  Also, for each $\x \in \domE$,
$i \in \{3,4\}$ and $\sigma \in \bar{\I}^{(i)}(\x)$, we shall let
$$
\frac{\partial^{l+m+n}\Y(\x,\sigma)}{\partial r^{l} \partial s^{m} 
\partial \sigma^{n}} :=
\frac{\partial^{l+m+n}\Y^{(i)}(\x,\sigma)}{\partial r^{l} \partial s^{m} 
\partial \sigma^{n}},
$$
if the above partial derivative of $\Y^{(i)}$ exists.
\end{definition}

Formerly, we let $\Y^{(i)}(\x,\sigma) := \Y^{(i)}(\x)(\sigma)$ and
$\Y(\x,\sigma) := \Y(\x)(\sigma)$; but a reexamination of all assertions
that we made made about $\Y^{(i)}(\x,\sigma)$ and $\Y(\x,\sigma)$ in the
past shows that they remain valid if we substitute our new definitions
for the old ones in those assertions.  For example, Cor.~\ref{5.5C} and
its proof remain valid if one replaces `$\partial^{n}[\Y(\x)(\sigma)]/
\partial\sigma^{n}$' by $\partial^{n}\Y(\x,\sigma)/\partial\sigma^{n}$'
and employs the new definition of the latter partial derivative.  This
new definition differs, however, from our former (usually tacit)
understanding of the meaning of $\partial^{n}\Y(\x,\sigma)/\partial\sigma^{n}$.
The point is that the domain of $\Y^{(i)}$, as defined above, is an open
subset of $R^{3}$; and (though the domain of $\Y$ is not an open subset of
$R^{3}$) the partial derivatives of $\Y$ are now defined in terms of partial
derivatives of $\Y^{(i)}$ and, therefore, employ sequences of points in
$R^{3}$ which may converge to a given point along any direction in $R^{3}$.
This has formal advantages when one employs the derivatives of $\Y$ at the
boundary of its domain.

\begin{definition}{Dfn.\ of $L^{(i)}(\sigma',\sigma)$ for each $\x \in \domE$
and $i \in \{3,4\}$\label{def73}}
For each $\sigma' \in \bar{\I}(\x)$ and $\sigma \in \I^{(i)}$, let
$$
L^{(i)}(\sigma',\sigma) := \frac{W(\sigma')-W^{(i)}(\sigma)}{\sigma'-\sigma}.
$$
\end{definition}

Employing the transformation defined by Eqs.\ (\ref{5.26a}) to (\ref{5.26i}),
Thm.~\ref{5.2B} [(\ref{5.27a}) and (\ref{5.27b})], the definition of
$p(\x,\theta)$ by Eq.\ (\ref{G5.3g}), the definition of $q(\x,\theta)$ by
Eq.\ (\ref{6.20e}) and the definition of $L^{(i)}(\sigma',\sigma)$ that we
just gave, one finds that Eqs.\ (\ref{5.30a}), (\ref{5.30b}) and (\ref{5.24b})
are expressible in the forms (in which `$\x$' is no longer suppressed)
\begin{eqnarray}
\Y^{(i)}_{1}(\x,\sigma) & = & W^{(i)}_{1}(\sigma) - \frac{2}{\pi} \int_{\Theta}
d\theta' p(\x,\theta') y_{2}(\x,\theta') W_{1}^{T}(\sigma(\x,\theta')) J
L^{(i)}_{1}(\sigma(\x,\theta'),\sigma) \nonumber \\
& & \text{for all } \x \in \domE \text{ and } \sigma \in
\check{\I}^{(i)}(x^{7-i}) \nonumber \\
& & \text{[after the extension defined by (\ref{5.31a})],}
\label{7.21a} \\
\Y^{(i)}_{2}(\x,\sigma) & = & W^{(i)}_{2}(\sigma) + \frac{2}{\pi} \int_{\Theta}
d\theta' q(\x,\theta') y_{1}(\x,\theta') W_{2}^{T}(\sigma(\x,\theta')) J
L^{(i)}_{2}(\sigma(\x,\theta'),\sigma) \nonumber \\
& & \text{for all } \x \in \domE \text{ and } \sigma \in
\check{\I}^{(i)}(x^{7-i}) \nonumber \\
& & \text{[after the extension defined by (\ref{5.31a})],} 
\label{7.21b}
\end{eqnarray}
and
\begin{eqnarray}
\bar{\bnu}(\x,\x_{0},\tau)^{-1} \bar{\F}(\x,\tau) & = & I - \frac{2}{\pi}
\int_{\Theta} d\theta' q(\x,\theta') y_{1}(\x,\theta')
\frac{W_{2}(\sigma(\x,\theta')) J}{\sigma(\x,\theta')-\tau} \nonumber \\
& & \text{for all } \x \in \domE \text{ and } \tau \in C - \bar{\I}(\x).
\label{7.22a}
\end{eqnarray}
Furthermore, from Eqs.\ (\ref{7.1a}), (\ref{7.2a}) and (\ref{7.22a}),
\begin{eqnarray}
H(\x) & = & H^{M}(\x_{0}) + 2(z-z_{0})\Omega - \frac{4i}{\pi} \int_{\Theta}
d\theta' q(\x,\theta') y_{1}(\x,\theta') W_{2}^{T}(\sigma(\x,\theta'))
\nonumber \\
& & \text{for all } \x \in \domE.
\end{eqnarray}

When proving the following theorem, one should bear in mind that
$\sigma(\x,\theta)$, $p(\x,\theta)$ and $q(\x,\theta)$ are analytic
functions of $(\x,\theta)$ throughout $\domE \times \Theta$.

\begin{abc}
\begin{theorem}[Differentiability properties of $\Y^{(i)}$, $\bar{\F}$
and $H$ when $\bv \in K^{3}$]
\label{7.1D} \mbox{ } \\
If $\bv \in K^{3}$, then
\begin{equation}
\begin{array}{l}
\partial\Y^{(i)}(\x,\sigma)/\partial r, \;
\partial\Y^{(i)}(\x,\sigma)/\partial s, \;
\partial^{2}\Y^{(i)}(\x,\sigma)/\partial r \partial s, \\
\partial^{2}\Y^{(i)}(\x,\sigma)/\partial \sigma^{2}, \;
\partial^{2}\Y^{(i)}(\x,\sigma)/\partial r \partial \sigma, \text{ and }
\partial^{2}\Y^{(i)}(\x,\sigma)/\partial s \partial \sigma \\
\text{exist and are continuous functions of $(\x,\sigma)$ throughout } \\
\{(\x,\sigma):\x \in \domE, \sigma \in \check{\I}^{(i)}(x^{7-i})\}.
\end{array}
\label{7.23}
\end{equation}
Also, upon letting $\grave{\F}$ denote the restriction of $\bar{\F}$ to
$$
\dom{\grave{\F}} := \{(\x,\tau):\x\in\domE, \tau \in C-\grave{\I}(\x)\},
$$
one has
\begin{equation}
\begin{array}{l}
\partial\grave{\F}(\x,\tau)/\partial r, \;
\partial\grave{\F}(\x,\tau)/\partial s, \text{ and }
\partial^{2}\grave{\F}(\x,\tau)/\partial r \partial s \\
\text{exist and are continuous functions of } (\x,\tau) \\
\text{throughout } \dom{\grave{\F}}; \text{ and, for each } \x \in \domE,
\text{ these } \\
\text{partial derivatives are holomorphic functions } \\
\text{of $\tau$ throughout } C - \grave{\I}(\x).
\end{array}
\label{7.24a}
\end{equation}
Furthermore,
\begin{equation}
H \text{ is } \bC^{1,1} \text{ on } \domE.
\label{7.24b}
\end{equation}
\end{theorem}
\end{abc}

\proof
{}From Thm.~\ref{7.3B}, statement (\ref{7.12e}) and the fact that $L^{(i)}$
is $\bC^{2}$, one concludes from Eq.\ (\ref{7.21b}) that
$\partial\Y^{(i)}_{2}(\x,\sigma)/\partial r$, 
$\partial\Y^{(i)}_{2}(\x,\sigma)/\partial s$,
$\partial^{2}\Y^{(i)}_{2}(\x,\sigma)/\partial r \partial s$,
$\partial^{2}\Y^{(i)}_{2}(\x,\sigma)/\partial \sigma^{2}$,
$\partial^{2}\Y^{(i)}_{2}(\x,\sigma)/\partial r \partial \sigma$ and
$\partial^{2}\Y^{(i)}_{2}(\x,\sigma)/\partial s \partial \sigma$ exist
and are continuous functions of $(\x,\sigma)$ throughout
$\{(\x,\sigma):\x \in \domE, \sigma \in \check{\I}^{(i)}(x^{7-i})\}$.
Then, from Eq.\ (\ref{7.21a}), one obtains like conclusions for
$\Y^{(i)}_{1}(\x,\sigma)$, whereupon the statement (\ref{7.23}) follows.

Statements (\ref{7.24a}) and (\ref{7.24b}) follow from Thm.~\ref{7.3B},
statement (\ref{7.12e}), the known differentiability and holomorphy
properties of $\bar{\bnu}(\x,\x_{0},\tau)^{-1}$ on $\dom{\grave{\F}}$,
and the theorem\footnote{Where?} on the holomorphy of functions given 
by Cauchy-type integrals.
\cheers

The following two lemmas will be used to prove Thm.~\ref{7.4D}.

\begin{abc}
\begin{lemma}[$d(\bar{\bnu}(\x_{0},\x,\tau) \grave{\F}(\x,\tau))$]
\label{7.2D} \mbox{ } \\
Recall that $\bar{\bnu}(\x,\x_{0},\tau)^{-1}=\bar{\bnu}(\x_{0},\x,\tau)$.
If $\bv \in K^{3}$, then the first partial derivatives of 
\begin{equation}
\frac{\bnu^{+}(\x_{0},\x,\sigma') \Y_{1}(\x,\sigma') W_{2}^{T}(\sigma') J}
{\sigma'-\tau}
\label{7.25a}
\end{equation}
with respect to $r$ and with respect to $s$ are summable over $\sigma'
\in \bar{\I}(\x)$; and
\begin{eqnarray}
d\left[\bar{\bnu}(\x_{0},\x,\tau) \grave{\F}(\x,\tau)\right] & = & 
- \frac{1}{\pi i} \int_{\bar{\I}} d\sigma'
\frac{d\left[\bnu^{+}(\x_{0},\x,\sigma')\Y_{1}(\x,\sigma')\right]
W_{2}^{T}(\sigma')J}{\sigma'-\tau} \nonumber \\
& & \text{for all } (\x,\tau) \in \dom{\grave{\F}}.
\label{7.25b}
\end{eqnarray}
\end{lemma}
\end{abc}

\begin{abc}
\proof
We shall tacitly employ statements (\ref{7.23}) and (\ref{7.24a}) of
Thm.~\ref{7.1D} in some steps of this proof.  We shall supply the proof
only for 
$\partial[\bar{\bnu}(\x_{0},\x,\tau)\grave{\F}(\x,\tau)]/\partial r$
and leave the proof for the partial derivative with respect to $s$ for
the reader.  The summability over $\bar{\I}(\x)$ of the partial derivative
with respect to $r$ of (\ref{7.25a}) is seen from the facts that
\begin{eqnarray}
\bnu^{+}(\x_{0},\x,\sigma') & = & \M^{+}(\sigma'-r)\M^{+}(\sigma'-s)
\left[\M^{+}(\sigma'-r_{0})\M^{+}(\sigma'-s_{0})\right]^{-1} \\
\text{and} & & \nonumber \\
\frac{\partial\bnu^{+}(\x_{0},\x,\sigma')}{\partial r} & = & 
-\frac{1}{2}\M^{+}(\sigma'-s) \left[\M^{+}(\sigma'-r)
\M^{+}(\sigma'-r_{0})\M^{+}(\sigma'-s_{0})\right]^{-1}
\end{eqnarray}
are both summable over $\bar{\I}(\x)$, and a summable function times a
continuous function over a bounded interval is summable.

In the proofs of this lemma and the next lemma, we shall employ the
shorthand notations
\begin{equation}
\begin{array}{rcl}
f(\x,\sigma') & := & \Y_{1}(\x,\sigma') W_{2}^{T}(\sigma') J, \\
g(\x,\sigma') & := & \bar{\bnu}(\x_{0},\x,\tau) \bar{\F}(\x,\tau),
\label{7.26a}
\end{array}
\end{equation}
whereupon Eq.\ (\ref{5.24b}) becomes
\begin{eqnarray}
g(\x,\tau) & = & I - \frac{1}{\pi i} \int_{\bar{\I}} d\sigma'
\bnu^{+}(\x_{0},\x,\sigma') \frac{f(\x,\sigma')}{\sigma'-\tau} \nonumber \\
& = & I - g_{1}(\x,\tau) - f(\x,r) g_{2}(\x,\tau),
\label{7.26b}
\end{eqnarray}
where
\begin{eqnarray}
g_{1}(\x,\tau) & := & \frac{1}{\pi i} \int_{\bar{\I}} d\sigma'
\bnu^{+}(\x_{0},\x,\sigma') \frac{f(\x,\sigma')-f(\x,r)}{\sigma'-\tau},
\label{7.26c} \\
\text{and} & & \nonumber \\
g_{2}(\x,\tau) & := & \frac{1}{\pi i} \int_{\bar{\I}} d\sigma'
\frac{\bnu^{+}(\x_{0},\x,\sigma')}{\sigma'-\tau}.
\label{7.26d}
\end{eqnarray}

We shall first deal with the term $f(\x,r)g_{2}(\x,\tau)$.
It is easy to show that 
\begin{equation}
g_{2}(\x,\tau) = \bnu(\x_{0},\x,\tau) - 1.
\label{7.26e}
\end{equation}
Therefore, for all $(\x,\tau) \in \dom{\grave{\F}}$,
\begin{equation}
\frac{\partial g_{2}(\x,\tau)}{\partial r} = - \frac{1}{2(\tau-r)}
\bar{\bnu}(\x_{0},\x,\tau).
\label{7.26f}
\end{equation}
Also, note that
\begin{eqnarray}
\frac{1}{\pi i} \int_{\bar{\I}} d\sigma' \frac{\partial
\bnu^{+}(\x_{0},\x,\sigma')/\partial r}{\sigma'-\tau} & = &
- \frac{1}{2\pi i} \int_{\bar{\I}} d\sigma'
\frac{\bnu^{+}(\x_{0},\x,\sigma')}{(\sigma'-r)(\sigma'-\tau)} \nonumber \\
& = & - \frac{\bar{\bnu}(\x_{0},\x,\tau)}{2(\tau-r)}.
\label{7.26g}
\end{eqnarray}
So, from Eqs.\ (\ref{7.26d}), (\ref{7.26f}) and (\ref{7.26g}),
\begin{eqnarray}
\frac{\partial}{\partial r} \left[ f(\x,r)g_{2}(\x,\tau) \right] & = &
\frac{1}{\pi i} \int_{\bar{\I}} d\sigma'
\frac{\partial\left[\bnu^{+}(\x_{0},\x,\sigma')f(\x,r)\right]/\partial r}
{\sigma'-\tau} \nonumber \\
& & \text{for all } (\x,\tau) \in \dom{\grave{\F}}.
\label{7.26h}
\end{eqnarray}
That takes care of the term $f(\x,r)g_{2}(\x,\tau)$. 

We shall next deal with the term $g_{1}(\x,\tau)$.  From statement
(\ref{7.23}) in Thm.~\ref{7.1D} and from Eq.\ (\ref{7.26a}), one can
see that
\begin{equation}
\frac{\partial}{\partial r} \left\{ \M^{+}(\sigma'-r)\M^{+}(\sigma'-s)
\left[f(\x,\sigma')-f(\x,r)\right]\right\}
\end{equation}
exists and is a continuous function of $(\x,\sigma')$ throughout
$\{(\x,\sigma'):\x \in \domE, \sigma' \in \bar{\I}(\x)\}$.  [We leave
details for the reader.]  No loss of generality will be incurred if we
tentatively introduce a closed and bounded convex neighborhood $\N$ of
the point $\x_{0}$ in the space $\domE$, whereupon it is seen that
$$
\{(\x,\sigma'):\x \in \N, \sigma' \in \bar{\I}(\x)\}
$$
is a bounded closed subspace of $R^{3}$; and, therefore,
\begin{eqnarray}
\lefteqn{M(\N) := } \\
& & \sup \left\{ \left|\left| \frac{\partial}{\partial r}
\left\{ \M^{+}(\sigma'-r)\M^{+}(\sigma'-s) \left[f(\x,\sigma')-f(\x,r)\right]
\right\}\right|\right|:\x \in \N,\sigma' \in \bar{\I}(\x)\right\} \nonumber
\end{eqnarray}
is finite; and the integrand in the expression for $g_{1}(\x,\tau)$ that is
given by Eq.\ (\ref{7.26c}) satisfies
\begin{equation}
\left|\left|\frac{\partial}{\partial r} \left[\bnu^{+}(\x_{0},\x,\sigma')
\frac{f(\x,\sigma')-f(\x,r)}{\sigma'-\tau}\right]\right|\right| \le
\left[\sqrt{|\sigma'-r_{0}||\sigma'-s_{0}|}\;|\sigma'-\tau|\right]^{-1}
M(\N).
\end{equation}
Since the right side of the above inequality is summable over $\bar{\I}(\x)$
and is independent of $\x$, a well-known theorem\footnote{See
Ref.~\ref{McShane}} on differentiation of a Lebesgue integral with respect 
to a parameter tells us that $\partial g_{1}(\x,\tau)/\partial r$ exists
(which, it happens, we already know) and is given by
\begin{equation}
\frac{\partial g_{1}(\x,\tau)}{\partial r} = \frac{1}{\pi i} \int_{\bar{\I}}
d\sigma' \frac{\frac{\partial}{\partial r} \left\{ \bnu^{+}(\x_{0},\x,\sigma')
\left[f(\x,\sigma')-f(\x,r)\right]\right\}}{\sigma'-\tau}
\label{7.27d}
\end{equation}
for all $\x \in \N$ and $\tau \in C - \grave{\I}(\x)$, where we have used the
fact that the contribution to $\partial g_{1}(\x,\tau)/\partial r$ due to
differentiation of the integral with respect to the endpoint $r \in
\{a^{3},b^{3}\}$ of the integration interval $\bar{\I}^{(3)}(\x)$ vanishes,
because the integrand in Eq.\ (\ref{7.26c}) vanishes when $\sigma'=r$.

However, since $\N$ can always be chosen so that it covers any given point
in $\domE$, Eq.\ (\ref{7.27d}) holds for all $(\x,\tau) \in \dom{\grave{\F}}$;
and upon combining (\ref{7.27d}), (\ref{7.26h}) and (\ref{7.26b}), one
obtains
\begin{eqnarray}
\frac{\partial g(\x,\tau)}{\partial r} & = & - \frac{1}{\pi i} \int_{\bar{\I}}
d\sigma' \frac{\partial[\bnu^{+}(\x_{0},\x,\sigma')f(\x,\sigma')]/\partial r}
{\sigma'-\tau} \nonumber \\
& & \text{for all } (\x,\tau) \in \dom{\grave{\F}},
\end{eqnarray}
which is the coefficient of $dr$ in Eq.\ (\ref{7.25b}).
\cheers
\end{abc}

Before we give the next lemma, note that application of the Plemelj relations
to Eq.\ (\ref{5.24b}) yields
\begin{eqnarray}
\frac{1}{2}[\F^{+}(\x,\sigma)+\F^{-}(\x,\sigma)] & = & 
- \Y_{1}(\x,\sigma) W_{2}^{T}(\sigma) J \nonumber \\
& & \text{for all } \x \in \domE \text{ and } \sigma \in \I(\x), 
\label{7.28a} \\
\text{and} & & \nonumber \\
\frac{1}{2}\bnu^{+}(\x_{0},\x,\sigma) [\F^{+}(\x,\sigma)-\F^{-}(\x,\sigma)]
& = & I - \frac{1}{\pi i} \int_{\bar{\I}} d\sigma' \bnu^{+}(\x_{0},\x,\sigma')
\frac{\Y_{1}(\x,\sigma')W_{2}^{T}(\sigma')J}{\sigma'-\sigma} \nonumber \\
& & \text{for all } \x \in \domE \text{ and } \sigma \in \I(\x).
\label{7.28b}
\end{eqnarray}
We could also have obtained Eq.\ (\ref{7.28a}) directly from Eq.\
(\ref{5.25a}); and Eq.\ (\ref{7.28b}) could have been obtained simply by
substituting (\ref{7.28a}) into Eq.\ (\ref{5.14c}).

\begin{abc}
\begin{lemma}[Differentiability properties of $\F^{\pm}$ when $\bv \in K^{3}$]
\label{7.3D} \mbox{ } \\
As in the preceding lemma, suppose that $\bv \in K^{3}$ and $\bar{\F}$ is
the solution of the HHP corresponding to $(\bv,\bar{\F}^{M})$.  Then the
following three statements hold:
\begin{romanlist}
\item 
The partial derivatives $\partial \F^{\pm}(\x,\sigma)/\partial r$,
$\partial \F^{\pm}(\x,\sigma)/\partial s$ and 
$\partial^{2}\F^{\pm}(\x,\sigma)/\partial r \partial s$ exist and are
continuous functions of $(\x,\sigma)$ throughout 
$\{(\x,\sigma):\x \in \domE, \sigma \in \I(\x)\}$.
\item 
The $1$-form
\begin{equation}
\frac{d[\bnu^{+}(\x_{0},\x,\sigma') \Y_{1}(\x,\sigma')] W_{2}^{T}(\sigma') J}
{\sigma'-\sigma}
\label{7.29a}
\end{equation}
is, for each $\x \in \domE$ and $\sigma \in \I(\x)$, summable over
$\bar{\I}(\x)$ in the PV sense.
\item 
For all $\x \in \domE$ and $\sigma \in \I(\x)$,
\begin{eqnarray}
\lefteqn{d \left\{ \frac{1}{2}\bnu^{+}(\x_{0},\x,\sigma) [\F^{+}(\x,\sigma)
-\F^{-}(\x,\sigma)] \right\} = } \nonumber \\
& & -\frac{1}{\pi i} \int_{\bar{\I}} d\sigma'
\frac{d[\bnu^{+}(\x_{0},\x,\sigma')\Y_{1}(\x,\sigma')]W_{2}^{T}(\sigma')J}
{\sigma'-\sigma}.
\label{7.29b}
\end{eqnarray}
\end{romanlist}
\end{lemma}
\end{abc}

\begin{abc}
\proofs
\begin{romanlist}
\item 
This follows from statement (\ref{7.23}), Eq.\ (\ref{7.28a}) and Eq.\ 
(\ref{5.25b}).
\cheers

The proofs of parts (ii) and (iii) will be supplied only for the
coefficients of $dr$ in Eqs.\ (\ref{7.29a}) and (\ref{7.29b}).  The
proofs for the coefficients of $ds$ are left to the reader.
\item 
As functions of $\sigma'$, $W_{2}^{T}(\sigma')$ is $\bC^{3}$,
$\Y_{1}(\x,\sigma')$ is $\bC^{2}$ and $\partial\Y_{1}(\x,\sigma')/\partial r$
is $\bC^{1}$ on $\bar{\I}(\x)$; and $\bnu^{+}(\x_{0},\x,\sigma')$ and
$\partial\bnu^{+}(\x_{0},\x,\sigma')/\partial r$ are summable over
$\bar{\I}(\x)$.  Therefore, for a sufficiently small $\epsilon > 0$,
\begin{equation}
\frac{\frac{\partial}{\partial r} \left[ \bnu^{+}(\x_{0},\x,\sigma')
\Y_{1}(\x,\sigma') \right] W_{2}^{T}(\sigma') J}{\sigma''-\sigma}
\label{7.30a}
\end{equation}
is summable over $\bar{\I}(\x) - ]\sigma-\epsilon,\sigma+\epsilon[$.
Moreover, since the numerator of (\ref{7.30a}) is a $\bC^{1}$ function of
$\sigma'$, it is well known that (\ref{7.30a}) is summable over
$[\sigma-\epsilon,\sigma+\epsilon]$ in the PV sense. 

Therefore, (\ref{7.30a}) is summable over $\bar{\I}(\x)$ in the PV sense.
\cheers
\item 
In terms of the shorthand notations (\ref{7.26a}), Eq.\ (\ref{7.28b}) is
expressible in the form
\begin{equation}
\frac{1}{2}[g^{+}(\x,\sigma)+g^{-}(\x,\sigma)] = I - \frac{1}{\pi i}
\int_{\bar{\I}} d\sigma' \bnu^{+}(\x_{0},\x,\sigma')
\frac{f(\x,\sigma')}{\sigma'-\sigma},
\label{7.30b}
\end{equation}
where Thm.~\ref{7.1D} furnishes the following properties of $f(\x,\sigma')$:
\begin{equation}
\begin{array}{l}
\partial f(\x,\sigma')/\partial r, \;
\partial f(\x,\sigma')/\partial s, \;
\partial^{2}f(\x,\sigma')/(\partial \sigma')^{2}, \\
\partial^{2}f(\x,\sigma')/\partial r \partial s, \; 
\partial^{2}f(\x,\sigma')/\partial r \partial \sigma' \text{ and } 
\partial^{2}f(\x,\sigma')/\partial s \partial \sigma' \\
\text{exist and are continuous functions of } (\x,\sigma') \\
\text{throughout } \{(\x,\sigma'):\x \in \domE, \sigma' \in \bar{\I}(\x)\}.
\end{array}
\label{7.30c}
\end{equation}
Let us introduce the additional shorthand notations
\begin{eqnarray}
f_{0}(\x,\sigma',\sigma) & := & 
\frac{f(\x,\sigma')-f(\x,\sigma)}{\sigma'-\sigma},
\label{7.30d} \\
f_{1}(\x,\sigma',\sigma) & := & f_{0}(\x,\sigma',\sigma)
-f_{0}(\x,r,\sigma), 
\label{7.30e} \\
g_{1}(\x,\sigma) & := & \frac{1}{\pi i} \int_{\bar{\I}} d\sigma'
\bnu^{+}(\x_{0},\x,\sigma') f_{1}(\x,\sigma',\sigma),
\label{7.30f} \\
g_{2}(\x,\sigma) & := & \frac{1}{\pi i} \int_{\bar{\I}} d\sigma'
\frac{\bnu^{+}(\x_{0},\x,\sigma')}{\sigma'-\sigma} \\
\text{and} & & \nonumber \\
g_{3}(\x,\sigma) & := & \frac{1}{\pi i} \int_{\bar{\I}} d\sigma'
\bnu^{+}(\x_{0},\x,\sigma').
\end{eqnarray}
Then Eq.\ (\ref{7.30b}) is expressible in the form
\begin{equation}
\frac{1}{2}[g^{+}(\x,\sigma)+g^{-}(\x,\sigma)] = I - g_{1}(\x,\sigma)
- f(\x,\sigma) g_{2}(\x,\sigma) - f_{0}(\x,r,\sigma) g_{3}(\x,\sigma).
\label{7.30i}
\end{equation}

Let us first consider the above terms that contain $g_{2}$ and $g_{3}$.
Upon interchanging $\x$ and $\x_{0}$ in Eq.\ (\ref{6.5a}), one obtains
\begin{equation}
g_{2}(\x,\sigma) = -1;
\end{equation}
and the usual contour integration technique yields
\begin{equation}
g_{3}(\x,\sigma) = \frac{1}{2}(r+s-r_{0}-s_{0}).
\end{equation}
Therefore, by using Eq.\ (\ref{6.10d}) and the fact that
$$
\partial\bnu^{+}(\x_{0},\x,\sigma')/\partial r = 
-\frac{1}{2}(\sigma'-r)^{-1} \bnu^{+}(\x_{0},\x,\sigma'), 
$$
the reader can prove that
\begin{eqnarray}
\frac{\partial g_{2}(\x,\sigma)}{\partial r} & = & 
\frac{1}{\pi i} \int_{\bar{\I}} d\sigma' 
\frac{\partial\bnu^{+}(\x_{0},\x,\sigma')/\partial r}{\sigma'-\sigma}
\\ \text{and} & & \nonumber \\
\frac{\partial g_{3}(\x,\sigma)}{\partial r} & = & 
\frac{1}{\pi i} \int_{\bar{\I}} d\sigma' 
\partial\bnu^{+}(\x_{0},\x,\sigma')/\partial r,
\end{eqnarray}
whereupon
\begin{eqnarray}
\frac{\partial[f(\x,\sigma)g_{2}(\x,\sigma)]}{\partial r} & = & \frac{1}{\pi i}
\int_{\bar{\I}} d\sigma'
\frac{\partial[f(\x,\sigma)\bnu^{+}(\x_{0},\x,\sigma')]/\partial r}
{\sigma'-\sigma},
\label{7.31e} \\
\text{and} & & \nonumber \\
\frac{\partial[f(\x,\sigma)g_{3}(\x,\sigma)]}{\partial r} & = & \frac{1}{\pi i}
\int_{\bar{\I}} d\sigma'
\frac{\partial}{\partial r} \left[f_{0}(\x,r,\sigma)\bnu^{+}(\x_{0},\x,\sigma')
\right].
\label{7.31f}
\end{eqnarray}
That completes the analysis of the terms in Eq.\ (\ref{7.30i}) that contain
$g_{2}$ and $g_{3}$.  

We next consider $g_{1}$.  From (\ref{7.30c}) to (\ref{7.30e}), one sees that
\begin{equation}
\begin{array}{l}
\text{For each } \sigma \in \I(\x), \partial f_{1}(\x,\sigma',\sigma)/\partial r
\text{ and } (\sigma'-r)^{-1} f_{1}(\x,\sigma',\sigma) \\
\text{exist and are continuous functions of } (\x,\sigma') \\
\text{throughout } \{(\x,\sigma'):\x \in \domE, \sigma' \in \bar{\I}(\x)\}.
\end{array}
\end{equation}
Therefore, as regards the integrand in the definition (\ref{7.30f}) of
$g_{1}(\x,\sigma)$, one readily deduces (by an argument similar to the one
used in the proof of the preceding lemma) that, corresponding to each closed
and bounded neighborhood $\N$ of the point $\x_{0}$ in the space $\domE$,
and each $\sigma \in \I(\x)$, there exists a positive real number
$M(\N,\sigma)$ such that
\begin{eqnarray}
\left|\left|\frac{\partial}{\partial r} \left[\bnu^{+}(\x_{0},\x,\sigma')
f_{1}(\x,\sigma',\sigma)\right]\right|\right| & \le &
\frac{M(\N,\sigma)}{\sqrt{|\sigma'-r_{0}||\sigma'-s_{0}|}} \nonumber \\
& & \text{for all } \x \in \N \text{ and } \sigma' \in \bar{\I}(\x)
- \{r_{0},s_{0}\}.
\end{eqnarray}
The remainder of the proof employs the same theorem on differentiation of a
Lebesgue integral with respect to a parameter that was used in the proof of
the preceding lemma.  The result is
\begin{equation}
\frac{\partial g_{1}(\x,\sigma)}{\partial r} = \frac{1}{\pi i} \int_{\bar{\I}}
d\sigma' \frac{\partial}{\partial r} \left[ \bnu^{+}(\x_{0},\x,\sigma')
f_{1}(\x,\sigma',\sigma) \right].
\label{7.32c}
\end{equation}
Upon combining the results given by Eqs.\ (\ref{7.31e}), (\ref{7.31f})
and (\ref{7.32c}), one obtains with the aid of Eqs.\ (\ref{7.30b}),
(\ref{7.30d}) to (\ref{7.30f}), and Eq.\ (\ref{7.30i}),
\begin{eqnarray}
\frac{\partial}{\partial r} \frac{1}{2} [g^{+}(\x,\sigma)+g^{-}(\x,\sigma)]
& = & -\frac{1}{\pi i} \int_{\bar{\I}} d\sigma'
\frac{\partial[\bnu^{+}(\x_{0},\x,\sigma')f(\x,\sigma')]/\partial r}
{\sigma'-\sigma} \nonumber \\
& & \text{for all } \x \in \domE \text{ and } \sigma \in \I(\x).
\end{eqnarray}
\cheers
\end{romanlist}
\end{abc}

The point of the preceding two lemmas is the following crucial theorem.

\begin{abc}
\begin{theorem}[Limits of $d\bar{\F}$ when $\bv \in K^{3}$]
\label{7.4D} \mbox{ } \\
Suppose $\bv \in K^{3}$ and $\bar{\F}$ is the solution of the HHP
corresponding to $(\bv,\bar{\F}^{M})$.  Then, the following three
statements hold:
\begin{romanlist}
\item 
For each $\x \in \domE$ and $\sigma \in \I(\x)$, $d\grave{\F}(\x,\sigma\pm\zeta)$
converges as $\zeta \rightarrow 0$ ($\Im{\zeta}>0$) and
\begin{equation}
\lim_{\zeta \rightarrow 0} d\grave{\F}(\x,\sigma\pm\zeta) = d\F^{\pm}(\x,\sigma).
\end{equation}
Note: The existences of $d\grave{\F}(\x,\tau)$ and $d\F^{\pm}(\x,\sigma)$ are
guaranteed by Thm.~\ref{7.1D} [statement (\ref{7.24a})] and by
Lem.~\ref{7.3D}(i), respectively.
\item 
$\bar{\F}(\x,\tau)$ converges as $\tau \rightarrow r_{0}$ and as $\tau 
\rightarrow s_{0}$ [$\tau \in C - \bar{\I}(\x)$]; and
$\bar{\bnu}(\x_{0},\x,\tau) \bar{\F}(\x,\tau)$ converges as $\tau \rightarrow
r$ and as $\tau \rightarrow s$.
\item 
For each $i \in \{3,4\}$,
\begin{equation}
(\tau-x^{i}) \frac{\partial \grave{\F}(\x,\tau)}{\partial x^{i}}
\end{equation}
converges as $\tau \rightarrow r_{0}$ and as $\tau \rightarrow s_{0}$, while
\begin{equation}
\bar{\bnu}(\x_{0},\x,\tau) (\tau-x^{i}) \frac{\partial \grave{\F}(\x,\tau)}
{\partial x^{i}}
\end{equation}
converges as $\tau \rightarrow r$ and as $\tau \rightarrow s$.
\end{romanlist}
\end{theorem}
\end{abc}

\begin{abc}
\proofs
\begin{romanlist}
\item 
We shall prove statement (i) for the coefficient of $dr$ in
$d\grave{\F}(\x,\tau)$ and leave the proof for the coefficient of $ds$ to
the reader.

Employ the shorthand notation
\begin{equation}
f(\x,\sigma') := \Y_{1}(\x,\sigma') W_{2}^{T}(\sigma') J
\label{7.34a}
\end{equation}
in the integrand of Eq.\ (\ref{5.24b}), which then becomes
\begin{eqnarray}
\bar{\bnu}(\x_{0},\x,\tau) \bar{\F}(\x,\tau) & = & I - \frac{1}{\pi i}
\int_{\bar{\I}} d\sigma' \bar{\bnu}(\x_{0},\x,\sigma')
\frac{f(\x,\sigma')}{\sigma'-\tau} \nonumber \\
& & \text{for all } (\x,\tau) \in \dom{\bar{\F}},
\label{7.34b}
\end{eqnarray}
whereupon, from Eq.\ (\ref{7.25b}) in Lem.~\ref{7.2D}, and from Eq.\
(\ref{7.26g}), 
\begin{eqnarray}
\lefteqn{-\frac{\bar{\bnu}(\x_{0},\x,\tau)}{2(\tau-r)} \bar{\F}(\x,\tau)
+ \bar{\bnu}(\x_{0},\x,\tau) \frac{\partial\bar{\F}(\x,\tau)}{\partial r}
= } \nonumber \\
& & \mbox{ } - \Phi(\x,\tau) + \frac{\bar{\bnu}(\x_{0},\x,\tau)}{2(\tau-r)}
f(\x,r) \nonumber \\
& & \text{for all } (\x,\tau) \in \dom{\grave{\F}},
\label{7.34c}
\end{eqnarray}
where
\begin{eqnarray}
\Phi(\x,\tau) & := & \frac{1}{\pi i} \int_{\bar{\I}} d\sigma'
\bnu^{+}(\x_{0},\x,\sigma') \frac{\phi(\x,\sigma')}{\sigma'-\tau}
\label{7.34d} \\
\text{and} & & \nonumber \\
\phi(\x,\sigma') & := & \frac{\partial f(\x,\sigma')}{\partial r}
- \frac{f(\x,\sigma')-f(\x,r)}{2(\sigma'-r)}.
\label{7.34e}
\end{eqnarray}
{}From Eq.\ (\ref{7.34e}) and the properties of $f(\x,\sigma')$ given by
statement (\ref{7.30c})
\begin{equation}
\begin{array}{l}
\partial\phi(\x,\sigma')/\partial\sigma' \text{ exists and is a continuous
function of } \\
(\x,\sigma')
\text{ throughout } \{(\x,\sigma'):\x \in \domE, \sigma' \in \bar{\I}(\x)\}.
\end{array}
\end{equation}
Therefore, $\bar{\bnu}(\x_{0},\x,\sigma') \phi(\x,\sigma')$ obeys a
H\"{o}lder condition of index $1$ on each closed subinterval of $\I(\x)$;
and it follows from the theorem in Sec.~16 in Muskhelishvili's 
treatise\footnote{See footnote \ref{Musk}.} that (\ref{7.34d}) satisfies
\begin{equation}
\Phi^{\pm}(\x,\sigma) := \lim_{\zeta \rightarrow 0} \Phi(\x,\sigma\pm\zeta)
\text{ exists for all } \sigma \in \I(\x).
\label{7.34g}
\end{equation}
Moreover, from the Plemelj relations [Eq.\ (17.2) in Sec.~17 of
Muskhelishvili's treatise],
\begin{equation}
\Phi^{\pm}(\x,\sigma) = \pm \bnu^{+}(\x_{0},\x,\sigma) \phi(\x,\sigma)
+ \frac{1}{\pi i} \int_{\bar{\I}} d\sigma' \bnu^{+}(\x_{0},\x,\sigma')
\frac{\phi(\x,\sigma')}{\sigma'-\sigma}.
\label{7.34h}
\end{equation}
[The existence of the above PV integral is demonstrated in Sec.~12 of 
Muskhelishvili's treatise.]  From Eq.\ (\ref{7.34c}), condition (3) in
the definition of the HHP [the one about the existence of $\F^{\pm}(\x)$]
and statement (\ref{7.34g}),
\begin{equation}
\lim_{\zeta \rightarrow 0} \frac{\partial\grave{\F}(\x,\sigma\pm\zeta)}
{\partial r} \text{ exists for each } \x \in \domE \text{ and }
\sigma \in \I(\x);
\label{7.34i}
\end{equation}
and, with the aid of Eqs.\ (\ref{7.28a}), (\ref{7.34a}) and (\ref{7.34h}),
\begin{equation}
\lim_{\zeta \rightarrow 0} \frac{1}{2} \left[
\frac{\partial\grave{\F}(\x,\sigma+\zeta)}{\partial r} + 
\frac{\partial\grave{\F}(\x,\sigma-\zeta)}{\partial r} \right] =
- \frac{\partial f(\x,\sigma)}{\partial r}
\end{equation}
and
\begin{eqnarray}
\lefteqn{\lim_{\zeta \rightarrow 0} \frac{1}{2} \frac{\partial}{\partial r}
\left[ \bar{\bnu}(\x_{0},\x,\sigma+\zeta) \grave{\F}(\x,\sigma+\zeta)
+ \bar{\bnu}(\x_{0},\x,\sigma-\zeta) \grave{\F}(\x,\sigma-\zeta) \right] }
\hspace{10em} \nonumber \\
& = & -\frac{1}{\pi i} \int_{\bar{\I}} d\sigma' \bnu^{+}(\x_{0},\x,\sigma')
\frac{\phi(\x,\sigma')}{\sigma'-\sigma}.
\label{7.34k}
\end{eqnarray}
However, from Eq.\ (\ref{6.10d}),
$$
\frac{1}{\pi i} \int_{\bar{\I}} d\sigma' \bnu^{+}(\x_{0},\x,\sigma')
\frac{f(\x,r)}{(\sigma'-r)(\sigma'-\sigma)} = 0.
$$
Therefore, from Eq.\ (\ref{7.34e}), Eq.\ (\ref{7.34k}) becomes
\begin{eqnarray}
\lefteqn{\lim_{\zeta \rightarrow 0} \frac{1}{2} \frac{\partial}{\partial r}
\left[ \bar{\bnu}(\x_{0},\x,\sigma+\zeta) \grave{\F}(\x,\sigma+\zeta)
+ \bar{\bnu}(\x_{0},\x,\sigma-\zeta) \grave{\F}(\x,\sigma-\zeta) \right] }
\hspace{10em} \nonumber \\
& = & -\frac{1}{\pi i} \int_{\bar{\I}} d\sigma' 
\frac{\partial[\bnu^{+}(\x_{0},\x,\sigma')f(\x,\sigma')]/\partial r}
{\sigma'-\sigma}.
\label{7.34l}
\end{eqnarray}

Next, from Eqs.\ (\ref{7.28a}) and (\ref{7.34a}),
\begin{equation}
\frac{1}{2} \left[ \frac{\partial\F^{+}(\x,\sigma)}{\partial r}
+ \frac{\partial\F^{-}(\x,\sigma)}{\partial r} \right] = 
- \frac{\partial f(\x,\sigma)}{\partial \sigma};
\label{7.35a}
\end{equation}
and, from Eq.\ (\ref{7.29b}) in Lem.~\ref{7.3D},
\begin{eqnarray}
\lefteqn{\frac{1}{2} \frac{\partial}{\partial r} \left\{
\bnu^{+}(\x_{0},\x,\sigma) [\F^{+}(\x,\sigma) -\F^{-}(\x,\sigma)] \right\} = } 
\nonumber \\
& & -\frac{1}{\pi i} \int_{\bar{\I}} d\sigma'
\frac{\partial[\bnu^{+}(\x_{0},\x,\sigma')f(\x,\sigma')]/\partial r}
{\sigma'-\sigma}.
\label{7.35b}
\end{eqnarray}
A comparison of the above Eqs.\ (\ref{7.35a}) and (\ref{7.35b}) with
Eqs. (\ref{7.35a}) and (\ref{7.34l}), together with the fact that
$$
\lim_{\zeta \rightarrow 0} \frac{\partial\bar{\bnu}(\x_{0},\x,\sigma\pm\zeta)}
{\partial r} = \frac{\partial\bnu^{\pm}(\x_{0},\x,\sigma)}{\partial r},
$$
now yields
\begin{equation}
\lim_{\zeta \rightarrow 0} \frac{\partial\grave{\F}(\x,\sigma\pm\zeta)}
{\partial r} = \frac{\partial\F^{\pm}(\x,\sigma)}{\partial r}.
\label{7.35c}
\end{equation}
Statements (\ref{7.34i}) and (\ref{7.35c}) complete the proof of part (i)
of our theorem for $\partial\bar{\F}(\x,\tau)/\partial r$.
\cheers

\item 
Since
\begin{equation}
\bnu^{+}(\x_{0},\x,\sigma) = \frac{\M^{+}(\sigma-r)\M^{+}(\sigma-s)}
{\M^{+}(\sigma-r_{0})\M^{+}(\sigma-s_{0})},
\label{7.36a}
\end{equation}
one has 
\begin{equation}
\bnu^{+}(\x_{0},\x,\sigma)f(\x,\sigma)=0 \text{ when } \sigma=r
\text{ and when } \sigma=s.
\end{equation}
Therefore, from statement $1^{0}$ in Sec.~29 of Muskhelishvili's treatise,
and from our Eq.\ (\ref{7.34b}),
\begin{equation}
\bar{\bnu}(\x_{0},\x,\tau)\bar{\F}(\x,\tau) \text{ converges as }
\tau \rightarrow r \text{ and as } \tau \rightarrow s \;
[\tau \in C - \bar{\I}(\x)].
\label{7.36c}
\end{equation}
Furthermore, from Eqs.\ (\ref{7.26d}), (\ref{7.26e}) and (\ref{7.34b}),
\begin{eqnarray}
\lefteqn{\bar{\F}(\x,\tau) = \bar{\bnu}(\x,\x_{0},\tau) I 
+ [\bar{\bnu}(\x,\x_{0},\tau)-1]f(\x,r_{0}) } \nonumber \\
& & \mbox{ } - \frac{\bar{\bnu}(\x,\x_{0},\tau)}{\pi i} \int_{\bar{\I}}
d\sigma' \bnu^{+}(\x_{0},\x,\sigma') \left[
\frac{f(\x,\sigma')-f(\x,r_{0})}{\sigma'-r} \right]. 
\label{7.36d}
\end{eqnarray}
{}From statement (\ref{7.30c}), $\partial f(\x,\sigma')/\partial \sigma'$
exists and is a continuous function of $\sigma'$ throughout $\bar{\I}(\x)$.
Therefore, as one can see from Eq.\ (\ref{7.36a}),
\begin{equation}
\bnu^{+}(\x_{0},\x,\sigma) [f(\x,\sigma')-f(\x,r_{0})] = 0
\text{ when } \sigma = r_{0};
\end{equation}
and it then follows from Eq.\ (\ref{7.36d}) and the same statement $1^{0}$
in Sec.~29 of Muskhelishvili that was used before that
\begin{equation}
\bar{\F}(\x,\tau) \text{ converges [to $-f(\x,r_{0})$] as } \tau \rightarrow
r_{0}.
\label{7.36f}
\end{equation}
Similarly, one proves that
\begin{equation}
\bar{\F}(\x,\tau) \text{ converges [to $-f(\x,s_{0})$] as } \tau \rightarrow
s_{0}.
\label{7.36g}
\end{equation}
Statements (\ref{7.36c}), (\ref{7.36f}) and (\ref{7.36g}) together 
constitute part (ii) of our theorem.
\cheers

\item 
We shall prove this part of our theorem for $i=3$, and the proof for $i=4$
is left to the reader.

We start with the definition (\ref{7.34d}) of $\Phi(\x,\tau)$.  The proof
that we have just given for part (ii) of this theorem is also applicable
to $\Phi(\x,\tau)$.  Specifically, the proof of part (ii) remains valid
if one makes all of the following substitutions in its wording and
equations:
\begin{eqnarray*}
\end{eqnarray*}
Therefore, the conclusion of part (ii) of our theorem remains valid if one
makes the substitution `$\bar{\bnu}(\x_{0},\x,\tau)\bar{\F}(\x,\tau)$'
$\rightarrow$ `$\Phi(\x,\tau)$'.  So, for all $(\x,\tau) \in \dom{\bar{\F}}$,
\begin{equation}
\begin{array}{l}
\Phi(\x,\tau) \text{ converges as } \tau \rightarrow r \text{ and as }
\tau \rightarrow s, \text{ and } \\
\bar{\bnu}(\x,\x_{0},\tau) \Phi(\x,\tau) \text{ converges as }
\tau \rightarrow r_{0} \text{ and as } \tau \rightarrow s_{0}.
\end{array}
\label{7.37}
\end{equation}
When the above statement (\ref{7.37}) is applied to Eq.\ (\ref{7.34c}),
one obtains the statement in part (iii) of our theorem for the case $i=3$.
\cheers
\end{romanlist}
\end{abc}

Note: The meanings that we assigned above to `$f(\x,\sigma')$',
`$\phi(\x,\sigma')$' and `$\Phi(\x,\tau)$' will not be used in the
remainder of these notes.  They were temporary devices for the purpose
of abbreviating the proofs of the preceding theorem and two lemmas.

\setcounter{equation}{0}
\setcounter{theorem}{0}
\subsection{The HHP solution $\bar{\F}$ is a member of $\S_{\bar{\F}}^{3}$
when $\bv \in K^{3}$}

\begin{abc}
\begin{theorem}[$\partial\grave{\F}/\partial x^{i} = \Gamma_{i} 
\grave{\F}$]
\label{7.1E} \mbox{ } \\
When $\bv \in K^{3}$, $\bar{\F}$ is the solution of the HHP corresponding
to $(\bv,\bar{\F}^{M})$ and $H$ is the function defined by Eq.\ (\ref{7.1a})
in Prop.~\ref{7.1A}, then [from Thm.~\ref{7.1D}] $d\grave{\F}(\x,\tau)$
and $dH(\x)$ exist; and, for each $i \in \{3,4\}$,
\begin{equation}
\frac{\partial \grave{\F}(\x,\tau)}{\partial x^{i}} = \Gamma_{i}(\x,\tau)
\grave{\F}(\x,\tau) \text{ for all } (\x,\tau) \in \dom{\grave{\F}},
\end{equation}
where
\begin{equation}
\Gamma_{i}(\x,\tau) := \frac{1}{2(\tau-x^{i})} \frac{\partial H(\x)}
{\partial x^{i}} \Omega.
\end{equation}
\end{theorem}
\end{abc}

\begin{abc}
\proof
{}From GSSM~\ref{4.3D}(ii), $\bar{\F}(\x,\tau)^{-1}$ exists for all
$(\x,\tau) \in \dom{\bar{\F}}$; and, for the continuous extension of $Y$
that is defined by Cor.~\ref{5.5C} (also, see the beginning of
Sec.~\ref{Sec_7}C) and Eq.\ (\ref{5.22b}), $Y(\x,\sigma)^{-1}$ exists
for all $\x \in \domE$ and $\sigma \in \bar{\I}(\x)$.  From Thm.~\ref{7.1D},
$d\bar{\F}(\x,\tau)$, $dY(\x,\sigma)$ and $dH(\x)$ exist and are continuous
functions of $(\x,\tau)$, $(\x,\sigma)$ and $\x$ throughout $\dom{\bar{\F}}$,
$$
\dom{Y} := \{(\x,\sigma):\x \in \domE, \sigma \in \bar{\I}(\x)\}
$$
and $\domE$, respectively; and, for each $\x \in \domE$, $d\bar{\F}(\x,\tau)$
is a holomorphic function of $\tau$ throughout $C - \grave{\I}(\x)$.  It
then follows, with the aid of conditions (1) through (3) in the definition
of the HHP, Eq.\ (\ref{7.1a}) in Prop.~\ref{7.1A}, and Thm.~\ref{7.4D}(i)
that, for each $\x \in \domE$,
\begin{eqnarray}
Z_{i}(\x,\tau) & := & (\tau-x^{i}) \frac{\partial \bar{\F}(\x,\tau)}
{\partial x^{i}} \bar{\F}(\x,\tau)^{-1} \text{ exists and is a holomorphic }
\nonumber \\
& & \text{function of $\tau$ throughout } C-\grave{\I}(\x), 
\label{7.39a} \\
Z_{i}(\x,\tau) & = & \frac{1}{2} \frac{\partial H(\x)}{\partial x^{i}}
\Omega + O(\tau^{-1}) \text{ in at least } \nonumber \\
& & \text{one neighborhood of } \tau = \infty,
\label{7.39b} \\
Z_{i}^{\pm}(\x,\sigma) & & \text{exists for each } \sigma \in \I(\x) \\
\text{and} & & \nonumber \\
Z_{i}^{+}(\x,\sigma) & = & Z_{i}^{-}(\x,\sigma) \nonumber \\
& = & (\sigma-x^{i}) \frac{\partial Y(\x,\sigma)}{\partial x^{i}}
Y(\x,\sigma)^{-1} + Y(\x,\sigma) \frac{1}{2} \frac{\partial H^{M}(\x)}
{\partial x^{i}} \Omega Y(\x,\sigma)^{-1} \nonumber \\
& & \text{for all } \sigma \in \bar{\I}(\x),
\label{7.39d}
\end{eqnarray}
where we have used the fact that the defining equation in condition (3)
for the HHP corresponding to $(\bv,\bar{\F}^{M},\x)$ is expressible in
the form
\begin{equation}
\F^{\pm}(\x,\sigma) = Y^{(j)}(\x,\sigma) \F^{M\pm}(\x,\sigma)
[\tilde{v}^{(j)}(\sigma)]^{-1} \text{ for all } \sigma \in \I^{(j)}(\x);
\label{7.39e}
\end{equation}
and we have used the fact that, since $\bar{\F}^{M\pm} \in \S_{\F^{\pm}}$,
\begin{equation}
\frac{\partial\F^{M\pm}(\x,\tau)}{\partial x^{j}} = \Gamma_{i}^{M}(\x,\tau)
\F^{M\pm}(\x,\tau) \text{ for all } \tau \in \bar{C}^{\pm} - 
\{r,s,r_{0},s_{0}\}.
\label{7.39f}
\end{equation}

We next define a continuous extension of $Z_{i}(\x)$ [which we also denote
by $Z_{i}(\x)$] to the domain $C - \{r,s,r_{0},s_{0}\}$ by letting
\begin{equation}
Z_{i}(\x,\sigma) := Z_{i}^{\pm}(\x,\sigma).
\end{equation}
Then, from the statement (\ref{7.39a}) and the theorem of Riemann that we
have already used in a different context, 
\begin{equation}
Z_{i}(\x,\tau) \text{ is a holomorphic function of } \tau 
\text{ throughout } C - \{r,s,r_{0},s_{0}\}.
\label{7.40b}
\end{equation}
However, from Eq.\ (\ref{7.39a}) and Thms.~\ref{7.4D}(ii) and~(iii),
\begin{equation}
\begin{array}{l}
\bar{\bnu}(\x,\x_{0},\tau) Z_{i}(\x,\tau) \text{ converges as }
\tau \rightarrow r_{0} \text{ and as } \tau \rightarrow s_{0},
\nonumber \\
\text{and } \bar{\bnu}(\x_{0},\x,\tau) Z_{i}(\x,\tau)
\text{ converges as } \tau \rightarrow r \text{ and as } 
\tau \rightarrow s.
\end{array}
\label{7.40c}
\end{equation}
Also, from Eq.\ (\ref{7.39d}) and the continuity on $\bar{\I}(\x)$
of $dY(\x,\sigma)$ and $Y(\x,\sigma)^{-1} = \Omega Y(\x,\sigma)^{T} \Omega$,
\begin{equation}
Z_{i}(\x,\sigma) \text{ converges as } \sigma \rightarrow r_{0},
\sigma \rightarrow s_{0}, \sigma \rightarrow r, \sigma \rightarrow s.
\label{7.40d}
\end{equation}
Combining (\ref{7.40b}), (\ref{7.40c}) and (\ref{7.40d}), one obtains,
by reasoning that should now be familiar to us, $Z_{i}(\x,\tau) = 
Z_{i}(\x,\infty)$, whereupon the conclusion of our theorem follows from
Eqs.\ (\ref{7.39a}) and (\ref{7.39b}).
\cheers
\end{abc}

\begin{abc}
\begin{corollary}[$d\grave{\F}=\Gamma\grave{\F}$]
\label{7.2E} \mbox{ } \\
For each $(\x,\tau) \in \dom{\grave{\F}}$,
\begin{equation}
d\grave{\F}(\x,\tau) = \Gamma(\x,\tau) \grave{\F}(\x,\tau),
\label{7.41a}
\end{equation}
where
\begin{eqnarray}
\Gamma(\x,\tau) & := & \frac{1}{2} (\tau-z+\rho\star)^{-1} dH(\x) \Omega
\nonumber \\
& = & \sum_{i} dx^{i} \Gamma_{i}(\x,\tau).
\label{7.41b}
\end{eqnarray}
\end{corollary}
\end{abc}

\proof
Obvious.
\cheers

\begin{theorem}[$\A \Gamma = \frac{1}{2} \Omega dH \Omega$]
\label{7.3E} \mbox{ } \\
Suppose $\bv \in K^{3}$ and $\bar{\F}$ is the solution of the HHP corresponding
to $(\bv,\bar{\F}^{M})$.  Then
\begin{equation}
\A \Gamma = \frac{1}{2} \Omega dH \Omega,
\label{7.42}
\end{equation}
where $H$, $\A$ and $\Gamma$ are defined by Eqs.\ (\ref{7.1a}), (\ref{7.3a})
and (\ref{7.41b}), respectively.
\end{theorem}

\proof
The proof will be given in three parts:
\begin{arablist}
\begin{abc}
\item 
We start by obtaining two results that we shall need for the proof.  For each
$H' \in \S_{H}$, note that [Eq.\ (\ref{1.12a})]
\begin{equation}
\Re{H'} = -h'
\end{equation}
and that the defining differential equation (\ref{1.12b}) for $\Im{H'}$ in
terms of $\Re{H'}$ is expressible in the following form, since $\star\star
= 1$:
\begin{equation}
h' \Omega d(\Re{H'}) = \rho \star (i \Im{H'}).
\label{7.43b}
\end{equation}
Recall that $h'$ is symmetric and $\det{h'} = \rho^{2}$.  So,
\begin{equation}
(h'\Omega)^{2} = \rho^{2} I.
\label{7.43c}
\end{equation}
{}From Eq.\ (\ref{7.43c}), Eq.\ (\ref{7.43b}) is equivalent to the equation
\begin{equation}
h'\Omega d(i\Im{H'}) = \rho \star d(\Re{H'});
\end{equation}
and, therefore, Eq.\ (\ref{7.43b}) is equivalent to the equation
\begin{equation}
h'\Omega dH' = \rho \star dH'.
\label{7.43c'}
\end{equation}
Furthermore, the above Eq.\ (\ref{7.43c'}) yields
$$
\A' \Gamma' =
[(\tau-z)\Omega+\Omega h' \Omega] \frac{1}{2} (\tau-z+\rho\star)^{-1}
dH' \Omega = \frac{1}{2} (\tau-z+\rho\star)^{-1}
[(\tau-z) \Omega dH' \Omega + \rho \star \Omega dH' \Omega].
$$
So, Eq.\ (\ref{7.43c'}) implies
\begin{equation}
\A' \Gamma' = \frac{1}{2} \Omega dH' \Omega
\text{ for each } H' \in \S_{H}.
\label{7.43d}
\end{equation}
The reader should have no difficulty in proving that, conversely, Eq.\ 
(\ref{7.43d}) implies (\ref{7.43c'}).  We shall use the above result
later in our proof.
\end{abc}

\begin{abc}
\item 
We now obtain a second result that we shall need for the proof.  From
Eq.\ (\ref{7.3a}) in Thm.~\ref{7.2A},
\begin{equation}
[\F^{\mp}(\x,\sigma)]^{\dagger} \A(\x,\sigma) \F^{\pm}(\x,\sigma) =
\A(\x_{0},\sigma) \text{ for all } \sigma \in \I(\x).
\label{7.44a}
\end{equation}
Now, recall that $\A(\x_{0},\sigma) = \A^{M}(\x_{0},\sigma)$ in our
gauge.  Also, recall that
\begin{equation}
[\bar{\F}'(\x,\tau^{*})]^{\dagger} \A'(\x,\tau) \bar{\F}'(\x,\tau) =
\A^{M}(\x_{0},\sigma) \text{ for all } \bar{\F}' \in \S_{\bar{\F}}.
\label{7.44b}
\end{equation}
Therefore, we obtain the following result by using Eqs.\ (\ref{7.39e})
[condition (3) in the definition of the HHP corresponding to
$(\bv,\bar{\F}^{M},\x)$, (\ref{7.44a}), (\ref{4.2}) [in Thm.~\ref{4.1A}]
and (\ref{7.44b}) [for $\bar{\F}' = \bar{\F}^{M}$]:
$$
Y^{\dagger}(\x,\sigma) \A(\x,\sigma) Y(\x,\sigma) = \A^{M}(\x,\sigma)
\text{ for all } \sigma \in \I(\x).
$$
However, recall that $Y(\x,\sigma)$ is now a continuous function of 
$\sigma$ throughout $\bar{\I}(\x)$.  Therefore,
\begin{equation}
Y^{\dagger}(\x,\sigma) \A(\x,\sigma) Y(\x,\sigma) = \A^{M}(\x,\sigma)
\text{ for all } \sigma \in \bar{\I}(\x).
\label{7.44c}
\end{equation}
We shall use this result below.
\end{abc}

\begin{abc}
\item 
{}From the definition of $\A$ and $\Gamma$, each component of $\A(\x,\tau)
\Gamma(\x,\tau)$ is a holomorphic function of $\tau$ throughout
$C - \{r,s\}$ and has no essential singularity at $\tau = r$ and at
$\tau = s$.  In fact, if there are any singularities at these points,
they are simple poles.  That much is obvious.

{}From Eqs.\ (\ref{7.41a}), (\ref{7.39e}) and (\ref{7.39f}),
\begin{eqnarray}
\A(\x,\sigma) \Gamma(\x,\sigma) & = & \A(\x,\sigma) [d\F^{\pm}(\x,\sigma)]
[\F^{\pm}(\x,\sigma)]^{-1} \nonumber \\
& = & \A(\x,\sigma) \left[dY(\x,\sigma) + Y(\x,\sigma) \Gamma^{M}(\x,\sigma)
\right] [Y(\x,\sigma)]^{-1} \nonumber \\
& & \text{for all } \sigma \in \I(\x).
\end{eqnarray}
The above equation becomes, after using Eq.\ (\ref{7.44c}),
$$
\A(\x,\sigma) \Gamma(\x,\sigma) = \left\{ \A(\x,\sigma) dY(\x,\sigma)
+ [Y(\x,\sigma)^{\dagger}]^{-1} \A^{M}(\x,\sigma) \Gamma^{M}(\x,\sigma)
\right\} [Y(\x,\sigma)]^{-1},
$$
which becomes, after using Eq.\ (\ref{7.43d}) with $H' = H^{M}$,
\begin{equation}
\A(\x,\sigma) \Gamma(\x,\sigma) = \left\{ \A(\x,\sigma) dY(\x,\sigma)
+ [Y(\x,\sigma)^{\dagger}]^{-1} \frac{1}{2} \Omega dH^{M}(\x) \Omega 
\right\} [Y(\x,\sigma)]^{-1}.
\end{equation}
{}From Thm.~\ref{7.1D} and the fact that $\det{Y(\x,\sigma)} = 1$, the
right side of the above equation is a continuous function of $\sigma$
throughout $\bar{\I}(\x)$.  Therefore, $\A(\x,\tau) \Gamma(\x,\tau)$
is extendable to a holomorphic function of $\tau$ throughout $C$; and
it follows that
$$
\A(\x,\tau) \Gamma(\x,\tau) = [\A(\x,\tau)\Gamma(\x,\tau)]_{\tau=\infty}
= \frac{1}{2} \Omega dH(\x) \Omega.
$$
\end{abc}
\end{arablist}
\cheers

\begin{corollary}[$h \Omega dH = \rho \star dH$]
\label{7.4E} \mbox{ } \\
Suppose $H$ is defined as in the preceding theorem.  Then
\begin{equation}
h \Omega dH = \rho \star dH.
\label{7.46}
\end{equation}
\end{corollary}

\proof
Multiply both sides of Eq.\ (\ref{7.42}) through by $2\Omega(\tau-z+\rho\star)$
on the left, and by $\Omega$ on the right; and then set $\tau=z$.
\cheers

In the following theorem we assert a result that is stronger than simply
that the HHP solution $\bar{\F}$ is a member of $\S_{\bar{\F}}^{3}$ when 
$\bv \in K^{3}$.

\begin{gssm}[HHP solution $\bar{\F} \in \S_{\bar{\F}}^{\Box}$]
\label{7.5E} \mbox{ } \\ \vspace{-3ex}
\begin{romanlist}
\item 
For each $\bv \in K^{\Box}$, where $\Box$ is $n \ge 3$, $n+$ ($n \ge 3$),
$\infty$ or `an', and, for each $\bar{\F}_{0} \in \S_{\bar{\F}}^{\Box}$,
there exists exactly one solution $\bar{\F}$ of the HHP corresponding to
$(\bv,\bar{\F}_{0})$.

\item 
Let $H$ be defined in terms of $\bar{\F}$ by Eq.\ (\ref{7.1a}), and let
$\E := H_{22}$.  Then $\E \in \S_{\E}$, and $H$ is identical with the
unique member of $\S_{H}$ that is constructed from $\E$ by using Eqs.\ 
(\ref{1.11a}) to (\ref{1.12c}).

\item 
Furthermore, $\bar{\F}$ is identical with the member of $\S_{\bar{\F}}$ that
is defined in terms of $H$ in Sec.~\ref{Sec_1}F  [and whose existence and
uniqueness for a given $H \in \S_{H}$ is asserted in Thm.~\ref{Thm_3}.]

\item 
Finally, let $\hat{\Q}$ denote the member of $\S_{\hat{\Q}}$ that is 
defined in terms of $\bar{\F}$ by Eq.\ (\ref{G2.14a}), and let $\bV$
denote the member of $\S_{\bV}$ that is defined in terms of $\hat{\Q}$
by Eq.\ (\ref{G3.1}).  Then $\bV \in \S_{\bV}^{\Box}$ and, therefore
(by definition), $\E \in \S_{\E}^{\Box}$, $H \in \S_{H}^{\Box}$ and
$\bar{\F} \in \S_{\bar{\F}}^{\Box}$.
\end{romanlist}
\end{gssm}

\proofs
\begin{romanlist}
\begin{abc}
\item 
Let $\bV_{0}$ denote the member of $\S_{\bV}$ that corresponds to
$\bar{\F}_{0}$.  Since $\bar{\F}_{0} \in \S_{\bar{\F}}^{\Box}$, there
exists (by definition of $\S_{\bar{\F}}^{\Box}$) $\bw \in B(\I^{3},\I^{(4)})$
such that
\begin{equation}
\bV_{0} \bw \in k^{\Box} \subset K^{\Box};
\end{equation}
and, since $K^{\Box}$ is a group,
\begin{equation}
\bv \bV_{0} \bw \in K^{\Box}.
\label{7.47b}
\end{equation}
{}From GSSM \ref{Thm_13}, there exists exactly one solution $\bar{\F}$ of the
HHP corresponding to $(\bv\bV_{0}\bw,\bar{\F}^{M})$; and, from GSSM \ref{4.2D},
it then follows that $\bar{\F}$ is also a solution of the HHP corresponding to
$(\bv,\bar{\F}_{0})$.  Finally, from GSSM \ref{4.3D}(iv), there is no other
solution of the HHP corresponding to $(\bv,\bar{\F}_{0})$.
\cheers
\end{abc}

\begin{abc}
\item 
{}From the premises of this theorem, $\bv \in K^{3}$.  Therefore, from 
statement (\ref{7.24b}) in Thm.~\ref{7.1D},
\begin{equation}
dH \text{ exists and is continuous }
\end{equation}
throughout $\domE$; and since
\begin{eqnarray}
(d^{2}H)(\x) & = & dr ds \left[ \frac{\partial^{2}H(\x)}{\partial r \partial s}
- \frac{\partial^{2}H(\x)}{\partial s \partial r} \right] \\
\text{and} & & \nonumber \\
(d\star dH)(\x) & = & -dr ds \left[
\frac{\partial^{2}H(\x)}{\partial r \partial s}
+ \frac{\partial^{2}H(\x)}{\partial s \partial r} \right], 
\end{eqnarray}
it is true that 
\begin{equation}
d^{2}H \text{ exists and vanishes }
\end{equation}
and
\begin{equation}
d\star dH \text{ exists and is continuous }
\end{equation}
throughout $\domE$.  Also, Eq.\ (\ref{7.46}) in Cor.~\ref{7.4E} asserts that
\begin{equation}
\rho \star dH = h \Omega dH,
\label{7.48d}
\end{equation}
where we recall from Eq.\ (\ref{7.1b}) in Prop.~\ref{7.1A} that
\begin{equation}
h := - \Re{H} = h^{T}
\label{7.48e}
\end{equation}
and, from Thm.~\ref{7.3A},
\begin{equation}
\det{h} = \rho^{2} \text{ and } f := \Re{\E} = -g_{22} < 0,
\label{7.48f}
\end{equation}
where $g_{ab}$ denotes the element of $h$ in the $a$th row and $b$th
column.  Since $\star\star=1$, Eq.\ (\ref{7.48d}) is equivalent to the
equation 
\begin{equation}
\rho dH = h \Omega \star dH
\label{7.48g}
\end{equation}
from which we obtain
\begin{equation}
\rho dH^{\dagger} \Omega dH = dH^{\dagger} \Omega h \Omega \star dH.
\label{7.48h}
\end{equation}
Upon taking the hermitian conjugates of the terms in the above equation,
and upon noting that $\Omega^{\dagger} = \Omega$, $h^{\dagger} = h$,
\begin{equation}
(\omega\eta)^{T} = - \eta^{T} \omega^{T} \text{ and }
\omega \star \eta = - (\star \omega) \eta \text{ for any }
n \times n \text{ matrix $1$-forms, }
\end{equation}
one obtains
\begin{equation}
-\rho dH^{\dagger} \Omega dH = dH^{\dagger} \Omega h \Omega \star dH.
\label{7.48j}
\end{equation}
{}From Eqs.\ (\ref{7.48j}) and (\ref{7.48h}),
\begin{equation}
dH^{\dagger} \Omega dH = 0.
\label{7.48k}
\end{equation}
[The above result (\ref{7.48k}) was first obtained by the authors in a
paper\footnote{I.~Hauser and F.~J.~Ernst, J.\ Math.\ Phys.\ {\bf 21},
1116-1140 (1980).  See Eq.\ (37).} which introduced an abstract geometric
definition of the Kinnersley potential $H$ and which derived other 
properties of $H$ that we shall not need in these notes.]

We next consider the $(2,2)$ matrix elements of Eqs.\ (\ref{7.48d}) and
(\ref{7.48k}).  With the aid of Eqs.\ (\ref{7.48e}) and (\ref{7.48f}), 
one obtains
\begin{equation}
\rho \star d\E = i(g_{12} d\E + f dH_{12})
\label{7.49a}
\end{equation}
and 
\begin{equation}
dH_{12}^{*} d\E - d\E^{*} dH_{12} = 0.
\label{7.49b}
\end{equation}
{}From Eq.\ (\ref{7.49a}),
\begin{eqnarray*}
f d(\rho \star d\E) - \rho d\E \star d\E & = & 
if (dg_{12} d\E + df dH_{12} - d\E dH_{12}) \\
& = & if \left[ -\frac{1}{2}(dH_{12}+dH_{12}^{*})d\E
+ \frac{1}{2}(d\E + d\E^{*}) dH_{12} - d\E dH_{12} \right] \\
& = & \frac{if}{2} (- dH_{12}^{*} d\E + d\E^{*} dH_{12}).
\end{eqnarray*}
Therefore, from Eq.\ (\ref{7.49b}),
\begin{equation}
f d(\rho \star d\E) - \rho d\E \star d\E = 0.
\label{7.49c}
\end{equation}
Furthermore, from Eqs.\ (\ref{1.12c}), (\ref{7.1b}) and (\ref{7.48f}),
\begin{equation}
\E(\x_{0}) = -1 \text{ and } \Re{\E} < 0.
\label{7.49d}
\end{equation}
Therefore,
\begin{equation}
\E \in \S_{\E},
\end{equation}
since $\E$ satisfies the Ernst equation (\ref{7.49c}) and the requisite
gauge conditions (\ref{7.49d})

Next, let
\begin{equation}
\chi := \Im{\E} \text{ and } \omega := g_{12}/g_{22}.
\end{equation}
Then, by taking the imaginary parts of the terms in Eq.\ (\ref{7.49a}),
one deduces
\begin{equation}
d\omega = \rho f^{-2} \star d\chi.
\label{7.50b}
\end{equation}
Furthermore, Eqs.\ (\ref{7.48e}) and (\ref{7.48f}) enable us to express
$h$ in the form
\begin{equation}
h = A \left( \begin{array}{cc}
\rho^{2} & 0 \\ 0 & 1
\end{array} \right) A^{T},
\end{equation}
where
\begin{equation}
A := \left( \begin{array}{cc}
1 & \omega \\ 0 & 1
\end{array} \right) \left( \begin{array}{cc}
1/\sqrt{-f} & 0 \\ 0 & \sqrt{-f}
\end{array} \right).
\end{equation}
Finally, the imaginary parts of the terms in Eq.\ (\ref{7.48g}) give us
\begin{equation}
\rho d(\Im{H}) = - h J \star dh.
\label{7.50e}
\end{equation}
A comparison of Eqs.\ (\ref{7.1b}), (\ref{7.1d}) and (\ref{7.50b}) to
(\ref{7.50e}) with the definition of $\S_{H}$ that is given in
Sec.~\ref{Sec_1}B demonstrates that $H$ is precisely that member of $\S_{H}$
that is computed from $\E$ by using Eqs.\ (\ref{1.11a}) to (\ref{1.12c}).
\cheers
\end{abc}

\begin{abc}
\item 
{}From statement (\ref{7.24a}) in Thm.~\ref{7.1D},
\begin{equation}
d\bar{\F}(\x,\tau) \text{ exists for all } \x \in \domE \text{ and }
\tau \in C - \grave{\I}(\x) = [C - \bar{\I}(\x)] - \{r_{0},s_{0}\};
\label{7.51a}
\end{equation}
and, from Cor.~\ref{7.2E},
\begin{equation}
d\bar{\F}(\x,\tau) = \Gamma(\x,\tau) \bar{\F}(\x,\tau) \text{ for all }
\x \in \domE \text{ and } \tau \in C - \grave{\I}(\x).
\end{equation}
Furthermore, from GSSM \ref{4.3D}(v),
\begin{equation}
\bar{\F}(\x_{0},\tau) = I \text{ for all } \tau \in C.
\end{equation}
Finally, consider that $\bar{\I}(\x) = \bar{\I}^{(3)}(\x)$ when $s=s_{0}$,
and $\bar{\I}(\x) = \bar{\I}^{(4)}(\x)$ when $r=r_{0}$; and, from condition 
(1) in the definition of the HHP, $\bar{\F}(\x,\tau)$ is a holomorphic
function of $\tau$ throughout $C - \bar{I}(\x)$.  Therefore,
\begin{equation}
\begin{array}{l}
\bar{\F}((r,s_{0}),\tau) \text{ and } \bar{\F}((r_{0},s),\tau) \\
\text{are continuous functions of $\tau$ at } \\
\text{$\tau=s_{0}$ and at $\tau=r_{0}$, respectively. }
\end{array}
\label{7.51d}
\end{equation}
{}From the above statements (\ref{7.51a}) to (\ref{7.51d}) and from the
definition of $\S_{\bar{\F}}$ in Sec.~\ref{Sec_1}F, it follows that
$\bar{\F} \in \S_{\bar{\F}}$.
\cheers
\end{abc}

\begin{abc}
\item 
{}From condition (3) in the definition of the HHP and from GSSM \ref{4.1C},
there exists $\bw' \in B(\I^{(3)},\I^{(4)})$ such that
\begin{equation}
\bv = \bV \bw' \bV_{0}^{-1}.
\label{7.52a}
\end{equation}
Therefore,
\begin{equation}
\bV = \bv \bV_{0} (\bw')^{-1}.
\end{equation}
However, from the proof of part (i) this theorem [see Eq.\ (\ref{7.47b})],
there then exists $\bw \in B(\I^{(3)},\I^{(4)})$ such that
\begin{equation}
\bV(\bw' \bw) = \bv \bV_{0} \bw \in K^{\Box}.
\end{equation}
Therefore, since $\bw' \bw \in B(\I^{(3)},\I^{(4)})$, it follows from the
definition of $\S_{\bV}^{\Box}$ given by Eq.\ (\ref{G3.8b}) that
$\bV \in \S_{\bV}^{\Box}$.  Hence, by definition, $\E \in \S_{\E}^{\Box}$,
$H \in \S_{H}^{\Box}$ and $\bar{\F} \in \S_{\bar{\F}}^{\Box}$.
\cheers
\end{abc}
\end{romanlist}

\begin{corollary}[$k^{\Box} = K^{\Box}$]
\label{7.6E} \mbox{ } \\
Suppose that $\Box$ is $n \ge 3$, $n+$ ($n \ge 3$), $\infty$ or `an'.
Then
\begin{equation}
k^{\Box} := \{ \bV \bw:\bV \in \S_{\bV}, \bw \in B(\I^{(3)},\I^{(4)}),
\bV \bw \in K^{\Box}\} = K^{\Box}.
\end{equation}
\end{corollary}

\begin{abc}
\proof
{}From its definition
\begin{equation}
k^{\Box} \subset K^{\Box}.
\label{7.54a}
\end{equation}

Now, suppose $\bv \in K^{\Box}$.  Since
\begin{equation}
\bV^{M} = (\delta^{(3)},\delta^{(4)}),
\label{7.54b}
\end{equation}
where
\begin{equation}
\delta^{(i)}(\sigma) := I \text{ for all } \sigma \in \I^{(i)},
\label{7.54c}
\end{equation}
we know that
\begin{equation}
\bar{\F}^{M} \in \S_{\bar{\F}}^{an} \subset \S_{\bar{\F}}^{\Box}.
\label{7.54d}
\end{equation}
Therefore, from the preceding theorem, there exists $\bar{\F} \in
\S_{\bar{\F}}^{\Box}$ such that $\bar{\F}$ is the solution of the HHP
corresponding to $(\bv,\bar{\F}^{M})$; and, if $\bV$ denotes the member
of $\S_{\bV}^{\Box}$ that corresponds to $\bar{\F}$, Eq.\ (\ref{7.52a})
in the proof of the preceding theorem informs us that $\bw' \in
B(\I^{(3)},\I^{(4)})$ exists such that $\bv = \bV \bw'$.  So $\bv \in
k^{\Box}$.

We have thus proved that
\begin{equation}
K^{\Box} \subset k^{\Box},
\end{equation}
whereupon (\ref{7.54a}) and (\ref{7.54b}) furnish us with the conclusion
$k^{\Box} = K^{\Box}$.
\cheers
\end{abc}

\setcounter{equation}{0}
\setcounter{theorem}{0}
\subsection{Proof of our generalized Geroch conjecture}

\begin{definition}{Dfn.\ of $Z^{(i)}$\label{def74}}
Let $Z^{(i)}$ denote the subgroup of $K(\x_{0},\I^{(i)})$ that is given by
\begin{equation}
Z^{(i)} := \{\delta^{(i)},-\delta^{(i)}\},
\end{equation}
where $\delta^{(i)}$ is defined by Eq.\ (\ref{7.54c}).
\end{definition}

\begin{proposition}[Center of $K$]
\label{7.1F} \mbox{ } \\
The center of $K(\x_{0},\I^{(i)})$ is $Z^{(i)}$.  Hence the center of
$K$ is $Z^{(3)} \times Z^{(4)}$.
\end{proposition}

\proof
Left for the reader.  Hint:  See the proof of Lem.~\ref{7.2F}(i).
\cheers

\begin{abc}
\begin{definition}{Dfn.\ of $[\bv]$ for each $\bv \in K^{3}$\label{def75}}
For each $\bv \in K^{3}$, let $[\bv]$ denote the function such that
\begin{equation}
\dom{[\bv]} := \S_{\bar{\F}}^{3}
\end{equation}
and, for each $\bar{\F}_{0} \in \S_{\bar{\F}}^{3}$,
\begin{equation}
[\bv](\bar{\F}_{0}) := \text{ the solution of the HHP corresponding to }
(\bv,\bar{\F}_{0}).
\end{equation}
Note that the existence of $[\bv]$ is guaranteed by GSSM \ref{4.2D}
and GSSM \ref{Thm_13}(iii).
\end{definition}
\end{abc}

\begin{definition}{Dfn.\ of $\K^{\Box}\s$ when $\Box$ is $n \ge 3$,
$n+$ ($n \ge 3$), $\infty$ or `an'\label{def76}}
Let 
\begin{equation}
\K^{\Box} := \{[\bv]:\bv \in K^{\Box}\}.
\end{equation}
\end{definition}

The following lemma concerns arbitrary members $\bv$ and $\bv'$ of $K$,
and arbitrary members $\bar{\F}_{0}$ and $\bar{\F}$ of $\S_{\bar{\F}}$.
Therefore, the lemma could have been given as a proposition in
Sec.~\ref{Sec_4}D.  However, we have saved it for now, because the
lemma is directly applicable in the proof of the next theorem.

\begin{lemma}[Properties of $K$]
\label{7.2F} \mbox{ } \\ \vspace{-3ex}
\begin{romanlist}
\item 
Suppose that $\bv \in K$, $\bar{\F}_{0} \in \S_{\bar{\F}}$ and $\bar{\F}
\in \S_{\bar{\F}}$.  Then $\bar{\F}$ is the solution of the HHP corresponding
to $(\bv,\bar{\F}_{0})$ if and only if $\bV^{-1} \bv \bV_{0} \in 
B(\I^{(3)},\I^{(4)})$, where $\bV_{0}$ and $\bV$ are the members of $\S_{\bV}$
corresponding to $\bar{\F}_{0}$ and $\bar{\F}$, respectively.

In particular, the solution of the HHP corresponding to $(\bv,\bar{\F}_{0})$
is $\bar{\F}_{0}$ if and only if $\bV_{0}^{-1} \bv \bV_{0} \in
B(\I^{(3)},\I^{(4)})$; and the solution of the HHP corresponding to
$(\bv,\bar{\F}^{M})$ is $\bar{\F}^{M}$ if and only if $\bv \in
B(\I^{(3)},\I^{(4)})$.

\item 
In addition to the premises of part (i) of this lemma, suppose that
$\bv' \in K$.  Thereupon, if $\bar{\F}$ is the solution of the HHP
corresponding to $(\bv,\bar{\F}_{0})$, and $\bar{\F}'$ is the solution
of the HHP corresponding to $(\bv',\bar{\F})$, then $\bar{\F}'$ is the
solution of the HHP corresponding to $(\bv' \bv,\bar{\F}_{0})$.

If $\bar{\F}$ is the solution of the HHP corresponding to $(\bv,\bar{\F}_{0})$,
then $\bar{\F}_{0}$ is the solution of the HHP corresponding to 
$(\bv^{-1},\bar{\F})$.
\end{romanlist}
\end{lemma}

\proofs
\begin{romanlist}
\item 
This theorem follows from GSSM \ref{4.1C} and the properties of members of
$\S_{\bar{\F}}$ that are given in GSSM \ref{Thm_3} [specifically, the
properties $\bar{\F}(\x,\infty)=I$ and (iv)] and GSSM \ref{3.7G}.  The
reader can easily fill in the details of the proof.
\cheers

\item 
This follows from the obvious facts that the equations
\begin{eqnarray*}
Y^{(i)}(\x,\sigma) & := & \F^{+}(\x,\sigma) \tilde{v}^{(i)}(\sigma)
[\F_{0}^{+}(\x,\sigma)]^{-1} \\
& = & \F^{-}(\x,\sigma) \tilde{v}^{(i)}(\sigma)
[\F_{0}^{-}(\x,\sigma)]^{-1}
\end{eqnarray*}
and
\begin{eqnarray*}
Y^{\prime(i)}(\x,\sigma) & := & \F^{\prime +}(\x,\sigma) \tilde{v}^{\prime(i)}(\sigma)
[\F^{+}(\x,\sigma)]^{-1} \\
& = & \F^{\prime -}(\x,\sigma) \tilde{v}^{\prime(i)}(\sigma)
[\F^{-}(\x,\sigma)]^{-1}
\end{eqnarray*}
imply
\begin{eqnarray*}
Y^{\prime(i)}(\x,\sigma) Y^{(i)}(\x,\sigma) & = & 
\F^{\prime +}(\x,\sigma) \tilde{v}^{\prime(i)}(\sigma) \tilde{v}^{(i)}(\sigma)
[\F_{0}^{+}(\x,\sigma)]^{-1} \\
& = & \F^{\prime -}(\x,\sigma) \tilde{v}^{\prime(i)}(\sigma)
\tilde{v}^{(i)}(\sigma)
[\F_{0}^{-}(\x,\sigma)]^{-1}
\end{eqnarray*}
and
\begin{eqnarray*}
[Y^{(i)}(\x,\sigma)]^{-1} & = & \F_{0}^{+}(\x,\sigma)
[\tilde{v}^{(i)}(\sigma)]^{-1} [F^{+}(\x,\sigma)]^{-1} \\
& = & \F_{0}^{+}(\x,\sigma)
[\tilde{v}^{(i)}(\sigma)]^{-1} [F^{+}(\x,\sigma)]^{-1}
\end{eqnarray*}
for all $i \in \{3,4\}$ and $\sigma \in \I^{(i)}$.
\cheers
\end{romanlist}

In the following theorem we assert a result that is stronger than the
assertion that we conjectured at the end of part II of these notes, where
$\Box$ was $3$.

\begin{gssm}[Generalized Geroch conjecture]
\label{7.3F} \mbox{ } \\ 
Suppose that $\Box$ is $n \ge 3$, $n+$ ($n \ge 3$), $\infty$ or `an',
and $\bv \in K^{\Box}$.  Then the following statements hold:
\begin{romanlist}
\item 
The mapping $\bv$ is the identity map on $\S_{\bar{\F}}^{\Box}$ if
and only if $[\bv] \in Z^{(3)} \times Z^{(4)}$.
\item 
The set $\K^{\Box}$ is a group of permutations of $\S_{\bar{\F}}^{\Box}$.
The mapping $\bv \rightarrow [\bv]$ is a homomorphism of $K^{\Box}$ onto
$\K^{\Box}$; and the mapping $\{\bv \bw: \bw \in Z^{(3)} \times Z^{(4)}\}
\rightarrow [\bv]$ is an isomorphism [viz., the isomorphism of
$K^{\Box}/(Z^{(3)}\times Z^{(4)})$ onto $\K^{\Box}$].
\item 
The group $\K^{\Box}$ is transitive.
\end{romanlist}
\end{gssm}

\proofs
\begin{romanlist}
\begin{abc}
\item 
The statement that $[\bv]$ is the identity mapping on $\S_{\bar{\F}}^{\Box}$
means that each $\bar{\F}_{0} \in \S_{\bar{\F}}$ is the solution of the
HHP corresponding to $(\bv,\bar{\F}_{0})$; and, from Lem.~\ref{7.2F}(i),
this is equivalent to the following statement:
\begin{equation}
\text{For each } \bV_{0} \in \S_{\bV}^{\Box},
\bV_{0}^{-1} \bv \bV_{0} \in B(\I^{(3)},\I^{(4)}).
\label{7.58a}
\end{equation}
Since $k^{\Box}=K^{\Box}$ (Cor.~\ref{7.6E}), each $\bv' \in K^{\Box}$ is
also a member of $k^{\Box}$, and this means that there exist $\bV' \in
\S_{\bV}$ and $\bw' \in B(\I^{(3)},\I^{(4)})$ such that $\bv' = \bV' \bw'$.
Therefore, from statement (\ref{7.58a}),
$$
\begin{array}{l}
\text{For each } \bv' \in K^{\Box}, \text{ there exists }
\bw' \in B(\I^{(3)},\I^{(4)}) \\
\text{such that } \bw' (\bv')^{-1} \bv \bV' (\bw')^{-1} \in
B(\I^{(3)},\I^{(4)}).
\end{array}
$$
So, since $B(\I^{(3)},\I^{(4)})$ is a group,
\begin{equation}
\text{For each } \bv' \in K^{\Box}, (\bv')^{-1} \bv \bv' \in
B(\I^{(3)},\I^{(4)}).
\label{7.58b}
\end{equation}
In particular, since $\bV^{M} \in K^{\Box}$ [see Eqs.\ (\ref{7.54b}) to
(\ref{7.54d})] and $(\bV^{M})^{-1} \bv \bV^{M} = \bv$,
\begin{equation}
\bv \in B(\I^{(3)},\I^{(4)}).
\end{equation}
Therefore, there exist
\begin{equation}
\alpha_{0}^{(i)}:\I^{(i)} \rightarrow R^{1} \text{ and }
\alpha_{1}^{(i)}:\I^{(i)} \rightarrow R^{1}
\end{equation}
such that
\begin{equation}
v^{(i)} = I \alpha_{0}^{(i)} + J \alpha_{1}^{(i)};
\label{7.58e}
\end{equation}
and, since $\bv \in K^{\Box}$ and $\det{v^{(i)}} = 1$,
\begin{equation}
\alpha_{0}^{(i)} \text{ and } \alpha_{1}^{(i)} \text{ are } \bC^{\Box}
\end{equation}
and
\begin{equation}
(\alpha_{0})^{2} + (\alpha_{1})^{2} = 1.
\label{7.58g}
\end{equation}

Also, the function whose domain is $\I^{(i)}$ and whose values are given by
$$
u^{(i)}(\sigma) = I \cosh[\mu^{+}(\x_{0},\sigma)] + \sigma_{3} 
\sinh[\mu^{+}(\x_{0},\sigma)]
$$
is a member of $K^{an}(\x_{0},\I^{(i)}) \subset K^{\Box}(\x_{0},\I^{(i)})$.
Upon letting $\bv' = (u^{(3)},u^{(4)})$ in Eq.\ (\ref{7.58b}), and upon
using Eq.\ (\ref{7.58e}), one obtains
$$
I \alpha_{0}^{(i)} + J \alpha_{1}^{(i)} [u^{(i)}]^{2} \in B(\I^{(i)});
$$
and this is true if and only if
\begin{equation}
\alpha_{1}^{(i)}(\sigma) \sinh[2\mu^{+}(\x_{0},\sigma)] = 0
\text{ for all } \sigma \in \I^{(i)}.
\label{7.58h}
\end{equation}
However, $\alpha_{1}^{(i)}$ is continuous.  Therefore, the condition
(\ref{7.58h}) can hold if and only if $\alpha_{1}^{(i)}$ is identically
zero, whereupon (\ref{7.58e}) and (\ref{7.58g}) yield $v^{(i)} = \pm
\delta^{(i)}$.  Hence $\bv \in Z^{(3)} \times Z^{(4)}$ is a necessary
and sufficient condition for $[\bv] = $ the identity map on
$\S_{\bar{\F}}^{\Box}$.
\cheers
\end{abc}

\begin{abc}
\item 
Suppose $\bV \in K^{\Box}$ and suppose $\bar{\F} \in \S_{\bar{\F}}^{\Box}$
and the corresponding member of $\S_{\bV}^{\Box}$ is $\bV$.  From the
definition of $\S_{\bar{\F}}^{\Box}$ and Cor.~\ref{7.6E}, there exists
$\bw \in B(\I^{(3)},\I^{(4)})$ such that 
$$
\bV \bw \in k^{\Box} = K^{\Box}.
$$
Therefore, since $\bv \in K^{\Box}$ and $K^{\Box}$ is a group,
$$
\bv^{-1} \bV \bw \in K^{\Box} = k^{\Box}.
$$
Therefore, from the definition of $k^{\Box}$, there exist $\bV_{0}
\in \S_{\bV}$ and $\bw' \in B(\I^{(3)},\I^{(4)})$ such that
$$
\bv^{-1} \bV \bw = \bV_{0} \bw'.
$$
So, since $B(\I^{(3)},\I^{(4)})$ is a group,
$$
\bV^{-1} \bv \bV_{0} \in B(\I^{(3)},\I^{(4)}).
$$
It then follows from Lem.~\ref{7.2F}(i) that $\bar{\F}$ is the solution
of the HHP corresponding to $(\bv,\bar{\F}_{0})$, where $\bar{\F}_{0}$
is the member of $\S_{\bar{\F}}^{\Box}$ that corresponds to $\bV_{0}$.

We have thus shown that every member $\bar{\F}$ of $\S_{\bar{\F}}^{\Box}$
is in the range of $[\bv]$; i.e., 
\begin{equation}
[\bv] \text{ is a mapping of } \S_{\bar{\F}}^{\Box} \text{ onto }
\S_{\bar{\F}}^{\Box}.
\label{7.59a}
\end{equation}

Next, suppose $\bar{\F}_{0}$ and $\bar{\F}'_{0}$ are members of 
$\S_{\bar{\F}}^{\Box}$ such that
$$
\bar{\F} := [\bv](\bar{\F}_{0}) = [\bv](\bar{\F}'_{0}).
$$
Then, $\bar{\F}$ is the solution of the HHP's corresponding to
$(\bv,\bar{\F}_{0})$ and to $(\bv,\bar{\F}'_{0})$, whereupon
Lem.~\ref{7.2F}(ii) informs us that $\bar{\F}'_{0}$ is the solution
of the HHP corresponding to $(\bv^{-1}\bv,\bar{\F}_{0})$.  Hence,
$\bar{\F}'_{0} = \bar{\F}_{0}$.

We have thus shown that $[\bv]$ is one-to-one.  Upon combining this result
with (\ref{7.59a}), we obtain
\begin{equation}
\begin{array}{l}
\text{For each } \bv \in K^{\Box}, [\bv] \text{ is a permutation of }
\S_{\bar{\F}}^{\Box} \\
\text{\{ i.e., $[\bv]$ is a one-to-one mapping of $\S_{\bar{\F}}^{\Box}$
onto $\S_{\bar{\F}}^{\Box}$ \}. }
\end{array}
\end{equation}

Furthermore, the reader can easily show from Lem.~\ref{7.2F}(ii) that, if
\begin{equation}
[\bv'] \circ [\bv] := \text{ the composition of the mappings }
[\bv'] \text{ and } [\bv],
\end{equation}
then
\begin{equation}
[\bv'] \circ [\bv] = [\bv' \bv].
\end{equation}
Lemma \ref{7.2F}(ii) also yields
\begin{equation}
[\bv]^{-1} = [\bv^{-1}].
\end{equation}
Therefore, since $K^{\Box}$ is a group, $\K^{\Box}$ is a group with
respect to composition of mappings.

The remainder of the proof is straightforward and is left to the reader.
\cheers
\end{abc}

\item 
Let $\bar{\F}_{0}$ and $\bar{\F}$ be any members of $\S_{\bar{\F}}^{\Box}$
such that the corresponding members of $\S_{\bV}^{\Box}$ are $\bV_{0}$
and $\bV$, respectively.  By definition of $\S_{\bV}^{\Box}$, there exist
members $\bw_{0}$ and $\bw$ of the group $B(\I^{(i)},\I^{(4)})$ such that
$$
\bV_{0} \bw_{0} \text{ and } \bV \bw \text{ are members of }
k^{\Box} = K^{\Box}.
$$
Then, from Lem.~\ref{7.2F}(i), $\bar{\F}$ is the solution of the HHP
corresponding to $(\bv,\bar{\F}_{0})$, where
$$
\bv := \bV \bw (\bw_{0})^{-1} \bV_{0}^{-1},
$$
and is clearly a member of $K^{\Box}$.  So, for each $\bar{\F}_{0}
\in \S_{\bar{\F}}^{\Box}$ and $\bar{\F} \in \S_{\bar{\F}}^{\Box}$,
there exists $[\bv] \in \K^{\Box}$ such that $\bar{\F} = [\bv](\bar{\F}_{0})$;
and that is what is meant by the statement that $\K^{\Box}$ is
transitive.
\cheers
\end{romanlist}

\setcounter{section}{0}
\setcounter{equation}{0}
\renewcommand{\thesection}{\Alph{section}}
\renewcommand{\theequation}{\Alph{section}\arabic{equation}}
\renewcommand{\thetheorem}{\Alph{section}.\arabic{theorem}}

\part{Appendices}

\section{Table of notational correspondences}
\begin{small}
\begin{tabular}{|l|l|} \hline
Symbol or convention in IVP3 and IVP4 & 
Symbol or convention in current paper \\ \hline \hline
$u,v$ & In effect, we let $u=r$ and $v=s$ in \\
& the current paper. \\ 
$(r,s)$ & $\x := (r,s)$ \\ 
$-1 \le r < 1$ & $r_{1} < r < r_{2}$ \\
$-1 < s \le 1$ & $s_{2} < s < s_{1}$ \\ 
$\{(r,1):-1 \le r < 1\}$ and &
$\{(r,s_{0}):r_{1} < r < r_{2}\}$ and \\
$\{(-1,s):-1 < s \le 1\}$ are &
$\{(r_{0},s):s_{2} < s < s_{1}\}$ are \\
the null lines on which the & the null lines on which the \\
initial data are prescribed. & initial data are prescribed. \\ 
In IVP papers, $\x_{1}=\x_{0}=(-1,1)$. & $\x_{1}$ and $\x_{0}$ are
distinct points. \\ 
$[-1,r] \cup [s,1]$ & $\grave{\I}(\x) := \grave{\I}^{(3)}(\x)
\cup \grave{\I}^{(4)}(\x)$ \\ 
No corresponding notations in IVP3-4 & $\I(\x)$ and $\bar{\I}(\x)$
\\ 
$D_{3r}$ & $C - \grave{\I}^{(3)}(\x)$ \\ 
$D_{2s}$ & $C - \grave{\I}^{(4)}(\x)$ \\ 
$D_{(r,s)}$ & $C - \grave{\I}(\x)$ \\ 
$D_{IV}$ & $\domE$ \\ 
$D$ & $\dom{\grave{\F}}$ \\ 
$\bchi_{3}$, $\bchi_{2}$ and $\bchi$ & $\grave{\bnu}_{3}$, 
$\grave{\bnu}_{4}$ and $\grave{\bnu}$
\\ 
No corresponding notations in IVP3-4 & $\bar{\bnu}_{3}$, $\bar{\bnu}_{4}$
and $\bar{\bnu}$ \\

$\bA(r,s,\tau)$ & $\Gamma(\x,\tau)$ \\ 
$S(u,v)$ and $\bS(r,s)$ & $\rho^{-1} h(\x)$ \\ 
$E := (\rho+ig_{12})/g_{22}$ & We use $\E = f + i\chi$ instead. \\ 
$H(u,v)$ and $\bH(r,s)$ & $H(\x)$ \\ 
$H(-1,1)=-I$, & $H(\x_{0}) = - \left( \begin{array}{cc}
\rho_{0}^{2} & 0 \\ 2iz_{0} & 1
\end{array} \right)$, \\ 
since, in IVP3-4, $\rho=1$ and $z=0$ & since we do not impose specific values \\
at $(r,s)=(-1,1)$. & on $r_{0}$ and $s_{0}$ in the current paper.  \\ 
$P(u,v,\tau)$ and $\bP(r,s,\tau)$ & $\grave{\F}(\x,\tau)$ \\ 
No corresponding notation in IVP3-4 & $\bar{\F}(\x,\tau)$ \\ 
$\bP_{3}$ and $\bP_{2}$ & $\grave{\F}^{(3)}$ and $\grave{\F}^{(4)}$ \\ 
No corresponding notations in IVP3-4 & $\bar{\F}^{(3)}$ and $\bar{\F}^{(4)}$ 
\\ 
$\bsQ_{3}$ and $\bsQ_{2}$ & $\grave{\J}^{(3)}$ and $\grave{\J}^{(4)}$ \\ 
No corresponding notations in IVP3-4 & $\bar{\J}^{(3)}$ and $\bar{\J}^{(4)}$ \\ 
$\Gamma_{3}$ and $\Gamma_{2}$ (contours in $C$) & $\Lambda_{3}$ and $\Lambda_{4}$ \\
$K_{3}(u,\sigma,\tau)$ and $\bK_{3}(r,\sigma,\tau)$ & $\grave{K}_{3}(r,\sigma,\tau)$
\\ 
$K_{2}(v,\sigma,\tau)$ and $\bK_{2}(s,\sigma,\tau)$ & $\grave{K}_{4}(s,\sigma,\tau)$
\\ 
$K(u,v,\Gamma_{3},\Gamma_{2},\sigma,\tau)$ and 
$\bK(r,s,\Gamma_{3},\Gamma_{2},\sigma,\tau)$ & $\grave{K}(\x,\Lambda,\sigma,\tau)$ \\ 
No corresponding notations in IVP3-4 & $\bar{K}_{3}$, $\bar{K}_{4}$, $\bar{K}$ \\ 
\hline
\end{tabular}
\end{small}
\newpage

\setcounter{equation}{0}
\section{$V^{(3)}$ and $V^{(4)}$ for some simple metrics}

The matrix functions $V^{(3)}$ and $V^{(4)}$ play an essential role
in our formalism.  In the case of many familiar metrics, these functions
can be evaluated explicitly.  For example, in the case of Minkowski space,
where $\E = \E^{M}$, $\Delta$ is identically zero.  Therefore, the 
corresponding member of the set $\S_{\hat{\Q}}$ has the value 
$\hat{\Q}(\x,\tau) = I$ for all $(\x,\tau)$ in its domain, and the
corresponding $V^{(i)}$ has the value 
\begin{equation}
V^{M(i)}(\sigma) = I \text{ for all $\sigma \in \I^{(i)}$.}
\label{G2.28}
\end{equation}

As a second example, the member of $\S_{\E}$ that represents the
Kasner metric of index zero [restricted, of course, to the
domain $\domE$] has the value
\begin{equation}
\E^{K}(\x) = -(\rho/\rho_{0}).
\end{equation}
The corresponding member of $\S_{\hat{\Q}}$ is found to be
\begin{equation}
\hat{Q}^{K}(\x,\tau) = N(\x,\tau) [N(\x_{0},\tau)]^{-1},
\end{equation}
where
\begin{eqnarray}
N(\x,\tau) & := & \left( \begin{array}{cc}
	\zeta(\x,\tau) & 0 \\ 0 & \zeta(\x,\tau)^{-1}
	\end{array} \right), 
\label{G2.30b} \\
\zeta(\x,\tau) & := & \frac{\M(\tau-r)+\M(\tau-s)}{\sqrt{s-r}},
\end{eqnarray}
and
\begin{equation}
\begin{array}{rcl}
\M(\tau-r) & := & \text{ the branch of $(\tau-r)^{1/2}$ for which} \\
\M^{\pm}(\sigma-r) & = & \left\{ \begin{array}{l}
	\sqrt{\sigma-r} \text{ if $\sigma \ge r$} \\
	\pm i \sqrt{r-\sigma} \text{ if $\sigma \le r$.}
	\end{array} \right.
\end{array} 
\label{G2.30d}
\end{equation}
Recalling that $(\x_{0},\x_{1},\x_{2})$ is type~A, one obtains
\begin{equation}
\begin{array}{rcl}
V^{K(3)}(\sigma) & = & -i \sigma_{3} N^{+}(\x_{0},\sigma)
\text{ for all $\sigma \in \I^{(3)}$,} \\
V^{K(4)}(\sigma) & = & N^{+}(\x_{0},\sigma) 
\text{ for all $\sigma \in \I^{(4)}$.}
\end{array} 
\label{G2.31}
\end{equation}

Finally, consider the example of the general vacuum Weyl metric.  Here,
$\E^{W} \in \S_{\E}$ is expressible in the form
\begin{equation}
\E^{W} = f = - \exp (2\psi), \quad \psi(\x_{0}) = 0.
\end{equation}
One readily obtains
\begin{equation}
V^{W(i)}(\sigma) = \exp \left[ -\mu^{+}(\x_{0},\sigma) \sigma_{3}
f^{(i)}(\sigma) \right],
\end{equation}
where
\begin{eqnarray}
f^{(3)}(\sigma) & = & [\sgn(\sigma-r_{0})] \int_{r_{0}}^{\sigma}
dr \frac{\psi_{r}(r,s_{0})}{\sqrt{(\sigma-r)(\sigma-r_{0})}} , 
\label{G2.34a} \\
f^{(4)}(\sigma) & = & [\sgn(\sigma-s_{0})] \int_{s_{0}}^{\sigma}
ds \frac{\psi_{s}(r_{0},s)}{\sqrt{(\sigma-s)(\sigma-s_{0})}} , 
\label{G2.34b} 
\end{eqnarray}
and $\sgn(x)$ denotes the sign of $x$.  The integrals over $r$
and $s$ are Abel transforms such as those that occur in the
tautochrone problem of classical mechanics.\footnote{See Ref.~\ref{IVP12}.} 

\newpage

\setcounter{equation}{0}
\section{The Kinnersley--Chitre subgroups $K_{KC}$ and $\K_{KC}$}

First let us define
\begin{equation}
\hat{\I} := \{\sigma \in R^{1}: r_{1} < \sigma < s_{1} \}
\end{equation}
and let $K(\x_{0},\hat{\I})$ be defined exactly as we defined 
$K(\x_{0},\I^{(3)})$ and $K(\x_{0},\I^{(4)})$  except that the domain 
of each element of the group $K(\x_{0},\hat{\I})$ is $\hat{\I}$.

We next identify that subgroup of $K^{3}$ which represents the K--C
group of permutations of $\S_{\bar{\F}}^{3}$.  This subgroup will be denoted
by $K_{KC}$ and is the set of all $\bv \in K$ for which there
exists
\begin{equation}
\hat{v} \in K(\x_{0},\hat{\I})
\end{equation}
such that $\tilde{\hat{v}}$ is analytic and ($i=3,4$)
\begin{equation}
v^{(i)} = \text{ the restriction of $\hat{v}$ to $\I^{(i)}$.}
\end{equation}

It is important to note that the statement that $\tilde{\hat{v}}$
is analytic is clearly equivalent to the statement that the
functions $\hat{\alpha}$ and $\hat{\beta}$ that are real linear
combinations of $I$ and $J$ and that occur in the equation
\begin{equation}
\hat{v}(\sigma) = \hat{\alpha}(\sigma) + \mu^{+}(\x_{0},\sigma)
\sigma_{3} \hat{\beta}(\sigma)
\end{equation}
are analytic functions of $\sigma$ throughout $\hat{\I}$.  From a
basic theorem\footnote{See Ref.~\ref{Bochner}.}
on holomorphic functions, $\hat{\alpha}$ and
$\hat{\beta}$ have holomorphic extensions $\alpha$ and $\beta$
to a connected open neighborhood of $\hat{\I}$ in the space $C$
such that the neighborhood is symmetric with respect to the real
axis.  If one lets
\begin{equation}
v(\tau) := \alpha(\tau) + \mu(\x_{0},\tau) \sigma_{3} \beta(\tau)
\end{equation}
for all $\tau$ in this neighborhood of $\hat{\I}$, then
\begin{equation}
\det{v} = 1, \quad v^{*} = v,
\end{equation}
and $\hat{v}(\sigma)$ is the limit of $v(\tau)$ as $\tau \rightarrow
\sigma$ from the upper half-plane.  The HHP corresponding to 
$(\bv,\bar{\F}_{0},\x)$ now has the equivalent form
\begin{equation}
Y(\x,\tau) = \bar{\F}(\x,\tau) \tilde{\bv}(\tau) [\bar{\F}_{0}(\x,\tau)]^{-1}
\end{equation}
for all $\tau$ on a simple closed contour $\Lambda$ that is chosen
so that $\tilde{v}(\tau)$ is holomorphic on $\Lambda$ and within the
region $\Lambda_{INT}$ enclosed by $\Lambda$, while $\hat{\I}(\x)$
lies in $\Lambda_{INT}$.  The functions of $\tau$ given 
by $Y(\x,\tau)$ and $\bar{\F}(\x,\tau)$ are required to be holomorphic
on $\Lambda \cup \Lambda_{INT}$ and on $C - \Lambda_{INT}$, respectively,
and one also requires that $\bar{\F}(\x,\infty) = I$.  This completely
summarizes one form of the HHP that is used to effect K--C 
transformations. 

The K--C group of permutations of $\S_{\bar{\F}}$ will be denoted by $\K_{KC}$
and is, of course, the set of all $[\bv] \in \K^{3}$ such that
$\bv \in K_{KC}$.  Of course, since we have proved that $\K^{3}$
exists, $\K_{KC}$ exists.  Even this is not without significance, as 
with our HHP of 1980 we only established the existence of a K--C group
of transformations among solutions in which the ``axis'' $\rho=0$ is 
accessible.

\newpage

\setcounter{equation}{0}
\setcounter{theorem}{0}
\section{The mappings $\btheta(\x):\bar{\I}(\x)\rightarrow\Theta$ and 
$\bsigma(\x):\Theta\rightarrow\bar{\I}(\x)$}

In integrals such as those in Thm.~\ref{5.1B}, it is sometimes useful to
introduce a new variable of integration for the purpose of getting rid of
the singularities of the integrands at $\sigma' \in \{r_{0},s_{0},r,s\}$.
This is especially important when one has to consider derivatives of the
integrals with respect to $r$ and $s$.  

We shall begin by defining several mappings that will be used in formulating
the theorem that is to follow.

\begin{definition}{Dfns.\ of $\Theta$, $\btheta(\x)$ and $\bsigma(\x)$
\label{def77}}
Let $\Theta$ denote that union of arcs
\begin{equation}
\Theta := \left[0,\frac{\pi}{2}\right] + \left[\pi,\frac{3\pi}{2}\right]
\label{5.26a}
\end{equation}
whose assigned orientations are in the direction of increasing 
$\theta \in [0,\pi/2]$ and $\theta \in [\pi,3\pi/2]$.  For each 
$\x \in \domE$, let
\begin{equation}
\btheta(\x):\bar{\I}(\x) \rightarrow \Theta
\end{equation}
be a mapping such that
\begin{equation}
\btheta(\x)(\sigma) := \btheta(\x,\sigma),
\end{equation}
where 
\begin{equation}
\begin{array}{r}
0 \le \btheta(\x,\sigma) \le \frac{\pi}{2} \text{ and }
\cos[2\btheta(\x,\sigma)] := \frac{2\sigma-(r_{0}+r)}{r_{0}-r} \\[1ex]
\text{when } \sigma \in \bar{\I}^{(3)}(\x)
\end{array}
\end{equation}
and
\begin{equation}
\begin{array}{r}
\pi \le \btheta(\x,\sigma) \le \frac{3\pi}{2} \text{ and }
\cos[2\btheta(\x,\sigma)] := \frac{2\sigma-(s_{0}+s)}{s_{0}-s} \\[1ex]
\text{when } \sigma \in \bar{\I}^{(4)}(\x).
\end{array}
\end{equation}
Also let
\begin{equation}
\bsigma(\x):\Theta \rightarrow \bar{\I}(\x)
\end{equation}
be a mapping such that
\begin{equation}
\bsigma(\x)(\theta) := \bsigma(\x,\theta),
\end{equation}
where
\begin{eqnarray}
\bsigma(\x,\theta) & := & r_{0} \cos^{2}\theta + r \sin^{2}\theta
\text{ when } \theta \in \left[0,\frac{\pi}{2}\right] 
\label{5.26h}
\end{eqnarray}
and
\begin{eqnarray}
\bsigma(\x,\theta) & := & s_{0} \cos^{2}\theta + s \sin^{2}\theta
\text{ when } \theta \in \left[\pi,\frac{3\pi}{2}\right].
\label{5.26i}
\end{eqnarray}
\end{definition}

\begin{theorem}[On the transformation $\sigma' \rightarrow \theta'$]
\label{5.2B} \mbox{ } \\
The mapping $\btheta(\x)$ is monotonic and is a continuous bijection
(one-to-one and onto) of $\I(\x)$ onto $\Theta$, and $\bsigma(\x)$
is its inverse mapping.  Moreover, $\bsigma(\x)$ is analytic [which
means that it has an analytic extension to an open subset of $R^{1}$].
Finally, one obtains the following transformation of integral
operators:
\begin{eqnarray}
\frac{1}{\pi i} \int_{a^{3}}^{b^{3}} d\sigma' \nu^{+}(\sigma') 
& \rightarrow & \frac{2(r_{0}-r)}{\pi} \int_{0}^{\pi/2} d\theta'
\sin^{2}\theta' \sqrt{\frac{s_{0}-\bsigma(\x,\theta')}
{s-\bsigma(\x,\theta')}},
\label{5.27a} \\
\frac{1}{\pi i} \int_{a^{4}}^{b^{4}} d\sigma' \nu^{+}(\sigma') 
& \rightarrow & \frac{2(s_{0}-s)}{\pi} \int_{\pi}^{3\pi/2} d\theta'
\sin^{2}\theta' \sqrt{\frac{\bsigma(\x,\theta')-r_{0}}
{\bsigma(\x,\theta')-r}}, 
\label{5.27b} \\
\frac{1}{\pi i} \int_{a^{3}}^{b^{3}} d\sigma' [\nu^{+}(\sigma')]^{-1}
& \rightarrow & \frac{2(r_{0}-r)}{\pi} \int_{0}^{\pi/2} d\theta'
\cos^{2}\theta' \sqrt{\frac{s-\bsigma(\x,\theta')}
{s_{0}-\bsigma(\x,\theta')}}, \\
\frac{1}{\pi i} \int_{a^{4}}^{b^{4}} d\sigma' [\nu^{+}(\sigma')]^{-1}
& \rightarrow & \frac{2(s_{0}-s)}{\pi} \int_{\pi}^{3\pi/2} d\theta'
\cos^{2}\theta' \sqrt{\frac{\bsigma(\x,\theta')-r}
{\bsigma(\x,\theta')-r_{0}}},
\label{5.27d}
\end{eqnarray}
where $a^{i} := \inf{\{x^{i},x_{0}^{i}\}}$ and
$b^{i} := \sup{\{x^{i},x_{0}^{i}\}}$.	
\end{theorem}

\proof
The proof is straightforward.  The reader may find the following relations
helpful:
\begin{eqnarray}
\nu^{+}(\sigma) & = & \pm i \sqrt{\frac{r_{0}-\sigma}{\sigma-r}}
\sqrt{\frac{s_{0}-\sigma}{s-\sigma}} \text{ when } a^{3} < \sigma < b^{3}
\text{ and } \pm (r_{0}-r) > 0, \\
\nu^{+}(\sigma) & = & \pm i \sqrt{\frac{\sigma-r_{0}}{\sigma-s_{0}}}
\sqrt{\frac{s_{0}-\sigma}{\sigma-s}} \text{ when } a^{4} < \sigma < b^{4}
\text{ and } \pm (s_{0}-s) > 0, \\
\tan\theta & = & \sqrt{\frac{r_{0}-\bsigma(\x,\theta)}{\bsigma(\x,\theta)-r}}
\text{ when } \theta \in \left[0,\frac{\pi}{2}\right],
\label{5.28c} \\
\tan\theta & = & \sqrt{\frac{s_{0}-\bsigma(\x,\theta)}{\bsigma(\x,\theta)-s}}
\text{ when } \theta \in \left[\pi,\frac{3\pi}{2}\right].
\label{5.28d}
\end{eqnarray}
The above Eqs.\ (\ref{5.28c}) and (\ref{5.28d}) are derived directly from
Eqs.\ (\ref{5.26h}) and (\ref{5.26i}).

\cheers

As regards the transformed integral operators in (\ref{5.27a}) to
(\ref{5.27d}), note that
\begin{eqnarray}
\sqrt{\frac{s-\bsigma(\x,\theta')}{s_{0}-\bsigma(\x,\theta')}}
\text{ is a real positive-valued analytic} \nonumber \\
\text{function of } (\x,\theta') \text{ on } \domE \times
\left[0,\frac{\pi}{2}\right]
\label{5.29a}
\end{eqnarray}
and
\begin{eqnarray}
\sqrt{\frac{\bsigma(\x,\theta')-r}{\bsigma(\x,\theta')-r_{0}}}
\text{ is a real positive-valued analytic} \nonumber \\
\text{function of } (\x,\theta') \text{ on } \domE \times 
\left[\pi,\frac{3\pi}{2}\right].
\label{5.29b}
\end{eqnarray}
The above statements are obtained from Eq.\ (\ref{5.26h}), Eq.\ (\ref{5.26i})
and the fact that $\grave{\I}^{(3)}(\x) < \grave{\I}^{(4)}(\x)$.  Note:  To
say that a function is analytic on a closed or semi-closed interval\footnote{
For $p>1$ a direct product of $p$ one-dimensional intervals is involved.}
in $R^{p}$ means that the function has an analytic extension to an open
interval in the space $R^{p}$.

\newpage

\setcounter{equation}{0}
\section{Three involutions of the set $\S_{\bar{\F}}$ of $\bar{\F}$-potentials}

For the stationary axisymmetric vacuum spacetimes, the Kramer--Neugebauer 
involution\footnote{D.~Kramer and G.~Neugebauer, {\em On axially symmetric
stationary solutions of the Einstein field equations for the vacuum},
Comm.\ Math.\ Phys.\ {\bf 10}, 132 (1968).} does not generally preserve the
reality of the metric tensor.  This peculiar complication does not, however,
occur for the spacetimes that we are considering in this paper.  Let 
$\I_{KN}$ denote the mapping whose domain is $\S_{\bar{\F}}$ and whose 
values are given by
\begin{equation}
\I_{KN} \E := (\rho_{0}f)^{-1} (\rho - i g_{12}), 
\label{G6.6a}
\end{equation}
where $\rho_{0} := (s_{0}-r_{0})/2$ and $g_{12}(0) = 0$.  As is
well known, the right side of the above equation is also a member
of $\S_{\E}$, and
\begin{equation}
\I_{KN} \left( (\rho_{0}f)^{-1} (\rho - i g_{12}) \right) = \E.
\label{G6.6b}
\end{equation}
Thus, $\I_{KN}$ is an involution of $\S_{\E}$; i.e., it is a permutation
of $\S_{\E}$ such that $\I_{KN}^{\quad 2}$ is the identity map on $\S_{\E}$.

Note: The K--N involution is generally defined without restricting its domain
to a special gauge of $\E$-potentials and without the factor $\rho_{0}^{-1}$
in Eqs.\ (\ref{G6.6a}) and (\ref{G6.6b}).  However, it suits the purpose of
this paper to let $\I_{KN}:\S_{\E} \rightarrow \S_{\E}$.  Incidentally, we 
expect no confusion to result from our failure to use a more explicit
notation [e.g., $\I_{KN}(\x_{0},\x_{1},\x_{2})$].  We shall, in fact, take
even greater notational liberties by employing `$\I_{KN}$' as a generic 
operator on functions that are determined by $\E$.  For example, if $H \in
\S_{H}$, then $\I_{KN}H$ will denote the member of $\S_{H}$ such that
$\I_{KN}\E = (\I_{KN}H)_{22}$; and, if $\bar{\F} \in \S_{\bar{\F}}$, then
$\I_{KN}\bar{\F}$ will denote the member of $\S_{\bar{\F}}$ such that
\begin{equation}
(\I_{KN}\bar{\F})(\x,\tau) = I + (2\tau)^{-1} \left[
(\I_{KN}H)(\x) - H^{M}(\x_{0}) \right] \Omega + O(\tau^{-2})
\end{equation}
in a neighborhood of $\tau=\infty$.

For the stationary axisymmetric vacuum spacetimes, C.\ Cosgrove\footnote{
C.~M.~Cosgrove, {\em Relationships between the group-theoretic and
soliton-theoretic techniques for generating stationary axisymmetric
gravitational solutions}, J.\ Math.\ Phys.\ {\bf 21}, 2417-2447 (1980);
{\em Relationship between the inverse scattering techniques of 
Belinskii-Zakharov and Hauser-Ernst in general relativity}, J.\ Math.\ 
Phys.\ {\bf 23}, 615 (1982).} derived an ingenious result for the effect
of the K--N involution on the Kinnersley--Chitre potential $F$.  We have
used his result as a guide to obtain the effect of $\I_{KN}$ on $\hat{\Q} 
\in \S_{\hat{\Q}}$.  We shall not enter into the details of our work but
simply state that
\begin{equation}
(\I_{KN}\hat{\Q})(\x,\tau) = N(\x,\tau) J \hat{\Q}(\x,\tau) J^{-1} 
N(\x_{0},\tau)^{-1}, 
\label{G6.8}
\end{equation}
where $N(\x,\tau)$ is defined by Eqs.\ (\ref{G2.30b}) to (\ref{G2.30d}).  From 
the above Eq.\ (\ref{G6.8}) and Eqs.\ (\ref{G2.21g1}) and (\ref{G2.21g2}) for $V^{(3)}$
and $V^{(4)}$, we obtained
\begin{eqnarray}
\I_{KN} V^{(3)} & = & V^{K(3)} J \sigma_{3} V^{(3)} \sigma_{3} J^{-1},
\label{G6.9a} \\
\I_{KN} V^{(4)} & = & V^{K(4)} J V^{(4)} J^{-1}, 
\label{G6.9b}
\end{eqnarray}
where $V^{K(3)}$ and $V^{K(4)}$ are associated with the Kasner metric, and 
are given by Eqs.\ (\ref{G2.31}).  Multiplying Eqs. (\ref{G6.9a}) and 
(\ref{G6.9b}) from the right by $J[V^{(3)}]^{-1}$ and $J[V^{(4)}]^{-1}$, 
respectively, one can express these equations in the following equivalent 
form:
\begin{equation}
(\I_{KN} \bV) \bw \bV^{-1} = \bV^{K} \bw \left(
\sigma_{3} V^{(3)} \sigma_{3} [V^{(3)}]^{-1}, \delta^{(4)} \right),
\label{G6.10a}
\end{equation}
where $\bw$ and $\bdelta$ are the members of $B(\I^{(3)},\I^{(4)})$
for which
\begin{equation}
w^{(i)}(\sigma) = J \text{ and } \delta^{(i)}(\sigma) = I
\text{ for all $\sigma \in I^{(i)}$.}
\label{G6.10b}
\end{equation}

Using the methods employed in the proof of GSSM \ref{7.5E}, one can show
that $\I_{KN}\bar{\F}$ is the solution of the HHP corresponding to 
$(\bv,\bar{\F})$ if $\bv$ is the member of $K$ that is equal to the right
side of Eq.\ (\ref{G6.10a}).  However, since $V^{(3)}$ does not generally
commute with $\sigma_{3}$, $(\I_{KN}\bV) \bw \bV^{-1}$ is generally 
different for different choices of $\bar{\F} \in \S_{\bar{\F}}^{3}$. It is
easy to show that the same conclusion is obtained when one uses a member
$\bw$ of $B(\I^{(3)},\I^{(4)})$ other than the one defined by Eq.\ 
(\ref{G6.10b}).  Therefore,
\begin{equation}
\I_{KN} \notin \K^{3}.
\end{equation}

Let $\I_{3}$ denote that permutation of $\S_{\E}$ for which the values
of $\E \in \S_{\E}$ on the null lines through $\x_{0}$ transform as
follows:
\begin{equation}
(\I_{3}\E)(r,s_{0}) := \E^{*}(r,s_{0}), \quad
(\I_{3}\E)(r_{0},s) := \E(r_{0},s). 
\label{G6.12}
\end{equation}
{}From the above Eqs.\ (\ref{G6.12}) and from Eq.\ (\ref{G2.15a}), one
obtains (employing `$\I_{3}$' as a generic operator in the same way
that we employed `$\I_{KN}$')
\begin{eqnarray}
(\I_{3}\hat{\Q})((r,s_{0}),\tau) & = & \sigma_{3} \hat{\Q}((r,s_{0}),\tau)
\sigma_{3}, \\
(\I_{3}\hat{\Q})((r_{0},s),\tau) & = & \hat{\Q}((r_{0},s),\tau).
\end{eqnarray}
Therefore, from the expressions (\ref{G2.21g1}) and (\ref{G2.21g2}) for
$V^{(3)}$ and $V^{(4)}$,
\begin{equation}
\I_{3}(V^{(3)},V^{(4)}) = \left( \sigma_{3} V^{(3)} \sigma_{3},
V^{(4)} \right).
\end{equation}
It is clear that $\I_{3}$ is an involution of $\S_{\bar{\F}}$ and that, for
reasons similar to those we gave above for $\I_{KN}$,
\begin{equation}
\I_{3} \notin \K^{3}.
\end{equation}

Although neither $\I_{KN}$ nor $\I_{3}$ is a member of $\K^{3}$,
the reader can now show without difficulty that
\begin{equation}
\I_{KN} \I_{3} = \I_{3} \I_{KN}
\end{equation}
and
\begin{equation}
\I_{KN} \I_{3} \bV = \bV^{K} \bw \bV \bw^{-1},
\end{equation}
where $\bw$ is the member of $B(\I^{(3)},\I^{(4)})$ that is defined
in Eqs.\ (\ref{G6.10b}).  Hence, $\I_{KN} \I_{3} \bar{\F}$ is the solution of
the HHP corresponding to $(\bv,\bar{\F})$, where
\begin{equation}
\bv = (\I_{KN}\I_{3}\bV) \bw \bV^{-1} = \bV^{K} \bw.
\end{equation}
Thus, we have the following remarkable theorem.
\begin{gssm}[$\I_{KN}\I_{3} \in \K^{3}$] 
\label{Thm_16}
The product $\I_{KN} \I_{3} = \I_{3} \I_{KN}$ is a member of
$\K^{3}$ and is given by $[\bV^{K}\bw] = \I_{KN} \I_{3}$,
where $w^{(i)}(\sigma) = J$ for all $\sigma \in I^{(i)}$.
\end{gssm}
One can similarly show that $\I_{KN}$ and $\I_{3}$ are not members
of $\K^{\infty}$ and not members of $\K^{an}$, but $\I_{KN}\I_{3}$
is a member of these groups.

At this time we have no deeper understanding of the curious facts that we
have described in this appendix.  Nevertheless, we feel that there may
indeed be some deeper significance, and we would encourage our readers to 
try to discern what that deeper significance might be.  In addition, one
might ask how one would execute an $\I_{3}$ involution $\S_{\bar{\F}}
\rightarrow \S_{\bar{\F}}$ in practice.

\section*{Acknowledgement}
Research supported in part by grants PHY-93-07762 and PHY-96-01043
from the National Science Foundation to FJE Enterprises.  

\end{document}